\documentclass[a4paper,12pt]{article}

\usepackage[margin=1.2in]{geometry}
\usepackage[dvipdfmx]{graphicx}
\usepackage{amssymb}
\usepackage{color}
\usepackage{multirow}
\usepackage{arydshln}

\usepackage{amsmath}
\usepackage{breqn}
\usepackage{bm}

\usepackage{ascmac}
\usepackage{tcolorbox}

\usepackage{hyperref}
\usepackage{wrapfig}
\usepackage{url}
\usepackage{color}
\usepackage{threeparttable}
\usepackage{float}
\usepackage{subfig}

\usepackage[
    sorting=none,
    style=phys,
    eprint=true,
    refsegment=section
]{biblatex}
\DeclareSourcemap{
  \maps[datatype=bibtex]{
    \map[overwrite=false]{
      \step[fieldsource=note]
      \step[fieldset=addendum, origfieldval, final]
      \step[fieldset=note, null]
    }
  }
}
\usepackage{suffix}

\newcommand\chapterauthor[1]{\authortoc{#1}\printchapterauthor{#1}}
\WithSuffix\newcommand\chapterauthor*[1]{\printchapterauthor{#1}}

\makeatletter
\newcommand{\printchapterauthor}[1]{%
  {\parindent0pt\vspace*{0pt}% <- tuned from -25pt
  \linespread{1.1}\normalsize\centering#1% <- tuned from large, \scshape removed, \centering added
  \par\nobreak\vspace*{25pt}}% tuned from 35pt
  \@afterheading%
}
\newcommand{\authortoc}[1]{%
  \addtocontents{toc}{\vskip0pt}% <-tuned from -10pt
  \addtocontents{toc}{%
    \protect\contentsline{chapter}%
    {\raggedleft\mdseries\protect\footnotesize#1}{}{}}% tuned from scriptsize, \scshape removed, \hskip1.3em removed, \raggedleft added
  \addtocontents{toc}{\vskip5pt}%
}
\makeatother
\usepackage[noblocks]{authblk}
\title{
Extension of \\ 
the J-PARC Hadron Experimental Facility \\
\Large - Third White Paper -
}

\date{\today}

\author[ ]{
Taskforce on the extension of the Hadron Experimental Facility\hspace{-1.5ex}
}

\author[1]{\normalsize Kazuya~Aoki}
\author[2]{Hiroyuki~Fujioka}
\author[3]{Toshiyuki~Gogami}
\author[1,4,5]{Yoshimasa~Hidaka}
\author[6]{Emiko~Hiyama}
\author[1]{Ryotaro~Honda}
\author[7,8,9]{Atsushi~Hosaka} %Nishina?
\author[8]{Yudai~Ichikawa}
\author[1]{Masaharu~Ieiri}
\author[10]{Masahiro~Isaka}
\author[7]{Noriyoshi~Ishii}
\author[11]{Takatsugu~Ishikawa}
\author[1]{Yusuke~Komatsu}
\author[1]{Takeshi~Komatsubara}
\author[1]{GeiYoub~Lim}
\author[6]{Koji~Miwa}
\author[1]{Yuhei~Morino}
\author[3]{Tomofumi~Nagae}
\author[6]{Sho~Nagao}
\author[6]{Satoshi~N.~Nakamura}
\author[12]{Hajime~Nanjo}
\author[3]{Megumi~Naruki}
\author[7]{Hidekatsu~Nemura}
\author[1]{Tadashi~Nomura}
\author[7,1]{Hiroyuki~Noumi}
\author[11]{Hiroaki~Ohnishi}
\author[1]{Kyoichiro~Ozawa}
\author[13]{Fuminori~Sakuma}
\author[1]{Shinya~Sawada}
\author[14]{Takayasu~Sekihara}
\author[7]{Sang-In~Shim}
\author[1]{Koji~Shiomi}
\author[7]{Kotaro~Shirotori}
\author[15]{Yasuhisa~Tajima}
\author[1]{Hitoshi~Takahashi}
\author[1]{Toshiyuki~Takahashi}
\author[9,16]{Sachiko~Takeuchi} %Nishina?
\author[1,9,17]{Makoto~Takizawa} %Nishina?
\author[6,8]{Hirokazu~Tamura}
\author[8]{Kiyoshi~Tanida}
\author[1]{Mifuyu~Ukai}
\author[8]{Takeshi~O.~Yamamoto}
\author[9]{Yasuo~Yamamoto} %Nishina?

\affil[1]{\small\it Institute of Particle and Nuclear Studies(IPNS), High Energy Accelerator Research Organization (KEK), Tsukuba 305-0801, Japan}
\affil[2]{\it Tokyo Institute of Technology, Tokyo, 152-8551 Japan}
\affil[3]{\it Kyoto University, Kyoto, 606-8502 Japan}
\affil[4]{\it Graduate University for Advanced Studies (Sokendai), Tsukuba 305-0801, Japan}
\affil[5]{\it RIKEN iTHEMS, RIKEN, Wako 351-0198, Japan}
\affil[6]{\it Tohoku University, Sendai 980-8578, Japan}
\affil[7]{\it Research Center for Nuclear Physics (RCNP), Osaka University, Ibaraki 567-0047, Japan}
\affil[8]{\it Advanced Science Research Center (ASRC), Japan Atomic Energy Agency (JAEA), Tokai 319-1195, Japan}
\affil[9]{\it RIKEN Nishina Center, RIKEN, Wako 351-0198, Japan}
\affil[10]{\it Hosei University, Tokyo 102-8160, Japan}
\affil[11]{\it Research Center for Electron Photon Science (ELPH), Tohoku University, Sendai 982-0826, Japan}
\affil[12]{\it Osaka University, Toyonaka 560-0043, Japan}
\affil[13]{\it RIKEN Cluster for Pioneering Research, RIKEN, Wako 351-0198, Japan}
\affil[14]{\it Kyoto Prefectural University, Kyoto 606-8522, Japan}
\affil[15]{\it Yamagata University, Yamagata 990-8560, Japan}
\affil[16]{\it Japan College of Social Work, Kiyose 204-8555, Japan}
\affil[17]{\it Showa Pharmaceutical University, Machida 194-8543, Japan}

\begin{document}

\pagenumbering{roman}

\maketitle

\thispagestyle{empty}
\addtocounter{page}{-1}

\clearpage

%==================================================================%
\section*{\centering Preface}
%==================================================================%

% flatex input: [preface.tex]
Toward realization of the extension of the Hadron Experimental Facility at J-PARC, this White Paper presents the physics to be newly developed in the extended facility.

The extension project has been discussed extensively among particle and nuclear physics communities in Japan since the early stage of the J-PARC construction.
In the user community of the Hadron Experimental Facility, Hadron Hall Users’ Association (HUA), a committee for the study of the facility extension was formed in August, 2015, and made two White Papers as arXiv.1706.07916 [nucl-ex] and arXiv.1906.02357 [nucl-ex].
In September, 2020, we organized a Task Force (TF) under HUA aiming at early realization of the extension through further discussions on the important physics features at the extended facility.
The task force was composed of three groups based on three core lines of research conducted at J-PARC: strangeness nuclear physics (HIHR/K1.1-TF), hadron physics (K10-TF), and flavor physics (KL2-TF).
We held a series of workshops in the first half of 2021, and substantial interest and support were given to the extension project at the international level.
Physics case deeply discussed in the workshops is summarized as this third White Paper.
On August 10,11,17,25, 2021, the extension project was reviewed by an international committee of 'Focused review committee of Hadron Experimental Facility Extension' formed under J-PARC PAC, for which this third White Paper was used as an input document.
Detailed information on the extension project can be found in HUA's home page, where documents related to the project, links to the workshops and review, and supporting letters for the project are available:
\begin{center} \url{https://www.rcnp.osaka-u.ac.jp/~jparchua/en/hefextension.html}
\end{center}

We wish this third White Paper strongly pushes forward the extension project of the Hadron Experimental Facility at J-PARC.
% flatex input end: [preface.tex]

%==================================================================%

\clearpage

%==================================================================%
\begin{abstract}
The J-PARC Hadron Experimental Facility was constructed with an aim to explore the origin and evolution of matter in the universe through the experiments with intense particle beams.
In the past decade, many results on particle and nuclear physics have been obtained at the present facility.
To expand the physics programs to unexplored regions never achieved, the extension project of the Hadron Experimental Facility has been extensively discussed.
This white paper presents the physics of the extension of the Hadron Experimental Facility for resolving the issues in the fields of the strangeness nuclear physics, hadron physics, and flavor physics.
\end{abstract}
%==================================================================%

\clearpage

\tableofcontents

\clearpage

\pagenumbering{arabic}

%==================================================================%
\section{\centering Executive Summary}
%==================================================================%

% flatex input: [summary.tex]
\subsection{Introduction}\label{sec:intro}

The Japan Proton Accelerator Research Complex (J-PARC) is a
multi-purpose accelerator facility located in Tokai village,
Japan~\cite{Nagamiya:2006en,Nagamiya:2012tma}.
The aim of J-PARC is to promote a variety of scientific research
programs ranging from the basic science of particle, nuclear, atomic, and
condensed matter physics and life science to the industrial application
and future nuclear transmutation using intense particle beams.
Among them, the Hadron Experimental Facility focuses on particle
and nuclear physics to explore the origin and evolution of matter in the
universe, using the primary 30 GeV proton beam and secondary beams of
pions, kaons, and muons.
With the intense hadron beams, a wide variety of experiments are performed to
approach the open questions in the universe:
\begin{itemize}
 \setlength{\itemsep}{0cm}
 \item Is there new physics beyond the Standard Model?
\item How is the matter-antimatter asymmetry 
  that resulted in the matter-dominance universe generated?
%\item What is the dark matter?
 \item What is the origin of the hadron mass that weights
 99.9\% of the visible matter in the universe ?
 \item How are hadrons built from quarks and gluons?
 %\item How are the hadron properties determined from quarks and gluons?
% \item How strong are interactions between hadrons under broken flavor SU(3)?
%%% HT
%% %\item What is the origin of short-range repulsion and spin–orbit force in
%%% nuclear force that form atomic nuclei?
\item What is the origin of the short-range part of the
nuclear force which plays essential roles in formation of atomic nuclei?
%\item How is high-density nuclear matter created in neutron stars?
%%% HT
%%% \item What is the properties of high-density nuclear matter
%%% that may exist in compact objects in the universe ?
\item What are the properties of high-density nuclear matter
 that may exist in compact stars in the universe ?
%%%
%
%%% %\item Can we answer above questions and explain consistently based on the basic theory ?
\end{itemize}
%Toward solution of these fundamental questions, 
%%% HT To answer these questions straightforward from basic laws, 
%%% HT
%%% To answer these questions straightforward from basic laws, 
To answer these questions based on fundamental physical laws, 
%%% 
the following three core lines of research have been conducted at J-PARC.

%%% The first is {\bf strangeness nuclear physics} for  
The first is {\bf strangeness nuclear physics}. It aims at 
%%%
elucidation of the matter containing strange quarks.
Observation of a neutron-star merger event by gravitational wave at
LIGO and Virgo and subsequent multi-messenger astronomical
observations have provided information on the equation of 
states (EOS) of nuclear matter and the synthesis of heavy chemical
elements~\cite{ABB17}.
However, the dense nuclear matter deep inside the neutron stars still
remains unknown because the properties of the nuclear matter and
interactions among the constituent particles, hadrons, have not been fully
understood.
Today, aiming to clarify the whole picture of neutron stars, a wide range
of scientific programs have been developed from microscopic to macroscopic approaches.
In particular, since hyperons are predicted to play an important role in
such dense environment, the interactions involving hyperons
in the nuclear matter should be determined with microscopic approaches.
At J-PARC, the world's leading research on hypernuclear physics has been
conducted and provided important information, together with precise
measurements of hyperon-nucleon scattering, to understand baryon-baryon interactions 
extended to the strangeness sector.

The second is {\bf hadron physics}. It aims to understand
the structure of hadrons as composite systems of quarks and gluons.  
Due to non-perturbative nature of QCD, 
hadrons are considered to be formed through complex non-trivial dynamics of quantum fields.
The most important features are confinement and the
spontaneous breaking of chiral symmetry, leading to
the emergence of constituent quarks as quasi-particles 
%%% and the Nambu-Goldstone pions.
and pions as Nambu-Goldstone bosons.
Furthermore, various flavor contents with different masses result in a
variety of structures including exotic hadrons, and  
in the flavor dependent structure of hadronic matter there should be a key for 
understanding deep inside the neutron stars. 
There are many issues not yet solved systematically.  
The high-intensity hadron beams in the energy region of several GeV
at J-PARC enable us to explore precise spectroscopy of hadrons with $u, d, s, c$ 
quarks. 
A framework to explain and predict various hadron properties 
will be established.
%of the K10 project.  

\if0
%The second is {\bf hadron physics} to investigate the nature of quantum
%chromodynamics (QCD).
%Hadron, described based on QCD as complex system of quarks and gluons,
%provides quite unique information on hierarchy of the visible matter in
%the universe; how the elementary particles form a variety of materials
%such as the stars and human beings in the universe.
%Nowadays, QCD has succeeded in describing the interactions between
%quarks and gluons, however, low-energy phenomena - such as the formation
%of a hadron - are not clearly explained because perturbation theory
%does not work in the low energy regime.
%High-intensity hadron beams in the energy region of $\sim$ GeV available
%at J-PARC enable to explore the nature of non-perturbative QCD, {\it
%i.e.}, strong coupling phenomena in the low energy region, through various
%measurements such as meson properties in nuclear media.
\fi

The third is {\bf flavor physics}. It aims at discovery of new physics
beyond the Standard Model (SM).
Since no clear evidence for new physics is found in vigorous direct searches at LHC,
flavor physics in intensity frontier plays a particularly important role today.
The $K_L \to \pi^0 \nu \bar{\nu}$ decay, which directly breaks the CP symmetry,
is utilized as a probe.
Since the SM predicts the branching ratio to be $3.0 \times 10^{-11}$ with small theoretical uncertainties~\cite{Buras:2015qea}, it provides us with a hit of new physics that the measured
branching ratio differ from the branching ratio predicted in the SM.
%observation of a discrepancy from the prediction provides us with a hint of new physics.

Charged lepton flavour violation is a definite sign for new physics beyond the SM.
The COMET experiment, aiming to search for coherent neutrino-less
conversion of a muon to an electron of $\mu^- + N(A,Z) \to e^- + N(A,Z)$
in muonic atoms ($\mu - e$ conversion), provides us with a window on new physics with the world's highest sensitivity.
%of $B(\mu^- + Al → e^- + Al)
%< 10^{-16}$, which is a factor of about 10,000 better than the present
%published limit.

Since the first delivery of the proton beam to the Hadron Experimental
Facility in January 2009,  experiments were carried out
and many fruitful results have been obtained.
In order to expand the programs of particle and nuclear physics to the regions that have not been
explored, more beam lines are
indispensable.
In the present Hadron Experimental Facility, a single production target is placed and is shared with a limited number of secondary beam lines due
to the limited space of the present hall.
About 50\% of the primary protons from the J-PARC Main
Ring (MR) are used
to produce secondary particles at the target, and the remaining protons are
transported to the beam dump~\cite{Agari:2012ana}.

The extension of the facility by adding more production targets and installing new secondary beam lines will substantially expand our research opportunities with the following merits.

%The new beam lines will enable us to promote new measurements which are impossible in the present facility.
\begin{screen}
{\bf Enabling new measurements that are not possible at the present
facility}.
\end{screen}
For the strangeness and hadron physics, construction of a `high-intensity
high-resolution beam line' and a `high-momentum mass-separated beam line',
with unprecedented capabilities, are highly anticipated to solve unsettled
problems in nuclear physics.
In particular, a solution to the ``hyperon puzzle'' is expected by high-precision spectroscopy of $\Lambda$-hypernuclei that
provides ultra-precise $\Lambda$ binding-energy measurements in a wide mass range
for the study of density-dependent $\Lambda N$ and
$\Lambda NN$ interactions.
In rare kaon decays, the extension of the hall enables us to
optimize the flux of kaons and neutrons in the `neutral beam line'.
With an extraction angle of 5 degrees, which had originally been chosen to be 16 degrees in the existing KL beam line at the present hall due to space limitations~\cite{Yamanaka:2012yma}, we will be able to utilize
a more intense kaon beam and thereby dramatically increase the
experiments' sensitivity.

%More flexible operation will be realized at the hadron experimental facility.
\begin{screen}
%%% HT {\bf Realization of more flexible operation of the facility.}
{\bf More efficient and flexible operation of the facility.}
%%%
\end{screen}
The existing K1.8 beam line is optimized for studying nuclei with
two strange quarks, whereas the studies of nuclei with a single strange
quark are performed at the branch line (K1.8BR)~\cite{Agari:2012kid}.
These two beam lines share the upstream part, and cannot be operated at the same time.
With a new dedicated `low-momentum beam line', which provides kaon beams with
higher intensity and 
better quality ({\it e.g.}, better $K/\pi$ ratio),
experiments can be performed simultaneously.
More flexible operation can also be realized for high momentum beams: the
existing primary proton beam line (high-p) and a new `high-momentum
mass-separated beam line'.
These beam lines, operated
simultaneously, can accommodate the increasing demands from
a wide variety of physics programs.

The extension project further promotes
the three core research lines: strangeness nuclear physics, hadron
physics, and flavor physics.
With the world's highest-intensity beams available at the
Hadron Experimental Facility, we aim to achieve the following goals:
\begin{itemize}
{\bf
%%% HT 
%%% \item to elucidate neutron star matter from nuclear physics through
%%% solving the hyperon puzzle,
 \item to elucidate at microscopic level neutron stars' EOS, by solving the hyperon puzzle,
%  \item to elucidate neutron star matter microscopically through solving the hyperon puzzle,
%%%
 \item to reveal baryon structure built from quarks and gluons, utilizing spectroscopic studies of strange and charm baryons, {\rm and}
%\item to reveal baryon structure built by quarks and gluons through spectroscopic studies of strange and charm baryons, {\rm and}
 \item to investigate new physics beyond the Standard Model through rare kaon decays.
}
\end{itemize}
The capability of the facility will be enhanced and enable us to reach
the objective
\begin{itemize}
{\bf
 \item to increase the diversity of world class physics programs at J-PARC on the basis of newly 
 constructed unique beam lines.
 %, which \color{red} realizes \color{black} flexible beam-time operation.
}
\end{itemize}
To maintain the world's leading position of the Hadron
Experimental Facility in the several-GeV energy region, the
diversity is of vital importance.
The operation with more production targets is also essential
to realize sustainable researches in the accelerator-based science in the future.

%%% HT
%%% The extension project had been considered in the early stage of J-PARC construction
%%% for particle and nuclear physics.
%%% The nuclear physics community in Japan supports it with the highest
%%% priority, and the particle physics community also recognizes
%%% it as important.
%%% The project was discussed in `Master Plan 2020' by the Science Council
%%% of Japan, and was selected as one of 15 projects from 31 important
%%% projects in `Roadmap 2020' by Ministry of Education, Culture, Sports,
%%% Science and Technology (MEXT) of Japan.
The extension project has been long discussed since the early stage of the J-PARC construction.
This is because when the whole J-PARC project was approved in 2000, the size of the Hadron Facility was reduced to almost a half of the original design due to budgetary limitation.
The nuclear physics community in Japan has requested 
realization of the extension 
with the highest priority, and the particle physics community has also expressed its importance.
The project was selected as one of 31 important projects in
“Japanese Master Plan for Large Research Projects 2020”
by the Science Council of Japan, 
and then selected as one of 15 projects in “Roadmap 2020 for promoting large scientific research projects” by Ministry of Education, Culture, Sports, Science and Technology (MEXT) of Japan.
%%%%

In the user community, Hadron Hall Users' Association (HUA), extensive
discussions on the project have been made.
A committee was organized under HUA for planning of the extension of the hadron experimental facility and made the first White Paper on the
Hadron Experimental Facility extension project as arXiv.1706.07916
[nucl-ex]~\cite{HEFextWP1:2017}.
It includes the roles of the project, overview of the facilities, and how
particle and nuclear physics can utilize it to attack the relevant open questions.
Concering details of the beam lines and many experimental lines of research to be
carried out in the project, the second White Paper was issued as
arXiv.1906.02357 [nucl-ex]~\cite{Takahashi:2019xcq}, based on the discussions
in `International workshop on the project for the extended hadron
experimental facility' held in March 2018.
The plan of the project has been updated and revised since then,
by taking into account recent progress of research 
in the Hadron Experimental Facility together with the global situation 
in the field of particle and nuclear physics.
In 2020, we organized a task force under
HUA, and held a series of workshops in the first half of
2021 to deepen discussions on the important physics features at the
extended Hadron Experimental Facility.
In this report, we summarize the updates.

This report is organized as follows.
The present
status of the Hadron Experimental Facility is briefly introduced in subsection~\ref{sec:status}.
The scientific goals and expected achievements in the extension
project are summarized for each subject in Sec.~\ref{sec:goal} together
with the present situation: what is the problem to be solved, what has been
achieved at the present facility so far, and how to approach the goal at
the new project.
A facility overview of the extended Hadron Experimental Facility is
described in Sec.~\ref{sec:facility}, and the expected timeline of the project is
given in Sec.~\ref{sec:timeline}.
Finally, we review the global situation of accelerator-based physics in
Sec.~\ref{sec:situation} to clarify the position of the J-PARC Hadron
Experimental Facility in the world.
In Sec.~\ref{sec:HIHR} and later, we describe details of each
experimental program planned at the new beam lines in the extended
facility.

\subsection{Present Status of the Hadron Experimental
  Facility}\label{sec:status}

\begin{figure}[htbp]
 \centerline{\includegraphics[width=0.6\textwidth]{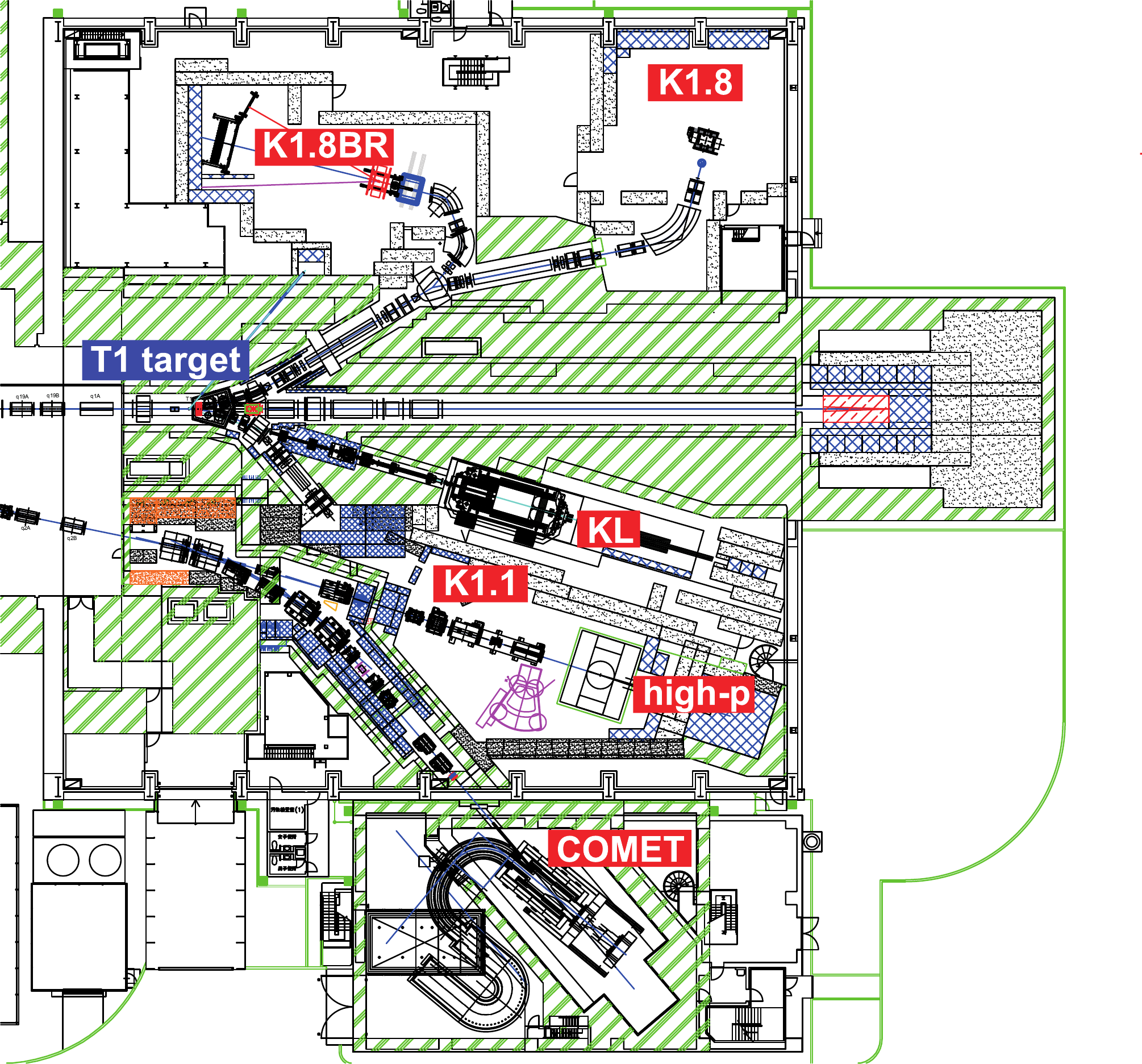}}
 \caption{Layout of the present experimental hall.
 The K1.8, K1.8BR, KL, and high-p beam lines are in operation.
 }
 \label{fig:HDhall}
\end{figure}

In the present Hadron Experimental Facility, a low-momentum
charged-kaon beam line (K1.8/K1.8BR), a neutral-kaon beam line (KL),
and a primary proton beam line (high-p) are being operated.
A new primary beam line for a muon-to-electron conversion experiment
(COMET) will also be ready for operation in 2022.
Figure~\ref{fig:HDhall} shows a layout of the present experimental
hall~\footnote{
To operate the K1.1 beam line in Fig.~\ref{fig:HDhall}, it is required to build the downstream part as well as
the experimental area at the south side of the
experimental hall.
There is spatial overlap between the high-p and K1.1 beam lines.
It takes more than nine months to get the experiment
ready for the changeover, during which other beam lines
in the hall cannot be operated due to the regulation of radiation protection.
}.
Primary protons are slowly extracted from the Main Ring
accelerator (MR) and transported to the experimental hall (MR-SX operation)~\cite{Koseki:2012wma}.
Kaons produced at the primary target (T1) are extracted to each secondary kaon beam line.
The primary protons are also delivered to the high-p and COMET
beam lines by branching off the protons in the switchyard at the upstream of the T1
target.

Table~\ref{table:exp} summarizes the experiments classified as ``completed'', ``ongoing'', ``forthcoming'',
and ``planned'' with their results.
Published results are also listed in the table.
Their highlights are as follows.
\begin{description}
 \setlength{\itemsep}{0cm}
\item[$\bigstar$ Search for penta-quark $\Theta^+$ via the
	   $\pi^-p\rightarrow K^-X$ reaction (E19)] \mbox{}\\
	   Upper limits on the production cross section and width
	   of $\Theta^+$ were obtained~\cite{Moritsu:2014bht}.
\item[$\bigstar$ Search for $^6_\Lambda$H via the $^6$Li$(\pi^-, K^+)$
	   reaction (E10)] \mbox{}\\
	   An upper limit on the production cross section of
	   $^6_\Lambda$H was obtained~\cite{Honda:2017hjf}.
	   The $\Sigma$-nucleus potential for $\Sigma$-$^{5}$He system was derived to be repulsive from analysis of the $^{6}$Li($\pi^{-}$, $K^{+}$) missing-mass spectrum \cite{Harada:2017spb}.
\item[$\bigstar$ $K^-pp$ bound state via the $d(\pi^+,K^+)$
	   reaction (E27)] \mbox{}\\
	   A $K^-pp$ like structure was
	   observed~\cite{Ichikawa:2014ydh}.
\item[$\bigstar$ $\gamma$-ray spectroscopy of $^4_{\Lambda}$He and
	   $^{19}_{\Lambda}$F (E13)] \mbox{}\\
	   A large charge symmetry breaking (CSB) in 4-body
	   $\Lambda$ hypernuclei was found and four $\gamma$-ray peaks from
	   $sd$-shell $\Lambda$ hypernucleus $^{19}_{\Lambda}$F were
	   observed~\cite{Yamamoto:2015avw,Yang:2017mvm}.
\item[$\bigstar$ $K^-pp$ bound state via the $^3$He$(K^-, n)$
	   reaction (E15)] \mbox{}\\
	   A $K^-pp$ bound state was observed~\cite{J-PARCE15:2020gbh}.
\item[$\bigstar$ $\Xi$ hypernucleus via the $^{12}$C$(K^-,
	   K^+)$ reaction (E05)] \mbox{}\\
	   A $^{12}_{\,\Xi}$Be hypernuclear state was
	   observed~\cite{Nagae:2019}.
\item[$\bigstar$ $\Lambda(1405)$ resonance via the
	   d$(K^-,n)$ reaction (E31)] \mbox{}\\
	   The $\bar K N \to \pi \Sigma$ scattering amplitude below the
	   threshold was obtained~\cite{Kawasaki}.
\item[$\bigstar$ Double-strangeness hypernuclei with hybrid
	   emulsion method (E07)] %\mbox{}\\
	   A double-$\Lambda$ hypernucleus $_{\Lambda\Lambda}$Be
	   (MINO event) and a coulomb-assisted nuclear bound state of
	   $\Xi^-$-$^{14}$N system (IBUKI event) were
	   observed~\cite{Ekawa:2018oqt, Hayakawa:2020oam}.
	   Recently, a deeply-bound $\Xi$ hypernuclear state was also observed~\cite{E07new}.
\item[$\bigstar$ Measurement of the $\Sigma p$ scatterings (E40)] \mbox{}\\
	   Differential cross sections of the $\Sigma^{-}p$ elastic
	   scattering were derived with drastically improved
	   precision~\cite{J-PARCE40:2021qxa}.
\item[$\bigstar$ Study of the $K_L \to \pi^0 \nu \bar{\nu}$ decay (E14
	   KOTO)] \mbox{}\\
       %The most stringent upper limit on the $K_L \to \pi^0 \nu \bar{\nu}$ decay
       %had been set to be $3.0\times10^{-9}$ at the 90\% confidence level with the
       %dataset taken in 2015. 
       A single-event sensitivity of $7.2\times10^{-10}$ was
       achieved with the dataset taken in 2016-2018. Three events were observed in the 
       signal region, which was consistent with the number of expected background events,
       $1.22\pm0.26$~\cite{KOTO:2020prk}. The physics data taking continues to improve the sensitivity
       by installing new counters to reduce background events.
       
	   %The highest sensitivity of 
	   %New upper limit of $4.9 \times 10^{-9}$ was set
	   %at the 90\% confidence level~\cite{KOTO:2020prk}.
\end{description}
A review paper of the nuclear physics programs at the Hadron Experimental
Facility was published~\cite{Ohnishi:2019cif}.

\begin{table}[htbp]
 \caption{Status of the experiments at the Hadron Experimental Facility with the publications.}
 \begin{center}
{\footnotesize
\begin{tabular}{lp{7cm}llp{3cm}}
\hline
 \multicolumn{2}{c}{\multirow{2}{*}{Experiment}} &
 \multirow{2}{*}{beam line} & Beam & \multirow{2}{*}{Status} \\
 & & & particle & \\
\hline 
 E03 & Measurement of X-rays from $\Xi$-atom & K1.8 & $K^-$ & completed \\
 E05 & Spectroscopic study of $\Xi$-hypernucleus, $^{12}_{\Xi}$Be, via the $^{12}$C$(K^-,K^+)$ reaction & K1.8 & $K^-$ & completed~\cite{Nagae:2019,Ichikawa:2020doq} \\
 E07 & Systematic study of double strangeness system with an emulsion-counter hybrid method & K1.8 & $K^-$ & completed~\cite{Ekawa:2018oqt,Hayakawa:2020oam,E07new} \\
 E08 & Pion double charge exchange on oxygen at J-PARC & K1.8 & $\pi^+$ & planned \\
 E10 & Production of neutron-rich $\Lambda$-hypernuclei with the double charge-exchange reactions & K1.8 & $\pi^-$ & completed~\cite{Sugimura:2013zwg,Honda:2017hjf} \\
 E13 & Gamma-ray spectroscopy of light hypernuclei & K1.8 & $K^-$ & completed~\cite{Yamamoto:2015avw,Yang:2017mvm} \\
 E18 & Coincidence Measurement of the Weak Decay of $^{12}_\Lambda C$ and the three-body weak interaction process & K1.8 & $\pi^+$ & forthcoming \\
 E19 & High-resolution search for $\Theta^+$ pentaquark in $\pi^- p \to K^- X$ reactions & K1.8 & $\pi^-$ & completed~\cite{Shirotori:2012ka,Moritsu:2014bht} \\
 E22 & Exclusive Study on the Lambda-N Weak Interaction in A=4 Lambda-Hypernuclei & K1.8 & $\pi^+$ & planned \\
 E26 & Direct measurements of $\omega$ mass modification in A$(\pi^-,n)\omega$ reaction and $\omega \to \pi^0 \gamma$ decays & K1.8 & $\pi^-$ & planned \\
 E27 & Search for a nuclear $\bar K$ bound state $K^-pp$ in the d$(\pi^+, K^+)$ reaction & K1.8 & $\pi^+$ & completed~\cite{Ichikawa:2014rva,Ichikawa:2014ydh} \\
 E40 & Measurement of the cross sections of $\Sigma p$ scatterings & K1.8 & $\pi^{\pm}$ & completed~\cite{J-PARCE40:2021qxa} \\
 E42 & Search for $H$-dibaryon with a large acceptance hyperon spectrometer & K1.8 & $K^-$ & completed \\
 E45 & 3-body hadronic reactions for new aspects of baryon spectroscopy & K1.8 & $K^-$ & forthcoming \\
 E70 & Proposal for the next E05 run with the $S$-$2S$ spectrometer & K1.8 & $K^-$ & forthcoming \\
 E75 & Decay Pion Spectroscopy of $^5_{\Lambda\Lambda}H$ Produced by $\Xi$-hypernuclear	Decay & K1.8 & $K^-$ & planned \\
\hdashline
 E15 & A search for deeply-bound kaonic nuclear states by in-flight $^3$He$(K^-, n)$ reaction & K1.8BR & $K^-$ & completed~\cite{Hashimoto:2014cri,Sada:2016nkb,Ajimura:2018iyx,J-PARCE15:2020gbh} \\
 E31 & Spectroscopic study of hyperon resonances below $\bar K N$ threshold via the $(K^-,n)$ reaction on deuteron & K1.8BR & $K^-$ & completed~\cite{Asano,Kawasaki} \\
 E57 & Measurement of the strong interaction induced shift and width of the 1st state of kaonic deuterium at J-PARC & K1.8BR & $K^-$ & planned \\
 E62 & Precision spectroscopy of kaonic helium 3 $3d \to 2p$ X-rays & K1.8BR & $K^-$ & completed~\cite{Okada:2016slo,Hashimoto:QNP2018} \\
 E72 & Search for a narrow $\Lambda^*$ resonance using the $p(K^-, \Lambda)\eta$ reaction with the hypTPC detector & K1.8BR & $K^-$ & forthcoming \\
 E73 & $^3_\Lambda H$ and $^4_\Lambda H$ mesonic weak decay lifetime measurement with $^{3,4}$He$(K^-, \pi^0)^{3,4}_\Lambda$H reaction & K1.8BR & $K^-$ & planned \\
 E80 & Systematic investigation of the light kaonic nuclei - via the in-flight $^4$He$(K^-, N)$ reactions & K1.8BR & $K^-$ & planned \\
\hline
\end{tabular}
}
 \end{center}  
 \label{table:exp}
\end{table}

\begin{table}[htbp]
 \begin{center}
{\footnotesize
\begin{tabular}{lp{7cm}llp{3cm}}
\hline
 \multicolumn{2}{c}{\multirow{2}{*}{Experiment}} &
 \multirow{2}{*}{beam line} & Beam & \multirow{2}{*}{Status} \\
 & & & particle & \\
\hline 
 E14 & Proposal for $K_L \to \pi^0 \nu \bar \nu$ Experiment at J-PARC & KL & $K^0_L$ & ongoing~\cite{Ahn:2016kja,Ahn:2018mvc,Ahn:2020orr,KOTO:2020prk} \\
\hdashline
 E16 & Electron pair spectrometer at the J-PARC 50-GeV PS to explore the chiral symmetry in QCD & high-p & $p$ & ongoing \\
 E50 & Charmed baryon spectroscopy via the $(\pi^-, D^{*-})$ reaction & high-p & $\pi^-$ & planned \\
 E79 & Search for an $I=3$ dibaryon resonance & high-p & $p$ & planned \\
 \hdashline
 E29 & Study of in medium mass modification for the $\phi$ meson using $\phi$ meson bound state in nucleus & K1.1 & $\bar p$ & planned \\
 E63 & Proposal of the 2nd stage of E13 experiment & K1.1 & $K^-$ & forthcoming \\
 \hdashline
 E36 & Measurement of $\Gamma(K^+ \to e^+ \nu)/\Gamma(K^+ \to \mu^+ \nu)$ and Search for heavy sterile neutrinos using the TREK detector system
  & K1.1BR & $K^+$ & completed~\cite{E36}\\
 \hdashline
 E21 & An Experimental Search for $\mu - e$ Conversion at a Sensitivity of $10^{-16}$ with a Slow-Extracted Bunched Beam & COMET & $\mu^-$ & forthcoming \\
 \hline
\end{tabular}
}
 \end{center}
  \end{table}

As of June 2021, a beam power of 64.5 kW was achieved with a 2.0 s
beam duration in a 5.2 repetition cycle, which corresponds to
7.0$\times$10$^{13}$ protons per pulse.
The present production target system T1, which is composed of a gold rod cooled by an indirect water-cooling system~\cite{Takahashi:2015,Watanabe:2020wor}, is allowable to 95 kW under the 5.2 repetition
cycle.
Further improvement of the accelerator beam power is being planned; the upgrade of the MR main-magnet power supplies will realize a MR-SX power over 100 kW by operation
with a higher repetition rate than that at the present.
To receive the beam of higher power,
a new production target system up to $>$150 kW is also being developed,
which employs a rotating-disk target with a direct helium-gas-cooling
system.

\subsection{Scientific Goals in the Extension Project}\label{sec:goal}
%With the new functional beam lines constructed in the extended Hadron
%Experimental Facility together with the already existing ones at present
%facility, we aim to pin down unsettled problems in the field of particle
%and nuclear physics.
%The physics goals at the extension project are summarized here, althogh
%details of the planned experiments are described in the following
%sections.

\subsubsection{Elucidation of neutron star matter microscopically through
solving the hyperon puzzle}
A so-called ``hyperon puzzle'' is the difficulty to reconcile the
astronomical observations of two-solar-mass neutron
stars~\cite{Demorest:2010bx, Antoniadis:2013pzd} with the
presence of hyperons in their interiors predicted by nuclear physics;
the hyperon presence makes the equation of state (EOS) softer and
thereby the maximum mass of neutron stars is incompatible with the observations.\footnote{The high momentum tail due to the nucleon-nucleon short-range correlations (SRC) also affects the EOS of neutron stars.
The observed dominance of the $pn$ SRC pairs compared to the $pp$ pairs is a clear consequence of the nucleon-nucleon tensor correlation.
In Jlab experiments, the high-momentum fraction was measured for both proton and neutron \cite{SRC_1}.
In $^{208}$Pb, the fraction of the high-momentum proton increased, whereas the fraction of the high-momentum neutron decreased.
In the neutron star, this SRC increases the average kinetic energy of protons, while it decreases that of neutrons.
This decrease of the neutron's energy makes the symmetry energy softer.
Therefore, the SRC is another source for softening the EOS \cite{SRC_2}.
%In the mean field calculation, such SRC should be newly introduced.
If the bare $NN$ interaction with the short-range repulsive core and the tensor force is used in microscopic approaches such as the variational method, the SRC effect is expected to be taken into account in the calculation.
%In such calculations, the SRC effect is included.
%In the BHF calculation, where raw two-body interaction can be used,  the treatment of SRC might not be included correctly compared with variational methods, because the short range repulsive force can be weakened in the G-matrix calculation.
%Theoretical calculations are ongoing to explain SRC scaling factor measured at Jlab.
In the neutron star density, the dominant source of repulsive interaction is three-body repulsive force. However, the SRC should also be discussed together in future.
}
The solution of this problem requires a mechanism 
providing an
additional repulsion between baryons to make the EOS stiffer.
Such additional repulsion would be described by two-body baryon-baryon interactions and three-body 
baryon-baryon-baryon interactions including hyperons,
which give essential effects in dense nuclear matter.
We need much more comprehensive information on the hyperon-nucleon
($YN$) and hyperon-hyperon ($YY$) interactions both in the free space
and in the nuclear medium.
To determine the strength of the hyperonic three-body repulsive forces, it is
vital to measure the $\Lambda$
binding energies ($B_{\Lambda}$) of $\Lambda$-hypernuclei precisely in a wide mass-number region.
The information on the density-dependent $\Lambda N$ interaction can be obtained from them.

At the Hadron Experimental Facility, experiments on hypernuclei have been proposed and performed to determine
the strengths of the baryon-baryon interactions 
extended to the strangeness sector.
The $\Lambda \Lambda$ and $\Xi N$ interactions have been and will be
derived with world's most accurate measurements of the
$\Xi/\Lambda\Lambda$ hypernuclei at the existing {\bf K1.8} beam line
using the $(K^-,K^+)$ reaction.
The $\Lambda N$ interactions in the free space will be precisely obtained by
$\Lambda p$ scattering experiments planned at the new {\bf K1.1} and {\bf
high-p} beam lines.
The strength of the $\Lambda N$-$\Sigma N$ coupling will also be precisely
obtained by spectroscopic studies of light neutron-rich hypernuclei and
precision measurements of charge symmetry breaking in $\Lambda$ hypernuclei.
The $\gamma$-ray spectroscopy of $\Lambda$ hypernuclei is a powerful tool to determine low-lying level structure of hypernuclei precisely.
The differential cross-section measurements of the
$\Sigma^- p \to \Lambda n$ and $\Lambda p \to \Sigma^{0}p$ scatterings are essential to determine the $\Lambda N$-$\Sigma N$ coupling in the free space, because the $\Lambda N$-$\Sigma N$ coupling could be modified in the nuclear medium.
These experiments have been and will be performed at K1.8 and K1.1.

To obtain information on the hyperonic three-body repulsive force, however, we need innovative measurements of $\Lambda$ hypernuclei with unprecedented precision in the wide mass-number region.
Such three-body effect affects the $\Lambda$ binding energy particularly in heavy $\Lambda$ hypernuclei.
The effect is expected to be from a few hundred keV to a few MeV, depending on the two-body $\Lambda N$ interaction.
In the heavy $\Lambda$ hypernuclei, the $\Lambda$ couples to a core nucleus of various one-hole excited states having a small energy difference with each other. It results in several hypernuclear states with narrow energy spacings of typically a few hundred keV.
In order to determine the $\Lambda$ binding energies by separating each state and to provide accurate information including the $\Lambda NN$ three-body effect, a new spectroscopic method should be introduced.
In fact, the past experiment at KEK could not separate each state due to a limited energy resolution of $\sim$2 MeV (FWHM).
New experiments proposed at the newly constructed {\bf HIHR}
beam line can break through the situation.
The binding energy can be determined with a resolution
of a few hundred keV (FWHM) from the light ($^4_{\Lambda}$He) to heavy
($^{209}_{\Lambda}$Pb) hypernuclei, by using the $(\pi,K^+)$
missing-mass spectroscopy by using high-intensity pion beams
and thin nuclear targets.
The systematic high-precision spectroscopy from light to heavy $\Lambda$
hypernuclei provides essential data to extract the repulsive hyperonic three-body force effect through theoretical studies based on the realistic $\Lambda N$ interaction determined from the $\Lambda p$ scattering experiments to be preformed at K1.1.
The $\gamma$-ray spectroscopy provides us with additional information to determine the level structure of $\Lambda$ hypernuclei, in particular for the heavy $\Lambda$ hypernuclei where the level spacing could be narrower than the HIHR resolution.
The $\gamma$-ray information will be also used to interpret the binding energy spectrum obtained by HIHR.
The density-dependent $\Lambda N$ interaction effects also appear in the mass-number dependence of the level spacing between the $s_{\Lambda}$ and $p_{\Lambda}$ states, because the averaged nuclear density which the $\Lambda$ feels in the $p_\Lambda$ orbital is expected to be different for different mass number.
The level spacing between $s_{\Lambda}$ and $p_{\Lambda}$ can be determined with a few keV accuracy from $\gamma$-ray spectroscopy using germanium detectors.

Only after combining the realistic $Y$$N$ and $Y$$Y$ interactions with the hyperonic three-body force strength obtained from the comprehensive studies in hypernuclear physics, a realistic EOS can be established to solve the hyperon puzzle.

\subsubsection{Revelation of baryon structure built by quarks and gluons
through spectroscopic studies of strange and charm baryons}
Quantum chromodynamics (QCD) has succeeded in describing the interactions between quarks and gluons, and hadron properties.
However, low energy phenomena such as the formation of hadrons are not
clearly explained yet, because perturbative calculations do not work in the
low energy regime.

At the present Hadron Experimental Facility, experimental approaches to investigate the in-medium properties of mesons are being made. They are expected to provide crucial information on the spontaneous breaking of chiral symmetry associated with the mass generation of hadrons, which is the most important feature of QCD at low energy.
Various investigations of in-medium meson properties have been and will be performed, such as the precise measurement of the spectral functions of vector mesons in medium at {\bf High-p} and systematic measurement of the light kaonic nuclei at {\bf K1.8BR}.
%The new information obtained at the present and new beam lines will make progress in understanding the low energy phenomena of QCD.

Spectroscopy of baryons (baryon spectroscopy) is another important approach to 
investigate how QCD works at low energy, which is to be explored at J-PARC.
%address this unsettled problem is to introduce a new degree of freedom inside hadrons, which is a powerful tool to describe hadron dynamics in effective theories of low-energy QCD.
In particular, a basic question "how quarks build hadrons" has yet to be answered clearly. 
It is vital to understand dynamics of effective degrees of freedom, that are constituent quarks and Nambu-Goldstone bosons, brought by the non-trivial gluon field at low energy.
Spectroscopy of baryons with heavy flavors provides a good opportunity to investigate interactions and correlations of the effective degrees of freedom in hadrons.
In excited baryons
%, particularly in a baryon 
containing a heavy quark,
the correlation between a quark pair inside the baryon is expected to
be enhanced.
%, and the quark pair will be acting as a new degree of freedom. 
It is referred as a diquark correlation.
There are longstanding arguments if the diquark correlations play roles on baryon structure and are related to the diquark condensations at highly dense quark matter,
though they are not settled yet.

At the energy of J-PARC, excited baryons with a charm quark or multi strange
quarks are appropriate.
%to explore the effective degrees of freedom inside hadrons.
An experiment to study the diquark correlation in charmed baryons (denoted as $Y_c^{*}$) is
planned at the ${\boldsymbol \pi} {\bf 20}$ beam line with secondary high-momentum
pions.

Spectroscopy of $\Xi$ baryons is planned at the new {\bf K10} %$\pi20$ 
with intense, separated kaons to investigate the diquark correlation in the strangeness sector, and to explore an unknown field of
excited $\Xi$ baryons (denoted as $\Xi^*$) \footnote{A pilot experiment on $\Xi$ baryons \cite{loi:xi-spectroscopy} is planned also at $\pi20$ with identifying kaons from pions of intensity more than 2 orders of magnitude, which will be able to provide basic information on the production cross sections as well as possible background processes prior to the further investigation at K10.}.

By introducing quarks with different flavors having different masses in a baryon, a relative motion between two quarks in a diquark and a collective motion of the diquark to the other quark are kinematically separated.
We will employ the the $\pi^-p\rightarrow D^{*-}Y_c^{*+}$ and $K^-p\rightarrow \Xi^*K^{(*)}$ reactions since the production cross sections reflect the above-mentioned diquark motions~\cite{shkim14,shim2}. It is noted that we can populate $Y_c^{*+}$ and $\Xi^*$ from the ground states up to highly excited states systematically since they are identified in the missing mass spectra. 
By measuring decay particles in coincidence with a populated state, we can measure the branching ratio (decay partial width) easily, which carries information on the internal structure of the excited baryon. 
This is an advantage in fixed-target experiments.
%is to identify both the formation and decay of the reaction.
%Thus, the production cross section and decay fraction provide us valuable information to clarify the quark-gluon dynamics inside.

Further investigations of the quark correlation can be performed at the
new K10 beam line, via the simplest $sss$ systems - $\Omega$ baryons.
The $\Omega$ baryons are unique because the pion coupling
is rather weak compared to the other hadrons including $u/\bar u$ and
$d/\bar d$ quarks, and thereby the quark-gluon dynamics will be directly
reflected to the exited state $\Omega^*$.
However, since the production cross sections of $\Omega^{*-}$ is expected to be
small in case of the $K^-p\rightarrow K^+K^{(*)0}\Omega^{*-}$ reaction, the
investigations can only be preformed using intense and separated kaon beams available at K10.
%It is worthy to consider a challenging experiment to measure an $\Omega^-$-$N$ bound state owing to abundant $\Omega$ production at K10.

\subsubsection{Investigating new physics beyond the Standard Model through
rare kaon decays}
The Standard Model (SM) in particle physics successfully describe how elementary particles 
interact with each other.
The last missing particle in the SM, a Higgs particle, 
was discovered in 2012 by the ATLAS and CMS experiments at the LHC~\cite{ATLAS:2012yve,CMS:2012qbp}.
However, there still remains many questions that cannot be explained by
the SM such as the matter-dominant universe and the low mass of the
Higgs boson, and thereby the searches for new physics beyond the SM have been pursued intensively.
Since direct production of new particles would not be discovered so easily in LHC,
the role of flavor physics experiments, which investigate new physics at a high energy 
scale through rare phenomena by using intensity frontier machines, become more important. 

At the {\bf KL} beam line in the present hall, the KOTO experiment searches for the rare kaon decay
$K_L \to \pi^0 \nu \overline{\nu}$. This decay directly violates CP symmetry. The branching ratio
 is highly suppressed, and is predicted to be $3.0\times10^{-11}$ with small theoretical uncertainties
 at a level of 2\%~\cite{Buras:2015qea}. Thus, this decay is one of the most sensitive probes to search for new physics.
The KOTO experiment set the most stringent upper limit on the $K_L \to \pi^0 \nu \overline{\nu}$  
branching ratio to be $3.0\times10^{-9}$ at the 90\% confidence level with the dataset taken in 2015~\cite{Ahn:2018mvc}.
The sensitivity of $7.2\times10^{-10}$ was achieved with the dataset taken in 2016-2018. 
Three events were observed in the signal region, which was consistent with the number of expected background events, $1.22\pm0.26$~\cite{KOTO:2020prk}. 
With newly installed counters to reduce background events, the KOTO experiment continues to take physics data and will reach a sensitivity better than
$10^{-10}$ in 3-4 years. However, the achievable sensitivity will be eventually saturated. 
It is desirable to design a new experiment that can observe the sufficient number of SM-predicted events for the measurement of the $K_L \to \pi^0 \nu \overline{\nu}$ decay. 

In the extended facility, the KOTO step-2 experiment plans to build a new neutral kaon beam line ({\bf KL2}) with 
a smaller extraction angle than that for the KOTO experiment to increase the $K_L$ flux and prepare 
a longer detector to extend the signal region and improve the signal acceptance.
The KOTO step-2 experiment aims to measure the branching ratio of the $K_L \to \pi^0 \nu \overline{\nu}$ decay, with a beam intensity of 100~kW for three snow-mass-years, 
in an accuracy of 27\% by observing 35 SM events with a signal-to-background ratio of 0.63. 

The breaking of time-reversal (T) symmetry
will also be investigated  at the {\bf K1.1BR} beam line by measuring the transverse polarization of the
muon from the $K^+ \to \pi^0 \mu^+ \nu$ decay.
T violation, connected to CP violation through the CPT theorem,
is an important key to solve the matter-antimatter asymmetry in the universe.

The observables from the kaon rare decays, together with the measurements
in $B$ factories such as the Belle II experiment, play an important role 
to investigate the flavor structure in new physics.

\subsection{Extended Hadron Experimental Facility and Its
  Realization}\label{sec:facility}
In the extension project, the construction of the following new beam lines will
expand the potential for particle and
nuclear physics programs at J-PARC.
\begin{itemize}
 \item {\bf HIHR beam line}:
       High-intensity and high-resolution charged $\pi$ meson beam line
       for high-precision spectroscopy of $\Lambda$-hypernuclei.
       The beam line will adopt state-of-the-art technology of the
       dispersion-matching method, and enable us to operate more than
       $10^8$ pions / spill up to 2 GeV/$c$ with the beam momentum
       resolution of $\delta p /p \sim 10^{-4}$.
       The $(\pi^\pm,K^+)$ missing-mass resolution of a few 100 keV (FWHM)
       will be achieved corresponding to the $\Lambda$-hypernuclei mass
       determination with several ten keV accuracy, which has never been
       realized so far.
 \item {\bf K10 beam line}:
       High-momentum (2--10 GeV/$c$) charged $K$ meson and anti-proton
       beam line for investigations of $S=-3$ strangeness physics and
       charm physics.
       The beam line will provide separated secondary beams with the
       higher momentum than any existing beam lines.
       The high-momentum particle separation will be realized by two-stage
       RF separators.
 \item {\bf K1.1 beam line}:
       Low-momentum ($<$1.2 GeV/$c$) charged $K$ meson beam line
       optimized for investigations of $S=-1$ strangeness physics.
       High-purity and high-intensity secondary particles will be available by
       using two-stage electrostatic separators.
       A branched beam line (K1.1BR) which focuses on experiments using stopped kaons
       will also be prepared.
 \item {\bf KL2 beam line}:
       High-intensity neutral $K_L$ meson beam line dedicated to measure 
       the branching ratio of the rare decay $K_L \to \pi^{0} \nu \bar \nu$.
       The extraction angle of 5 degrees will be
       adopted, instead of 16 degrees in the existing beam line,
       to increase the $K_{L}$ yield while keeping the ratio of
       kaons and neutrons which could become a source of background
       events.
\end{itemize}

\begin{figure}[htbp]
\centerline{\includegraphics[width=0.95\textwidth]{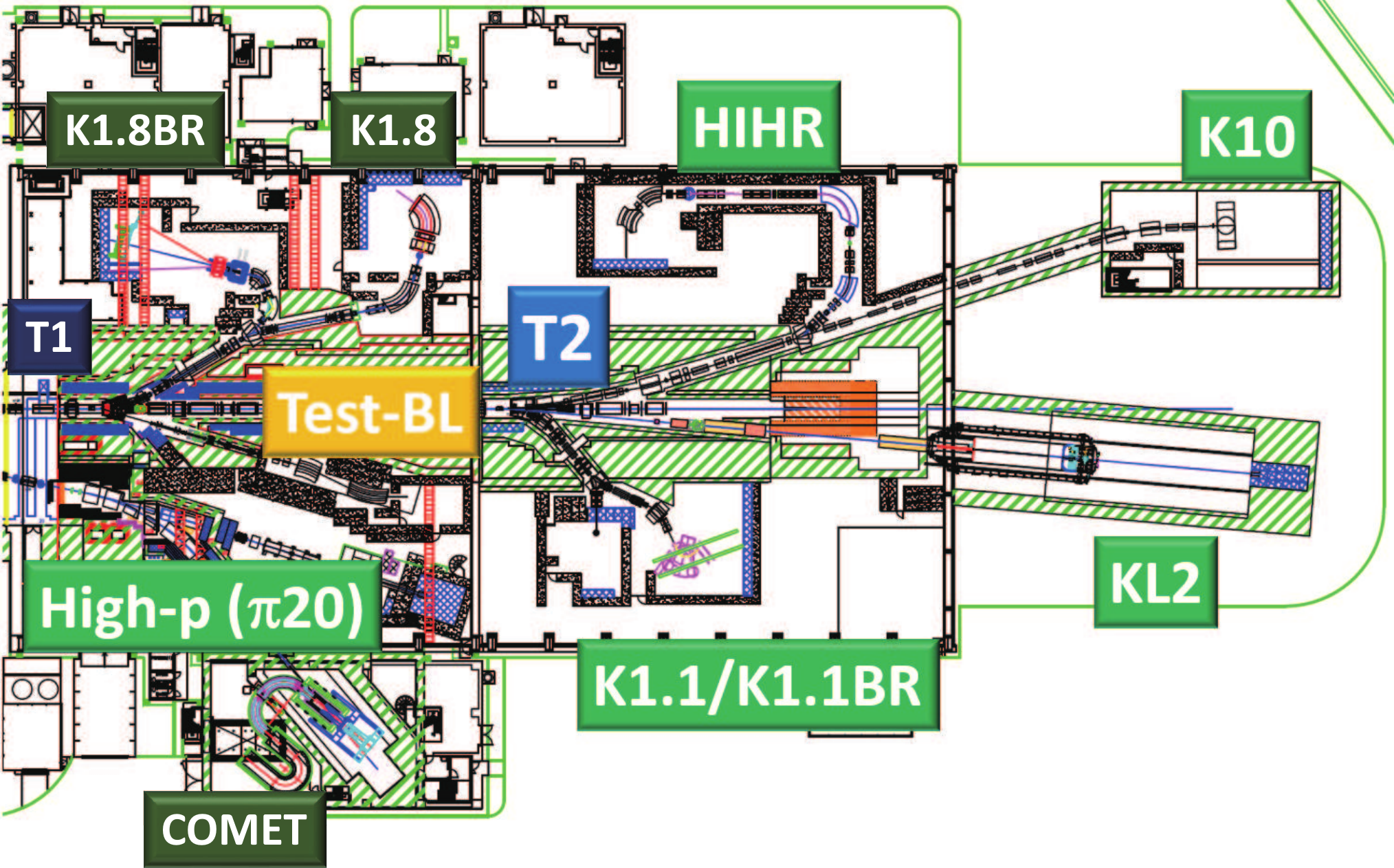}}
\caption{Layout of the extended Hadron Experimental Facility.
The T2 production target and the beam lines, HIHR, K1.1/K1.1BR, KL2,
and K10 will be newly constructed.
In addition, the $\pi20$ beam line and the test beam line are expected to be
realized in an early stage of the project.
}
\label{FigExtHEF}
\end{figure}

Figure~\ref{FigExtHEF} shows a layout of the extended Hadron
Experimental Facility together with the present facility.
The size of the experimental area will be twice larger and a new
production target (T2) will be placed.
The new beam lines will be connected from the T2 target station.
A test beam line is planned to be constructed in the experimental
area that has previously been used for the KL experiment.
In the extended experimental hall, five beam lines (plus the test
beam line) in total will be operated simultaneously.
The specifications of the beam lines for both the present and the
extended Hadron Experimental Facility are summarized in
Table~\ref{BLspec}.
By using the two production targets, effective utilization of
intense primary protons will be realized.

\begin{table}[htbp]
\caption{Specifications of beam lines in the present and the extended 
Hadron Experimental Facility.
Beam intensities at the present beam lines are typical and scaled to 
$\sim$ 80 kW beam on a 50\% loss production target.
The intensities of charged and neutral particles at the new beam lines are the designed values for a $\sim$ 50 kW and $\sim$ 100 kW beam on a 50\% loss production target, respectively.
This table is taken from~\cite{HEFextWP1:2017} with modifications.
}
\begin{center}
{\footnotesize
\begin{tabular}{l|llll}
\hline
 & Particles	& Momentum	& Intensity	& Characteristics	\\
\hline
\multicolumn{5}{l}{
beam lines in the present hadron experimental facility} \\
\hline
K1.8		& $K^{\pm}, \pi^{\pm}$ 		& 1.0 -- 2.0 GeV/{\it c}	
 &	$\sim$ 10$^6$ $K^-$ /spill (1.8)	& mass separated	\\
K1.8BR	& $K^{\pm}, \pi^{\pm}$		& $<$ 1.1 GeV/{\it c}
 &	$\sim$ 5$\times$10$^5$ $K^-$ /spill (1.0)	& mass separated	\\
KL		& $K_L$	& 2.1 GeV/{\it c} in ave.	& 10$^7$ $K_L$ /spill	
 & 16$^\circ$ extraction angle	\\
high-p    & $p$	&	& 10$^{10}$ $p$ /spill	& primary beam (30 GeV)	\\
$\pi20$   & $\pi^{\pm}$	&	$<$ 31 GeV/{\it c}	& 10$^7$ $\pi$ /spill
		& secondary beam \\
COMET	& $\mu^-$	&	&	& for $\mu^- \rightarrow e^-$  experiment \\
\hline
\multicolumn{5}{l}{
beam lines in the extended area} \\
\hline
K1.1		& $K^{\pm}, \pi^{\pm}$	& $<$ 1.2 GeV/{\it c}	
 & $\sim$ 4$\times$10$^5$ $K^-$ /spill (1.1)	& mass separated	\\
K1.1BR	& $K^{\pm}, \pi^{\pm}$	& 0.7 -- 0.8 GeV{\it c}
 & $\sim$ 1.5$\times$10$^5$ $K^-$ /spill 	& mass separated	\\
HIHR	& $\pi^\pm$	&	$<$ 2.0 GeV/{\it c}
 & $\sim$ 2$\times$10$^8$ $\pi$ /spill (1.2)    & mass separated	\\
	&	&	&	& $\times$10 better $\Delta p/p$	\\
K10	& $K^\pm, \pi^\pm, \overline{p}$		& $<$ 10 GeV/{\it c}
 & $\sim$ 7$\times$10$^6$ $K^-$ /spill	& mass separated \\
KL2	& $K_L$	& $\sim$ 5~GeV/{\it c} in ave.	
 & $\sim$ 4$\times$10$^7$ $K_L$ /spill	& 5$^\circ$ extraction angle \\
	&	&	&	& optimized $n/K_L$	\\
\hline
\end{tabular}
}
\end{center}
\label{BLspec}
\end{table}

In the original plan of the extension project,
the size of the experimental area was three times larger than the present area and two
new production targets (T2 and T3) were considered to be placed, as described in the
first and second white papers.
Thereafter we have reconsidered the extension plan to reduce the
cost, by decreasing both the size of
the experimental area and the number of the production targets as shown in Fig.~\ref{FigExtHEF}\footnote{
The number of the new secondary beam lines is unchanged from
the original plan.}.
A staging plan of the construction has also been considered aiming at
early realization of high-precision $(\pi,K^+)$ spectroscopy at the
HIHR beam line, precise measurement of $YN$ scsttering at K1.1,
and the highest sensitivity search for $K_L \to \pi^0 \nu \bar{\nu}$
at KL2.
The HIHR, K1.1, and KL2 beam lines will be construct
at first, and the K10 and K1.1BR beam lines will follow.
In parallel, the $\pi20$ beam line is expected to be constructed by
modifying the high-p beam line in an early stage of the project,
to conduct charmed baryon spectroscopy.
The basic information on $\Xi$ and $\Omega$ baryons, such as elementary 
cross section of $K^- p$ reaction, will also be obtained at $\pi20$.
This investigation of $\Xi$ and $\Omega$ baryons is
important to design the baryon spectroscopy of $\Xi^*$/$\Omega^*$
planned at K10.
The new test beam line is also expected to be built after the KL beam line is removed.
The realization of a test beam line has been strongly requested from 
the particle and nuclear physics communities as well as astrophysics
communities.

%The HIHR beam line enables to measure the density dependent $\Lambda$
%single-particle energy in $\Lambda$ hypernuclei with ultra
%high-resolution of 100 keV (FWHM) that has never been achieved.
%The $\Lambda NN$ three-body force, a key of the ``Hyperon Puzzle''
%solution, will be investigated by combining with comprehensive studies
%at the existing facility.

\subsection{Timeline of the Project}\label{sec:timeline}
The construction of the extended hadron hall 
will take 6 years in total including a period of 2.5 years 
of suspending the beam delivery to the existing beam lines.
The suspension is necessary to move the beam dump and to
extend the primary beam line to the new hall.
From the summer of 2021 to the autumn of 2022, the MR has been shut down
for the upgrade of magnet power supplies aiming to increase beam power
over 1 MW for the fast-extraction operation (FX) and to improve the 
operation in both the SX and FX.
Thus, to maximize the physics output from the Hadron Experimental
Facility, it is essential that current approved or planned experiments at the
existing beam lines are effectively conducted before the beam
suspension.

\begin{figure}[htbp]
 \centerline{\includegraphics[width=1\textwidth]{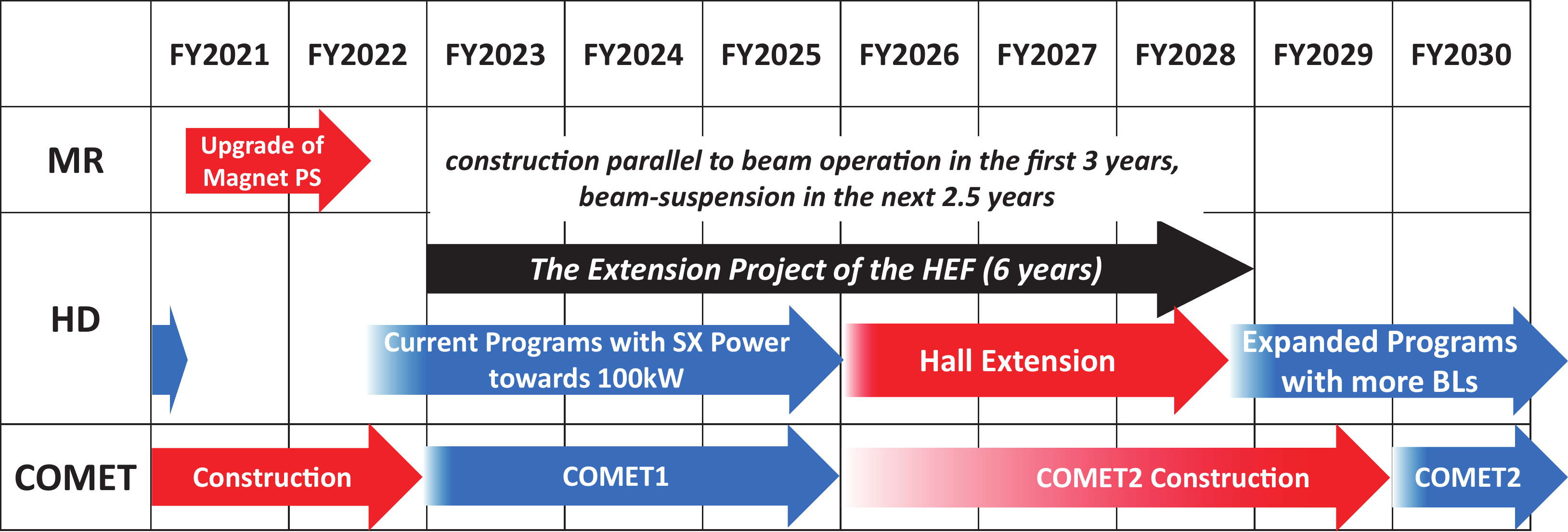}}
 \caption{
 Timeline of the extension project with the start of the funding in FY2023.
 }
 \label{fig:timeline}
\end{figure}

The timeline of the project is shown in Fig.~\ref{fig:timeline}.
for the case that the funding of the extension project starts in FY2023.
%The hall extension will start from FY2026 with beam suspension for 
%2.5 years.
Since the construction of the equipment parallel to the beam operation is possible 
in the first 3 years, most of the current programs planned with the SX power 
towards 100~kW will be completed before the beam suspension.
Then the beam operation will be suspended for 2.5 years from FY2026 to connect the existing and extended halls.

At the {\bf K1.8} beam line, experiments of $\Xi$ hypernuclear
spectroscopy (E70 and E75) will be performed by installing the $S-2S$
spectrometer in 2021-2022, and successor experiments will be followed for the 
systematic investigation of $S=-2$ systems.
Experiments related to kaonic atoms and nuclei will be conducted at the
{\bf K1.8BR} beam line to investigate the $\bar K N$ interaction close
to the mass threshold, such as the X-ray spectroscopy of the kaonic 
deuterium atom (E57) and the systematic measurement of the light kaonic
nuclei (E80), as well as measurements of the hypertriton lifetime (E73) and a 
narrow $\Lambda^*$ resonance (E72).
Most of these experiments will be realized by improving the beam line
and detector system.

At the {\bf high-p} beam line providing a primary 30 GeV proton beam,
the measurement of vector meson mass spectra in nuclei (E16) will continue by using $p + A$ reactions.
A part of experiments proposed at {\bf K1.1} - $\gamma$-ray spectroscopy of
light hypernuclei (E63) - would have a chance to be performed after the
completion of the first stage of E16, by rebuilding the layout
of the beam-line magnets and the experimental area at the south side of the hall.
There is spatial overlap between the high-p and the K1.1 beam lines.
Since the time and cost of the changeover between 
high-p and K1.1 will give considerable effects to the whole program of the facility,
new experiments planned at K1.1 are desired to be conducted at the newly constructed K1.1
beam line in the extended hall.

At the {\bf KL} beam line, study of the rare CP-violating kaon
decay (E14-KOTO) will continue to search for new physics at the sensitivity
of the $K^0_L \to \pi^0 \nu \bar{\nu}$ branching fraction of
$\mathcal{O}(10^{-11})$.
The {\bf COMET} beam line at the south experimental hall will also start
delivering 8 GeV protons in the bunched slow extraction mode in
2023, for the experiment to search for $\mu^-$-to-$e^-$ conversion (E21-COMET).
Before the beam suspension scheduled in 2026 
in the earliest case, the COMET phase I experiment
will be completed with three-year operation, and will be followed by 
the COMET phase II experiment with significant detector upgrades.

During the beam suspension, a new target system capable of the $>$150 kW beam
will be installed as the new production targets, which employ a directly-cooled rotating-target method.
Detector upgrades will be conducted at the K1.8, K1.8BR, and COMET
beam lines, and a new {\bf test beam line} will be built during the shutdown period. %if we have a chance.
The high-p beam line is also expected to be upgraded to the
${\boldsymbol \pi} {\bf 20}$ beam line by placing a thin production
target at the branching point in the switchyard.
Secondary high-momentum and mass-unseparated beams of pions, kaons, and
antiprotons up to 20 GeV/$c$ will be available, with which a charmed baryon spectroscopy experiment (E50) will be performed to investigate diquark correlation
in heavy baryons.

\subsection{Global Situation of Accelerator-Based
  Physics}\label{sec:situation}
Particle accelerators are important in the modern physics to explore the 
fundamental components of the matter in the universe.
Many accelerator facilities are in operation and planned in all over the world.

% ### HEP ###
\subsubsection{Particle physics}
There are two approaches to search for new physics beyond the Standard Model (SM)
through the accelerator-based experiments.

\vskip.5\baselineskip

% high-energy frontier
{\bf Energy Frontier:}\\
One approach is the exploration of the energy frontier, where direct
searches for new particles such as supsersymmetric
(SUSY) particles and dark matter (DM) candidates are performed.
The state-of-the-art accelerator in energy frontier is the Large Hadron
Collider (LHC) at CERN.
Now the ATLAS and CMS experiments at the LHC have searched for new
signatures of physics beyond the SM based on $\sim$ 150 fb$^{-1}$ of
data from $pp$ collisions at a center-of-mass energy of $\sqrt{s}$ = 13 TeV
collected during LHC Run 2 (2015--2018).
The research will continue in LHC Run 3 that will start at around 2022 with
$\sqrt{s}$ = 14 TeV, followed by High-Luminosity LHC upgrade (HL-LHC) which will be
in operation at the end of 2027 (LHC Run4).
The HL-LHC project aims to increase the potential for discoveries of
new physics by cranking up the performance of the LHC.

\vskip.5\baselineskip

% intensity frontier
{\bf Intensity Frontier:}\\
The other approach is the intensity frontier that provides us with indirect probes of new
physics effects.
The Belle II experiment at the SuperKEKB accelerator is a cutting-edge
experiment in the asymmetric $e^+e^-$ collider, which is designed to
make precise measurements of CP violating parameters in the $b$-quark sector ($B$ physics) and find new physics effects.
%The accelerator was achieved a new instantaneous luminosity world
%record, $2.4 \times 10^{34}$ cm$^{-2}$s$^{-1}$ in June 2020.
The main purpose of the LHCb experiment at the LHC is also to explore
new physics via the $B$ physics; LHCb has already delivered many
results and planned to continue running in LHC Run 3 and Run 4 with 
major upgrades.
The $c$-quark sector ($D$ physics) and tau leptons are being investigated by Belle II as well as the BESIII experiment at the BEPCII.
The upgraded versions of the BESIII/BEPCII facility have been proposed.
The $s$-quark sector ($K$ physics) has played an essential role in
the SM through the kaon decay measurements, and has given a large
impact.
The rare kaon decays $K_L \to
\pi^0 \nu \bar{\nu}$ and $K^+ \to \pi^+ \nu \bar{\nu}$ have been studied by the {\bf KOTO}
experiment at J-PARC and the NA62 experiment at CERN-SPS, respectively.
As a future plan at the CERN-SPS, the KLEVER experiment starting at around 2027 has been
considered to measure $K_L \to \pi^0 \nu \bar{\nu}$ with better
sensitivity than the goal of the KOTO experiment.

Toward new physics beyond the SM, many experiments are ongoing and planned with intensity frontier
accelerators.
Long baseline neutrino oscillation experiments aim to reveal CP violation in the
lepton sector and to resolve the neutrino-mass hierarchy problem.
The {\bf T2K} (J-PARC) and NO$\nu$A (Fermilab) experiments are ongoing,
and the {\bf Hyper Kamiokande} and DUNE experiments have been planned as the next
generation projects.
Lepton flavor violation (LFV) is naturally made in SM extensions,
but LFV of
charged leptons (CLFV) has never been observed.
International programs of CLFV search have proceeded and being planned
using muon decays.
The MEG2 experiment at PSI will start soon, aiming at a sensitivity
improvement of one order of magnitude compared to the predecessor
experiment MEG.
The {\bf COMET} and Mu2e experiments are also in preparation at J-PARC
and Fermilab, respectively, and the mu3e experiment is planned after
MEG2 at PSI.
There are {\bf muon $\mathbf{g-2}$} measurements both in Fermilab and J-PARC.
The anomalous magnetic dipole moment of muon will provide us with one of the
sensitive tests of the SM.
The $g-2$ experiment at Fermilab recently reported their first result and showed a difference from theories at a significance of 4.2 $\sigma$
by combining with the results from the previous experiment at BNL~\cite{Muong-2:2021ojo}.
The J-PARC $g-2$ experiment will be constructed and start soon.

\vskip.5\baselineskip

% ILC
{\bf International Linear Collider (ILC):}\\
The International Linear Collider (ILC), a next-generation experimental facility aiming at the
discovery of new physics beyond the SM through precise Higgs measurements,
is being proposed by the
international community of high energy physics.
Electron and positron beams will collide with a center-of-mass
energy of 250 GeV in the current baseline design.
The primary goal of the ILC is to produce a large number of Higgs boson
particles, as the Higgs factory.
The ILC construction is expected to start in 2020s, and the physics
experiments would start in 2030s.

% ### NP ###
\subsubsection{Nuclear physics}
The ultimate goal of nuclear physics is to reveal 
formation and evolution of the matter widely ranging from
hadrons as complex systems of quarks and gluons to neutron stars
described as ``giant nuclei''.
The whole of their aspects will be described in
quantum chromodynamics (QCD).
For this purpose, various approaches are adopted at a wide variety of
accelerator facilities.

\vskip.5\baselineskip

% proton accelerator, J-PARC
{\bf Proton Accelerator Facility:}\\
At high-intensity proton accelerator facilities, various secondary meson
beams such as pion and kaon as well as a primary proton beam are utilized
to investigate hadron structure and hadron-hadron interactions.
The {\bf Hadron Experimental Facility} of J-PARC is now the world's leading
facility in this field.
Nuclear physics programs at the facility have focused on strangeness
physics extended to double strangeness ($S=-2$) systems, which is
essential to understand the equation of state (EOS) of nuclear matter with strangeness.
The kaon-nucleon interaction has also been investigated by utilizing
high-intensity kaon beams, and an experimental study of spectral change
of vector mesons in nuclei has started in 2020 with a 30 GeV primary
proton beam.
%Details of the experimental programs are as described in Sec.~1.
An upgrade of the MR main-magnet power supplies being conducted
in 2021-2022
will realize delivery of more intense beams to the Hadron Experimental
Facility by over 100 kW operation of the MR-SX.
%At the CERN-SPS, the COMPASS experiment is continuing to study the
%structure of the proton and more specifically of its spin with high
%intensity muon and hadron beams.

\vskip.5\baselineskip

% HI and anti-proton facility, FAIR
{\bf Heavy-Ion and Anti-Proton Accelerator Facility:}\\
One of the biggest accelerator complex projects comparable with J-PARC is
the International Anti-proton Heavy Ion Research Facility (FAIR) led by
the Institute for Heavy Ion Research in Germany (GSI).
The research at FAIR will cover a wide range of topics from nuclear and
hadron physics to applications in condensed matter physics and biology.
Experimental programs dedicated to nuclear and hadron physics
consists of the NUSTAR, CBM, and PANDA experiments.
The NUSTAR project aims at exploration of nuclei
with large neutron or proton excess by using intense radioactive beams
with energies between 0.5 GeV/u to 1.5 GeV/u employing the Superconducting FRagment Separator (Super-FRS).
The goal of the CBM experiment is to explore the QCD phase diagram in
the region of high baryon densities using high-energy nucleus-nucleus
collisions with a beam energy range up to 11 GeV/u for the heaviest
nuclei.
The PANDA experiment is aiming to reveal the dynamics of quarks and gluons that exhibits non-perturbative behaviors.
HESR is a storage ring 
for antiprotons from 1.5 GeV/$c$ to 15
GeV/$c$ generated by a proton beam from the SIS100 accelerator.
The main topics are the production of hadrons, including exotic ones,
and the elucidation of their internal structure.
Another type of the important topics are
the hadron formation and hadron mass spectra in nuclei
for investigating the partial restoration of chiral symmetry in 
nuclear matter.
Besides, hypernucleus generation and its spectroscopic studies are
planned.
The beam will be accelerated and delivered to the experimental halls
after 2027.

% RIBF, FRIB, HIAF, J-PARC HI

Similar facilities focusing on expanding the chart of the nuclides
using intense ion beams are the Radioactive Isotope Beam Factory (RIBF) at
RIKEN, Japan, and the Facility for Rare Isotopes Beams (FRIB) at
Michigan State University in the United States.
The RIBF started operation in 2007. The final-stage accelerator, the
Superconducting Ring Cyclotron (SRC), is the largest and most powerful
cyclotron in the world at present.
The FRIB, the next-generation accelerator for conducting rare isotope
experiments, is currently under construction with completion scheduled
for 2022.
The High Intensity heavy ion Accelerator Facility (HIAF) in China has
also been constructed and its completion is expected in 2024.
The primary goal of the HIAF is the same as those of the RIBF and FRIB.
With the intense heavy ion beams with energies up to a few GeV/u
available at HIAF, experimental programs aiming at exploration of the
QCD phase diagram of nuclear matter as well as investigation of
hypernuclear production in heavy ion collision are planned.

At J-PARC, a future program with acceleration of heavy-ions
({\bf J-PARC HI}) has been discussed.
In this program, a heavy-ion injector consisting of a linac and a
booster ring will be newly constructed, and the two existing
synchrotrons 3 GeV RCS (Rapid-Cycling Synchrotron)~\cite{Hotchi:2012vma} and MR 
will be used to accelerate heavy-ions.
High-intensity heavy-ion beams up to uranium (U) with the energies of 1-19
GeV/u will be extracted to the extended Hadron Experimental Facility
with the world's highest beam rate of $10^{11}$ Hz.
%The major physics objectives of the J-PARC HI project are to determine
%the QCD phase structures the equation of state of dense nuclear matter.

\vskip.5\baselineskip

% HI collider, LHC, RHIC
{\bf Heavy-Ion Collider:}\\
Heavy-ion colliders enable us to access higher temperature
regions of the QCD phase diagram 
where a quark-gluon plasma (QGP) phase appears, which is a novel state of
matter wherein quarks and gluons are no longer confined in hadrons.
At the LHC, the ALICE experiment is optimized to study the collisions of
nuclei at the ultra-relativistic energies.
During the LHC Run 1 (2009--2013) and Run 2 (2015--2018), the ALICE
recorded Pb-Pb collisions at 2.76 TeV and 5 TeV, respectively, as well
as comparison data in $p$-$p$ and $p$-Pb collisions.
To extend its physics program in the future runs of Run 3 and 4, 
detector upgrades are ongoing.
For comprehensive exploration of the high baryon density region of the
QCD phase diagram with precision measurements, the STAR experiment at
the Relativistic Heavy Ion Collider (RHIC) at Brookhaven National
Laboratory (BNL) has preformed the Beam Energy Scan (BES) program with
Au-Au collisions.
The program allows precision measurements of the
intermediate-to-high baryon chemical potential $\mu_B$ region of the QCD
phase diagram, by covering a wide center-of-mass energy region from
$\sqrt{s_{NN}}$ = 3 to 19.6 GeV.
Recently, new and precise results on hadron-hadron femtoscopic
measurements have been reported in high energy $p$-$p$, $p$-$A$, and $A$-$A$
collisions vigorously from both the ALICE and STAR experiments; such as
the $\Lambda$-$\Lambda$ and $\Omega$-$p$ interactions above their
thresholds have been extracted via comparison with various theoretical
calculations.
The new experimental program at RHIC, sPHENIX, has been in
preparation to obtain detailed properties of the QGP matter 
through the new experimental approach of jet correlations and $\Upsilon$s
measurements.

\vskip.5\baselineskip

% electron accelerator, JLab
{\bf Electron Accelerator Facility:}\\
High-energy electron accelerators allow us to extract information on the
quark and gluon structure of nucleons.
The 12 GeV upgrade project of the Continuous Electron Beam
Accelerator Facility (CEBAF) at the Jefferson National Accelerator
Laboratory (JLab) has been completed.
This upgrade of the facility opens a new scientific project on the JLab
physics programs, {\it i.e.}, the further elucidation of the nucleon
structure and photo-production of hadrons.
In the upgrade, extensive improvements to the existing experimental
equipment at Hall A, Hall B, and Hall C have been done, and the new
fourth experimental hall, Hall D, has been constructed.
Investigations of the electromagnetic form factors at large momentum
transfer, direct measurement of nucleon-nucleon short-range correlation,
and hypernuclear physics programs with the $(e,e'K^+)$ reaction are
planned at Hall A.
At Hall B, mapping of the nucleon's 3-dimensional structure (nuclear
femtography) will be conducted through measurements of generalized
parton distributions and transverse momentum distributions, using the
CLAS12 large acceptance spectrometer system.
The planned experimental programs at Hall C focus on precision
determination of the nucleon and nuclear structure function.
One of the main goals at Hall D is to explore the origin of quark
confinement by an experimental search for exotic hadrons such as
glueballs and hybrid mesons with the new experimental apparatus GlueX.
%\textcolor{red}{[Mainz,SPring-8]}
%The Mainz Microtron (MAMI) in Germany also provides high intensity
%electron beam with an energy up to 1.6 GeV.

\vskip.5\baselineskip

% EIC
{\bf Electron-Ion Collider:}\\
For understanding of inner structure of protons and nuclei at
very high precision, Electron Ion Collider (EIC), a future
electron-proton and electron-ion collider, will be constructed at BNL
in this decade.
The main goals of the EIC program are providing tomographic 3D images of 
quark/gluon distribution in protons and nuclei, solving the proton
spin puzzle, and searching for gluon saturation and the color glass
condensate.
The operation of the EIC is expected to start in 2030.

\vskip.5\baselineskip

% other collider experiments
% LHC, B-factrory, C-factory
{\bf Proton-Proton and Electron-Positron Colliders:}\\
The hadron spectroscopy at the collider experiments have
played an important role in hadron physics.
At the LHC, a huge number of hadrons are produced and $\sim$ 60 new
hadrons have been discovered by the ATLAS, CMS and LHCb experimentss
in 11 years of LHC operation.
As represented by the observation of pentaquark particles in the
charm-sector ($P_c$) at the LHCb experiment, new exotic hadron
searches will continue in LHC Run 3 and 4.
The $e^+e^-$ collider experiments of
the SLAC National Accelerator, PEP-II/BaBar, and the KEKB/Belle have
discovered many excited states of hadrons including some exotic hadron
candidates of tetraquark, such as $X Y Z$ particles.
In the Beijing BES III experiment, many interesting results have been
obtained on not only exotic hadrons but also more exotic states, such as
the antiproton-proton bound states.
The Belle II experiment has started to take data.
New hadrons including exotic ones are expected to be discovered.

% ### summary ###
\subsubsection{Position of the J-PARC Hadron Experimental Facility}
J-PARC is a multi-purpose accelerator facility that is
unique in a variety of secondary-particle beams utilized for a wide
range of scientific programs.
In J-PARC, the Hadron Experimental Facility is
a unique facility in the world that enables us 
to conduct
particle and nuclear physics programs with the highest-intensity meson
(pion and kaon) beams in the several GeV energy region.
With this advantage, experiments will be performed,
focusing on (1) comprehensive investigation of hypernuclei extended to
$S=-2$ systems aiming at revelation of the strangeness matter EOS,
(2) hadron physics for the exploration of the low-energy QCD through 
the meson properties in nuclei, and (3) the search for the rare neutral-kaon decay to explore new physics beyond the Standard Model.
The increasing beam power after the MR power-supply upgrade in
2021-2022 will push forward with these research programs that can be
done nowhere else.

In such a situation, the extension project of the Hadron
Experimental Facility has great possibilities to open new
high-intensity frontier in particle and nuclear physics.
The new experimental programs at the extend facility are described in the
following sections, all of which cannot be implemented at the existing
facilities.
In particular, the systematic high-precision $(\pi,K^+)$ spectroscopy of
$\Lambda$-hypernuclei at the HIHR will clarify density dependence of the
$\Lambda N$ interaction in medium with unprecedented precision.
The measurements will give us a big stepping stone toward the elucidation of
neutron star matter in microscopic approach.
%The strange and charm baryon spectroscopies at $\pi20$ and K10 will
%reveal the diquark correlation in baryons with unique
%measurements that can be done nowhere else.
%The measurements will derive the quark-gluon dynamics inside.
Spectroscopy of strange baryons at K10
as well as that of charm baryons at $\pi 20$
provides crucial ingredients on the structure of hadrons as composite systems 
of quarks and gluons.
The diquark correlation in baryons will be revealed by unique
measurements that can be done nowhere else.
Experimental studies of the rare decay $K_L \to \pi^0 \nu \bar{\nu}$
with the highest intensity $K_L$ beam available at KL2 will continue to
lead flavor physics around the world.

\begin{figure}[htbp]
 \centerline{\includegraphics[width=1\textwidth]{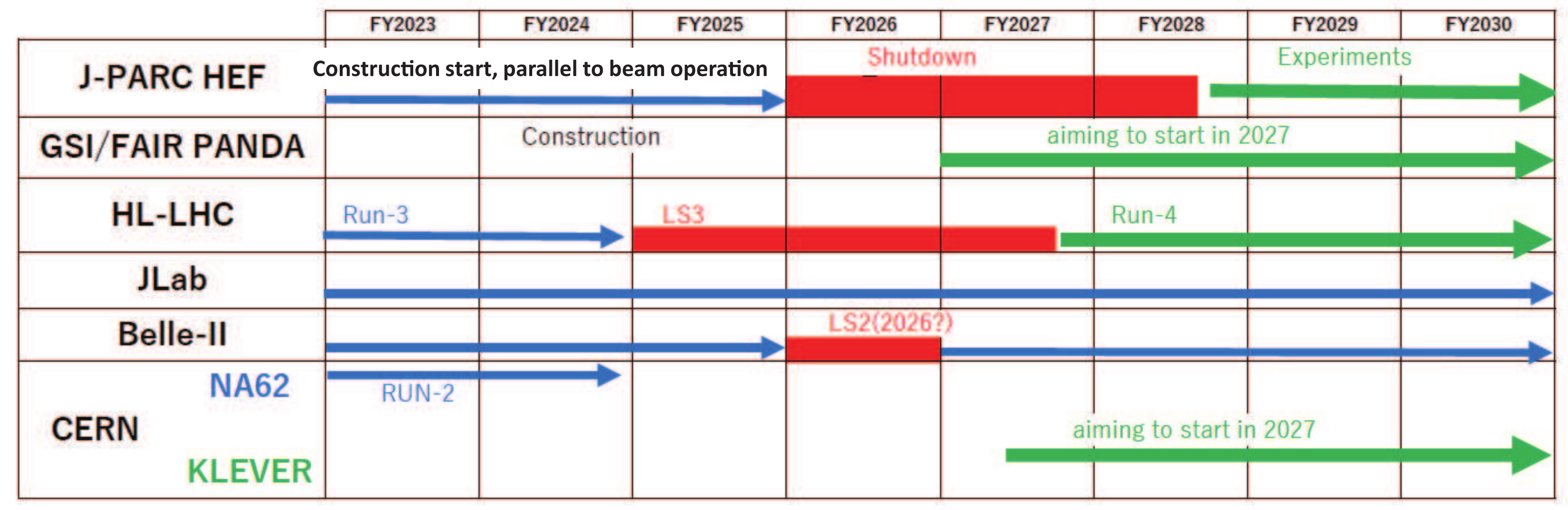}}
 \caption{
 Timeline of facilities in the world related to the Hadron
 Experimental Facility.
 }
 \label{fig:urgency}
\end{figure}

Contributions to particle and nuclear physics from the Hadron
Experimental Facility in J-PARC are essential, together with
other world’s frontier facilities, to the development of science.
As summarized in Fig.~\ref{fig:urgency}, the PANDA, HL-LHC, and KLEVER
experiments - potential competitors of experiments at J-PARC -
will start their operations around after 2027.
To maintain leading position of J-PARC in the field of particle and
nuclear physics utilizing several GeV secondary beams, therefore, early
realization of the extension project is essential.

% ### Competitiveness ###
\subsubsection{Global competitiveness of the extension project}
There are growing efforts worldwide in particle and nuclear physics,
and many projects are on going and planned.
Global competitiveness of the extension project is summarized
in each project of the beam lines.

\vskip.5\baselineskip

{\bf Hypernuclei at the HIHR Beam Line:}\\
{\bf JLab} has a campaign of hypernuclear physics programs to investigate
isospin and mass dependence of 
the $\Lambda$ binding energies in the 
$\Lambda$-hypernuclei with electron beams.
The energy resolution of the $(e,e'K^+)$ reaction spectroscopy,
$0.5$–$1.0$ MeV (FWHM)~\cite{TAN14}, is comparable to that in HIHR.
However, the $(e,e'K^+)$ reaction converts a proton to a $\Lambda$
while the $(\pi^+,K^+)$ reaction converts a neutron to a $\Lambda$, and
thus produced hypernuclei are different in nuclear species.
Thus, comparison between hypernuclei produced by
the $(\pi^+,K^+)$ and $(e,e'K^+)$ reactions 
from the same target will give
direct information on charge symmetry breaking.
So far, the resolution of $(\pi^+,K^+)$ experiments does not match
to that achieved by the $(e,e'K^+)$ reaction, and detailed comparison was impossible.
The JLab hypernuclear program could be a strong competitor as well as a
strong collaborator.
It should be noted that there are lots of backlog of accepted experiments
at JLab and installation/decommission of major spectrometers take a
long time (from a few months to a half year).
Frequent beam time assignment is difficult at JLab.

{\bf MAMI} has an electron microtron which can produce strangeness
and similar hypernuclear programs to JLab can be conducted.
However, it is unlikely that $(e,e'K^+)$ reaction spectroscopy 
experiments will be carried out on a large scale, 
due to limitation of the existing kaon spectrometer.
Decay pion spectroscopy of electro-produced $\Lambda$ hypernuclei 
was successfully carried out for $^4_\Lambda{\rm H}$, where 
the kaon spectrometer at MAMI served perfectly as a kaon tagger for this experiment~\cite{ESS15}.
Though measurements are limited only to the ground states of light
$\Lambda$ hypernuclei, 
MAMI already established high-resolution spectrometers for
90-130 MeV/$c$ pions in this measurement, and
achieved a remarkable energy resolution of 0.15 MeV (FWHM) for $^4_\Lambda{\rm H}$.
At HIHR, there are experimental plans to perform decay pion 
spectroscopy experiments for heavier ($p$-shell) hypernuclei.
MAMI is a strong competitor on measurement of the ground state mass of
light $\Lambda$ hypernuclei.

{\bf GSI-HypHI, RHIC-STAR {\rm and} LHC-ALICE} have started 
spectroscopic studies of hypernuclei with heavy ion beams.
HI hypernuclear spectroscopy attracts attention because it can produce
highly exotic hypernuclei, such as proton-rich or neutron-rich
hypernuclei that are out of reach by standard spectroscopic techniques.
Future experiments at {\bf GSI-FAIR} are planned also for light hypernuclei.
Although it is a promising and practically only program to access highly
exotic hypernuclei, they cannot be competitor for HIHR in terms of energy
resolution and signal-to-noise ratio.
{\bf FAIR-PANDA} is planning a spectroscopy of multi-strangenss hypernuclei 
with a $\gamma$-ray measurement by using anti-proton beams.
Ultimate physics goal of link between QCD and nuclear physics might be
shared with HIHR, but experimental technologies are totally different
and it cannot be a direct competitor for HIHR.

\vskip.5\baselineskip

{\bf Hyperon-Nucleon Scattering at the K1.1 Beam Line:}\\
Recently, {\bf LHC-ALICE} and {\bf RHIC-STAR} have strongly promoted "femtoscopy" measurements using baryon pairs (and meson-baryon pairs) produced in pp and heavier ion collisions \cite{STAR:2014dcy,ALICE:2018ysd,Morita:2014kza,ALICE:2019buq}.
In the femtoscopy measurements, information on low-energy hadron-hadron interactions is derived from two-particle correlation 
with small relative momenta.
In particular, the ALICE collaboration has vigorously pushed forward the
measurement with huge statistics collected in LHC-Run2, and will
continue the study in LHC-Run3 and Run4.
These measurements are suitable to deduce information on the $S$-wave interaction such as the scattering lengths and effective ranges 
for the baryon-baryon systems, including
multi-strangeness systems that are difficult 
to be studied in the direct scattering experiments.
In the proposed experiment at J-PARC, measurement of the very low 
energy YN scatterings is rather difficult due to less sensitivity in detection of low energy protons.
On the other hand, we can measure the differential information on cross
sections and spin observables for higher momentum regions 
dominated by $P$- and higher wave.
Direct scattering at higher momentum is also essential to study short range interactions.
Such differential information can be directly connected to phase shift
analysis which has never been performed for YN interactions.
Beside, the experimental condition is well controlled; various beam
momenta can be selected according to the physics purpose, and exclusive
measurement enables us to specify the reaction channel unambiguously 
from the hyperon production to the YN scattering.
A spin-polarized hydrogen target can be also used 
for further studies of spin-dependent interactions, 
while such studies are not possible in the femtoscopy method.
Thus, the scattering and the femtoscopy experiments 
are complementary to each other.
It should also be stressed that 
the differential information is long awaited by theorists because it 
is essential to construct baryon-baryon interaction models.

{\bf CLAS} collaboration at JLab also has a potential to measure YN
scatterings from the re-scattering of hyperons produced by the
photo-induced reaction~\cite{CLAS:2021gur}.
In their analysis, the $\Lambda$ momentum region is higher than 1 GeV/$c$.
Thus, the CLAS experiment is potential competitor of the J-PARC
experiments.

\vskip.5\baselineskip

{\bf Baryon Spectroscopy at the ${\boldsymbol \pi} {\bf 20}$ and K10 Beam Lines:}\\
{\bf LHCb} and {\bf Belle II} have an excellent potential to 
perform baryon spectroscopy in the charm sector.
Both experiments have reconstructed excited charm baryons mainly from
excited bottom baryons and bottom mesons decays, respectively, {\it i.e.},
via invariant mass reconstruction.
This approach is powerful for the new state search as demonstrated
so far.
On the other hand, determination of branching fractions and total cross
sections is not easy by using the invariant mass reconstruction.
In the strangeness sector, high-energy collider experiments can also produce
a wide variety of excited $\Xi$ and $\Omega$ baryons; however, identification
of excited strange baryons is difficult due to huge pion-multiplicity
environment.
At J-PARC, we will use a missing-mass technique to identify excited
states of strange and charm baryons using $(K^-,K^+)$, $(K^-,K^+K^{(*)0})$,
and $(\pi^-,D^{*-})$ reactions.
%Great advantage of these experiments are that we can determine total 
%cross section by missing mass as well as performing exclusive
%measurement with decay particle measurement that enables us to specify
%the reaction channel unambiguously.
Taking an advantage of the technique,
we do not have to detect any daughter particles from the baryon of interest
for determining the production cross section independently of its decay mode.
When we additionally detect a daughter particle and identify the decay mode,
we can highly suppress the background contributions.
Detailed investigation of both production and decay can realize
determination of decay-branching ratios and spin-parity.
It should be noted that direct production of highly-excited states would
be possible by bringing a high angular momentum into the production
reaction.
Thus, these experiments are in a complementary relation
in which totally different methods are utilized.

\vskip.5\baselineskip

{\bf Rare Neutral-Kaon Decay at the KL2 Beam Line:}\\
{\bf CERN-NA62} and its successor project, {\bf KLEVER}, are direct
competitors of the KOTO step-2 experiment.
NA62 has measured the branching ratio of the $K^+ \to \pi^+ \nu
\bar{\nu}$ decay to be $(10.6^{+4.0}_{-3.4} ({\rm stat.}) \pm 0.9 ({\rm syst.}))
\times 10^{-11}$ at 68\% confidence level (CL) using Run 1 data
collected during 2016-18~\cite{NA62:2021zjw}.
The indirect limit on the $K_L \to \pi^0 \nu \bar{\nu}$ decay 
from the NA62 result with 68\% CL is $6.4 \times 10^{-10}$.
NA62 will continue their data taking in Run 2 scheduled for 2021-24.
After completion of NA62, the KLEVER project is planned
to search for the $K_L \to \pi^0 \nu \bar{\nu}$ decay with a 
completely new detector system~\cite{Moulson:2019ifj}.
Although the kaon momentum utilized is different between the KOTO step-2
($\sim 5$ GeV/$c$) and the KLEVER ($\sim 40$ GeV/$c$), the target 
sensitivity of both experiments is about 60 events for the $K_L \to \pi^0 \nu \bar{\nu}$ decay at the SM prediction of the branching ratio.
KLEVER would aim to start data taking in LHC Run 4 (2027-).

\vskip.5\baselineskip

For the past decade, J-PARC has been a major player in hypernuclear
study and rare neutral-kaon decay search.
However, simply maintaining the current activities for the next decade would deteriorate the leading position of J-PARC.
Early realization of the extension project and pursuing new
possibilities are essentially important to keep our international
competitiveness.
% flatex input end: [summary.tex]

%==================================================================%

\printbibliography[segment=\therefsegment,heading=subbibliography]

\clearpage

%==================================================================%
\section{\centering Physics Programs at HIHR and K1.1 Beam Lines}\label{sec:HIHR}
\chapterauthor{
H.~Fujioka, T.~Gogami, E.~Hiyama, R.~Honda, Y.~Ichikawa, M.~Isaka,\\
K.~Miwa, T.~Nagae, S.~Nagao, S.~N.~Nakamura, H.~Noumi, T.~Takahashi,\\
H.~Tamura, K.~Tanida, M.~Ukai, T.~O.~Yamamoto, and Y.~Yamamoto\\
}
%==================================================================%

%%%%%% HIHR/K1.1 Motivation  (Tamura)
% flatex input: [HIHRK11-motivation.tex]

\subsection{The Main Physics Motivation: Elucidating Neutron Stars Microscopically Through Solving "Hyperon Puzzle"}

\subsubsection{Nuclear physics and neutron stars}

One of the main goals of nuclear physics is to elucidate the origin and the evolution of all the matter in the universe, by describing a variety of hadrons, atomic nuclei, and neutron stars starting from the common theoretical basis of QCD.

Nuclear physicists have long made efforts to understand properties and structure of various nuclei, starting from structureless nucleons and (semi-)phenomenological nuclear force without considering the inner structure of nucleons and the origins of nuclear force. 
Rather recently, studies have started to clarify the connection between quarks and gluons, hadrons, and nuclei based on QCD.
In this situation, understanding of matter inside neutron stars is an ultimate problem to test the validity and applicability of the current nuclear physics theories. Since experimentally-available nuclear data are limited to those around the nuclear saturation density ($\rho_0$), (semi-)phenomenological nuclear theories cannot be applied to nuclear matter with higher densities than $\rho_0$. In order to describe the high density matter in neutron stars, deeper understanding of nucleons (baryons) and nuclear force (baryon-baryon forces) are indispensable,
and then complicated many-body problems should be appropriately solved.
Astronomical data on neutron stars such as mass, radius, cooling speed, etc., plus information added by gravitational wave from neutrons star mergers, need to be reproduced microscopically from nuclear physics.

\begin{figure}
\begin{center}
\includegraphics[width=10cm]{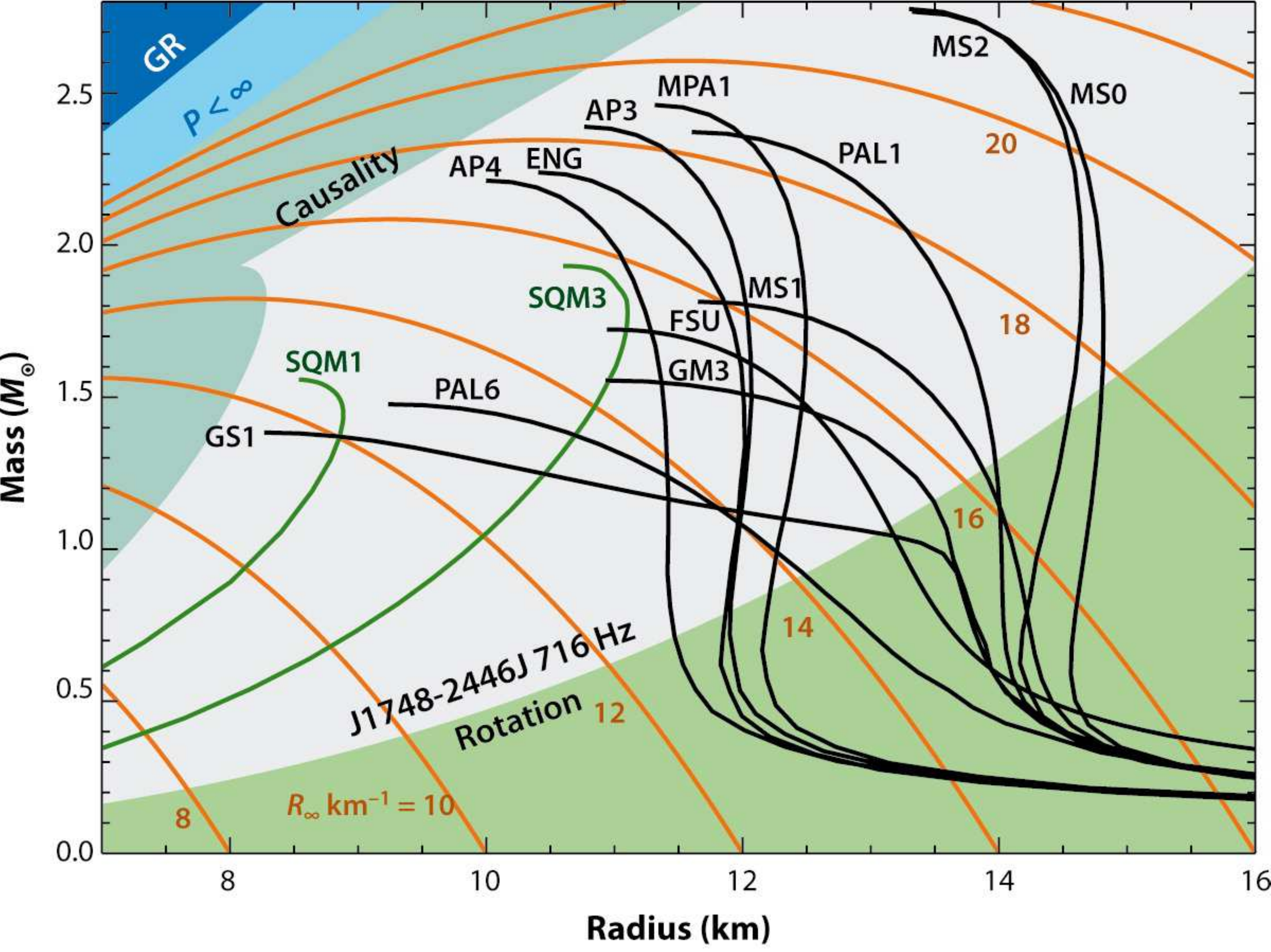}
\caption{Neutron star mass and radius relation (MR relation) for various EOS's \protect\cite{LAT12}.
}
\label{MR}
\end{center}
\end{figure}

The present status is far from satisfactory. 
Once one takes a certain nucleon-nucleon ($N$$N$) (and baryon-baryon ($B$$B$)) interaction model and a theoretical framework for calculating many-body nucleonic (or baryonic) systems, the Equation of State (EOS), the energy (or the pressure) per nucleon (baryon) as a function of the baryon density (or the baryon chemical potential) is determined, and the mass-radius relationship of neutron stars is uniquely derived from hydrostatic pressure balance inside a neutron star.
As shown in Fig.~\ref{MR}, however, predictions of mass-radius relationship widely diverge depending on variety of nuclear force (baryon-baryon interaction) models giving different EOS's. 
In astronomy, masses of many neutron stars have been measured, precisely in some cases, but their radii have not been measured reliably.
Currently, however, reliable measurement of neutron star radii is being made by an X-ray detector NICER installed at International Space Station. The NICER team has reported that the neutron star PSR J0030+0451 has a radius of $13.02^{+1.24}_{-1.06}$ km and a mass of $1.44^{+0.15}_{-0.14} M_{\odot}$ \cite{MIL19}. 
Such precise observational data of neutron star radii strictly constrain EOS. 
Recently, a gravitational wave from a neutron star merger
was detected for the first time \cite{ABB17}. 
The tidal deformability obatined from the gravitational wave data also constrains the neutron star radius \cite{ABB18}.
In this situation, nuclear physicists are required to develop a ``new generation of nuclear physics'' that can be quantitatively applied to high density matter in neutron stars.

\subsubsection{Hyperon puzzle of neutron stars}

\begin{figure}
\begin{center}
\includegraphics[width=0.9\textwidth]{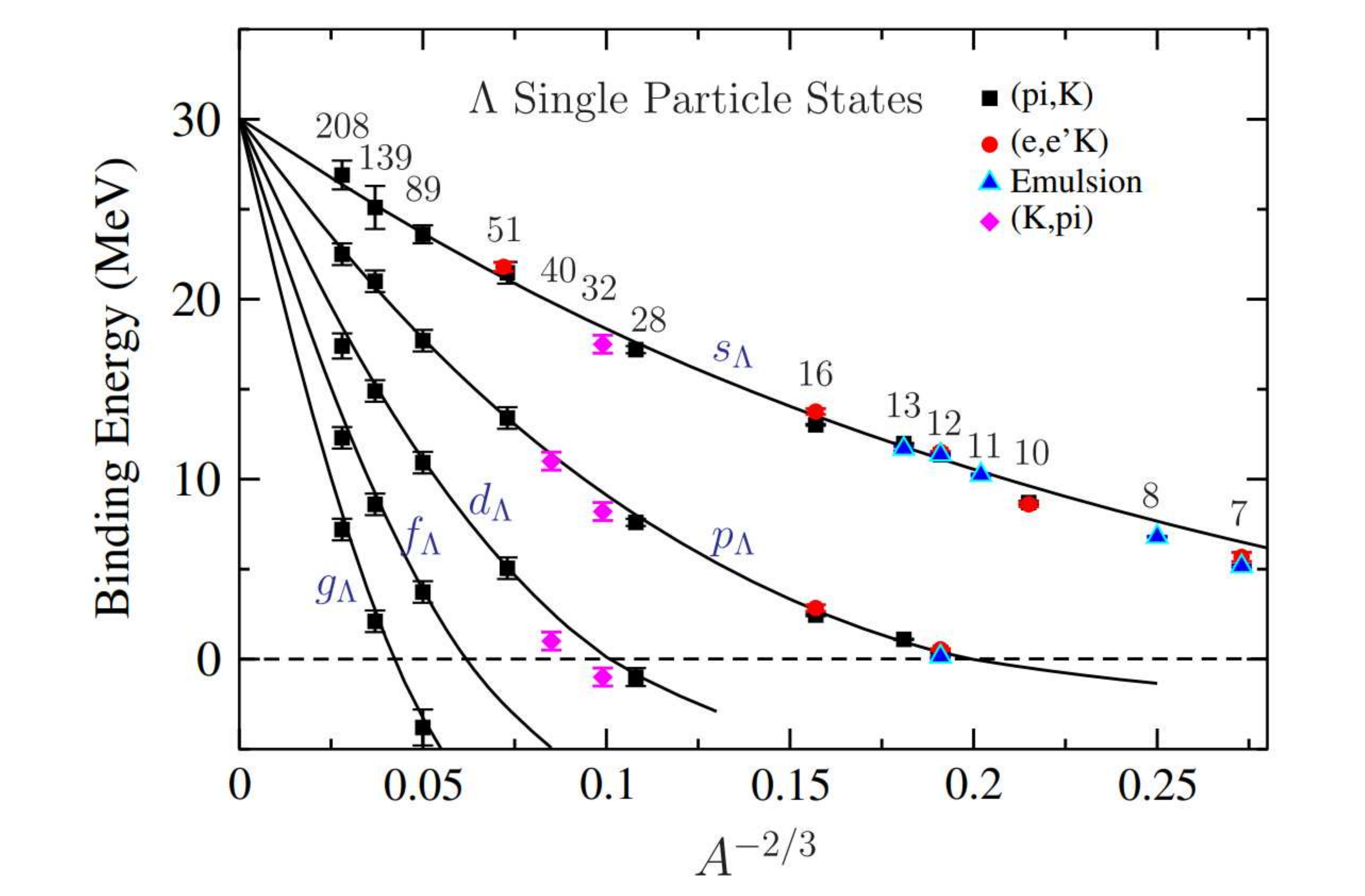}
\caption{Binding energies of $\Lambda$'s single-particle orbits, plotted as a function of $A^{-2/3}$ ($A$ is the mass number), measured via various hypernuclear experiments \protect\cite{GAL16}. The data for
heavy ($A>40$) hypernuclei shown in black points come from the $(\pi^+,K^+)$ spectroscopy experiment at KEK-PS with a limited accuracy \protect\cite{HAS96,HAS06}.} 
\label{Millener}
\end{center}
\end{figure}

\begin{figure}
\begin{center}
\includegraphics[width=12cm,height=8cm]{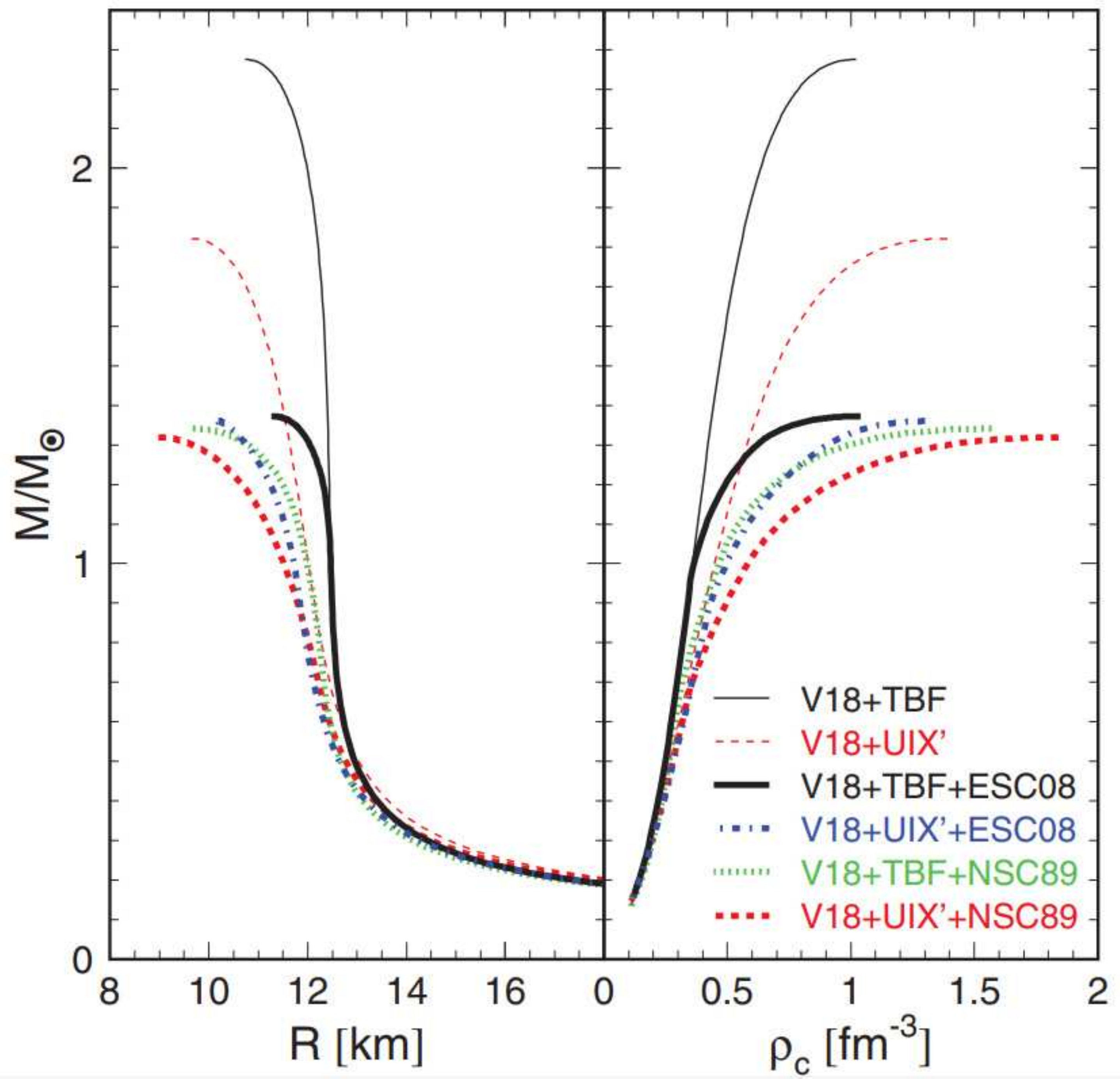}
\caption{Neutron star mass and radius(left)/central density(right) relations calculated for various EOS's from the V18 $N$$N$ interaction plus the $N$$N$$N$ three-body force (TBF or UIX'), without and with $Y$$N$ interactions from Nijmegen ESC08 or NSC89 models \protect\cite{SCH11}.
}
\label{SCHULZE}
\end{center}
\end{figure}

From observational data, the density at the center of neutron stars is expected to be larger than $\sim 3\rho_0$. 
In such a condition hyperons are naively expected to appear because the Fermi energy of neutrons exceed the mass difference between a hyperon and a neutron , and consequently a part of the neutrons are converted to hyperons via weak interaction. In the Fermi gas model which ignores the $B$$B$ interactions, $\Lambda$ hyperons appear at $\rho = \frac{1}{3\pi^2}(m_\Lambda^2 - m_n^2)^{3/2} \sim 6 \rho_0$. 
In calculations considering the attractive potential of a $\Lambda$ hyperon in nuclear matter, of which depth is determined to be 30 MeV from hypernuclear binding energy data (see Fig.\ref{Millener})
via $(\pi^+,K^+)$ reaction experiments at KEK \cite{HAS96,HAS06,GAL16}, $\Lambda$ hyperons appear at densities around $2\rho_0 - 3\rho_0$ \cite{VID00,SCH11}. On the other hand, appearance of hyperons makes the EOS drastically soft and massive neutron stars cannot be supported against gravitational collapse; those calculations show that the maximum mass of neutron stars does not exceed 1.5$M_{\odot}$ as shown in Fig.~\ref{SCHULZE}\cite{DJA10,VID00,SCH11}. 

Since 2010, three reliable observations of heavy neutron stars with masses around 2$M_{\odot}$, J1614-2230 with $(1.97 \pm 0.04) M_{\odot}$ \cite{Demorest:2010bx}, J0348-0432 with $(2.01 \pm 0.04) M_{\odot}$ \cite{Antoniadis:2013pzd}, 
and MSP J0740-6620 with $(2.14^{+0.10}_{-0:09}) M_{\odot}$ \cite{CRO20}, have been reported and given a great impact on nuclear physics. This discrepancy called "hyperon puzzle" reveals that the current nuclear physics has a serious defect.

\subsubsection{Density dependence of $BB$ interactions in medium --three body $BBB$ force}

\begin{figure}
\includegraphics[width=\textwidth]{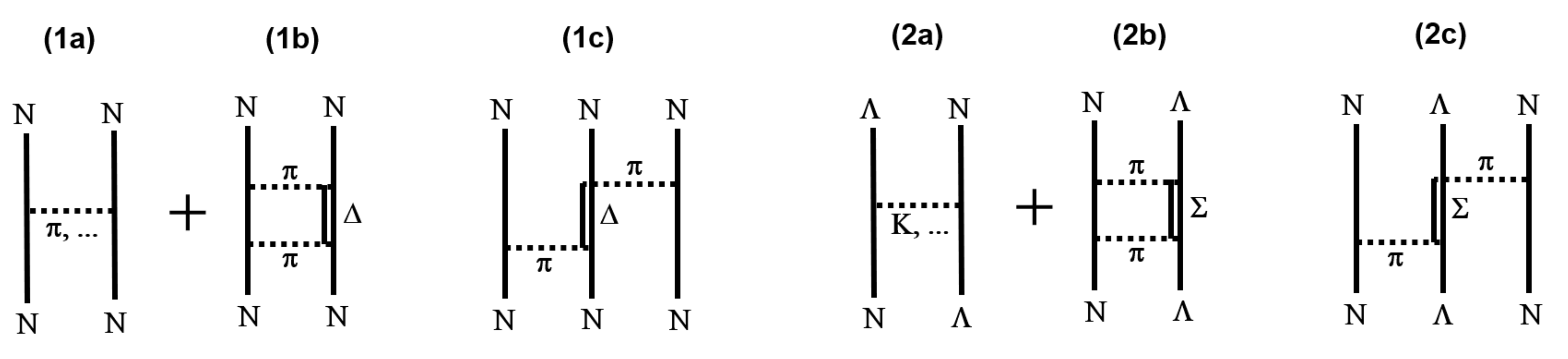}
\caption{Diagrams for $B$$B$ and $B$$B$$B$ interaction with baryon mixing. (1a) $N$$N$ two-body interaction via meson exchange. (1b) $N$$N$ two-body interaction via $\Delta$ excitation. (1c) $N$$N$$N$ three-body interaction induced by $\Delta$ excitation called Fujita-Miyazawa diagram. (2a) $\Lambda$$N$ two-body interaction via one boson exchange. (2b) $\Lambda$$N$ two-body interaction via $\Sigma$ excitation. (2c) $\Lambda$$N$$N$ three-body interaction induced by $\Sigma$ excitation.
}
\label{FM}
\end{figure}

Since a nucleon (baryon) is a compound system made of quarks,
it is likely that three-body and many-body nucleon (baryon) systems cannot be described 
by combinations of the two-body $N$$N$ ($B$$B$) force alone.
To explain such an effect the three-body force (3BF) is introduced.
The existence of the $N$$N$$N$ 3BF is established. 
With the realistic two-body $N$$N$ interaction model (AV18) \cite{WIR95-1,
%WIR95-2
WIR02} alone, 
which was precisely determined from rich $N$$N$ scattering data, 
ab initio calculations significantly underestimate $^3$H and $^3$He binding energies. 
This discrepancy is attributed 
to the two-pion exchange (Fujita-Miyazawa type) 3BF (see Fig.~\ref{FM} (1c)) with attractive nature \cite{FUJ57}.
The existence of the $N$$N$$N$ 3BF was also confirmed later by $d$-$p$ scattering experiments \cite{SEK02}.

\begin{figure}
\begin{center}
\includegraphics[width=0.9\textwidth]{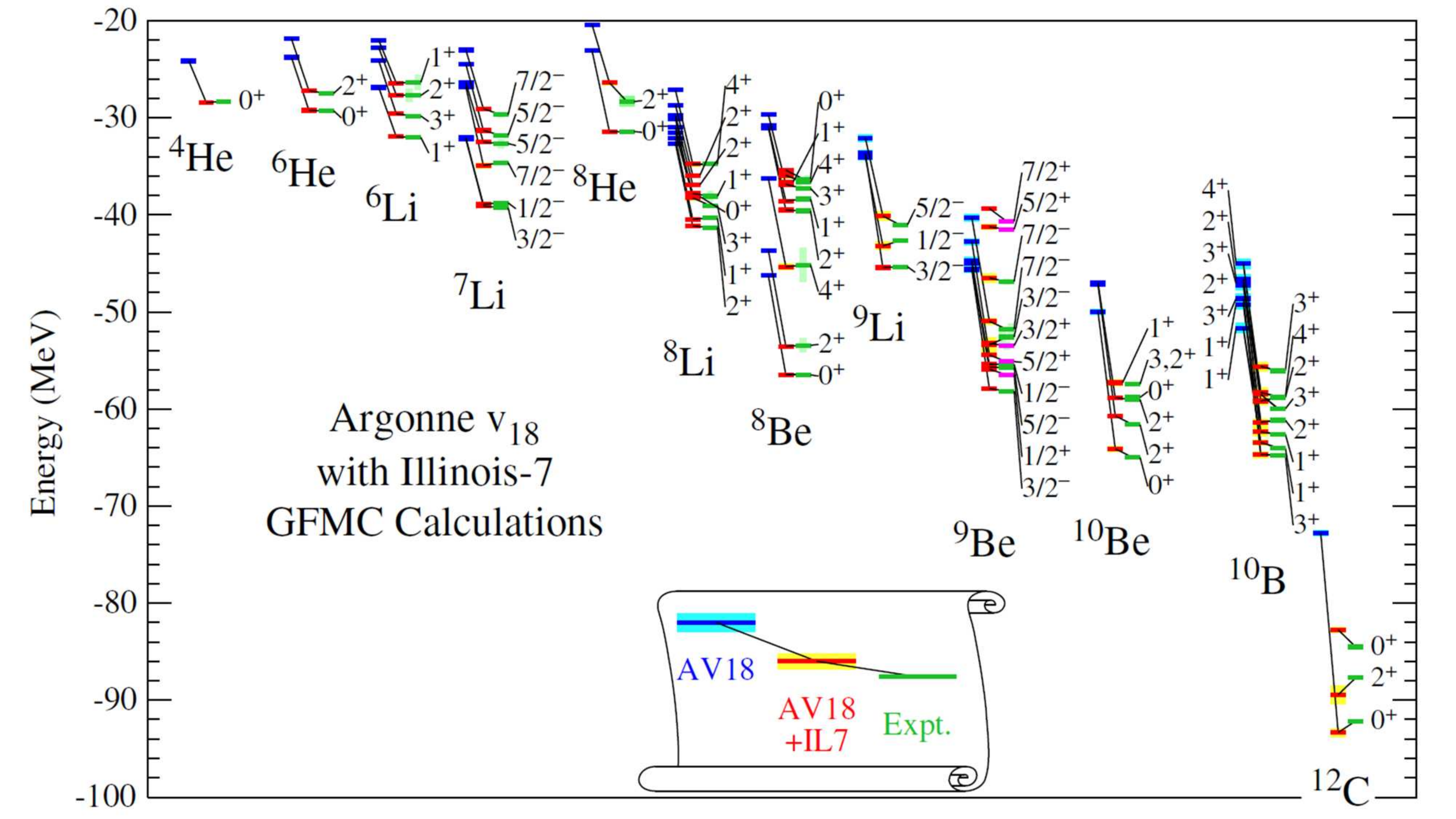}
\caption{Energies of light non-strange nuclear levels calculated from AV18 and from AN18+IL7 interactions via 
Green's function Monte Carlo method \protect\cite{CAR15}.}
\label{Carlson}
\end{center}
\end{figure}

On the other hand,
calculations with the realistic two-body $N$$N$ force (such as AV18) alone
predict the nuclear saturation density considerably higher than the real value \cite{AKM98}, 
indicating the existence of repulsive forces acting at high density around $\rho_0$.
Thus, a phenomenological three-body $NNN$ repulsive force is introduced. 
Such a force gives larger effects inside neutron stars; 
actually, massive neutron stars with $2 M_{\odot}$ can be supported \cite{AKM98} 
by adding the Urbana-type $N$$N$$N$ 3BF (UIX) \cite{PUD95}, 
which contains a phenomenological isospin-independent repulsive 3BF introduced to reproduce the nuclear saturation density, $\rho_0$, in addition to the Fujita-Miyazawa type 3BF.
Via the Green's function Monte Carlo method, the AV18+UIX force was found to reproduce the binding energies of light nuclei (A $\le$ 7) quite well \cite{PUD95,PUD97}. After revision of the 3BF (AV18+IL7), all the binding energies of nuclear levels up to A=12 were precisely reproduced as shown in Fig.~\ref{Carlson} \cite{WIR02,CAR15}.

To solve the hyperon puzzle, theorists proposed various ideas to stiffen the EOS at higher density \cite{OER17}. 
Among them, it is natural to consider repulsive 3BFs with hyperons (for the $Y$$N$$N$, $Y$$Y$$N$, and $Y$$Y$$Y$ channels)
similar to the repulsive $N$$N$$N$ 3BF, although there is currently no experimental evidence for such 3BFs with hyperons.
In addition, the $Y$$N$ two-body force could be significantly modified in dense nuclear matter 
due to $\Lambda$$N$-$\Sigma$$N$ coupling, of which effects (see Fig.~\ref{FM} (2b),(2c)) should be also clarified.
To solve the hyperon puzzle, some theorists conjecture a possible phase transition to deconfined quark matter at high density region inside neutron stars \cite{MAS13,BAY18}. 
To make such discussions more realistic, properties of baryonic matter at high density should be described well.  In other words, an exciting scinario of quark matter would be confirmed after the dense baryonic matter is quantitatively understood.

In nucleonic matter, rich and precise data of the two-body $N$$N$ scattering and a wide variety of nuclei allow us to separate the effects of the 3BF from those of the 2BF. 
The situation has been much different in the case with hyperons because of sparse $Y$$N$ scattering data caused by experimental difficulties. Until now, hypernuclear data have been used to obtain supplemental information on $Y$$N$ and $Y$$Y$ two-body interactions. The binding energy data ($B_\Lambda$) for $\Lambda$'s single-particle orbits in light to heavy $\Lambda$ hypernuclei as shown in Fig.~\ref{Millener}, and the level scheme data for $s$- and $p$-shell hypernuclei, provided the potential depth of $\Lambda$ in nuclear matter (-30 MeV) and the
strengths of the spin-dependent $\Lambda$$N$ interactions \cite{HAS06,GAL16}. Then, they have been then used to test and improve the $B$$B$ interaction models. Limited but important information has been also obtained for $\Sigma$$N$, $\Xi$$N$, and $\Lambda$$\Lambda$ interactions from hypernuclear data \cite{GAL16}. 

In order to investigate the $Y$$N$ interactions in dense nuclear matter
and to extract the $Y$$N$$N$ 3BF effects,
two-body $Y$$N$ interactions in free space need to be precisely determined without employing hypernuclear data. $\Lambda$$p$ scattering experiments with high statistics are necessary for a wide range of momenta up to $\sim$1.2 GeV/c, including differential cross sections and spin observables. In parallel, quality and variety of $\Lambda$ hypernuclear data should be also much improved.
Precise data on $\Lambda$ binding energies and hypernuclear structure for light to heavy $\Lambda$ hypernuclei including neutron-rich ones should be theoretically analyzed based on 
the two-body $Y$$N$ interactions. It enables us to extract the strength and properties of the $Y$$N$$N$ 3BF, which will be applied to update the EOS to solve the hyperon puzzle.

The main motivation of the HIHR/K1.1 beam lines at the extended Hadron Facility in J-PARC is to promote those experimental 
for the purpose of studies and to solve the hyperon puzzle.

\subsubsection{Present status and prospects of theoretical studies}

Recent theoretical development on the baryon-baryon interactions 
is closely related to the HIHR/K1.1 project.

The one-pion exchange potential from Yukawa's theory successfully reproduced 
the experimental nuclear ($N$$N$) force at a long range ($>2$ fm). 
The shorter range part of the nuclear force potential was phenomonologically 
made based on the rich $N$$N$ scattering data (in the Argonne models),
or constructed by extending the pion exchange picture 
to heavier pseudoscaler, vector and scalar mesons (in the one-boson exchamge models).
%The Argonne potentials are the most phenomenological and precise model which are widely
%used for various nuclear structure calculations.
%The parameters on the size of the short-range core and the strength of each potential term
%are determined by fitting rich $N$$N$ scattering data.
%In the AV19 potential, over 3000 experimental data points of $N$$N$ scattering 
%are well fitted with a reduced chi-square of 1.09 with 40 parameters. 
%This potential is called a "realistic model".
%This type of model is not available for YN and YY interactions due to lack of scattering data. 
In the one-boson exchange models,
parameters such as meson-baryon coupling constants and short-range (high-momentum) cutoff parameters
are determined by fitting the scattering data.
In 1990s, several models (Argonne AV18 \cite{WIR95-1,
%WIR95-2
WIR02}, Nijmegen I, II \cite{STO94}, and CD Bonn \cite{MAC01} models) 
succeeded in accurately describing all the $N$$N$ scattering data quite well with a limited number of parameters. 
Those models called ``realistic nuclear force''
triggered development of ab initio calculations of nuclear many-body systems starting
from the two-body $N$$N$ interaction.
It is to be noted that the findings of the attractive and repulsive components 
in the $N$$N$$N$ 3BF were achieved by virtue of the realistic $N$$N$ interaction model
as well as the progress of ab initio and other reliable theoretical methods for many-body systems.

~\\
{\bf  Approaches from one-boson exchange models }
~\\

The one-boson exchange models
were extended to the $B$$B$ interactions with hyperons ($B$ stands for octet baryons), 
where the coupling constants are assumed to follow the flavor SU(3) symmetry. 
by the Nijmegen group and the Bonn-Juelich group.
Here the parameters can be determined by fitting to sparse $Y$$N$ scattering data together with rich $N$$N$ data. 
However, since they are not well constrained by insufficient $Y$$N$ scattering data, 
hypernuclear data were also refered to in order to improve the Nijmegen models 
via the Brueckner-Hartree-Fock theory (G-matrix method) which connects the two-body interaction in the free space 
to the effective interaction in nuclear medium.
As a result, a recent Nijmegen models (ESC16) provides $\Lambda$, $\Sigma$, and $\Xi$ potential depths 
in nuclear matter almost consistent with experimental data \cite{NAG19}. 
In addition these ESC potentials reproduce most of the available $\Lambda$ hypernuclear binding energies over a wide mass number range from $^3_\Lambda$H to $^{208}_\Lambda$Pb in 1--2 MeV accuracy, and also 
the excitation energies of the $\Lambda$ hypernuclear excited states within a few hundreds keV \cite{YAM14}. 
However, calculations using the two-body forces from the Nijmegen interaction thus determined 
cannot support neutron stars heavier than $\sim$1.5$M_{\odot}$ \cite{YAM13}.

\begin{figure}
\begin{center}
\includegraphics[width=0.48\textwidth]{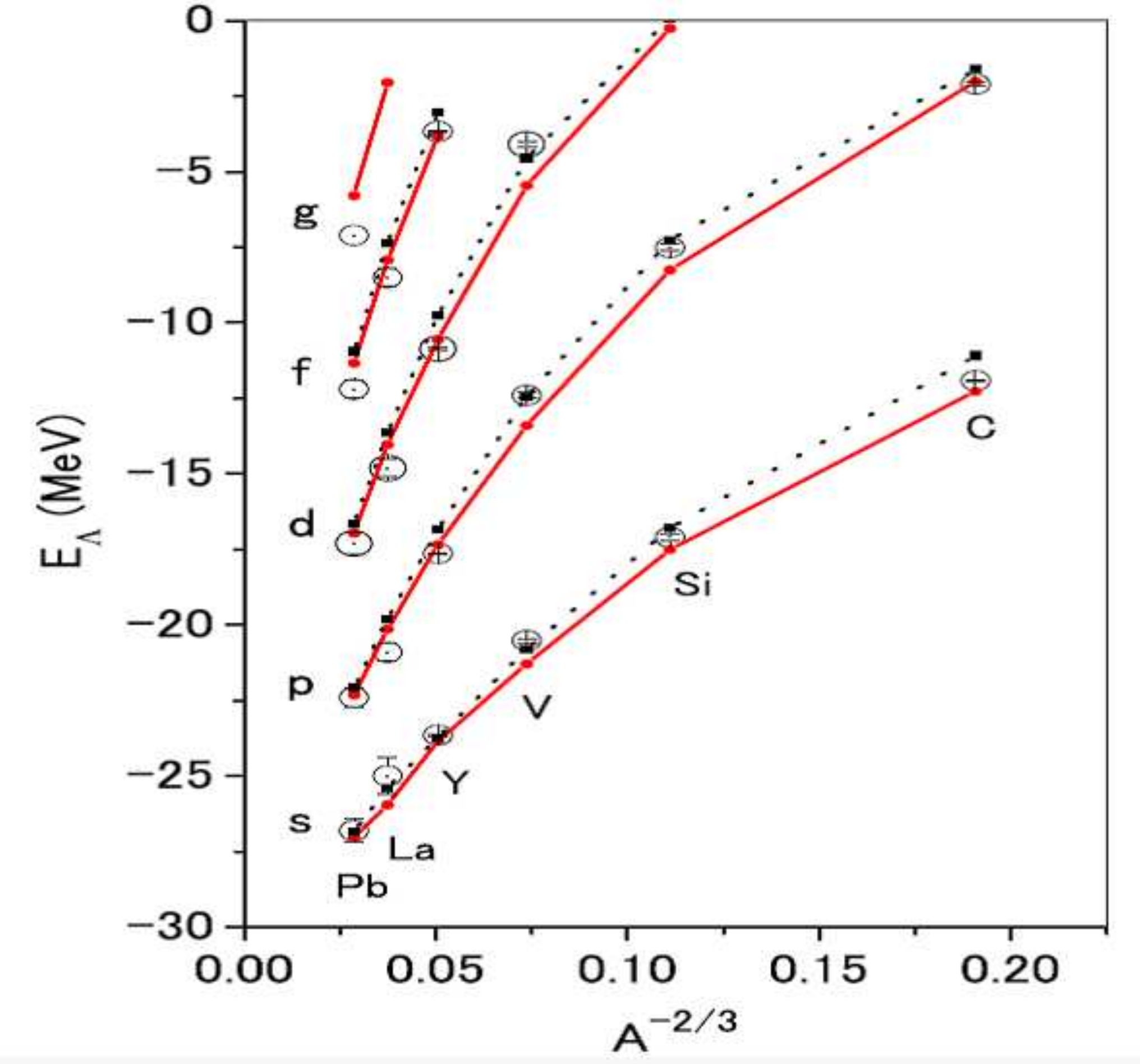}
\includegraphics[width=0.51\textwidth]{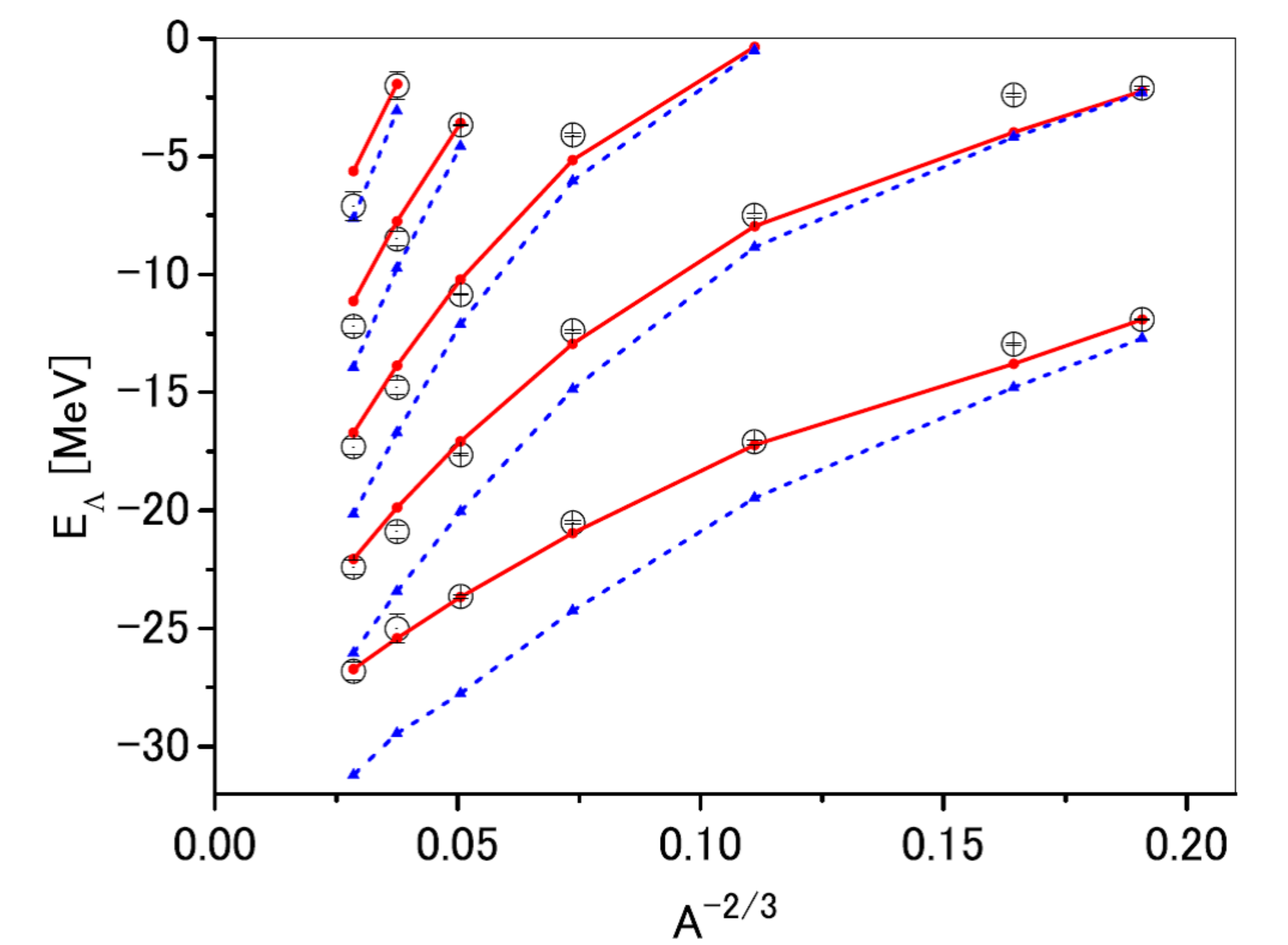}
\caption{Left: A(target mass number) dependence of the $B_\Lambda$ value for light to heavy hypernuclei calculated from Nijmegen ESC08c model without (dotted line) and with (red line) a universal $B$$B$$B$ 3BF \protect{\cite{YAM14}}.
Right: Same as the Left figure but calculated from Nijmegen ESC16 model without (dotted line) and with (red line) a universal $B$$B$$B$ 3BF \protect\cite{NAG19}.
The slope of the line is smaller With the 3BF (red lines) than without 3BF (dotted lines) due to larger repulsive effects of the 3BF for larger A hypernuclei.  Open circles show data points from the previous ($\pi^+.K^+$) experiment at KEK-PS \protect\cite{HAS96}.
}
\label{YAM_BL}
\end{center}
\end{figure}

Recently, Yamamoto et al. introduced a repulsive 3BF in the $N$$N$$N$ channel so as to reproduce the $^{16}$O-$^{16}$O scattering cross section data at 70 MeV/A which is sensitive to the EOS of dense matter with $\sim 2\rho_0$, and then
assumed that the same strength of the 3BF exists universally in all the $B$$B$$B$ channels \cite{YAM14}.
The calculated EOS with the ESC08c interaction plus the universal 3BF predicted 
the maximum mass of neutron stars larger than 2.0 $M_{\odot}$.
It also predicts slightly weaker A-dependence of the $B_\Lambda$ values for the single-particle $\Lambda$ orbits
in hypernuclei, compared to the prediction without 3BF, as shown in Fig.~\ref{YAM_BL} Left \cite{YAM13,YAM14}.
It is because the repulsive 3BF pushes up the hypernuclear energy more for heavier (denser) hypernuclei. 
It suggests that the hypernuclear binding energy data, if sufficiently precise, contain effects of the 
nuclear density dependence of the $\Lambda$$N$ interaction such as the $Y$$N$$N$ 3BF.
Such studies of density dependence are impossible for the $N$$N$ interaction because single-particle nucleon hole states
are not observed in nuclear excitation spectra except for shallow (low excitation) orbits near the Fermi surface.
This is a striking feature of $\Lambda$ hypernuclei, in which a $\Lambda$ is free from Pauli exclusion principle from
nucleons and behaves as a distinguishable particle in a nucleus. 

It is to be noted that the most recent version of the Nijmegen interaction, ESC16, gives A-dependence of $B_\Lambda$ significantlydifferent from the ESC08c as shown in Fig.~\ref{YAM_BL} Right. 
It is caused by large ambiguities in the two-body $\Lambda$$N$ interaction models available at present,
particularly in the p-wave (and higher waves) $\Lambda$$N$ interactions.

%According to AMD calculations of hypernuclear binding energies from Nijmegen's ESC03 and ESC09** interactions, 
%they predict slightly different A-dependence of the hypernuclear binding energies due to different $p$-wave contribution 
%as shown in Fig.~\ref{ISAKA} \cite{ISA}. It indicates that the $\Lambda$$N$ $p$-wave interaction has to be 
%determined by $\Lambda$$N$ scattering experiments in order to extract effects of the $Y$$N$$N$ 3BF separately 
%from the 2BF. 
%Thus, measurement of spin observables in the $\Lambda$$N$ scattering and high-statistics $\Lambda$$p$ 
%scattering experiments for higher $\Lambda$ momentum ($p_\Lambda=$ 0.8--1.2 GeV/c) are particularly important. 

Therefore, sufficiently-precise $Y$$N$ scattering data have to be measured before studying the 3BF. 
Namely, once the two-body $Y$$N$ interaction is well established,
precise data of $B_\Lambda$ values for a wide range of A will clarify the density dependence 
of $\Lambda$$N$ interaction, the strength of the $Y$$N$$N$ 3BF to be added to the two-body interaction.   

\vskip1.5\baselineskip

\begin{figure}
%\begin{center}
\includegraphics[width=0.5\textwidth]{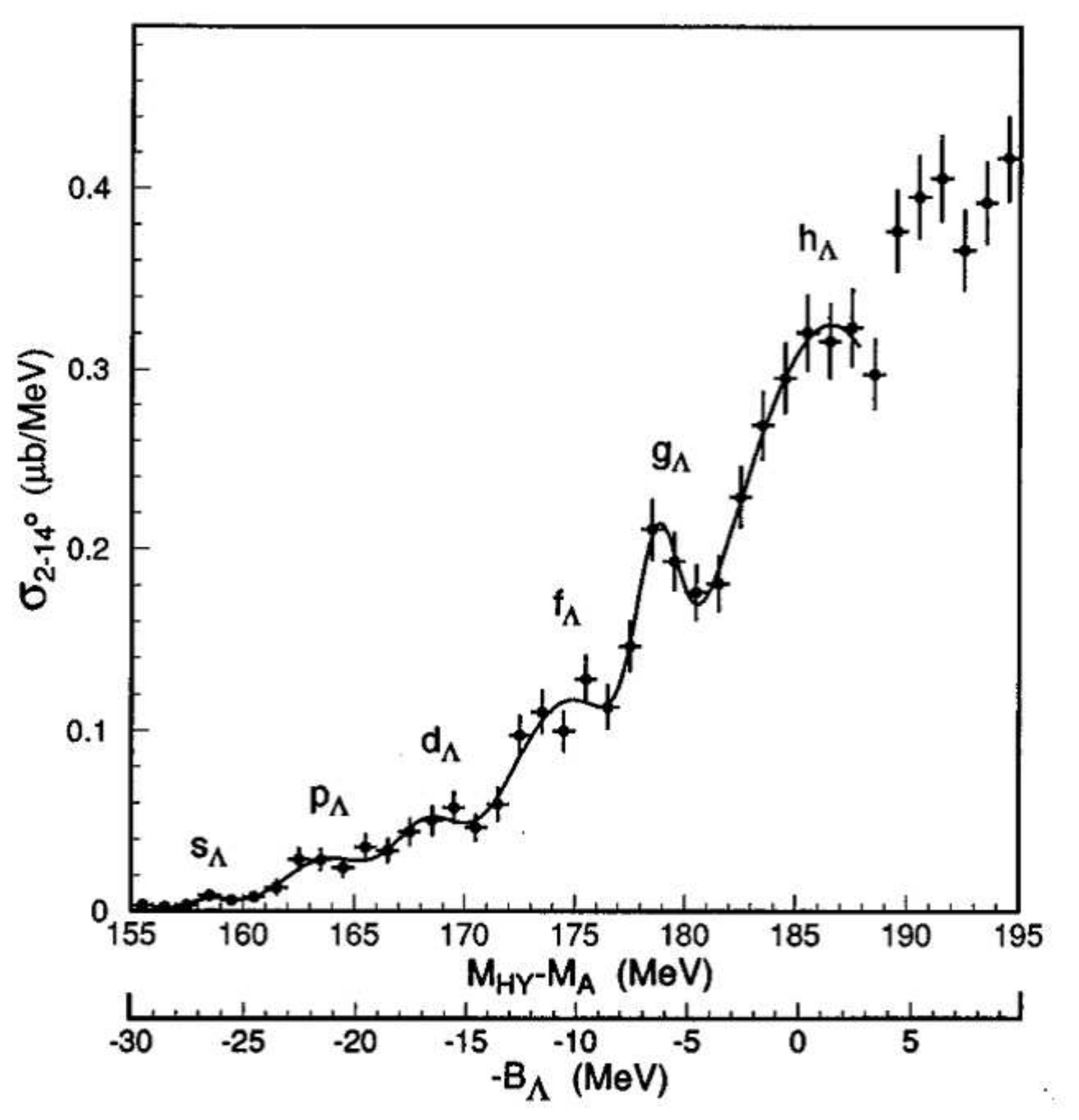}
\includegraphics[width=0.5\textwidth]{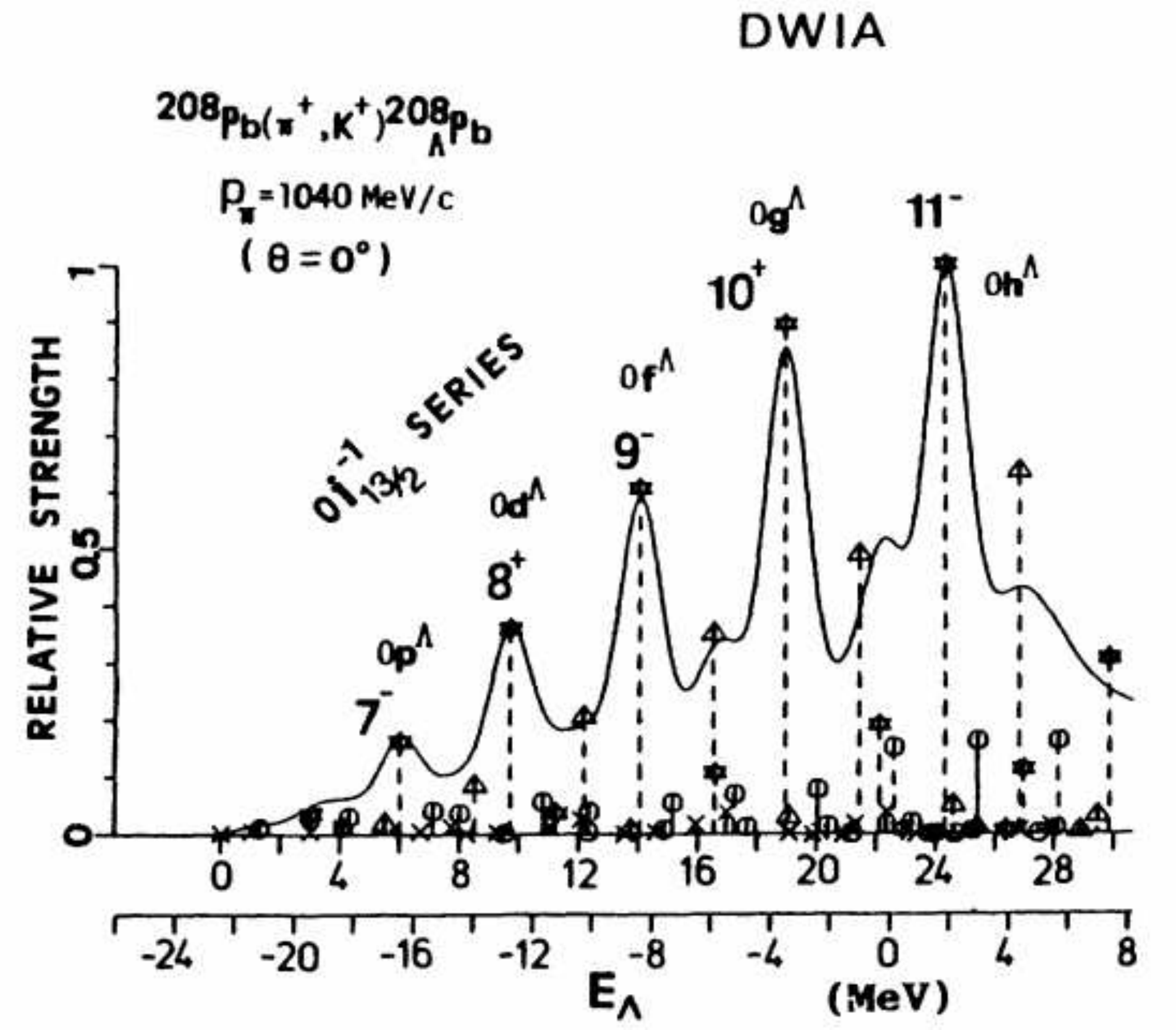}
\caption{
Left: A $^{208}_\Lambda$Pb energy spectrum taken by ($\pi^+$,$K^+$) reaction with a resolution of 2.5 MeV (FWHM) \protect{\cite{HAS96}}. The spectrum shows major shell ($s_\Lambda$, $p_\Lambda$, $d_\Lambda$,...) structure which is not clearly separated.
Rigfht: A calculated spectrum of  $^{208}$Pb ($\pi^+$,$K^+$) $^{208}_\Lambda$Pb reaction \protect{\cite{MOT88}}, suggesting that each bump in the experimental spectrum (a) consists of several core states and spin-spin/spin-orbit multiplets.
}
\label{Pbspectrum}
%\end{center}
 \end{figure}

~\\
{\bf  Chiral EFT approach}
~\\

Recently, another type of approach to $B$$B$ interactions is in progress in the framework of chiral effective field theory (chEFT).
In the chEFT framework, the $N$$N$ interaction is evaluated via power counting in terms of the momentum scale, where pion exchange diagrams are systematically calculated but short-range parts of the interaction are renormalized as low energy constants (LEC) that have to be determined from experimental data. The phase shift data of low energy ($E_{lab} < 300$ MeV) $N$$N$ scattering are reproduced quite well in the N$^3$LO calculation in a similar accuracy to the realistic $N$$N$ interaction models such as AV19. 
In this framework, the $N$$N$$N$ 3BF appears naturally in the N$^3$LO and higher order calculations, where two LEC's have to be determined from $^3$H binding energy and few-body scattering data.  In order to apply the chEFT method to $B$$B$/$B$$B$$B$ interactions including hyperons, studies have been made by Juelich-Bonn-Muenchen group and the NLO calculation has been made \cite{HAI13,LE20-2}. 
This interaction is also used to calculate 
the $\Lambda$ binding energies in hypernuclei \cite{HAI20}.
In addition, ab initio calculations 
with the non-core shell model have been carried out for $s$ and $p$-shell $\Lambda$ hypernuclei with the LO and NLO interactions \cite{WIR16,LE20}.
Due to insufficient $Y$$N$ scattering data, however, the LEC's cannot be determined 
well at present even if the flavor SU(3) symmetry is imposed to the parameters. 
Thus, a full calculation for N$^2$LO where the 3BF appears is not carried out yet, although the $Y$$N$$N$ 3BF effects are estimated based on a specific model \cite{PET16}. 
  Precise and comprehensive $Y$$N$ scattering data are eagerly awaited.

~\\
{\bf  Lattice QCD approach}
~\\

Recently, studies of $B$$B$ interactions with the lattice QCD method have been developed.
In particular, the HAL QCD approach succeeded in deriving the $B$$B$ interaction potentials reliably \cite{ISH07,AOK12}, 
which can be used to calculate nuclear many-body systems.
Since the lattice QCD method suffers from statistical errors from light quarks, 
recent results calculated on the physical pion mass 
can be well compared with experimental data for $Y$$N$ and $Y$$Y$ interactions.
With the $B$$B$ interactions determined from the HAL QCD method,
bound states are searched for 
in the $\Omega$$N$ and $\Omega$$\Omega$ systems \cite{IRI19, GON18}
and in light $\Xi$ hypernuclei \cite{HIY20}.

In addition, the HAL QCD method is being extended to the three-nucleon forces,
although the cost of the calculation is quite expensive. 
The first calculation in the triton channel with a heavy pion mass
shows a repulsive 3BF at a short distance \cite{DOI12}.

\subsubsection{Our scenario to solve the hyperon puzzle}

\begin{figure}
\includegraphics[width=\textwidth]{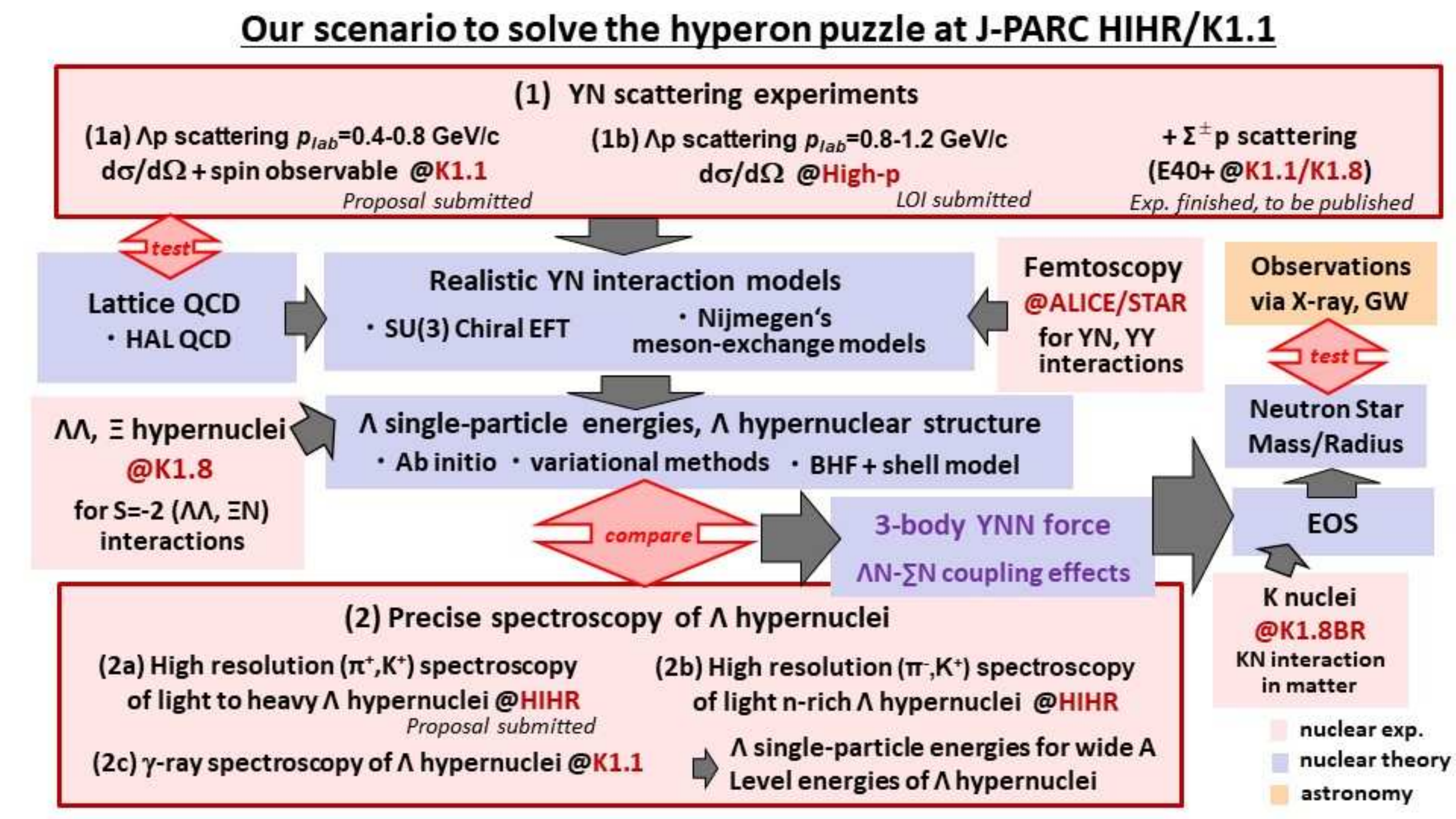}
\caption{Scenario to solve the hyperon puzzle and to elucidate matter in neutron stars microscopically.}
\label{SCENARIO}
\end{figure}

Figure \ref{SCENARIO} schematically shows our scenario
to solve the hyperon puzzle and elucidate matter in neutron stars.
We will collect high-quality $\Lambda$$p$ scattering data at the K1.1 and the High-p beam lines,
and together with the $\Sigma$$p$ scattering data already taken in the J-PARC E40 experiment
we provide $Y$$N$ scattering database to theorists.
They will construct upgraded theoretical models of $B$$B$ interactions.
Nijmegen's meson-exchange models will be improved to be a ``realistic model'' which
describes all the available scattering and hypernuclear data.
The extended chiral EFT models will proceed to the N$^2$LO calculation 
and the LEC's responsible for the three-body force will
be determined.
In this process, $S=-2$ data on $\Xi$ and $\Lambda$$\Lambda$ hypernuclei
collected at the K1.8 beam lines via E07 (the emulsion experiment; under analysis), E03 (the $\Xi$-atomic X-ray spectroscopy; under analysis), 
E70 (the ($K^-,K^+$) spectroscopy for $^{12}_\Xi$Be hypernucleus; schedule to run in 2022), and E75 ($^6_\Xi$Li and $^5_{\Lambda\Lambda}$H;
schedule after 2022) experiments will be combined.
In addition, high energy heavy ion collion experiments (ALICE in LHC and STAR in RHIC) 
have reently provided $B$$B$ correlation data via ``femtoscopy'' technique \cite{ACH19a,ACH19b}.
In the coming runs, ALICE will provide  
high statistics data on various channels including $\Lambda$$\Lambda$, $\Xi^-p$, $\Omega^-p$, etc.
Although this method can hardly provide information on 
$B$$B$ interactions for $p$- (and higher-$L$) wave and for spin-separated ($^3S_1$ and $^1S_0$) channels,
the data will be quite unique and valuable, particularly for the multi-strange ($S \le -2$) channels.

It is also worth mentioning that
the $B$$B$ interaction database and the realistic $B$$B$ interaction models will be
also used to test HAL QCD results for various $B$$B$ channels, which will appear in near future
on the physical pion mass.
Once the HAL QCD results are completely confirmed, they will be used to supplement
experimental $B$$B$ interaction data to construct realistic $B$$B$ interaction models.

In parallel, we also collect high-precision spectroscopic data of various $\Lambda$ hypernuclei
at the HIHR and K1.1 beam lines as shown below. 
At first, nuclear effects of the $\Lambda$$N$-$\Sigma$$N$ coupling force, which will be 
incorporated in the $B$$B$ interaction models based on the $Y$$N$ scattering data,
will be precisely tested by comparing ab initio calculations 
and precise data for light (neutron-rich) $\Lambda$ hypernuclei.
When they agree well, as was achieved in non-strange nuclei as shown in Fig.~\ref{Carlson},
the realistic $B$$B$ interaction models for $\Lambda$ hyperons are confirmed to be established. 

The $\Lambda$ binding energy data for 
$\Lambda$'s single-particle orbits at various densities in nuclear matter will be obtained.
Those data will be compared with 
$\Lambda$'s single-particle binding energies and level schemes calculated
from the realistic $B$$B$ interactions via various many-body approaches.
The standard method is to derive the effective interaction as a function of the fermi momentum
via the Brueckner-Hartree-Fock theory and derive
$\Lambda$'s single-particle energies in hypernuclei \cite{YAM13,NAG19}
and also calculate level structure with shell models
and variational methods such as antisymmetrized molecular dynamics (AMD) \cite{ISA16-1,ISA16-2}.
Another type of calculations to derive $\Lambda$ binding energies in finite nuclei 
with variational methods, such as Auxiliary Field Diffusion Monte Carlo method \cite{LON14},
is also being developed. 
Thus, the density-dependent binding energy difference
between the data and the calculations will be quantitatively obtained.
  %as the first evidence for the $\Lambda$$N$$N$ 3BF effects.

The obtained density dependence of the $\Lambda$ binding energies 
should be extrapolated to high density matter in neutron stars.
In Nijmegen models, the coupling constant of the phenomenological 3BF (multi-Pomeron exchange interaction),
which is assumed to be universal over the octet baryons at present,
will be adjusted for the $\Lambda$$N$$N$ channel, or the theoretical framework
will be improved, so as to reproduce the data.
Finally, when the updated theories can reproduce all the precise spectroscopic data,
the EOS of neutron stars will be reliably calculated.

In the framework of chiral EFT, 
the N$^2$LO calculations with the 3BF will be carried out and the LEC's will be 
determined via all the scattering data and light $\Lambda$ hypernuclear binding energy data.
Then this interaction will be applied to various hypernuclear data.
If the data and the calculations agree well, the interaction can be used to
calculate the EOS. 

Finally, the calculated EOS will be tested by observational data of
the neutron star mass and radius, which will be accumulated by the NICER detector and the XRISM satellite  
in coming years.
Gravitational wave data will also provide additional information on the stiffness of the neutron star mergers
in the future. At present, however, simulated waveform templates depend on the EOS and microscopic $B$$B$ 
interactions behind. When the $B$$B$ interaction is reliably determined by our measurement,
more realistic calculations of the templates will be possible and unique information 
will be able to be extracted from gravitaional wave data.

Through this project, the properties and the strength of the repulsive $\Lambda$$N$$N$ 3BF
will be clarified and the matter in neutron stars will be elucidated
microscopically. However, the origin of the 3BF
may not be easily understood from the quark level. 
However, the lattice QCD calculations for the 3BF will be available in the future and greatly helps 
us understand the mechanism of the $B$$B$$B$ three-body forces.
In the $B$$B$ interactions, the origins of the short-range repulsive core were previously studied 
in the quark cluster model \cite{OKA80,OKA00}, of which ideas are being tested and confirmed at present
by the $Y$$N$/$Y$$Y$ interaction data and the HAL QCD results. 
Similarly, we expect to reach deeper understanding of the 
mechanism of the $B$$B$$B$ forces which govern both atomic nuclei and neutron stars.

\subsubsection{Experimental plans for our scenario}

\begin{itemize}
\item[(1)]
{\bf $\Lambda$$N$ scattering experiments}

Compared with the $N$$N$ scattering, the $Y$$N$ scattering data are extremely poor 
in statistics, particularly for hyperon momenta higher than 0.5 GeV/$c$.
In most of the channels differential cross sections as well as spin observables are not measured. 
In order to apply the $B$$B$ interaction models to high density nuclear systems, 
high-statistics $Y$$N$ scattering data for a wide momentum range are indispensable.
In particular, differential cross sections and spin observables as well as higher momentum data up to $\sim$1.2 GeV/c
play important roles to determine the unknown interactions 
for the $p$-wave (and higher $L$ waves).
We propose to collect the following
new scattering data at K1.1 and High-p beam lines at J-PARC.
The data will make both the Nijmegen's meson-exchange model and the extended chiral EFT model with hyperons
much more reliable and realistic.

(1a) As shown in the next section and in \cite{Miwa_Proposal}, 
we propose to collect $\Lambda$$p$ scattering events by polarized $\Lambda$ beams from 0.4 to 0.8 GeV/$c$ . 
In the E40 experiment recently performed at J-PARC we have successfully 
obtained $\Sigma^-p$ and $\Sigma^+p$ differential cross section 
data with statistics $10^2$ times larger than the past experiments 
by introducing a new method to identify scattering events kinematically \cite{
%Miwa:2021
J-PARCE40:2021qxa}. 
By employing the same method and the same detector system around the target,
differential cross sections for the $\Lambda$$p$ elastic
scattering will be measured with better than 10\% statistical error 
with an angular step of $d$cos$\theta$= 0.1 for each of the 50 MeV/c $\Lambda$ momentum bins between 0.4 and 0.8 GeV/c. Analyzing power and depolarization are measured with a 10\%-level statistical error with an angular step of $d$cos$\theta$= 0.2 
for each 100 MeV/c momentum bin.
Here, a $\Lambda$ hyperon beam is produced by the $p(\pi^-,K^{0})\Lambda$ reaction with a 1.05 GeV/c pion beam
in which a produced $\Lambda$ is known to have almost 100\% polarization.
For details, see the next section and the proposal.

(1b) In addition, another experiment to measure the differential cross sections of $\Lambda$$p$ scattering for higher momenta
($p_{lab} \sim$ 0.8--1.2 GeV/$c$) is planned at the High-p beam line. 
$\Lambda$ hyperons are produced by $p(\pi^-,K^{*0})\Lambda$ reaction in a liquid 
hydrogen target with 8.5 GeV/c $\pi^-$ beams,
and production of $\Lambda$ hyperons is tagged by a large acceptance spectrometer developed for the E50 experiment.
The $\Lambda$$p$ scattering is identified by a cylindrical spectrometer around the target.
$\Lambda$$p$ scattering events will be collected with 10$^2$ times larger statistics than the previous experiments.
See \cite{Honda:2019} for details.

%\begin{figure}
%\includegraphics[width=\textwidth]{isaka.pdf}
%\caption{$B_\Lambda$ values calculated for various hypernuclear ground state by AMD method with Nijmegen interactions (ESC12 %and ESC14) without and with 3BF (MPP + TBA).
%The difference of the two interactions is $p$-wave strength \protect{\cite{ISA17}}.
%}
%\label{ISAKA}
% \end{figure}

\item[(2)]

{\bf Precise $\Lambda$ hypernuclear spectroscopy}

Precise spectroscopy of a wide variety of $\Lambda$ hypernuclei is essential 
to learn the nature of the $\Lambda$$N$ interaction in nuclear matter
and to apply it to higher density matter. Here we propose three types of experiments.

As described above, effects of a possible repulsive three-body $\Lambda$$N$$N$ interaction
(or density-dependent effects of the $\Lambda$$N$ interaction)
are expected to appear in A-dependence of the $\Lambda$ binding energy of hypernuclei.
The size of such effects cannot be theoretically predicted, but the calculation with Nijmegen ESC08c shown above
(Fig.~\ref{YAM_BL} Left) implies that the effect can be less than 1 MeV.
Thus, in order to quantitatively extract the strength of the $\Lambda$$N$$N$ interaction,
measurement of the $B_\Lambda$ values with accuracy of $\sim$0.1 MeV is desirable.

%As shown in Fig.~\ref{MILLENER}, 
The $B_\Lambda$ values for $\Lambda$'s single-particle
orbits ($s_\Lambda$, $p_\Lambda$, $d_\Lambda$, $f_\Lambda$,...)  
have been measured
by various reactions and the $\Lambda$'s potential depth
in the nuclear matter was determined to be -30 MeV, as shown in
Fig.~\ref{Millener}.
However, the data for heavy hypernuclei $A \ge 40$) only come from the
($\pi^+,K^+$) spectroscopy experiments carried out at KEK-PS with a mass resolution of
$\sim$2 MeV (FWHM) \cite{HAS96}. 
As shown in the $^{208}$Pb$(\pi^+, K^+)$$^{208}_\Lambda$Pb spectrum (see Fig.~\ref{Pbspectrum}), 
hypernuclear peaks are not well separated for those heavy hypernuclei \cite{HAS96}. 
It is because, when a $\Lambda$ occupies a certain single-particle orbit, the core nucleus
is split into various excited states, each of which appears as a peak at a slightly different energy.
Since the spacing between the core excited states is typically $\sim$0.5 MeV, energy resolution 
better than $\sim$0.5 MeV is necessary to separate those peaks.
In the KEK experiment, the expected spectrum shape 
composed of several core excited states was
calculated relying on theoretical cross sections
and convoluted with the experimental resolution of $\sim$2 MeV,
and then it was used to fit the measured spectrum to extract the $\Lambda$'s single-particle energy values. 
Thus, these $B_\Lambda$ values may have systematic errors around 1--2 MeV.

At the HIHR beam line, we plan to measure the spectrum with $<$400 keV (FWHM) resolution
to separate each of the hypernuclear peaks in order to
determine each single particle energy without such ambiguity.

In addition, since a hypernuclear level is split into a doublet (or a multiplet) by spin-spin and spin-orbit interactions between a $\Lambda$ and the core nucleus, 
the energies of both doublet states should be averaged to obtain a $\Lambda$'s single-particle energy,
if the doublet spacing is expected to be not negligible.
 Since a double spacing is typically of the order of a few 100 keV and both of the doublet ($\Lambda$-spin-flip and non-flip states) are usually not populated simultaneously, 
the spacing had better be determined by measuring $\gamma$ rays
and constructing a level scheme of the hypernucleus.
Fortunately, we have a technique and apparatus of $\gamma$-ray spectroscopy
of $\Lambda$ hypernuclei \cite{TOY15}. By using the $(K^-,\pi^-)$ reaction
with intense $K^-$ beams at the K1.1 beam line
$\gamma$-ray measurement will be carried out for some of the target hypernuclei
studied at HIHR.

Effects of baryon mixing such as $\Lambda$$N$-$\Sigma$$N$ and $\Xi$$N$-$\Lambda$$\Lambda$ significantly 
change the two-body $\Lambda$$N$, $\Sigma$$N$, $\Xi$$N$ and $\Lambda$$\Lambda$ interactions in nuclear matter 
through Pauli effect of intermediate nucleons and the three-body $Y$$N$$N$ interactions.
In particular, it is pointed out that the $\Lambda$$N$-$\Sigma$$N$ coupling via the tensor interaction
can generate a significant difference between the $^3S_1$ $\Lambda$$N$ interaction in the free space 
and that in the nuclear matter due to a nucleon Pauli effect \cite{HAI17}. 
Since such an effect can be large in dense nuclear matter in neutron stars, it should be experimentally clarified.
The baryon-changing baryon-baryon-meson coupling constants in the meson-exchange models 
should be determined to correctly treat these problems.

For this purpose, we also plan to measure high resolution spectra of light neutron-rich $\Lambda$ hypernuclei 
via $(\pi^-,K^+)$ reaction at HIHR beam line,
because the 3BF effect due to $\Lambda$-$\Sigma$ mixing is expected to be enhanced 
in large-isospin hypernuclei.
We propose to measure $^9_\Lambda$He and $^{10}_\Lambda$Li (and then $^{11}_\Lambda$Li and  $^{12}_\Lambda$Be) using $^9$Be and $^{10}$B (and  $^{11}$B and $^{12}$C) targets with accuracy of $\sim$0.1 MeV and  a resolution of $<$400 keV (FWHM).  

The information on the $\Lambda$$N$-$\Sigma$$N$ coupling
will be also obtained from the cusp spectrum of $d(\pi^+,K^+)\Lambda$$p$,$\Sigma$$p$ reaction
at HIHR, as well as $\Lambda$$p$$\to$$\Sigma^0$$p$ and $\Sigma^-$$p$$\to$$\Lambda$$n$
scattering experiments.
A $\Sigma^-$$p$$\to$$\Lambda$$n$
scattering experiment (E40) has been already conducted, while
a $\Lambda$$p$$\to$$\Sigma^0$$p$ scattering experiment will
be performed at K1.1 beam line.

The $Y$$N$ interaction models updated with high statistics $Y$$N$ scattering data
including the above channels at the K1.1 beam line will be used for
ab initio calculations of these light hypernuclei.
If the calculations agree with the precise experimental data,
the long-standing problem of $\Sigma$ mixing in $\Lambda$ hypernuclei
is solved and the interaction model can be used to investigate the $\Lambda$$N$$N$ 3BF effects.

In summary,

(2a) We propose to measure high-resolution ($<$400 keV FWHM) $(\pi^+,K^+)$ spectra for light 
to heavy $\Lambda$ hypernuclei ($^6_\Lambda$Li,
 $^7_\Lambda$Li, $^9_\Lambda$Be, $^{10}_\Lambda$B, $^{10}_\Lambda$B,
$^{11}_\Lambda$B, $^{12}_\Lambda$C, $^{28}_\Lambda$Si, $^{40}_\Lambda$Ca, $^{51}_\Lambda$V,
$^{89}_\Lambda$Y, $^{139}_\Lambda$La, and $^{208}_\Lambda$Pb).
With the momentum dispersion matching technique, hypernuclear excitation spectra with
a resolution of $<$400 keV (FWHM) will be achieved.
Details are described in the following sections and Ref.~\cite{Nakamura_Proposal}.

(2b) We will also propose to measure high resolution ($<$400 keV FWHM) $(\pi^-,K^+)$ spectra 
for light neutron-rich $\Lambda$ hypernuclei, $^9_\Lambda$He and $^{10}_\Lambda$Li (and then $^{11}_\Lambda$Li and  $^{12}_\Lambda$Be)
at the HIHR beam line to obtain $\Lambda$$N$-$\Sigma$$N$ coupling effects.

(2c) We propose $\gamma$-ray spectroscopy experiments for the same $\Lambda$ hypernuclei as (2a) using $(K^-,\pi^-)$ reaction at the K1.1 beam line
in order to clarify their level schemes and extract $\Lambda$'s single particle energies precisely.

\end{itemize}

% flatex input end: [HIHRK11-motivation.tex]

%%%%%% HIHR/K1.1 Motivation  (Tamura)

%%%%%% HIHR beam line (SNN)
% flatex input: [HIHRBL-HIHR2.tex]
%%%%%%%%%%%%%%%%%%%%%%%%%%%%%%%%
%%% Copied from P84 proposal
\subsection{High Intensity High Resolution (HIHR) Beam Line}
\label{SEC:HIHRbeamline}

\begin{figure}[hb]
\begin{center}
\includegraphics[width=0.9\hsize]{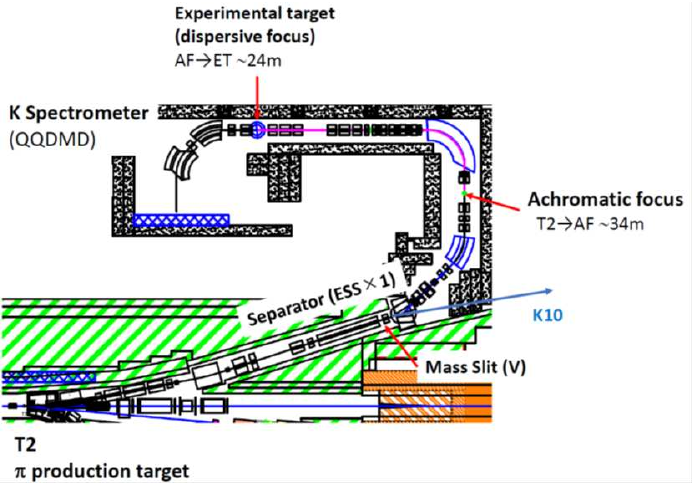}\vspace*{5mm}\\
\caption{Schematic illustration of the High Intensity High Resolution beam line with a kaon spectrometer.}
\label{FIG:3-1}
\end{center}
\end{figure}

The proposed experiment is to perform the high precision mass spectroscopy of $\Lambda$ hypernuclei 
produced by the $(\pi^+,K^+)$ reaction and will employ a newly designed momentum dispersion matching 
pion beam line and kaon spectrometer, the High Intensity High Resolution beam line (HIHR) which is schematically 
illustrated in Fig.~\ref{FIG:3-1}.

%%%%
% Fig3-1
%%%%%

The high intensity, high resolution beam line (HIHR) ion-optically realizes a strongly correlated beam in position and momentum. 
Combining a properly designed kaon spectrometer to HIHR, which is the so-called dispersion matching ion-optical technique, 
as described below, one can carry out high-precision spectroscopy in nuclear physics. In particular, $\Lambda$-single particle 
energies in various $\Lambda$ hypernuclei can be measured at an energy resolution of as high as a few hundred keV which 
enables us to determine Lambda binding energy with an accuracy of $<100$~keV.
As the dispersion matching beam line does not require a beam measurement, one can remove beam line detectors and 
reduce beam line materials that affect the energy resolution due to multiple scattering and momentum straggling effects. 
HIHR provides a solution to utilize full-intensity beams potentially available at J-PARC, while available beam intensity 
would be limited a capability of detectors used as beam counters in usual beam lines.
Detailed description of the momentum match technique will be given in the next section.

HIHR is composed of 4 sections as explained below. In the most upstream section, the secondary particles produced 
at the primary target are collected at the production angle of 3 degrees. The maximum momentum of the secondary 
beam is designed to be 2~GeV/$c$. The base design of the spectrometer follows it of K1.8 beam line in the current 
hadron hall 
%\cite{Agari:PTEP2012}.
\cite{Agari:2012kid}.
and the beam line layout of the extraction part would be essentially the same as that of K1.8. 
The secondary beam is focused vertically at the intermediate focal (IF) point. The secondary beam image is redefined here 
by the IF slit placed.

In the second section after the IF point, a pion beam is separated from the other particle beams. In this section, by 
using two sets of a pair of quadrupole magnets (Q-doublet), a so-called point-to-point ion-optics in the vertical direction is 
realized between the IF point and the MS point at the end of this section. An electrostatic separator (ESS) having a pair of 
parallel plate electrodes of 10 cm gap and 4.5 m long is placed between the Q-doublets.
Charged particles are kicked vertically by an electrostatic field produced in the ESS. One can compensate the vertical 
kick by the pitching magnets placed just before and after the ESS.
While the kicked angle by the ESS is proportional to the inverse of the particle velocity, the kicked angle by the magnets 
is proportional to the particle momentum, Namely, a particle beam compensated its kicked angle keeps the beam level and 
the other particle beams not fully compensated their kicked angles are focused at the different positions in vertical 
at the MS point. A mass slit (MS) is placed at the MS point so as to select a particle beam and block the other unwanted 
particle beams.
In the third section after the MS point, the secondary beam is focused achromatically in the horizontal and vertical 
directions at the IF2 point (Achromatic focus in Fig.~\ref{FIG:3-3}). The beam profile is defined again by the beam slits placed 
at the IF2 point. In the final section after the IF2 point, by placing a large bending magnet and a several quadrupole magnets, 
the beam is focused with a magnification of 1.1 and a dispersion of 11 [cm/\%] at the experimental target. 
Here, ion-optical aberrations to the second order are eliminated by three sextupole magnets.

Overall, this experimental design is for (1) the highest possible resolution ($<400$~keV FWHM) to determine $\Lambda$ 
binding energies with an accuracy of $<100$~keV, (2) the highest reachable yield with the high intensity $\pi^+$ beam, 
in the reaction spectroscopic study of hypernuclei. 

\subsubsection{Kinematics and key parameters of HIHR beam line and spectrometer}

%%%%%%
% Fig 3-2
\begin{figure}[htb]
\begin{center}
\includegraphics[width=0.7\hsize]{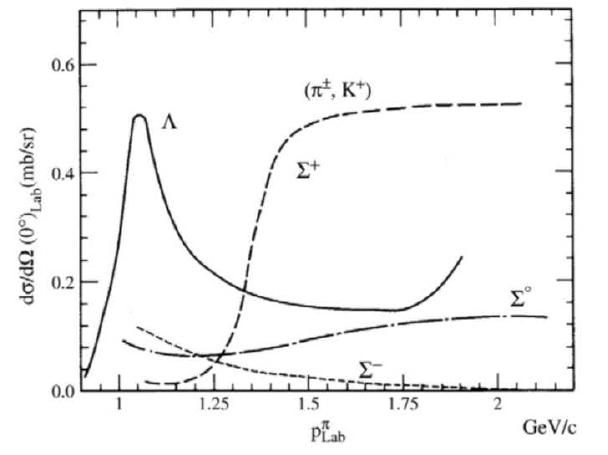}\vspace*{5mm}\\
\caption{The hyperon production cross section as a function of incident pion momentum \cite{HAS06}.}
\label{FIG:3-2}
\end{center}
\end{figure}
%%%%%%

%%%%%%
% Fig:3-3 
\begin{figure}[htb]
\begin{center}
\includegraphics[width=0.7\hsize]{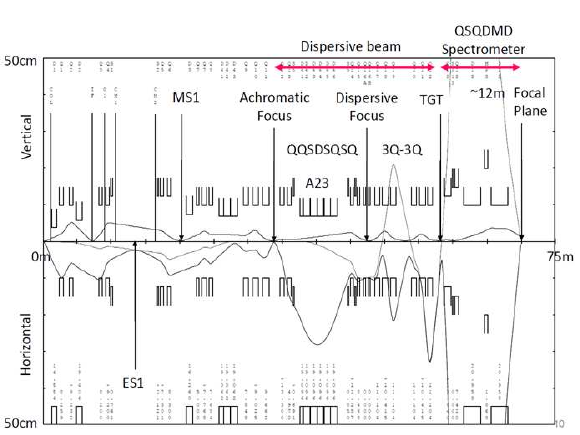}\vspace*{5mm}\\
\caption{The layout of HIHR magnets and calculated beam envelope with TRANSPORT \cite{
%NOU19
Takahashi:2019xcq}.}
\label{FIG:3-3}
\end{center}
\end{figure}
%%%%%%

The proposed kinematics is assuming use of a pion beam momentum of $p = 1.05$~GeV/$c$, where the cross section 
of the elementary process becomes maximum as shown in Fig.~\ref{FIG:3-2}.
For the $^A Z(\pi^+,K^+)_\Lambda^A Z$ reaction, the central momentum of the kaon spectrometer is determined to 
be 0.72~GeV/$c$. if the $\pi^+$ beam momentum is set at 1.05~GeV/$c$.

\begin{figure}[htb]
\begin{center}
\includegraphics[width=0.7\hsize]{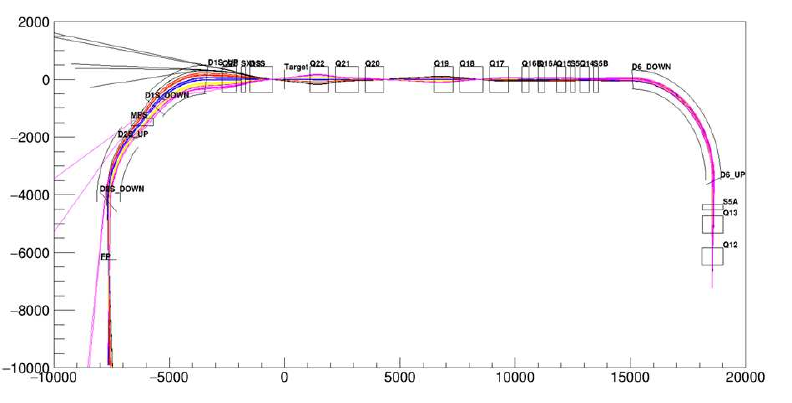}\vspace*{5mm}\\
\caption{GEANT4 model of the HIHR beam line after the achromatic focus point (IF2) and 
the kaon spectrometer. Unit of scale is millimeter.}
\label{FIG:3-4}
\end{center}
\end{figure}

The layout of HIHR magnets and calculated beam envelope are shown in Fig.~\ref{FIG:3-3}. 
Though detailed design of HIHR beam line highly depends on the design of primary target and beam extraction 
part which will be shared with the other beam lines in the hadron extension hall, optical features of the $\pi$ beam line 
and K spectrometer were studied with a GEANT4 Monte Carlo simulation (Fig.~\ref{FIG:3-4}) based on conceptual 
design \cite{
%NOU19
Takahashi:2019xcq} of HIHR calculated by an optics optimization code, TRANSPORT. 

The kinematic and key parameters for the current design of HIHR estimated by the GEANT4 simulation with 
the 2nd order matrix tune are summarized in Table~\ref{TAB:3-I}. 

%%%%%%
% Fig 3-4
%%%%%%

\begin{table}[h]
\caption{\label{TAB:3-I}Parameters of the HIHR beam line (GEANT4 with 2nd order matrix tune)}
\begin{center}
\begin{tabular}{|l|c|}
\hline
Beam $\pi^+$ centreal momentum (GeV/$c$) & 1.1\\
\hline
Total length (m) from achromatic focus to the experimental target & 24.8\\
\hline
Momentum acceptance (\%) & $\pm$ 1\\
\hline
Horizontal magnification & $-1.13$\\
\hline
Vertical magnification & $-0.88$\\
\hline
Dispersion (cm/\%) & 11.28 \\
\hline
\end{tabular}
\end{center}
\end{table}

\begin{table}[h]
\caption{\label{TAB:3-II}Parameters of the HIHR kaon spectrometer (GEANT with 2nd order matrix tune)}
\begin{center}
\begin{tabular}{|l|c|}
\hline
$K^+$ centreal momentum (GeV/$c$) & 0.71\\
\hline
Total length (m) & 11.4\\
\hline
Horizontal angular acceptance (mrad) & $\pm~60$\\
\hline
Vertical angular acceptance (mrad) & $\pm~100$\\
\hline
Momentum acceptace (\%) & $\pm~5$\\
\hline
Horizontal magnification & $-1.84$\\
\hline
Vertical magnification & $-0.54$\\\hline
Dispersion (cm/\%) & 10.72 \\
\hline
\end{tabular}
\end{center}
\end{table}

\subsubsection{beam line and K spectrometer dispersion match setting}

The kaon spectrometer matched to HIHR beam line comprises the QSQDMD configuration, 
where D, Q, S, and M stand for a dipole, a quadrupole, a sextupole, and a multipole magnet, respectively. 
A dispersion matching ion optical technique \cite{SJO60, FUJ97, FUJ99} is a well-established technique for long-living ions, 
but there has been no application to the secondary meson GeV beam line.
Applying this technique to the secondary $\pi^+$ beam line with high-resolution kaon spectrometer, 
a $\Lambda$ hypernuclear energy spectrum can be obtained in the $(\pi^+,K^+)$ reaction by measuring $K^+$ position 
distribution at the final focal plane of the kaon spectrometer. 

%%%%
% Fig 3-5
\begin{figure}[htb]
\begin{center}
\includegraphics[width=0.7\hsize]{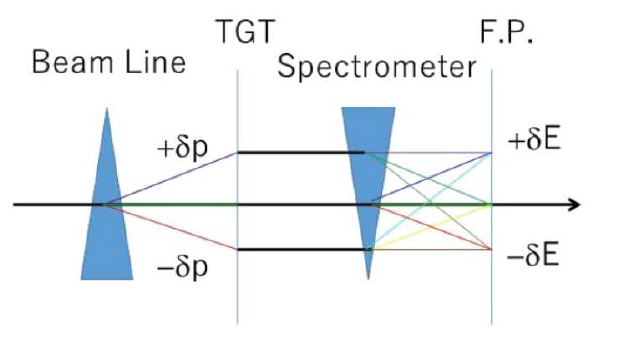}\vspace*{5mm}\\
\caption{Conceptual illustration of momentum dispersion matching.}
\label{FIG:3-5}
\end{center}
\end{figure}
%%%%

\begin{figure}[htb]
\begin{center}
\includegraphics[width=0.7\hsize]{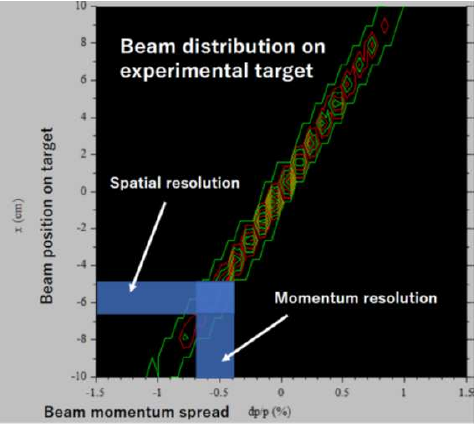}\vspace*{5mm}\\
\caption{Pion beam momentum spread is converted to beam position spread on the experimental target. 
Beam distribution was calculated with a beam optics code, TRANSPORT. 
Momentum spread of 1~\% is converted to spatial beam spread of 10~cm.}
\label{FIG:3-6}
\end{center}
\end{figure}

Figure~\ref{FIG:3-5} shows a conceptual illustration of the momentum dispersion matching technique. 
Secondary $\pi^+$ is transferred from the primary target with a momentum.bite of $\pm 1\%$, which corresponds to 
$\pm 11$~MeV/$c$. Therefore, it is essential for sub-MeV spectroscopy to measure their momentum, 
one by one if a conventional beam line is used. Limitation of beam rate for beam detectors does not allow us to use very 
high intensity beam. In the momentum dispersion matching beam line, the beam momentum spread is converted to beam 
position distribution on the target where $(\pi^+,K^+)$ reaction takes place (Fig.~\ref{FIG:3-6}). 
Kaon spectrometer’s dispersion and magnification are adjusted to cancel the beam line’s momentum dispersion.

%%%
% Fig3-6
%%%%

Let us explain how the dispersion match conditions are realized. 
The first order beam transfer matrix is given as follows with the initial 
coordinate $(x_0, \theta_0, \delta_0)$, position, angle and relative momentum with respect to the central momentum and
$(x_f, \theta_f, \delta_f)$ for them at the final point.

\begin{small}
\begin{equation}
\left ( \begin{array}{c} 
x_f\\
\theta_f\\
\delta_f\\ 
\end{array} \right ) =
\left ( \begin{array}{ccc} 
s_{11} & s_{12} & s_{16}\\
s_{21} & s_{22} & s_{26}\\
0&0&1\\
\end{array} \right ) 
\left ( \begin{array}{ccc} 
T & 0 & 0\\
0 & \theta/\theta_1 + 1 & 0\\
0&0& (K\theta + DQ)/\delta_0 + C\\
\end{array} \right ) 
\left ( \begin{array}{ccc} 
b_{11} & b_{12} & b_{16}\\
b_{21} & b_{22} & b_{26}\\
0&0&1\\
\end{array} \right ) 
\left ( \begin{array}{c} 
x_0\\
\theta_0\\
\delta_0\\ \end{array} \right ),
\end{equation}
\end{small}

\begin{eqnarray}
\theta_1 &=& b_{21} x_0 + b_{22} \theta_0 + b_{26}\delta_0,\\
K &=&  (\partial p_{scat}/\partial \theta)(1/p_{scat}),\\
C &=&  (\partial p_{scat}/\partial p_{beam})(p_{beam}/p_{scat}),\\
D &=& (\partial p_{scat}/\partial Q)(1/p_{scat}).
\end{eqnarray}

Here, $T, \theta$ and $Q$ are a cosine of an angle between a beam direction and a normal to a target plane, a scattering angle, 
and an excitation energy, respectively. Values of $K$ and $C$  are respectively a derivative of the scattered momentum ($p_{scat}$) 
to the scattering angle ($\theta$) and that to the beam momentum ($p_{beam}$). $D$ is a derivative of $p_{scat}$ to $Q$. 
The position $x_f$ at the final focal plane of kaon spectrometer can be expressed as:
\begin{equation}
x_f = (\partial x_f/\partial x_0)x_0 + (\partial x_f/\partial \theta_0)\theta_0 
+(\partial x_f/\partial \delta_0) \delta_0 + (\partial x_f/\partial \theta)\theta 
+s_{16}DQ. \label{Eq:HIHR6}
\end{equation}
The matching conditions are obtained as follows.
\begin{eqnarray}
(\partial x_f/\partial x_0) &=& s_{11} b_{11} T + s_{12} b_{21} \rightarrow {\rm minimize},\\
(\partial x_f/\partial \theta_0) &=& s_{11} b_{12} T + s_{12} b_{22} \rightarrow 0,\\
(\partial x_f/\partial \delta_0) &=& s_{11} b_{16} T + s_{12} b_{26} + s_{16}C \rightarrow 0,\\
(\partial x_f/\partial \theta) &=& s_{12} + s_{16} K \rightarrow 0.
\end{eqnarray}

Fixing reaction kinematics and a scattering angle, magnetic component of beam line and spectrometers are tuned 
to satisfy the above conditions. Degrees of freedom to be tuned in beam line and spectrometer systems are 6 and 3, respectively.
 Once the matching conditions are satisfied, the excitation energy $Q$ can be given as $x_f/(s_{16} D$ with an energy 
resolution of $(\partial x_f/\partial x_0 ) x_0$. In the real operation of the beam line, higher order terms of matrices 
should be taken into account and the momentum matching parameter optimization will be performed with 
an established parameter minimization algorism on the computer.

%%%%%% End SNN 20210724
% flatex input end: [HIHRBL-HIHR2.tex]

%%%%%% HIHR beam line (SNN)

\clearpage

%%%%%% HIHR main part (SNN)
% flatex input: [HIHR-Phys.tex]
%%%%%%%%%%%%%%%%%%%%%%%%%%%%%%%%
% SNN copied ProposalP84 
\subsection{High Precision Spectroscopy of $\Lambda$ Hypernuclei with the $(\pi^+, K^+)$ Reaction at HIHR}

\subsubsection{Introduction}

The major goal of nuclear physics is understanding the nature of many-body system 
whose dynamics is dictated by the strong interaction. There is a hierachy of three such systems in our universe: 
1) baryons/mesons, the bound system of quarks, 2) nuclei, the self-bound system of baryons and 
3) neutron stars, isospin-asymmetric nuclear matter bound by the gravitational force. 
The size scale of 1) and 2) is femtometers and that of 3) is ~10 km. Though they differ by 10$^{19}$ order of magnitudes, 
it is common that the strong interaction plays a key role to govern their structure and properties.  
The strong interaction in the low energy region where the QCD is not perturbative has been investigated in the framework of 
baryon potential models. The construction of NN potentials has taken advantage upon the availability of a substantial number 
of scattering data. High-precision potential models fit all these data with extreme accuracy. 
However, when only two-body forces are accounted for, light nuclei turn out to be under-bound and the saturation 
properties of infinite isospin-symmetric nuclear matter are not correctly reproduced. 
It indicates the need for a three-body interaction. The same theoretical framework could be extended to the strange sector. 
Though the strange quark is heavier than the u and d quarks, it is still lighter than the QCD cut-off ($\sim$1~GeV) 
unlike the heavy quarks $(c, t, b)$. Hence, the strange quark can be treated in the framework of the flavor SU(3) symmetry 
which is a natural extension of the isospin symmetry for ordinary nucleons. 
To devise a unified description of the baryonic interaction within the flavor SU(3) basis, one must then quantitatively 
understand the hyperon-nucleon (YN) and the hyperon-hyperon (YY) interactions. 
Investigation on double-$\Lambda$ and $\Xi$, $S=-2$ hypernuclei and direct measurement of YN scattering experiments 
made great progresses at the existing J-PARC hadron hall. 
However, the bare YN interaction and effective YN interaction in nuclei are different 
due to quantum many-body effects; spectroscopic investigation of $\Lambda$ hypernuclei, nuclear many-body systems 
containing one $\Lambda$ particle, provides a unique and, currently, the only practical tool to provide 
effective YN interaction in various nuclei with an excellent precision. 

While several models of YN forces were proposed in the past, recent observation of neutron stars with masses 
of  $2M_\odot$ (two solar mass) poses several important questions. 
There is a very simple argument to justify the appearance of strange degrees of freedom in the inner core of a neutron star. 
In the pure neutron matter case, whenever the chemical potential becomes sufficiently large to match the chemical potential 
of a hyperon in the same matter, the hyperon becomes stable since it is a distinguishable particle, and creates its own Fermi sea, 
thereby lowering the kinetic energy of the system. This results in a so-called "softening" of the equation of state (EOS), 
due to a decrease of the Fermi pressure. In turn, a soft EOS predicts a lower sustainable mass for a neutron star. 
So far, existing YN interactions typically predict a maximum mass no larger than $1.5~M_\odot$, 
in strong contrast with the astrophysical observations (the "hyperon puzzle"). 
The key for solving this apparent contradiction is a more repulsive YN interaction, 
which increases the hyperon chemical potential, moving the onset of hyperons at higher densities. 
Most models nowadays agree on this aspect, but additional constraints are needed, 
and they can only be inferred from accurate spectroscopic data of hypernuclei. 
The relation between hypernuclei and matter inside a neutron star is not straightforward. 
It is clear that the hyperon chemical potential at high density cannot be effectively constrained 
from the existing hypernuclear data with a few MeV resolution, unless very accurate measurements on hypernuclei 
are performed in wide mass range. Tiny changes in the binding energy, and consequently on the determination of 
the YN/YNN force, can have dramatic consequences on the EOS of matter at supra-saturation conditions. 
Another interesting aspect of the YN interaction is the charge symmetry breaking (CSB). 
Recently decay-pion spectroscopy of electro-produced $^4_\Lambda$H, which was originally proposed at JLab \cite{TAN10}, 
was successfully carried out at Mainz \cite{ESS15} and the excitation energy of $^4_\Lambda$He $1^+$ state was also 
successfully measured by the gamma-spectroscopy at J-PARC \cite{
%TOY15
Yamamoto:2015avw}. 
These new experimental data strongly support the fact that the A=4 hypernuclear isospin-doublet has a large CSB 
for the ground states ($0^+$) and small CSB for the $1^+$ excited states. 
Although the origin of such a large CSB is not fully understood, it is clear that the discussion based on 
two-body YN/NN interaction only does not suffice and that the inclusion of $\Lambda$N-$\Sigma$N coupling 
and three-body force are essential. 
Behavior of $\Lambda$ in symmetric nuclear matter and neutron-rich environments would be quite different 
and  $\Lambda$N-$\Sigma$N coupling and 3-body force are important for the discussion of nuclear matter property \cite{HAI20}.
It is quite important to investigate these effects for the heavier and neutron richer hypernuclear systems, 
but these effects tend to be smaller \cite{GAL13} and thus high resolution is an important key.

Though a systematic study of $A \leq 208~~\Lambda$ hypernuclei with a few MeV accuracy has already been carried out 
with the conventional $(\pi^+, K^+)$ reaction spectroscopy at BNL-AGS and KEK-PS \cite{PIL91,HAS95,HAS96,HAS98,HOT01}, 
the accuracy of existing hypernuclear data is not adequate to the features of the baryonic force models needed to properly address 
the hyperon puzzle, and thus higher precision data are necessary. 
We propose to measure $\Lambda$ binding energies of $\Lambda$ hypernuclei in wide mass region via 
the $(\pi^+, K^+)$ reaction at the HIHR beam line which will be realized in the extended hadron experimental hall of J-PARC. 
Precise hypernuclear spectroscopy in wide mass region is crucially important to constrain or provide necessary information
 to construct a reliable interaction which can be used to calculate EOS of the neutron stars. 
Such high precision hypernuclear spectroscopy is only possible at momentum dispersion match $\pi$ beam line, HIHR and 
cannot be performed at other facilities.

\subsubsection{Experiment of the $(\pi^+, K^+)$ hypernuclear spectroscopy at HIHR}

\begin{figure}[hb]
\begin{center}
\includegraphics[width=0.9\hsize]{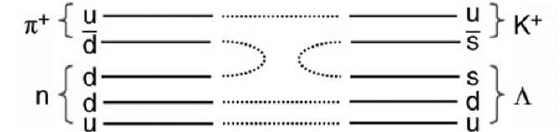}\vspace*{5mm}\\
\caption{The elementary process of the $n(\pi^+, K^+)\Lambda$. 
A neutron in a nucleus is converted by this reaction to a $\Lambda$ hyperon. 
Some fraction of produced $\Lambda$ is sticked to the nucleus to produce 
a $\Lambda$ hypernucleus.}
\label{FIG:2-1}
\end{center}
\end{figure}

The proposed experiment investigates $\Lambda$ hypernuclei in wide mass range 
with the $^A Z(\pi^+,K^+) _\Lambda^A~Z$ reaction (Fig.~\ref{FIG:2-1}).

%%%%
%Fig2-1
%%%%%

There are various methods for hypernuclear spectroscopy. It can be categorized in the reaction spectroscopy 
such as the missing mass spectroscopy with the  $(\pi^+,K^+)$ reaction and others which include decay product 
studies such as gamma and decay-pion spectroscopies. The production information is complementary 
to the information obtained by decay product studies.  
Reaction spectroscopy with the missing mass technique is quite powerful way of investigation of hypernuclei. 
It provides reaction cross sections and detailed information of energy levels of hypernuclear states. 

The hypernuclear reaction spectroscopy started with the $(K^-,\pi^-)$ reaction. An s-quark in the beam $K^-$ is 
exchanged with a d-quark in a neutron to produce a $\Lambda$ in the $(K^-,\pi^-)$ reaction. 
On the other hand, the $(\pi^+,K^+)$ reaction produces $s$ and $\overline{s}$ quark pair associatively and the a $\Lambda$ and 
a $K^+$ are produced simultaneously. These reactions convert a neutron to a $\Lambda$, and thus both $(K^-,\pi^-)$ and 
$(\pi^+,K^+)$ reactions produced the same hypernuclei with the same targets while they are exothermic and endothermic reactions.
Typical cross sections of $\Lambda$ hypernuclei for these reactions are 1~mb/sr and 10~$\mu$b/sr for 
$(K^-,\pi^- )$ and  $(\pi^+,K^+)$ reactions, respectively.

%%%%
% Fig.2-2
\begin{figure}[htb]
\begin{center}
\includegraphics[width=0.8\hsize]{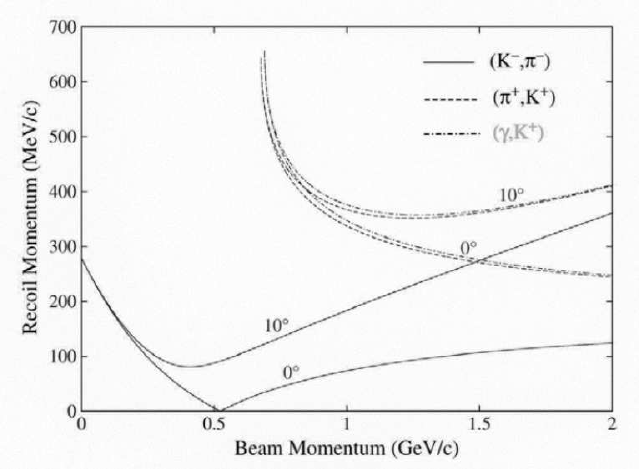}\vspace*{5mm}\\
\caption{Beam momentum dependence of hypernuclear
 recoil momentum with a $^{12}$C target for various reactions \cite{HAS06}.}
\label{FIG:2-2}
\end{center}
\end{figure}
%%%%

Figure~\ref{FIG:2-2} shows recoil momentum transfer to the produced hypernuclei 
as a function of beam momentum. 
The $(K^-,\pi^- )$ reaction gives small momentum transfer, and thus a neutron is converted to a $\Lambda$ 
without changing angular momentum. It populates predominantly substitutional states; the produced $\Lambda$ occupies 
the orbit with the same angular momentum as the converted neutron. Contrary to the  $(K^-,\pi^- )$ reaction, 
$(\pi^+,K^+)$ and $(\gamma,K^+) / (e,e'K^+)$ give large momentum transfer to hypernculei and 
they can populate high-spin states with a neutron hole in large angular momentum orbits and a $\Lambda$ 
in small angular momentum states. This feature allows $(\pi^+,K^+)$ reaction to have a large cross section to populate 
deeply bound $\Lambda$ states like a ground state by converting a neutron with a high angular momentum to a 
$\Lambda$ in $s$ orbit while the cross section of such a state is highly suppressed for the $(K^-,\pi^- )$ reaction 
due to small angular momentum transfer.
The  $(e,e'K^+)$  reaction has similar features to populate deep bound $\Lambda$ hypernuclei as the $(\pi^+,K^+)$ reaction, 
because both are endothermic reactions and momentum transfer is large. However, its hypernuclear production cross section is 
extremely small, 0.1~$\mu$b/sr which is 100 times less than it of the $(\pi^+,K^+)$ reaction. 
In order to compensate small production cross section, very high intensity primary electron 
beam (100~$\mu$A=$6.3 \times 10^{14}~e^-/sec$) is used. 
High quality electron beam allows to avoid measuring momentum and position of electron beams and combination of high-resolution spectrometers for kaon and scattered electrons make it possible to measure hypernuclear mass with a resolution of 0.5 MeV (FWHM) , however accidental coincidence background originating from the high intensity primary electron beam causes a serious problem about signal-to-noise ratio \cite{TAN14, GOG18}.
So far, the $(\pi^+,K^+)$ reaction spectroscopy has limitations for resolution as well as hypernuclear yields. 
Since $\pi^+$ beam is produced as the secondary particles and its large momentum spread makes analysis of 
beam momentum inevitable. Intensity of $\pi^+$ beam is limited by operational rate limit of detector system on the beam line. 
Due to these reasons, the energy resolution and beam intensity for the $(\pi^+,K^+)$ hypernuclear spectroscopy have 
been limited to be a few MeV and several million $\pi^+$s per a second, respectively.
Introduction of HIHR will remove those limitations from the  $(\pi^+,K^+)$ $\Lambda$ hypernuclear spectroscopy and 
it will realize the energy resolution of $<0.4$~MeV (FWHM) which is equivalent or better than it of $(e,e'K^+)$ spectroscopy 
and such a good energy resolution enables us to determine binding energies of various $\Lambda$ hypernuclei with 
an accuracy of $<100$~keV. Furthermore, there is no limitation for $\pi^+$ beam intensity. 
Detail of HIHR will be given in following sections. 
It should be noted that the $(\pi^+,K^+)$ reaction populates spin non-flip states in forward angles while the 
$(\gamma,K^+) /(e,e'K^+)$ reaction populates both spin-flip and non-flip states because photon has spin 1. 
This spin states selectivity of the $(\pi^+,K^+)$ reaction spectroscopy may simplify the analysis of heavy hypernuclei of 
which density of energy levels becomes high. Since the $(\pi^+,K^+)$ reaction converts a neutron to a $\Lambda$ and 
the $(e,e'K^+)$ reaction converts a proton, different hypernuclei which belong to the same iso-spin multiplet will 
be studied with the same target. They are highly complementary, and it is important to have the same level of accuracy 
for these reaction spectroscopic techniques in order to discuss subtle effect of charge symmetry breaking effects 
of the $\Lambda$N interaction.
Table~\ref{TAB:2-I} summarizes the features of $(K^-,\pi^- ), (\pi^+,K^+)$ and $(e,e'K^+)$ reaction spectroscopy of $\Lambda$ hypernuclei.
After the first proposal of the $(\pi^+,K^+)$ reaction spectroscopy by H.Thiessem \cite{THI80}, pioneering experiments were 
performed at BNL AGS \cite{MIL85,PIL91} and then intensive research was carried out at KEK-PS with 
SKS spectrometer \cite{HAS95, HAS96, HAS98, HOT01}.

\begin{table}
\begin{center}
\caption{Reactions for $\Lambda$ hypernuclear spectroscopy}
\label{TAB:2-I}
\begin{tabular}{cccccc}
\hline\hline
Reaction & Resolution & Cross section & Momentum & Elementary & Signal to  \\
            & (MeV, FWHM) & ($\mu$b/sr) for & Transfer & process & Noise ratio\\
\hline 
$(K^-, \pi^-)$ & $>2$ & 1000 & Small & $n (K^-, \pi^-) \Lambda$ & $\bigcirc$ \\
\hline
$(\pi^+, K^+)$ & $1.5$ & 10 & Large & $n (\pi^+,K^+) \Lambda$ & $\bigcirc$ \\
& {\bm $\mathbf \rightarrow 0.4$ (HIHR)}\\
\hline
$(e,e^\prime K^+)$ & $0.5$ & 0.1 & Large & $p (e,e^\prime K^+) \Lambda$ & $\bigcirc$ \\
\hline\hline
\end{tabular}
\end{center}
\end{table}

\begin{figure}[htbp]
\begin{center}
\includegraphics[width=0.7\hsize]{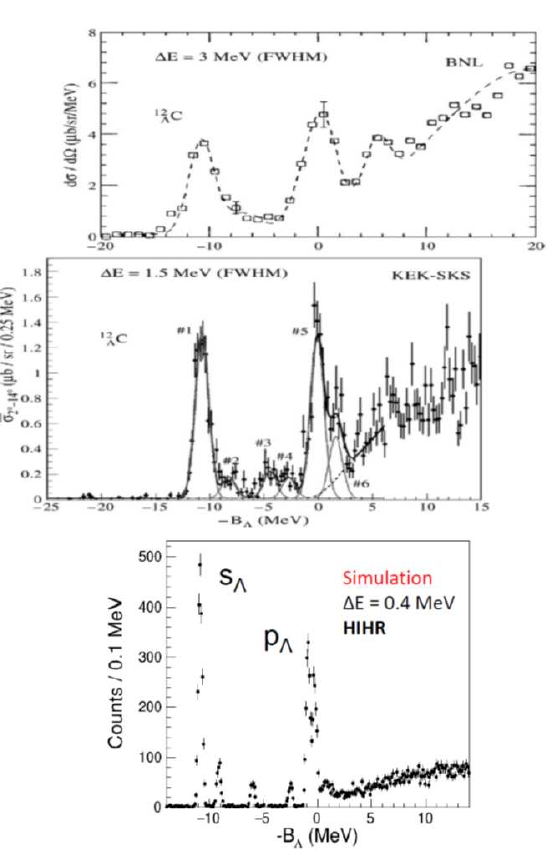}\vspace*{5mm}\\
\caption{$\Lambda$ binding energy spectra of $^{12}_\Lambda$C measured at BNL-AGS, KEK-SKS and HIHR(expectation). 
Simulation for HIHR was carried out based on theoretical calculation for major peaks \cite{MOT10} and quasi-free $\Lambda$ 
production events, background events were generated to be consistent with the experimental data \cite{HOT01}.}
\label{FIG:2-3}
\end{center}
\end{figure}

Figure~\ref{FIG:2-3} shows $\Lambda$ binding energy spectra of $^{12}_\Lambda$C measured with the
$(\pi^+,K^+)$ reaction at BNL-AGS, KEK-SKS and J-PARC-HIHR (expectation). 
The energy resolutions are respectively 
3~MeV, 1.5~MeV and 0.4~MeV (expectation) for BNL, KEK and HIHR. Comparing the spectra from BNL and KEK, 
one can easily see that a factor of two improvement of the energy resolution provides much richer information. 
Further improvement of energy resolution from KEK to HIHR by a factor of more than three, determination 
precision of energies for major peaks which correspond to $\Lambda$ in $s$, $p$-orbits is improved drastically as well as it will 
enable us to separate sub-peaks originating from core nucleus excited states. Since heavier hypernuclei have more complex 
nuclear structure, better energy resolution contributes to improve sensitivity of small cross section peaks.

\begin{figure}[htb]
\begin{center}
\includegraphics[width=0.7\hsize]{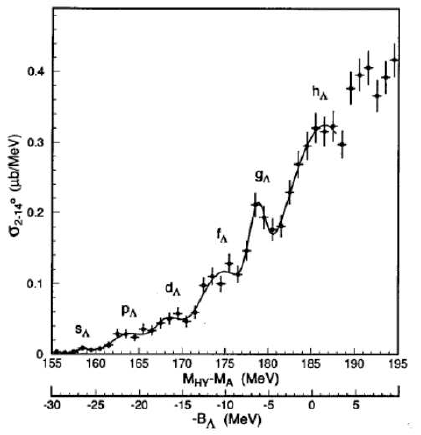}\vspace*{5mm}\\
\caption{$\Lambda$ binding energy spectra of $^{208}_\Lambda$Pb measured at KEK-SKS \cite{HAS96}. 
Major neutron hole series of $(i_{13/2})^{-1}$ with a $\Lambda$ in $s, p, d, f, g, h$ orbits are barely 
observed as $s_\Lambda, p_\Lambda, d_\Lambda, f_\Lambda, g_\Lambda$ and $h_\Lambda$ 
peaks with an energy resolution of 2.3~MeV.}
\label{FIG:2-4}
\end{center}
\end{figure}

Figure~\ref{FIG:2-4} shows $^{208}_\Lambda$Pb energy spectrum measured with SKS at KEK \cite{HAS96} and 
peaks corresponding to the states with $\Lambda$ in $s,p,d,f,g,h$-orbits are barely observed for a major neutron hole series 
of  $(i_(13/2) )^{-1}$ but a clear observation of these peaks was hampered by overlapping with sub-major neutron series 
from a neutron hole in $(h_(9/2) )^{-1}$ where a neutron was converted to a $\Lambda$.
Due to small cross section of $^{208}_\Lambda$Pb, a thicker target (3.418~g/cm$^2$)  was used at SKS and 
it results in energy resolution of 2.3~MeV (FWHM), though a thinner carbon target (0.859~g/cm$^2$) was used 
to achieve so far best energy resolution of 1.45~MeV as the $(\pi^+,K^+)$ reaction. 
Sub-MeV resolution is necessary to observe clear peaks for heavy hypernuclei and a high intensity $\pi^+$ beam 
enables us to use a thin target enough not to deteriorate resolution; these requirements can be met only at HIHR.

In the following section, the physics justification of the measurement will be given. 

%%%
% Fig 2.3
%%%

%%%
% Fig 2.4
%%%

\subsubsection{Neutron stars and the hyperon puzzle}

Neutron stars (NS) are the most compact and dense stars in the universe, with typical masses 
$M \sim 1.4~M_\odot$ and radii $R \sim 10$~km. Their central densities can be several times larger 
than the nuclear saturation density, $\rho_0 = 0.16$~fm$^{-3}$. Since the Fermi energy of fermions at such 
densities is in excess of tens of MeV, thermal effects have little influence on the structure of NS. 
Therefore, they exhibit the properties of cold matter at extremely high densities, very far from being realized 
in present terrestrial experiments. In the era of multi-messenger astronomical observations, NS offers a 
unique opportunity to test a broad class of theories, from nuclear physics to general relativity, including the 
recent observation of gravitational waves \cite{ABB17} and X-ray hotspot measurement by NICER 
which constraints on mass-radius ratio \cite{PAN21, RAA21}. 
These recent progresses in astronomical observations have deepened our understanding of {\bf macroscopic} 
features of NS, such as mass, radius, and stiffness. Therefore, {\bf microscopic understanding becomes more 
important than ever; we should answer what kind of physics determines the size and mass of NS and why NS is so stiff.} 

%%%%
% Fig 2-5
\begin{figure}[htb]
\begin{center}
\includegraphics[width=0.7\hsize]{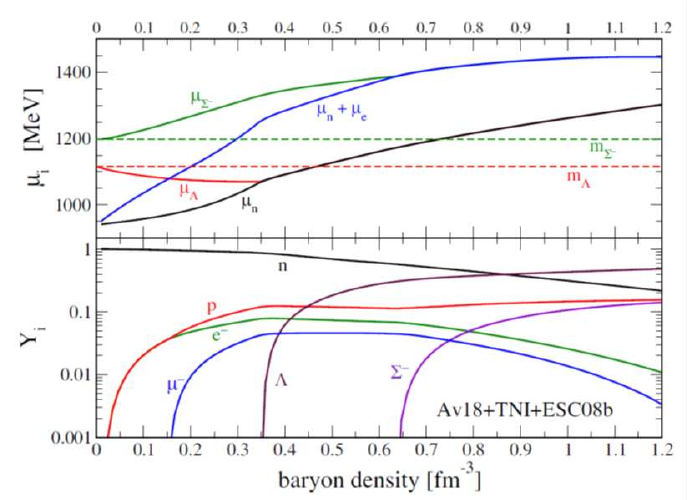}\vspace*{5mm}\\
\caption{Chemical potentials $\mu$, and concentrations $Y$ of the stellar constituents 
in hyperonic matter as a function of the baryon density \cite{BOM16}.}
\label{FIG:2-5}
\end{center}
\end{figure}
%%%%

From the surface to the interior of a NS, stellar matter undergoes a number of transitions. From electron and neutron-rich ions 
in the outer envelopes, the composition is believed to change into a degenerated gas of neutrons, protons, electrons, and 
muons in the outer core. At densities larger than $\sim 2\rho_0$, new hadronic degrees of freedom or exotic phases are 
likely to appear. Figure~\ref{FIG:2-5} shows the chemical potentials and concentrations of stellar constituents in beta-stable 
hyperonic matter as a function of baryon density, obtained from a recent theoretical calculation employing modern baryonic 
potentials \cite{BOM16}. 
The appearance of hyperons in the core of a NS was already advocated in 1960 \cite{AMB60}. In the degenerate dense matter 
forming the inner core of a NS, Pauli blocking would prevent hyperons from decaying by limiting the phase space 
available to nucleons. When the nucleon chemical potential is large enough, the conversion of nucleons into hyperons 
becomes energetically favorable. This results in a reduction of the Fermi pressure exerted by the baryons and a softening 
of the equation of state (EOS). As a consequence, the maximum mass determined by the equilibrium condition between 
gravitational and nuclear forces is reduced. The value of about $1.5~M_\odot$ for the maximum mass of a NS, inferred 
from neutron star mass determinations \cite{THO99}, was considered the canonical limit, and it was compatible with most 
EOS of matter containing strangeness. However, the recent measurements of the large mass values of the millisecond pulsars 
J1614-2230 ($1.97(4)~M_\odot$) 
%\cite{DEM10},
\cite{Demorest:2010bx} PSR J0348+0432 ($2.01(4)~M_\odot$) \cite{ANT13}
% bug?? maybe Antoniadis:2013pzd ?? : 
% Corrected in bibtex SNN (2021728)
and 
MSP J0740+6620 ($2.14(20)~M_\odot$)
%\cite{CRO19]}
\cite{CRO20} % maybe correct
require a much stiffer equation of state.  

This seems to contradict the appearance of strange baryons in high-density matter given what is known at present about 
the hyperon-nucleon interaction. This apparent inconsistency between NS mass observations and theoretical calculations 
is a long-standing problem known as “hyperon puzzle”. Its solution requires better understanding of the 
hyperon-nucleon (YN) interaction in a wide range of systems from light to medium and heavy hypernuclei as well 
as more accurate theoretical calculation frameworks.

Currently there is no general agreement among the predicted results for the EOS and the maximum mass of NS including 
hyperons. This is due to incomplete knowledge of the interactions governing the system (both two- and three-body forces 
for hypernuclei) as well as difficulties of theoretical treatment of many-body systems. There are theoretical calculations 
which were applied to the hyperonic sector and resulted in the appearance of hyperons at around $2-3~\rho_0$, and 
a strong softening of EOS, implying a sizable reduction of the maximum mass \cite{VID11, 
%HJS11
SCH11, MAS12}. On the other hand, 
other approaches suggest much weaker effects arising from the presence of strange baryons in the core of the 
star \cite{BED12,WEI12,MIY13,LOP14}. 

Rich nucleon-nucleon (NN) scattering data allow one to derive satisfactory models of two-body nuclear forces, 
either purely phenomenological \cite{WIR95-1,
%WIR95-2
WIR02} or built on the basis of an effective field theory \cite{MAC96, EPE05, EKS13, GEZ13}. 
In contract to NN scattering, quite limited scattering data are available for hyperon-nucleon (YN) scattering, though a 
$\Sigma$N scattering experiment was recently performed at J-PARC \cite{MIW11} and a new $\Lambda$p scattering experiment 
is planned. Furthermore, there is no scattering data for $\Lambda$-neutron an additional information on the three-nucleon interaction is inferred from saturation properties of 
symmetric nuclear matter (Urbana IX force \cite{PUD95}), the resulting PNM EOS turns out to be stiff enough to be compatible 
with astrophysical observations \cite{GAN12, GAN14}.

\begin{figure}[htb]
\begin{center}
\includegraphics[width=0.6\hsize]{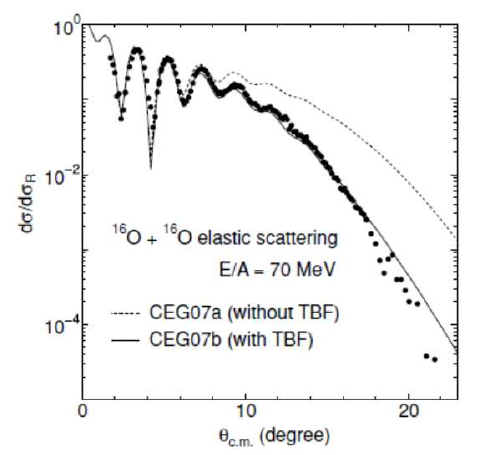}\vspace*{5mm}\\
\caption{Differential cross sections for $^{16}$O+$^{16}$O elastic scattering at $E/A = 70$~MeV \cite{FUR09}. }
\label{FIG:2-6}
\end{center}
\end{figure}

Recent analysis of $^{16}$O+$^{16}$O scattering data shows that the established meson exchange 
potential model (Nijmegen ESC08c \cite{NAG14}) cannot reproduce the cross section at large scattering angles where 
density of the system becomes large and inclusion of 3-body force (TBF) solves the problem as shown in Fig.~\ref{FIG:2-6} \cite{FUR09}. 

%%%%
% Fig 2-6
%%%%

%%%%
% Fig 2-7
\begin{figure}[htb]
\begin{center}
\includegraphics[width=0.6\hsize]{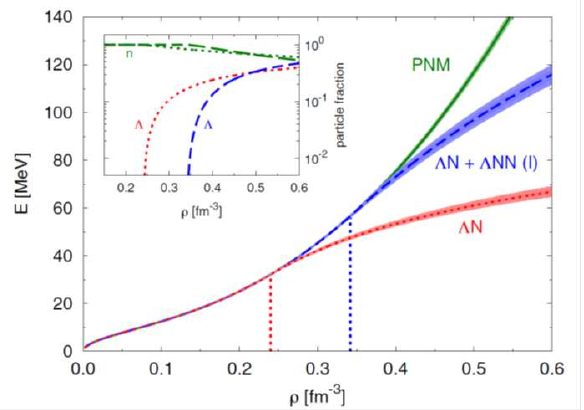}\vspace*{5mm}\\
\caption{Energy as a function of baryon density \cite{LON15}. The vertical dotted lines indicate the 
$\Lambda$ threshold densities. In the inset, neutron 
and $\Lambda$ fractions correspond to the two hyper-neutron matter EOSs.}
\label{FIG:2-7}
\end{center}
\end{figure}
%%%%

It is a quite promising scenario to solve the hyperon puzzle that inclusion of such 3-body repulsive force in the $\Lambda$NN 
interaction makes EOS harder or prevents cross over of chemical potentials of neutron and $\Lambda$ to suppress the 
appearance of $\Lambda$ in NS.

Figure~\ref{FIG:2-7} shows one example of EOS of neutron star calculated by quantum Monte Carlo method \cite{LON15}. 
The EOS of NS without hyperons (Pure Neutron Matter; green) is hard but inclusion of $\Lambda$ makes the EOS soft ($\Lambda$N; red) 
where density of system becomes higher ($\rho >1.5~\rho_0 \sim 0.24$ fm$^{-3}$),  and inclusion of 3-body repulsive force (blue), 
stiffness of EOS recovers. Inset plot shows neutron and $\Lambda$ fractions. One can see that the 3-body repulsive 
force pushes the $\Lambda$ appearance threshold up.

Thus, there is a general indication that 3-body repulsive forces become quite significant at high density, and it harden 
the EOF of neutron stars. The binding energies of light hypernuclei with high statics YN scattering data are necessary 
to construct realistic YN interaction models, but they are not enough to solve the hyperon puzzle. 
Additional information must necessarily be inferred from the properties of medium and heavy hypernuclei 
in order to extrapolate to the infinite-mass limit for discussion of highly massive asymmetric nuclear matter such as 
neutron stars and strange hadronic matters ($n_u \sim n_d \sim n_s$). 

Such information is essential to extrapolate the hyperon behavior to the infinite limit and thus to reliably predict neutron star 
properties.

\subsubsection{Theoretical models of structure calculation of $\Lambda$ hypernuclei }

Except for very light $\Lambda$ hypernuclei which ab-inito or detailed cluster calculation can handle, 
the structure of $\Lambda$ hypernuclei is generally studied by employing the weak-coupling 
approximation that assumes the wave function of a $\Lambda$ hypernucleus to be decomposed into a core nucleus and 
a $\Lambda$ hyperon. In this picture, the hypernuclear Hamiltonian consists of the Hamiltonian for the core-nucleus, 
the $\Lambda$ kinetic energy and the sum of $\Lambda$N interaction terms that can be derived with various theoretical frameworks. 

Realistic nuclear interaction models have been constructed based on rich nucleon-nucleon (NN) scattering data 
but scarce hyperon-nucleon (YN) and completely no YY scattering data limits construction of reliable baryonic interaction models.
Therefore, the analysis of hypernuclear binding energies with the limited number of available hyperon-nucleon (YN) scattering 
data play key role to formulate various baryon-baryon interaction models. Traditionally, interaction models based on meson exchange 
picture such as Nijmegen ESC \cite{RIJ10, NAG14, 
%NAG15
NAG19} and Bonn-Jülich \cite{REU94, HAI05} are widely used. 
Recently SU(3) chiral effective field theory (ChEFT) \cite{HAI13} attracts attention since it allows improvement of precision 
of the calculation in systematical way by power counting and two- and three-body forces are derived in a consistent way.  
Recent progresses of lattice QCD calculation with massive CPU power computers enable us also to derive the baryonic 
interaction potential models \cite{AOK10, AOK11, ISH12} and coming closer to a level where they give additional constraints 
on the baryonic interaction, however, direct calculation of light baryonic systems by lattice QCD is still challenging task.

Ab-initio calculations and microscopic cluster calculations enable us to use these bare interactions to calculate the binding energies 
of light hypernuclear systems \cite{WIR18, HIY09}.
However, analysis of heavier hypernuclei needs effective interactions which have been derived from the bare $\Lambda$N force 
by various calculation techniques such as G-matrix methods based on Bruckner theory \cite{RIJ99} because heavier nuclear systems 
are too complicated for those microscopic calculations, and it is practically impossible to take all quantum many-body effects 
into account.

In order to perform theoretical analysis of binding energy information of hypernuclei in wide mass range, following information 
are necessary: 1) reliable two-body baryonic interaction, 2) systematic method to convert bare two-body interaction to the 
effective interaction which takes quantum many body effects adequately into account, and 3) nuclear structure calculation 
with the derived effective interaction model.

\begin{figure}[htb]
\begin{center}
\includegraphics[width=0.7\hsize]{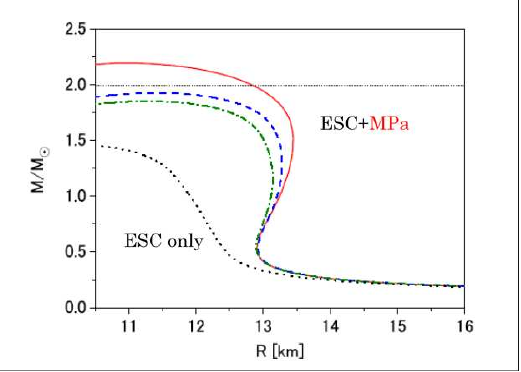}\vspace*{5mm}\\
\caption{Neutron star masses as a function of the radius \cite{YAM14}. Dotted black curve is for ESC 
only without three body forces and it was shown that inclusion of hyperon makes EOS very soft. 
Red solid line shows the case with multi-body forces, MPa. Blue 
and green lines are different parameter set of multi-body forces \cite{YAM14}.}
\label{FIG:2-8}
\end{center}
\end{figure}

Let us give one example of such calculations: a meson exchange baryon-baryon interaction, ESC08c \cite{RIJ10}, which takes 
two-meson and meson pair exchanges as well as various one boson exchanges into account with broken SU(3) symmetry, 
is selected as the bare baryon-baryon interaction. As already recognized in the non-strange sector by the study of $^{16}$O+$^{16}$O 
scattering data \cite{FUR09}, the ESC08c model needs to be supported by a 3/4-body repulsive force originating from multi-Pomeron 
exchange (MPa) that can be universally extended to the strange sector. Finally, effective YN and YY interaction models 
were constructed from ESC08c+MPa by Gaussian parametrization of the G-matrices theory in nuclear matter and 
EOS of neutron star was constructed. Then, Tolmann-Oppenheimer-Volkoff (TOV) equation was solved for the hydrostatic 
structure and mass-radius relations of neutron stars was obtained \cite{YAM14}.
Figure~\ref{FIG:2-8} shows neutron star masses as a function of the radius. Dotted black curve is for ESC only without 
three body forces and it was shown that inclusion of hyperon makes EOS very soft. Inclusion of three-body force makes 
EOS stiffer and maximum mass of NS becomes larger; the case for MPa, two solar-mass NS can be supported. 

%%%
% Fig 2-8
%%%
% Fig 2-9
\begin{figure}[htb]
\begin{center}
\includegraphics[width=0.6\hsize]{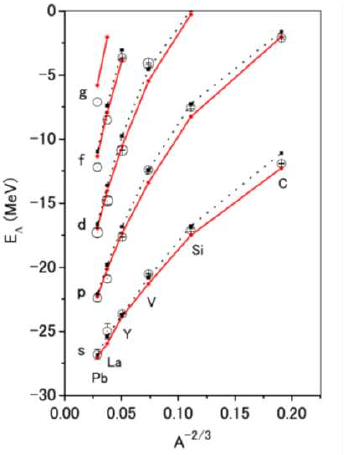}\vspace*{5mm}\\
\caption{Energy spectra of various $\Lambda$ hypernuclei. Experimental data were marked by 
open circles and theoretical calculation results with 3-body force (MPa) is red solid line and without 
3-body force (ESC) is black dots \cite{YAM14}. }
\label{FIG:2-9}
\end{center}
\end{figure}
%%%

\begin{figure}[htb]
\begin{center}
\includegraphics[width=0.6\hsize]{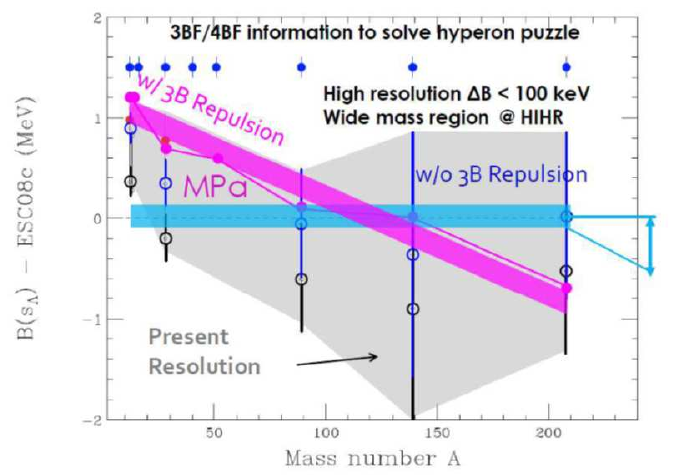}\vspace*{5mm}\\
\caption{Expected effects of 3-body repulsive $\Lambda$NN force for various $\Lambda$ hypernuclei \cite{YAM14}. 
Open black circles are previous data obtained by previous  $(\pi^+, K^+)$ spectroscopy of $\Lambda$ 
hypernuclei. Recently possible ambiguity of 0.5~MeV was pointed out \cite{GOG16, BOT17} for all 
these data and blue circles shows the experimental data with this ambiguity. 
Expected precision of experiments at HIHR was shown by error bars of closed circles.}
\label{FIG:2-10}
\end{center}
\end{figure}

 Obtaining $k_f$-dependent local potential derived from the G-matrices calculation for the interaction which recovered 
the stiffness of NS, hypernuclear energy spectra were calculated with Skyrme Hatree-Fock wave functions (Fig.~\ref{FIG:2-9}). 
Figure~\ref{FIG:2-10} shows the effects of three-body force which were obtained from the $\Lambda$ binding energies 
in $s$-orbit with multi-body force (ESC+MPa) and without it (ESC only). Energy differences of the experimental data 
obtained by previous $(\pi^+,K^+)$  spectroscopy and calculated values without multi-body force were plotted, too. 
Though 3-body force plays a significant role for NS, the effects for hypernuclear energies are small, typically less than 1~MeV. 
It is clear that the experimental data have too large errors as well as ambiguities to constraint theoretical models. 
Experiments at HIHR will provide $\Lambda$ binding energies in wide mass region with an accuracy of $< 100$~keV. 
It should be noted that similar discussions for $\Lambda$ binding energies for $p,d,f,...$-orbits are possible and 
interaction models are further constrained with those information when reliable $\Lambda$ binding energies are obtained at HIHR.

%%%
% Fig2-10
%%%%

There are variety of theoretical calculations depending on choice of bare interactions and treatment of quantum many-body problem.  
The effective G-matrix potential from the Nijmegen potential including many-body force was combined 
with a microscopic calculation, Hyper-AMD to calculate the binding energies of  $^{12}_\Lambda$B, $^{13}_\Lambda$C, 
$^{16}_\Lambda$O, $^{28}_\Lambda$Si, and $^{51}_\Lambda$V \cite{ISA16-1, ISA16-2, ISA17}. 
The different version of Nijmegen potential with Pomeron and Odderon exchanges (ESC16) with the G-matrix calculation \cite{NAG19} was also carried out.

\begin{center}
\begin{figure}[htbp]
\begin{minipage}[b]{0.5\linewidth}
%\centering
%\includegraphics[width=0.9\hsize]{HIHRfig/HIHR-2-11a.pdf}\vspace*{5mm}\\
\includegraphics[width=0.9\hsize]{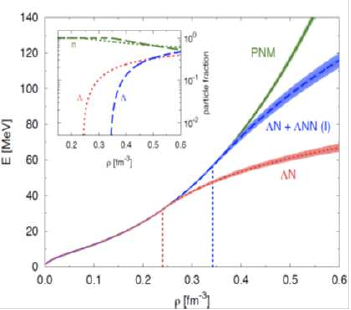}\vspace*{5mm}\\
{\small (a) Equations of state. The vertical dotted lines indicate the $\Lambda$ threshold densities. 
In the inset, neutron and $\Lambda$ fractions correspond to the two hyper-neutron matter EOSs.}
\end{minipage}\hspace*{5mm}
\begin{minipage}[b]{0.5\linewidth}
%\centering
\includegraphics[width=0.9\hsize]{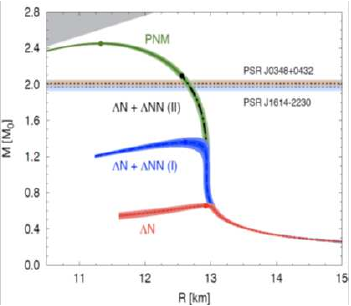}\vspace*{5mm}\\
{\small (b) Mass-radius relations given by AFDMC. Closed circles represent the predicted 
maximum masses. Horizontal bands at $2 M_\odot$ are
 the observed masses of the heavy neutron stars
 \cite{Demorest:2010bx,ANT13}.}
 %\cite{DEM10, ANT13}.}
 \end{minipage}
\caption{EOS and neutron star mass-radius relations calculated by AFDMC \cite{LON15}. }
\label{FIG:2-11}
\end{figure}
\end{center}

\begin{center}
\begin{figure}[htbp]
\begin{minipage}[b]{0.5\linewidth}
%\centering
\includegraphics[width=0.9\hsize]{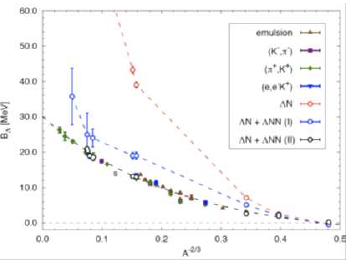}\vspace*{5mm}\\
{\small (a) Experimental $B_\Lambda$ values in s wave and AFDMC calculation results with 
2-body $\Lambda$N interaction alone, and two different 
parametrizations of the 3-body YNN interaction (updated from \cite{LON14}).}
\end{minipage}\hspace*{5mm}
\begin{minipage}[b]{0.5\linewidth}
%\centering
\includegraphics[width=0.9\hsize]{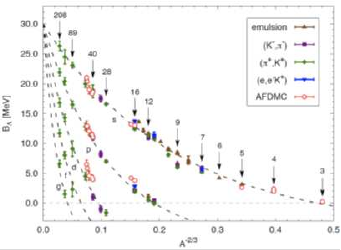}\vspace*{5mm}\\
{\small (b) Experimental results for $\Lambda$ in $s, p, d, f$ and $g$ orbits. 
Red open circles are the AFDMC results obtained including the most recent 2-body plus 3-body hyperon-nucleon 
phenomenological interaction model (updated from \cite{PED15}).}
\end{minipage}
\caption{$\Lambda$ separation energies as a function of $A^{-2/3}$ calculated by AFDMC \cite{LON15}. }
\label{FIG:2-12}
\end{figure}
\end{center}

Recently, the auxiliary field diffusion Monte Carlo (AFDMC) technique for strange systems has made substantial progresses. 
By using this microscopic ab-initio approach, an accurate analysis of the $\Lambda$ separation energy of light- and 
medium-heavy hypernuclei has been carried out \cite{LON14, PED15} using a phenomenological interaction \cite{BOD84, USM95-1, USM95-2, USM95-3, USM95-4, IMR14} 
in which the two-body potential has been fitted on the existing $\Lambda$p scattering data.
As shown in Fig.~\ref{FIG:2-11}(a), when only the two-body $\Lambda$N force is considered (red curve), the calculated hyperon separation energies tend to 
disagree with the experimental data (green curve) as the density increases. This has a sizable effect on the predicted NS structure, Fig.~\ref{FIG:2-11}(b).  
The inclusion of the three-body $\Lambda$NN force in this scheme leads to a satisfactory description of the hyperon separation energies in a wide mass range 
and for the $\Lambda$ occupying different single particle state orbitals ($s, p$ and $d$ wave), as shown in Figs. \ref{FIG:2-12}(b). 
The resulting EOS spans the whole regime extending from the appearance of a substantial fraction of hyperons at $\sim 2\rho_0 \simeq 0.32$~fm$^{-3}$ 
to the absence of $\Lambda$ particles in the entire density range of the star, as shown in Fig.~\ref{FIG:2-12}(a). 

Recently SU(3) Chiral Effective Field Theory (ChEFT), which based on chiral symmetry of QCD Lagrangian, attracts wide
 attention \cite{POL06, PET16, KOH18, HAI20}. ChEFT established a prescription to extend the interaction model to include many-body forces. 
Performed next-to-leading order (NLO) calculations already included some components of the 3-body force (the ones that can be reduced to 2-body terms), 
but next-to-next-to-leading order (NNLO) calculations are necessary for the inclusion of a genuine 3-body interaction, and more experimental inputs 
such as high statistics YN scattering data to be available at K1.1 beam line of J-PARC extended hadron hall and high precision binding energy information 
of light hypernuclei, are required to constraint the Low-Energy Constants which are necessary parameters to make ChEFT predictable.

%%%%%%%
% Fig2-11
%%%%%%%
% Fig2-12
%%%%%%%

Although the above calculations are based on different bare potentials: meson exchange Nijmegen potential, phenomenological one or ChEFT, 
and they adopt different approaches how to apply these potentials to hypernuclei: G-matrix calculation, microscopic ab-initio quantum Monte Carlo 
calculation, the general tendencies of results are consistent.
They predict that inclusion of 3-body repulsive force makes relatively small differences in the $\Lambda$ separation energies of hypernuclei but it gives 
dramatically different results for the properties of the infinite medium like a NS: hardening of EOS or even suppression of the hyperon 
appearance \cite{YAM14, LON15, GER20}.  

However, while 2-body baryonic force models based on different theoretical frameworks are expected to be reasonably accurate 
from new data of YN scattering at K1.1 J-PARC and information of light hypernuclei, detailed information on the 3-body hyperon-nucleon interaction 
is still missing. This lack of knowledge is to be attributed to a poor experimental information for medium to heavy hypernuclei, 
which are the key to infer properties of the infinite hyper-nuclear matter. Therefore, in order to properly assess the role of hyperons in 
NSs and reconcile theoretical predictions with astrophysical observations, $i.e.$ solve the hyperon puzzle, precise experimental investigation on light, 
medium and heavy targets is of paramount importance. In order to provide such accurate experimental data efficiently, realization of HIHR beam line is essential. 

\subsubsection{Choice of targets, additional physics outputs}
Other proposals for wide variety of targets will follow after the success of the proposed experiment but let us limit the choice targets for 
the first campaign of hypernuclear research at HIHR for now.
As shown in Fig.~\ref{FIG:2-10}, 3-body force effect depends on mass number of hypernuclei, choice of targets should cover wide mass range from 
light to heavy targets. As discussed in section~\ref{SEC:targetthick}, thin targets are necessary to realize a high-resolution experiment. 
Therefore, self-supporting solid targets with a thickness of about millimeters were chosen for the first campaign of hypernuclear experiments at HIHR. 
Surely, the main goal of the proposed experiment is to solve the hyperon puzzle by systematic study of $\Lambda$ hypernuclei in wide mass range, 
but physics outputs of high resolution spectroscopy of $\Lambda$ hypernuclei are not limited to that.   
So far $^{12}$C target was used as a reference for the $(\pi^+,K^+)$ spectroscopy of hypernuclei due to its easy handling. 
Because the proposed experiment would be the first experiment at HIHR, beam line commissioning including dispersion matching parameters’ optimization 
will be carried out.

Li, Be, B targets can be prepared as self-supporting targets and they are theoretically within scope of precise few body calculation techniques, 
such as microscopic cluster model \cite{HIY09} and no-core shell model calculation which was extended to $^7_\Lambda$Li  for hypernuclei \cite{LE20} 
and to $^{12}$C for normal nuclei \cite{NAV00}. 
Importance of $\Lambda$N-$\Sigma$N coupling is widely recognized for discussion of Charge Symmetry Breaking (CSB) 
of $\Lambda$ hypernuclei  \cite{HIY01,GAZ16,NOG19}. For heavier and neutron richer hypernuclear systems, such $\Lambda$N-$\Sigma$N coupling and 
3-body force become more important, implying that the behavior of $\Lambda$ in symmetric nuclear matter and neutron-rich environments would 
be quite different. It was pointed out that the $\Lambda$N-$\Sigma$N coupling is quite important also for the discussion of NS \cite{
%HAI07
HAI17}.
Since there is high resolution ($< 1$~MeV resolution) spectroscopic experiments with electron beams for $^7_\Lambda$He \cite{NAK13,GOG16-2}, $^9_\Lambda$Li \cite{GOG21}
 and $^{10}_\Lambda$Be \cite{GOG16} at JLab, and thus measurement of their isospin multiplet partners, $^7_\Lambda$Li, $^9_\Lambda$Be and
$^{10}_\Lambda$B are important to study with same or better energy resolution at HIHR; such precise data will enable us to discuss CSB and 
the $\Lambda$N-$\Sigma$N coupling in detail.
The region of medium mass hypernuclei is also interesting. Precision measurement of them will clarify the single-particle behavior of a $\Lambda$ hyperon 
in the nuclear system, investigation of baryonic many-body system with strangeness degree of freedom, and the effective $\Lambda$-N interaction. 
Dynamics of a core nucleus, which is a many-body system of protons and neutrons, couples to a $\Lambda$ hyperon motion which is free 
from Pauli effects of nucleons, and then new symmetric states, “genuine hypernuclear states”, will appear \cite{MOT21, ISA13}. 
Deformation of nuclei can be studied by measuring these states.

 Though recent progresses of microscopic calculations such as Hyper-AMD \cite{ISA17} and AFDMC \cite{LON14} make it possible to discuss medium-heavy 
hypernuclei, the mean-field picture is also important.

%%%
%Fig2-13
\begin{figure}[htb]
\begin{center}
\includegraphics[width=0.85\hsize]{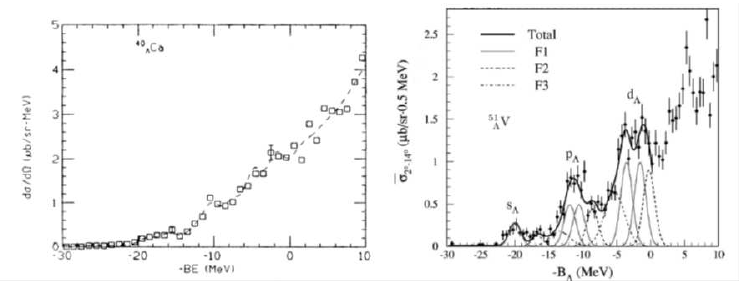}\vspace*{5mm}\\
\caption{$^{40}$Ca \cite{PIL91} and $^{51}$V \cite{HOT01} spectra obtained by the $(\pi^+, K^+)$ reaction. }
\label{FIG:2-13}
\end{center}
\end{figure}
%%%

\begin{figure}[htb]
\begin{center}
\includegraphics[width=0.6\hsize]{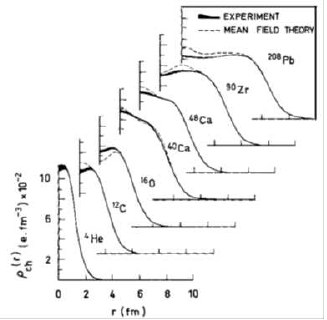}\vspace*{5mm}\\
\caption{Charge density distribution of doubly closed-shell nuclei \cite{FRO87}.}
\label{FIG:2-14}
\end{center}
\end{figure}

In the investigation of hadronic many-body systems with strangeness, there is a fundamental question, “to what extent does a $\Lambda$ hyperon keep its 
identity as a baryon inside a nucleus?” \cite{YAM84}. Spectroscopic data in heavier hypernuclei can help to answer this question. Indeed, the relevance of 
the mean-field approximation in nuclear physics is one of the prime questions related to the role that the sub-structure of nucleons plays in the nucleus. 
The mean-field dynamics dominates the structure of medium to heavy nuclei and $\Lambda$ hypernuclei prove the existence of single-particle motion 
from the deepest $s$-orbit up to large $\Lambda$ valence orbits. The existing data from $(\pi^+, K^+)$ reactions obtained at KEK, however, do not 
resolve the fine structure in the missing mass spectra due to limited energy resolution, and theoretical analyses suffer from those uncertainties 
as shown in Fig.~\ref{FIG:2-13}. The improved energy resolution at HIHR, which is comparable to the spreading widths of the excited hypernuclear states, 
will provide wide variety of information. For example: 1) differentiation between the effects of the static spin-orbit potential and dynamical self-energies 
due to core polarization; 2) access to collective motion of the core nucleus, namely deformation of the core nucleus, utilizing the $\Lambda$ as a probe; 
3) modify energy levels of a core nucleus by adding a $\Lambda$ as an impurity and so on.
Effective masses of a $\Lambda$ hyperon in the nuclear potential will be obtained, which appear to be closer to that of the free value in contrast 
to the case of ordinary nuclei. Therefore, the proposed precision measurement of the single particle levels can address the degree of 
non-locality of the 
effective $\Lambda$-Nucleus potential and can be compared, for example, with the advanced mean field calculations based on the quark-meson 
coupling (QMC) model \cite{TSU98} and on DDRH \cite{KEI00}. This can be related to the nature of the $\Lambda$N and $\Lambda$NN interactions, 
and to the $\Lambda$N short range interactions \cite{MOT88}. In a more exotic way, the binding energies were discussed in terms of the 
distinguishability of a $\Lambda$ hyperon in the nuclear medium, which will result in a different $A$ dependence of the binding energy 
as suggested by Dover \cite{DOV87}.
  Lead target is a holy grail for the $\Lambda$ hypernuclear spectroscopy. The $^{208}$Pb nucleus is doubly magic nucleus and heaviest 
target which can be easily available as a self-supporting plate. Its charge density distribution is nearly constant for a very large fraction (~70\%) 
of the nuclear volume as shown in Fig.~\ref{FIG:2-14} \cite{FRO87}, and its properties were expected to reflect those of uniform nuclear matter.
List of targets for the proposed experiment will be compiled in section~\ref{SEC:targetthick} with the requesting beamtime.

%%%
%Fig 2-14
%%%

\subsubsection{Experimental setup}

Details of HIHR beam line were already given in section~\ref{SEC:HIHRbeamline}.
Detection of 0.7~GeV/$c$ $K^+$ is possible with the well-established experimental techniques. 
Detector package based on it of SKS in K1.8 \cite{TAK12} can be adopted, namely a plastic scintillation counter for 
time-of-flight measurement, aerogel and lucite Cherenkov counters for kaon identification at the trigger level and 
a drift chamber system to measure positions and angles of $K^+$s at the focal plane of the spectrometer. 
As discussed in section~\ref{SEC:HIHR-Res}, position resolution of 0.2~mm (rms) which was achieved by the SKS detector 
is good enough for the tracking devices at HIHR.

%%%%%%
% Fig 3-7
\begin{figure}[htb]
\begin{center}
\includegraphics[width=0.6\hsize]{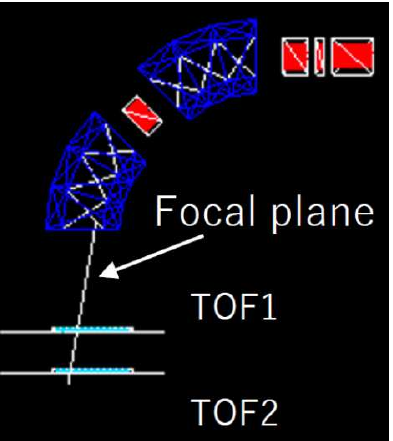}\vspace*{5mm}\\
\caption{A set of plastic scintillation walls are placed just downstream of the focal plane 
in GEANT4 simulation model (detailed discussion about GEANT4 simulation will be given 
in section~\ref{SEC:HIHR-GEANT4}).}
\label{FIG:3-7}
\end{center}
\end{figure}
%%%%%%

\begin{figure}[htb]
\begin{center}
\includegraphics[width=0.6\hsize]{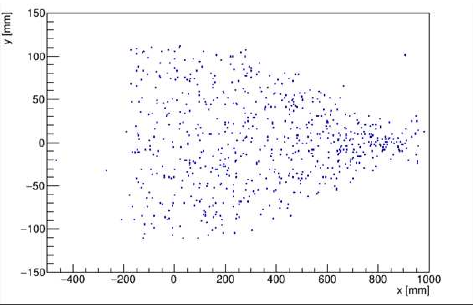}\vspace*{5mm}\\
\caption{Hit position distribution at the front TOF wall.}
\label{FIG:3-8}
\end{center}
\end{figure}

\begin{figure}[htb]
\begin{center}
\includegraphics[width=0.6\hsize]{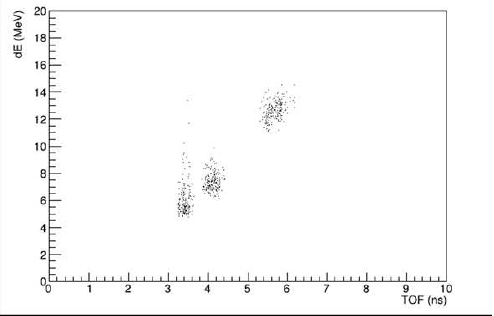}\vspace*{5mm}\\
\caption{Time of flight between two TOF counters vs. energy deposit in TOF1nt TOF wall. }
\label{FIG:3-9}
\end{center}
\end{figure}

Figure~\ref{FIG:3-7} shows a set of plastic scintillation walls placed just downstream of the kaon spectrometer’s focal plane 
with a distance of 1 meter. Particle hit position distribution was estimated with GEANT4 simulation. From the hit position distribution 
shown in Fig.~\ref{FIG:3-8}, TOF walls should cover at least 1200~mm$\times$200~mm area.  
Scatter plot for time of flight between TOF1 and TOF2, and energy deposit in TOF1 for $0.69-0.73$~GeV/$c$
$\pi^+, K+$ and $p$ are given in Fig.\ref{FIG:3-9}. Central values of TOF for $\pi^+, K^+$ and $p$ are respectively 3.39, 4.09 and 5.66~ns. 
The momentum acceptances of them make their widths as 0.11, 0.11 and 0.19~ns (rms). 
Assuming a standard resolution of 0.15~ns for TOF counters, more than 3.7$\sigma$ separation of $K^+$ from $\pi^+$ and $p$ is possible. 
GENAT4 simulation shows a conventional counter technique can separate them clearly. 
At the trigger level of the experiment, aerogel Cherenkov counters are used for pion rejection and 
lucite Cherenkov counters will separate kaons from protons.

%%%%%%
% Fig 3-8
%%%%%%

%%%%%%
% Fig 3-9
%%%%%%

\subsubsection{Expected Yield of Hypernuclei}
The hypernuclear spectroscopy with the $(\pi^+, K^+)$ reaction at the novel dispersion match beam line HIHR aims 
to investigate hypernuclei in wide mass range with a sub MeV resolution while maintaining good hypernuclear yield as well as signal/accidental ratio.
In this section, yield of hypernuclei is estimated based on the current conceptual design of HIHR.

\subsubsection{Primary target and $\pi^+$ beam extraction}
\begin{figure}[htb]
\begin{center}
\includegraphics[width=0.6\hsize]{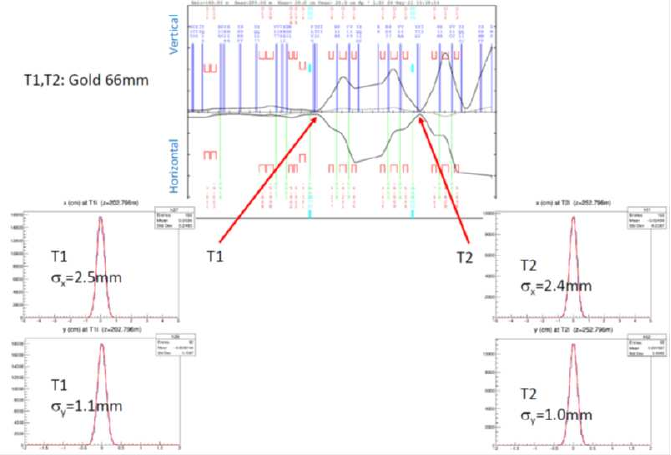}\vspace*{5mm}\\
\caption{Optics study of the beam transport between T1 and T2 primary targets. 
Taking multiple scattering into account, beam spot size of $2.5^H \times 1.0^V$~mm$^2$ can be maintained. }
\label{FIG:4-1}
\end{center}
\end{figure}

A beam spot size of the primary target (T1, 66~mm length gold target) in the existing the hadron hall 
is $2.5^H \times 1.0^V$~mm$^2$. 
It is known that optimization of the primary beam line magnetic components, the beam spot size on the beam optics study 
shows that beam size on the T2 target can be maintained as it on T1 even taking multiple scattering effect into account (Fig.~\ref{FIG:4-1}).

%%%%%%
% Fig 4-1
%%%%%

\begin{figure}[htb]
\begin{center}
\includegraphics[width=0.6\hsize]{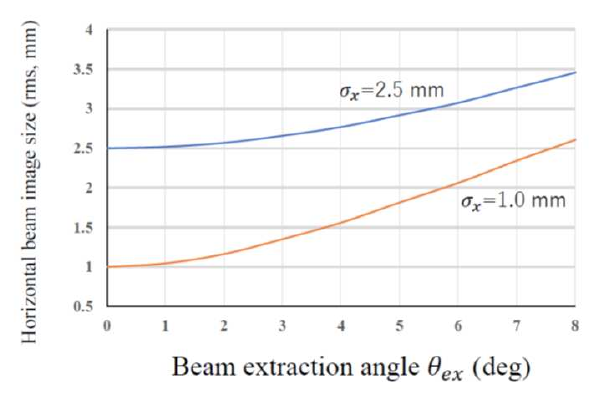}\vspace*{5mm}\\
\caption{Beam extraction angle ($\theta_{ex}$) dependence of horizontal beam image size.  
Energy resolution of HIHR is highly affected by horizontal beam image size but not by vertical one. 
Platinum target of 60~mm length is assumed as the primary target. 
Blue line is for the same beam spot size on existing T1 target ($\sigma_x=2.5$~mm, $\sigma_y=1$~mm) 
and orange line is for ($\sigma_x= 1.0$~mm, $\sigma_y=2.5$~mm) option. }
\label{FIG:4-2}
\end{center}
\end{figure}

The horizontal beam image $f(x)$ is expressed by the convolution of the horizontal primary beam distribution assumed as 
a gaussian on the target and a flat distribution with a target image length $T \cos \theta_{ex}$. Here, $\theta_{ex}$ is a beam extraction angle.  
\[
f(x)= \int^{(T \sin \theta_{ex})/2}_{(-T \sin \theta_{ex})/2} \exp\left( \frac{-(x-u)^2}{2(x_{ptarg})^2} \right) \cos \theta_{ex}~du.
\]
Let us calculate root mean square of $f(x)$ assuming a platinum target of 60~mm length ($T=60$~mm) of which material thickness 
is equivalent to it of 66~mm gold. 
Figure~\ref{FIG:4-2} shows the $\theta_{ex}$ dependence of root mean square of horizontal beam image distribution $f(x)$. 
Though smaller extraction angle $\theta_{ex}$ and beam size ($x_{ptarg}$) on the target are favored in terms of beam image size 
which affects directly the energy resolution of hypernuclear spectroscopy, careful optimization is necessary to have a reasonable $\pi^+$ 
yield, beam spot size and not to disturb other beam lines in the extended hadron hall. 
It should be noted that vertical beam image size is almost independent of target length and beam extraction angle. 

%%%%
% Fig4-2
%%%%

\begin{figure}[htb]
\begin{center}
\includegraphics[width=0.6\hsize]{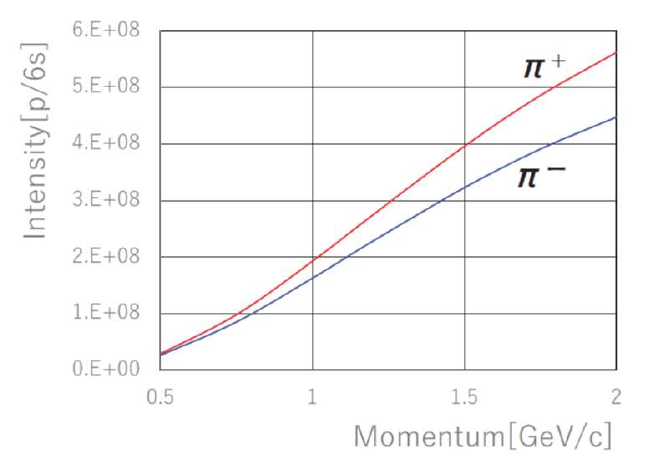}\vspace*{5mm}\\
\caption{Expected beam intensity as a function of beam momentum for 30-GeV, 50-kW primary 
proton beam is irradiated on a 60-mm thick platinum target and pion beam is extracted with an angle of 6~degrees. 
The calculation was performed with the Sanford-Wang formula \cite{SanfordWang1967,SAN60-2}.}
\label{FIG:4-3}
\end{center}
\end{figure}

\begin{figure}[htb]
\begin{center}
\includegraphics[width=0.6\hsize]{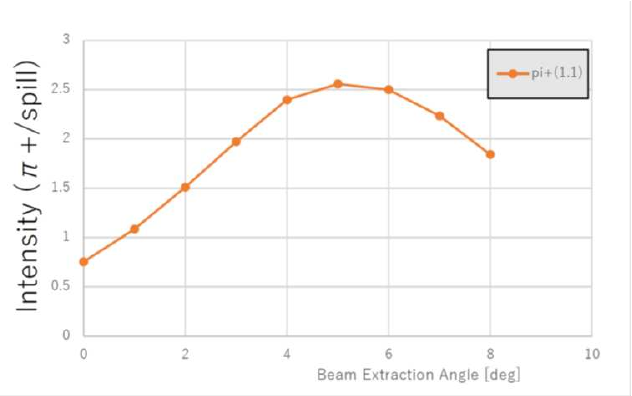}\vspace*{5mm}\\
\caption{Expected beam intensity as a function of beam extraction angle for 30-GeV, 50-kW 
primary proton beam is irradiated on a 60-mm thick platinum target, 
calculated by using the Sanford-Wang formula \cite{SanfordWang1967,SAN60-2}. }
\label{FIG:4-4}
\end{center}
\end{figure}

Figure~\ref{FIG:4-3} shows expected beam intensities in the case that a 30-GeV, 50-kW primary proton beam is irradiated on 
a 60-mm thick platinum target. About $2.5 \times 10^8$ $\pi^+$s per beam spill for 1.1~GeV/$c$ $\pi^+$ with the beam extraction 
angle of $\theta_{ex}= 6$~degrees. 
Figure~\ref{FIG:4-4} shows the extraction angle dependence of $\pi^+$ intensity. 
The beam extraction angle should be optimized in a range of $3-6$ degrees taking other beam lines conditions into account. 
For now, extraction angle of 3~degrees for 50-kW primary beam and $\pi^+$ intensity of $2 \times 10^8$  $\pi^+$/spill 
at the experimental target are assumed for following hypernuclear yield estimation. 
It should be noted that no rate limitation exists for detector system of HIHR and beam with a higher pion rate is desirable 
if future improvement of accelerator makes it possible.

%%%%
% Fig4-3
%%%%

%%%%
% Fig4-4
%%%%

\subsubsection{Solid angle estimation of kaon spectrometer}

%%%%
% Fig4-5
\begin{figure}[htb]
\begin{center}
\includegraphics[width=0.6\hsize]{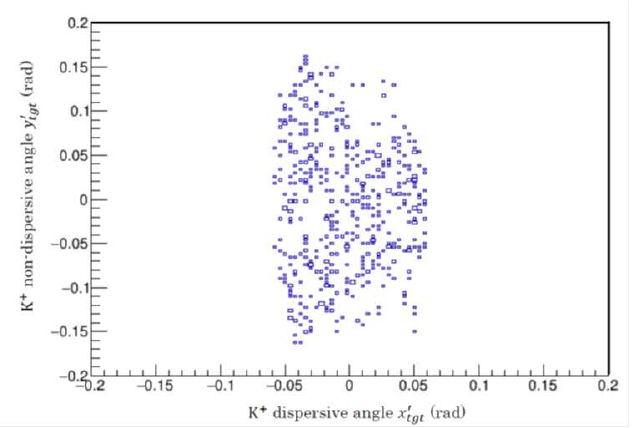}\vspace*{5mm}\\
\caption{Angle distribution of $K^+$ with a momentum of 0.725~GeV/$c$ at the experimental target. 
Events are selected for the particles reached 
to the focal plane. $x_{tgt}'$ and $y_{tgt}'$ are dispersive and non-dispersive angles, respectively.}
\label{FIG:4-5}
\end{center}
\end{figure}
%%%%

Solid angle of the kaon spectrometer of HIHR was estimated based on GEANT4 Monte Carlo simulation. 
Effective width and gap of dipole magnets (D1S, D2S) of the kaon spectrometer were assumed as 100~cm $\times$ 20~cm, respectively. 

Figure~\ref{FIG:4-5} shows angular distribution of 0.725~GeV/$c$ $K^+$s passed through the kaon spectrometer. 
Solid angle of the spectrometer was estimated as $\Omega = 4 \pi N_{fp}/N_{gen}$, where $N_{fp}$ is the number of $K^+$s 
which reach to the focal plane and $N_{gen}$ is the number of uniformly generated $K^+$s in the simulation.

\begin{figure}[htb]
\begin{center}
\includegraphics[width=0.6\hsize]{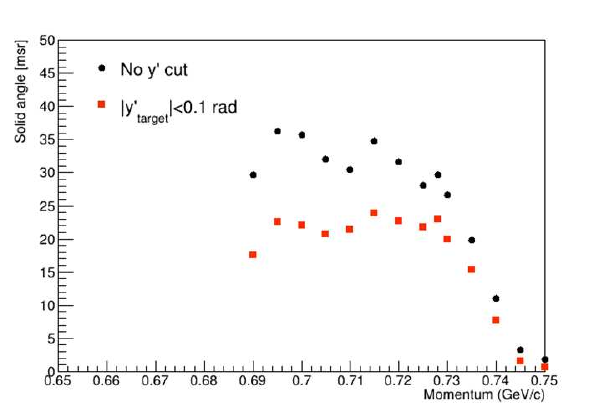}\vspace*{5mm}\\
\caption{Figure 4-6: Solid angle of the kaon spectrometer as a function of kaon momentum. 
Red dots show the solid angle with a selection of $|y_{target}' |<0.1$~rad and block ones without the selection. 
On the experimental target, uniform distribution of reaction points 
for $|x_{target} |<50$~mm was assumed for this simulation.}
\label{FIG:4-6}
\end{center}
\end{figure}

Figure~\ref{FIG:4-6} shows momentum dependence of the solid angles of the kaon spectrometer with and without 
vertical angular selection at the experimental target. There would be a room for optimization, but resolution for the events 
with larger $|y_{target}' |$  tends to be worse. Therefore $|y_{target}' | < 0.1$~rad is now assumed to make the solid angle 
flat over kaon momentum acceptance. Applying this cut, solid angle is about 20~msr over entire momentum acceptance.

%%%%
% Fig4-6
%%%%

\subsubsection{Yield estimation of $^{12}_\Lambda$C}
Reaction spectroscopy of $\Lambda$ hypernuclei normally utilize a carbon target as a reference and commissioning of setup 
because easy handling of solid target and nuclear structure of $^{12}_\Lambda$C has been well studied.

It is assumed that primary proton of 30~GeV, 50~kW bombarded the target of platinum 60~mm length with a beam spot size of 
$2.5 \times 1.0$~mm$^2$. Beam repetition cycle of 5.2~s and the meson extraction angle of 3~degrees was assumed. 
The experimental target of $^{12}$C of 400~mg/cm$^2$ is assumed to achieve about 400~keV energy resolution 
as discussed in the next section. 
 Cross section of $^{12}$C$(\pi^+, K^+)^{12}_\Lambda$C is assumed as 8.1~$\mu$b/sr which is an average of $2-14$~degrees 
for kaon emission angles \cite{HOT01}. In the proposed experiment, kaon emission angle of 0~degrees will be measured and 
thus cross section is expected to be slightly larger than the assumed value. 
Key parameters for $^{12}_\Lambda$C yield estimation are summarized in Table~\ref{TAB:4-I}.

\begin{table}
\begin{center}
\caption{Yield estimation of $^{12}_\Lambda$C hypernucleus at HIHR}
\label{TAB:4-I}
\begin{tabular}{|l|c|}
\hline\hline
    & HIHR@J-PARC \\
\hline
Primary proton beam & 30~GeV, 50~kW\\
& repetition cycle 5.2~s\\
\hline
Central momentum of $\pi^+$ beam & 1.1~GeV/$c$\\
\hline
$\pi^+$ beam rate at target & $3.85 \times 10^7$~/s\\
                                    & (200~M $\pi^+$/spill\\
\hline
Reaction & $^{12}$C$(\pi^+, K^+)^{12}_\Lambda$C\\
\hline
Cross section of ground state of $^{12}_\Lambda$C & 8.1~$\mu$b/sr\\
\hline
Central momentum of $K^+$ & 0.71~GeV/$c$\\
\hline
Solid angle of kaon spectrometer & $>20$~msr\\
\hline
Kaon survival ratio & 0.12\\
& 11.4~m for QSQDMD\\
\hline
\hline
Estimated count rate of $^{12}_\Lambda$C$_{gs}$ & 53.1 counts/hr\\
\hline\hline
\end{tabular}
\end{center}
\end{table}

\subsubsection{Resolution estimation}

Detailed design of HIHR and target will be optimized in collaboration with J-PARC beam line group. 
The energy resolution estimation was carried out with an optics code TRANSPORT and Monte Carlo package GEANT4. 

\subsubsection*{TRANSPORT, optics code study}

The following factors contribute to the total energy resolution of the $(\pi^+, K^+)$ reaction spectroscopy experiment.
\begin{enumerate}
\item Beam momentum resolution.\\
Though a separate discussion of beam momentum resolution is not necessary to evaluate the total energy resolution of
$\Lambda$ hypernuclei for a dispersion matching experiment, the beam momentum resolution plays an important role 
in calibration as well as a possible option of non-dispersion matching experiment. Therefore, it is useful to estimate beam 
momentum resolution based on the first order beam optics. Assuming beam image size at IF2 as $\sigma_x = 1$~mm, 
magnification $(x_{tgt} |x_0)=1.3$ and dispersion $(x_{tgt} |\delta)=11.28$~cm/\%, momentum resolution is estimated 
as $dp/p = 2.4 \times 10^{-4}$ for a 1.1~GeV/$c$ $\pi^+$ beam. It corresponds to 250~keV/$c$ (FWHM). 

\item Kaon spectrometer momentum resolution.\\
Similar to the beam momentum resolution, separate discussion of kaon spectrometer is not necessary to estimate 
the total energy resolution of HIHR. Position resolution of 0.2~mm (rms) and tracking efficiency of 99.8\% is assumed 
as they are achieved for SKS tracking chamber at K1.8 beam line of J-PARC \cite{TAK12}. 
Dispersion $(x_{fp} |\delta)=10.72$~cm/\% and a point source are assumed, momentum resolution of 
$dp/p = 1.9 \times 10^{-4}$ for a 0.71~GeV/$c$ $K^+$ beam. It corresponds to 32~keV/$c$ (FWHM). 
Beam size on the target will contribute in addition to this value. With magnification of $(x_{fp} |x_{tgt})=1.84$ and reaction point uncertainty 
of 1.3~mm, momentum resolution including this contribution is estimated as 370~keV/$c$ (FWHM). 
Momentum dispersion match operation needs a thin solid target but additional tracking device after the target will make it 
possible to optically separate beam and kaon spectrometers for non-dispersion matching operation. 
It should be noted that high resolution (sub-MeV) spectroscopy is also possible for a thick gaseous target with 
this non-dispersion matching operation of HIHR at the cost of maximum $\pi^+$ intensity.

\item Energy loss and straggling in the target.\\
Since we have no information of reaction point in a solid target (typically the thickness of the target is less than 
a millimeter for a material thickness of $100-400$~mg/cm$^2$), energy loss of charged particles can be corrected only 
as an average. Its distribution including straggling will contribute the final mass resolution. 
Most probable energy loss ($\Delta_p$) and straggling ($w$) were calculated with Landau-Vavilov formula and compiled in 
Tab.~\ref{TAB:5-I}. Angular distribution affects the effective thickness of the target, however, horizontal angular acceptance 
of the kaon spectrometer is $\pm50$~mr and this effect is estimated as $\{1/\cos(5 \times 10^{-2})\}-1|\simeq 1.3 \times 10^{-3}$
 which can be safely neglected.
\end{enumerate}

With the above estimation into account, the second order ion optics calculation with TRANSPORT was performed.

\begin{table}
\begin{center}
\caption{Most probable energy loss and straggling for typical solid targets with a thickness of 50/100/200~mg/cm$^2$ . 
They correspond to the cases for reaction at the center of 100/200/400~mg/cm$^2$ thick target. Unit is keV.}
\label{TAB:5-I}
\begin{tabular}{|c|c|c|c|c|c|}
\hline\hline
Target & Thickness& \multicolumn{2}{|c|}{1.1~GeV/$c$ $\pi^+$} & \multicolumn{2}{|c|}{0.71~GeV/$c$ $K^+$} \\
\cline{3-6}
 & (mg/cm$^2$) & $\Delta_p$ (MPV) & $w$ (FWHM) & $\Delta_p$ (MPV) & $w$ (FWHM)\\
\hline \hline
$^{10}$B & 200 & 136 & 31 & 124 & 36\\
\hline
$^{12}$C & 100 & 65 & 16 & 59 & 18\\
\hline
$^{12}$C & 200 & 136 & 31 & 124 & 36\\
\hline
$^{12}$C & 400 & 283 & 62 & 260 & 72\\
\hline
$^{28}$Si & 400 & 258 & 62 & 232 & 72\\
\hline
$^{51}$V & 400 & 221 & 56 & 196 & 65\\
\hline
$^{89}$Y & 400 & 203 & 55 & 176 & 63\\
\hline
$^{139}$La & 400 & 182 & 51 & 156 & 59\\
\hline
$^{208}$Pb & 400 & 162 & 49 & 135 & 57\\
\hline\hline
\end{tabular}
\end{center}
\end{table}

%%%
%Fig 5-1
\begin{figure}[htb]
\begin{center}
\includegraphics[width=0.7\hsize]{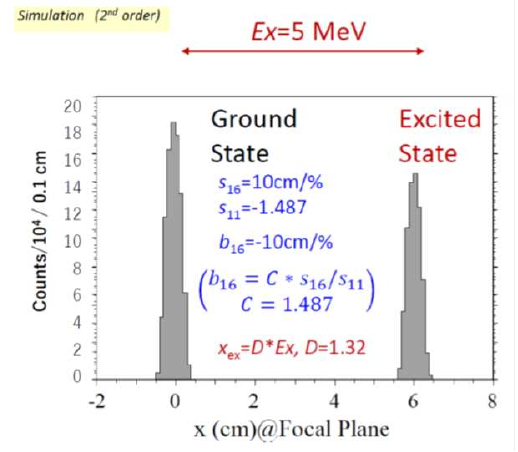}\vspace*{5mm}\\
\caption{Horizontal position distribution at the focal plane calculated by the second order optical calculation.  }
\label{FIG:5-1}
\end{center}
\end{figure}
%%%%

Figure~\ref{FIG:5-1} shows horizontal position distribution of $K^+$s emitted at the focal plane calculated 
by the second order optical calculation. Kaon emission angle at the experimental target was assumed as $\theta_K=0$ and 
energy loss and straggling effects in the target were not included in this calculation. Therefore, this result reflects purely beam line 
and spectrometer optical features. The first peak corresponds to the ground state of $^{12}_\Lambda$C and 
the second peak is an artificially created excited state with $E_x= 5$~MeV. 
Peak width was fitted as $\sigma = 0.161$~cm and it can be converted to the energy resolution (optics) as:
\[\Delta E_{opt} = 2\sqrt{2 \ln 2} \times 0.161~{\rm cm} \times \frac{5~\rm MeV}{6.02~\rm cm} = 315~{\rm keV (FWHM)}.\]
This resolution shows ideally achievable value for a very thin target with complete dispersion matching conditions are satisfied. 

Let us summarize results of energy resolution studies with the optics code TRANSPORT in Table~\ref{TAB:5-II}. 
 Taking  energy loss and straggling effects into account, energy resolution is estimated as better than 0.4~MeV 
for the ground state of $^{12}_\Lambda$C.

\begin{table}
\begin{center}
\caption{Summary of energy resolution study with TRANSPORT}
\label{TAB:5-II}
\begin{tabular}{|c|c|}
\hline\hline
 & TRANSPORT\\
 & Optical code\\
\hline
Reaction  & $^{12}$C$(\pi^+, K^+)^{12}_\Lambda$C\\
\hline
$\pi^+$ beam line momentum resolution (FWHM) & 0.25~MeV/$c$\\
\hline
$K^+$ spectrometer momentum resolution (FWHM) & 0.37~MeV/$c$\\
\hline\hline
Energy resolution of  & 0.32~MeV\\
\hline
Energy loss and straggling & 0.09~MeV\\
in target (100~mg/cm$^2$ $^{12}_\Lambda$C) & \\
\hline\hline
Mass resolution of $^{12}_\Lambda$C & {\bf 0.33~MeV}\\
(FWHM) &\\
\hline\hline
\end{tabular}
\end{center}
\end{table}

\subsubsection{GEANT4, Monte Carlo simulation study}\label{SEC:HIHR-GEANT4}

As shown in Fig.~\ref{FIG:3-4}, the HIHR dispersion matching beam line after IF2, 
an experimental target and the kaon spectrometer were modeled in GEANT4 simulation code. 
Optics code is useful for quick optimization of magnet parameters, and it is suitable to examine 
the rough behavior of the system. On the contrary, Monte Carlo simulation takes longer time and 
tunes of beam element parameters are not straightforward as the optics code does, but it is necessary to 
investigate the behavior of a complex system in which various factors are intertwined.  
Ideal magnetic fields for dipole, quadrupole and sextupole magnets were now assumed in the model and 
they should be replaced by more realistic fields map calculated by 3D finite element method such as TOSCA 
or measured magnetic field after detailed design or fabrication of magnets.

%%%%
%Fig5-2
\begin{figure}[htb]
\begin{center}
\includegraphics[width=0.7\hsize]{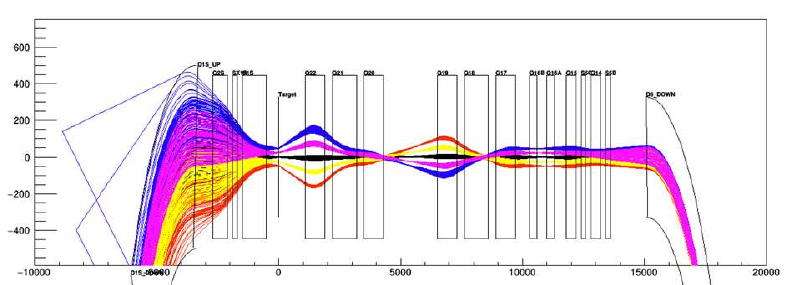}\vspace*{5mm}\\
\caption{ Example of beam trajectories simulated by GEANT4. Beam color (red, yellow, black, magenta and blue) 
corresponds to $\pi^+$ 
beam momenta of 1.045, 1.0475, 1.050, 1.0525 and 1.055~GeV/$c$, respectively. }
\label{FIG:5-2}
\end{center}
\end{figure}
%%%%

Figure~\ref{FIG:5-2} shows GEANT4 simulated beam trajectories. It can be seen that momentum spread of beam
$\pi^+$ was converted to spatial distribution at the experimental target ($z=0$) as expected.

In the simulation, following processes were taken into account. Passing through the experimental target, beam 
$\pi^+$ loses its energy, experiences multiple scattering and finally react with a target nucleus through 
the $^{12}$C$(\pi^+, K^+)^{12}_\Lambda$C reaction. In this section, excitation energies of 0, 5, 10, 15~MeV were 
considered (they are nothing to do with real $^{12}_\Lambda$C structure) and $K^+$ emission angles were uniformly 
distributed in the angular acceptance.

%%%%
% Fig 5-3
\begin{figure}[htb]
\begin{center}
\includegraphics[width=0.7\hsize]{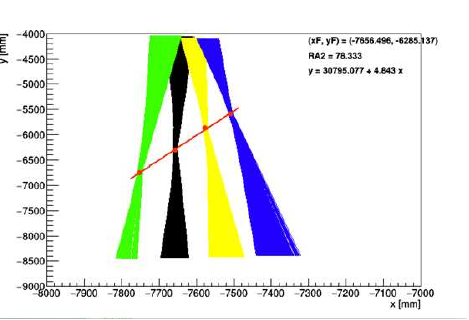}\vspace*{5mm}\\
\caption{Example of $K^+$ trajectories simulated by GEANT4. Beam color (green, black yellow and blue) 
corresponds to the excitation energy of 15, 10, 5, 0~MeV for $^{12}$C$(\pi^+, K^+)^{12}_\Lambda$C reaction. 
Measurement of beam position gives excitation energy.  }
\label{FIG:5-3}
\end{center}
\end{figure}
%%%%

%%%%
% Fig 5-4
\begin{figure}[htb]
\begin{center}
\includegraphics[width=0.9\hsize]{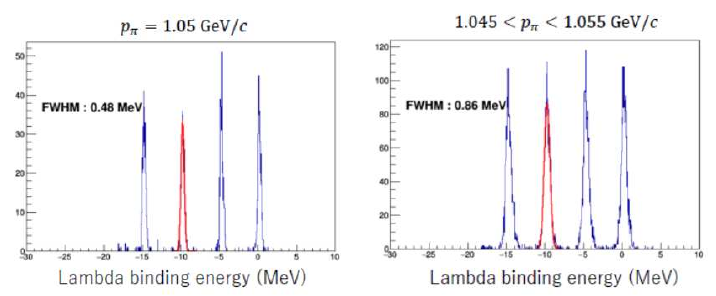}\vspace*{5mm}\\
\caption{Lambda binding energy spectrum for $p_\pi=1.05$~GeV/$c$ (left) and 
$1.045< p_\pi<1.055$~GeV/$c$ (right).}
\label{FIG:5-4}
\end{center}
\end{figure}
%%%%

Figure~\ref{FIG:5-3} shows that $K^+$s with the same excitation energy converge at the focal plane and 
it is different from the case that $K^+$s with the same momentum from a point source converge at the focal plane 
of a normal spectrometer. 

The $K^+$ horizontal position distribution was converted to Lambda binding energy ($B_\Lambda$). 
Figure~\ref{FIG:5-4}(left) shows $-B_\Lambda$ for beam pion momentum was selected as $p_\pi =1.05$~GeV/$c$, 
and (right) shows it for $1.045< p_\pi <1.055$~GeV/$c$. Energy resolution for $p_\pi =1.05$~GeV/$c$, was 0.48~MeV (FWHM) 
and it for $1.045 < p_\pi <1.055$~GeV/$c$ was 0.86~MeV (FWHM).  It is known that deterioration of energy resolution 
in this simulation was caused by an imperfect alignment of focal planes for different beam momenta. 
Ideally intrinsic resolution of HIHR beam optics system is expected to approach the ideal value, 315~keV (FWHM) as shown 
by TRANSPORT calculation in the previous section. Further GEANT4 simulation study on beam elements tunes with 
higher order optical elements for better dispersion matching condition is in progress. 

It should be noted that the resolution was discussed here assuming use of the kaon spectrometer as a “hardware spectrometer” which 
has a clean focal plane and gives energy or momentum information as a dispersive position information. 
Recent progresses of computer power enable us to obtain resolution of $\Delta p/p = 2 \times 10^{-4}$  for a so-called 
“software spectrometer”, which has no clear focal plane, by using 6 orders polynomial function of position and angle information of 
$x, x', y, y'$ at reference plane \cite{GOG18}. Introduction of such techniques would enable us to compensate higher order aberration 
which cannot be compensated perfectly by hardware and contribute to improve energy resolution.

By using GEANT4 simulation, how target thickness affects energy resolution was studied. 
Target thickness affects energy resolution in following two ways and these effects are automatically included 
in the result of GEANT4 simulation.

\begin{enumerate}
\item Effective thickness changes due to $K^+$ emission angle. This effect is already discussed in the previous section.  and it is safely neglected.
\item Energy loss depends on the reaction position. If the reaction happens near to front surface of the target, 
$p(\pi^+)$ is larger than it is expected for the case the reaction happens at the center of the target, 
since energy loss in the target is smaller. The same thing happens for emitted $K^+$. If the reaction happens near to front, 
actual $p(K^+)$ is also larger than it measured by the kaon spectrometer. If reaction happens near to back surface of the target, 
the opposite effects will be observed. In addition to energy loss effects, energy straggling occurs even if the reaction point does not move.
These effects can be easily estimated with GEANT4 simulation. Preparing relatively thick target in the model and reaction points were 
uniformly distributed along the incoming beam direction, then spatial distribution of $K^+$s at the focal plane was measured and 
converted to the energy scale. This study shows that 1~mm change of reaction point along the beam direction 
in the $^{12}$C target corresponds to 78.5~keV change of missing mass of $^{12}_\Lambda$C. 
Since 200~mg/cm$^2$ of $^{12}$C (density 1.8~g/cm$^3$) is 1.1~mm thick, energy loss and straggling give 87.2~keV deterioration 
to the energy resolution.
\end{enumerate}

\begin{figure}[htb]
\begin{center}
\includegraphics[width=0.9\hsize]{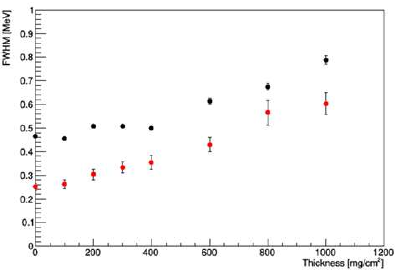}\vspace*{5mm}\\
\caption{Target thickness dependence of the energy resolution for $^{12}$C target. 
Black points and red points are respectively without and with angular acceptance 
selection of $|x_{target}' |<20$~mr and $|y_{target}' |<20$~mr.}
\label{FIG:5-5}
\end{center}
\end{figure}

\label{SEC:targetthick}
Final result of target thickness dependence of energy resolution for $^{12}$C (density 1.8~g/cm$^3$) with  
$p_\pi=1.05$~GeV/$c$ was shown in Fig.~\ref{FIG:5-5}.  Black points show the resolution for all generated events 
and red points for the events with $K^+$ angular acceptance cuts of $|x_{target}'|<20$~mr and $|y_{target}' |<20$~mr.  
Red points give better resolution because momentum matching conditions are less satisfied for the events with larger 
emission angles. It indicates that further study of optics tuning is essentially important as well as limitation of 
angular acceptance results in improvement of resolution (250~keV (FWHM) for the thin target limit) at the cost of yield. 
Looking black points, it can be seen that up to 400~mg/cm$^2$ which corresponds to 2~mm-thick $^{12}$C , 
intrinsic resolution of optics system dominates under currently achieved matching condition. 
For target thickness over 400~mg/cm$^2$ energy straggling and other factors in target begin to contribute 
to the energy resolution. Limitation of $K^+$ angular acceptance makes intrinsic optical resolution good enough 
to see almost pure target thickness dependence as shown by red points.

%%%
% Fig5-5
%%%

\subsubsection{Summary of energy resolution study}\label{SEC:HIHR-Res}

\begin{center}
\begin{figure}[htb]
\begin{minipage}[b]{0.5\linewidth}
%\centering
\includegraphics[width=0.9\hsize]{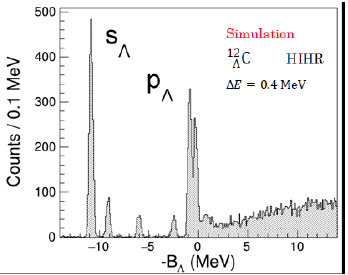}\vspace*{5mm}\\
\end{minipage}\hspace*{5mm}
\begin{minipage}[b]{0.5\linewidth}
%\centering
\includegraphics[width=0.9\hsize]{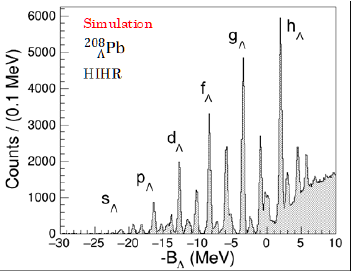}\vspace*{5mm}\\
\end{minipage}
\caption{ Expected $\Lambda$ binding energy spectra for $^{12}_\Lambda$C (left) and 
$^{208}_\Lambda$Pb (right). Resolution of 0.4~MeV (FWHM) and number of the events 
for the ground states of 1000 and 100 were assumed for $^{12}_\Lambda$C 
and $^{208}_\Lambda$Pb, respectively.}
\label{FIG:5-6}
\end{figure}
\end{center}

\begin{figure}[htb]
\begin{center}
\includegraphics[width=0.9\hsize]{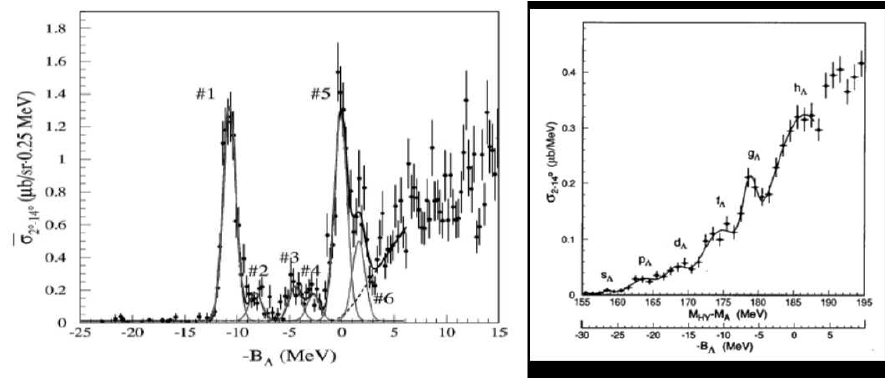}\vspace*{5mm}\\
\caption{Previously measured $\Lambda$ binding energy spectra for $^{12}_\Lambda$C (left) \cite{HOT01} and
 $^{208}_\Lambda$Pb (right) \cite{HAS96}. Resolutions were 1.45~MeV (FWHM) and
 2.3~MeV (FWHM) for $^{12}_\Lambda$C and $^{208}_\Lambda$Pb, respectively.}
\label{FIG:5-7}
\end{center}
\end{figure}

The energy resolution estimation was carried out with an optics code TRANSPORT and a Monte Carlo package GEANT4. 
Optics code has advantages of quick matrix tuning and easier investigation of optical features of the beam line and 
spectrometer systems. However, detailed study with reactions in the target cannot be carried out. 
On the contrary to the optics code, Monte Carlo simulation can easily consider various reactions in materials on the 
beam line but tuning of beam elements is not easy as the optics code. During conceptual design works, both optics code 
and simulation studies will be complementarily performed.
When a reliable design of magnets is finalized and detailed magnetic field maps are implemented in the simulation code, 
a thorough optimization of optical parameters will be performed. After such an optimization, adjustments for minor 
change of experimental condition will be quickly done with a dedicated GEANT4 simulation code to be developed for 
this purpose.
Though further study and optimization of beam line components are necessary, current study shows that the 
dispersion match technique enables an energy resolution of better than 400 keV (FWHM) for less than 400 mg/cm2 targets at HIHR. Limitation of acceptance with a thin target might enable further improvement of the resolution at the cost of hypernuclear yield. Target thickness and acceptance limitation will be optimized depending on the experimental requirements, such as resolution or yield.
Figure~\ref{FIG:5-6} shows expected $\Lambda$ binding energy spectra for $^{12}_\Lambda$C (left) and $^{208}_\Lambda$Pb (right). 
They should be compared with previously measured spectra at KEK (Fig.~\ref{FIG:5-7}).

%%%%
%Fig 5-6
%%%%
% Fig5-7
%%%%

\subsubsection{Beamtime request}
The targets listed in Table~\ref{TAB:6-I} are required to achieve the original goal of this proposal, 
but choice of targets is not limited, because HIHR aims to be the hypernuclear factory for wide variety of experiments.
Since HIHR is a brand-new beam line, commission time for hardware development will be requested separately.
Assuming the energy resolution of HIHR is 400~keV (FWHM) as discussed in section~\ref{SEC:HIHR-Res}, 
more than 100 events of a peak can determine the peak centroid with a statistical error better than 
$400~{\rm keV} / \sqrt{100} \simeq 40~{\rm keV}$. Typically, cross sections of excited bound states of $\Lambda$ 
binding energy of heavy hypernuclei are several times larger than they for the ground states, for example, cross sections of
$\Lambda$ in $s$-orbit (ground state) and $p$-orbit for the $^{208}$Pb$(\pi^+, K^+)^{208}_\Lambda$Pb reaction are 
respectively 0.2 and 0.8~$\mu$b/sr. Therefore, statistical uncertainty of binding energies for excited bound states 
would be a few 10~keV. 
Background shape ambiguity, accuracy of energy scale calibration would contribute to systematic errors, 
but overall uncertainties of binding energies of hypernuclei can be controlled to be less than 100~keV.
Carbon target is normally used as a reference for reaction spectroscopy of hypernuclei. 
Optimized parameters of magnetic elements for dispersion matching condition will be investigated with the carbon runs. 
During beam line tuning, limitation of acceptance will be applied to have clean events with the best energy resolution 
which is optically achievable and thus higher statistics is essentially important. Reference data for various target 
thickness (100, 200, 400 mg/cm$^2$) will be accumulated.
It should be noted that $(\pi^+, K^+)$ reaction hypernuclear spectroscopy cannot use 
hyperon ($\Lambda$ and $\Sigma^0$) masses as the absolute mass calibration sources due to unavailability of 
neutron target, and thus $^{12}_\Lambda$C mass will serve as one of energy references. 
All previous $(\pi^+, K^+)$ experiments used the $^{12}_\Lambda$C mass as the mass calibration, 
but it was pointed out that old $^{12}_\Lambda$C mass measured by nuclear emulsion almost half a century ago 
would have a large error \cite{GOG16, BOT17}. Recent state-of-art hybrid emulsion experiment, J-PARC E07, 
has lots of  $^{12}_\Lambda$C events and analysis of these events will provide a new $^{12}_\Lambda$C mass reference 
in near future. Decay pion spectroscopy experiment at HIHR is now planned \cite{FUJ21, NAG18} and it will serve as 
an independent calibration source. 

For light to medium heavy targets, hypernuclear production cross section for ground state is relatively large 
and thus beamtime for 100 events accumulations with 200~mg/cm$^2$ thick targets are requested.
For heavy targets, cross sections of ground states are small and energy level density is large. Therefore luminosity-oriented 
beamtime to measure ground state which needs high statistics and resolution-oriented beamtime to separate complicate 
excitation energy levels which have larger cross sections than the ground state, are separately requested.
Beamtimes for light to medium-heavy hypernuclei need 724~hours (30~days) and heavy targets need 1764~hours (73~days), 
ground total of 2488~hours (104~days) for the first campaign of experiments at HIHR. 
Divided beamtime allocation is desirable rather than a continuous one, because time for analysis of accumulated data 
is important to feed back its result to the next beamtime.

\begin{table}
\begin{center}
\caption{Summary of requesting beamtime for 50~kW proton beam power. Differential cross sections at
$\theta_K \sim 0$ were estimated by using data of prior $(\pi^+, K^+)$  experiments \cite{PIL91, HAS94, HAS96,  HOT01, HAS06}. }
\label{TAB:6-I}
\begin{tabular}{|c||c|c|c|c|c|}
\hline\hline
 Hypernuclei & Assumed g.s.  & Target thickness & Expected & Requested  & Beam\\
 & cross section & (mg/cm$^2$)                     & Yield (/h) & number of   & time (h)\\
 &  ($\mu$b/sr) &&&events for g.s. &\\
\hline\hline
$^{12}_\Lambda$C & 8.1 & 100 & 13.3 & 1000 & 79\\
\hline
$^{12}_\Lambda$C & 8.1 & 200 & 26.6 & 2000 & 79\\
\hline
$^{12}_\Lambda$C & 8.1 & 400 & 53.1 & 2000 & 39\\
\hline
$^{6}_\Lambda$Li & 1.9 & 200 & 12.7 & 100 & 8\\
\hline
$^{7}_\Lambda$Li & 1.9 & 200 & 10.9 & 100 & 10\\
\hline
$^{9}_\Lambda$Be & 0.2 & 200 & 1.1 & 100 & 98\\
\hline
$^{10}_\Lambda$B & 0.9 & 200 & 3.5 & 100 & 30\\
\hline
$^{11}_\Lambda$B & 0.9 & 200 & 3.2 & 100 & 33\\
\hline
$^{28}_\Lambda$Si & 0.5 & 400 & 1.4 & 100 & 75\\
\hline
$^{40}_\Lambda$Ca & 0.5 & 400 & 0.94 & 100 & 112\\
\hline
$^{51}_\Lambda$V & 1.2 & 400 & 1.8 & 100 & 59\\
\hline
$^{89}_\Lambda$Y & 0.6 & 200 & 0.53 & 100 & 199\\
\hline\hline
Sub-total & & & & & 724\\
(light to  & & & & & (30 days)\\
 midium heavy &&&&&\\
targets) & &&&&\\
\hline\hline
$^{139}_\Lambda$La & 0.3 & 200 & 0.085 & 20 & 236\\
\hline
$^{139}_\Lambda$La & 0.3 & 400 & 0.17 & 80 & 471\\
\hline
$^{208}_\Lambda$Pb & 0.3 & 200 & 0.057 & 20 & 352\\
\hline
$^{208}_\Lambda$Pb & 0.3 & 400 & 0.11 & 80 & 705\\
\hline\hline
Sub-total & & & & & 1764\\
(heavy)  & & & & & (73 days)\\
\hline\hline
Grand &&&&& 2488\\
total  &&&&& (104 days)\\
\hline\hline
\end{tabular}
\end{center}
\end{table}

\subsubsection{Summary of spectroscopic study of $\Lambda$ hypernuclei with the ($\pi^+, K^+)$ reaction at HIHR}
Based on a newly designed momentum matching pion beam line with a specially designed 
kaon spectrometer system, HIHR at the J-PARC hadron extension hall, a campaign of $\Lambda$
 hypernuclear $(\pi^+, K^+)$ reaction spectroscopy experiments was proposed.
Unprecedent energy resolution of $<400$~keV (FWHM) as a hypernuclear reaction spectroscopy 
with high statistics $>100$~events for ground states of various $\Lambda$ hypernuclei will 
enable us to determine $\Lambda$ binding energies with an accuracy better than 100~keV. 
Precise determination of hypernuclear binding energies in wide mass region enables us to clarify 
the existence of the $\Lambda$NN three-body repulsive force which is a key to solve 
the puzzle of heavy neutron stars (hyperon puzzle). HIHR will be a unique hypernuclear 
factory which provides various high precision data to construct a reliable baryonic interaction model 
and such experiments cannot be performed at other facilities.

%%%%% End SNN 20210726

% flatex input end: [HIHR-Phys.tex]

%%%%%% HIHR main part (SNN)

\clearpage

%%%%%% DCX (SNN)
% flatex input: [DCX.tex]
\subsubsection{Spectroscopic study of neutron-rich $\Lambda$ hypernuclei with the double charge exchange (DCX) reaction}

HIHR is designed primarily for the study of $\Lambda$ hypernulei with the $(\pi^+, K^+)$ reaction, but the dispersion match 
technique can be applied to the $\pi^-$ beam.  The double charge exchange reaction, $^A Z(\pi^-, K^+)^A_\Lambda (Z-2)$, enables us to investigate
neutron rich hypernuclei. Isospin asymmetric systems serve as an ideal test ground for study of $\Lambda N - \Sigma N$ coupling effect 
which is one of key issues to solve the hyperon puzzle. 
It should be noted that inclusion of the coherent $\Lambda - \Sigma$ coupling effect which can be treated as the effective three-body force with a hyperon 
is essentially important to theoretically reproduce experimentally-measured energies of the $s$-shell hypernuclei \cite{AKA00}.

As it is well known, some of neutron rich nuclei near the neutron drip line exhibit interesting properties such 
as neutron halos and skins and they are characterized by having loosely bound neutrons \cite{TAN85, KOB88}.
Adding a $\Lambda$ hyperon to such systems,  the glue like role of $\Lambda$ makes the system tighter bound and 
changes their basic properties such as radius and surface diffuseness \cite{HIY04}.

\begin{figure}[htb]
\begin{center}
\includegraphics[width=0.7\hsize]{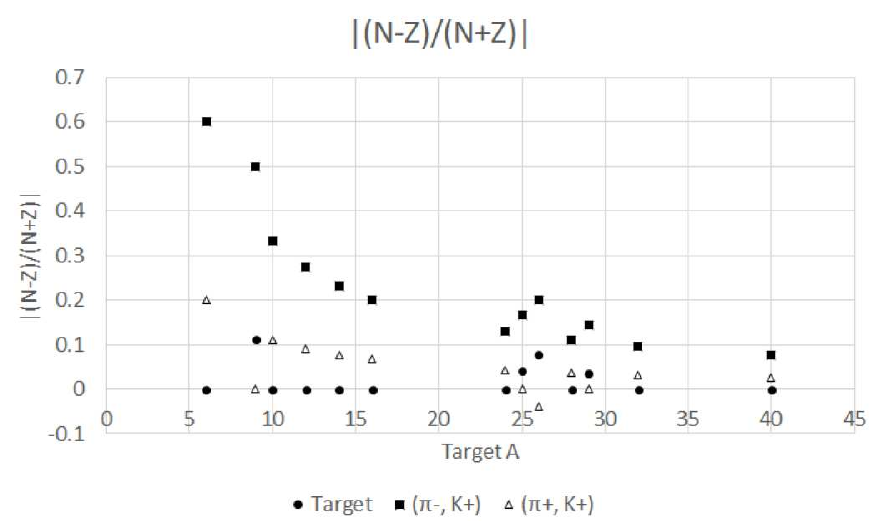}\vspace*{5mm}\\
\caption{The isospin asymmetric parameter, $|N-Z|/(N+Z)$, for various $\Lambda$ hypernuclei produced 
via the $(\pi^-, K^+)$ and $(\pi^+, K^+)$ reactions. Closed circle, closed square, and open triangle show the asymmetric 
parameters for the target nuclei and hypernuclei produced with the $(\pi^-, K^+)$ and $(\pi^+, K^+)$ reactions, respectively.
 \cite{
 %HON18
 Takahashi:2019xcq}.}
\label{FIG:DCX-1}
\end{center}
\end{figure}

Figure~\ref{FIG:DCX-1} shows isospin asymmetry of $\Lambda$ hypernuclei produced by the $(\pi^\pm, K^+)$ reactions.
It tells us that the $(\pi^-, K^+)$ (DCX) reaction produces neutron richer hypernuclei with light targets. 
Though cross sections of DCX are small, it enables us to access the new region of neutron-rich hypernuclei which are beyond neutron drip line for normal nuclei.  

%%%%
% Fig.H-1
%%%%

\begin{figure}[htb]
\begin{center}
\includegraphics[width=0.7\hsize]{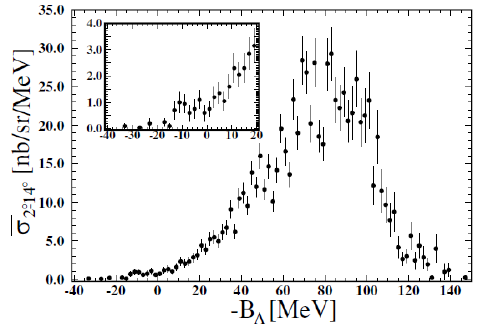}\vspace*{5mm}\\
\caption{Missing mass spectrum of the $^{10}{\rm B}(\pi^-, K^+)$ reaction. Inset shows an expanded view near the threshold \cite{SAH05}.}
\label{FIG:DCX-2}
\end{center}
\end{figure}

Though clear peaks of neutron rich hypernuclei has not been observed yet with the DCX reactions, 
experiments with $(\pi^-, K^+)$ at KEK-PS \cite{SAH05} and $(K^-_{stop},  \pi^+)$ at DA$\Phi$NE \cite{AGN06, AGN12} were already carried out.
For the $^{10}{\rm B}(\pi^-, K^+)^{10}_\Lambda$Li reaction, events in the bound region were observed (Fig.~\ref{FIG:DCX-2}) and
the number of the events was converted to the cross section, $11.3 \pm 1.9$~nb/sr \cite{SAH05}. 
Recently heavy hyper-hydrogen ($^6_\Lambda$H) search experiment was carried out with the $^6$Li$(\pi^-, K^+) X$ reaction and 
an upper limit of cross section, 0.56~nb/sr, was obtained\cite{
%HON17
Honda:2017hjf}.

Assuming 1.20~GeV/$c$ momentum for $\pi^-$ and 0.88~GeV/$c$ for $K^+$ which were the same setting as KEK-PS-E521 \cite{SAH05},
$\pi^-$ intensity at HIHR is $2 \times 10^8 \pi$/spill (Fig.~\ref{FIG:4-3}) which is 50 times stronger than KEK-PS K6 beam line 
and almost the same as $\pi^+$ intensity of 1.05~GeV/$c$ at HIHR.
With an adequate adjustment of momentum dispersion match conditions for $\pi^-$, the same energy resolution of 0.4~MeV is expected
for the $(\pi^-, K^+)$ reaction at HIHR. Therefore, yield estimation needs to take cross section difference between $(\pi^+, K^+)$ 
and $(\pi^-, K^+)$ DCX reactions into account.

Table~\ref{TAB:HIHR-DCX} compares important parameters between prior DCX experiment at KEK-PS, ($\pi^+, K^+)$ and DCX at HIHR.
Assuming a cross section of 10~nb/sr for both $^{9}{\rm Be}(\pi^-, K^+)^{9}_\Lambda$He and $^{10}{\rm B}(\pi^-, K^+)^{10}_\Lambda$Li reactions, 
$^{9}_\Lambda$He yield of 0.092~h$^{-1}$ and 
$^{10}_\Lambda$Li yield of 0.078~h$^{-1}$ are expected. In order to achieve similar precision to $(\pi^+, K^+)$ reaction spectroscopy by accumulating 100 counts, 
1091 hours (= 45 days) and 1282 hours (= 53 days) of beamtimes are necessary
respectively for $^{9}_\Lambda$He and $^{10}_\Lambda$Li 
with 50~kW proton beam on the primary target. 
In order to study wider variety of neutron rich hypernuclei in reasonable beamtime with DCX reaction at HIHR, usage of higher beam intensity or introduction 
of thicker target at the cost of resolution could be considered. 
Further optimization on the experimental conditions is in progress.

%%%%
% Fig.H-2
%%%%

\begin{small}
\begin{table}
\begin{center}
\caption{Important parameters for $(\pi^\mp, K^+)$ reaction spectroscopy of $\Lambda$ hypernuclei.}
\label{TAB:HIHR-DCX}
\begin{tabular}{ccccccc}
\\
\hline\hline
Reaction & Target & $p_\pi$  & $p_K$  & Resolution & 
Cross section& Yield\\
            & thickness & (GeV/$c$) & (GeV/$c$) & (MeV, FWHM) &
 (nb/sr) &  (/h)\\
&(g/cm$^2$)\\
\hline\hline 
\multicolumn{7}{c}{KEK PS K6 \cite{SAH05}}\\ 
\hline
%$^{10}{\rm B}(\pi^+, K^+)^{10}_\Lambda$B & 
%\hline
$^{10}{\rm B}(\pi^-, K^+)^{10}_\Lambda$Li &  3.5 & 1.20 & 0.88 & 2.5 & 
$11.3 \pm 1.9$& \\
\hline\hline
\multicolumn{7}{c}{HIHR}\\ \hline
$^{9}{\rm Be}(\pi^+, K^+)^{9}_\Lambda$Be &  0.4 & 1.05 & 0.72 & 0.4 & 
$2.4 \times 10^2$ & 2.2 \\
\hline
$^{10}{\rm B}(\pi^+, K^+)^{10}_\Lambda$B &  0.4 & 1.05 & 0.72 & 0.4 & 
$9 \times 10^2$ & 7.0 \\
\hline
$^{9}{\rm Be}(\pi^-, K^+)^{9}_\Lambda$He &  0.4 & 1.20 & 0.88 & 0.4 & 
10$^*$ & 0.092 \\
\hline
$^{10}{\rm B}(\pi^-, K^+)^{10}_\Lambda$Li &  0.4 & 1.20 & 0.88 & 0.4 & 
10$^*$& 0.078 \\
\hline\hline 
\multicolumn{7}{r}{$^*$assumed value.}
\end{tabular}
\end{center}
\end{table}
\end{small}

%
%[AKA00] Y. Akaishi et al., Phys. Rev. Lett., 84, 3539 (2000).
%[TAN85] I. Tanihata, et al., Phys. Rev. Lett. 55 (1985) 2676.
%[KOB88] T. Kobayashi, et al., Phys. Rev. Lett. 60 (1988) 2599.
%[HIY04] E. Hiyama, Few-Body Systems 34 (2004) 79.
%[HON18] R, Honda, Write-ups for workshop on the project for the extended hadron experimental facility of J-PARC, by HUA arXiv:1906.02357.
%[SAH05] P. K. Saha et al., Phys. Rev. Lett, 94, 052502 (2005).
%[AGN06] M. Agnello et al., Phys. Lett. B, 640, 145 (2006).
%[AGN12] M. Agnello et al., Phys. Rev. Lett, 108, 042501 (2012).
%[HON17] R. Honda et al., Phys. Rev. C, 96, 014005 (2017).

% flatex input end: [DCX.tex]

%%%%%% DCX (SNN)

%%%%%% HIHR other experiments (Tamura)
%\input{HIHR-others}
% flatex input: [HIHR-others_rev.tex]

\subsection{Other Experiments Planned at HIHR}

\subsubsection{Cusp spectroscopy for $\Sigma$$N$ interaction}

The $\Sigma$$N$ interaction has recently been studied via high-statistics
$\Sigma^+$$p$ and $\Sigma^-$$p$ scattering experiments (J-PARC E40) \cite{MIW11}.
This experiment will make a great contribution to
construction of the realistic $Y$$N$ interaction models.
However, in this type of experiment,  $\Sigma$$N$ scattering cross sections
at very low energies ($p_Y^{lab} <$ 400 MeV/$c$) 
are difficult to measure.
In order to study low-energy $\Sigma$$N$ interaction,
another type of  experiment is proposed
to measure a high-resolution
missing-mass spectrum
for the $d(\pi^+,K^+)$ reaction
around the $\Sigma$$N$ threshold.

\begin{figure}
\centerline{\includegraphics[width=17cm]{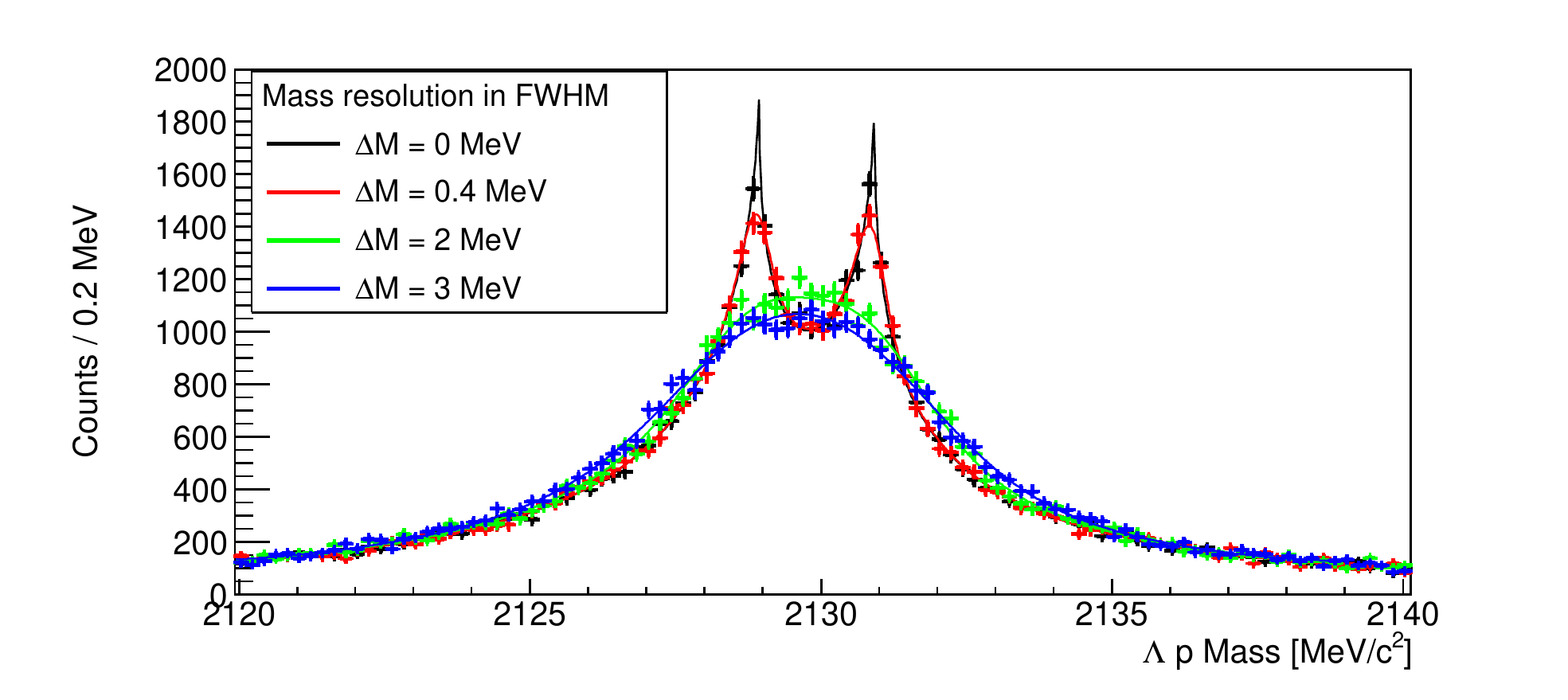}}
\caption{Calculated spectra with the scattering length of $A_{\Sigma} = 2.06 - i4.64$ fm. 
The black line shows the original spectrum and 
the colored lines show the smeared spectra with
experimental resolutions of $\Delta M = 0.4$ (red, HIHR), 2 (green, COSY HIRES\cite{cusp_HIRES}), 
and 3 (J-PARC E27 
%\cite{cusp_E27}
\cite{Ichikawa:2014rva}) MeV in FWHM.
The points with statistical error bars show the expected yield when we assume the total yield as 
$5 \times 10^{4}$ events. 
}
\label{fig:cusp}
\end{figure}

A sharp enhancement (``cusp'') in the $\Lambda p$ invariant mass spectra of 
$K^- d \to \pi^- \Lambda p$, $\pi^+ d \to K^+ \Lambda p$, and $pp \to K^+ \Lambda p$
reactions at the $\Sigma$$N$ threshold was previously observed in several experiments, but 
the obtained peak position and width were not consistent with each other~\cite{cusp_Machner}. 
It was discussed that the mass spectrum can be expressed in terms of the $\Sigma N$ scattering length of 
the spin-triplet $T$(isospin) $= 1/2$ channel \cite{cusp_Dalitz1,cusp_Dalitz12}. 
However, the past experiments did not allow extraction of the scattering length 
due to their limited mass resolution. 
Figure~\ref{fig:cusp} shows the calculated mass spectra when we choose the scattering length as 
$A_{\Sigma} = 2.06 - i4.64$ fm. 
The black line shows the original calculated spectrum without smearing due to the mass resolution. 
The points with statistical error bars show
the expected yield when we assume the total yield as 
$5 \times 10^{4}$ events. 
Two prominent peaks are expected to appear due to $\sim$2 MeV difference 
of the $\Sigma N$ thresholds (2128.9 MeV for $\Sigma^+ n$ and 2130.9 MeV for $\Sigma^0 p$), because the cusp structure should be seen for both 
$\Sigma^+ n$ and $\Sigma^0 p$ thresholds.
The yield ratio should be 
related to the amplitude of the elementary processes of $K^- p \to \pi^- \Sigma^{+}$ and $K^- n \to \pi^- \Sigma^{0}$ and their interference. 
Figure~\ref{fig:cusp} shows the case for the yield ratio as $\Sigma^+ n : \Sigma^0 p = 1 : 1$. 
The colored lines show the smeared spectra with the experimental resolution 
of $\Delta M = 0.4$, 2, and 3 MeV in FWHM. 
The resolution of $\Delta M = 2$ MeV corresponds to the best value of the past experiment 
achieved by HIRES at COSY \cite{cusp_HIRES}. 
The prominent two peak structure in the original spectrum disappears for $\Delta M \geq 2$ MeV, 
and much better resolution is necessary to determine the scattering length from the cusp spectrum. 
Moreover, it is difficult to distinguish the mass spectra 
with $\Delta M \geq 2$ MeV from 
a simple Breit-Wigner distribution. 
Therefore, the mass resolution is a key for the $\Sigma N$ cusp measurement. 
The inconsistency of the peak positions and widths among the past experiments may originate from the yield ratio between the $\Sigma^+ n$ and $\Sigma^0 p$ channels, which is different depending on the reactions and the beam momenta.

In the proposed experiment, a $d(\pi^+,K^+)$ missing mass spectrum will be
measured with a pion beam momentum of 1.7 GeV/$c$ 
with the unprecedented high resolution ($< 0.4 $ MeV (FWHM))
by the HIHR beam line \cite{
%ICH19
Takahashi:2019xcq}.
By tagging the $\Lambda$$p$ final state,
events in the isospin $T=1/2$ channel is separated from those in the $T=3/2$ channel, and
the real and imaginary parts of the $\Sigma$$N$ scattering length for the 
spin-triplet $T=1/2$ channel 
will be accurately obtained for the first time. 
By using the HIHR beam line, not only the $\Sigma N$ cusp but also 
other cusp spectra can be measured with a high resolution.  
In particular, measurement of the $\bar{K} N$ (isospin $T = 1$) cusp is promising by using the $p(\pi^{\pm}, K^+)\Lambda \pi^{\mp}$ 
reaction. 

\subsubsection{Precise decay pion spectroscopy for hypernuclear ground state mass}

In the precise measurement of $\Lambda$ binding energies 
in hypernuclei proposed at the HIHR beam line,
precise and reliable energy calibration of the hypernuclear excitation spectrum is 
of great importance. 
In the $(e,e'K^+)$ spectroscopy at JLab, $p(e,e'K^+)$$\Lambda$,$\Sigma^0$ reactions
with a hydrogen target are used for energy calibration of the spectrometer system.
This method cannot be used for the $(\pi^+,K^+)$ reaction
because of the absence of a neutron target.
Instead, the ground-state binding energy of $^{12}_\Lambda$C, which was 
reported to be $10.76 \pm 0.19\,\mathrm{MeV}$
by old emulsion experiments,
has been used for the energy calibration.
However, it has been pointed out~\cite{GOG16,BOT17} that
the ground-state binding energies of other $p$-shell hypernuclei thus obtained are systematically shifted by $\approx 0.54\text{--}0.6\,\mathrm{MeV}$
from those obtained in old emulsion experiments as well as those measured by the FINUDA experiment using the $(K^-_{\rm stop}, \pi^-)$ reaction. 
It should be noted that the FINUDA experiment 
could make use of the two-body decay of $K^+$ into $\mu^+\nu_\mu$ and $\pi^+\pi^0$ for calibration.
Therefore, 
an independent method to measure the hypernuclear ground-state binding energies is called for, in order to obtain precise and reliable standard values of some $p$-shell hypernuclei.

The $B_\Lambda$ values of hypernuclear ground states can be reliably determined 
by measuring the pion momentum in the two-body mesonic decays,
$^A_\Lambda$Z$_{gs}$ $\to$  $^A$(Z+1) $+ \pi^-$, as was successfully performed
for $^4_\Lambda$H \cite{ESS15}.
At the HIHR beam line, the ground states of light hypernuclei are produced
by intense pion beams on a thin target, and the momentum spectrum
of their decay pions are measured with 
an additional magnetic spectrometer around the target. 
One of the promising ideas is to use a compact superconducting solenoid magnet, in which a low-momentum decay pion moves along a helical orbit.
The momenta of the decay pions range between $90\,\mathrm{MeV}/c$ and $135\,\mathrm{MeV}/c$.

For the calibration purpose, $^9_\Lambda$Be or $^{13}_\Lambda$C
hypernuclei are best suited.
The first reason is that the ground state is not split into the doublet because the spin of the even-even core nucleus ($^8\mathrm{Be}$ or $^{12}\mathrm{C}$)  is 0.
Under this condition, one
can safely relate the lowest-energy peak in the hypernuclear mass spectrum
to the ground state; 
this is not the case when both the upper and lower state in the ground-state doublet can be populated.
The second reason is that the excitation energy of the first excited state 
of the daughter nucleus ($^{9}\mathrm{C}$ or $^{13}\mathrm{N}$) is 
sufficiently large.
Otherwise, an experimental determination of the absolute pion momentum for the $^A_\Lambda$Z$_{gs}$ $\to$  $^A$(Z+1)$_{gs}+ \pi^-$ decay will be more or less complicated.
In addition, $^{12}_\Lambda\mathrm{C}$ may be used for the energy calibration as well, taking the advantage of a large formation cross section.
The formation of the upper $2^-$ state of the $^{12}\mathrm{C}$ ground-state doublet will be suppressed because of the non-spin-flip nature of the $(\pi^+,K^+)$ reaction. It is noted that the $^{12}\mathrm{C}$ ground state with the spin-parity $1^-$ decays dominantly to the first excited state of $^{12}\mathrm{N}$, but not to the ground state~\cite{MOT88b}.

In Refs.~\cite{MOT88b, MOT94}, excitation spectra for pionic decay from the lower and upper states of various hypernuclear ground-state doublets are shown.
Since the spectrum is totally different between the lower and upper states, spin-parity assignment of the hypernuclear ground states is feasible. This technique was already applied in the FINUDA experiment~\cite{AGN09}, resulting in the first spin assignment of $^{15}_\Lambda$N to be $3/2^+$. 
One of the interesting objects to be investigated is a neutron-rich  hypernucleus $^{12}\mathrm{Be}$~\cite{HOM15}, which can be populated in $^{12}$C$(\pi^-,K^+)$ reaction.
Whereas the core nucleus $^{11}\mathrm{Be}$ has a positive-parity ground state, as is known as parity inversion, the inclusion of a $\Lambda$ hyperon may revert the parity of the ground state.

This measurement is also essential in studies of charge symmetry breaking (CSB) effects,
because the CSB effect in the $B_\Lambda$ values 
for $p$-shell and heavier $\Lambda$ hypernuclei
is expected to be of the order of 100 keV.
By comparing the $B_\Lambda$ values measured in the $(\pi^*,K^*)$ reaction at HIHR with those of the corresponding mirror hypernuclei measured
in the $(e,e'K^+)$ reaction, 
as well as by comparing their level schemes to be measured in $\gamma$-ray 
spectroscopy as mentioned above, 
the CSB effects will be investigated beyond the $A=4$ hypernuclei
and the origin of this phenomenon will be understood
in terms of the $B$$B$ interactions in nuclear matter.

\subsubsection{Search for $\eta$ and $\eta'$ nuclear bound states}

The HIHR beam line will be also used in non-strangeness experiments
for high resolution missing mass spectroscopy using intense pions beams below 2 GeV/c.

Meson-nucleus bound states as well as mesonic atoms
will provide us with clear information on hadron-hadron interactions and
hadron properties in nuclear matter related to partial restoration of chiral symmetry.
In J-PARC, $K^-$ nuclear bound states and $K^-$ atoms have been 
intensively studied, and are further planned, mainly at the K1.8BR beam line with $K^-$ beams.

In the extended hadron facility at J-PARC,
$\eta$ and $\eta'$ nuclear bound states, which are not observed yet \cite{CHR88,TAN16,TAN18,TOM20}
probably due to small cross sections,
will be searched for via the $(\pi,N)$ reaction \cite{
	%ITA19
	Takahashi:2019xcq} 
taking advantage of intense ($> 10^8$/spill) pion beams 
and an excellent ($< 400$ keV (FWHM)) energy resolution.
In particular, behavior of $\eta$' meson in nuclear matter has 
attracted physicists' attention because
the heavy mass of $\eta$' is believed to be explained by the UA(1) anomaly.
In the $\eta'$ nucleus search experiment,
we employ the $^{12}$C($\pi^+,p$) reaction with a 1.8 GeV/c pion beam and
high energy protons from
$\eta'NN \to NN$ decay will be tagged around the target
to overcome huge background.

\subsubsection{Pure neutron system and neutron-rich nuclei via $(\pi^-,\pi^+)$ reaction}

Another type of non-strangeness experiments suitable for the HIHR beam line
is a study of pure-neutron systems and very neutron-rich light nuclei via
the double charge exchange $(\pi^-,\pi^+)$ reaction.

A candidate of a tetraneutron resonance state, $^4$n, were 
reported at RIBF via a heavy-ion double charge exchange reaction, 
$^4$He($^8$He,$^8$Be)$^4$n \cite{KIS16}.
In order to confirm its existence and investigate the structure, 
a missing mass measurement of 
the $^4$He$(\pi^-,\pi^+)$$^4$n reaction at HIHR beam line is proposed \cite{FUJ16}.
When an intense pion beam ($\sim 1.6\times 10^8$ pions/spill)
at 980 MeV/c is irradiated to a 2.0 g/cm$^2$-thick $^4$He target,
$\sim$100 events of $^4$n will be collected in two weeks for a cross section of 1nb/sr.
With a mass resolution ($< 400$ keV (FWHM))
better than the previous RIBF experiment (1 MeV (FWHM)),
the width as well as the accurate energy of the state can be obtained.

In addition, the $(\pi^-,\pi^+)$ reaction can be used to study heavy hydrogens
($^4$H, $^5$H, $^6$H, $^7$H) and other neutron-rich nuclei. 
Among the heavy hydrogens, $^6$H is not observed yet and should be searched for at HIHR.

%\bibitem{MOT88b} T. Motoba, K. Itonaga, and H. Band\={O}, Nucl. Phys. A \textbf{489}, 683 (1988).
%\bibitem{MOT94} T. Motoba and K. Itonaga, Prog. Theor. Phys. Suppl. \textbf{117}, 361  (1994).

%\bibitem{AGN09} M.~Agnello \textit{et al.} (FINUDA Collaboration) and A.~Gal, Phys. Lett. B \textbf{681}, 139 (2009).
%\bibitem{HOM15} H. Homma, M. Isaka, M. Kimura, Phys. Rev. C \textbf{91}, 014314 (2015) 
%\bibitem{TAN18} Y.~K.~Tanaka \textit{et al.}, Phys. Rev. C \textbf{97}, 015202 (2018).
%\bibitem{TOM20} N.~Tomida \textit{et al.} (LEPS2/BGOegg Collaboration),  Phys. Rev. Lett. \textbf{124}, 202501 (2020). 

% flatex input end: [HIHR-others_rev.tex]

%\input{HIHR-others}

\clearpage

%%%%% K1.1 beamline (T.Takahashi)
% flatex input: [K11-BL.tex]
\subsection{Design of K1.1 and K1.1BR Beam Lines}
\begin{figure}
\begin{center}
\includegraphics[width=0.75\hsize]{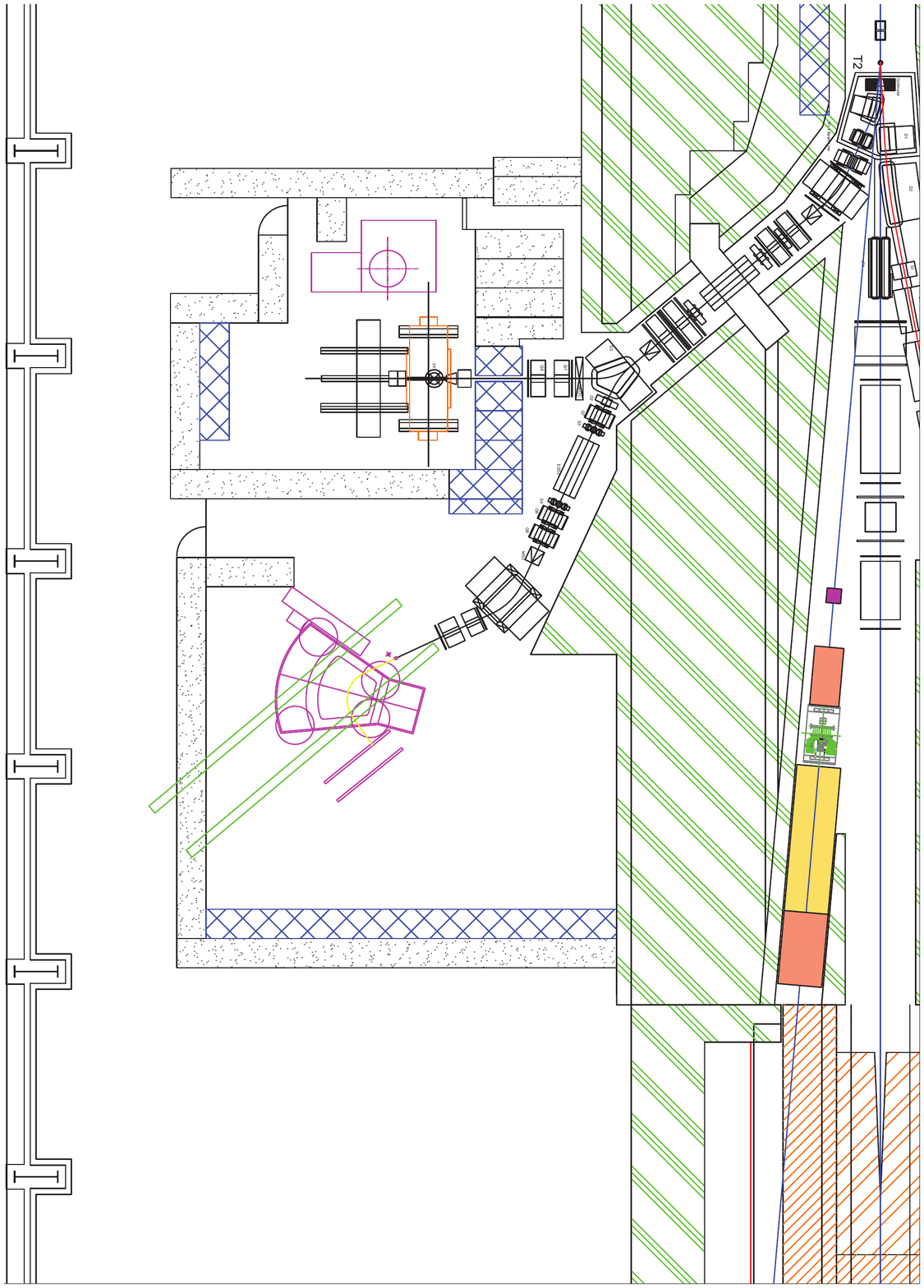}
\end{center}
\caption{Layout of the K1.1 and K1.1BR beamliines.}
\label{fig:K11BLopt2}
\end{figure}

The K1.1 and K1.1BR are mass-separated charged beam lines for low-momentum
kaons of $\sim$1.1 and $\sim$0.8 GeV/{\it c}, respectively.
As shown in Fig \ref{fig:K11BLopt2},
the beam line is extracted at  6$^\circ$ and use the cross fields type separators.
Total length of 28.32 and 20.78 m for K1.1 and K1.1BR, respectively.
Almost all elements such as magnets, electrostatic separators, slits, and 
power supplies  already exist for use of the K1.1BR and K1.1 in the current
Hadron Experimental Hall.  
The most upstream magnet D1, which will be installed in the T2 vacuum chamber,
will be newly fabricated to conform with the configuration around 
the production target (T2).
Then the distance between the T2 to K1.1D1 becomes shorter than that of
the beam line at the current Hadron Hall (2.0 m $\rightarrow$ 1.2m).
This optimization to the K1.1 beam line enhances the beam line acceptance about twice.
Field (at pole ) values are listed in Table \ref{tbl:K11MagField}.
The D2 maximum field limits the maximum momentum of $\sim$1.2GeV/{\it c}.

% remove since this is design in the original plan
%\begin{figure}
%\begin{center}
%\includegraphics[width=0.75\hsize]{K11BLoption1.pdf}
%\end{center}
%\caption{Layout of the K1.1 and K1.1BR beamliines (option1).}
%\label{fig:K11BLopt1}
%\end{figure}

\begin{table}
\caption{Field (at pole) values of the K1.1 beam line elements.}
\label{tbl:K11MagField}
\begin{center}
\begin{tabular}{l|r|r|r}
\hline\hline
Elements	& Max. field [kG]	& K1.1 (1.1GeV/{\it c})	 [kG]	& K1.1BR (0.8GeV/{\it c}) [kG]	\\
\hline\hline
D1			& (new)				& 12.3948						&  9.0144						\\
Q1			&  7.75				&  5.9733						&  4.2744						\\
Q2			& 11.65				& -9.0733						& -6.5176						\\
D2			& 15.512			& 14.2068						& 10.33216						\\
IF(H,V)		&					&								&								\\
Q3			& 1223				&  7.0645						&  5.92312						\\
O1			&  4.09				&								&								\\
Q4			& 12.34				& -8.0711						& -5.65928						\\
S1			&  5.22				& -1.3024						& -0.62264						\\
ESS1		&					&								&								\\
S2			&  5.22				& -1.3024						& -0.63856						\\
Q5			& 12.07				& -9.0749						& -7.4448						\\
Q6			& 12.07				&  8.047						& 8.53512						\\
MomS		& 					&								&								\\
MS1		&					&								&								\\
D3			& 21.18				& -14.6772						& 14.825512					\\
\hline
O2			&					&								&								\\
Q7			& 11.87				& -5.104						& 								\\
S3			& 2.56				& -1.1848						&								\\
ESS2		&					&								&								\\
S4			& 2.56				&  0.3611						&								\\
Q8			& 11.87				& -9.4507						&								\\
Q9			& 11.4				&  9.4533						&								\\
MS2		&					&								&								\\
D4			& 21.46				& 17.927						&								\\
Q10		&  8.2				& -6.7281						&								\\
Q11		& 11.5				& -6.1168						&								\\
\hline
HFOC		&					&								&								\\
Q7-BR		&					&								& -7.56032						\\
Q8-BR		&					&								&  8.392336					\\
\hline\hline
\end{tabular}
\end{center}
\end{table}

\begin{figure}
\begin{center}
\includegraphics[width=0.95\hsize]{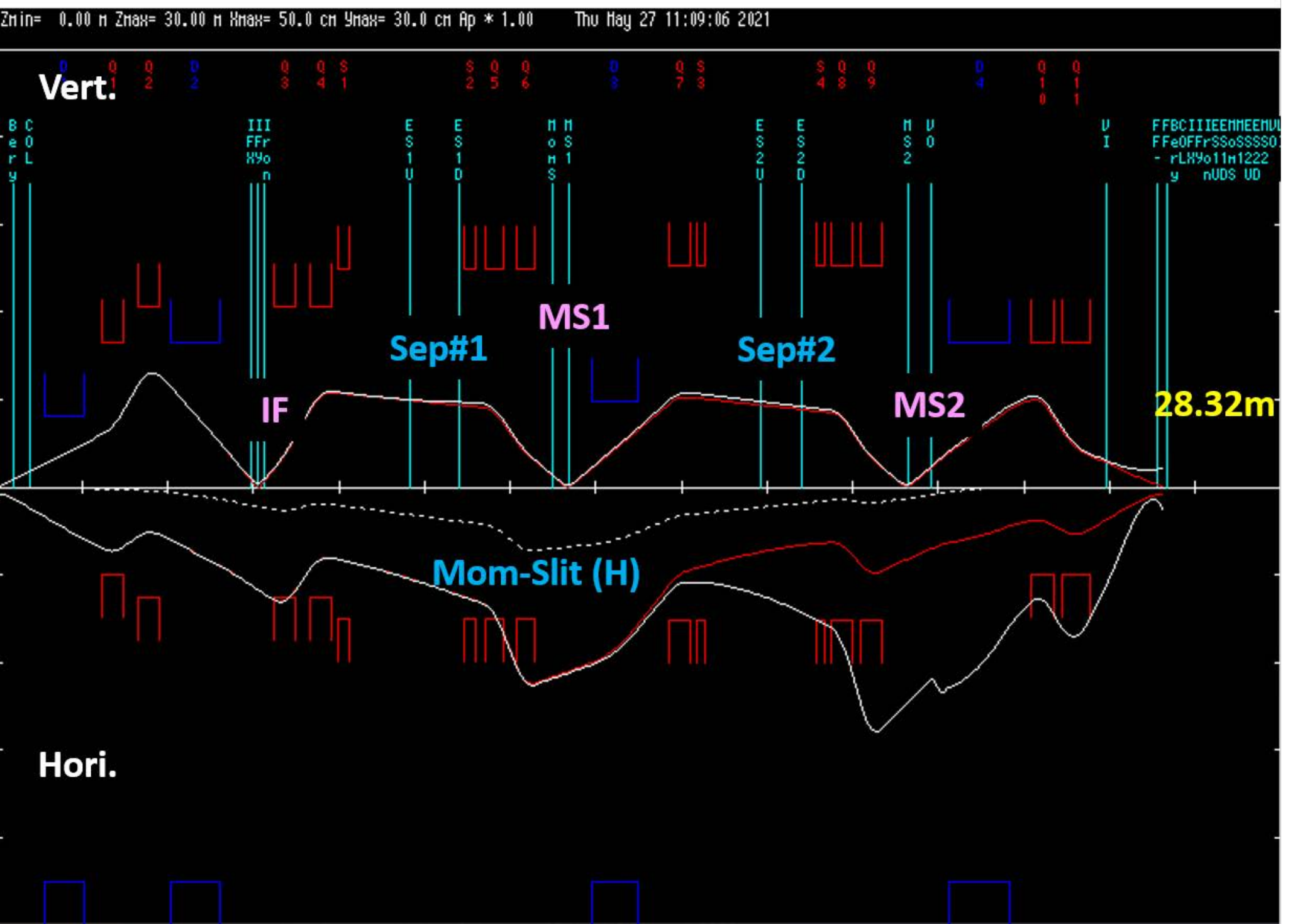}
\end{center}
\caption{Beam envelope of the K1.1 beam line optics.
Beam source is assumed to be an ellipse shape with $\pm$7.7 mm (horizontal) 
and $\pm$1.7 mm (vertical) and spreads of $\pm$30mrad in horizontal, 
$\pm$15mrad in vertical, and momentum spread of $\pm$3\%.
Red curves show the 1st order optics, while white one show
the one including the 2nd order.
}
\label{fig:K11optics}
\end{figure}

Figure \ref{fig:K11optics} shows the beam envelope of the K1.1 beam line
calculated by {\it TRANSPORT}.
Beam source is assumed to be an ellipse shape with $\pm$7.7 mm (horizontal) 
and $\pm$1.7 mm (vertical) and spreads of $\pm$30 mrad in horizontal, 
$\pm$15mrad in vertical, and momentum spread of $\pm$3\%.
This source size is taken into account of the primary proton beam size of
1.0 mm (H) $\times$ 2.5 mm (V) in rms and the length of the prodction target
of 66 mm and the beam extraction angle of 6.0 degree.
Top half of the figure shows vertical one, while bottom half shows horizontal.
Red and  white solid curves show optics only the 1st order and including the 
2nd order, respectively. Dashed one shows one due to the momentum spread.
The beam is focused vertically at the following 4 points, IF (Intermediate Focus), 
MS1 (Mass Slit 1), MS2, and FF (Final Focus), for particle-mass separation. 
At IF vacuum window is installed to separate accelerator-class vacuum with
the secondary beam line class vacuum. 
To minimize the effects of the vacuum window and redefine the vertical beam size, 
the beam is vertically focused at IF.
From IF to MS1 is the 1st mass-separator section.
The particles with different mass, namely different velocity, are focused at
the different vertical position on MS1 by using an 1.9m long electrostatic field 
and compensation magnetic field.
Mass slit with the tapered shape is installed at MS1 to prevent
the unwanted particles passing through.
Momentum slit is installed the upstream of the mass slit 1, where 
momentum dispersion is large, to define momentum bite of the secondary beam.
From MS1 to MS2 is the 2nd mass-separator section similar to the 1st 
mass-separator section. 
After MS2, beam analyzer is located.
At FF, beam is focused horizontally, vertically, and achromatically.

\begin{figure}
\begin{center}
\includegraphics[width=0.5\hsize]{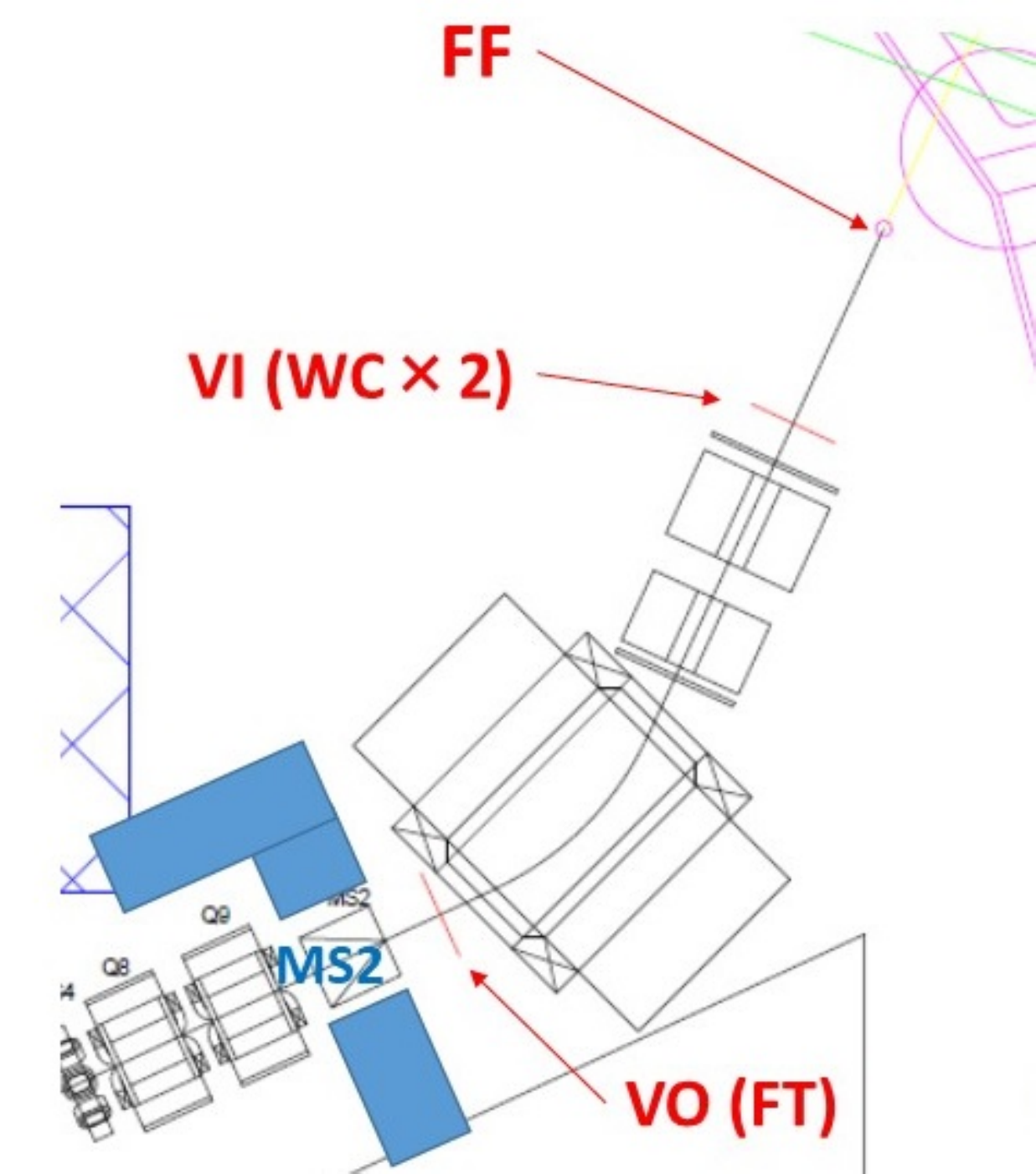}
\end{center}
\caption{K1.1 beam analyzer.}
\label{fig:K11BeamAnalyzer}
\end{figure}

Figure \ref{fig:K11BeamAnalyzer} shows the K1.1 beam analyzer
with {\it DQQ} configuration.
It is necessary to measure at least 5 independent observables 
in order to determine particle's momentum.
In the beam analyzer, particle trajectory on both horizontal
and vertical positions and their directions at the downstream
part (VI) with trackers such as two sets of wire chambers, 
At the upstream part (VO), it is enough to measure the horizontal
position to determine particle momentum. 
Scintillating fiber tracker (FT) is used 
Coefficients of  VO $\rightarrow$ VI transport are as follows;
$R_{11}$=1.38544, $R_{33}$=-0.66853 (horizontal and vertical magnifications),
$R_{12}$=0.43615 cm/rad.
and $R_{16}$=2.108 cm/\$ (dispersion).
In the 1st order optics calculation, momentum resolution is expressed as follows,
\begin{equation}
 \Delta p/p = \frac{\sqrt{1 + R_{11}^2}}{R_{16}} \Delta X,
\end{equation}
where $\Delta X$ is the resolution of the horizontal position measurements.
If we assume $\delta X$=0.2 mm (RMS), $\Delta p/p$= 1.6$\times$10$^{-4}$ (RMS),
namely 3.8$\times$10$^{-4}$ (FWHM), is expected.
However  $R_{12}$ can not be adjusted to small or zero like QQDQQ configuration
analyzer.
The multiple scattering by the FT at VO affects the momentum resolution.
In case of  0.5g/cm$^2$ material at VO, the multiple scattering gives
$\delta\theta \sim$1.2mrad (RMS). 
Then the momentum resolution by the multiple scattering effect is estimated
as follows,
\begin{eqnarray}
\Delta p/p 	& = & \frac{\sqrt{1 + R_{11}^2}}{R_{16}} \times R_{12}\Delta\theta \\
			& = & 4.2 \times 10^{-4} {\rm (RMS)} \nonumber \\
			& \rightarrow & 1.0 \times 10^{-3} {\rm (FWHM)} \nonumber
\end{eqnarray}
The multiple scattering effect significantly contributes the momentum resolution of the beam analyzer.

\begin{figure}
\begin{center}
\includegraphics[width=0.95\hsize]{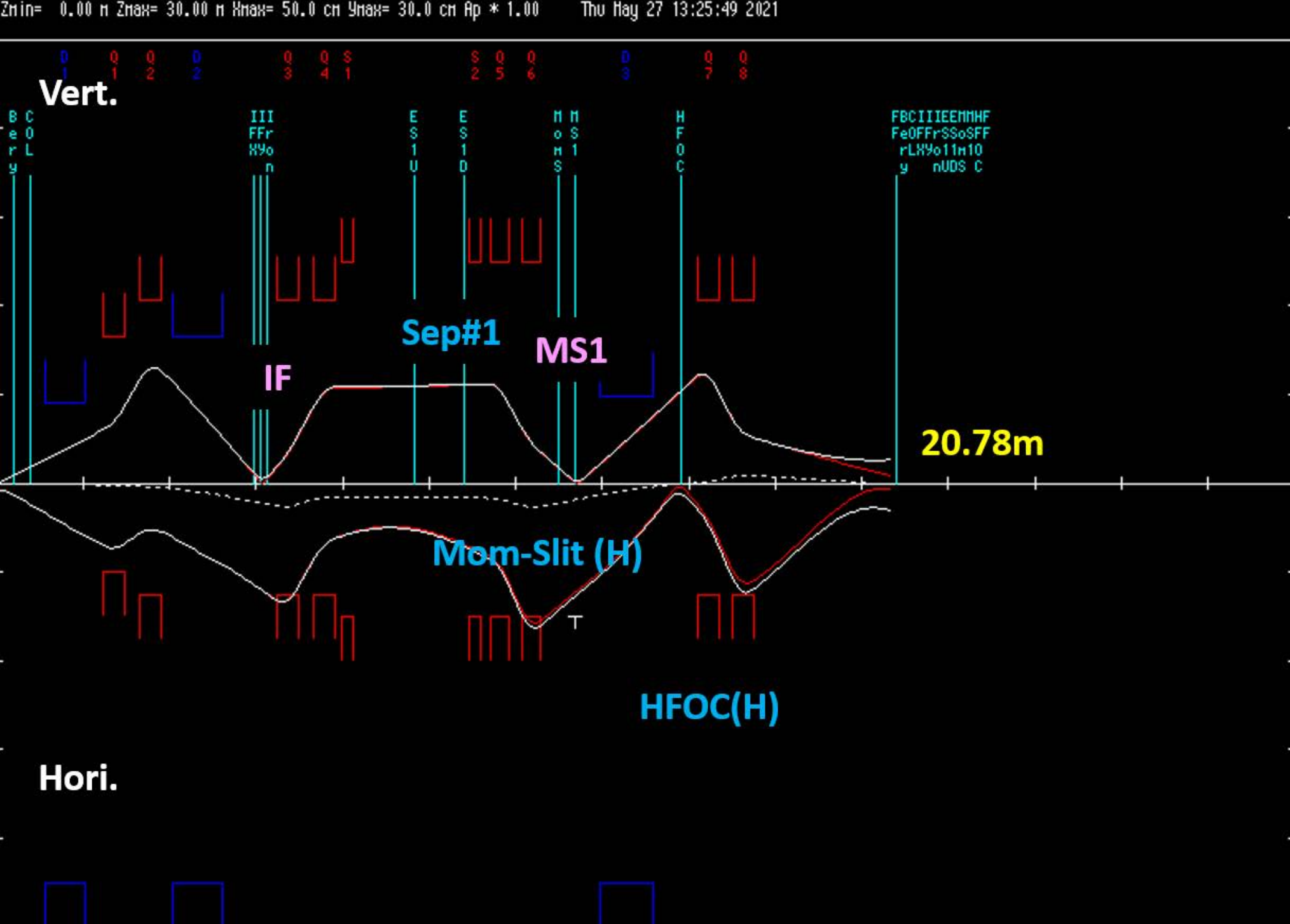}
\end{center}
\caption{Beam envelope of the K1.1BR beam line optics
Beam source size and spread are same as those of K1.1 beam line.
}
\label{fig:K11BRoptics}
\end{figure}

Figure \ref{fig:K11BRoptics} shows the beam envelope of t
he K1.1BR beam line calculated by {\it TRANSPORT}.
The same beam source as the K1.1 beam line is assumed.
In the K1.1BR optics, the beam is horizontally focused after
D3 (HFOC).

Kaon yield and purity were estimated using {\it TURLE} .
The assumption of the simulation is as follows;
5.0$\times$10$^{13}$ protons\footnote{ 5.0$\times$10$^{13}$
corresponds to 48kW (30GeV) with 5.0 s repetition cycle.
Note that number of proton on the T2 target 
depends on not only MR beam power but also the T1 target.}
 on the 50\% loss production target,
production cross section by Sanford-Wang parameterization with kinematic reflection,
%\cite{swp}
and no cloud-pion effect.
Table \ref{tbl:yieldK11} shows 1.1GeV/{\it c} $K^-$ yield and purity for each
slit opening at K1.1 beam line, while table \ref{tbl:yieldK11BR} shows
those of 0.8 GeV/{\it c} $K^\pm$ at K1.1BR beam line.
IFH and MomS are fully open.

\begin{table}
\caption{Yield and purity of the 1.1 GeV/{\it c} $K^-$ at K1.1 beam line.}
\label{tbl:yieldK11}
\begin{center}
\begin{tabular}{rr|r|rr|rr}
\hline
\multicolumn{2}{c|}{slit opening}		& \multicolumn{1}{c|}{acceptance}	& 
	 \multicolumn{2}{c|}{$E$=70kV/cm}		&  \multicolumn{2}{c}{$E$=60kV/cm} \\
IFV 	& MS1/MS2				& \multicolumn{1}{c|}{ [msr \%] }		& 
 	yield	& purity		& yield	& purity		\\
\hline
$\pm$1.5mm	& $\pm$1.0mm	& 1.19	& 219k	& 98\%	& 220k	& 97\%	\\
$\pm$2.0mm  & $\pm$1.0mm    & 1.28  & 234k  & 98\%  & 234k  & 95\%  \\
$\pm$3.0mm	& $\pm$1.0mm	& 1.31	& 242k	& 98\%	& 242k	& 95\%	\\
$\pm$1.5mm	& $\pm$1.5mm	& 1.70	& 312k	& 71\%	& 312k	& 16\%	\\
$\pm$2.0mm  & $\pm$1.5mm    & 1.94  & 357k  & 41\%  & 357k  & 10\%  \\
$\pm$3.0mm	& $\pm$1.5mm	& 2.14	& 393k	& 32\%	& 394k	& 9.2\%	\\
$\pm$1.5mm	& $\pm$2.0mm	& 1.95 	& 358k	& 14\%	& 358k	& 4.0\%	\\
$\pm$2.0mm  & $\pm$2.0mm    & 2.38  & 437k  & 6.2\% & 437k  & 2.8\% \\
$\pm$3.0mm,	& $\pm$2.0mm	& 2.79	& 511k	& 4.9\%	& 515k	& 1.5\%	\\
\hline
\end{tabular}
\end{center}
\end{table}

\begin{table}
\caption{Yield and purity of the 0.8 GeV/{\it c} $K^\pm$ at K1.1BR beam line 
with $E$=60 kV/cm.}
\label{tbl:yieldK11BR}
\begin{center}
\begin{tabular}{rrr|r|rr|rr}
\hline
\multicolumn{3}{c|}{slit opening}		& \multicolumn{1}{c|}{acceptance}	& 
 	\multicolumn{2}{c|}{$K^+$}		&  \multicolumn{2}{c}{$K^-$} \\
IFV	& MS1	& HFOC				& \multicolumn{1}{c|}{ [msr \%] }		& 
 	yield	& purity		& yield	& purity		\\
\hline
$\pm$1.5mm	& $\pm$1.0mm	& $\pm$10mm	& 1.36	& 131k	& 91\%	&  79k	& 88\%	\\
$\pm$2.0mm  & $\pm$1.0mm    & $\pm$10mm & 1.50  & 145k  & 83\%  &  87k  & 78\%  \\
$\pm$3.0mm	& $\pm$1.0mm	& $\pm$10mm	& 1.62	& 156k	& 54\%	&  94k	& 45\%	\\
$\pm$1.5mm	& $\pm$1.5mm	& $\pm$10mm	& 1.85	& 179k	& 67\%	& 108k	& 60\%	\\
$\pm$2.0mm  & $\pm$1.5mm    & $\pm$10mm & 2.21  & 213k  & 59\%  & 128k  & 51\%    \\
$\pm$3.0mm	& $\pm$1.5mm	& $\pm$10mm	& 2.55	& 245k	& 39\%	& 148k	& 32\%	\\
$\pm$1.5mm	& $\pm$2.0mm	& $\pm$10mm	& 2.09	& 201k	& 47\%	& 121k	& 38\%	\\
$\pm$2.0mm  & $\pm$2.0mm    & $\pm$10mm & 2.60  & 250k  & 43\%  & 151k  & 35\%    \\
$\pm$3.0mm	& $\pm$2.0mm	& $\pm$10mm	& 3.21	& 310k	& 31\%	& 187k	& 24\%	\\
\hline
\end{tabular}
\end{center}
\end{table}
% flatex input end: [K11-BL.tex]

%%%%% K1.1 beamline (T.Takahashi)

\clearpage

%%%%% K1.1 physics (K. Miwa)
% flatex input: [K11-Phys.tex]
\subsection{$\Lambda$p Scattering Experiment at the K1.1 Beam Line}
\subsubsection{Background of YN interaction and YN scattering experiment}
Nucleus is a many-body system of nucleons which are bound by a nuclear force.
The nuclear force is crucially important interaction which determines the structures from nuclei up to compact stars.
The nuclear force shows the attractive nature in the middle and long ranges and the averaged attractive force binds the nuclei.
In order to make the nuclei and compact stars be a stable bound state, 
the balance between the attractive force in the middle-long range and repulsive force in the short range is very important,
because the repulsive force in the short range prevents nuclei and compact stars from collapsing due to its attractive force.
The nuclear force is a strong interaction between the color-less nucleon and can be described theoretically by not taking into account the role of quarks.
For example, boson exchange models try to explain the nuclear force by considering the possible meson exchange including  pseudo scalar, vector, scalar mesons
and its pair or multi-meson exchange diagrams.
Experimentally, the nuclear force has been intensively studied by plenty of proton-proton and neutron-proton scattering experiments.
Very precise scattering data such as differential cross section and spin observables exist and these data were essential to construct so-called "realistic models" \cite{
%Machleidt:2001
MAC01, 
%Stoks:1994
STO94, 
%Wiringa:1995
WIR95-1} for nuclear force
which reproduce these experimental observables with a reduced $\chi^{2} \sim 1$.
Such modern interaction models are used to understand the structure of the nuclei and three-body nuclear force.

In order to understand the role of quark flavor  in the nuclear force, we should extend the nuclear force to the baryon-baryon (BB) interaction including hyperon-nucleon (YN) and hyperon-hyperon (YY) interactions,
because new interaction-multiplets are expected to show very different features in the short range region where two baryons overlap with each other.
The quark cluster model predicts the quite repulsive core in the 10 and $8_{s}$ multiplets due to the Pauli effect in quark level 
and the attractive core in the flavor singlet channel due to the color-magnetic interaction \cite{Oka:1986, Fujiwara:2007}.
These predictions are now reproduced by lattice QCD simulations which become a powerful theoretical tool to derive the YN and YY potentials from the first principle in QCD \cite{
%Aoki:2012
AOK12, Inoue:2012, Nemura:2018}.
The BB interaction is an essential test to describe the meson exchange picture with a uniform treatment assuming the SU(3) flavor symmetry \cite{
%Rijken:1999
RIJ99, 
%Nagels:2019
NAG19, 
%Haidenbauer:2005
HAI05}.
In order to test these theoretical models of two-body BB interaction, important experimental inputs are the binding energy of few-body bound system including hyperon such as hypertriton
and two-body scattering data between hyperon and proton.
As we mentioned, the $pp$ and $np$ scattering data played an essential role to establish the realistic nuclear force models.
On the other hand, the hyperon-proton scattering experiment was quite difficult experimentally due to the low intensity of the hyperon beam and its short lifetime.
36 cross section data of the $\Lambda p$ and $\Sigma p$ scatterings  \cite{Sechi-zorn:1968, Alexander:1968, Kadyk:1971, Hauptman:1977, Engelmann:1966, Eisele:1971, Stephen:1970, Kondo:2000, Kanda:2005} were actually indispensable to construct the present BB interaction models.
However the quality and quantity of these data are insufficient to impose a strict constraint on the theoretical models.
Therefore, historically, the BB interaction has been investigated from the hypernuclear structure because their binding energies and level structures reflect the YN interaction \cite{
%Hashimoto:2006
HAS06}.
The effective two-body interaction potential in the hypernuclei was obtained by the G-matrix calculation using the bare two-body interaction \cite{Yamamoto:2010}.
Then, the calculated energy levels of the hypernulear system were compared with the hypernuclear data and the bare two-body interaction was updated so as to reproduce the experimental data.
This theoretical and experimental strategy seems to work well at least for the many hypernucler phenomena.
However, we know that a neutron star with two-solar mass \cite{
%Demorest:Nature2010
Demorest:2010bx} can not be supported by the present two-body YN interaction due to softening of the equation of state by the appearance of hyperons in the high density region in the neutron star.
The many-body repulsive interactions such as YNN three-body force are expected to play an essential role in the such high-density region to support the massive neutron star with hyperon.
Such many-body effect is expected to appear in the heavy $\Lambda$ hypernuclei as the mass difference % of  a few 100 keV  level 
in the ground state.
A new project to measure the $\Lambda$ hypernuclear bounding energy with ultra precise resolution are just launching to explore the $\Lambda NN$ three-body interaction by the ($\pi^{+}, K^{+}$) spectroscopy in the extended hadron experimental facility by utilizing a high-intensity high-resolution (HIHR) beam line \cite{Nakamura_Proposal}.
In such a situation, it is crucially important to establish the realistic two-body YN interaction from the two-body system, that is, YN scattering data.
That is because the current theoretical treatment to derive YN two-body interaction from the hypernuclear structure is already suffered from uncertainties from the many-body effect in the hypernuclei.
Therefore, we have to really change the strategy for deriving the YN two-body interaction, that is, the realistic two-body YN interaction should be constructed based on the two-body scattering data.

Recently, we have realized a high-statistics $\Sigma p$ scattering in the J-PARC E40 experiment where 
$\sim$5,000 scattering events for both $\Sigma^{-}p$ and $\Sigma^{+}p$ channels were identified \cite{Miwa:2020, 
%Miwa:2021
J-PARCE40:2021qxa}.
We have introduced a new experimental technique to overcome the experimental difficulty of a hyperon-proton scattering experiment.
A liquid hydrogen target was used as both hyperon production and hyperon proton scattering targets to identify the hyperon production and the scattering events kinematically 
without any imaging data to identify the scattering topology.
High intensity $\pi$ beams were handled to accumulate $\Sigma$ beams as much as possible and 
more than 100 times more $\Sigma$ beams ($\sim$17 M $\Sigma^{-}$ and $\sim$70 M $\Sigma^{+}$) than that in the past KEK experiment were accumulated.
The differential cross sections measured in E40 together with a past measurement (KEK-PS E289 data for 400 $< p$ (MeV/$c$) $<$ 700 \cite{Kondo:2000}) and theoretical calculations are shown in
Figure 17 in reference\cite{
%Miwa:2021
J-PARCE40:2021qxa}.
%we successfully measured the differential cross sections of the $\Sigma^{-}p$ elastic scattering for the momentum region from 470 to 850 MeV/$c$ at J-PARC.
The statistical error of 10\% level was achieved  with a fine angular step of $d\cos \theta = 0.1$ by identifying  the largest statistics of about 4,500 $\Sigma^{-}p$ elastic scattering  events from 1.72 $\times$ $10^{7}$ $\Sigma^{-}$ particles.
The differential cross sections show clear forward peaking structure and the forward and backward ratio is large particularly in the higher momentum regions.
Although the experimental inputs of the two-body hyperon-proton scattering were quite limited up to now, the success of the $\Sigma^{-}p$ scattering is a remarkable step to provide accurate data to improve the $BB$ interaction models and to establish the realistic $BB$ interactions with both theoretical and experimental efforts.
%Analysis to derive the differential cross sections of the $\Sigma^{-}p \to \Lambda n$ reaction and the $\Sigma^{+}p$ elastic scattering is on going.
%Because all channels are related to each other within the flavor SU(3) symmetry, these data also impose important constraint on the two-body $BB$ interaction theories.
We apply the same experimental method to the $\Lambda p$ scattering experiment to derive the differential cross section and spin observables with good accuracy.
By synthesizing all the experimental information ($\Sigma N$ channel in E40 and $\Lambda N$ channel in the proposed experiment), better understanding of the $BB$ interactions will be achieved in near future.

%\begin{figure}[]
%  \centerline{\includegraphics[width=0.5\textwidth]{figure_k1.1/showAllSigmaMPdSdW3.pdf}}
%  \caption{ Differential cross sections obtained in the E40 experiment (black points) taken from \cite{
%  %Miwa:2021
%  J-PARCE40:2021qxa} . The error bars and boxes show  statistical and   systematic uncertainties, respectively.
%  Red points are averaged differential cross section of 400 $<$ $p$ (MeV/$c$) $<$ 700 taken in KEK-PS E289 (the same points are plotted in all of the four momentum regions).
%  The dotted (magenta), dot-dashed (blue) and solid (yellow) lines represent the calculated cross sections by the Nijmegen ESC08c model  based on boson-exchange picture, the fss2 model including  QCM and the extended $\chi$EFT model, respectively.
%  }
%  \label{showAllSigmaMPdSdW3}
%\end{figure}

\subsubsection{Formalism of the spin-dependent YN interaction}\label{subsec_spindepYN}
In this proposed experiment, the important new aspect is to measure the spin observables with a polarized $\Lambda$ beam.
The hyperon-proton scattering is the scattering between two spin 1/2 particles and the spin observables are substantial tool for the study of the spin-dependent YN interaction\cite{Ishikawa:2004}.

The $T$ matrix for the elastic scattering between particle $a$ and $b$ is represented in terms of spin-independent, spin-spin, symmetric $LS$ (SLS), antisymmetric $LS$ (ALS) and tensor components as
\begin{equation}
M = V_{c} + V_{\sigma}(\bm{s_{a}}\cdot \bm{s_{b}})+V_{SLS}(\bm{s_{a}}+\bm{s_{b}})\cdot \bm{L} + V_{ALS}(\bm{s_{a}}-\bm{s_{b}}) \cdot \bm{L} + V_{T}([\bm{s_{a}} \otimes \bm{s_{b}}]^{(2)} \cdot \bm{Y_{2}}(\hat{\bm{r}}))
\end{equation}
where $\bm{L}$ is the $a$-$b$ relative orbital angular momentum, $\bm{r}$ is the $a$-$b$ relative coordinate and $V$'s are form-factor functions
for the spin-independent central interaction $V_{c}$, the spin-spin interaction $V_{\sigma}$, the SLS interaction $V_{SLS}$, the ALS interaction $V_{ALS}$, and the tensor interaction $V_{T}$.
In order to describe the scattering observables, new scalar amplitudes, vector amplitudes and tensor amplitudes are defined as follows,
\begin{equation}
U_{\alpha} \equiv <\bm{k}_{f}|V_{c}|\bm{k}_{i}>,  U_{\beta} \equiv <\bm{k}_{f}|V_{\sigma}|\bm{k}_{i}> 
\end{equation}
for the scalar amplitudes,
\begin{equation}
S_{ALS} \equiv <\bm{k}_{f}|V_{ALS}L_{1}|\bm{k}_{i}> , S_{SLS} \equiv <\bm{k}_{f}|V_{SLS}L_{1}|\bm{k}_{i}> 
\end{equation}
for the scalar amplitudes and
\begin{equation}
T_{j} = \frac{1}{2} <\bm{k}_{f}|V_{T}Y_{2j-1}|\bm{k}_{i}> 
\end{equation}
for the tensor amplitudes for $j=1, 2, 3$. For later convenience, the following $T_{\alpha}$ and $T_{\beta}$ are also used,
\begin{equation}
T_{\alpha} = \frac{1}{\sqrt{6}}T_{1}+T_{3},  T_{\beta} = \frac{1}{\sqrt{6}}T_{1}-T_{3}, 
\end{equation}
which give
\begin{equation}
T_{2} = -\tan \theta (\frac{1}{2}T_{\alpha}+T_{\beta}).
\end{equation}
The differential cross section is described by the sum of the all contributions as follows,
\begin{equation}
\Bigl(\frac{d\sigma}{d\Omega} \Bigl)=\frac{1}{4}{\rm Tr}(MM^{\dag})=|U_{\alpha}|^{2}+\frac{3}{16}|U_{\beta}|^{2}+\frac{1}{2}(|S_{SLS}|^{2} + |S_{ALS}|^{2})+\frac{1}{4}|T_{1}|^{2}+\frac{1}{2}(|T_{2}|^{2}+|T_{3}|^{2}).
\end{equation}
Therefore, in order to determine the each component separately, a lot of spin observables are necessary.
In the YN scattering case, the measurable spin observables are limited. 
However, such spin observables data are quite important to pin down the each spin-dependent component.
The analyzing power $A_{y}(Y)$ for the polarized hyperon beam is described as,

\begin{equation}
A_{y}(Y) = -\frac{1}{\sqrt{2}\sigma ( \theta )} {\rm {Im}} \left \{  (U_{\alpha} + \frac{1}{4}U_{\beta} )^{*} S_{SLS}+ (U_{\alpha} - \frac{1}{4}U_{\beta} )^{*} S_{ALS} -\frac{1}{2}T_{\alpha}^{*} (-S_{ALS}+S_{SLS}) \right\},
\end{equation}
where $\sigma(\theta)$ represents the differential cross section.
If we regard that  the component including the tensor amplitude ($T_{\alpha}$) is the 2nd order value, the analyzing power is sensitive to the LS forces.
In the NN interaction, the ALS is not allowed for the isospin symmetry and the ALS is the purely the new interaction appeared in the YN sector.

By measuring the change of the polarization of hyperon before and after the scattering, the depolarization can be measured.
The depolarization is described as

\begin{dmath}
%D_{y}^{y} \approx \frac{1}{\sigma(\theta)}{\rm Re} \left\{  |U_{\alpha}|^{2}-\frac{1}{16}|U_{\beta}|^{2} + \cdots \right\},
D_{y}^{y}  =  \frac{1}{\sigma(\theta)}{\rm Re} \left\{  \frac{1}{2 \sqrt{3}} \left ( U_{0}+\frac{1}{\sqrt{3}} U_{1}\right)^{*}U_{1} + \frac{1}{2} \left ( U_{0}-\frac{1}{\sqrt{3}} U_{1}\right)^{*} \left (\frac{1}{\sqrt{6}}T_{1}+T_{3}  \right ) -S_{1}^{*}S_{2}   +  \frac{1}{2}|S_{3}|^{2}  - \frac{1}{\sqrt{6}}T_{1}^{*} \left (\frac{1}{\sqrt{6}} T_{1}-T_{3} \right ) - \frac{1}{2}|T_{2}|^{2} \right\}.
\end{dmath}
Although the equation is complicated form of several amplitudes, $D_{y}^{y}$ is expected to be rather sensitive to tensor force \cite{Haidenbauer:1992}.
It is still difficult to determine each spin-dependent component separately from these limited observables. 
However, accumulation of these differential measurements can contribute to impose strong constraints on the present YN interaction models.

As for the type of the existing YN scattering observables, total cross section data exist even though the accuracy is not sufficient.
Therefore theoretical calculations of all models become similar for the total cross section because the existing cross section data are used to fit parameters in the theoretical models as shown in next section.
However, since there are almost no spin observable data except for the analyzing power measured in KEK \cite{Kurosawa:2006},
the theoretical prediction are quite different among the different theoretical frameworks.
Therefore the spin observable measurement is quite important to test and improve the theoretical frameworks.

\subsubsection{Theoretical studies for the $\Lambda p$ scattering}
\begin{figure}[h!]
\begin{center}
\includegraphics[width=16cm]{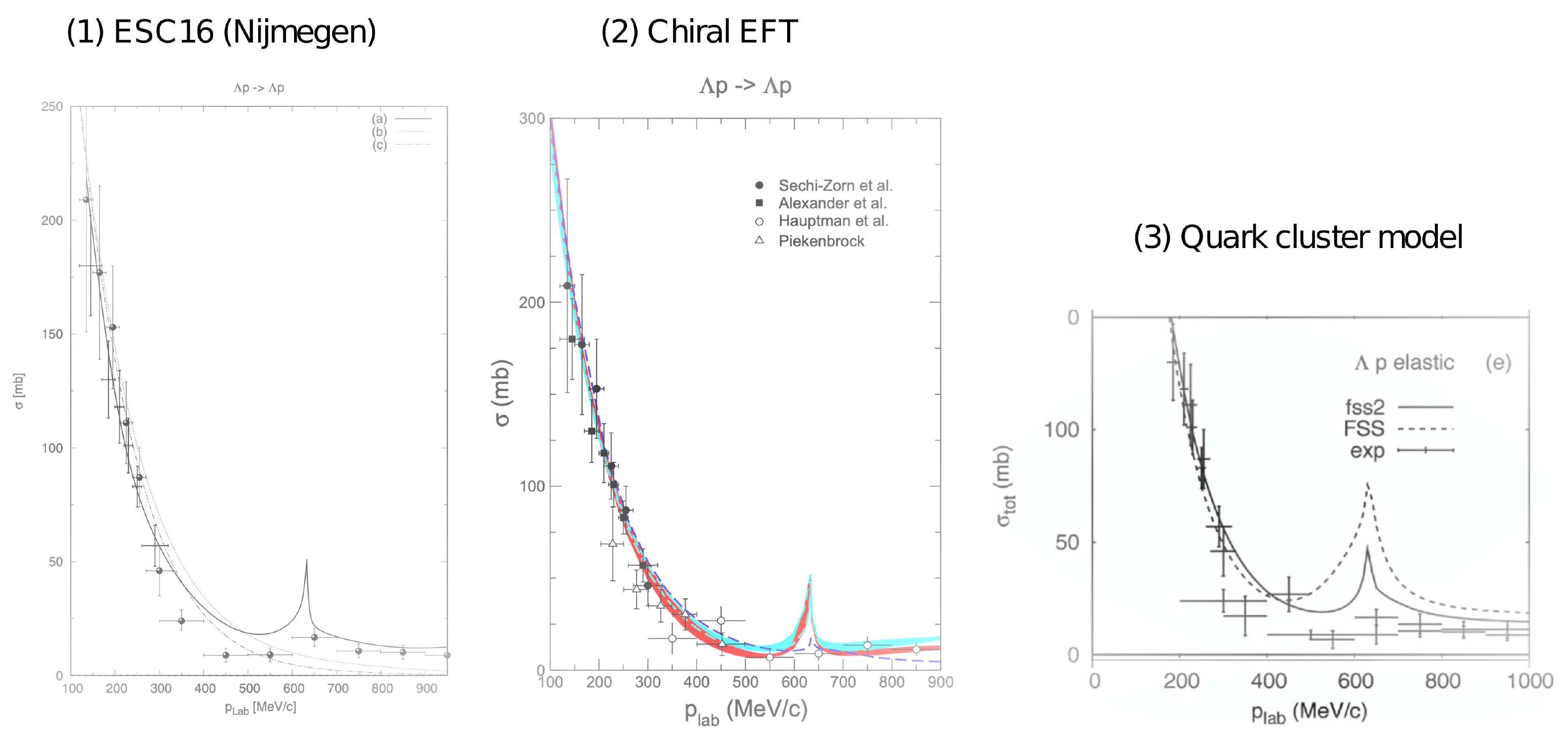}
\caption[]{Total cross section of the $\Lambda p$ elastic scattering and theoretical calculation. (1)Calculation by ESC16 model \cite{
%Nagels:2019
NAG19}. The solid line (a) shows the calculation including the higher wave contribution, whereas dotted lines (b) and (c) are obtained by the effective range approximation with some scattering length and effective range. 
(2) Calculation by the chiral EFT model with contributions up to the next leading order \cite{
%Haidenbauer:2013
HAI13, 
%Haidenbauer:2020
HAI20}. The red and cyan bands represent the result for NLO13 and the alternative version NLO19, respectively. The dotted line show the result for the Nijmegen NSC97f potential. (3) Calculation by the quark cluster model, that is, fss2 (solid curve) and FSS (dashed curve) \cite{Fujiwara:2007}.}
\label{figs_LambdaP_totalCS}
\end{center}
\end{figure}

In Figure \ref{figs_LambdaP_totalCS}, the total cross sections of the $\Lambda p$ elastic scattering are plotted with theoretical models such as the Nijmegen Extended Soft-Core (ESC) 16 model \cite{
%Nagels:2019
NAG19},
the chiral Effective Field Theory (EFT) extended to YN sector \cite{
%Haidenbauer:2013
HAI13, 
%Haidenbauer:2020
HAI20} and the quark cluster model \cite{Fujiwara:2007}.
In the low energy below 400 MeV/$c$, there exist several cross section measurements with hydrogen bubble chambers and these are the essential inputs to determine the S-wave contribution of the $\Lambda p$ channel.
On the other hand, in the higher momentum range higher than 400 MeV/$c$, the cross section data were reported by one bubble chamber experiment
and the total cross sections for each momentum region were determined from only $\sim$20 scattering events.
Due to such limited statistics, there is no differential information such as differential cross section and spin observables.

\begin{table}
\begin{center}
\caption{The relationship between the isospin basis and the flavor SU(3) basis for the $NN$ and $YN$ with $S=-1$ channels}
\label{relation_YN_SU3_table}
\begin{tabular}{cccc}
\\
\hline\hline
$S$ & $BB$ channel (I) & $^{1}E$ or $^{3}O$ & $^{3}E$ or $^{1}O$  \\
\hline\hline 
0 & $NN (I=0)$ & $--$ & $(10^{*})$ \\
  & $NN (I=1)$ & (27) & $--$ \\
\hline\hline 
  & $\Lambda N (I=1/2)$ & $\frac{1}{\sqrt{10}}[(8_{s})+3(27)]$ &  $\frac{1}{\sqrt{2}}[-(8_{a})+(10^{*})]$\\
$-1$    & $\Sigma N (I=1/2)$ & $\frac{1}{\sqrt{10}}[3(8_{s})-(27)]$ &  $\frac{1}{\sqrt{2}}[(8_{a})+(10^{*})]$\\
    & $\Sigma N (I=3/2)$ & (27) &  (10)\\
\hline\hline
\end{tabular}
\end{center}
\end{table}

The $NN$ and $YN$ interactions should be understood in a unified way using (broken) SU(3) symmetry.
Table \ref{relation_YN_SU3_table}  shows the relationship between the isospin basis and the flavor SU(3) basis for the $NN$ and $YN$ channels with strangeness $S=-1$.
These channels are related each other through the SU(3) flavor multiplet and $\Lambda N$ and $\Sigma N$ channels are discussed together theoretically.
Now, the $\Sigma p$ scattering cross section data will be drastically updated by the E40 experiment.
%This should have theoretical impact even for the description of the $\Lambda N$ interaction.

All theories predict the sizable cusp (enhancement) of the cross section at the $\Sigma N$ threshold due to the strong $\Lambda N - \Sigma N$ ($I=1/2$) $^{3}S_{1}-^{3}D_{1}$ coupling caused by the tensor force of the pion exchange potential.
In the quark cluster model, the importance of the ALS which causes the $\Lambda N ^{3}P_{1} - \Lambda N ^{1}P_{1} $  and $\Lambda N - \Sigma N (I=1/2) ^{3}P_{1} - ^{1}P_{1} $ transitions 
is discussed because the additional ALS force due to the quark-quark interaction is predicted in the quark picture.
Such ALS effect should appear in the Analyzing power (polarization) in the $\Lambda p$ scattering at this $\Lambda N-\Sigma N$ threshold region.

In the following subsection, we summarize the characteristics of  several theoretical models.

\begin{itemize}
    \item Nijmegen ESC model
\end{itemize}

Nijmegen extended-soft-core (ESC) models whose latest version is ESC16 \cite{
%Nagels:2019
NAG19}, describe $NN$, $YN$ and $YY$ interactions in a unified way using broken SU(3) symmetry.
The potentials consists of local and nonlocal potentials due to one-boson exchange which are the members of nonets of pseudoscalar, vector, scalar and axial-vector mesons,
two psudoscalar exchange, meson-pair exchange and diffractive exchange.
The meson-baryon coupling constants are calculated by imposing the SU(3) symmetry.
In the Nijmegen model, there have been difficulties to reproduce the sufficiently repulsive short-range interaction in $\Sigma N (I=3/2, ^{3}S_{1})$ channel and 
$\Sigma N (I=1/2, ^{1}S_{0})$ channel which are explained quite naturally as the Pauli effect in the quark level in the (10) and (8$_{s}$) multiplets in the SU(3) representation  in the quark level in the quark cluster model.
In the latest ESC16 version, such "forbidden state" effect are taken into account phenomenologically by making an effective Pomeron potential as the sum of 
pure Pomeron exchange and of a Pomeron-like representation of the Pauli repulsion.
The other difficulty was  small LS splittings in the $\Lambda$ hypernuclei
which are characterized by the LS potential of ($V_{SLS}-V_{ALS}$) type.
In conventional OBE model, ALS becomes smaller compared with SLS which leads the large LS splitting in the $\Lambda$ hypernuclei.
In this model, in order to reproduce the small LS splitting, some prescriptions have been performed such as meson pair exchange of the axial-vector pairs. 
These treatments of the ALS force should be tested with the Polarization (is equal to the Analyzing power) of the $\Lambda p$ scattering 
where the ALS and SLS should make a sizable contribution.

\begin{itemize}
    \item Quark cluster model
\end{itemize}

In the quark cluster model \cite{Fujiwara:2007}, the interaction Hamiltonian for quarks consists of the phenomenological confinement potential, the color Fermi-Breit interaction
with explicit flavor-symmetry breaking, and the effective-meson exchange potentials of pseudoscalar, scalar and vector mesons.
The six quarks are put in the Gauss potential which is characterized by a size parameter $b$.
These six quarks are imposed to be anti-symmetric under the exchange between any quark combination.
The fermion nature of quark is taken into account and its effect is characterized by the size parameter $b$.
Large repulsive nature in the (10) and (8$_{s}$) multiplets in the SU(3) representation can be naturally predicted by the Pauli effect in the quark level.
The color Fermi-Brit interaction between quarks gives another source of ALS, SLS and tensor forces in addition to the contribution from meson exchange potential.
They predict the sizable ALS comparable to SLS and explain the small $\Lambda$ hypernuclear LS splitting.

\begin{itemize}
    \item Chiral Effective Field Theory (Chiral EFT)
\end{itemize}

Chiral EFT has turned out to be a  powerful tool for the derivation of nuclear forces.
Its most notable feature is that there is an underlying power counting which allows one to improve calculations systematically by going to 
higher orders in a perturbative expansion.
In addition, it is possible to derive two- and three- nucleon forces in a consistent way.
This method is applied for the YN interaction and the $\Lambda N$ and $\Sigma N$ interactions are obtained at next-to-leading order where 
contributions from one- and two-pseudoscalar meson exchange diagrams and from four-baryon contact terms are considered \cite{
%Haidenbauer:2013
HAI13, 
%Haidenbauer:2020
HAI20}.
The chiral YN potential contains meson exchanges and a series of contact interactions with an increasing number of derivatives.
For the former meson exchange, contributions from the pseudoscalar octet ($\pi, K, \eta$) are taken into account.
The latter represent the short-range part of the interaction and are parametrized by low-energy constants (LEC), which need to be fixed by a fit to data.
The present Chiral EFT framework is constrained  from SU(3) symmetry in order to reduce the number of free parameter
and there are five, eight and ten independent LECs for the $S$-waves at LO, the $S$-waves and $S$-$D$ transition at NLO and for the $P$-waves at NLO.
%Even with such constraint,  the number of the LEC parameters increases at the higher leading order.
The present YN scattering data are insufficient to determine the LEC parameters higher than the next-leading order.
One of the the most important experimental tasks is to provide higher precision data for the YN channels to impose the constraint to determine these parameters at NLO more accurately.
Because the existing YN experimental cross sections are concentrated on the lower energy where the $S$-wave contribution is dominant,
the experimental constraints for the P-wave parameters are quite weak.
Therefore the differential cross sections and the polarization observables above $p_{\Lambda} > 400$ MeV/$c$ are essential even for determining the LEC at NLO.
At the NNLO, the number of LEC does not increase for the two-body interaction \cite{Haidenbauer:priv}.
Therefore, accumulation of YN scattering data can contribute to extend Chiral EFT to  NNLO.

\begin{figure}[h!]
\begin{center}
\includegraphics[width=8cm]{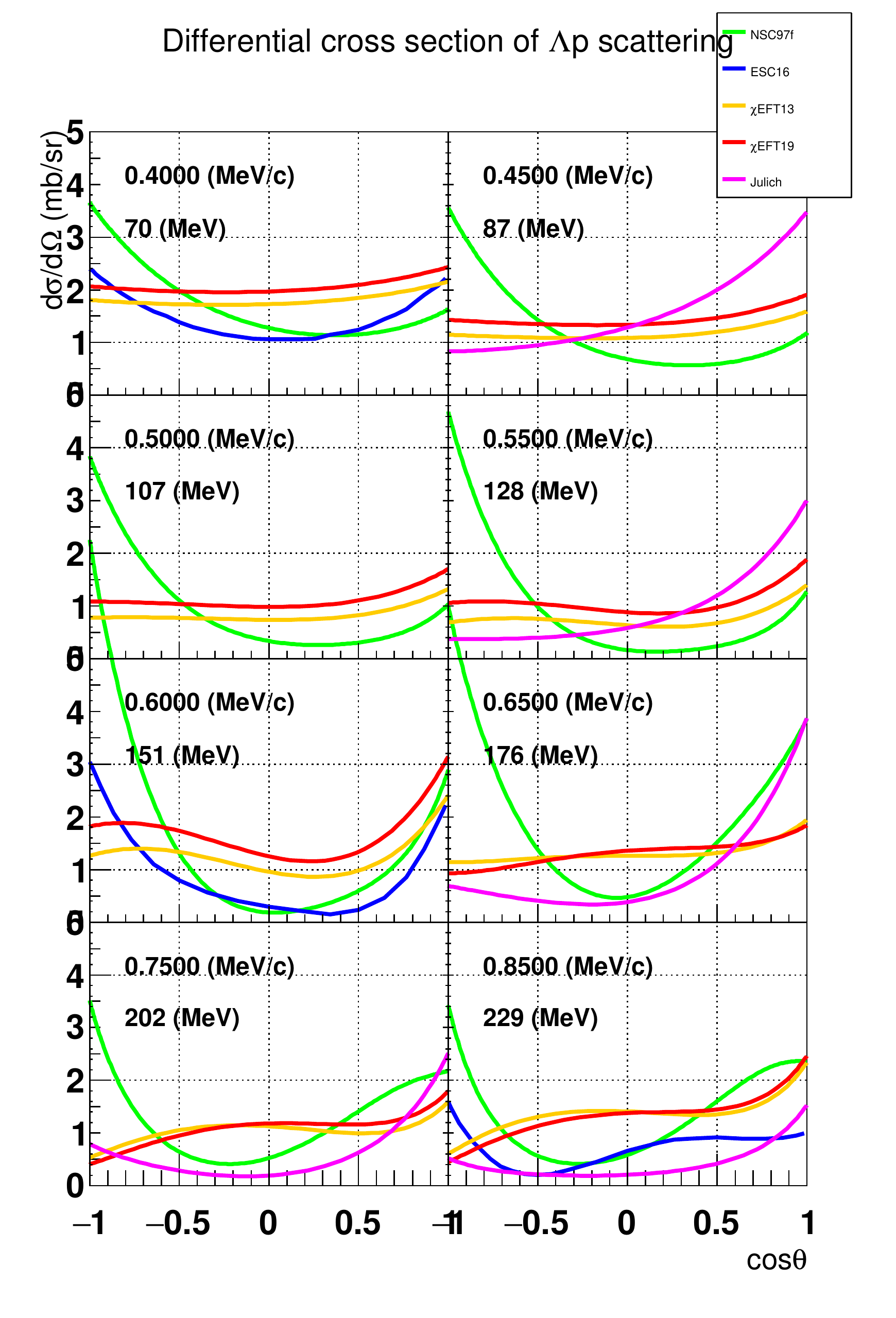}
\caption[]{Differential cross sections calculated by theoretical models for the momentum range between 0.4 and 0.85 GeV/$c$.
Two types of Nijmegen models (NSC97f and ESC16 \cite{
%Nagels:2019
NAG19}) and J\"{u}lich model \cite{
%Haidenbauer:2005
HAI05} are presented as the typical example of the boson exchange picture.
The two results by chiral EFT (chiral EFT13 and 19) \cite{
%Haidenbauer:2013
HAI13, 
%Haidenbauer:2020
HAI20} are calculated with the different sets of the LEC parameters both of which reasonably reproduce the YN scattering cross section.}
\label{fig_showDiffCS_Theory}
\end{center}
\end{figure}

\begin{figure}[!t]
\begin{center}
\begin{minipage}{6.5cm}
\includegraphics[width=0.98\textwidth]{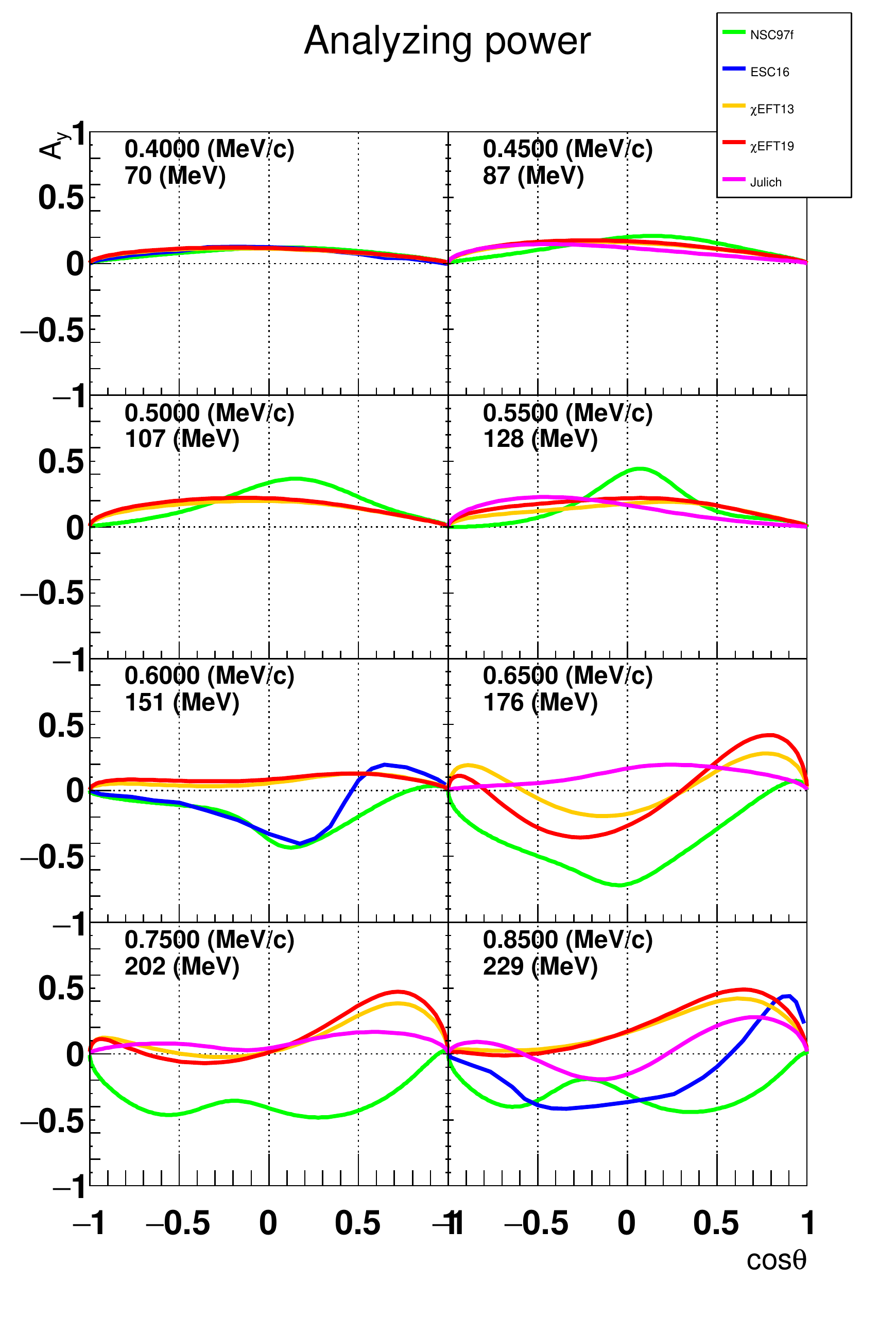}
\end{minipage}
\hspace{0.2cm}
\begin{minipage}{6.5cm}
\includegraphics[width=0.98\textwidth]{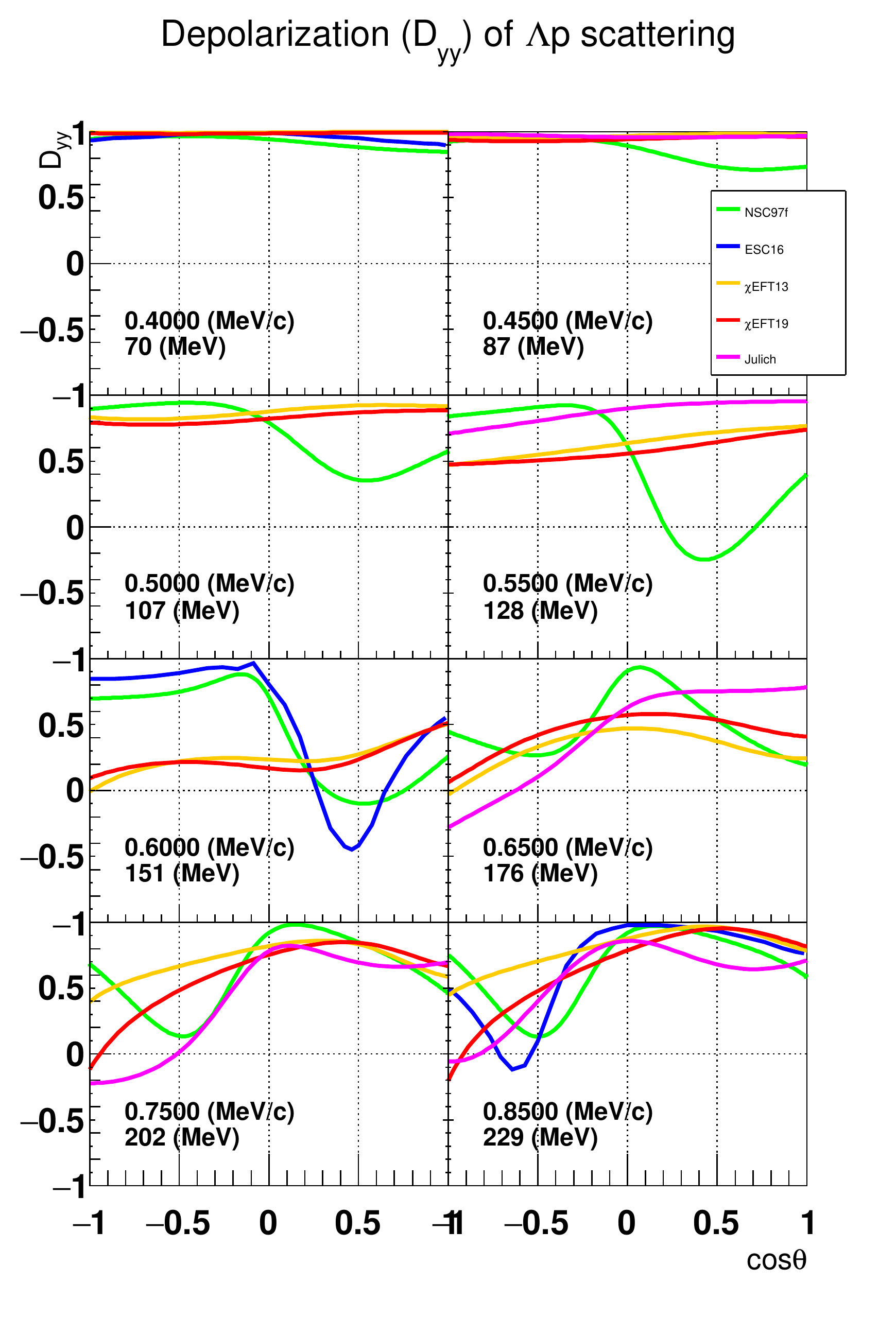}
\end{minipage}
\caption{Theoretical calculation of the Analyzing power ($A_{y}$) (left) and the Depolarization ($D_{y}^{y}$) (right) for the polarized $\Lambda$ beam.
The each colored line is same with figure \ref{fig_showDiffCS_Theory}}
\label{figs_showLR_Asymmetry_Theory}
\end{center}
\end{figure}

Figure \ref{fig_showDiffCS_Theory} shows the differential cross sections calculated by each model for the momentum range between 0.4 and 0.85 GeV/$c$.
Two types of Nijmegen models (NSC97f and ESC16 \cite{
%Nagels:2019
NAG19}) and J\"{u}lich model \cite{
%Haidenbauer:2005
HAI05} are presented as the typical example of the boson exchange picture.
Two results by chiral EFT (chiral EFT13 \cite{
%Haidenbauer:2013
HAI13} and 19 \cite{
%Haidenbauer:2020
HAI20} )  are calculated with the different sets of the LEC parameters both of which reasonably reproduce the YN scattering cross section.
However, the strength of the $\Lambda N$-$\Sigma N$ coupling potential, which is closely related to the attraction in $\Lambda N$ interaction in $\Lambda$ hypernuclei, in chiral EFT13 is much larger than that in chiral EFT19.
As for the total cross section, there are experimental data even for these momentum regions and the each model's parameters were determined to reproduce the existing total cross section.
However the differential cross section in each model shows quite different angular distribution due to the lack of the experimental input for the angular dependence.
%Such angular dependence might result from the higher-wave contributions and the charged kaon exchange in the $\Lambda p$ reaction for the backward scattering like the isospin exchange term in the $np$ scattering.
It is true that there were no experimental inputs to impose any constraint on such contributions.
For this purpose, the accurate differential cross section data in the $P$-wave region are essential.
The polarization measurements are also essential to determine the LS contribution which were never determined from the YN scattering data.
Figure \ref{figs_showLR_Asymmetry_Theory} shows the Analyzing power (Left) and the Depolarization (Right) for the polarized $\Lambda$ beam \cite{Haidenbauer:2021, Haidenbauer:priv, Rijken:private}.
In the lower momentum around 0.4 GeV/$c$ where the $P$-wave contribution is less significant, there is no large model difference.
However, the model difference appears clearly in the higher momentum region, where the $P$- and higher waves contribute in the reaction.
Especially, the difference in the $\Sigma N$ threshold (0.633 GeV/$c$ for $\Sigma^{+} n$ and 0.642 GeV/$c$ for $\Sigma^{0} p$) becomes significant
due to the different treatments of tensor force ($^{3}S_{1}-^{3}D_{1}$ transition) and ALS ($^{3}P_{1}-^{1}P_{1}$ transition).
Such effect should be reflected in all measurements, that is, the differential cross section, Analyzing power and  Depolarization. 

In present, the experimental situation is changing.
The J-PARC E40 experiment successfully detected the $\Sigma^{+} p$, $\Sigma^{-} p$ elastic scatterings and the $\Sigma^{-} p \to \Lambda n$ reaction
in the momentum range between 0.4 and 0.8 GeV/$c$.
The differential cross sections for these channels will be derived in near future \cite{
%Miwa:2021
J-PARCE40:2021qxa}.
Such data are the first experimental inputs to impose the remarkable constraint on the theoretical models.
As shown in Table \ref{relation_YN_SU3_table}, the $\Lambda N$ and $\Sigma N$ channels are related with each other through the flavor SU(3) representation.
Because we measured both the $\Sigma^{+} p (I=3/2)$, $\Sigma^{-} p$ (superposition of $I=1/2, 3/2$) elastic scatterings,
the knowledge for all multiplets not included in the $NN$ channels will be updated.
Then, the theoretical prediction for the $\Lambda N$ channel should become less uncertain compared with the present situation.
It is worth comparing the experimental results of the $\Lambda p$ scattering with the updated theoretical calculations to test the theoretical framework with the flavor SU(3) symmetry.

\subsubsection{Impact on the neutron star physics}

Hypernuclear physics are studied based on the two-body $\Lambda N$ interaction, especially the Nijmegen models are widely used to study the hypernuclear structure.
Due to the lack of the YN scattering data, such two-body models are partially tuned to reproduce the hypernuclear phenomena such as level splittings of $\Lambda$ hypernuclei and 
$\Sigma$ nuclear potential.
In the last section, we have reviewed that there is a large disagreement among models even for the $\Lambda N$ interaction which is intensively studied from $\Lambda$ hypernuclei.
The present urgent issue to be solved is the so-called "hyperon puzzle" in neutron star.
This is the unsolved question how one can reconcile the softening of the equation of state (EOS) due to the appearance of hyperons with the observed two-solar-mass neutron star.
Because $\Lambda$ hyperon plays an essential role for the softening of the EOS, the understanding of the $\Lambda$ potential in the high density region is indispensable.
Currently, the importance of the three-body interaction including hyperon such as $\Lambda NN$ is widely discussed.

\begin{figure}[h!]
\begin{center}
\includegraphics[width=10cm]{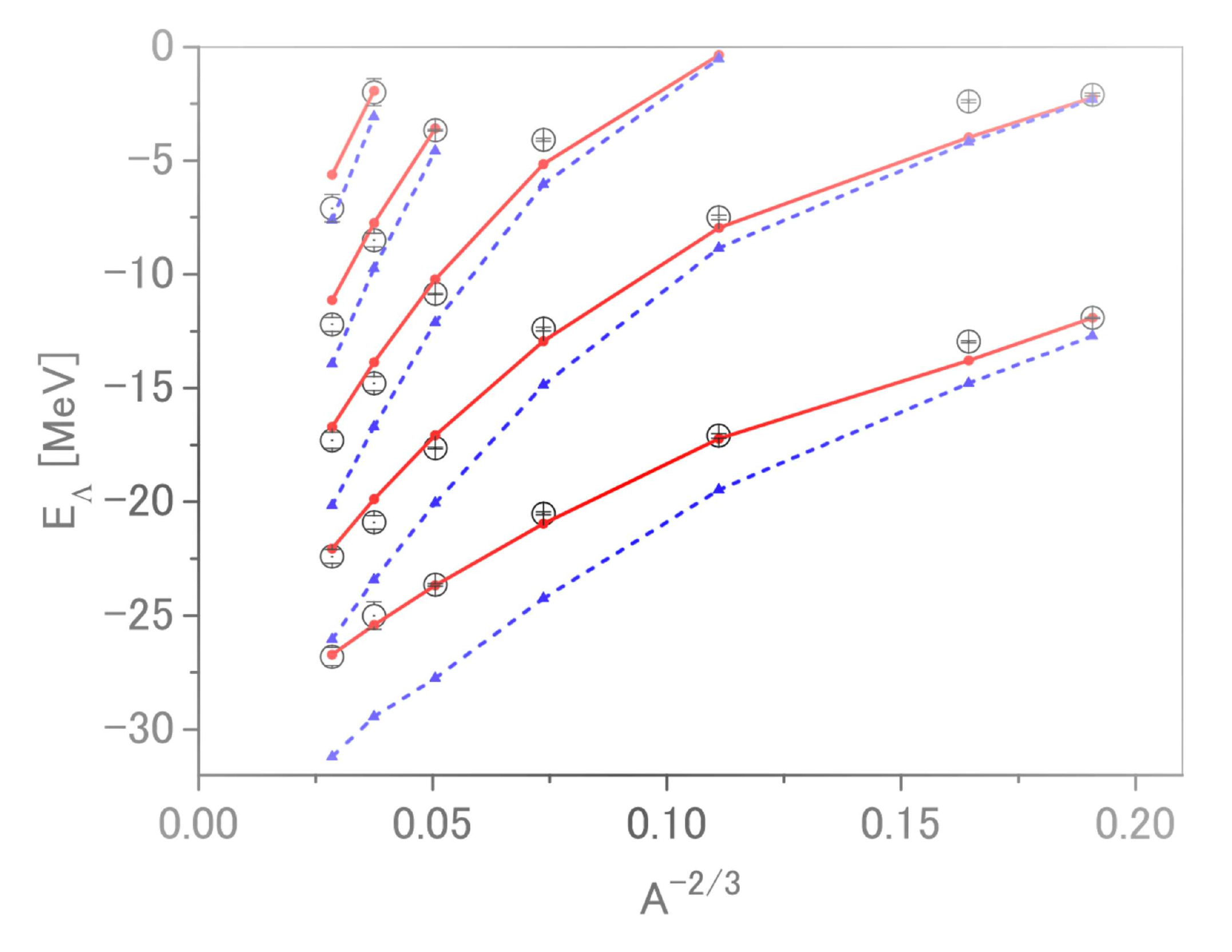}
\caption[]{
Energy spectra of  $^{13}_{\Lambda}$C, $^{16}_{\Lambda}$O, $^{28}_{\Lambda}$Si, $^{51}_{\Lambda}$V, $^{89}_{\Lambda}$Y,
$^{139}_{\Lambda}$La, $^{208}_{\Lambda}$Pb as a function of $A^{-2/3}$ calculated with the ESC16 model \cite{
%Nagels:2019
NAG19}.
Solid and dashed lines show calculated value by the $G$-matrix folding model derived from ESC16 with and without the three-body interaction, respectively.
Open circles and error bars denote the experimental values.
}
\label{fig_LambdaBindingEnergy}
\end{center}
\end{figure}

Yamamoto {\it et al.} pointed out that such $\Lambda NN$ three-body effect appears in the $\Lambda$ binding energies in heavy $\Lambda$ hypernuclei \cite{
%Yamamoto:2014
YAM14}.
Two-body $\Lambda N$ potential in the nuclei obtained by the G-matrix calculation from the Nijmegen ESC model was used.
The three-body interaction, composed of the multipomeron exchange repulsive potential (MPP) and the phenomenological three-body attraction (TBA), is added as the density dependent potential.
These three-body interaction's parameters were adjusted to reproduce the angular distribution of $^{16}$O $+$ $^{16}$O elastic scattering at $E/A = 70$ MeV
with use of the G-matrix folding potential, and values of the saturation density and the energy per nucleon there in nuclear matter.
One of the possible scenarios to explain two-solar-mass neutron star is that such density dependent three-body repulsion (MPP) contributes additional repulsive force to support the gravity of the neutron star.
Such three-body interaction should give a sizable contribution to the energy spectra of $\Lambda$ hypernuclei.
They calculated it systematically for wide mass number $\Lambda$ hypernuclei ($^{13}_{\Lambda}$C, $^{16}_{\Lambda}$O, $^{28}_{\Lambda}$Si, $^{51}_{\Lambda}$V, $^{89}_{\Lambda}$Y,
$^{139}_{\Lambda}$La, $^{208}_{\Lambda}$Pb) in the framework of  the latest Nijmegen model (ESC16)  with and without the three-body force as shown in Figure \ref{fig_LambdaBindingEnergy} where
its energy spectra is reproduced well by including the three-body force \cite{
%Nagels:2019
NAG19}.
They concluded that the density dependent three-body force works to reproduce better energy spectra of heavy system.
In the high-density region, this three-body term dominated by MPP is expected to lead the stiff EOS of the hyperon-mixed neutron-star matter.

\begin{figure}[h!]
\begin{center}
\includegraphics[width=14cm]{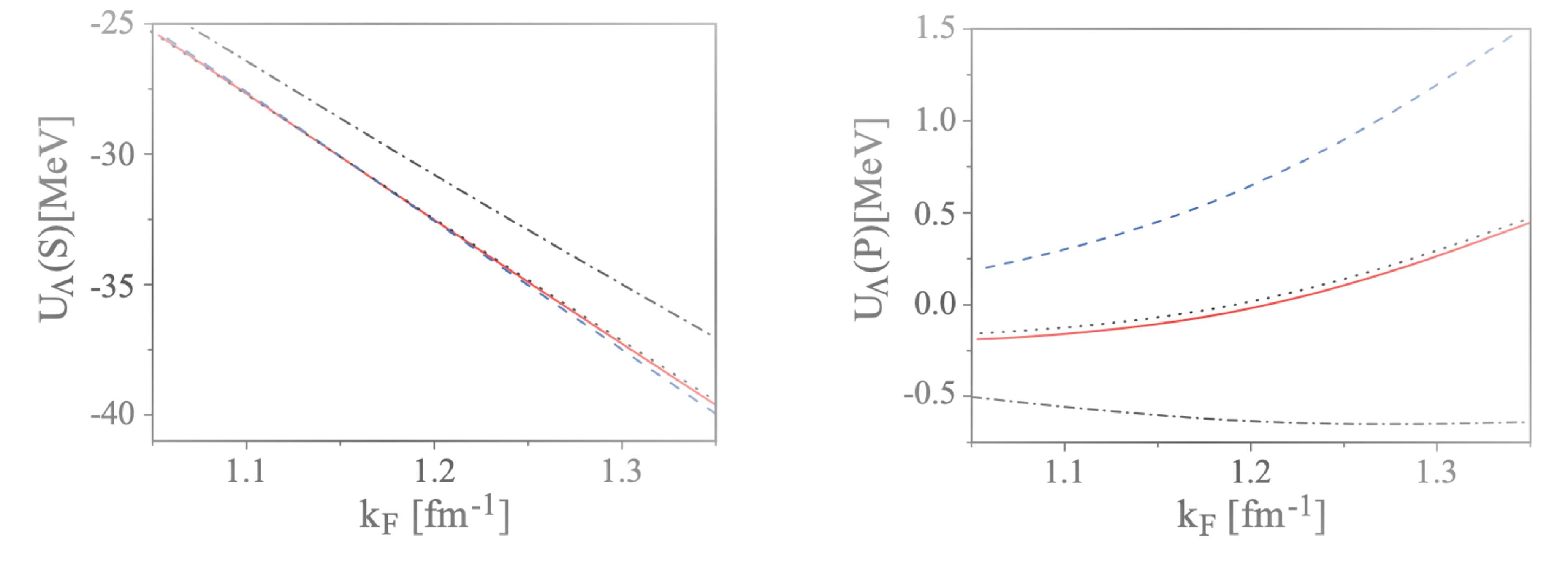}
\caption[]{
$S-$state (left) and $P-$state (right)  contributions to $U_{\Lambda}$ as a function of $k_{F}$ taken from ref. \cite{
%Isaka:2017
ISA17}.
Red solid, blue dashed, black dotted, and black dot-sashed curves are for ESC14, ESC12, ESC08a, and ESC08b, respectively.
The latest ESC16 is similar to ESC14.
}
\label{fig_U_Lambda}
\end{center}
\end{figure}

Isaka {\it et al.} also analyzed the $\Lambda$ binding energies with the various Nijmegen two-body interaction models with AMD framework \cite{
%Isaka:2017
ISA17}.
They pointed out that the three-body force strength reproducing the $\Lambda$ binding energies depends on the two-body interaction model and 
there are considerable difference even within various ESC and NSC models.
Such difference comes due to the potential of $P$-state contribution.
Figure \ref{fig_U_Lambda} shows the density dependence of the $S$- and $P$-state contributions to $\Lambda$ potential as a function of $k_{F}$ \cite{
%Isaka:2017
ISA17}.
In case of ESC14 which is similar to ESC16, the $P$-state contributions are small. 
On the other hand, in case of ESC12 and NSC96f, the $P$-state contributions are substantially repulsive.
Therefore, in the former model,  the $\Lambda$ binding energies are well reproduced by adding the strong MPP repulsion which can support the two-solar mass neutron star.
However, in the later model, the $P$-state repulsive contribution can reasonably work to reproduce the $\Lambda$ binding energy and there is no room to introduce the strong MPP repulsion.
The difference in the $P$-wave contribution even within the same Nijmegen framework is quite reasonable 
because these models predict the different behavior in differential cross section as shown in Figure \ref{fig_showDiffCS_Theory}.
In the NSC97f, the larger $P$-wave contribution might make the differential cross section larger than that in ESC16.

We believe that comparison between the high-resolution experimental data of the $\Lambda$ binding energy and the Nijmegen model calculation with the three-body interaction
is a crucial way to derive the $\Lambda NN$ three-body interaction.
In order to reinforce such the strategy, the accurate $\Lambda p$ scattering data are really awaited.
In the $NN$ sector, the realistic model of the nuclear force is established by updating the theoretical frameworks so as to reproduce a plenty of two-body scattering observables.
Now, we also have to construct the realistic $YN$ interaction model and both of theoretical and experimental efforts are necessary.
For the experimental side, various scattering observables as many as possible with a reasonable precision should be measured as the basic information for constructing a real bridge to the many-body system.

Chiral EFT is a promising theoretical framework to describe the nuclei and neutron matter.
One of the biggest advantages of this model is that the three-body forces appear naturally and automatically in a consistent implementation of the framework.
In the power counting in the present YN Chiral EFT, such three-body forces arise first at next-to-next-to-leading order (N$^{2}$LO) in the chiral expansion \cite{
%Petschauer:2016
PET16}.
If the decuplet baryons are included as explicit degrees of freedom, the three-body forces with decuplet excitation appear already at NLO.
In this calculation, two types of B-M-B$^{*}$  and B$^{*}$-B-B-B couplings appear  where B, M and B$^{*}$ represent octet baryons, pseudo scalar mesons and decuplet baryons, respectively.
The LECs in the B-M-B$^{*}$ coupling can be estimated through decuplet saturation  \cite{Petschauer:2017}.
The two-pion-exchange $\Lambda NN$ \cite{Petschauer:2017} and $\Lambda NN$-$\Sigma NN$ \cite{
%Kohno:2018
KOH18} forces  are already applied to study a density-dependent effective potential for the BB interactions by integrating one nucleon degree of freedom in the medium.
The repulsive effect of the $\Lambda NN$ interaction is calculated to be about 5 MeV at normal density ($\rho_{0}$) and about 20 MeV at 2$\rho_{0}$ in symmetric nuclear matter \cite{
%Kohno:2018
KOH18}.
A similar repulsive contribution is also obtained in pure neutron matter.
The $\Lambda NN$-$\Sigma NN$ coupling effect works to cancel the repulsive effect of the $\Lambda NN$ interaction at normal density in symmetric nuclear matter.
However, in the higher densities, the cancellation is incomplete and the net three-body force contribution is repulsive.
In pure neutron matter, the effect of the $\Lambda NN$-$\Sigma NN$ coupling effect is very small and, therefore, does not weaken the repulsion from $\Lambda NN$ force\cite{
%Kohno:2018
KOH18}.
In the present calculation, the three-body forces with B$^{*}$-B-B-B couplings is not included yet, because there are two LECs which should be determined from experimental data.
However, the number of LECs are limited.
These parameters can be fixed from the $0^{+}$ and $1^{+}$ states of $^{4}_{\Lambda}$He\cite{Haidenbauer:priv}.
A $\Lambda$-deuteron scattering experiment, which is also possible in future at J-PARC, can also contribute to fix these LECs.
In general, such $\Lambda$-deuteron or $\Lambda$-$\alpha$ scattering measurement includes rich information to fix these LECs compared with the binding energy of a few-body hypernuclei.
The structure of single-$\Lambda$ hypernuclei is studied using the chiral EFT up to NLO by J\"{u}lich-Bonn-Munich group \cite{Haidenbauer_Vidana}, where the three-body interaction is not included yet.
A qualitatively good agreement with data is obtained for the chiral EFT19 interaction over a fairly large range of mass number values.
This shows the potential of the chiral EFT model and it is worth providing scattering observables experimentally to update the chiral EFT model.

\subsubsection{$\Lambda p$ scattering experiment at the K1.1 beam line}

\begin{figure}[h]
\begin{center}
\includegraphics[width=13cm]{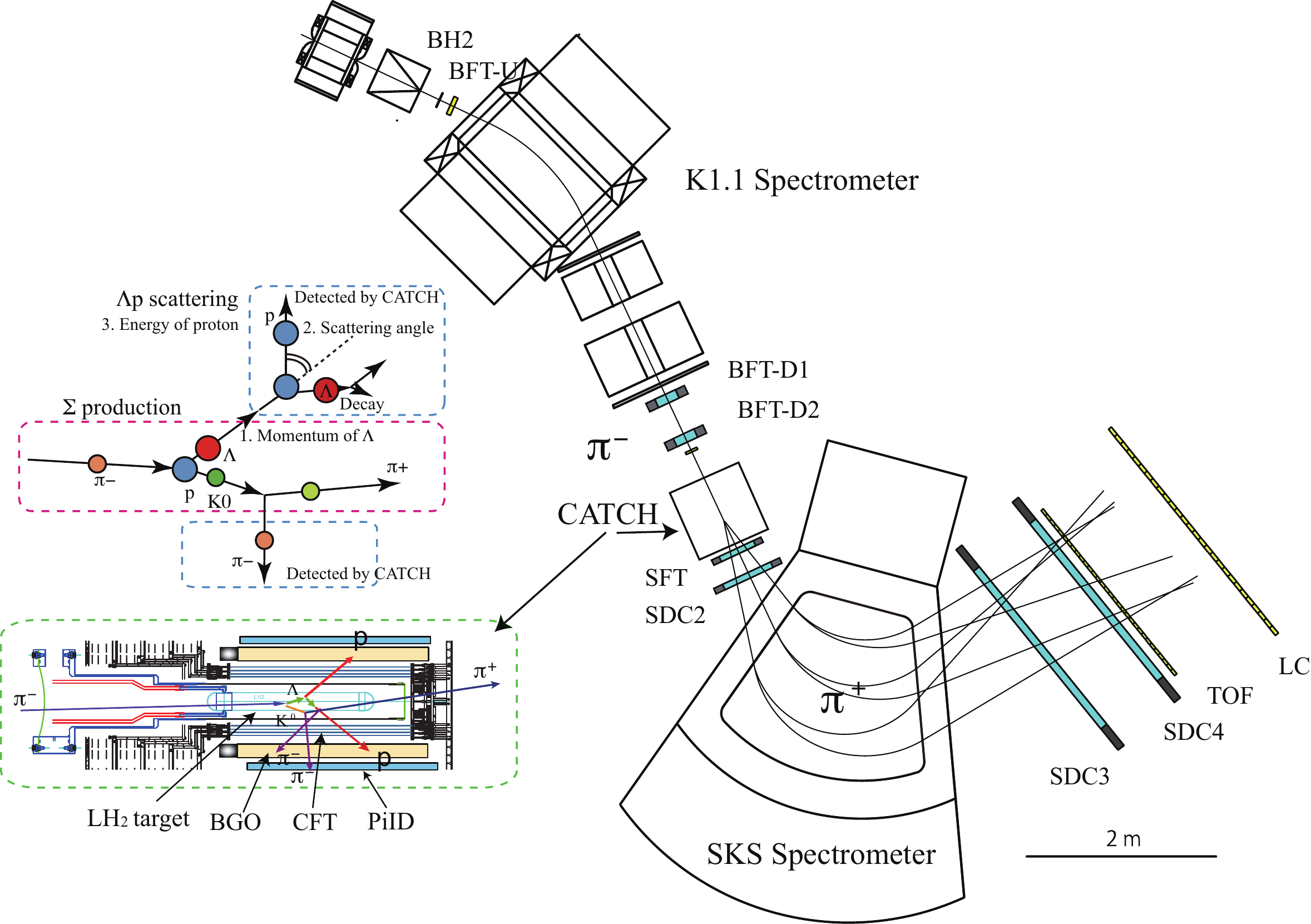}
\caption[]{
Experimental setup for the proposed $\Lambda p$ scattering experiment.
The schematic drawing of how to identify $K^{0}_{S}$ and $\Lambda p$ scattering is also shown.

}
\label{figs_K1.1withCATCH}
\end{center}
\end{figure}

We propose a new experiment for a  $\Lambda p$ scattering by using a momentum-tagged $\Lambda$ beam produced by the $\pi^{-} p \to K^{0}_{S} \Lambda$ reaction at the K1.1 beam line.
Figure \ref{figs_K1.1withCATCH} shows the experimental setup.
Experimental concept is almost the same with the E40 experiment for the $\Sigma p$  scattering.
1.05 GeV/$c$ $\pi^{-}$ beam is irradiated to the liquid hydrogen (LH$_{2}$) target with 30 cm thickness.
We design experiment to use 30 M/spill $\pi^{-}$ beam which is 1.5 times higher intensity than that in E40 in order to accumulate as much $\Lambda$ beam as possible.
The SKS spectrometer was used as a forward spectrometer to detect $\pi^{+}$ \cite{
%Takahashi:2012
TAK12}.
Because the momentum resolution of SKS is expected to be $\Delta p/p=10^{-3}$ (FWHM) which is more than 10 times better than that of KURAMA,
the missing mass resolution for $\Lambda$ will be improved and this also results in the improvement of the S/N ratio for $\Lambda$ identification.
In order to detect the $\Lambda p$ scattering and $\pi^{-}$ from $K^{0}_{S}$ decay, 
CATCH, which consists of a cylindrical fiber tracker, BGO calorimeter and PiID,  is used as shown in enlarged figure in Figure \ref{figs_K1.1withCATCH}.
CATCH surrounds the LH$_{2}$ target and detect recoil proton and scattered $\Lambda$ by detecting the $\pi^{-}p$ decay.
%In this section, we describe the experimental details.

\begin{itemize}
    \item Advantage of the $\pi^{-}p \to K^{0} \Lambda$  at $p_{\pi} = 1.05$ GeV/$c$
\end{itemize}

\begin{figure}[!t]
\begin{center}
\begin{minipage}{6.5cm}
 \includegraphics[width=0.98\textwidth]{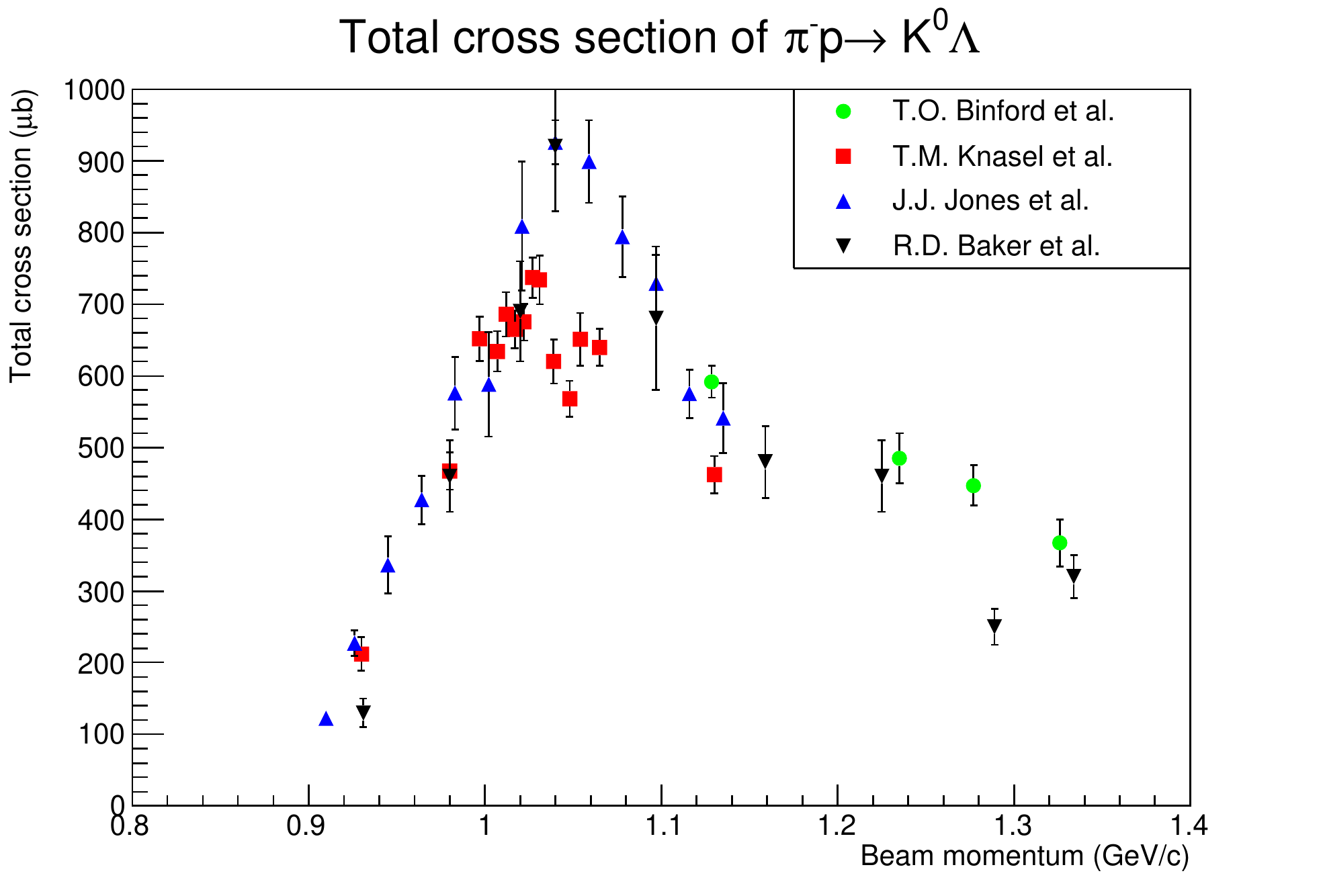}
\end{minipage}
\hspace{0.2cm}
\begin{minipage}{6.5cm}
 \includegraphics[width=0.98\textwidth]{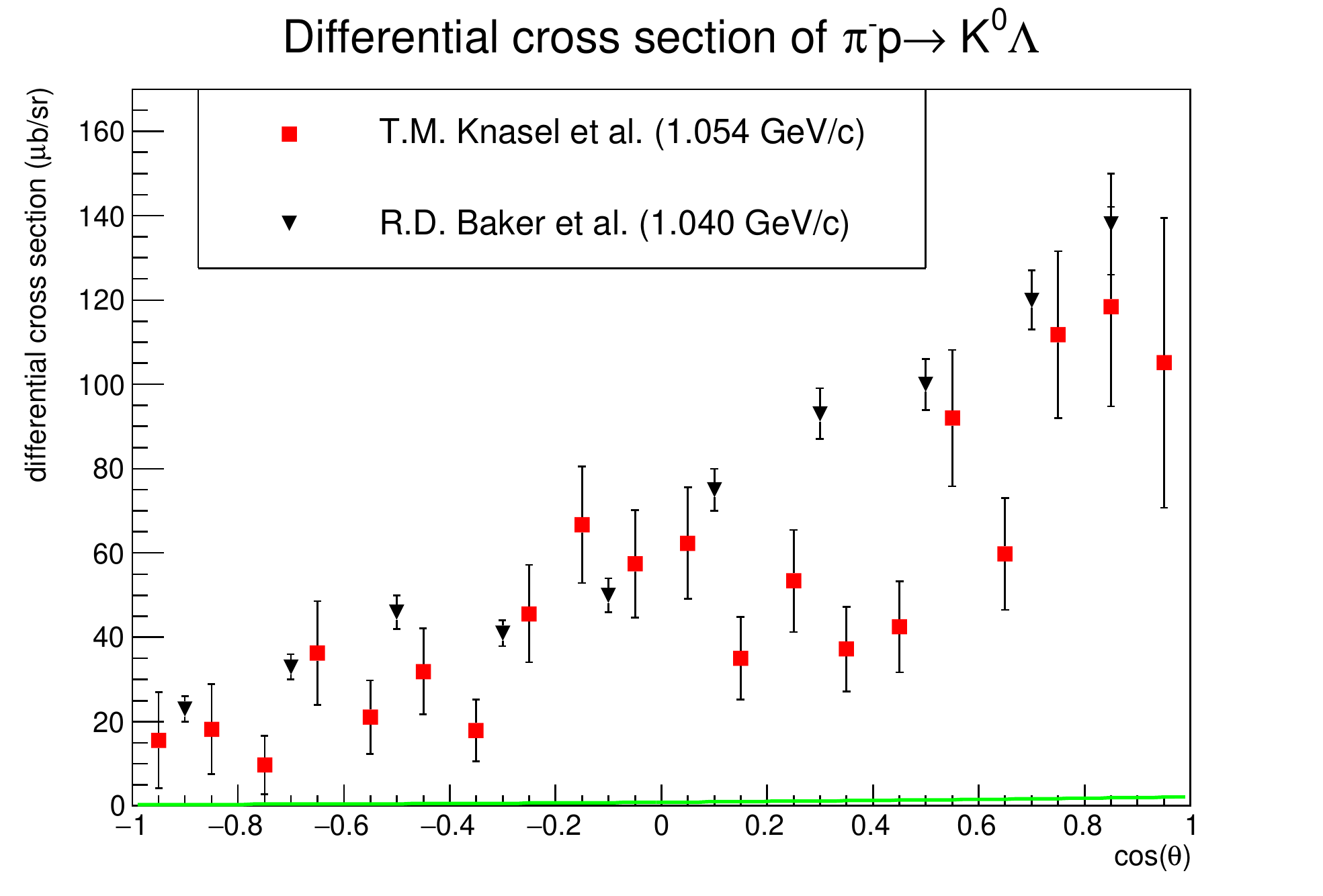}
\end{minipage}
\caption{(Left) Momentum dependence of the total cross section of the $\pi^{-}p\to K^{0}\Lambda$ reaction. (Right) Differential cross section around $p$=1.05 GeV/$c$.\cite{Jones:1971, Knasel:1971, Binford:1971, Baker:1978}} 
\label{figs_showCrossSectionK0Lambda}
\end{center}
\end{figure}

In order to realize the high-statistics $\Lambda p$ scattering experiment, 
it is the very important to accumulate as much $\Lambda$ beam as possible.
Therefore, the $\Lambda$ production cross section is one of the most important experimental parameters.
Figure \ref{figs_showCrossSectionK0Lambda} (left) shows beam momentum dependence of the total cross section.
The total cross section becomes maximum around $\sim$900 $\mu$b at 1.05 GeV/$c$.
Figure \ref{figs_showCrossSectionK0Lambda} (right) shows the differential cross section.
There are two measurements that are not consistent with each other, but the measurement by R.D. Baker {\it et al.} \cite{Baker:1978} seems to be more reliable.
In this simulation study, we have assumed that the $\Lambda$ production cross section is 900 $\mu$b and the angular distribution by R.D. Baker {\it et al.} are used. 

\begin{figure}[t]
\begin{center}
\includegraphics[width=12cm]{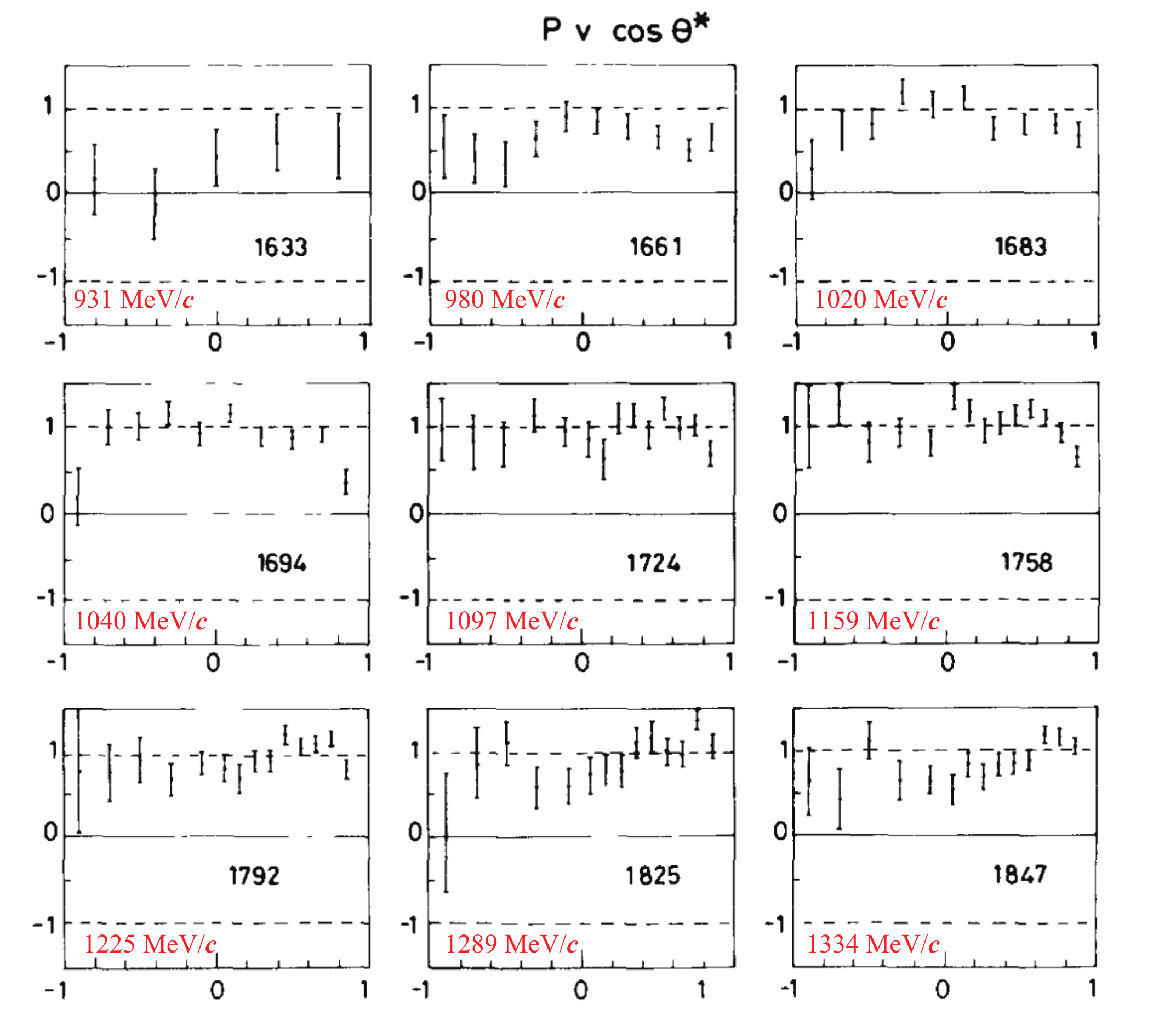}
\caption[]{
Induced polarization of the $\Lambda$  by the $\pi^{-} p \to K^{0} \Lambda$ reaction for the momentum of  $0.931 \le p$ (GeV/$c$) $ \le 1.334$.
Horizontal axis show the $\cos \theta_{CM}$, where $\theta_{CM}$ is the scattering angle of $K^{0}$ in the CM frame.
High polarization can be obtained for all momentum ranges. 
}
\label{figs_LambdaPolarization}
\end{center}
\end{figure}

Second important characteristics of this $\Lambda$ production around 1.05 GeV/$c$ is that $\Lambda$ particles are produced with almost 100\% polarization 
for the ($\pi^{-}, K^{0}$) production plane.
Figure \ref{figs_LambdaPolarization} shows the $\Lambda$ polarization in the momentum range of $0.931 \le p$ (GeV/$c$) $ \le 1.334$
measured by R.D. Baker {\it et al.}\cite{Baker:1978} .
This polarization can be measured by the angular dependence of the proton from the $\Lambda$ decay.
As shown in this figure, the $\Lambda$ polarization is sufficiently high to measure the spin observables.
%The detail of the $\Lambda$ beam polarization is described section*,.
The $K^{0}$ angle can be covered from 1 to 0 in $\cos \theta_{CM}$, where $\theta_{CM}$ is the scattering angle of $K^{0}$ in the CM frame.
Therefore almost 100\% polarized $\Lambda$ can be used as hyperon beam automatically.
The production of the polarized $\Lambda$ beam enables us to measure not only the differential cross section but also the spin observables such as Analyzing power ($A_{y}$) and 
depolarization ($D_{y}^{y}$).

\subsubsection{Goal of $\Lambda p$ scattering experiment at K1.1}
%We propose a new experiment to measure the differential cross section and spin observables (Analyzing power and depolarization) with a polarized $\Lambda$ beam
%in the momentum range between 0.4 and 0.8 GeV/$c$ at the K1.1 beam line.
%$\Lambda$ beam is tagged by the $\pi^{-}p \to K^{0} \Lambda$ reaction at $\pi^{-}$ beam momentum of 1.05 GeV/$c$.
%In this momentum, there are two experimental advantages.
%The first one is that the $\Lambda$ production cross section becomes maximum around $\sim$1 mb.
%The other one is that the produced $\Lambda$ is polarized for the ($\pi^{-}, K^{0}$) reaction plane with almost 100\% polarization.
%These features enable us to derive not only the differential cross section but also the spin observables.
%One of the experimental challenges is the identification of the $\pi^{-}p \to K^{0} \Lambda$ reaction, 
%because the detection of $K^{0} \to \pi^{+} \pi^{-}$ with an enough acceptance is  not straightforward.
%However we have established identification method of the $\pi^{-}p \to K^{0} \Lambda$ reaction with E40 by-product data
%which is described in Chapter \ref{Study_in_E40_data}.

The $\Lambda p$ scattering event is identified by the CATCH detector system, which was developed for the $\Sigma p$ scattering experiment, 
by detecting the scattered $\Lambda$ and recoil proton.
In this experiment, $\pi^{-}$ from $K^{0}$ decay is also detected by CATCH as shown in the enlarged figure in Figure \ref{figs_K1.1withCATCH}.
Experimental method is basically the same with one developed for the $\Sigma p$ scattering.
A liquid hydrogen target is used as the $\Lambda$ production and $\Lambda p$ scattering targets and the $\Lambda p$ scattering events are kinematically identified 
by detecting particles in the final state without any imaging information.
High intensity $\pi^{-}$ beam of 30 M/spill will be used to accumulate high intensity $\Lambda$ beam.
In this experiment, we want to accumulate 100 M $\Lambda$ beam in total with separated two periods.
We request 34-days beam time including 29-days production run and 5-days commissioning and calibration runs as first stage to measure the differential cross section and the Analyzing power of the $\Lambda p$ scattering by accumulating 50M momentum-tagged $\Lambda$ beam.
As second stage, we request additional 34-days beam time including the production and commissioning/calibration runs with the same ratio to measure the depolarization and to improve the accuracy of measurements of the differential cross section and Analyzing power.
Here, we assume that spill structure is 5.2 s cycle with a beam duration of 2 s and the $\pi^{-}$ beam intensity is 30 M/spill.
The experimental and analysis details are written in reference \cite{Miwa_Proposal}.

\begin{figure}[]
\begin{center}
\includegraphics[width=12cm]{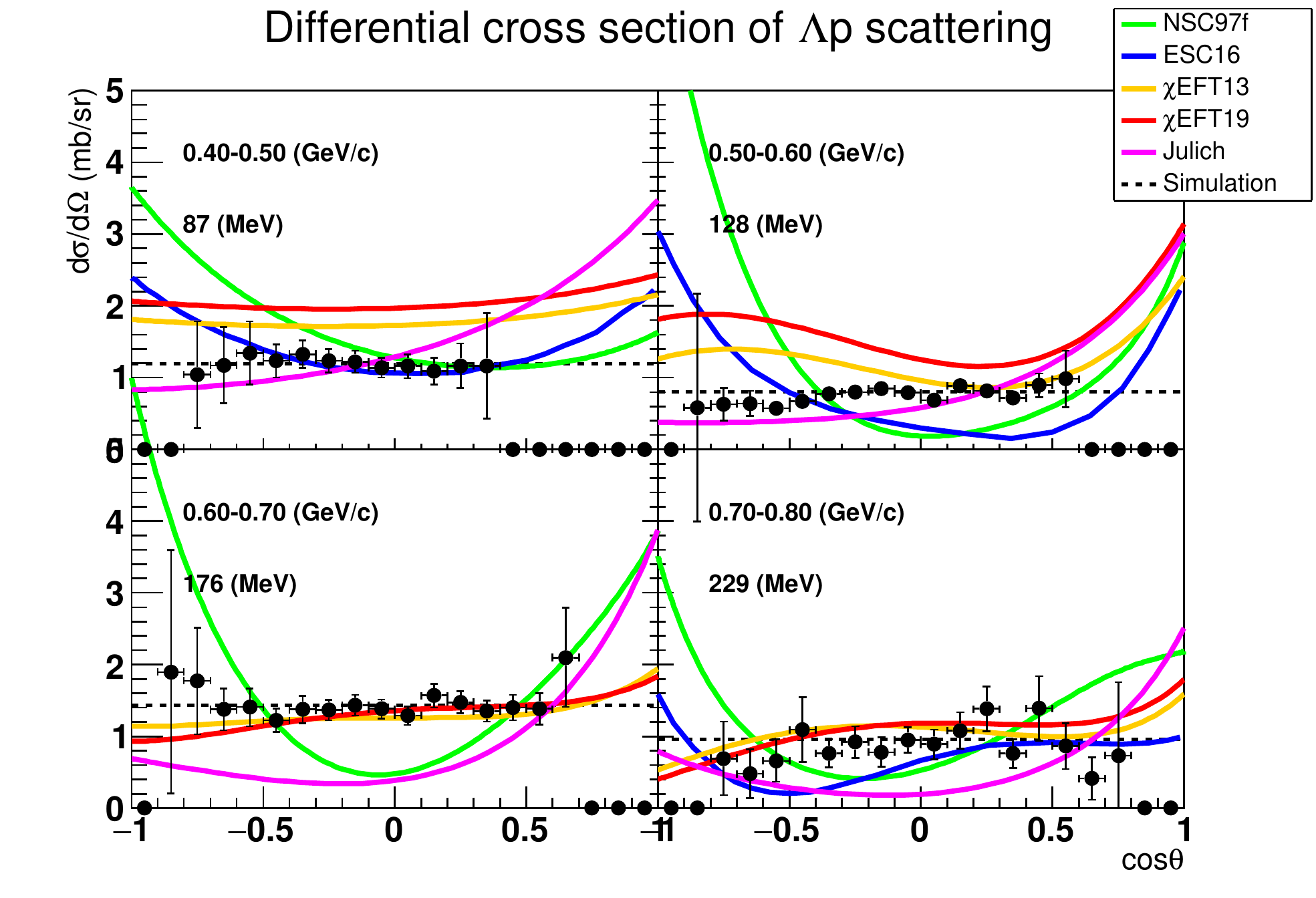}
\caption[]{
Simulated results for the differential cross section measurement of the $\Lambda p$ scattering in the momentum range from 0.4 to 0.8 GeV/$c$ with 0.1 GeV/$c$ momentum interval for 50M $\Lambda$ beam.
Theoretical calculations are also presented.
}
\label{fig_showAllLambdaPdSdW}
\end{center}
\end{figure}

\begin{figure}[]
\begin{center}
\includegraphics[width=12cm]{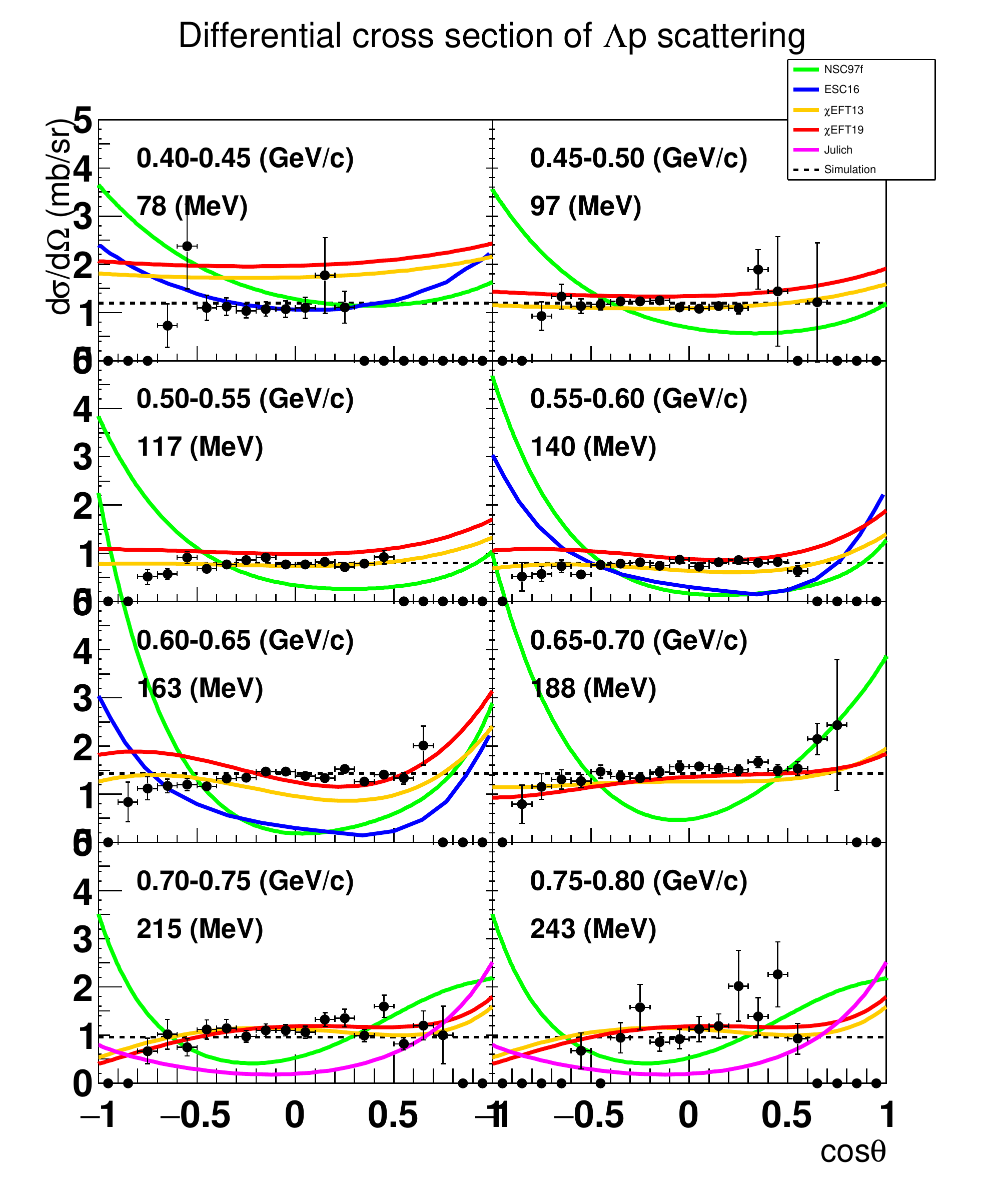}
\caption[]{
Simulated results for the differential cross section measurement of the $\Lambda p$ scattering in the momentum range from 0.4 to 0.8 GeV/$c$ with 0.05 GeV/$c$ momentum interval with 100M $\Lambda$ beam.
Theoretical calculations are also presented.
}
\label{fig_showDSDW0_8MomDiv}
\end{center}
\end{figure}

The differential cross section is derived at least for four different momentum ranges (100 MeV/$c$ momentum step)  for 50M $\Lambda$ beam as shown in Figure \ref{fig_showAllLambdaPdSdW}.
The statistical error for angular region of 0 $<$ $\cos \theta$ $<$ 0.1 as a typical angular bin is expected to be 10\%.
In this simulation, $\Lambda p$ total cross sections for each momentum are assumed to be 15, 10, 20 and 12 mb, respectively, based on the past experimental data and theoretical calculation.
The angular dependence is simply assumed to be flat for checking easily whether the analysis procedure works well or not.
In the present analysis, we request two protons in the final state to suppress the background due to the background contamination in the $\Lambda$ production identification.
The angular acceptance for the differential cross section is limited due to this cut.
We hope the angular acceptance becomes much wider by loosing the two-proton cut by improving the $\Lambda$ identification method in future study.
In this proposal, we are showing conservative but assured result in present study.
%We will cover as wider as possible to measure the differential cross section from both the scattered $\Lambda$ and the recoil proton.
%The angular dependence which is important to determine the higher wave contributions can be measured with the enough accuracy to improve the theoretical models.
As shown in Figure \ref{fig_showAllLambdaPdSdW} for 50 M $\Lambda$ beam, we can select best theoretical picture among the Nijmegen model (NSC97f and ESC16), J\"{u}lich boson-exchange model and chiral EFT(13 and 19) very clearly.
In present, we are requested to select better model within the same theoretical framework (for example selection between chiral EFT13 and 19 or selection among NSC97f and ESC16 and other ESC versions).
Selection of better model in the same theoretical framework could be possible even for 50 M $\Lambda$ beam depending on the difference in the theoretical prediction.
However, differential cross section measurement with narrower momentum step is quite effective to discriminate the better framework in the same theoretical model, 
because the momentum dependence of the differential cross section becomes quite large as shown in Figure \ref{fig_showDiffCS_Theory} toward the $\Lambda$ beam momentum of the $\Sigma N$ threshold ($\sim$630 MeV/$c$) due to the tensor $^{3}S_{1}$-$^{3}D_{1}$ coupling.
Therefore, the differential cross section measurement with narrower momentum range of  50 MeV/$c$ momentum step is very important to measure this momentum dependence.
By accumulating 100M $\Lambda$ beam in total, we will measure differential cross sections with  10\% statistical error for each 50 MeV/$c$ $\Lambda$ beam momentum step from 0.4 to 0.8 GeV/$c$ as shown in Figure \ref{fig_showDSDW0_8MomDiv}.
As mentioned in later,  we need also 100M $\Lambda$ beam for the $D_{y}^{y}$ measurement.

\begin{figure}[]
\begin{center}
\includegraphics[width=12cm]{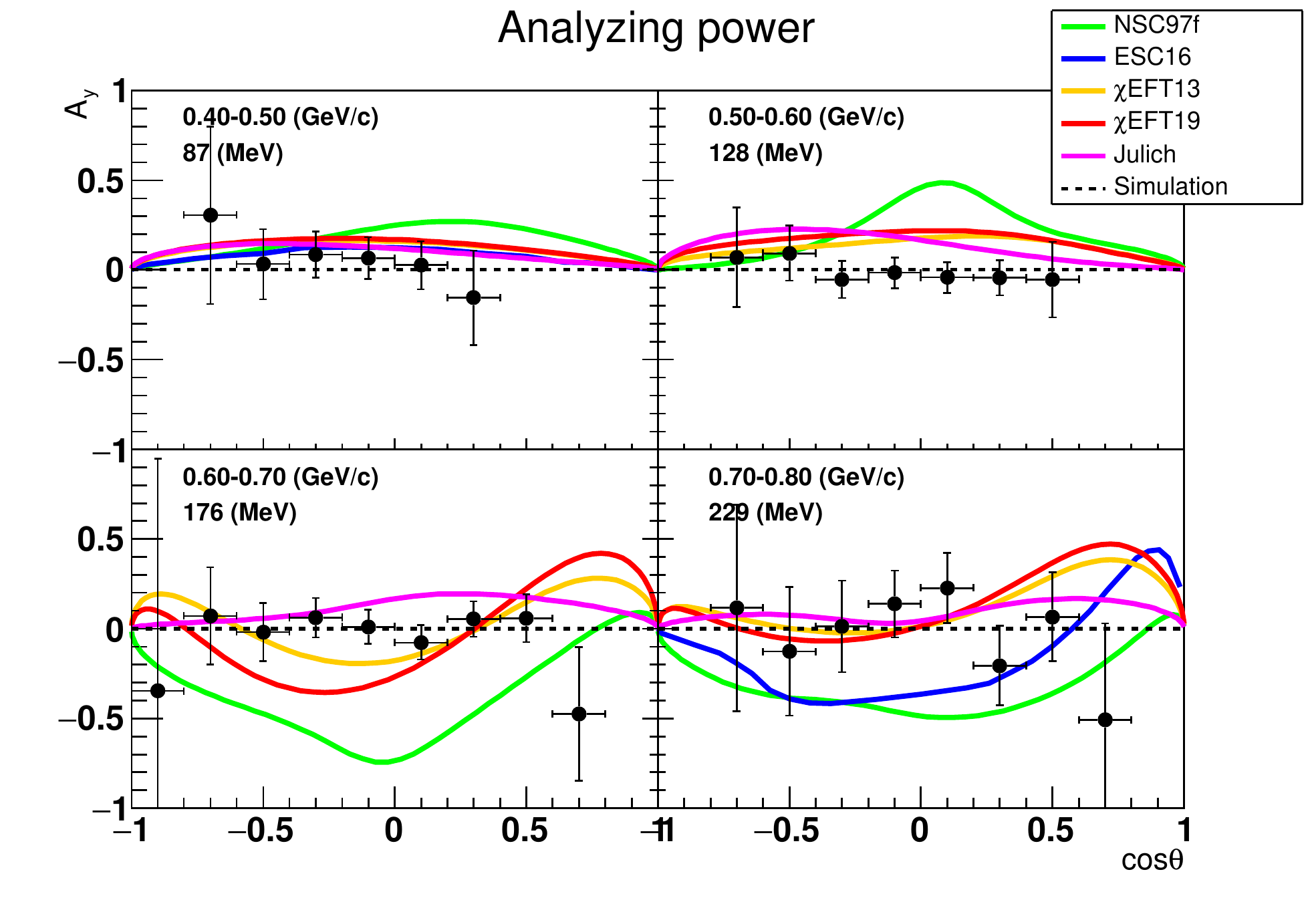}
\caption[]{
Simulated results for $A_{y}$ measurement of the $\Lambda p$ scattering in the momentum range from 0.4 to 0.8 GeV/$c$ with 0.1 GeV/$c$ momentum interval for 100M $\Lambda$ beam.
Theoretical calculations are also presented.
}
\label{fig_showLR_Asymmetry_c2}
\end{center}
\end{figure}

\begin{figure}[]
\begin{center}
\includegraphics[width=12cm]{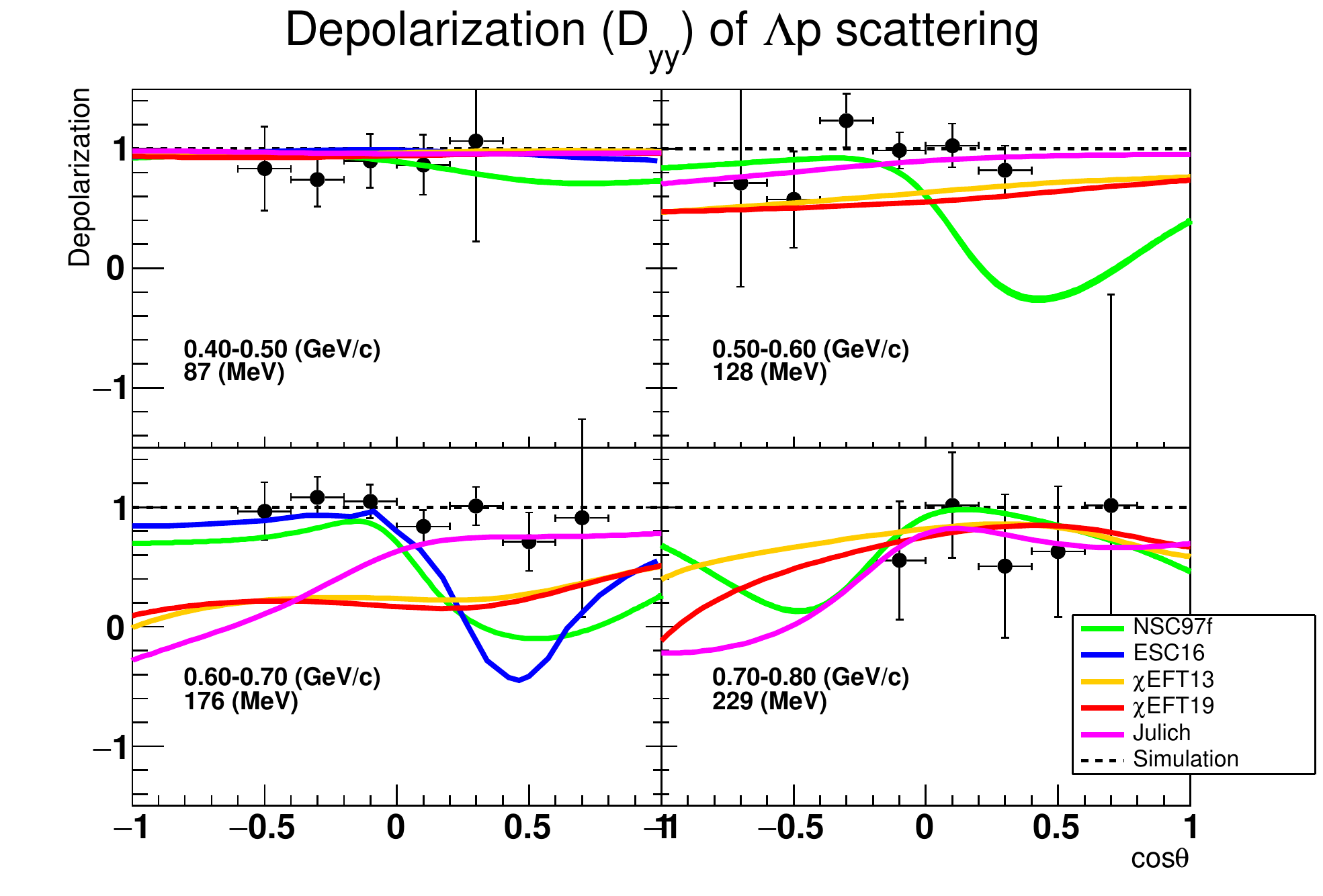}
\caption[]{
Expected $D^{y}_{y}$ accuracy for 100M $\Lambda$ beam.
In this simulation,  no spin transfer ($D_{y}^{y}=0$) and no induced polarization ($P=0$) in the $\Lambda p$ scattering are assumed for simplicity.
The beam polarization of $P_{\Lambda} = 1$ is also assumed based on the past measurement.
Theoretical calculations are also shown together.
Due to the limitation of the statistics, errors in the high momentum region (0.7-0.8 GeV/$c$) are still large.
However,  in the middle momentum region (0.5 to 0.7 GeV/$c$) model difference between the chiral EFT and Nijmegen models can be clearly separated.

}
\label{fig_showAllDyy}
\end{center}
\end{figure}

Figure \ref{fig_showLR_Asymmetry_c2} shows the sensitivity of Analyzing power ($A_{y}$) which were obtained from the left-right asymmetry for the $\Lambda p$ scattering for 100M $\Lambda$ beam.
The expected statistical error is 8\% for angular step of $d\cos \theta = 0.2$.
In this simulation,  $A_{y} = 0$ is assumed to check the experimental validity easily.
The normal technique used in the $NN$ case where the beam polarization is flipped and the number of the scattered particle are counted with a single detector setup or
the identical two detectors  were placed in the left-right symmetric position, can not be used, 
because the $\Lambda$ beam is produced  in the tilted direction for the CATCH detector.
Therefore the differential cross sections for the left scattered case and the right scattered case were derived and 
$A_{y}$ is derived from these left- and right-differential cross sections.
We confirmed that $A_{y}$ values were reasonably obtained by such analysis method in the simulation.
In the momentum range higher than 0.5 GeV/$c$, there are large differences among three theoretical models.
In order to discriminate these models, 10\% accuracy is necessary and 
such new data are quite important to improve the theoretical models because sizable difference can appear in spin observables even thought the difference in the differential cross section is small.
For example, there is large difference in $A_{y}$ at 0.65 GeV/$c$ between chiral EFT13 and 19, whereas there is very small difference in the differential cross section.
$A_{y}$ is sensitive for the anti-symmetric LS force (ALS).
In the calculation by chiral EFT13, $A_{y}$  values can be varied by adding ALS with a difference of 0.1 - 0.2 depending on the size of ALS \cite{Haidenbauer:priv}.
Therefore the present accuracy of less than 10\% is also necessary to constrain the ALS force from the $A_{y}$ measurement.

Figure \ref{fig_showAllDyy} shows the simulated results for the depolarization ($D_{y}^{y}$) measurement for 100M $\Lambda$ beam.
For this measurement, the polarization of the scattered $\Lambda$  has to be measured from the up-down asymmetry of the decay proton from $\Lambda$ with respect to the scattering plane.
The up-down symmetry in CATCH is rather better compared to the left-right symmetry because CATCH has the symmetric acceptance with respect to the central axis of the system.
However, there are some detector asymmetry depending on the scattering vertex position and scattering angle.
Such systematic effect due to the asymmetry in CATCH is also studied in this simulation.
In this simulation, $D_{y}^{y}=1$ which means no spin flip occurred in the scattering is assumed for the simplicity.
We have confirmed the $D_{y}^{y}$ can be reproduced reasonably by using the up-down asymmetry technique although there are some deviations due to the asymmetric effect.
This is the first trial to measure the polarizations in both initial and final states and this is an important step to measure further measurements such as rotation parameters.
As mentioned in Section \ref{subsec_spindepYN},  $D_{y}^{y}$ is also the mixed combination of several spin-dependent and spin-independent amplitudes.
However, $D_{y}^{y}$ is expected to be closely related to the tensor force.
Although the accuracy is limited (15\% level), the discrimination of theoretical models is possible for momentum range from 0.5 to 0.7 GeV/$c$ as shown in Figure \ref{fig_showAllDyy} .
These data can also impose constraint on theories from different aspects.
Accumulation of these spin observables and differential cross section are essential to improve these theoretical models and to establish the realistic YN interaction model in future.

\subsubsection{International situation and future prospect at J-PARC}
Recently, femtoscopic measurements are widely applied to baryon pairs produced in $pp$ and heavier ion collisions at LHC and RHIC \cite{
%STAR:2015
STAR:2014dcy, 
%ALICE:2019
ALICE:2018ysd, 
%Morita:2019
Morita:2014kza,
%ALICE:2020
ALICE:2019buq}.
These measurements are very strong method to deduce the scattering length and effective range of the BB interactions including multi-strangeness systems. 
Because the femtoscopy is applied for the baryon pairs with small relative momentum, the $S$-wave information will be updated from these studies.
In the proposed setup at J-PARC, measurement of  the low energy YN scattering is rather difficult due to the difficulty in detection of low energy protons.
However, in the present experiment, we can measure the differential information for higher momentum region.
Therefore, these experiments are complemental relation.
It should also be stressed that the differential information is quite awaited from theoretical side.

CLAS collaboration also has a potential to measure YN scattering from the re-scattering of hyperons produced by the photo-induced reaction \cite{
%Prince:2019
Price:2019eoc}.
In their analysis, the $\Lambda$ momentum region is higher than 1 GeV/$c$.
Therefore, relation between CLAS and this experiment is also complemental.
However, CLAS is a potential competitor.

In J-PARC, Honda {\it et al.} \cite{Honda:2019} are considering a $\Lambda p$ scattering experiment at the High-p beam line in the secondary-particle mode.
In the experiment at the High-p beam line,  a wide momentum range (0.4 $<$ $p$(GeV/$c$) $<$ 2.0) of $\Lambda$ particle can be identified as $\Lambda$ beam thanks to a large acceptance of the forward spectrometer and higher incident $\pi^{-}$ beam momentum.
This experiment is also important to measure the energy dependence of the $\Lambda N$ cross section up to high momentum range (2 GeV/$c$), because the cross section at high energy is related to the short range interaction which becomes important at high density environment like the inner region of neutron star.
Although there is an overlap  in the $\Lambda$ beam momentum region with the present experiment, there are unique features in both experiments.
\begin{itemize}
\item K1.1 : spin observable measurement around $\Sigma N$ threshold region is a unique feature at the K1.1 beam line 
\item High-p : wide $\Lambda$ momentum range up to 2 GeV/$c$ and high statistics are unique feature at the High-p beam line
\end{itemize}
Therefore, both experiments can contribute the understanding of the YN two-body interaction.

In the proposed experiment at the K1.1 beam line, by accumulating several scattering observables, 
the data points ($d\sigma / d \Omega$, $A_{y}$ and $D_{y}^{y}$) become comparable with the number of phase shift values up to $F$ or $G$ waves.
There is a possibility to derive the phase shift value from the global fit of all scattering observables like the $NN$ case.
In future, we should also consider this possibility for the construction of the realistic BB interaction model.

Once the two-body $\Lambda N$ interaction is established, we should also measure the $\Lambda d$ scattering to study the $\Lambda NN$ three-body force.
Experimentally, the $\Lambda d$ scattering experiment can be performed by exchanging the liquid hydrogen target to a liquid deuterium target.
Such data are essential input to fix the LECs for the chiral three-body force.

% flatex input end: [K11-Phys.tex]

%%%%% K1.1 physics (K. Miwa)

%%%%%% K1.1 other experiments (Tamura)
% flatex input: [K11-others.tex]
\subsection{Other Experiments Planned at K1.1}

\subsubsection{$\gamma$-ray spectroscopy of $\Lambda$ hypernuclei}

$\gamma$-ray spectroscopy with germanium (Ge) detectors provides us 
with precise excitation energies and
level scheme data in accuracy of the order of 1 keV. 
We performed a series of experiments at KEK-PS, BNL-AGS, and J-PARC 
with dedicated Ge detector arrays (Hyperball, Hyperball-2, and Hyperball-J) 
and revealed level structure of $^4_\Lambda$He, 
 $^4_\Lambda$Li,  $^9_\Lambda$Be, 
 $^{11}_\Lambda$B,  $^{12}_\Lambda$C,  $^{15}_\Lambda$N, 
 $^{16}_\Lambda$O, and  $^{19}_\Lambda$F 
 by observing and assigning 25 $\gamma$ transitions \cite{TAM00,HAS06,Yamamoto:2015avw,
 %YAN18
 Yang:2017mvm}.
 With these data, the strengths of the $\Lambda$$N$ 
 spin-spin, spin-orbit, and tensor interactions
have been determined well \cite{MIL07}, and the charge symmetry breaking (CSB) 
effect in the $A=4$ hypernuclei \cite{Yamamoto:2015avw}
as well as hypernuclear shrinking effect in $^7_\Lambda$Li \cite{TAN01}
were also discovered.

At the K1.1 beam line, $\gamma$-spectroscopy experiments will be continued particularly 
for the purposes of \\
(1) Study of $\Lambda$$N$$N$ three-body force via $^3_\Lambda$H excited states,\\
(2) Further studies of CSB effects using $p$-shell $\Lambda$ hypernuclei, \\
(3) Study of density dependence of $\Lambda$ binding energies from $s_\Lambda$ $\to$
$p_\Lambda$ transitions, and\\ 
(4) Study of baryon modification in nuclear matter via $\Lambda$'s $B(M1)$ measurement 

In addition, $\gamma$-ray spectroscopy experiments 
are also necessary for the precise determination of
$\Lambda$'s single-particle energies in the ($\pi^+,K^+$) reaction spectroscopy experiments
to be carried out at HIHR. 
The $\Lambda$'s single-particle level is split into spin-doublet states as well as several core-excited states
with level spacings of the order of a few 100 keV. 
The $\gamma$-ray experiments will unveil level schemes of those hypernuclei studied at HIHR, 
which allows precise extraction of the $\Lambda$'s single-particle energies.
\\

\noindent{\bf (1) Study of $\Lambda$$N$$N$ three-body force via $^3_\Lambda$H excited states}

In our studies of the $\Lambda$$N$$N$ 3BF and
the $\Lambda$$N$-$\Sigma$$N$ coupling, which are the main problems for neutron star EOS,
the excited states of $^3_\Lambda$H would provide decisive information if they are observed.

\begin{figure}
\begin{center}
\includegraphics[width=0.8\hsize]{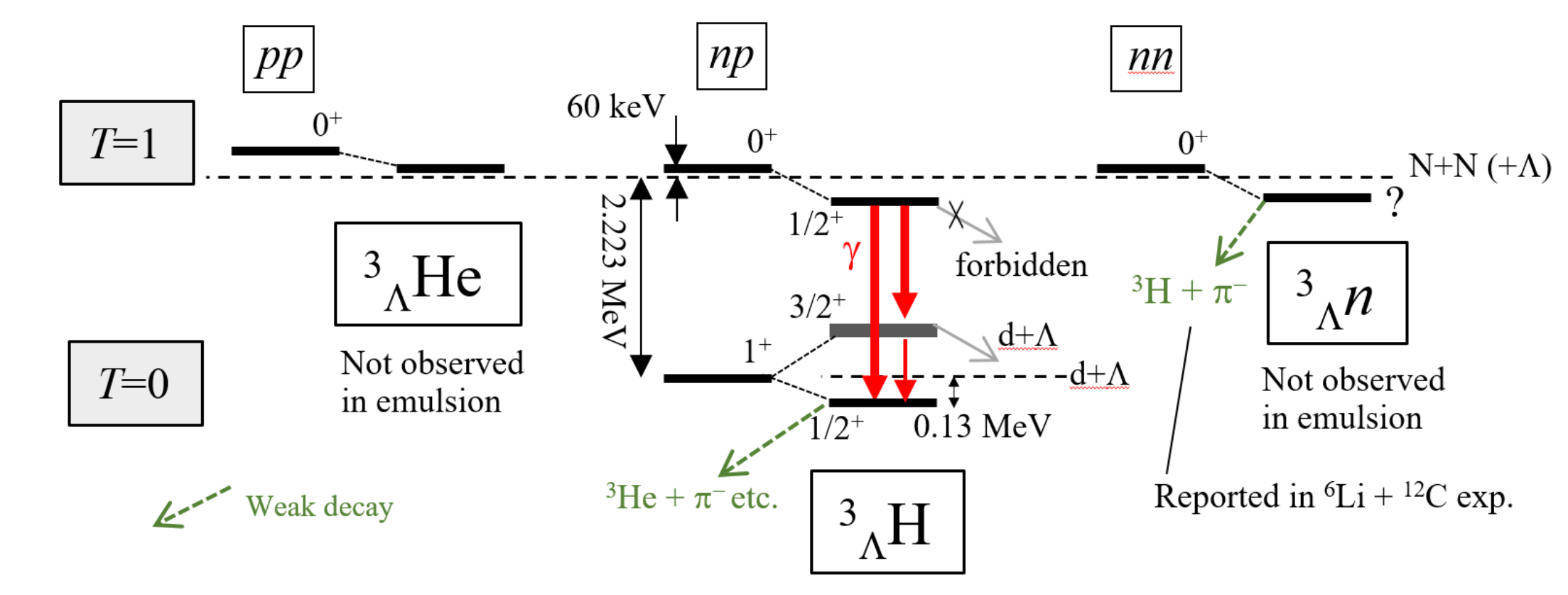}
\end{center}
\caption{Expected level scheme of $A=3$ $\Lambda$ hypernuclei.}
\label{fig:A3level}
\end{figure}

Figure \ref{fig:A3level} shows the expected level scheme of $A=3$ $\Lambda$ hypernuclei.
We hope to study the $3/2^+;T=0$ state, the spin-doublet partner of the ground state ($1/2^+;T=0$) of $^3_\Lambda$H.
It is expected to be excited by a few 100 keV to 1 MeV from the ground state. 
If the energy of this state is lower than or slightly above the $d+\Lambda$ threshold, 
we will be able to observe the $3/2^+ \to 1/2^+$ transition.

We also hope to observe transitions from the isospin 1 ground state ($1/2^+;T=1$) 
which is expected to be located around $\sim$2 MeV excitation from the $^3_\Lambda$H ground state. Although this state should be located above the $d+\Lambda$ threshold
but the strong decay to $d+\Lambda$ is greatly suppressed due to isospin conservation.
Therefore, if the energy of the $1/2^+;T=1$ state is lower than or slightly higher than
the $p+n+\Lambda$ threshold, $\gamma$ transitions to the ground state ($T=0$) doublet
are expected to be observed.
If the $^3_\Lambda n$ hyeprnucleus ($nn\Lambda$ state) is bound as reported by 
the HypHI experiment \cite{RAP13}, these transitions will be observed and the precise energy of the $^3_\Lambda$H $T=1$ state will be deduced. It would provide invaluable information
on the $T=1$ $\Lambda$$N$$N$ three-body force.

In the experiment we employ the $^7$Li($K^-,\pi^-$)$^7_\Lambda$Li$^*(3/2^+)$, 
$^7_\Lambda$Li$^*(3/2^+) \to ^3_\Lambda$H$^*$ +$^4$He reaction at $p_K \sim 1$ GeV/$c$.
Since the $^7_\Lambda$Li$^*(3/2^+)$ state is so-called a substitutional state having
a huge production cross section,
observation of $\gamma$-rays from these $^3_\Lambda$H excited states
may be possible even if these states are particle unbound 
and the $\gamma$-ray yields are quite small, thanks to
the intense kaon beam at the K1.1 beam line and the large acceptance ($\sim$ 100 msr)
of the SKS spectrometer

\noindent{\bf (2) Further studies of CSB effects using $p$-shell 
$\Lambda$ hypernuclei} \cite{
%YAM19
Takahashi:2019xcq} 

Even after our confirmation of a large CSB effect in $A=4$ hypernuclei \cite{Yamamoto:2015avw},
the origin of this effect has not been understood yet. 
It is expected that the CSB effect for heavier ($p$-shell) $\Lambda$ hypernuclei 
will provide a clue to understand its mechanism.
By using the $(\pi^-,K^0$) reaction, which will be used in the proposed $\Lambda$$p$ scattering
experiment, we can produce mirror $\Lambda$ hypernuclei with respect to the hypernuclei already 
well studied via $(K^-,\pi^-)$ and/or $(\pi^+,K^+)$ reactions with $\gamma$-ray spectroscopy.
We will study $^7_\Lambda$He, $^{12}_\Lambda$B, $^{16}_\Lambda$N, etc.,
and compare them with  
$^7_\Lambda$Li ($T=1$), $^{12}_\Lambda$C  and $^{16}_\Lambda$O data.
Because of a lower detector efficiency for $(\pi^-,K^0$) reaction, we need to occupy
an intense pion beam line for several months. 
The K1.1 line + SKS spectrometer is ideal apparatus for these experiments.\\

\noindent{\bf (3) Study of density dependence of $\Lambda$ binding energies 
from $p_\Lambda \to s_\Lambda$ transitions}

In the precise $(\pi^+,K^+)$ spectroscopy at HIHR, we aim at
measuring density dependent effects
of the $B_\Lambda$ values for the $s_\Lambda$, $p_\Lambda$, $d_\Lambda$, $f_\Lambda$ orbits
in a wide variety of $\Lambda$ hypernuclei.
In the $\gamma$-ray spectroscopy, 
we can obtain extremely precise (a few keV accuracy) data of 
the $p_\Lambda$-$s_\Lambda$ energy spacings via $E1$($p_\Lambda \to s_\Lambda$)
transitions. In heavy hypernuclei such as 
$^{139}_\Lambda$La and $^{208}_\Lambda$Pb, 
both $p_\Lambda$ and $s_\Lambda$ wave functions are confined 
in the center of the nucleus with a density of $\sim \rho_0$, while in lighter hypernuclei such as
$^{28}_\Lambda$Si and $^{40}_\Lambda$Ca 
the $p_\Lambda$ orbit partly overlaps with less dense part near the nuclear surface.
Thus, $A$-dependence of the $p_\Lambda$-$s_\Lambda$ spacing is expected to provide
information on the density dependence of the $\Lambda$$N$ interaction that can be described by the $\Lambda$$N$$N$ 3BF. The measured $\gamma$-ray spectra will be also used to 
determine the low-lying level schemes of the hypernuclei, 
which is necessary to precisely extract the 
single-particle $\Lambda$ binding energies from the high precision ($\pi^+,K^+$) spectra taken at HIHR.
\\

\noindent{\bf (4) Study of baryon modification in nuclear matter via $\Lambda$'s $B(M1)$}

In the approved experiment E63 \cite{TAM12,TAM_LOI}, 
we plan to measure the $\Lambda$-spin-flip
M1 transition probability $B(M1)$ in $^7_\Lambda$Li,
in order to investigate possible change of the magnetic moment of a $\Lambda$ in nuclear medium
due to a possible modification of baryon structure as well as baryon mixing effects.
The experiment employs the $(K^-,\pi^-)$ reaction at the K1.1 beam line.
If a deviation of the $\Lambda$'s $g$-factor from the free space value 
is observed, further studies are necessary with different hypernuclei to clarify its origin
by investigating nuclear density dependence and isospin dependence of the deviation.
Since a measurement for each hypernucleus requires a beam time more than one month,
the independent K1.1 beam line is necessary.\\

\subsubsection{Weak decays of $\Lambda$ hypernuclei}

Weak decays of $\Lambda$ hypernuclei were mainly
studied at KEK-PS and DA$\Phi$NE.
Experimental data on the lifetime and the $n$/$p$-induced rates of the
nonmesonic weak decay (NMWD) ($\Lambda$$n$, $\Lambda$$p$ $\to$ $n$$n$, $n$$p$) 
confirmed a short-range nature of NMWD described by heavy meson exchanges \cite{GAL16}.
However, the three-body non-mesonic decay, $\Lambda$$N$$N$$\to$ $N$$N$$N$, 
is not clearly understood, and the stage-2 approved experiment (E18) will be carried 
out when the K1.1 line is available.
In addition, the following experiments are planned at K1.1.\\

\noindent{\bf (1) Mesonic and nonmesonic weak decays via  $(\pi^-,K^0)$ 
and $(\pi^+,K^+)$ reactions}

The $(\pi^-,K^0$) reaction method will allow us to extend hypernuclear weak decay studies
to the neutron-rich side \cite{FEL19a,
%FEL19b
Takahashi:2019xcq}, as is also planned for $\gamma$-ray spectroscopy.
Using this reaction, a test of the $\Delta I=1/2$ rule in NMWD,
which has been considered to be quite important but
not successful due to experimental difficulties, 
will become feasible.
The $\Delta I=1/2$ rule is a long-standing problem 
to be solved from low-energy QCD dynamics.
Some theories suggest that this rule does not hold in NMWD.
To test it we need to measure the $p$-induced NMWD rate ($\Gamma_p$) in $^4_\Lambda$H as well as 
 the $n$-induced NMWD rate ($\Gamma_n$) in $^4_\Lambda$He 
precisely.
By using intense pion beams at K1.1 and the $(\pi^-,K^0)$ reaction technique, 
the $^4_\Lambda$H measurement will be feasible 
in a rather short beam time.
This method is also useful in precise determination of the $^3_\Lambda$H lifetime
of which experimental data are in a puzzling situation at present \cite{ALI21}.

At the K1.1 beam line, weak decay studies 
of various hypernuclei ($^3_\Lambda$H, $^4_\Lambda$H, and 
neutron-rich $p$-shell hypernuclei) via $(\pi^-,K^0)$ reaction
are proposed \cite{FEL19a,
%FEL19b
Takahashi:2019xcq}.
The pion spectra in the mesonic weak decay will be used
for spin-parity assignment of hypernuclear ground states,
as was confirmed by the FINUDA experiments at DA$\Phi$NE \cite{FEL15}.
In addition, coincidence measurement of weak decay particles
and $\gamma$ rays will provide a new method to study the spin-flip $\Lambda$'s B(M1)
values of heavy hypernuclei \cite{TAM01}.\\

\noindent{\bf (2) Beta decay of $\Lambda$ hypernuclei} \cite{TAM_LOI}

Weak decay of $\Lambda$ in a nucleus is a unique probe to investigate
possible modification of baryon structure in nuclear medium.
To avoid strong interaction effects in the final state, 
measurement of the $\Lambda$'s beta decay, $\Lambda \to p e^- \bar{\nu}$, is desirable.
Because the $u$ quark distribution of a $\Lambda$ hyperon 
is expected to be more spread
than the $s$ quark distribution in nuclear matter, the beta-decay rate is expected to be 
significantly reduced due to a smaller overlap between $s$ and $u$ quarks 
in nuclear matter than in the free space.
An experiment is planned at K1.1 to measure
the lifetime and the branching ratio of $^5_\Lambda$He hypernucleus
and to obtain the beta-decay rate precisely \cite{TAM_LOI}.
The hypernuclei are produced by $^6$Li$(\pi^+,K^+)$ reaction with the SKS spectrometer,
and beta-decay electrons are detected by a 4$\pi$ calorimeter around the target. 
Since the beta-decay branching ratio of a $\Lambda$ is small ($8.3 \times 10^{-4}$),
reduction of huge background from the main weak decay modes (mesonic and nonmesonic decays)
is a key of the measurement. A recent simulation study shows that the background can be
sufficiently reduced,
and the beta-decay rate can be measured in a statistical accuracy of $\sim$4\%
in two months of beam time.

% flatex input end: [K11-others.tex]

%%%%%% K1.1 other experiments (Tamura)
%\input{K11-others_rev}

\printbibliography[segment=\therefsegment,heading=subbibliography]

\clearpage

%==================================================================%
%\section{\centering Physics objectives at the K10 beam line}
\section{\centering Physics Objectives at \boldmath$\pi20$ and K10 Beam Lines}
\chapterauthor{
K.~Aoki, Y.~Hidaka, A.~Hosaka, N.~Ishii, T.~Ishikawa, Y.~Komatsu,\\
Y.~Morino, M.~Naruki, H.~Nemura, H.~Noumi, H.~Ohnishi, K.~Ozawa,\\
F.~Sakuma, T.~Sekihara, S.~I.~Shim, K.~Shirotori, H.~Takahashi,\\
S.~Takeuchi, and M.~Takizawa\\
}
%==================================================================%

% flatex input: [./k10docu/k10main.tex]
%\newpage
%\section{Physics objectives at the K10 beam line}
%\textcolor{red}{Whether "K10" is the name of the beamline? or the project name? It is mixed at this moment. I think that the "Physics program at K10 beamline." or "physics goals at K10 beamline" will be appropriate for the section title. (Hiroaki)}\\
%%overview
\newcommand{\cred}{\color{black}}%red}}
\newcommand{\cblue}{\color{black}}%blue}}

% flatex input: [./k10docu/k10PhysBG.tex]

Hadrons are composite particles of quarks interacting with gluons. 
Hence the goal of hadron physics is to answer the question 
``how quarks build hadrons".  
By this we should be able to explain various properties of observed hadrons, 
to predict new phenomena, and to tell how best to look for. 
The resulting knowledge can be used as the basis 
of various extended studies where hadrons appear such as dense 
hadronic matter.  
Furthermore, QCD is the only gauge theory that shows non-perturbative nature of dynamics and can be accessible by experiments. Thus the exploring QCD should be helpful to %the understanding of 
understand what happens in various other physics systems, which can be described by the quantum field theories with the strong couplings.
%{\color{red} (Takizawa san: Furthermore, the QCD is the only gauge theory that shows non-perturbative 
%nature of dynamics and can be accessible by experiments.  
%Thus the exploring QCD should be helpful to the understanding 
%of what happens in various other physics systems.)}

Spectroscopy of hadrons (in the following this is referred to as ``hadron spectroscopy")
has been carried out %studied 
for many years, 
mostly driven by experimental discoveries of new states~\cite{Zyla:2020zbs}.
Recent lattice QCD calculations have been making crucial contributions to establish 
that QCD is indeed the theory of the strong interaction for hadrons, 
and opened a way even to nuclear systems from QCD.
Nevertheless, there are still unsolved or unsettled issues.  
They are important not only for their own sake but also due to their wide 
application to related fields.  
For instance, the question of the high density hadronic matter
has been considered to be one of the most challenging problems in 
the strong interaction physics, QCD. 

\begin{figure}[h]
\begin{center}
\includegraphics[width=1 \linewidth]{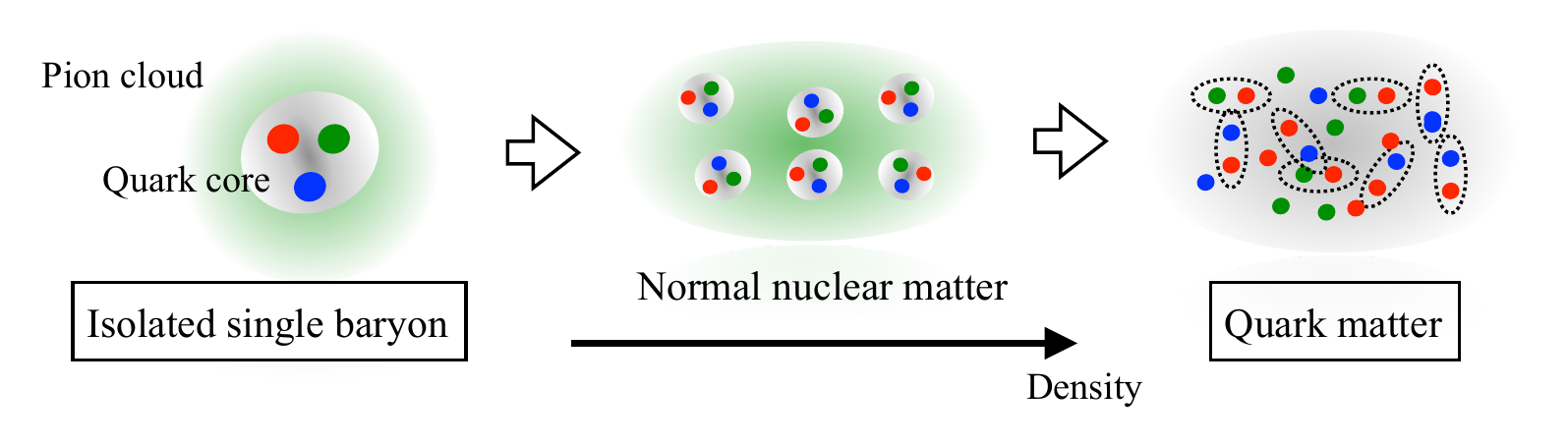}
\end{center}
\vspace{-5mm}
\caption{Schematic picture for the isolated nucleon, nuclear matter, 
and quark matter.  
}
\label{fig_NtoQmatter}
\end{figure}

{\cred %For example, a
At five times or more than the normal nuclear matter densities, the nuclear matter 
is expected to turn into the quark matter where exotic structure such as color superconductivity 
could be formed. 
Such a change in the matter structure should affect the properties of the neutron stars that have 
the high density region in the core part.  
As illustrated in Fig.~\ref{fig_NtoQmatter}, at such densities the core parts of the nucleon 
start to overlap, where the dynamics of the quarks and their correlations (of primary importance is the diquarks) becomes relevant.  
We then expect that such properties could be learned by baryon spectroscopy.  
Although the information that can be obtained by the spectroscopy (at zero density) 
is not exactly relevant, it provides the information 
at the initial point that extrapolates to finite density regions.  
%In K10 project, 
In the project at the $\pi20$ and K10 beam lines,
we will perform %study 
high statistics baryon spectroscopy 
with keeping these scopes.  }

%-------------------------------------
%-------------------------------------

%--------------------------------------
%\subsection{General Introduction} 
%\subsection{From quark to baryon} 
\subsection{What We Know About Baryons} 
%--------------------------------------

Let us start with looking at a simple example of atoms 
that are built by electrons (and a nucleus).
An obvious fact is that the mass of the electron 
is much larger than the energies that are relevant 
to various atomic phenomena, the binding energies
and excitation energies of electrons.  
Thus the creation of electron and anti-electron (positron) 
pair is suppressed, 
and therefore, the electrons inside an atom are essentially the same 
as isolated electrons, enabling the electrons 
as good building blocks, 
an ``ordinary path" of how electrons build atoms.  
A similar situation occurs in nuclear systems.  

\begin{figure}[h]
\begin{center}
\includegraphics[width=1 \linewidth]{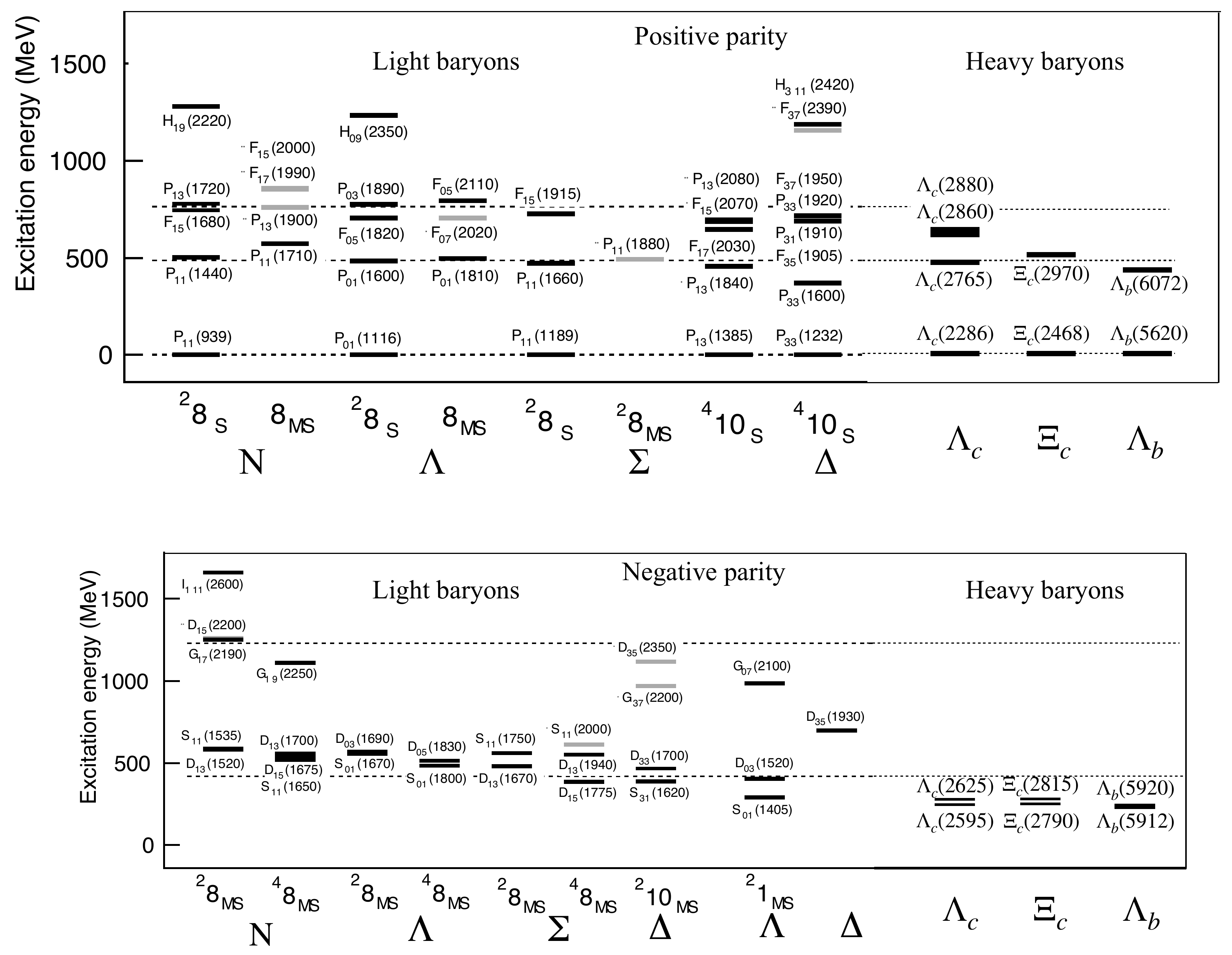}
\end{center}
\vspace{-5mm}
\caption{Excitation energies of various baryon states.
Plots for $u, d, s$ flavors are from Ref.~\cite{Takayama:1999kc}, 
with  heavy baryon data are added.
}
\label{fig_systematics}
\end{figure}

The situation changes drastically for hadrons.  
Indeed 
the current $u, d$ quarks in the QCD Lagrangian are 
far lighter than the proton~\cite{Zyla:2020zbs}. 
This makes it very difficult to draw an ordinary path 
with the current quarks used as good building blocks.   
If we want to create the mass of the proton about 940 MeV 
from the current $u,d$-quark masses
of order a few MeV, 
we need additional steps of many $q \bar q$ creations. 
Obviously such a process leads to enormous complications.  
If so, is the resulting hadron spectrum % so: removed (T. Ishikawa)
chaotic? 
In fact, the data seem very systematic % seems -> seem (T. Ishikawa)
in the region of excitation energies up to around 1 GeV, 
as shown in Fig.~\ref{fig_systematics}.
This implies the emergence of effective degrees of freedom 
that build the observed hadrons in a rather simple manner.  

%-------------------------------------
%-------------------------------------

This is the place where the non-perturbative dynamics of QCD comes in.  
It is known that the QCD vacuum $|0\rangle$ is populated by the non-zero 
expectation value of the quark condensate for the light $q = u, d, s$ quarks, 
$\langle 0 | \bar q q | 0 \rangle \neq 0$.  
Consequently chiral symmetry is spontaneously broken and the 
light quarks acquire a finite mass of order 
%several hundreds MeV, (TI)
of several hundred MeV,
%the emergence of constituent quarks (TI)
constituent quarks emerge as quasi-particles.
The constituent quarks may interact effectively by exchanging gluons, 
which is often treated one gluon exchange. 
Furthermore, light pions are generated as the Nambu-Goldstone bosons
formed by $u, d$ quarks and their antiquarks.  
Kaons of $s, u$ (or $d$) quark content are also regarded as the Nambu-Goldstone bosons
whose masses are however significantly heavier than the masses of the pions
due to heavier mass of the $s$ quark~\footnote{
The current quark masses are $m_u \sim 2$ MeV, $m_d \sim 5$ MeV and $m_s \sim 100$ MeV.
Consequently, the mass of the pion is $\sim$ 140 MeV while the mass of the kaon $\sim$ 
500 MeV.~\cite{Zyla:2020zbs}}.  
As we will discuss in Section~\ref{sec:physpi20k10}, 
%later {\color{green}\fbox{?} subsections},
the difference in the $u,d$ and $s$ quarks 
and hence in the pions and kaons affects significantly the properties of baryons of different flavor contents.  

One of possible mechanisms of causing the non-trivial vacuum 
is the instanton scenario~\cite{Diakonov:2002fq}.  
The instanton is a topological object of gluons as a classical solution of 
the SU(3) color gauge theory.  
The presence of such an object leads to the zero modes of quarks 
that leads to the finite quark condensate as explained by the Banks-Casher 
relation~\cite{Banks:1979yr}.  
The instanton also induces the $U_A(1)$ breaking interaction 
for quarks~\cite{tHooft:1976rip,tHooft:1976snw}.
Such an interaction explains well the 
$\eta$-$\eta^\prime$ mass difference~\cite{Kobayashi:1970ji}, 
one of the phenomenological consequences of the $U_A(1)$ anomaly.  
The interaction may be also effective for quarks in baryons.  
However, the role of such an interaction for baryons is not fully explored.
We will discuss this issue in Section~\ref{sec:charmedbaryons}.

In view of the above discussions,  the constituent quarks and pions can be 
building blocks of baryons.  
The quarks are confined in 
the ``quark core" region with its radius smaller than $\sim$ 0.5 fm, 
and the pion forms the ``meson cloud" around it for baryons containing $u,d$ quarks
as illustrated in the left of Fig.~\ref{fig_NtoQmatter}.  
The pion cloud may be suppressed when $s$ or heavy $c, b$ quarks replace the $u,d$ quarks inside the quark core.  
The constituent quarks may interact each other by the effective interactions
due to the non-perturbative effect of QCD.
{%\color{green}\fbox{?}
They may be modeled by a form of one gluon exchange (OGE), instanton induced (III), %one (III) 
and even meson exchange type interactions.} 
These interactions generate yet an important quasi-particle degrees of freedom
through their spin-dependent components, that is the diquark.  
%The diquarks are utilized in many discussions of hadron physics, but their detailed nature is not yet well established.  
An attractive correlation is expected for flavor-anti-symmetric quark pairs of $ud$, $us$, $ds$, etc.  
Due to its spin dependent nature the correlation is  strong for the above light quark
pairs, and weakened 
for heavier quark pairs.  
Though the diquarks are utilized in many discussions of hadron physics, their precise correlation nature is not yet established, which however will be an important 
inputs for the physics of the hadronic matter.   %{\color{green}\fbox{?}The last sentence seems to give an equivalent message repeatedly to the sentence that starts with "The diquarks are ...".}

Discussions on the spin dependent interactions are given in Section~\ref{sec:spin-dep-int}.
These interactions are primarily two-body, and 
%plays (TI)
play the main role 
when discussing the baryon spectroscopy. %at K10}.  
It is noted, however, that three-body interactions are also possible 
which is a further interesting subject of spectroscopy in this project. %at K10. and finite density problems.  

\subsection{What We Will Explore at \boldmath$\pi20$ and K10\label{sec:physpi20k10}} 
%--------------------------------------

Based on the above general descriptions, 
there are several important aspects in hadron physics.  
First,  different favor combinations of  quarks with different masses would cause 
non-trivial flavor dependent dynamics, although 
the fundamental interaction of quarks and gluons are flavor blind.  
The quarks are classified into light $u, d, s$ quarks and heavy 
$c, b$ quarks as their masses are compared with the QCD scale parameter $\Lambda_{\rm QCD}$.  
The position of the $s$ quark is a bit subtle, as it is not as light as the $u,d$ quarks and neither 
as heavy as the $c$ quarks, 
which causes interesting observations in hadrons containing 
%strange (TI)
the $s$ quarks.  

Second,  large spin-dependent forces exist.  
This contrasts with, for example, atomic systems dictated by electromagnetic interaction.  
As anticipated in the previous section, 
the sufficiently attractive spin-spin forces cause 
strong correlation in pairs of quarks, diquarks, in its spin and flavor anti-symmetric channels
in orbitally ground state that plays a crucial role not only 
in spectroscopy but also in high density hadronic matter.  
In the sufficiently high density region, the color super-conducting phase 
is expected to be formed that governs various matter properties~\cite{Fukushima:2010bq}.  
For spectroscopy, we consider also internal structure of the diquarks with orbital excitation.  
%which corresponds to the $\rho$ mode excitations as we will discuss later.  
Different flavor combinations, in particular those of $u, d, s$ quarks 
may results in different properties.  
Strongly correlated diquarks are also considered to be important 
for exotic hadrons of multiquark contents.  
Hence the detailed knowledge of the spin dependent interaction for $u, d, s$ quarks 
is a very important issue that can be studied at $\pi 20$ and K10.  

Spin-orbit ($LS$) force is another issue in hadron spectroscopy.  
In the light flavor sector the $LS$ force is suppressed 
as indicated by the mass difference of the  nucleon excited states 
$N(1535)$ of spin-parity $J^P = 1/2^-$ and $N(1520)$ of $3/2^-$.  
In contrast, for the heavy baryon sector, as the pair of $\Lambda_c(2595)1/2^-$-$\Lambda_c(2625)3/2^-$ and 
that of $\Lambda_b(5912)1/2^-$-$\Lambda_b(5920)3/2^-$ implies the importance of the $LS$ force. 
The amounts of the splittings that decrease as the mass increases, as well as their ordering 
for $1/2^-$ and $3/2^-$ states 
seem to be consistent with what we expect from the OGE and III.  
Because the spin-dependent force is one of the central issue of this project, we discuss it 
in some detail in Section~\ref{sec:spin-dep-int}
%Appendix zzz.  

\begin{figure}[h]
\begin{center}
\includegraphics[width=0.9 \linewidth]{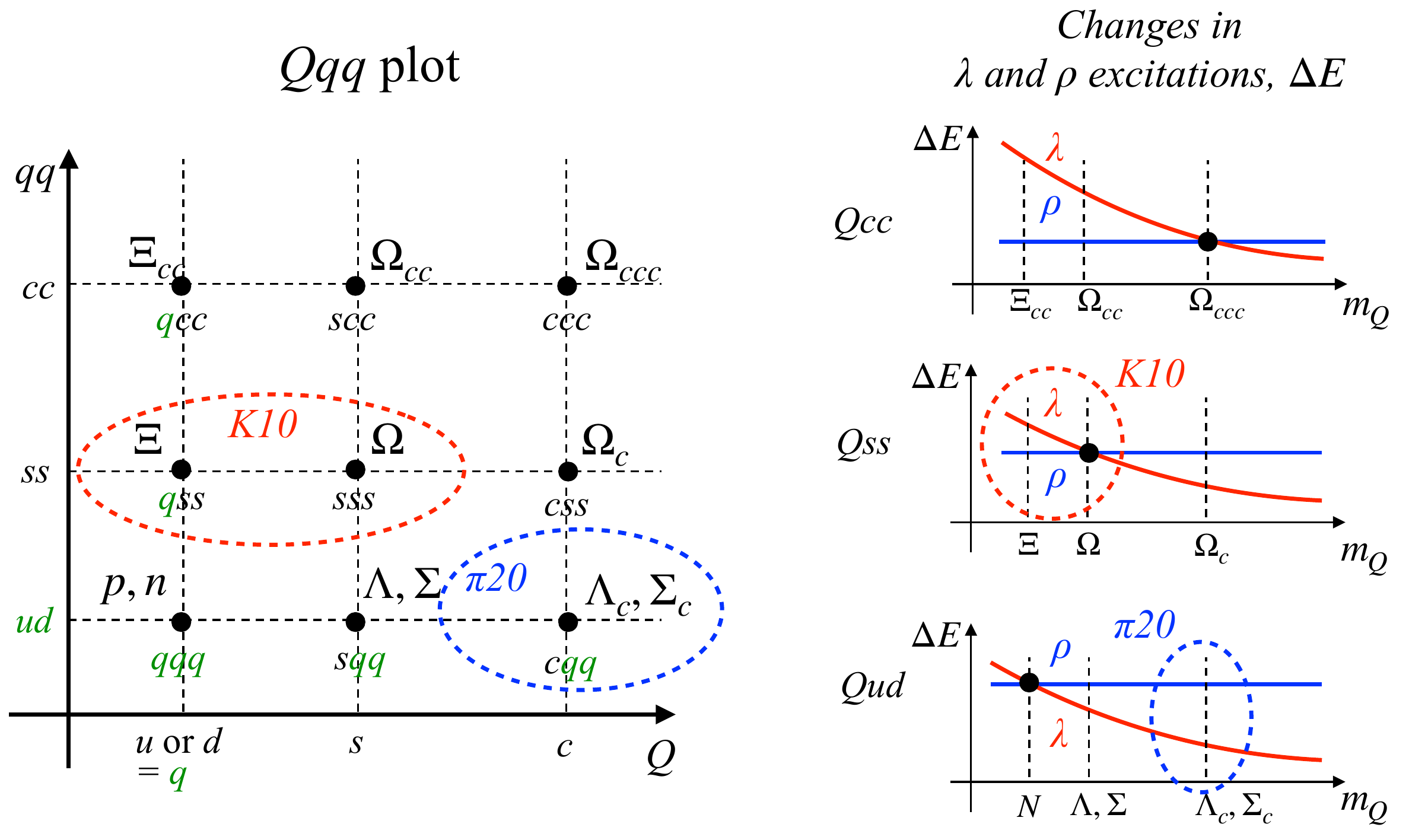}
\end{center}
\vspace{-5mm}
\caption{Left: various $Qqq$ baryons indicating what we can study at $\pi 20$ and K10 at J-PARC.  
Right: Character changes of the $\lambda$ and $\rho$ modes for $Qcc$, $Qss$ and $Qud$ baryons
as functions of the mass of the $Q$-quark, $m_Q$.  }
\label{fig_baryonspecies}
\end{figure}

Physics studies at $\pi 20$ and K10 can cover much part of the above issues.  
The main object is the baryon spectroscopy.  
Hence in Fig.~\ref{fig_baryonspecies} we have shown baryons which we can study 
at $\pi 20$ and K10 at J-PARC ($Qqq$ plot on the left panel)~\footnote{
In fact, there are more baryons whose flavor contents are all different such as $usc$. }.  
In the figure %strange 
the $s$ quark is shown to be  different from the $u,d$ quarks, although 
it is classified as a light quark. 
As shown in there, 
%in 
at K10 we can study $\Xi$ and $\Omega$ baryons, while 
at $\pi 20$ (singly) charmed baryons are studied.  

%-------------------------------------
%-------------------------------------

On the right panel of Fig.~\ref{fig_baryonspecies} typical behavior of the two internal modes, 
$\lambda$ and $\rho$ modes, of baryons are shown for the three cases, $Qcc$, $Qss$ and $Qud$.    
Here the $\lambda$ mode is the relative motion between $Q$ and the other pair of quarks, and
the $\rho$ mode is the internal excitation of the pair of quarks.  The behaviors of the  $\lambda$ and $\rho$ modes are discussed  furthermore in the cases of $Qqq$ $(Qud)$ and $Qss$ in Sections~\ref{sec:baryonshf} and \ref{sec:baryonsqss}. 
They show up in orbital excitations which generate large structure of energy levels.  
Then fine structure due to the spin-dependent forces may follow.  
The different combinations of quark flavors results in qualitatively different 
behaviors of the excitation energies $\Delta E$ of the $\lambda$ and $\rho$ modes.  
As discussed in some detail in the next subsections, for charmed baryons of such as $cud$
the $\lambda$ mode is lower than the $\rho$ mode, while for $\Xi$ baryons of such as $uss$
the ordering changes.   
Such separation of the $\lambda$ and $\rho$ modes puts constraints in spin configurations of the diquarks, 
which can be efficiently used to extract the information of spin-dependent forces.  
%For charmed baryons,  the $\lambda$ mode is the relative motion of the $ud$ diquark 
%and the charm quark, 
%the level structure reflects the properties of the $ud$ diquark which is useful to extract
%the information of the spin-dependent forces for the diquark.  
In fact, this is the one of the main purposes of the $\pi 20$ project.  
In K10 from the $\Xi$-spectroscopy we expect also to extract the properties 
of the diquarks containing a strange quark, $su$ and $ds$ diquarks.
%In this case, the role of the $\lambda$ and $\rho$ modes are interchanged, but still we can use 
%this mode separation to extract the information of the spin-dependent forces for
%the $su$ and $ds$ diquarks.  

$\Omega$ baryons are unique in that all quarks are the same, $sss$.
The symmetry in flavor structure can impose constraints
in spin and orbital motion that suppresses some spin-dependent forces.  
Furthermore, the $sss$ structure suppreses the pion cloud unlike 
the protons and neutrons~\footnote{
An intuitive picture, for instance, of the neutron which contains a virtual component of
the proton and negatively charged pion can explain consistently the negative charge radius 
of the neutron~\cite{Thomas:2007bc}.}.
Such a simplification of the $\Omega$ with suppression of spin-dependent forces and pion cloud
may enable us to access directly to the dynamics of the quark core region of baryons.  

To summarize we will study charmed baryons, $\Lambda_c$ and $\Sigma_c$,  
at $\pi 20$ and $\Xi$ and $\Omega$ baryons at K10.
Using the QCD effective theories and lattice simulations we 
will establish a framework that explains 
observed spectrum consistently from the light to heavy flavored hadrons.
The results will be useful not only for further studies of hadron spectroscopy 
but also for other related fields, in particular for physics of high density hadronic matter.
In this way the $\pi 20$  and K10 will contribute to  the grand challenge of  
the J-PARC extension project. 

\subsubsection{Spectroscopy of charmed baryons}\label{sec:charmedbaryons}

Since the charm ($c$) quark is much heavier than $u$ and $d$ quarks, 
in baryons with one $c$ quark, internal motions of $ud$ quarks can be distinguished from that of the $c$ quark. 
One can find that the orbital excitation energy of a $ud$ pair to $c$  ($\lambda$ mode) 
is lower than that between $u$ and $d$ ($\rho$ mode), 
as shown in the bottom of the right panel of Fig.~\ref{fig_baryonspecies}.
This is a kinematical effect under the confinement potential known as the isotope shift. 
In addition, the spin-spin interaction with the heavy charm quark $c$ is suppressed,
as the color magnetic interaction between two quarks is proportional 
to the inverse of the relevant quark masses. 
The spin-singlet $ud$ state is of particular interest since its interaction is attractive. 
Hence, the $ud$ diquark correlation is developed. 
Therefore, baryons with a charm quark provide a good opportunity to study the $ud$ diquark correlation.
We also expect to decompose the origin of the quark interactions by 
analyzing the high statistics data in terms of QCD based effective theories.

Concerning the $U_A(1)$ anomaly, an interesting scenario was proposed in Ref.~\cite{Kim:2020imk}
where the $\lambda$-$\rho$ inversion was pointed out 
for the negative parity $\Xi_c$ baryons
($\Xi_c(\rho)$ is lighter than $\Xi_c(\lambda)$).  
For this inversion mechanism, the $0^- (^3P_0)$ $us$ (or $ds$) diquark plays an essential role.
The corresponding $ud$ $0^-$ diquark forms a $\rho$-mode $\Lambda_c$ baryon of 
$J^P = 1/2^-$.  
This state has a unique feature that the decay into $\pi \Sigma_c$ is forbidden
due to the selection rule associated with the conservation of spin and parity~\cite{Nagahiro:2016nsx}.  
If so, there is a good chance to observe such a $\Lambda_c$ state because of 
its narrow width~\footnote{
Mixing with different states might cause a narrow decay width.}.  

%This together with the dynamical origin of the hadron masses associated 
%with the spontaneous breaking of chiral symmetry and $U_A(1)$ anomaly 
%will be useful for the study of finite density hadronic matter.  

%It is widely believed nowadays that the constituent quarks emerge 
%as effective degrees of freedom to form hadrons as a result of spontaneous symmtery breaking of chiral symmetery. 
%The masses of the constituent quarks, and hence hadrons, are dynamically generated. Properties of hadrons would change as they are put in a didffent evironment in temperature and/or density. 
%Systematic behaviers of baryon excited states over different flavors have to be investigated in order to extract dynamical nature of the qurak-quark interaction that should be related to the dynamics of chiral condensateion and/or $U_A(1)$ anomaly in QCD. 

%-------------------------------------
\begin{figure}[h]
\begin{center}
\includegraphics[width=0.8 \linewidth]{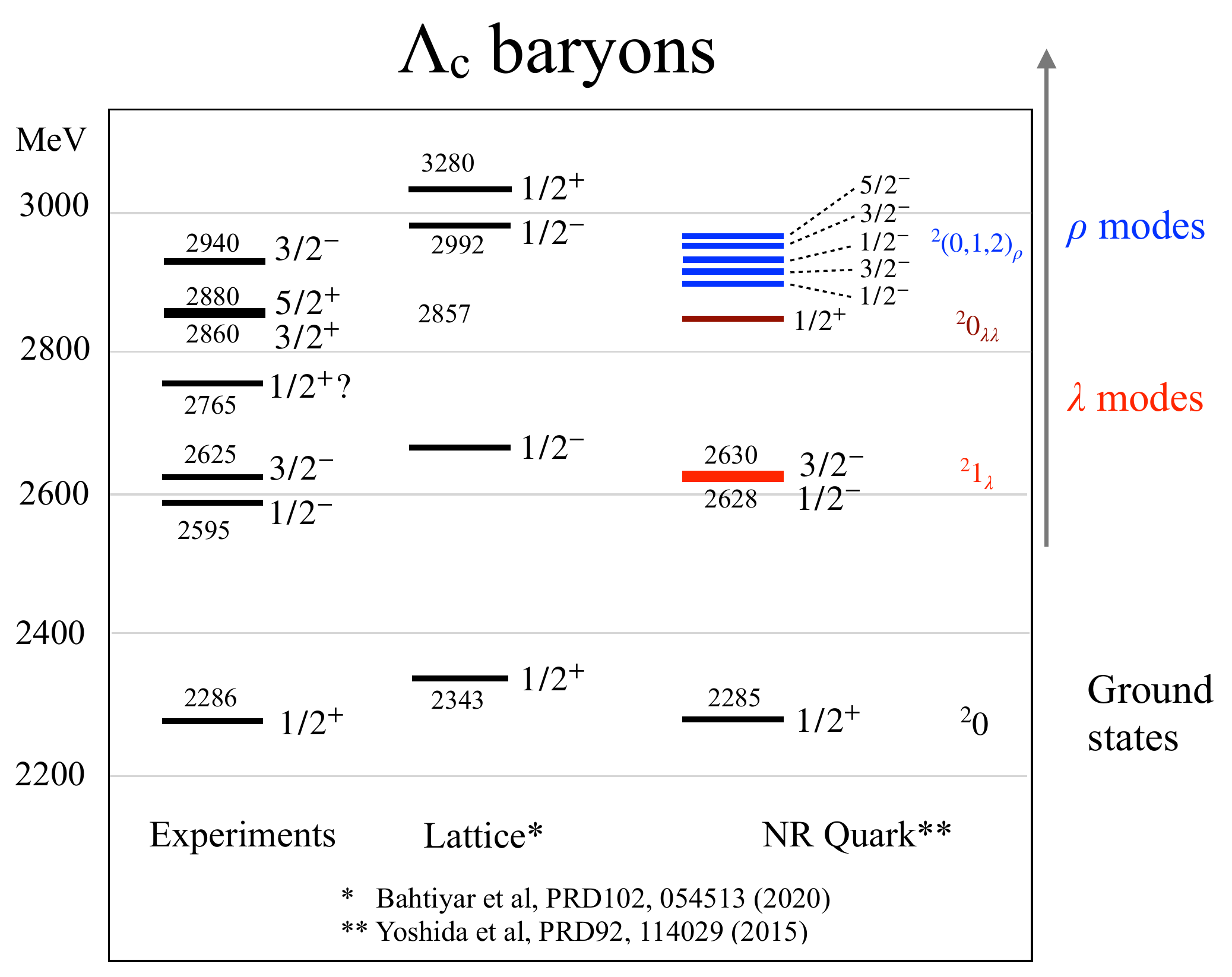}
\end{center}
\vspace{-5mm}
\caption{Comparison of $\Lambda_c$ baryon masses from experiments~\cite{Zyla:2020zbs}, 
lattice simulations~\cite{Can:2015exa} and non-relativistic quark model~\cite{Yoshida:2015tia}.  
Orange shaded is the region of orbital excitations by the quark model.}
\label{fig_Lcbaryons}
\end{figure}
%-------------------------------------

In Fig.~\ref{fig_Lcbaryons} masses of $\Lambda_c$ charmed baryons are shown and compared among
those of  experiments, 
lattice simulations~\cite{Can:2015exa} and non-relativistic quark model~\cite{Yoshida:2015tia}. 
%The large structure of the lowering the $\lambda$ mode is expected from the results of the quark model calculations.  
The quark model calculations show the lower excited states corresponding to the $\lambda$ modes
expected by the kinematical effect.
Then, large splitting between the $\lambda$ and $\rho$ modes is expected to appear,
which makes a major structure in the mass spectrum.
However, identification of higher lying states of various spins and parities is not clear, 
except for the first negative parity pair of $1/2^-, 3/2^-$, though the quark model 
does not reproduce sufficient $LS$ splitting (see the discussion below).  
To fill these gaps is the primary goal of the study of charmed baryon spectroscopy.  

\begin{figure}[ht]
\begin{center}
\includegraphics[width=0.8 \linewidth]{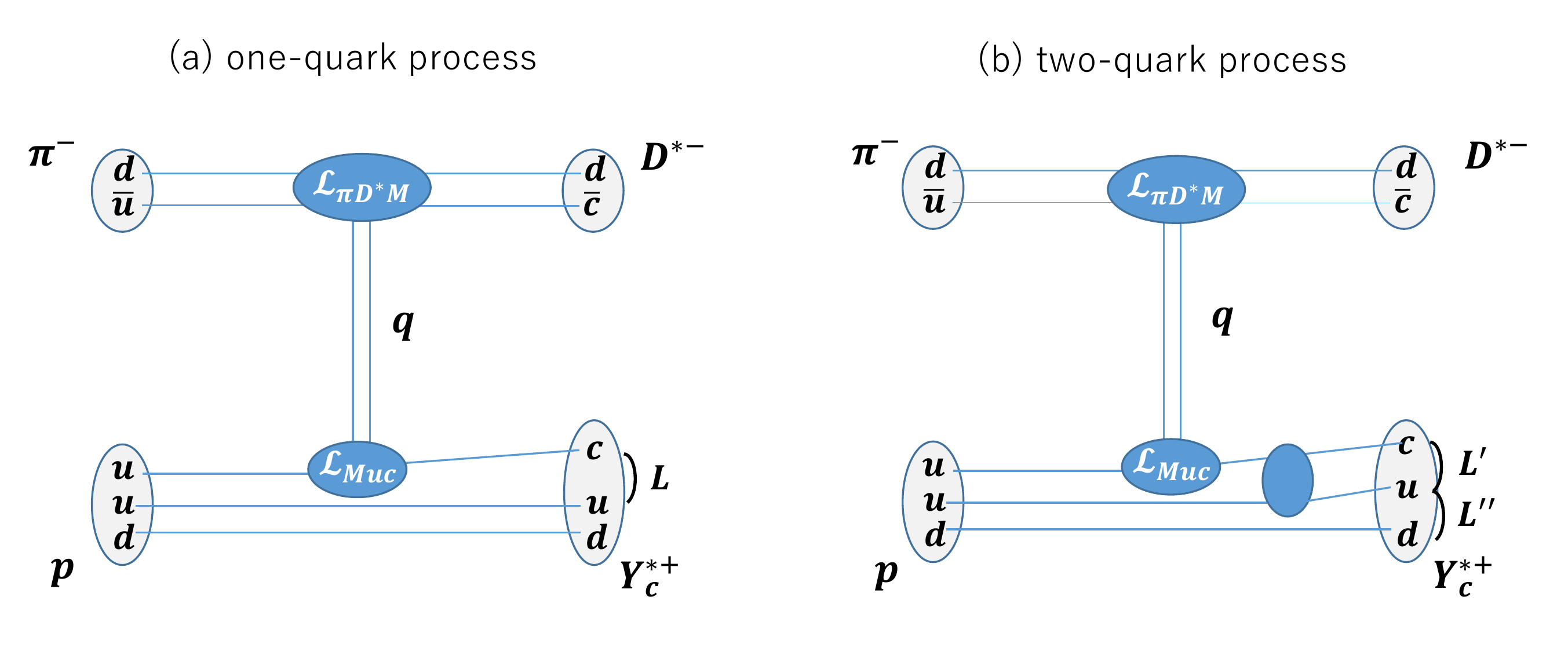}
\end{center}
\vspace{-5mm}
\caption{Reaction diagrams of the $\pi^-p\rightarrow D^{*-}Y_c^{*+}$ reaction. 
(a) one-quark process: a meson with a (3-)momentum transfer of $\boldsymbol{q}$ is exchanged in the $t$ channel. 
The meson is coupled to $u$ in the proton and $c$ in the charmed baryon and 
introduces an angular momentum $L$ between $ud$ and $c$. 
(b) two-quark process: after the meson exchange, introduced $L$ is shared among 3 quarks in the final state. 
}
\label{fig:reaction-diagram}
\end{figure}

%At High-p facility, 
We have proposed charmed baryon spectroscopy at the J-PARC high-momentum beam line ($\pi$20), as approved as E50~\cite{E50exp}.
To produce charmed baryons, 
missing-mass technique is employed via the $\pi^-p\rightarrow D^{*-}Y_c^{*+}$ reaction, where $Y_c^{*+}$ represent excited charmed baryons with a $c$ quark. A series of the $Y_c^{*+}$ states will be observed in the missing mass spectrum, as shown in Fig.~\ref{k10_charm_fig2} in Section~\ref{sec:charm-spectroscopy}. The production rates of the excited states are expected to depend on their internal configurations of spins and orbital angular momenta (parity). 
In the $\pi^-p\rightarrow D^{*-}Y_c^{*+}$ reaction, a $u$ quark in the proton is converted into a $c$ quark in the charmed baryon $Y_c^{*+}$, 
that we call one quark process here, is expected to 
takes place dominantly~\cite{shim2}. In the one quark process, 
the angular momentum will be introduced between $c$ and $ud$, 
as shown in Fig.~\ref{fig:reaction-diagram} (a).
Thus, the $\lambda$-mode excited states are favored to be populated 
in the one quark process.
The reaction cross section is estimated by calculating the so-called sticking probability of $c$ to $ud$, written as\\
$$\left|\langle\varphi_f(\boldsymbol{r})|2\sigma_-\exp{(-i\boldsymbol{qr})}|\varphi_i(\boldsymbol{r})\rangle\right|^2
\propto\left(q/\alpha\right)^{2L}\exp{(-q^2/\alpha^2)},$$\\
where $\varphi_{i(f)}$, $L$, $q$, and $\alpha$ represent an initial $u$ (a final $c$) wave function, 
the angular momentum introduced in the charmed baryons, the momentum transfer 
of the reaction, and a typical size parameter of baryon ($\sim$0.4 GeV/$c$), respectively.
Here, $\sigma_-$ is a spin operator required in the case of a vector meson exchange in the reaction. 
In the right-hand side of the above equation, the wave functions of the initial and final states for harmonic oscillator potential are employed for analytic calculations.
Since the reaction involves a large momentum transfer of $q\sim$1.4 GeV/$c$, 
the sticking probability, thus the cross section, becomes very small due to 
the exponential factor.
On the other hand, taking the ratio of the cross section of the excited state with $L$ to that of the ground state, the exponential factor is almost cancelled, and the ratio do not go down owing to the factor of $\left(q/\alpha\right)^{2L}$. 
The intense pion beam provided by the $\pi$20 beam line compensates the small cross sections.
In the end, relatively high possibility to populate the higher excited states with a finite $L$ is rather advantageous. 

We expect two-quark involved process, as shown in the right panel of 
Fig~\ref{fig:reaction-diagram} (b) \cite{shim2}.
In the two-quark process, the momentum transfer will be shared by $c$ and $u$ or $d$ in the final baryon. 
Then, both $\lambda$- and $\rho$-mode states are populated.
Though the magnitude of the two-quark process is of interest to understand 
the reaction mechanism, 
the production ratio of $\Lambda$ to $\Sigma$, which is about 2 to 1~\cite{Crennell:1972km}, 
may give a good reference for the ratio of the one-quark process to 
the two-quark one, which is to be about 5 to 1 in charm sector.
We expect to populate finite magnitude of the $\rho$-mode states as well.  

%-------------------------------------
%-------------------------------------

As an example, let us consider 
$\Lambda_c(2625)$ and $\Lambda_c(2595)$ states, which are well established as 3 stars in PDG rating. Their spin-parities are assigned as 3/2$^-$ and 1/2$^-$ as they are believed to be so-called $LS$ partners with having an orbital angular momentum of $L=1$ between $c$ and $ud$ and the $ud$ spin-parity of 0$^+$ with isospin 0 (flavor anti-symmetric state). The production rates of 
$\Lambda_c(2625)$ and $\Lambda_c(2595)$ should be $L+1:L=2:1$, respectively \cite{shim2}. By measuring the production rates, we could determine their spin-parities and settled that they are indeed the $LS$ partners of the $\lambda$ mode with having a spin-singlet $ud$ state. 
%{\cblue(relation of the ratio to the concreteness of the $\lambda/\rho$ mode separation, the form factors of the $\lambda$-mode states, what does the comparison between $L=0,1,2,...$ tell us?)}
The theoretical calculation suggests good separation of the $\lambda$ and $\rho$ modes in $\Lambda_c$ \cite{Yoshida:2015tia}.
We will be able to confirm it by measuring the production ratios of the spin doublets in the expected $\lambda$-mode excited states, which are  described by the relative motion of a $c$ quark to the $ud$ diquark. As we discussed above, the production cross section is described by the overlap integral of the initial ($p$) and final ($\Lambda_c$) wave functions at the momentum transfer ($q$) of the relevant reaction. Thus, the total cross sections of the $\lambda$-mode states with having different orbital angular momenta $L$'s would provide detailed information on the %collective 
motion of the $ud$ diquark.

The $\rho$-mode states have not yet been identified clearly. As we discussed above, the $\Lambda_c$ state of $J^P=1/2^-$ having $ud$ diquark of $0^-$ is a good target to observe. Its mass and width provide how the $U_A(1)$ anomaly works in baryons~\cite{Kawakami:2020sxd}. 

The decay branching ratios also carry information on internal motions of $ud$ quarks.
%in the other excited states. 
 In particular, the $\lambda$ mode excitation with the $ud$ diquark, is expected to favor a decay into a $D$ meson and a nucleon ($N$) compared with decays into a pion and a low-lying charmed baryon ($Y_c^{*\prime}$) if the decay channel is open. 
 We will measure their decay branching ratios (decay partial widths) by detecting decay particles in coincidence with $Y_c^{*+}$ produced by the $\pi^-p\rightarrow D^{*-}Y_c^{*-}$ reactions. 
%{\cblue(relation of the decay rate to ``solidness of the $us$-diquark")}

%{\cred
%As for a $\lambda$-mode $\Lambda_c^*$ state that is validated once in the production process, 
%we expect that the decay partial width of the $\Lambda_c^*\rightarrow D^0p$ process would tell us how the $ud$-diquark correlation is formed firmly in the charmed baryon.
%The transition probability of the decay process, $\Lambda_c^*\rightarrow D^0p$, is described similarly to the production cross section. Namely, it is a transition form factor described with the initial $\Lambda_c^*$ and final $P$-wave functions at the momentum $q^\prime$ of the final state at the rest frame of the initial particle.
%We could investigate the wave functions of the $\lambda$-mode $\Lambda_c^*$ state in the production and decay processes at a large and small transfer momenta, respectively. 
%}
%{\cblue(relation of the decay rate to ``solidness of the $us$-diquark")}

\begin{figure}[h]
\begin{center}
\includegraphics[width=0.8 \linewidth]{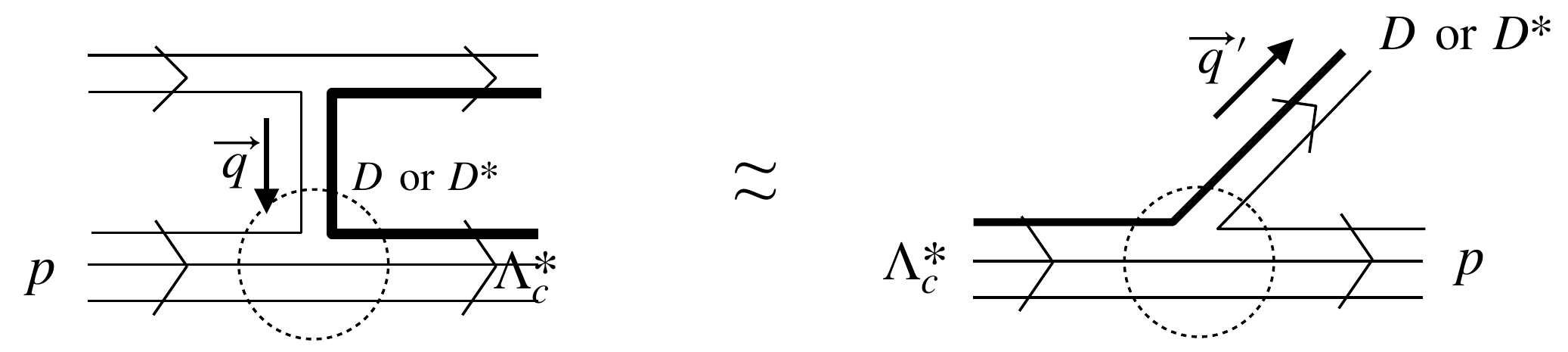}
\end{center}
\vspace{-5mm}
\caption{Similar vertex structures appearing in the production and decay reactions.}
\label{fig_MBBverices}
\end{figure}

{\cred
Once $\lambda$-mode is confirmed for a $\Lambda_c^*$ baryon in the production process, 
we expect that the decay partial width of the $\Lambda_c^*\rightarrow D^0p$ process would tell us how the $ud$-diquark correlation is formed firmly in the $\Lambda_c^*$.
Moreover, considering a similar quark line structure for the vertices of the decay and production reaction
such as $\Lambda_c^*\leftrightarrow D^0p$, one would expect that the production and decay rates 
are related by the transition form factor of different momenta, $\vec q$ and $\vec q^\prime$, entering the vertices, 
as shown in Fig.~\ref{fig_MBBverices}.
We could investigate the wave functions of the $\lambda$-mode $\Lambda_c^*$ state in the production and decay processes at a large and small transfer momenta, respectively. 
}
%-------------------------------------
%-------------------------------------

\subsubsection{Spectroscopy of $\Xi$ baryons\label{sec:physxi}}

\begin{figure}[h]
\begin{center}
\includegraphics[width=0.8 \linewidth]{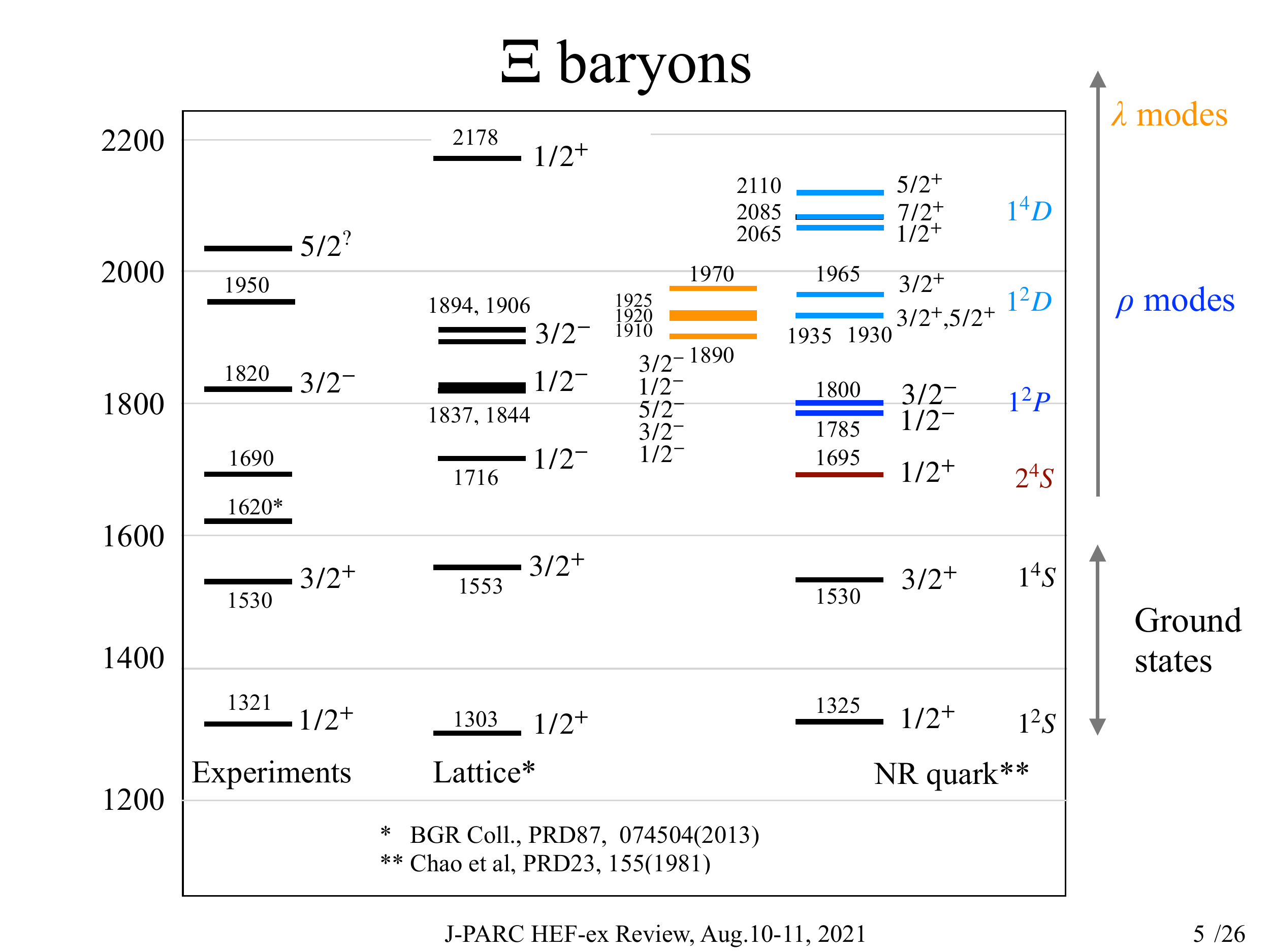}
\end{center}
\vspace{-5mm}
\caption{Comparison of $\Xi$ baryon masses from experiments~\cite{Zyla:2020zbs}, 
lattice simulations~\cite{Engel:2013ig} and non-relativistic quark model~\cite{Chao:1980em}.  
Orange shaded is the region of orbital excitations by the quark model.}
\label{fig_Xibaryons}
\end{figure}

\begin{figure}[ht]
\begin{center}
\includegraphics[width=0.5 \linewidth]{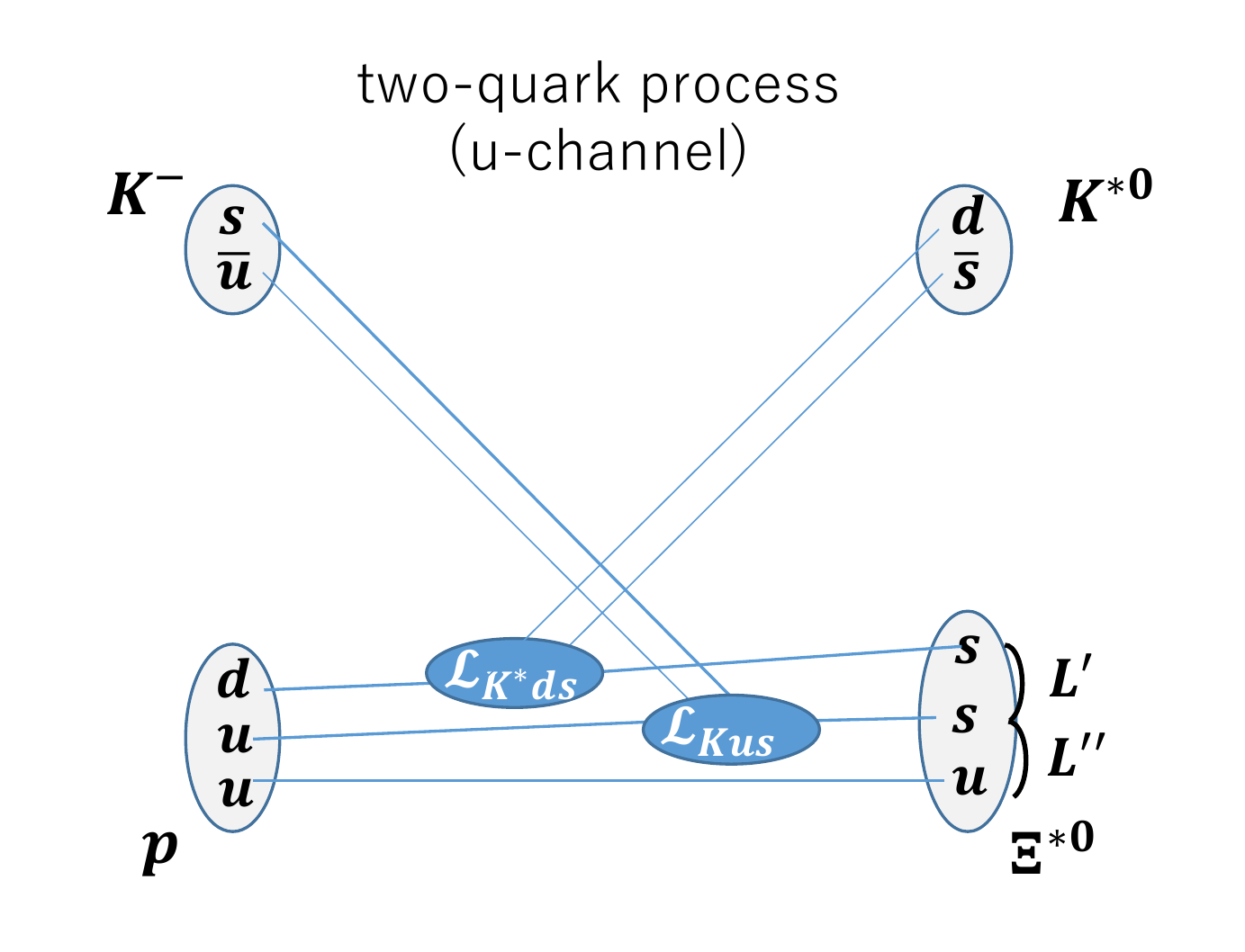}
\end{center}
\vspace{-5mm}
\caption{Reaction diagrams of the $K^-p\rightarrow K^{*0}\Xi^{*0}$ reaction. 
The two quarks are involved in the $u$-channel reaction, and angular momenta between two $s$ quarks and $s$ and $u$ are introduced in the final state. 
}
\label{fig:reaction-diagram2}
\end{figure}

\begin{figure}[ht]
\begin{center}
\includegraphics[width=0.8 \linewidth]{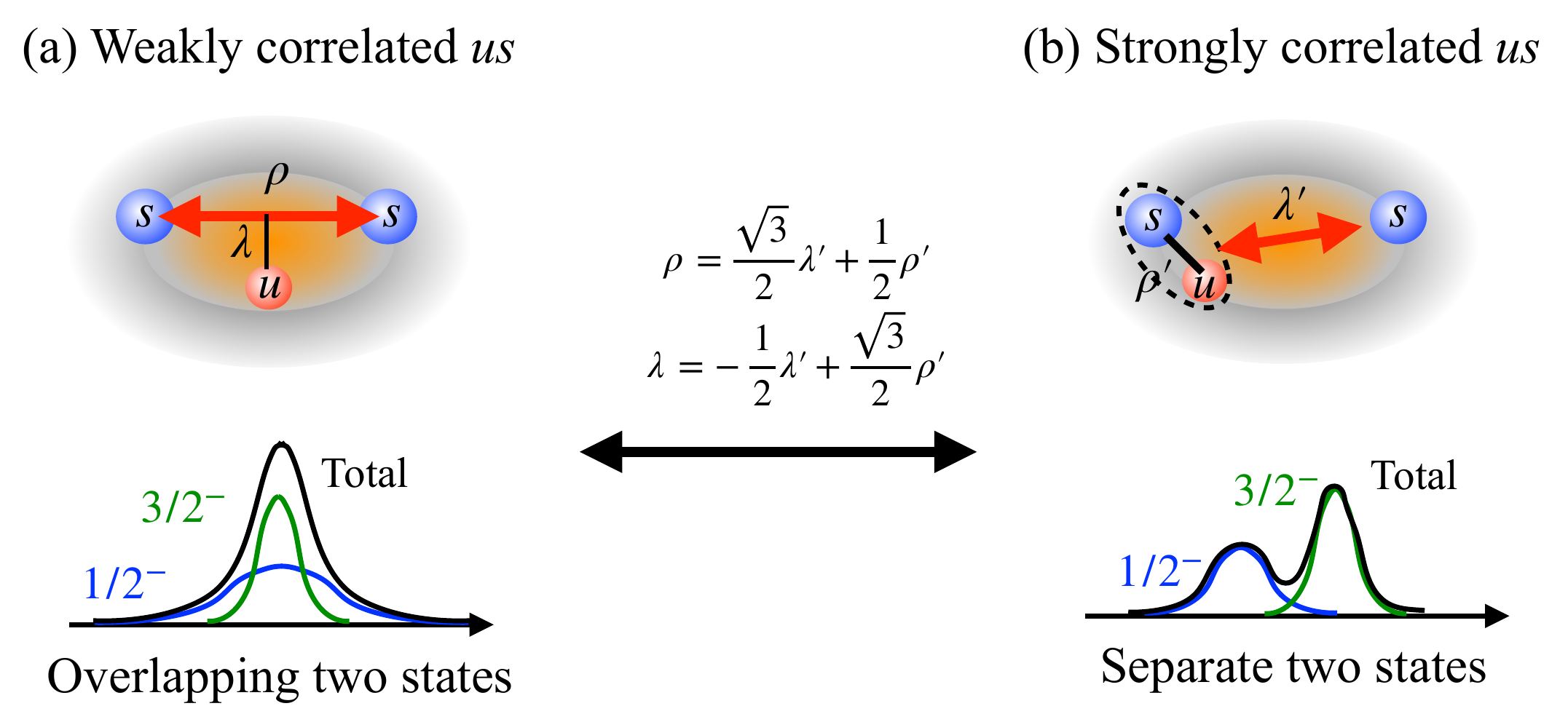}
\end{center}
\vspace{-5mm}
\caption{$\rho$-mode excited state of $\Xi$ ($P$-wave). (a) In the case of weakly-correlated $sq$ ($q=u, d$), the orbital angular momentum of 1 is carried between the spin-antisymmetric $ss$ pair and hence it does not contribute to the $LS$ splitting. (b) In the case of strongly-correlated $sq$, the orbital angular momentum of 1 is partly carried between the correlated $sq$ and the other $s$, and thus the finite $LS$ splitting is expected.
}
\label{fig_ls-xi}
\end{figure}

%At K10 t
The high intensity kaon beam produces abundantly $\Xi$ baryons from the ground state 
up to 1-GeV excitations and even higher, as demonstrated in Fig.~\ref{k10_xi_fig2} in Section~\ref{sec:xi-spectroscopy}.  
Currently ten excited states are listed in PDG~\cite{Zyla:2020zbs},
while only two states are known with established spin and parity.  
In Fig.~\ref{fig_Xibaryons}, states of masses lower than 2.1 GeV are plotted, 
where $\Xi(1530)$ is rated as four-star, $\Xi(1620)$ one-star, and the others three-star.  
In comparison with the lattice and quark model  results, the nature of the observed states 
is  uncertain.  
However, as shown in Fig.~\ref{fig_baryonspecies}, 
the $\rho$ mode, the relative motion of the two strange quarks, is expected to be lowered
which can be used efficiently for the study of $\Xi$ baryons~\footnote{
For $\Xi$ baryons of $uss$ or $dss$ content, the $\lambda$ mode is defined 
to be the relative motion between the $ss$ pair and $u$ (or $d$) quark, and 
the $\rho$ mode the relative motion between the two $ss$ quarks as shown in the left of 
Fig.~\ref{fig_ls-xi}.}. 
Furthermore, from the production experimental point of view, 
two light quarks in the protons are to be converted to two strange quarks 
in $\Xi^*$; namely, a two quark-process is needed unlike the singly charmed baryon productions 
where only one light quark is converted into the charm quark. 
The two-quark process 
%is likely to be more involved but contains more information in the internal structure of baryons and also 
can populate both $\rho$ and $\lambda$ modes, as illustrated in Fig.~\ref{fig:reaction-diagram2}.  

%-------------------------------------
%-------------------------------------

%-------------------------------------
%The 
%-------------------------------------
%$\Xi$ baryons prodive a good place to investigate $ds$ and $su$ diquark correlations, which play a role in the color-super conducting phase in highly dense cold quark matter, as discussed above.
%Similar to the pion-induced charmed baryon production associated with a $D^{*-}$ meson, as mentioned above, we could produce excited $\Xi$ baryons systematically via the $K^-p\rightarrow K^*\Xi^*$ reactions. Again, similar to the charmed baryons, we could kinematically separate the relative motions between two strange quarks ($\rho$ mode) and corrective motions of two strange quarks to the other light quark ($\lambda$ mode) in excited $\Xi$ baryons ($\Xi^*$). Interestingly, the $rho$-mode excitation energy is lowered to the $\lambda$-mode one in the case of $\Xi^*$.
%In the production reaction, two light quarks in the protons are to be converted to two strange quarks in $\Xi^*$; namely, two quark-process must be involved. Both the $\rho$ and $\lambda$ modes are expected to be populated. 

Now, following the quark model we would expect the lowest excitation 
is the $P$-wave excitation of the $\rho$ mode.
Due to the Pauli principle, the spin of the $ss$ pair is zero and hence 
the total spin of the $\Xi$ is expected to form a spin doublet 
$J^P = 1/2^-, 3/2^-$ together with the spin 1/2
of the other $u$ or $d$ quark.  
However, the current experimental data indicates only the state of $3/2^-$.  
There must be the other partner of $1/2^-$.
{%\cblue
%(To be modified. $\rho/\lambda$ to $\rho^\prime/\lambda^\prime$ coordinate transfer and its relation to the $LS$ force should be discussed with illustrations of the coordinates and the expected spectra of $3/2^-$ and $1/2^-$ states with respect to the LS splitting.)
Its location is expected to be close to the $3/2^-$ state if we naively apply the empirical 
fact of the small $LS$ force from the nucleon excitations, for instance
$N(1520)$ of $3/2^-$ and $N(1535)$ of $1/2^-$.  
}
{\cred
In the lowest $\rho$-mode excitations, the $ss$ pair carries the orbital angular momentum of 1, 
and the spin of the $ss$ pair is 0 because of the Pauli principle. 
In this case, the two-body $LS$ force vanishes, and thus does not contribute to the energy splitting between the doublet states (Fig.~\ref{fig_ls-xi}(a)). 
Naively, one would expect the widths for the $1/2^-$ state is wider than that of the $3/2^-$ state, when we consider 
their decays into a $J^P=0^-$ meson and a $1/2^-$ baryon because of the partial wave nature of the decaying particles. }
{\cblue
Experimentally the overlapping two peaks could be identified with sufficient statistics.  
This would be true if the $s$ quark is very heavy when the $us$ or $ds$ diquark correlation is neglected. }
{\cred 
In reality with a finite $s$ quark mass 
the spin-antisymmetric $us$ (or $ds$) diquark may become non-negligibly correlated, 
the $LS$ splitting may be finite through $\lambda^\prime$ orbital excitation, 
as shown in Fig.~\ref{fig_ls-xi}(b). 
Here, the relation between the two Jacobi coordinates is employed as\\
$$
\begin{pmatrix}
\lambda\\
\rho\\
\end{pmatrix}
=
\begin{pmatrix}
-1/2&\sqrt{3}/2\\
\sqrt{3}/2&1/2\\
\end{pmatrix}
\begin{pmatrix}
\lambda^\prime\\
\rho^\prime
\end{pmatrix}.
$$}

%------------------------------------------
%------------------------------------------

{%\cblue
As discussed in Section~\ref{sec:spin-dep-int}, the $LS$ splitting depends on the masses of the quarks and hence 
the location of the $1/2^-$ state carries important information on the nature or the origin 
of the $LS$ force.  
Furthermore, the decay widths carries dynamical information of excitations. 
The width of the $1/2^-$ state would be wide 
if it belongs to the flavor octet~\cite{Arifi:2021private}.  
This could be the reason that the state has not been yet recognized.  
}
The identification of the $1/2^-$ state is one of the important issue.% in K10.
The problem of the missing $1/2^-$ state exists also in the $\Omega$ baryons, 
where we will come back to the problem in more detail.  

Not only the interests in the spectroscopy, the $\Xi$ baryons can provides a platform 
for the study of diquark correlations of the $ds$ and $su$ diquarks.  
The ground state splitting between the octet and decuplet $\Xi$'s already carries 
much information.  
In addition, the $\rho$ mode excitation of the $ss$ pair leaves the relative motion between 
the $s$ and $u$ (or $d$) quarks in $S$-wave, 
from which we study some detailed properties of the diquark correlation.

%are expected to be developed in the spin-singlet flavor-antisymmtric $ds$/$su$ state in the $rho$-mode excited $\Xi$ states. Several excited $\Xi$ states have been reported so far. However, little of their properties such as spin-parities and their decay branching ratios are known. For example, the $\Xi(1820)$ has been reported as its spin-parity of 3/2$^-$, which is a candidate of the $\rho$ mode excitation with havng a configuration of $ds$/$su$(0$^+$) and the orbital angular momentum of 1 between the $ds$/$su$ state and the other $s$.
%If so, lts $LS$ partner of the spin-parity 1/2$^-$ state should exist. The size of the $LS$ splitting in $\Xi$ baryons are not known, which provide information of flavor dependence of spin-dependent interactions between quarks that should be characterized by the non-trivial QCD dynamics.  We could identify the excitation mode of $\Xi(1820)$ together with that of its partner if it exists by measuring their production rates and decay patterns.

To summarize shortly, we will study masses and widths of various $\Xi$ baryons with the determination of  
spins and parities.
How the spin and parity are determined is discussed in Section~\ref{sec:angulardist}.  
Moreover, high lying $\Xi$ excited states such as $\Xi(2370)$ or higher will be used as 
a doorway to produce $\Omega$ baryons~\footnote{
There is a report that $\Xi(2370)$  has a significant amount of branching ratio of $9\%\pm4$\%
decaying into $\Omega\;\bar{K}$~\cite{Kinson:1980vs}.}.
% with the branching ratio of 9\%. Therefore, productions of highly excited states are expected to be a doorway to produce $\Omega$ baryons.

% flatex input end: [./k10docu/k10PhysMultiSBaryon.tex]

%blue}}
%%Spectroscopy of Omega baryons
% flatex input: [./k10docu/k10PhysOmega.tex]
\subsubsection{Spectroscopy of $\Omega$ baryons\label{sec:physomega}}

The $\Omega$ baryon is a system of three strange quarks, $sss$.
%-------------
%The features in the mass spectrum expected by equal masses have been discussed in the previous section.  
%-------------
In flavor SU(3) symmetry group it is assigned as a member of the decuplet representation, 
which was realized by the quark model whose building blocks are the $u, d$ and $s$ quarks.  
%Here in what follows, these quarks are the ``constituent" ones as we discussed in the previous section.  
Historically nine short-lived baryons of the spin $3/2^+$ were known, four $\Delta$'s, three $\Sigma^*$'s and 
two $\Xi$'s fitting into the decuplet members, and the last piece $\Omega$ was predicted 
in the quark model~\cite{GellMann:1964nj}.
The confirmation of $\Omega^-$ particle  in
1964~\cite{Barnes:1964pd}
%1973~\cite{Alvarez:1973vb,Barnes:1964pd} 
supported the validity of the quark model 
based on the SU(3) flavor symmetry.
%This has been considered as an evidence of the constituent quarks.

In reality, the SU(3) symmetry is broken by the mass of the $s$ quark, which is heavier 
than the almost equal masses of $u, d$ quarks. 
The pattern of SU(3) breaking could predict the mass of $\Omega$ under the equal mass spacing 
between the decuplet members.  
The broken SU(3) also leads to the obvious mass difference of the pion ($m_\pi \sim 140$ MeV) 
and kaon ($m_K \sim 500$ MeV).  
This difference together with the singly flavored nature leads to the unique features of the $\Omega$ baryons.  

%-------------------------------------
\begin{figure}[h]
\begin{center}
\includegraphics[width=0.8 \linewidth]{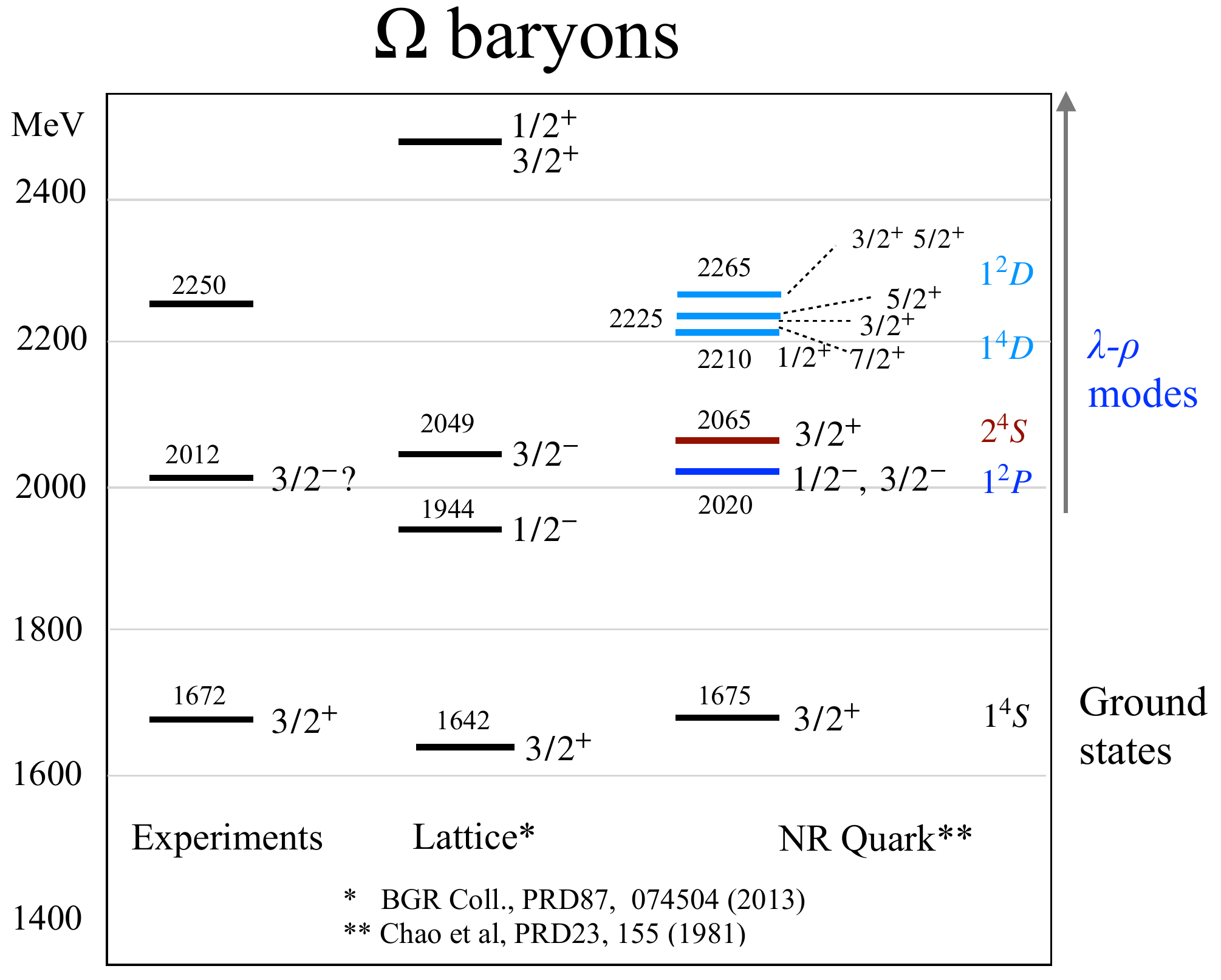}
\end{center}
\vspace{-5mm}
\caption{Comparison of $\Omega$ baryon masses from experiments~\cite{Zyla:2020zbs}, 
lattice simulations~\cite{Engel:2013ig} and non-relativistic quark model~\cite{Chao:1980em}.  
Orange shaded is the region of orbital excitations by the quark model.}
\label{fig_Omegabaryons}
\end{figure}
%-------------------------------------

The current status $\Omega$ baryon spectroscopy is shown in Fig.~\ref{fig_Omegabaryons}.
Indeed the experimental status is poor, and so it is not yet the stage of comparison between data and theory.  
Nevertheless, we would like to discuss the $\Omega$ spectroscopy in some detail.  
First the $sss$ combination put some restrictions on the wave function due to the Pauli's exclusion principle.  
For example, the fact that spin of the ground state $\Omega$ is 3/2 rather than 1/2 is precisely from that requirement.  
Similar restrictions apply to excited states.
Some consequences from the restrictions may manifest in the fine structures of excited states. 
This will be discussed in Section~\ref{sec:1p-qm}(A).  

Second is associated with the heavier mass of the kaon than that of the pion.  
In fact, the pion is very light as it is the NG boson, and therefore, it in many cases 
plays a role of radiation like a photon with the axial-vector nature.  
%Thus many excited baryons emit pions when they decay.  
On the other hand, the kaon when coupled to excited states, it plays not only 
a role of radiation but also a role of a constituent forming hadronic molecule.  
This could be the reason that the $\Lambda(1405)$ can develop the molecular 
structure of $\bar K N$.  
A similar situation may occur in excited states of $\Omega$'s, which is also 
the subject in Section~\ref{sec:1p-molecule}(B).  

Third, the $\Omega$ baryons do not couple to the pion due to isospin symmetry.  
Therefore, without the meson cloud which exists in the light baryon structure, 
we may be able to directly access to the quark core region of the baryons.
Furthermore, we may expect that an $\Omega$ baryon is 
smaller in spatial extension as compared to the light flavored baryons
which is surrounded by meson cloud.  
We will discuss a related topic in Section~\ref{sec:2s-state}(C). 
A possible experimental study to access baryon size from scattering cross sections is discussed in Section~\ref{sec:ynscattering}.%3.8.4.  

%In the following we would like to discuss several unsolved/unsettled issues in hadron physics.  
%To solve these problems, we consider it very important to approach the problems from various directions.
%High-intensity high-momentum negative kaon beams with high purity
%will contribute to it by the study of $\Omega$ baryon spectroscopy.  

% flatex input end: [./k10docu/k10PhysOmega.tex]

%%Spectroscopy of Omega baryons
% flatex input: [./k10docu/k10PhysGoal.tex]
%--------------------------------------
%\paragraph{{\it 1$P$ states -- quark orbital excitation} :}\hfill
%--------------------------------------
%--------------------------------------
%\subsubsection{1$P$ states -- quark orbital excitation\label{sec:1p-qm}}
\subsubsection*{(A) 1$P$ states -- quark orbital excitation\label{sec:1p-qm}}
%--------------------------------------

The confined constituent quarks can be excited orbitally.
The lowest excited state is expected to be in $P$-wave with an orbital angular momentum $L = 1$.
In the spin-flavor SU(6) classification they form the 70-dimensional representation 
that are decomposed into spin-flavor multiplets $^{2S+1}{\bf N} = \ ^2 8, ^2 10, ^4 8$ and $^2 1$.
Here $S$ is the total intrinsic spin of the three quarks, 
$S = 1/2, 3/2$, and ${\bf N}$ the dimension of the flavor representation.
Among them, $^2 8$ and $^4 8$ are assigned to $N$, 
$\Lambda$, $\Sigma$ and $\Xi$ baryons, 
and $^2 1$ to $\Lambda(1405)$ and $\Lambda(1520)$.
Then the $^2 10$ representation includes the $P$-wave $\Omega$  
together with $\Delta, \Sigma$ and $\Xi$ excited states.
Because of their intrinsic spin $S=1/2$, they form the spin doublet, 
$J = S + L = 1/2, 3/2$, and their parity is negative due to the $P$-wave.
It is emphasized that in the standard picture, there must be 
$2S+1$ or $2L+1$ fold degeneracies depending on whether  $S < L$ or $S > L$, respectively.  

The $\Omega(2012)$ which has been recently reported from Belle is a candidate of 
one of the multiples~\cite{Yelton:2018mag}.
The mass 2012 MeV is about 500 MeV above the expected ground state when possible spin-spin interaction contribution is removed.
Therefore, the state is expected to be one of doublets of $^2 10$.  
Its decay width is as narrow as $6.4^{+2.5}_{-2.0} \pm1.6$ MeV. 
The decay $\Omega(2012) \to \Xi \bar K $ was studied in the quark model, 
resulted in a decay width about 12 MeV for $J^P = 1/2^-$ and about 6 MeV for $J^P = 3/2^-$. 
The width of the $J^P = 3/2^-$ state is narrower due to its $D$-wave nature of the decaying 
two-body channel in the final state.  
Therefore, the quark model seems to prefer $3/2^-$ state.

Assuming a two-body $LS$ interaction, we need both $L = 1$ and $S = 1$, 
where $L$ and $S$ are the orbital angular momentum and spin of an $ss$ pair.
However, such a combination is not allowed due to the Pauli principle,  
when spin, flavor,  orbital and color wave functions are combined.  
Therefore, we expect that the $LS$ splitting of $1/2^-$ and $3/2^-$ is suppressed 
for these $\Omega$ states.  

As explained in Section~\ref{sec:spin-dep-int}
the two-body spin-orbit interaction is derived from the OGE and III interactions.
In many cases they contribute with opposite signs, while in others
only the OGE survive due to flavor anti-symmetric nature of III and its suppression 
for heavy quarks systems.  
Such relations are summarized in Table~\ref{table_LS} for various baryon states, 
where expectation not only for $P$-wave but also for $D$-wave excitations, 
where in the latter the OGE from the III may exclusively contribute to the $\Omega$'s.  
In Ref.~\cite{Liu:2019wdr}, a phenomenological $LS$ interaction was introduced
resulting in significant splittings.  
One possible origin of such an interaction would be three-body nature.  
The clarification of the $LS$ splitting is crucial to the systematic understanding 
for the baryon spectroscopy.  

%-----------------------
\begin{table}[h]
\caption{
Expected contributions of $LS$ interaction to various baryons.  
In the second line $P$ and $D$ denote the total orbital angular momentum of quarks 
inside baryons.
Positive values $A$ and $B$ represent matrix elements of the $LS$ interaction
which depend on baryon states, and their   
computations were performed, for instance, in Ref.~\cite{Takeuchi:1998mv}.
The symbol ``?" indicates that there is no experimental data. 
}
\label{table_LS}
\begin{center}
\begin{tabular}{c c c c c c c c c}
\hline
 & \multicolumn{2}{c}{$\Omega^*$} 
 & \multicolumn{2}{c}{$N^*$} &\multicolumn{2}{c}{$\Lambda^*$} & \multicolumn{2}{c}{$\Lambda^*_c$}\\
Orbit & $P$ & $D$ & $P$ & $D$ & $P$ & $D$ & $P$ & $D$\\
\hline
OGE & -- & $+A$ & $+A$ & $+A$ & $+A$ & $+A$ & $+A$ & $+A$\\
III      & -- & --     & $-B$ & $-B$ & $-B$ & $-B$ & -- & -- \\
Sum & 0 & $+A$ & $\sim 0$ & $\sim 0$ & small & small & $+A$ & $+A$ \\
Exp (MeV)  & ? & ? & $\sim 0$ & $\sim 0$ & 36 & ? & 34 & ? \\
\hline
\end{tabular}
\end{center}
\end{table}
%-----------------------

%--------------------------------------
%\paragraph{{\it 1$P$ states -- moleculars} :}\hfill
%--------------------------------------
%--------------------------------------
%\subsubsection{1$P$ states -- moleculars\label{sec:1p-molecule}}
\subsubsection*{(B) 1$P$ states -- moleculars\label{sec:1p-molecule}}
%--------------------------------------

Another interpretation has been proposed for $\Omega(2012)$; 
since its mass is located approximately 10 MeV below the $\Xi^*\bar{K}$ threshold, it could be a $\Xi^*\bar{K}$ molecular state~\cite{Valderrama:2018bmv,Ikeno:2020vqv}.
The formation of such a molecular state is expected near the threshold region of two or more 
particles with some heavy mass.  
The $\Xi^*$ and $\bar{K}$ would be such.  
Though the kaon is heavier than the pion, it also shares the properties of the 
NG boson when the chiral interaction is effective.  
In this regards, the system shows an interesting feature; 
the direct interaction for $\Xi^*\bar{K}$ is absent, but the offdiagonal interaction to the coupling 
to $\Omega\eta$  with higher mass can drive an attraction. 
The decay width was also estimated by chiral counting, expecting values consistent with 
the data~\cite{Valderrama:2018bmv}.
The mechanism itself is interesting, and deserves study.  
A caveat in this picture is that $1/2^-$ state is not easy to be explained.  

To establish the nature of $\Omega(2012)$ is an urgent issue in the present project.
The determination of basic properties of such as spin and parity, decay width, and also a search of spin partner provides an important information to explore the $\Omega$ spectroscopy.

%\paragraph{{\it 2$S$ states, physics of the Roper-like states} :} \hfill
%--------------------------------------
%\subsubsection{2$S$ states, physics of the Roper-like states\label{sec:2s-state}}
\subsubsection*{(C) 2$S$ states, physics of the Roper-like states\label{sec:2s-state}}
%--------------------------------------

Baryon resonances possessing the same spin-parity as that of the ground states are of special interest. 
By now many such states are observed in a wide range of flavor contents; 
$N(1440) 1/2^+$, $\Lambda(1600) 1/2^+$, $\Sigma(1660) 1/2^+$, $\Delta(1600) 3/2^+$ and $\Xi_c(2970) 1/2^+$.
The nucleon resonance $N(1440)$ has been known for long time as the Roper resonance, while $\Xi_c(2970)$ has been established only recently~\cite{Belle:2020tom}.
Furthermore, $\Lambda_c(2765)$ 
and newly found $\Lambda_b(6072)$~\cite{Sirunyan:2020gtz, Aaij:2020rkw} are also expected to be their siblings.
Interestingly all of their masses are about 500 MeV above their corresponding ground states
as shown clearly in Fig.~\ref{fig_systematics}. 
%In the light flavor sector of $u, d, s$, many of these levels appear below the first $Pp$-wave excitations
%such as $N(1535)$ and $N(1520)$.  
%These features are not explained easily by the conventional quark model. 
%One would be tempted to draw a speculation that yet unknown flavor independent dynamics exists.
%On the other hand, negative parity resonances are systematically lowered 
%as the flavor contents become heavier as suggested 
%by the lowering of the $\lambda$ mode in the quark model.

In the nucleon sector the problem has been known as too low mass of the Roper resonance.
In the quark model the $1/2^+$ state is realized as a radial (nodal) excitation 
whose mass appears at around 1800 MeV which is obviously too high 
as compared to the observed one of 1440 MeV. 
Related to this the quark model predicts the wrong mass ordering; 
the Roper resonance appears higher than the negative parity state, 
for instance $N(1535) 1/2^-$. A lowering mechanism has been proposed 
due to the meson cloud around the quark core of $u$, $d$ quarks~\cite{Suzuki:2009nj,Burkert:2017djo}. 
Turning to the heavy quark sector, due to the lowering of the $\lambda$-mode, 
the observed states of positive and negative parities appear 
as in the expected order with their mass values qualitatively 
consistent with the quark model predictions. 
Thus another option to explain the equal excitation masses is 
flavor dependent mechanism so as to cancel the natural flavor dependence 
in the quark model. 
These two interpretations for the Roper resonance and siblings must be converged.

To answer this question, information of $\Xi$ and $\Omega$ resonances are useful. 
The mass of the strange quark is between those of light and heavy quarks, and therefore more strange baryons interpolate the light and heavy flavor dynamics, filling the missing link between the two. 
As anticipated, with the suppression of the meson cloud 
the $\Omega$ baryons provide an ideal platform to extract the dynamics of the constituent quarks. 
This is particularly expected in the study of various transitions such 
as radiative decays $\Omega^* \to \Omega \gamma$ and 
kaon decays $\Omega^* \to \Xi \bar K$ since the decay rates are sensitive to the forms of the wave functions. 

The decay of the Roper-like state through one kaon emission 
$\Omega^* \to \Xi \bar K$ may contain interesting  physics.  
The decay occurs by the chiral interaction between the NG kaon and quarks, 
\begin{eqnarray}
{\cal L}_{\phi qq} = \frac{g_A^q}{f_\pi} \ 
\bar q \gamma_\mu \gamma_5 \lambda_a q \ \partial^\mu\phi^a
\end{eqnarray}
where $\phi^a$ is the flavor SU(3) NG boson field and 
$g_A^q \sim 1$~\cite{Weinberg:1990xm}  the axial coupling constant of the quark.
In the previous publications, the Lagrangian was expanded non-relativistically and 
took up to the first order in the velocity $v \sim p/m$ of the quarks 
where $p$ is the momentum of a quark inside the baryon, 
\begin{eqnarray}
{\cal L}_{\pi qq} \sim  \ \frac{g_A^q}{f_\pi} 
\left( \vec \sigma \cdot \vec q
+ \frac{\omega_\pi}{2m_q}(\vec \sigma\cdot \vec q - 2 \vec \sigma \cdot \vec p)
\right)
\end{eqnarray}
The problem of the truncation up to this order is in the fact that 
the leading term of order $v^0$ vanishes exactly due to the operator structure of 
$\vec \sigma \cdot \vec q$ where $\vec q$ is the momentum carried by the kaon, 
and the orthogonality between the wave functions of the radially excited and ground states
in the long wave length limit.  
Such a kind of forbidden process was originally pointed out in the 
radiative decay of the nucleon Roper resonance~\cite{Kubota:1976ft}.  
The next to leading order term of $v^1$ also turns out to be small.  
Actual computation shows that these terms (and similar studies) 
predicted only small decay widths
which contradicts the observed data~\cite{Arifi:2021orx,JuliaDiaz:2004qr}.  

Thus next-next to leading order term of $v^2$ is important.  
Explicitly 
\begin{eqnarray}
{\cal L} \sim \frac{g_A^q}{f_\pi} 
\left(
\frac{m_\pi^2}{8m_q^2}\vec \sigma \cdot \vec q + 
\frac{1}{4m_q^2}
\vec \sigma\cdot (\vec q - 2 {\vec p}) \times
(\vec q \times {\vec p})
\right)
\end{eqnarray}
The second term is the dominant term which results in the transition matrix element
as proportional to $\langle p^2 \rangle \sim 1/ \langle r^2 \rangle$, 
which survives in the long-wave length limit with no forbidden selection rule applies.  
Physically, we can interpret that the transition rate is dictated by the Fermi motion of the confined quarks 
and hence the inversely proportional to the size of the quark core of baryons.  
It was shown that such higher order term
contributed significantly to the decay width of the heavy baryons
and improved the agreement with the data~\cite{Arifi:2021orx}.  
A preliminary calculation show the decay width of the Roper-like $\Omega$ 
is around 100 MeV after the inclusion of the $v^2$ contributions, while the terms 
of order $v^0$ and $v^1$ give only around 20 MeV.  
The small value in the lower order terms is qualitatively consistent with the previous studies.  

After discussing this much about the Roper-like $\Omega$, it is very important to observe 
the state and to measure the width.  
The width is interesting in that it carries the information of the internal motion of 
constituent quarks in the core region which is related to the size of that region.

\clearpage
% flatex input: [./k10docu/k10pi20bl.tex]
\subsection{High-Momentum Secondary Beam Line --- $\pi$20}

Since early 2020, the High-p beam line (B line) has been operated for an experimental study on spectral changes of light vector mesons in nuclei, E16, with a primary beam branched from the existing slow-extraction beam line (A line).

As we have discussed, baryon spectroscopy with various flavors is vital to reveal the non-trivial flavor dependent dynamics in baryons. The charmed baryon spectroscopy is important and closely related to the spectroscopy of multi-strangeness baryons, $\Xi$'s and $\Omega$'s at K10, as we review characteristic natures of baryons with a charm quark and those with two strange quarks in Sections~\ref{sec:charmedbaryons}, \ref{sec:physxi}, and \ref{sec:physomega} as well as Sections~\ref{sec:baryonshf} and \ref{sec:baryonsqss}.
We have already proposed a spectroscopic study of charmed baryons (E50)~\cite{E50exp} and received a stage-1 approval. E50 needs a high-intensity pion beam at 20 GeV$/c$ at the $\pi$20 beam line. 

\begin{figure}[htbp] 
 \centerline{\includegraphics[width=0.9\textwidth]{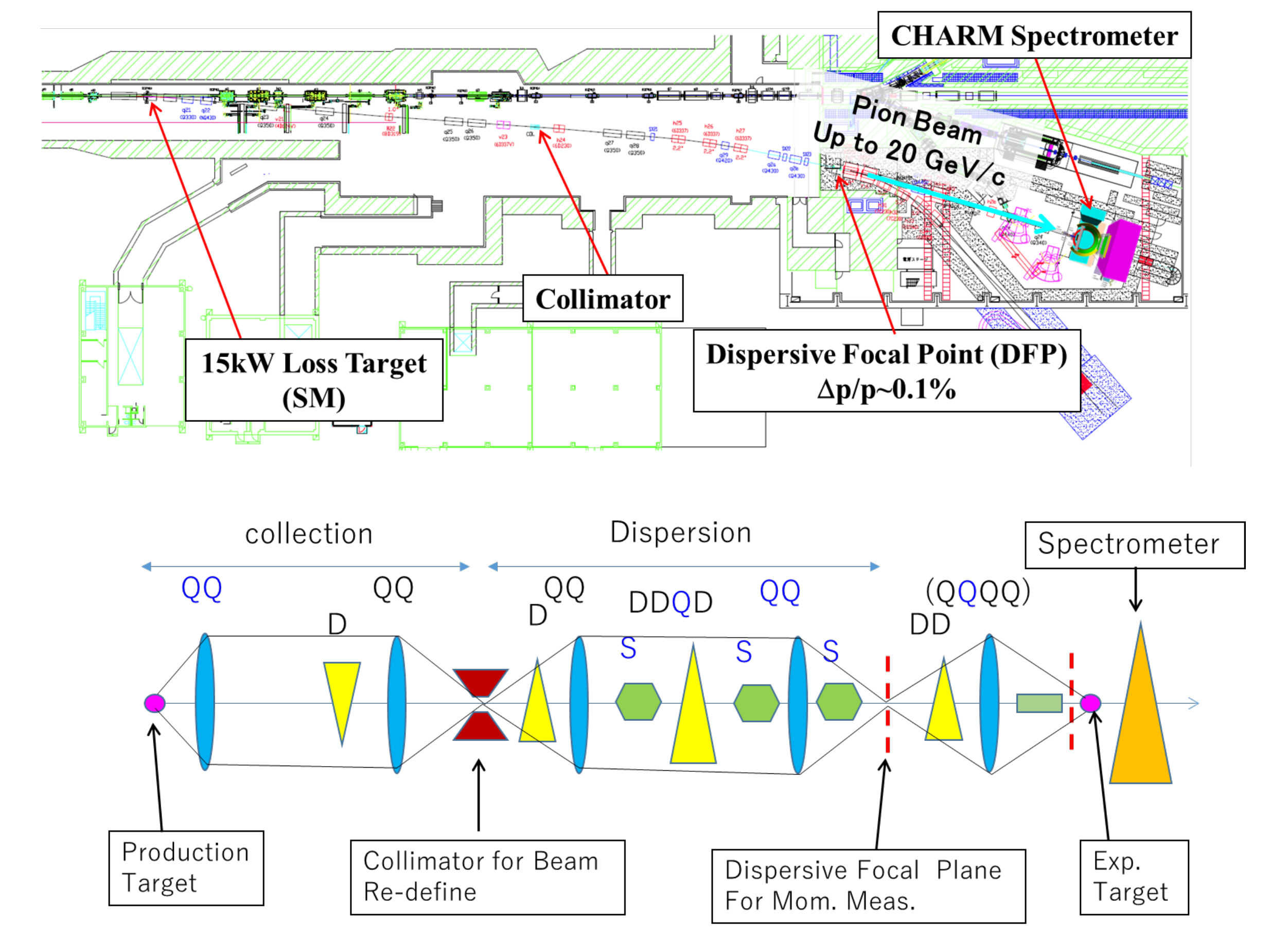}}
 \caption{Plan view of the $\pi20$ beam line (top) and schematic illustration of the beam line configuration (bottom). \label{fig:pi20beamline}}
\end{figure}

\begin{figure}[htbp] 
 \centerline{\includegraphics[width=0.9\textwidth]{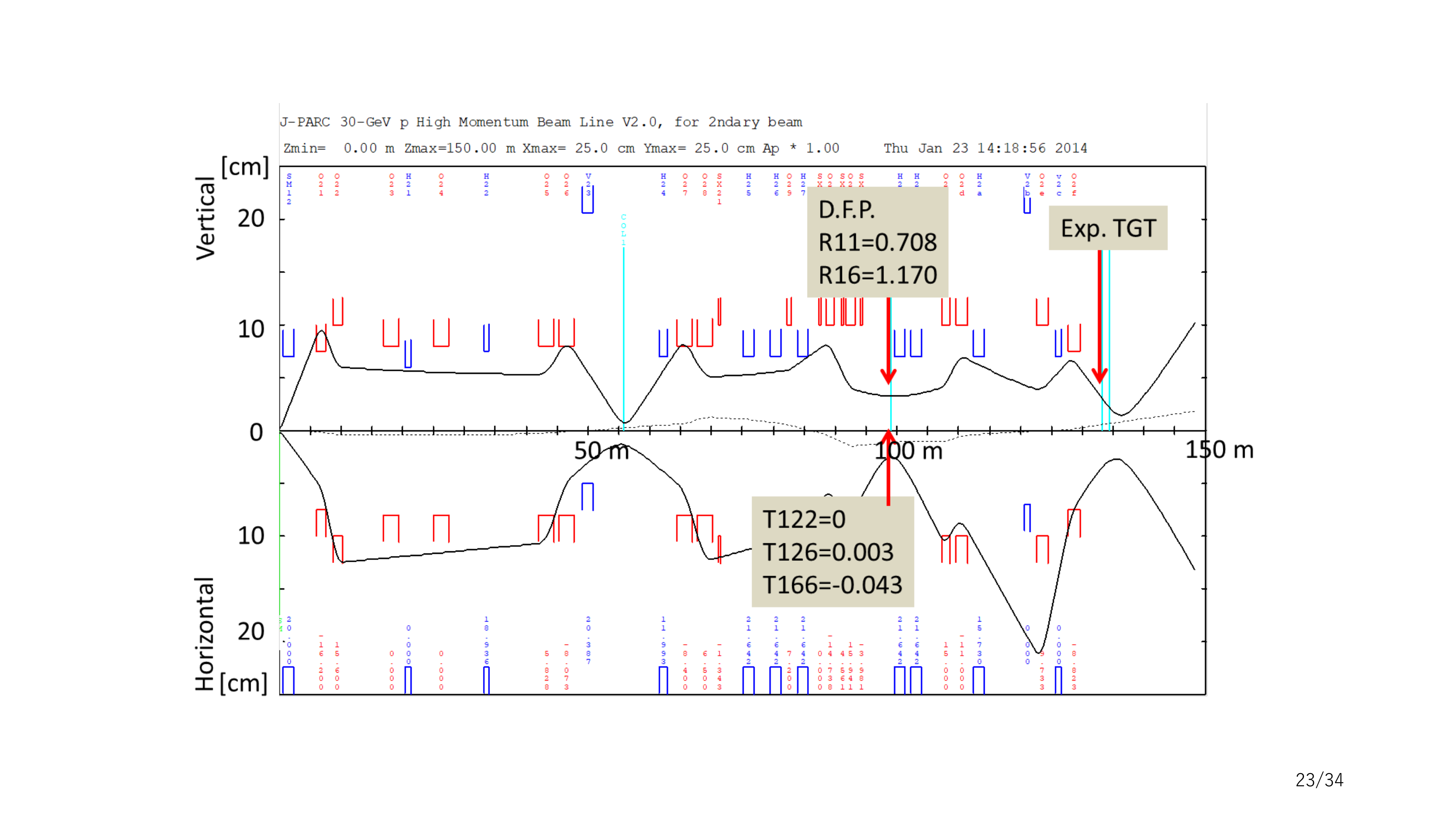}}
 \caption{Beam envelope calculated by the TRANSPORT beam optics code to the second order for the $\pi20$ beam line. \label{fig:pi20beamenvelope}}
\end{figure}

\begin{figure}[htbp] 
 \centerline{\includegraphics[width=0.9\textwidth]{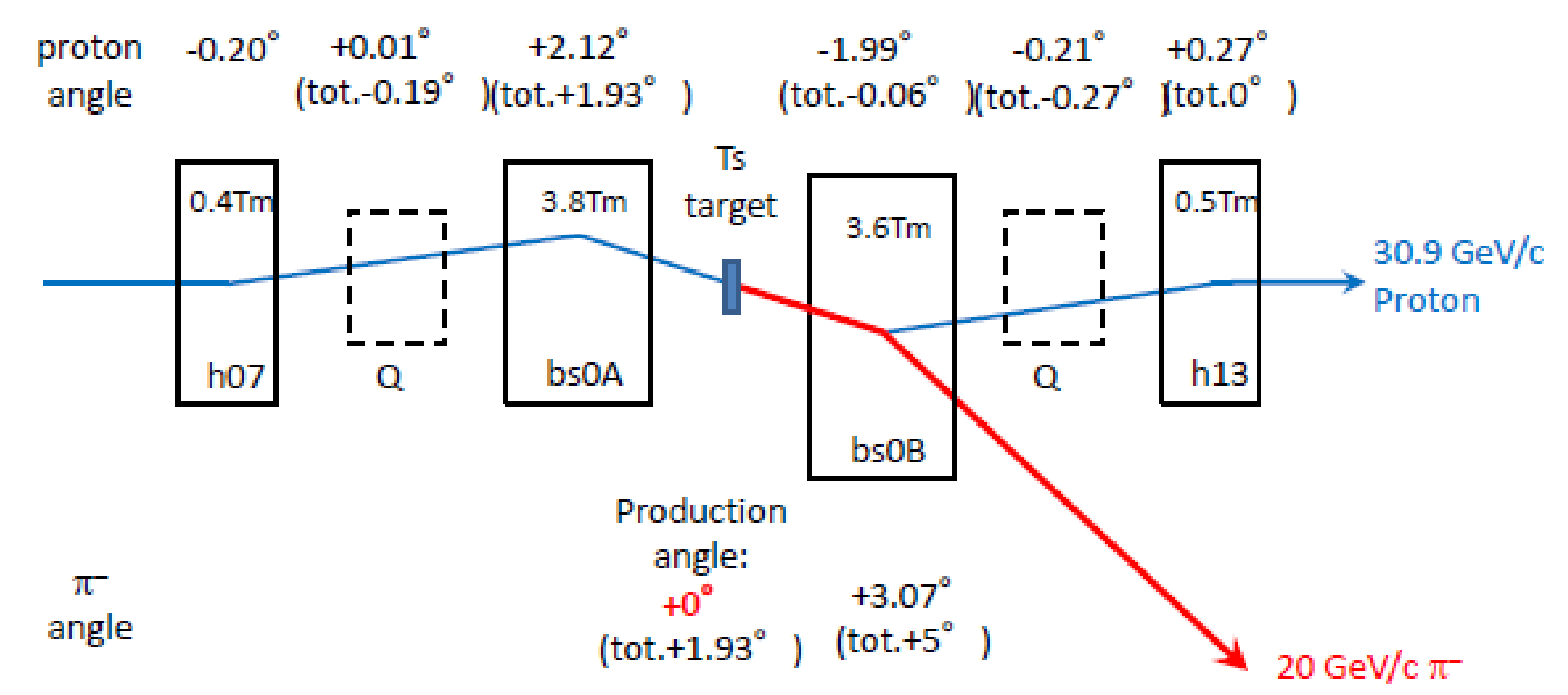}}
 \caption{Proposed layout at the production target for the $\pi20$ beam line. For a negative secondary beam, the primary beam trajectory is swung by two dipole magnets (h07 and bs0A) to the left hand side before the target and swung back by another two dipole magnets (bs0B and h13) to the A line after the target. The layout is optimized so that the negative secondary beam of 20 GeV/$c$ produced at zero degree is extracted to the $\pi20$ beam line. This is the so-called beam swinger optics~\cite{Tanaka:1994rr}.  \label{fig:pi20bso}}
\end{figure}

The B line is designed so that the secondary beams produced at the branching point in the A line can be transported without major modification of the beam-line configuration (Fig.~\ref{fig:pi20beamline}), except that the most upstream part around the primary target and additional structure/equipment required for radiation safety have yet to be provided.
The beam envelope calculated by the TRANSPORT code~\cite{TRANSPORT} is shown in Fig.~\ref{fig:pi20beamenvelope}. 
The secondary beam with a negative charge produced at zero degree at the production target is departed from the primary beam course (A line) to the $\pi20$ beam line by using the so-called beam swinger optics~\cite{Tanaka:1994rr}, as illustrated in Fig.~\ref{fig:pi20bso}. 
For a negative secondary beam, the primary beam trajectory is swung by two dipole magnets (h07 and bs0A) to the left hand side before the target and swung back by another two dipole magnets (bs0B and h13) to the A line after the target. The layout is optimized so that the negative secondary beam of 20 GeV/$c$ produced at zero degree is extracted to the $\pi20$ beam line.

\begin{figure}
 \centerline{\includegraphics[width=0.9\textwidth]{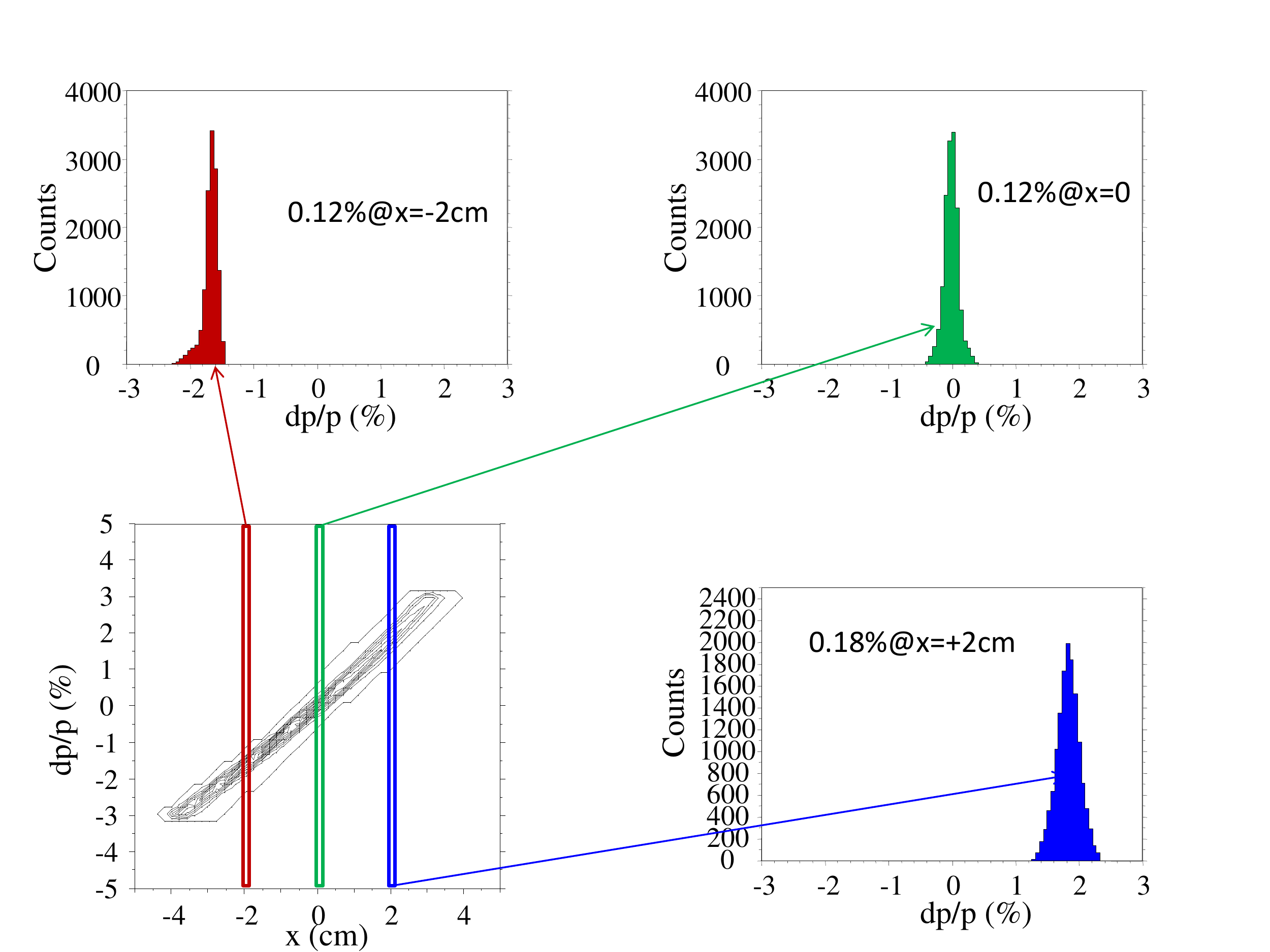}}
 \caption{Correlation of horizontal position ($x$) to momentum (dispersion $dp/p(\%)$ of beam particles at the dispersive focal plane. \label{fig:pi20dispersion}}
\end{figure}

\begin{figure}
 \centerline{\includegraphics[width=0.8\textwidth]{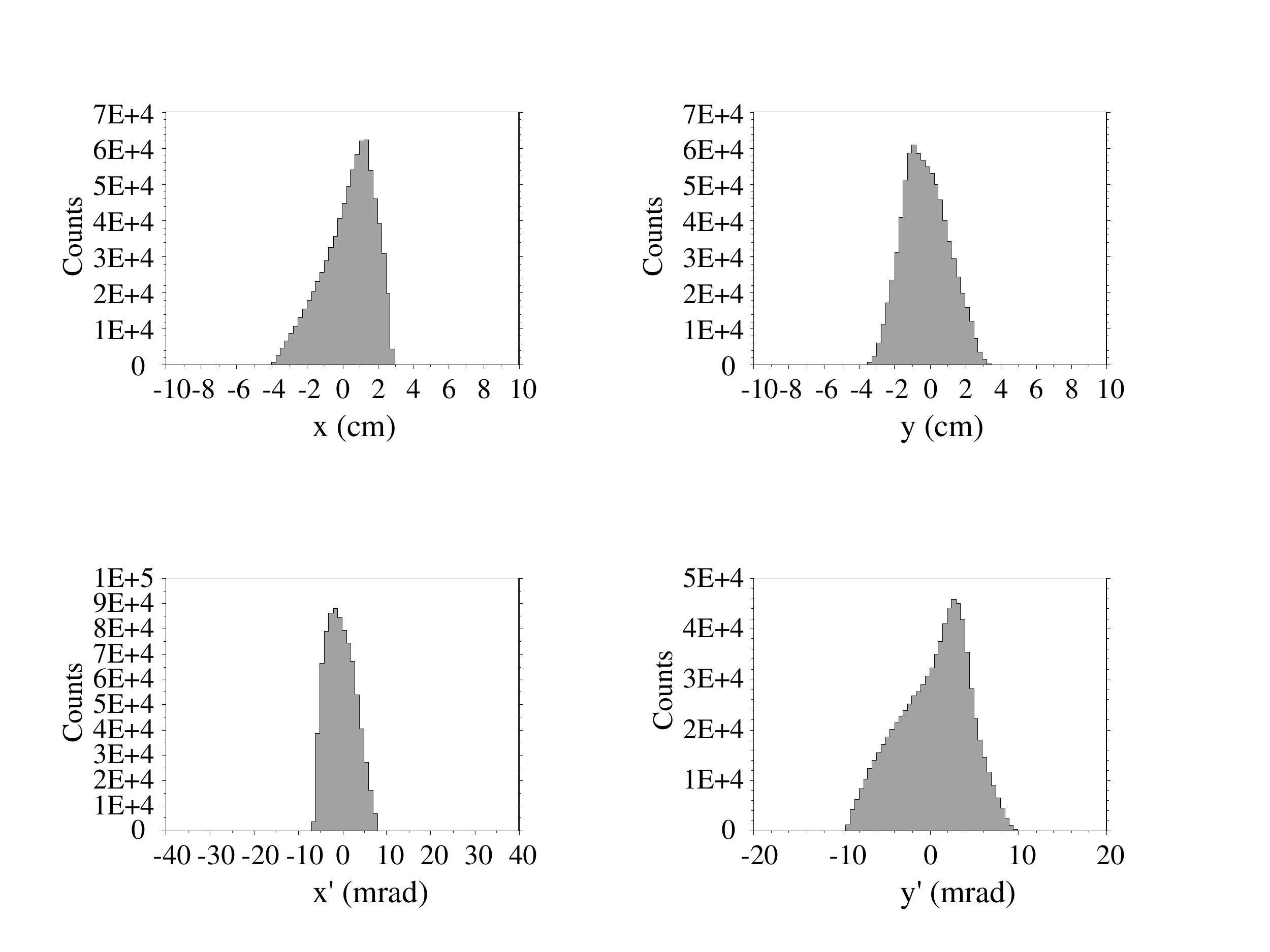}}
 \caption{Beam profile estimated by TURTLE at the experimental target. \label{fig:pi20profileff}}
\end{figure}

\begin{figure}
 \centerline{\includegraphics[width=0.9\textwidth]{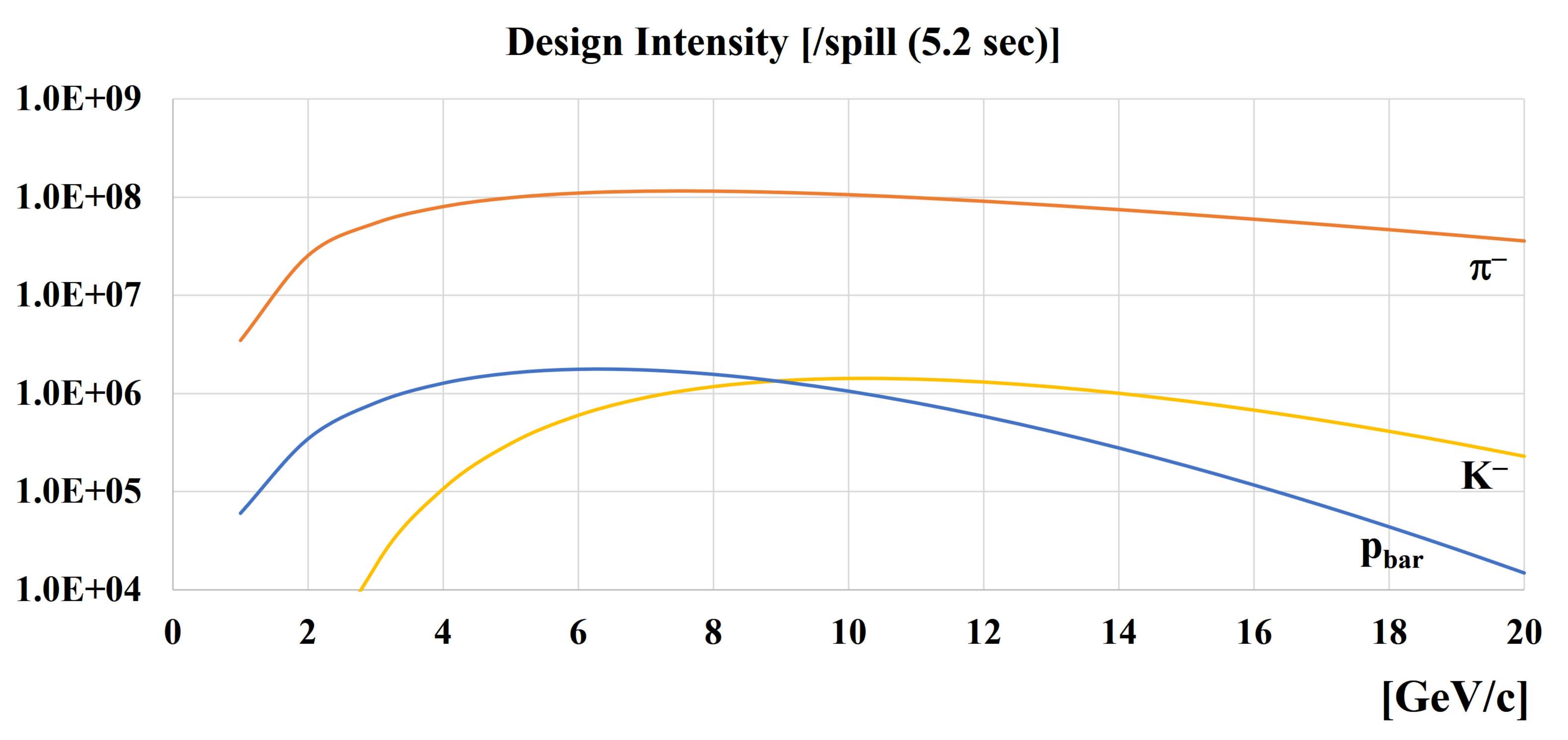}}
 \caption{Expected intensities of negative pions, kaons, and antiprotons per spill (5.2-second spill cycle) as functions of their momenta in the case of 30-kW primary protons on a 60-mm-long Platinum target (15-kW loss) estimated by Sangford and Wang's formula \cite{SanfordWang1967}. \label{fig:pi20intensity}}
\end{figure}

The secondary beam is collected by the first doublet of quadrupole magnets in the $\pi20$ beam line and focused at a collimator placed 50 m downstream from the production target to define the beam image at the production target. After the collimator, the beam is focused again at the dispersive focal plane with a large dispersion of 1.170 \%/cm.
Here, 4 sextupole magnets are employed to eliminate/minimize major geometric and chromatic aberrations to the second order.
A correlation of horizontal position ($x$) to momentum (dispersion $dp/p$(\%)) of beam particles at the dispersive focal plane is calcualted by TURTLE \cite{TURTLE}, as shown in Fig.~\ref{fig:pi20dispersion}. The figure demonstrates that a momentum resolution as good as 0.1\% can be realized with a spatial resolution of 1 mm for a beam particle. 
A beam profile estimated by TURTLE at the experimental target is shown in Fig.~\ref{fig:pi20profileff}.
Figure~\ref{fig:pi20intensity} shows expected intensities of negative pions, kaons, and antiprotons per spill (5.2-second spill cycle) as functions of their momenta in the case of 30-kW primary protons on a 60-mm-long Platinum target (15-kW loss) estimated by Sangford and Wang's formula \cite{SanfordWang1967}.

At the present hadron experimental facility, no secondary charged beams greater than 2 GeV/$c$ are available. The upgrade of the B line to utilize high-momentum secondary beams is strongly desired.



%----------------------------------------------------

% flatex input end: [./k10docu/k10pi20bl.tex]

%%Spectroscopy of Omega baryons
%%Conceptual design for the K10 beam line
\clearpage
% flatex input: [./k10docu/k10beamline.tex]
\subsection{Conceptual Design for the K10 Beam Line}

High-intensity high-momentum negative kaon beam with high purity
is a crucial ingredient for the proposed experiment. Therefore, the K10 beam line under discussion in the project "Hadron hall extension" is the only place where the proposed experiment can be performed. 

%The K10 beam line is designed to serve high-momentum separated secondary beams.

Two options are under consideration for the particle separation at 
the K10 beam line;
an ordinary electrostatic (ES)-separator option and an RF-separator option. 
The ES-separator option can be applied up to $4 \sim 6$~GeV/$c$. 
In contrast, the RF-separator option is suitable for higher momentum.
By making a common design at the front-end section
of both options of the beam line,
we can switch the separation method without accessing the high-radiation area.
The production angle of secondary beams is chosen to be 3~degrees,
which is smaller than those of the K1.8 and K1.1 beam lines,
because the production cross-sections of $K^-$ and $\bar{p}$ 
at 3~degrees are about five times larger than those at 6~degrees,
according to the empirical formula by Sanford and Wang\cite{SanfordWang1967}.
%\begin{figure}[tbp]
% \centering
% \includegraphics[width=0.8\textwidth]{k10/t2station.pdf}
% \caption{????æ¨???è¿?????10???¼ã?????¤ã?³ã?????¤ã?¢ã??????}
% \label{K10FrontLayout}
%\end{figure}
%????æ¨???è¿????????¤ã?¢ã????????ref{K10FrontLayout}??¤º????
%K10???¼ã?????¤ã?³ã??æ¬¡ã???¼ã??????ç¬?æ¨????¹ã???¼ã?·ã?§ã??(T2) ?§ç??????????
%1æ¬¡ã???¼ã?????¤ã?³ã???????´ã?¸å????ºã????????
%??å¯¾å?´ã????1.1???¼ã?????¤ã?³ã??è¨?½®????????
%ä¸?æµ????»ç??³ã????1æ¬¡ã???¼ã?????¤ã?³ã??K1.1???¼ã?????¤ã?³ã???»ç??³ã??¹²æ¸?????????????
%æ³???±ã????????å¿?è¦?????????
%\subsection{RF-separator option}
Since the separation by an ordinary ES separator is proportional to $1/p^3$,
the separation is much difficult for the momentum higher than
4~GeV/$c$ ($K^-$) or 6~GeV/$c$ ($\bar{p}$).
Therefore, we need RF separators to handle such a high-momentum region, which is essential for the proposed experiment.

\begin{figure}[b]
  \centering
  \includegraphics[width=0.6\textwidth]{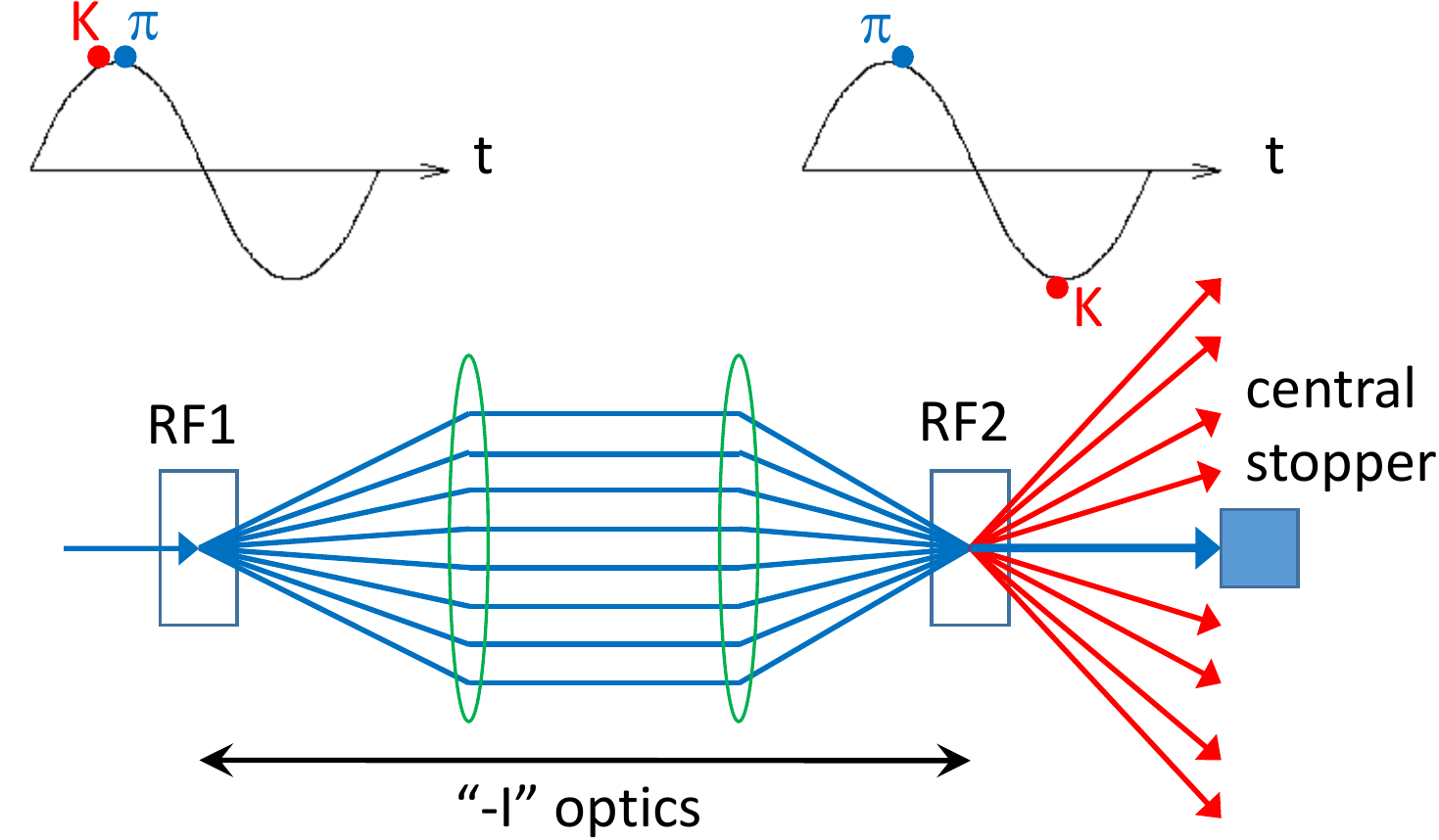}
  \caption{The principle of the particle separation using two RF cavities.}
  \label{fig:RF2Principle}
\end{figure}

The principle of the particle separation by using RF cavities
is schematically shown in Fig.\ref{fig:RF2Principle}.
In this method, two RF cavities (RF1 and RF2) are located in the beam line,
and the optics between the two cavities are set so that
the transport matrix is equal to ``$-I$''.
If the RF amplitude of the two cavities are the same,
the sum of the beam deflection from the two cavities is
\begin{eqnarray}
D &=& -A \sin\left(\omega t\right) + A \sin\left(\omega t + \Delta \phi\right) \\
  &=& 2A \sin\frac{\Delta \phi}{2} \cos\left(\omega t + \frac{\Delta \phi}{2}\right),
\end{eqnarray}
where $\omega t$ is the phase at the first cavity,
and $\Delta \phi$ the phase difference between the two cavities.
The minus sign in the first line comes from the ``$-I$'' optics
between the two cavities.
The amplitude $A$ is given by
\begin{equation}
A = \frac{eEl}{pc\beta},
\end{equation}
where $e$, $p$, and $\beta$ are the charge, momentum, and velocity
of the particle, respectively, and $E$ and $l$ denote
the field gradient and the effective length of the cavity, respectively.
When the phase at the second cavity is set to be same as
that of the first cavity for an unwanted particle, namely $\pi$,
the deflection of the particle is canceled
in whichever phase it passes the first cavity,
and it is absorbed with a central stopper downstream.
On the other hand, the phase of the second cavity for particles
with the other mass and velocity ($K^-$ or $\bar{p}$) differs by
\begin{eqnarray}
\Delta \phi^u_w &=& \frac{2 \pi f L}{c}\left(\frac{1}{\beta_w} - \frac{1}{\beta_u}\right) \\
                &\sim& \frac{\pi f L}{c} \frac{m^2_w - m^2_u}{p^2 c^2},
\end{eqnarray}
they are deflected by $2 A \sin \frac{\Delta \phi^u_w}{2}$ in maximum
depending on the phase at the first cavity,
and pass outside of the central stopper.
Here, $f$ is the RF frequency, $L$ the distance between the two cavities,
$p$ the momentum, and $\beta_w$/$\beta_u$ and $m_w$/$m_u$ are
the velocities and masses of the wanted/unwanted particles, respectively.

\begin{figure}[tbp]
  \centering
  \includegraphics[width=0.8\textwidth]{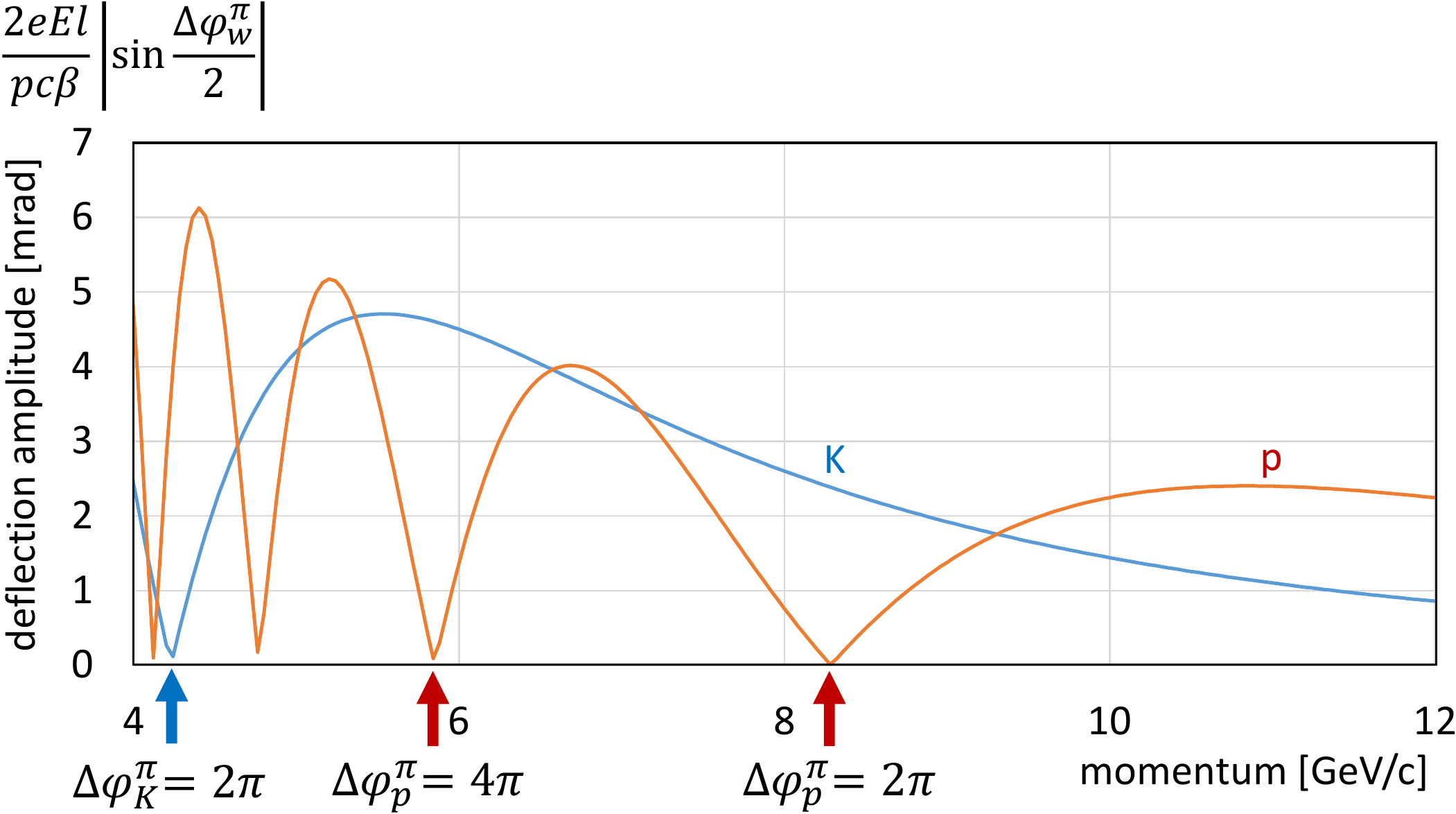}
  \caption{Momentum dependence of the deflection amplitude
           for $K^-$ (blue) and $\bar{p}$ (red) in the case that
           the RF phase of the two cavities are same for $\pi^-$.
           Since the deflection for $\pi^-$ is always canceled,
           usable is the momentum region in which
           the deflection for $K^-$ or $\bar{p}$ is large.}
  \label{fig:RF2Separation}
\end{figure}

For example, assuming $\pi^-$ as an unwanted particle,
the momentum dependence of the deflection amplitude for $K^-$ and $\bar{p}$
in the case of $f = 2.857$~GHz and $L = 16.8$~m is plotted
in Fig.\ref{fig:RF2Separation}.
The deflection for the wanted particle is also canceled
in the momentum range corresponding to
\begin{equation}
\Delta \phi^u_w = 2 n \pi \,\,\, (n = 1, 2, 3, ...),
\end{equation}
whereas the deflection get maximum in the range satisfying
\begin{equation}
\Delta \phi^u_w = (2 n - 1) \pi \,\,\, (n = 1, 2, 3, ...).
\end{equation}

Another operation mode can be considered where
the phase of the second cavity is tuned to be same as that of the first cavity
for a wanted particle ($K^-$ or $\bar{p}$).
Then the deflection of the wanted particle is always canceled,
while the unwanted particle ($\pi^-$) is deflected
depending on the phase difference and is eliminated with a slit.

\begin{figure}[tbp]
  \centering
  \includegraphics[width=\textwidth]{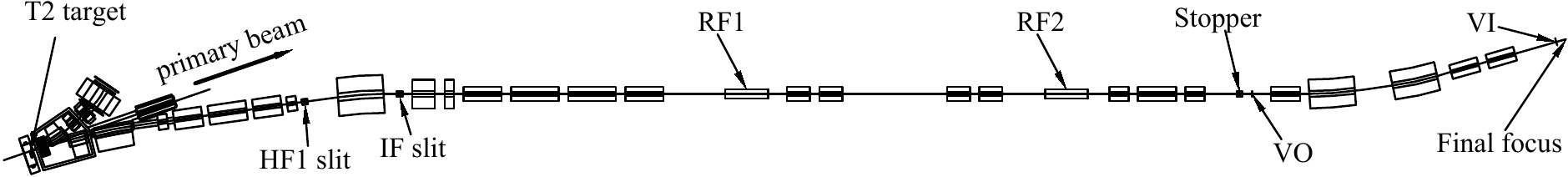}
  \caption{Layout of the RF option of the K10 beam line.}
  \label{fig:K10RFLayout}
\end{figure}
\begin{figure}[tbp]
  \centering
  \includegraphics[angle=-90,width=0.9\textwidth]{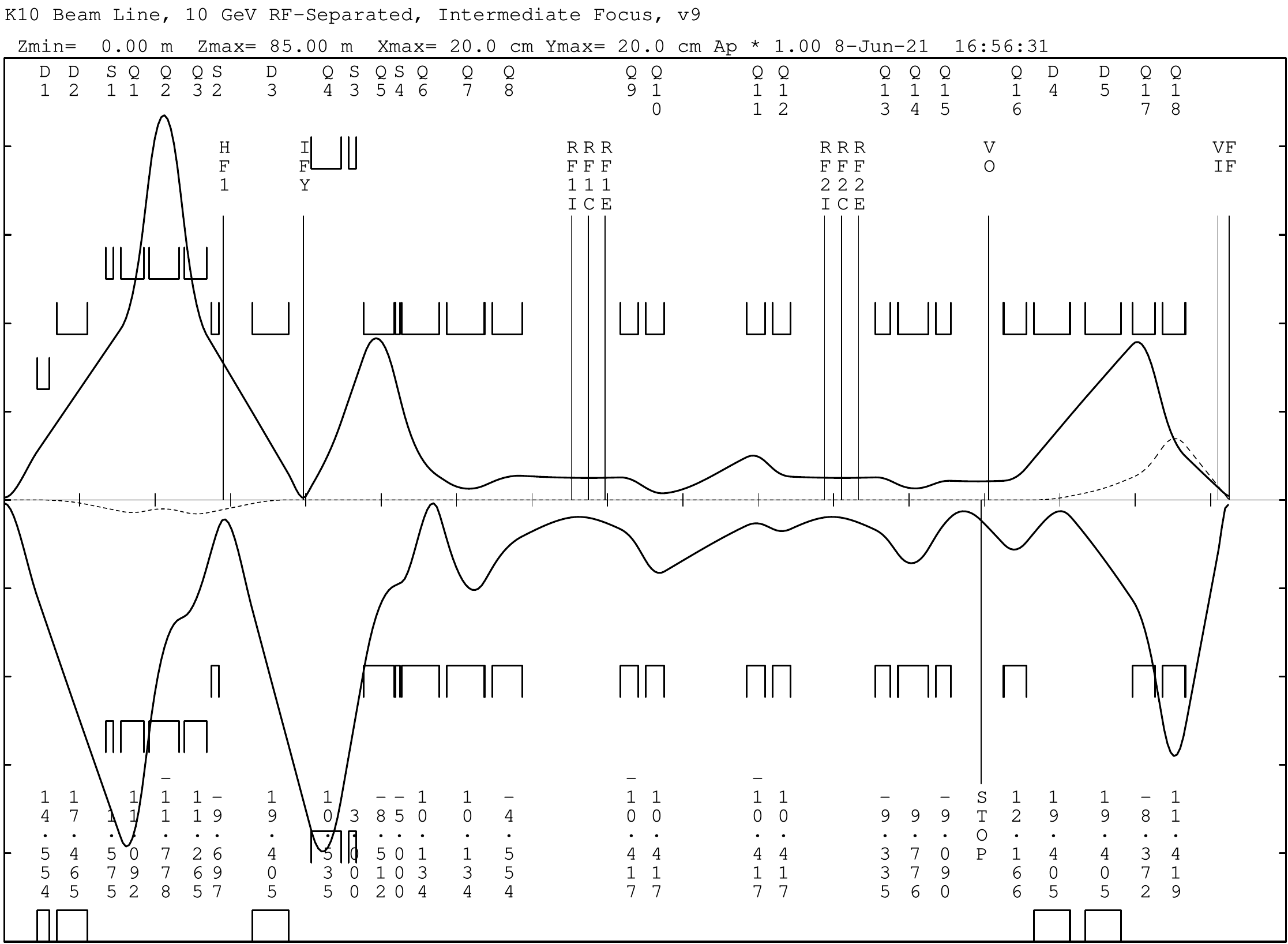}
  %k10-rf-envelope.pdf}
  \caption{First-order beam envelope of the RF option of the K10 beam line.The upper and lower show the vertical and horizontal directions, respectively. The dash line indicate the horizontal dispersion.}
  \label{fig:K10RFEnvelope}
\end{figure}

The layout of the RF-separator option of the K10 beam line is
presented in Fig.\ref{fig:K10RFLayout}, and
the beam envelope calculated with the TRANSPORT code\cite{TRANSPORT}
is shown in Fig.\ref{fig:K10RFEnvelope}.
%The layout and beam envelope of the RF option of the K10 beam line
%are shown in Fig.\ref{fig:K10RFLayout} and Fig.\ref{fig:K10RFEnvelope},
%respectively.
The total length of the beam line is $80.7 \sim 81.2$~m
depending on the configuration of the beam spectrometer.
The RF frequency and the distance between the cavities are
set to 2.857~GHz and 16.8~m, respectively.
The effective length of the cavities is 2.25~m.

The beam line consists of three sections; the front-end section,
the separation section, and the analyzing section.
In the front-end section, secondary beams generated at the production target
are extracted from the primary beam line,
and focused vertically at the intermediate (IF) slit
to reduce so-called ``cloud $\pi$''.
The optics are also tuned to make beams almost achromatic at the IF slit.
In the separation section, the optics is tuned to obtain
a parallel and narrow beam at the RF cavity, and
the transport matrix of ``$-I$'' is realized between the cavities
by using four quadrupoles with same field gradient.
After analyzed with a beam spectrometer,
the beam is focused to an experimental target
both in horizontal and vertical directions.

As for the beam spectrometer, we have two types of the configuration.
Type~I consists of two dipole magnets and three quadrupoles
with the order of QDDQQ, as shown in Fig.~\ref{fig:K10RFEnvelope}.
In the transport matrix of the section from VO to VI,
the condition of the point-to-point focus is satisfied
in the horizontal direction ($R_{12}=0$),
and the magnification ($R_{11}$) and the dispersion ($R_{16}$)
are $-1.627$ and $-0.635$~cm/\%, respectively.
By assuming the position resolution of tracking devices located at VO and VI
is $\sigma_x = 300$~$\mu$m, the expected momentum resolution is
$\sigma_p = \frac{\sqrt{1+R_{11}^2}}{|R_{16}|} \sigma_x = 0.090$~\%.

\begin{figure}[tbp]
  \centering
  \includegraphics[angle=-90,width=0.9\textwidth]{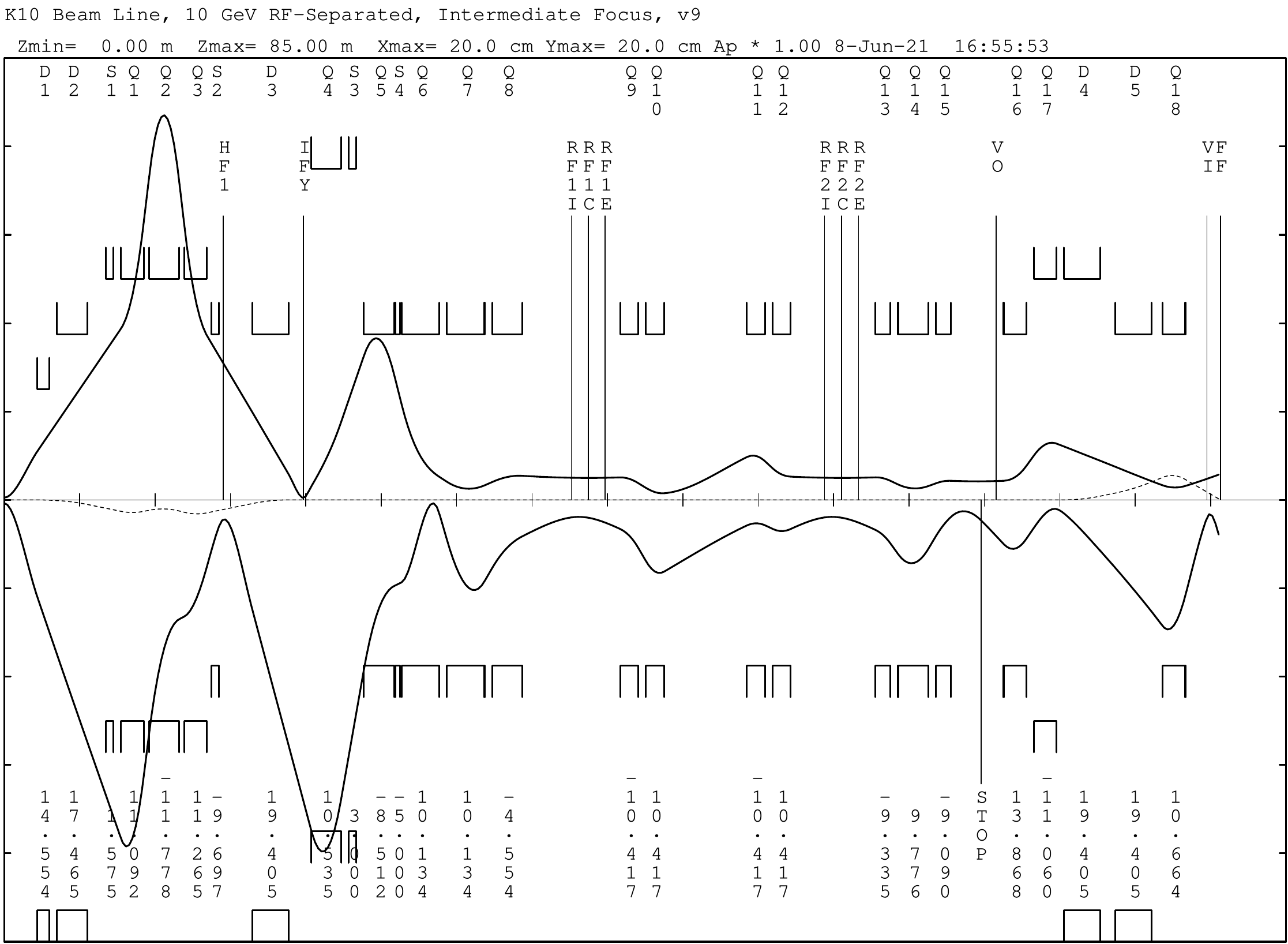}
  %k10-rf-envelope2.pdf}
  \caption{Same as Fig.\ref{fig:K10RFEnvelope} but for Type~II of the beam spectrometer.}
  \label{fig:K10RFEnvelope2}
\end{figure}

Another design, Type~II, has a QQDDQ configuration (Fig.~\ref{fig:K10RFEnvelope2}).
It also realizes the point-to-point focus ($R_{12}=0$) from VO to VI,
and has the magnification of $R_{11} = -0.228$ and
the dispersion of $R_{16} = -0.354$~cm/\%.
The expected momentum resolution, therefore, is 0.087~\%.

Compared with Type~II, the vertical beam size
in the analysing section in Type~I is so large that
it causes a transmission problem when the phase of the second cavity
is tuned to $\pi^-$'s ($\pi$-tuned mode).
Type~I spectrometer, therefore, can be used only with a slit
in a $K$-tuned or a $\bar{p}$-tuned mode.
On the other hand, Type~II can be operated with a central stopper
in a $\pi$-tuned mode as well as a slit in a $K$- or $\bar{p}$-tuned mode.

\begin{table}[tbp]
 \begin{center}
  \caption{Expected $K^-$ intensity per spill and purity ($K^-$:$\pi^-$) of
           the RF option of the K10 beam line.
           The beam loss of 25~kW at the production target and
           the spill repetition of 5.2~s were assumed.
           The production cross-section was calculated
           by using Sanford and Wang formula.
           Decay muons and so-called ``cloud-$\pi$'' were not included.
           Slit conditions were varied to achieve a moderate purity
           for each case.
           The field gradient of RF cavities was also varied
           in the range from 5 to 9~MV/m.}
  \label{tbl:K10RFResult}
  \vspace{5mm}
  \begin{tabular}{cccc} \hline \hline
  & Type~I (QDDQQ) & Type~II (QQDDQ) & Type~II (QQDDQ) \\
  & $K$-tuned mode & $\pi$-tuned mode & $K$-tuned mode \\ \hline
  5 GeV/$c$ & $3.9 \times 10^6$ (1:7.0) & $5.4 \times 10^6$ (1:3.5) & $6.1 \times 10^6$ (1:7.8) \\
  6 GeV/$c$ & $5.2 \times 10^6$ (1:5.0) & $7.3 \times 10^6$ (1:2.6) & $8.4 \times 10^6$ (1:5.4) \\
  7 GeV/$c$ & $6.0 \times 10^6$ (1:4.8) & $8.3 \times 10^6$ (1:2.1) & $9.4 \times 10^6$ (1:5.1) \\
  8 GeV/$c$ & $5.8 \times 10^6$ (1:5.7) & $7.9 \times 10^6$ (1:2.1) & $9.1 \times 10^6$ (1:6.1) \\
  9 GeV/$c$ & $4.9 \times 10^6$ (1:7.3) & $6.7 \times 10^6$ (1:2.1) & $8.0 \times 10^6$ (1:7.6) \\
  10 GeV/$c$ & $4.0 \times 10^6$ (1:9.4) & $4.7 \times 10^6$ (1:2.5) & $6.4 \times 10^6$ (1:10.0) \\ \hline \hline
  \end{tabular}
 \end{center}
\end{table}

\begin{figure}[tbp]
  \centering
  \begin{tabular}{ccc}
  Type~I ($K$-tuned mode) & Type~II ($\pi$-tuned mode) & Type~II ($K$-tuned mode) \\
  \includegraphics[width=0.3\textwidth]{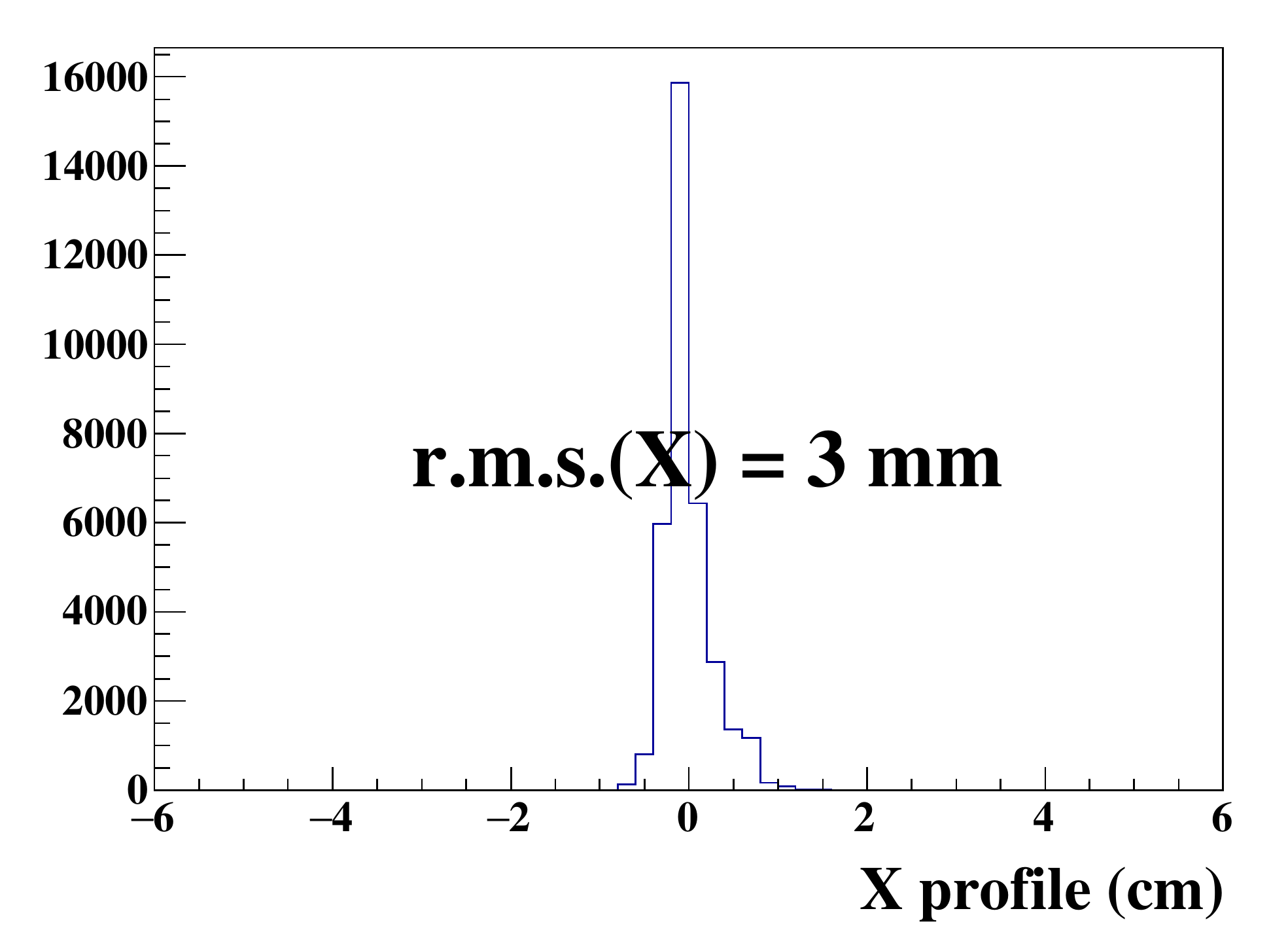} &
  \includegraphics[width=0.3\textwidth]{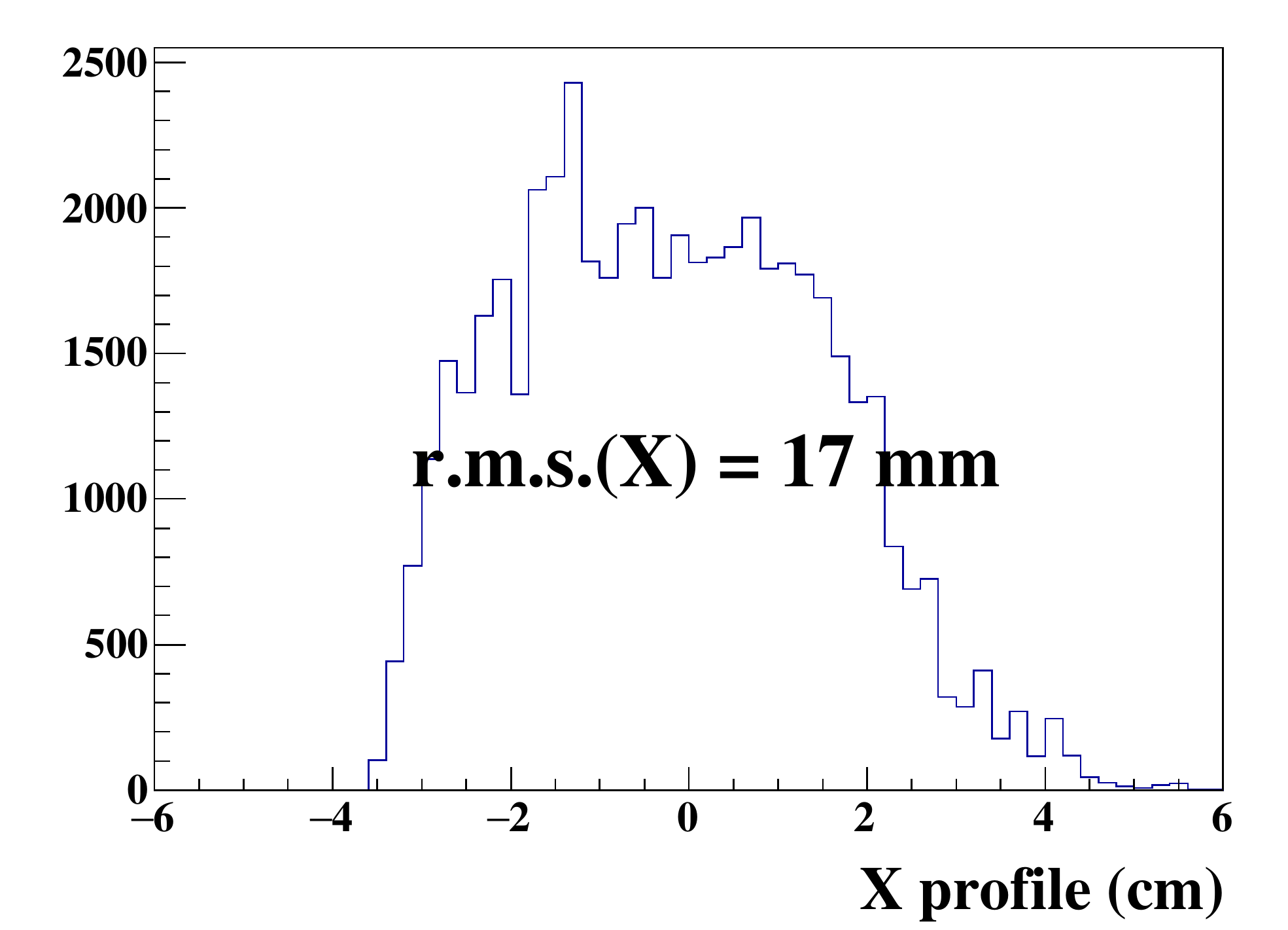} &
  \includegraphics[width=0.3\textwidth]{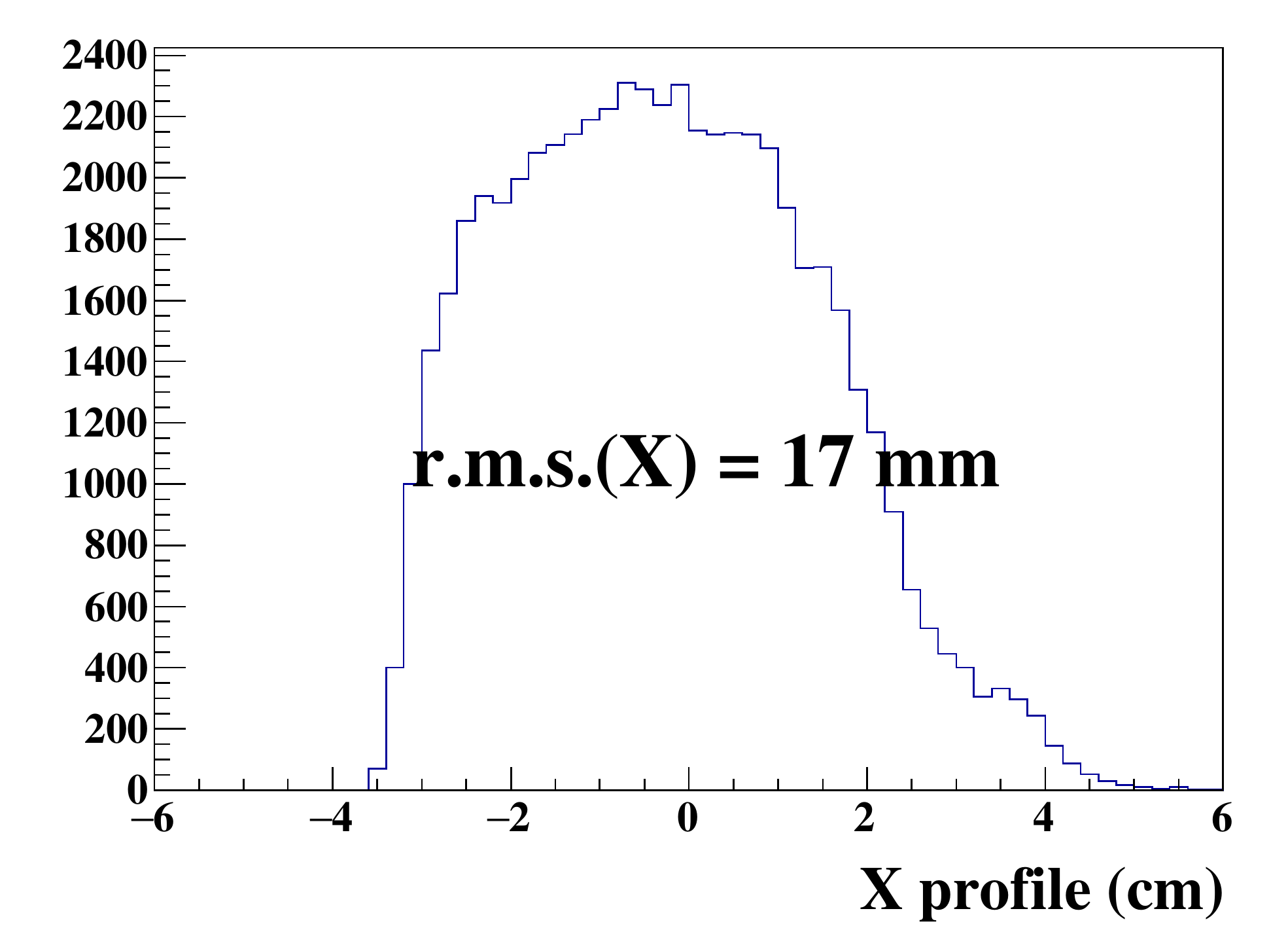} \\
  \includegraphics[width=0.3\textwidth]{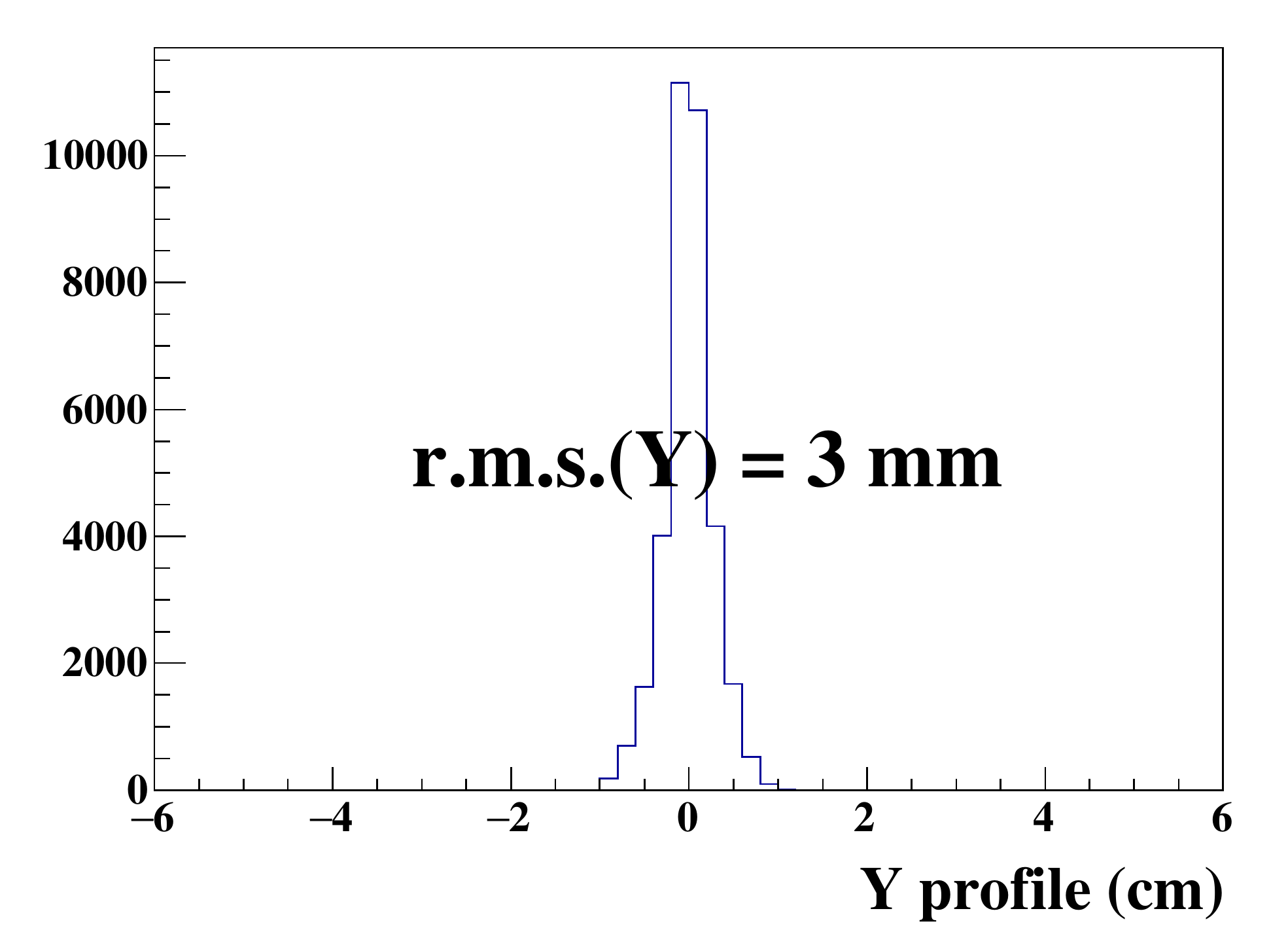} &
  \includegraphics[width=0.3\textwidth]{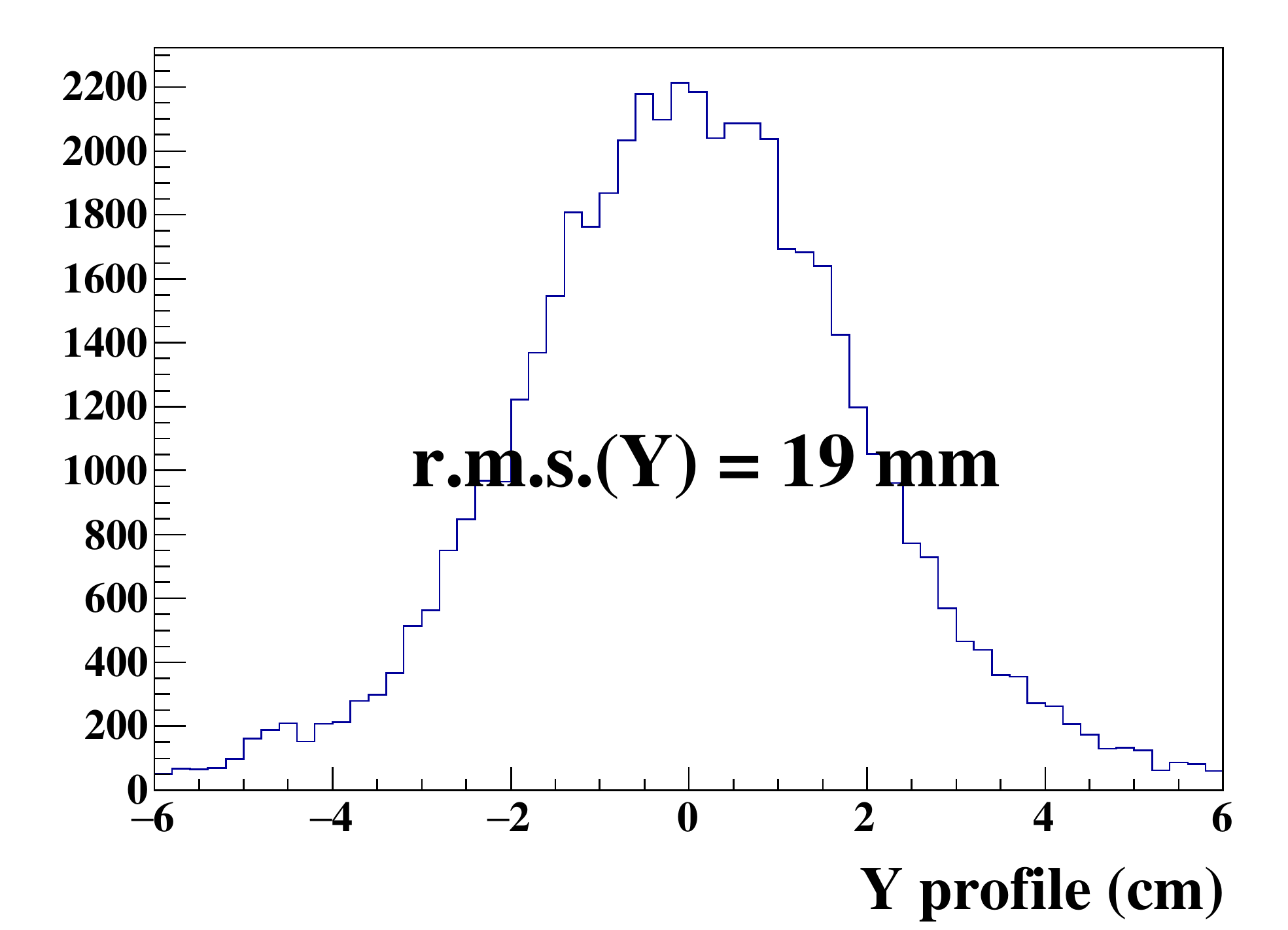} &
  \includegraphics[width=0.3\textwidth]{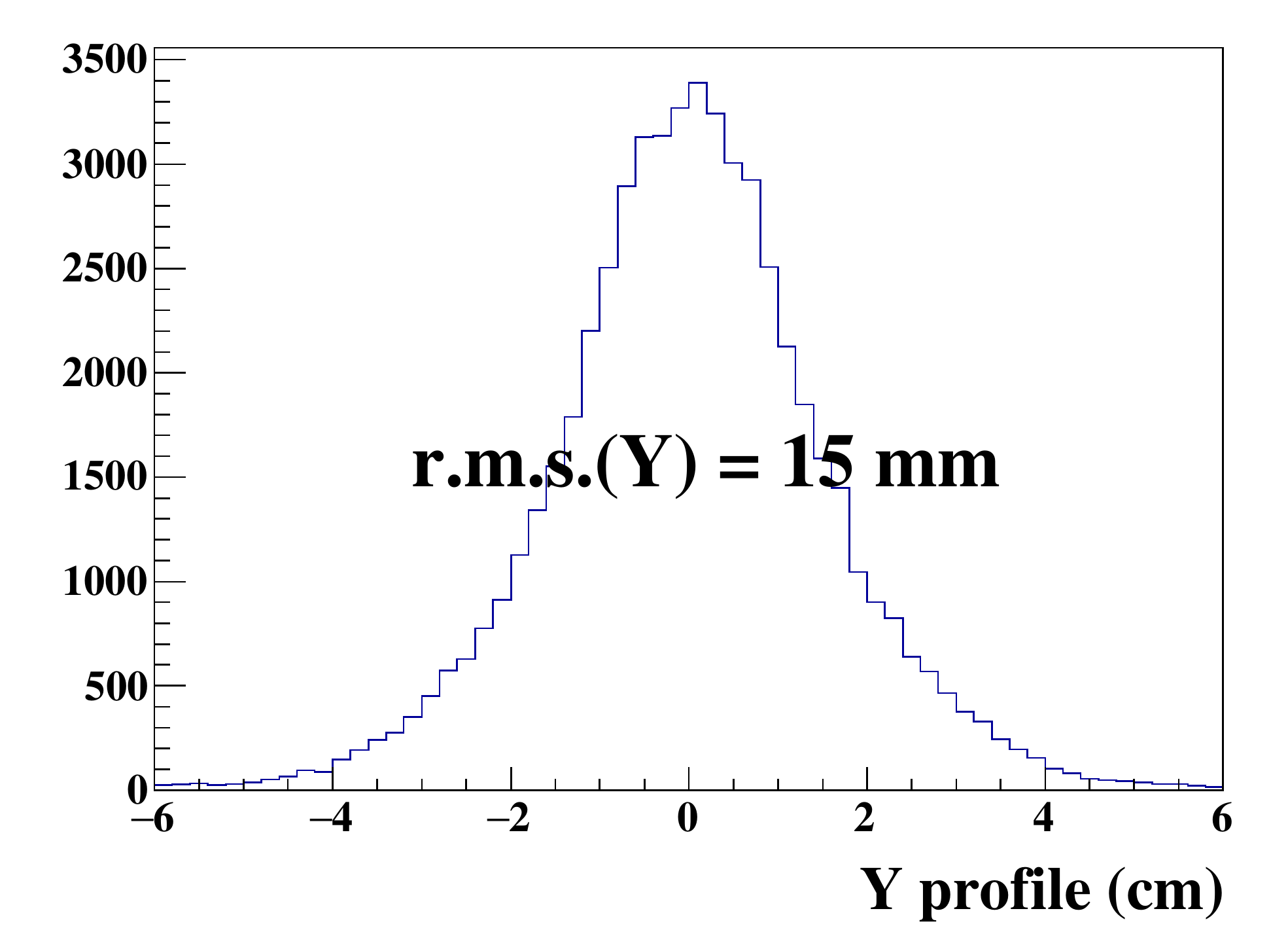}
  \end{tabular}
  \caption{Expected beam profile at the final focusing point for each type of
           the beam spectrometer and the operation mode.}
  \label{fig:K10RFProfileFF}
\end{figure}

Table~\ref{tbl:K10RFResult} lists the result of
the beam simulation using DecayTURTLE\cite{TURTLE}
for each type of the beam spectrometer and the operation mode,
assuming the beam loss of 25~kW at the production target and
the spill repetition of 5.2~s.
The horizontal and vertical beam profiles are shown
in Fig.~\ref{fig:K10RFProfileFF}.
Type~II has an advantage in terms of intensity and purity,
whereas Type~I provides a much smaller beam spot size
at the experimental target.

% flatex input end: [./k10docu/k10beamline.tex]

%%Conceptual design for the K10 beam line
% flatex input: [./k10docu/k10ESseparator.tex]
\subsubsection*{ES-separator option for K10 beam line}

\begin{figure}[hppb]
  \centering
  \includegraphics[width=\textwidth]{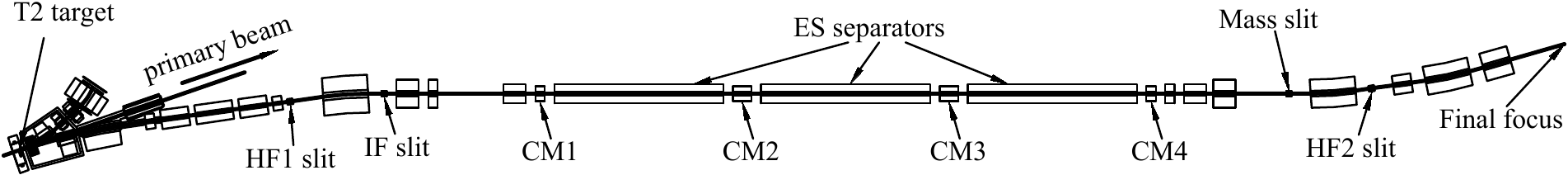}
  \caption{Layout of the ES option of the K10 beam line.}
  \label{fig:K10ESLayout}
\end{figure}

\begin{figure}[bp]
  \centering
  \includegraphics[width=0.9\textwidth]{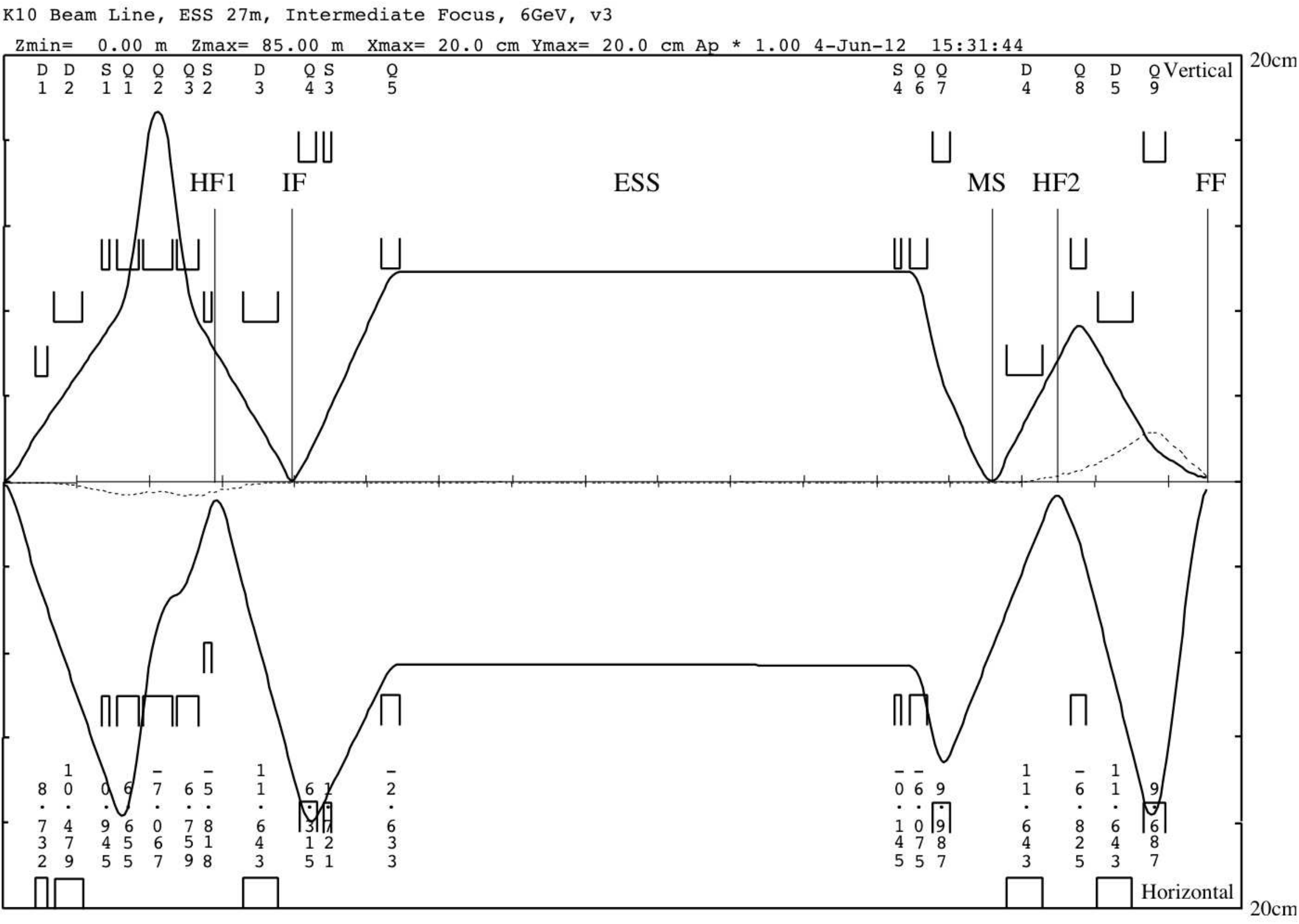}
  \caption{First-order beam envelope of the ES option of the K10 beam line.
           The upper and lower show the vertical and horizontal directions,
           respectively.
           The dash line indicate the horizontal dispersion.}
  \label{fig:K10ESEnvelope}
\end{figure}

The layout of the ES-separator option of the K10 beam line is
presented in Fig.\ref{fig:K10ESLayout}, and
the beam envelope calculated with the TRANSPORT code
is shown in Fig.\ref{fig:K10ESEnvelope}.
The total length of the beam line is 82.8~m,
which is almost same as that of the RF option.
%?»ã?????¼ã?¿ã?????£ã???²ã???????????¼ã??è»??????????»ã?»ã?????¼ã?¿ã???ä¸?æµ????³é????
%ç½???????CM1ï½?CM4??????4?°ã???æ¥µé?»ç??³ã?????£ã???æ­£ã??????
%?¹å???³ª??????£ã??äº?æ¬¡ç?å­???????ä¸?æµ????¹ã??????MS????????????

The front-end section from the production target to the IF is completely same
as that in the RF option.
The separation of the secondary particles is performed
by using three ES separators.
Each of the ES separators have the effective length of 9~m,
the electrode gap of 10~cm, and the electrostatic field of 75~kV/cm.
In the separation section, parallel beams are made
in both the horizontal and vertical directions 
to pass through such long ES separators,
and then the beams are focused vertically at the mass slit (MS).
They are analyzed with the beam spectrometer in the analyzing section,
and finally focused both horizontally and vertically at an experimental target.
In addition, there are two horizontal focal points in the beam line (HF1, HF2),
where slits are located to determine the momentum bite and
to increase the beam purity.

The beam intensity and purity of the beam line was estimated
by using a lay-tracing code DecayTURTLE.
The assumption for the beam loss and spill repetition is
same as that for the RF option.
The result is summarized in Table~\ref{tbl:K10ESResult}.


\begin{table}[hb]
 \begin{center}
  \caption{The expected intensity and purity of the ES option
           of the K10 beam line.
           Assumed condition of the primary protons is same as Table~\ref{tbl:K10RFResult}.}
  \label{tbl:K10ESResult}
  \vspace{1mm}
  \begin{tabular}{cccc} \hline \hline
    & 4 GeV/$c$ $K^-$ & 4 GeV/$c$ $\bar{p}$ & 6 GeV/$c$ $\bar{p}$ \\ \hline
  acceptance [msr-\%] & 0.33 & 1.2 & 0.55 \\
  intensity [/spill] & $1.6 \times 10^6$ & $1.5 \times 10^7$ & $7.5 \times 10^6
$ \\
  purity ($K^-:\pi^-$ or $\bar{p}:\pi^-$) & 1.1:1 & 81:1 & 1:3.4 \\ \hline \hline
  \end{tabular}
 \end{center}
\end{table}

% flatex input end: [./k10docu/k10ESseparator.tex]

%%Conceptual design for the K10 beam line

\clearpage
%%Experiment
\subsection{Baryon-Spectroscopy Experiment at \boldmath$\pi$20 and K10}\label{sec:k10-exp}
% flatex input: [./k10docu_exp/k10-exp-intro_v3.tex]
%%%%%%%%%%%%%%%%%%%%%%%%%%%%%%%%%%%%%%%%%%%%%%%%%%%%%%%%%%%%%%%%%%%%%%%%%%%%%%%%%%%%%

At the High-p ($\pi 20$) and K10 beam lines, 
we conduct spectroscopy of $\Lambda_c$ and $\Sigma_c$ baryons (charmed baryons)
as well as that of $\Xi$ and $\Omega$ baryons (multi-strangeness hyperons).
At $\pi 20$, excited $\Lambda_c$ ($\Sigma_c$) baryons, $\Lambda_c^*$'s ($\Sigma_c^*$'s),
are produced in the two-body $\pi^-\;p\to D^{*-}\; \Lambda_c^{*+}$
($\pi^-\;p\to D^{*-}\; \Sigma_c^{*+}$) reaction,
and they are identified from the $p\left(\pi^-, D^{*-}\right)$ missing mass\footnote{
The reaction of interest is often denoted by $T(I,S)R$
for scattering experiments in Nuclear Physics,
where $I$, $T$, $S$, and $R$ stand for the incident, target, scattering, 
and recoil particles, respectively.
We use this notation also in Hadron Physics
by using $S$ for a set of detected particles, and $R$ a set of undetected particles.
When $R$ is comprised of a single particle,
its mass can be calculated from four-momenta $p$ of $I$, $T$, and $S$:
$
m_R^2 = p_R^2 = \left(p_I+p_T-p_S\right)^2
$.
We call $m_R$  the $T(I,S)$ missing mass.
}.
Similarly, excited $\Xi$ baryons, $\Xi^*$'s, are produced in $K^-\;p\to K^{(*)}\; \Xi^{*}$,
and they are identified from the $p\left(K^-, K^{(*)}\right)$ missing mass.
Only low-lying $\Xi^*$'s are studied at $\pi20$ 
owing to the limited kaon beam intensity, and highly-excited $\Xi^*$'s will be investigated at K10.
At K10, we additionally explore excited $\Omega$ baryons, $\Omega^*$'s, 
in terms of $p(K^-,K^+\;K^{(*)0})$ for $K^-\;p \to \Omega^{*-}\; K^+\; K^{(*)0}$.

A spectrometer system was originally designed for the J-PARC E50 experiment
at $\pi20$ aiming at spectroscopy of $\Lambda_c^*$'s and $\Sigma_c^*$'s~\cite{E50exp}.
This spectrometer was expected to have a wide angular coverage with high 
momentum-resolution, 
and high particle-identification power in a wide momentum range
up to $\sim 20$ GeV$/c$ for particles emitted in experiments at $\pi 20$.
The spectrometer shows high performance not only for spectroscopy of $\Xi^*$'s,
but also for that of $\Omega^*$'s.
The configuration of the spectrometer is described in sub-subsection \ref{sec:k10-spectrometer}.
Expected $p\left(\pi^-, D^{*-}\right)$ missing-mass spectra for $\Lambda_c^*$ and $\Sigma_c^*$ production
are shown in subsubsections \ref{sec:charm-spectroscopy}.
The $p\left(K^-, K^{(*)}\right)$ and $p(K^-,K^+\;K^{(*)0})$ spectra appear 
in sub-subsections \ref{sec:xi-spectroscopy} and \ref{sec:omega-spectroscopy}
for $\Xi^*$ and $\Omega^*$ production, respectively.
Since $\pi^-$-induced $\Omega^*$ production requires more than two final-state particles,
the relevant analyses are more complicated than that for $\Lambda_c^*$'s, $\Sigma_c^*$'s, or $\Xi^*$'s.
The details of the $\Omega^*$-related performance of the spectrometer
are described in sub-subsections~\ref{sec:omega-spectroscopy-suppl1},
\ref{sec:omega-spectroscopy-suppl2}, and \ref{sec:omega-spectroscopy-suppl3} as supplemental information.

%\label{sec:charm-spectroscopy-suppl1}
%\label{sec:xi-spectroscopy-suppl1}
%\label{sec:omega-spectroscopy-suppl1}
%\label{sec:omega-spectroscopy-suppl2}
%\label{sec:omega-spectroscopy-suppl3}

%%%%%%%%%%%%%%%%%%%%%%%%%%%%%%%%%%%%%%%%%%%%%%%%%%%%%%%%%%%%%%%%%%%%%%%%%%%%%%%%%%%%%
% flatex input end: [./k10docu_exp/k10-exp-intro_v3.tex]

%%Experiment
% flatex input: [./k10docu_exp/k10-spectrometer_v7.tex]
%%%%%%%%%%%%%%%%%%%%%%%%%%%%%%%%%%%%%%%%%%%%%%%%%%%%%%%%%%%%%%%%%%%%%%%%%%%%%%%%%%%%%
\subsubsection{Spectrometer system}

\begin{figure}[t]
  \begin{center}
  \includegraphics[width=13.5cm,keepaspectratio,clip]{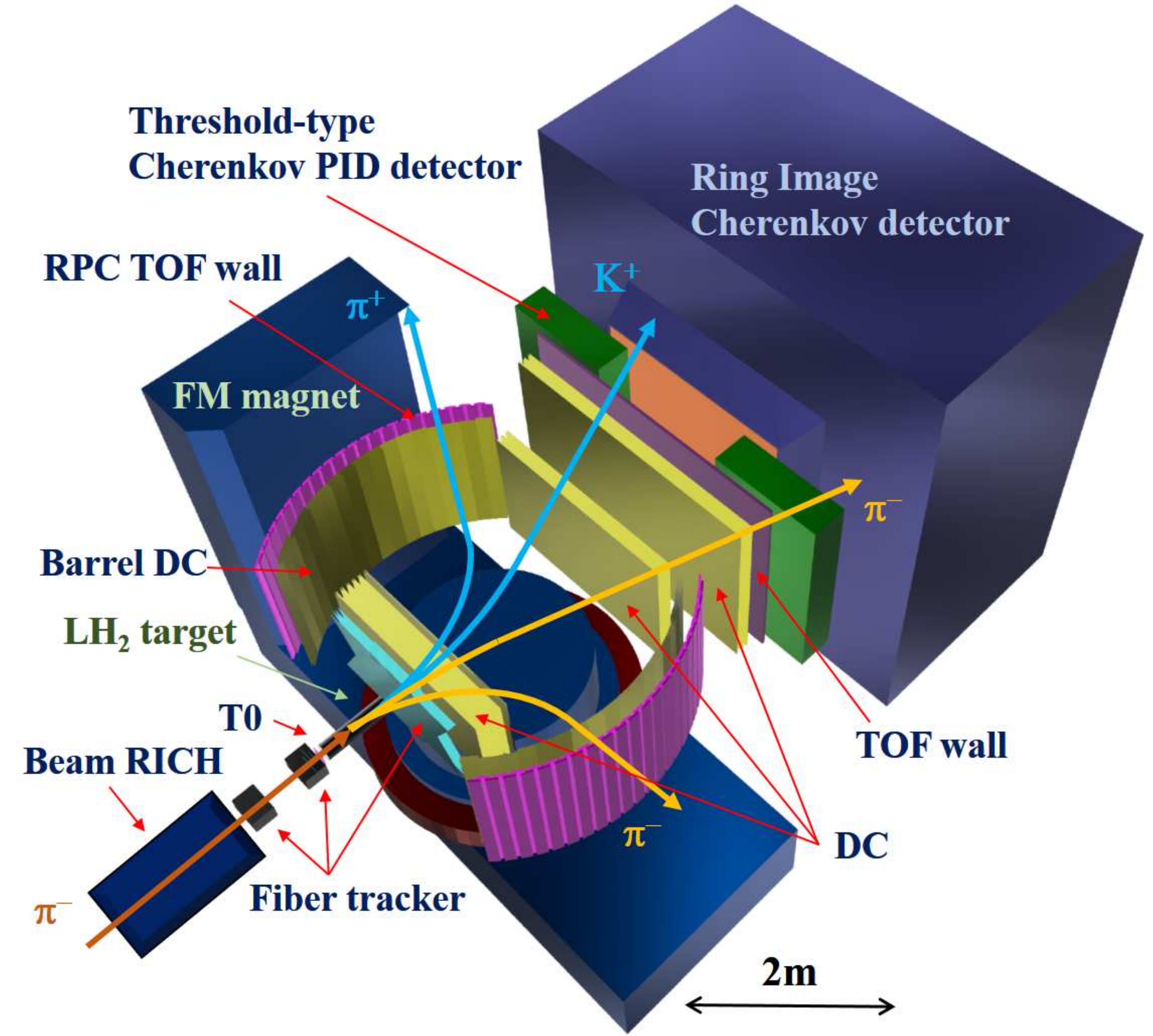}
  \end{center}
  \caption{ 
  Schematic view of the spectrometer to be constructed for the E50 experiment at the $\pi20$ beam line.
  It consists of a large dipole magnet, tracking detectors, time-of-flight detectors, and particle-identification detectors.
  Some of tracking detectors and time-of-flight detectors are placed inside the gap of the magnet (internal detectors).
  The liquid hydrogen target is placed just in front of the entrance of the dipole magnet.
  }
  \label{k10spec_fig1}
\end{figure}

The spectrometer for the E50 physics program~\cite{E50exp} (E50 spectrometer)
satisfies requirements of all the planned spectroscopy experiments at $\pi$20 and K10.
Figure~\ref{k10spec_fig1} shows the schematic view of the E50 spectrometer.
In the E50 experiment, charmed baryons ($Y^{*+}_c$'s: a generic term of $\Lambda_c^*$'s and
$\Sigma_c^*$'s) are produced in the $\pi^{-}\;p \rightarrow D^{*-}\;Y^{*+}_c$ reaction.
The $Y^{*+}_c$ mass is determined from the $p(\pi^-,D^{*-})$ missing mass
by detecting the final-sate $K^+\;\pi^-\;\pi^-$ particles from the $D^{*-}$ decay.
The daughter particles from the $Y^{*+}_c$ decay are also planned to be detected.
The E50 spectrometer is designed for detecting all the charged particles 
produced in the $\pi^{-}\;p \rightarrow D^{*-}\;Y^{*+}_c$ reaction at an incident pion momentum of 20 GeV/$c$.
Emitted at forward angles are high-momentum particles from the $D^{*-}\to\bar{D}^0\pi^-$ decay 
followed by $\bar{D}^0\to K^+\;\pi^-$ as well as daughter particles from the $Y^{*+}_c$ decay.
The momenta of the emitted particles range from 0.2 to 16 GeV$/c$.
The E50 spectrometer has been designed as a multi-purpose system at $\pi20$,
and it can be be also used  for spectroscopy of  $\Xi^*$'s and $\Omega^*$'s at K10.

The E50 spectrometer consists of a large dipole magnet, 
tracking detectors, time-of-flight detectors, and particle-identification detectors.
The stationary target (liquid hydrogen) is placed just in front of the entrance of the dipole magnet.
To determine the $Y_c^{*+}$ mass from the $p(\pi^-,D^{*-})$ missing mass,
it is necessary to measure the four-momenta of the final-sate $K^+\;\pi^-\;\pi^-$ particles.
Since high-momentum $K^+$ and $\pi^-$ from the $\bar{D^0}$ decay are emitted at forward angles,
placed are scintillating-fiber hodoscopes~\cite{aram19}
behind the target, several drift chambers (DC's), and ring-imaging Cherenkov detector (RICH)
for the trajectory determination (momentum analysis), 
time-of-flight (TOF) measurement, and particle identification, respectively.
Additional tracking and time-of-flight detectors are placed inside the gap of the magnet (internal detectors)
to detect the slow $\pi^-$'s emitted at forward angles from the $D^{*-}$ decay (not from the $\bar{D}^0$ decay)
and daughter particles from the $Y^{*+}_c$ decay.
These detectors should have a wide angular coverage
since the slow $\pi^-$'s cannot reach the exit of the dipole magnet and since the daughter particles are emitted in a wide angle range.
Placed in front of the magnet yokes are the barrel-shaped DC's and high time-resolution 
TOF detector wall consisting of resistive-plate chambers (PRCs) 
for detecting the slow $\pi^-$'s and horizontally-emitted daughter particles from the $Y^{*+}_c$ decay.
The RPC's are also placed on the pole pieces for detecting vertically-emitted $Y^{*+}_c$ daughter particles.
The scintillating-fiber hodoscopes and DC's located behind the target are used 
not only for high-momentum $K^+\;\pi^-$ particles at forward angles but also for all the other produced particles.
To get high-counting-rate capability, the scintillating-fiber 
hodoscopes are planned to be used just behind the target.
They are also placed at upstream of the target for measuring the incident momentum and profile, or $(x,y)$ intensity map.
To get the reference timing for all the detectors, a fine-segmented acrylic Cherenkov counter (time-zero counter)~\cite{akai19} 
is placed just in front of the target.
Another RICH counter, Beam RICH~\cite{naru20}, is used for identifying 
incident particles in the secondary beam located at most upstream.
The details of the E50 spectrometer are described elsewhere~\cite{E50spec}.

We plan to use a trigger-less streaming data acquisition (DAQ) system~\cite{E50daq-2, E50daq-1},
which selects the events corresponding to the reaction of interest 
without any hardware trigger (trigger-less).
All the detector signals are digitized in front-end electronics,
and the digitized data fragments are continuously transferred (streaming) to a personal computers (PCs).
The events of interest are selected using software on PC's by combining the streaming data fragments.
In a streaming DAQ system,
a complicated hardware trigger system is not necessary,
which makes a busy time for processing an event corresponding to a trigger.
We are developing a streaming DAQ system primarily for the E50 experiment
as well as special time-to-digital converters (TDC's) 
for a continuous time measurement without any external trigger.
We have tested and successfully demonstrated a prototype streaming DAQ system~\cite{E50daq-1} 
using high-intensity electrons and positrons converted from a bremsstrahlung photon beam 
at the Research Center for Electron Photon Science, Tohoku University~\cite{ishi10}.
The streaming DAQ system allows us to make the DAQ efficiency close to 100\% without any trigger bias.

%%%%%%%%%%%%%%%%%%%%%%%%%%%%%%%%%%%%%%%%%%%%%%%%%%%%%%%%%%%%%%%%%%%%%%%%%%%%%%%%%%%%%

% flatex input end: [./k10docu_exp/k10-spectrometer_v7.tex]
\label{sec:k10-spectrometer}
%%Experiment
% flatex input: [./k10docu_exp/k10-exp-multistrange_v3.tex]
%%%%%%%%%%%%%%%%%%%%%%%%%%%%%%%%%%%%%%%%%%%%%%%%%%%%%%%%%%%%%%%%%%%%%%%%%%%%%%%%%%%%%
\subsubsection*{Requirements for observing $\Xi^*$'s and $\Omega^*$'s}

In spectroscopy of $\Xi^*$'s and $\Omega^*$'s, 
we have to primarily measure the excitation spectrum of $\Xi^{(*)}$'s and $\Omega^{(*)}$'s, 
and determine the angular distribution of a daughter particle from their decay.
$\Xi^{(*)}$'s and $\Omega^{(*)}$'s are produced 
in the $K^{-}\;p \rightarrow K^{(*)}\; \Xi^{(*)}$\footnote{
Note that $K^{*}\;\Xi^{(*)}$ includes
$K^{(*)0}\;\Xi^{(*)0}$ and $K^{(*)+}\;\Xi^{(*)-}$.
} 
and $K^{-}\;p \rightarrow \Omega^{(*)-}\;K^+\;K^{(*)0}$ reactions 
at incident kaon momenta ranging from 5 to 8 GeV$/c$ and 7 to 10 GeV$/c$, respectively.
Since the $\Xi^{(*)}$- and $\Omega^{(*)}$-produced
events are identified by the $p(K^-, K^{(*)})$ and the $p(K^-, K^+K^{(*)0})$ missing mass, respectively.
For detecting the emitted particles 
in association with the hyperon of interest,
high acceptance is required for detecting the final-state $K^+$ and $K^{*0}$, 
$K^+\;K^{*0}$ and $K^+\;K^0$, where
$K^{*0}$ and $K^{0}$ are identified from the $K^+\;\pi^-$ and $\pi^+\;\pi^-$ decays.
Additionally, high momentum-resolution is necessary 
for each of the emitted particles to get enough high $\Xi^*$- and 
$\Omega^*$-mass-resolution (5--7 MeV in $\sigma$)
for observing $\Xi^*$'s and $\Omega^*$'s separately, 
and determining their widths directly from the mass spectra.
The daughter particles from the $\Xi^*$ and $\Omega^*$ decays should be detected 
in a wide angular coverage for determining the spin-parities 
of $\Xi^*$'s and $\Omega^*$'s.

High-intensity high-momentum $K^-$ beam is necessary to produce $\Xi^{(*)}$'s and $\Omega^{(*)}$'s effectively. 
At such incident kaon momenta, 
high-momentum particles are likely to be emitted at forward angles in the laboratory frame when we use a stationary target. 
A spectrometer system with a dipole magnet covering the forward direction is suitable
to detect these particles to get high mass-resolutions for $\Xi^{(*)}$'s and $\Omega^{(*)}$'s. 
The requirements for the spectrometer system are as follows:
\begin{itemize}
 \item wide angular coverage of polar angles up to $30^\circ$ for detecting the final-state $K^+$, $K^+\;\pi^-$, $K^+\;K^+\;\pi^-$ and $K^+\;\pi^+\;\pi^-$ particles,
 \item wide angular coverage of polar angles up to $45^\circ$ for detecting the daughter particles from the $\Xi^*$ and $\Omega^*$ decays,
 \item high momentum-resolution of a few $10^{-3}$ ($\sigma$),
 \item multi-layer tracking system to detect multiple particles, and
 \item high particle-identification power (the efficiency is higher than 97\%, and mis-identification fraction is lower than 1\% for 0.2--8.0 GeV/$c$ particles).
\end{itemize}
Those requirements for spectroscopy of $\Xi^*$'s and $\Omega^*$'s are satisfied by 
the E50 spectrometer.
It is thus a possible candidate of a spectrometer at K10.

%%%%%%%%%%%%%%%%%%%%%%%%%%%%%%%%%%%%%%%%%%%%%%%%%%%%%%%%%%%%%%%%%%%%%%%%%%%%%%%%%%%%%
% flatex input end: [./k10docu_exp/k10-exp-multistrange_v3.tex]

%%Experiment
% flatex input: [./k10docu_exp/k10-charm-spectroscopy_v3.tex]
%%%%%%%%%%%%%%%%%%%%%%%%%%%%%%%%%%%%%%%%%%%%%%%%%%%%%%%%%%%%%%%%%%%%%%%%%%%%%%%%%%%%%
\subsubsection{Charmed baryons ($Y_c^*$: $\Lambda_c^*$ and $\Sigma_c^*$)}

\begin{figure}[t]
  \begin{center}
  \includegraphics[width=15cm,keepaspectratio,clip]{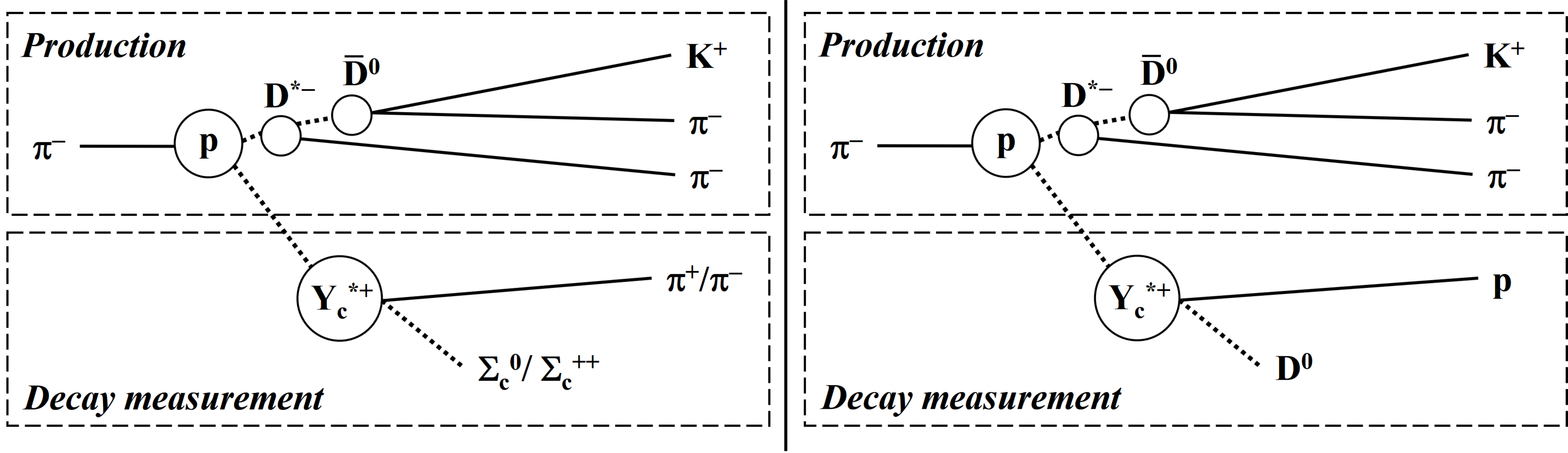}
  \end{center}
  \caption{
  Schematic view of the production mechanisms of $Y_c^*$'s in the $\pi^-\;p$ reaction, and their decay.
  The solid lines represent the initial- and final-state particles to be directly detected. 
  The dotted lines show unstable particles to be reconstructed from kinematic variables of the detected particles.
  $Y_c^*$'s are expected to decay into 
$\Sigma_c^{0}\;\pi^{+}$,
$\Sigma_c^{++}\;\pi^{-}$
 (left), and $p\;D^{0}$ (right).
  }
  \label{k10_charm_fig1}
\end{figure}

At the $\pi 20$ beam line, 
we investigate primarily the masses and decays of excited charmed baryons 
($Y^{*+}_c$'s: a generic term of $\Lambda_c^*$'s and $\Sigma_c^*$'s) 
in the $\pi^-\; p \to D^{*-}\; Y_c^{*+}$ reaction at an incident pion momentum of 20 GeV/$c$.
Figure~\ref{k10_charm_fig1} shows 
the schematic view of the production mechanisms of $Y_c^*$'s in the $\pi^-\; p$ reaction, and their decay.
The decay chain of the $D^{*-}$ meson, $D^{*-} \rightarrow \bar{D}^0\;\pi^{-}$ (branching ratio: 67.7\%) 
and $\bar{D}^0 \rightarrow K^{+}\;\pi^{-}$ (branching ratio: 3.95\%), 
is used for determining the mass of the produced $Y_c^*$'s.
Detected with the E50 spectrometer are $K^{+}$'s and $\pi^{-}$'s with momenta
from 2 to 16 GeV/$c$ from the $\bar{D}^0$ decay 
and $\pi^{-}$'s with momenta from 0.5 to 1.7 GeV/$c$ from the $D^{*-}$ decay.
The $Y_c^*$ mass is determined from the $p\left(\pi^-, D^{*-}\right)$ missing mass (production measurement).
The daughter particles from the $Y^{*+}_c$ decay are also planned to be detected 
(decay measurement)
in the
 $Y^{*+}_c \rightarrow \Sigma_c^{0}\;\pi^{+}$, 
 $Y^{*+}_c \rightarrow \Sigma_c^{++}\;\pi^{-}$, 
and $Y^{*+}_c \rightarrow p \;D^{0}$ channels.
The four momentum of the produced $Y^{*+}_c$ is known after it is identified 
by using the $p(\pi^-, D^{*-})$ missing mass.
The masses of the daughter particles,
$\Sigma_c^{0}$,
$\Sigma_c^{++}$,
and $D^{0}$,
from the $Y_c^{*+}$ decay can be obtained by the 
 $p(\pi^-, D^{*-}\pi^+)$, 
 $p(\pi^-, D^{*-}\pi^-)$, and
 $p(\pi^-, D^{*-}p)$  missing masses, respectively.
Thus, required is only 
detection of $\pi^\pm$ or $p$
from the
 $Y^{*+}_c \rightarrow \Sigma_c^{0}\;\pi^{+}$, 
 $Y^{*+}_c \rightarrow \Sigma_c^{++}\;\pi^{-}$, 
or $Y^{*+}_c \rightarrow p \;D^{0}$ decay
with momenta ranging from 0.2 to 4.0 GeV/$c$.

In the proposal for the E50 experiment~\cite{E50exp},
we show the expected $Y_c^{*+}$-mass spectra estimated by a Monte-Carlo simulation
based on Geant4 with the E50 spectrometer configuration. 
Here, the conditions and results of this simulation are described briefly
since the conditions are common except for the production cross sections of $Y_c^{*+}$'s.
The  estimated cross section of $Y_c^{*+}$ production 
is 10$^{-4}$ smaller in the $\pi^-\; p \to D^{*-}\; Y_c^{*+}$ reaction
than that of hyperon ($Y^*$) production (10--100 $\mu$b for  $\pi^{-}\;p \rightarrow K^{*}\;Y^{*}$).%~\cite{Hosaka}, 
To estimate the  $Y_c^{*+}$ yield,
we assumed 1 nb for the cross section of the ground-state $\Lambda_c^{+}$ baryon,
6.0$\times$10$^7$/spill for the beam intensity, 
and 4 g/cm$^2$  (57 cm) for the thickness of the liquid hydrogen target. 
The acceptance for detecting the daughter particles from the $D^{*-}$ decay 
was found to be 60\%--70\% for the excitation energy up to 1 GeV
by assuming the angular distribution of $D^{*-}$ emission corresponding to the $t$-channel process.
The momentum resolution of 0.2\%($\sigma$) was obtained at 5 GeV/$c$,
giving the invariant-mass resolution of 5.5 MeV($\sigma$) and 0.6 MeV($\sigma$)
for reconstructing $\bar{D}^0$ and $D^{*-}$, respectively.
The mass resolution for the ground-state $\Lambda_c^+$ was  8.3 MeV($\sigma$).
A better mass-resolution is expected for the higher-mass $Y_c^{*+}$'s
since the momenta of the $K^+\;\pi^-\;\pi^-$ particles from the $D^{*-}$ decay are lower for these $Y_c^{*+}$'s.
In the decay measurement, a wide angular coverage of $\pi$ emission
is expected $\cos\theta_{\pi} \ge -0.9$
for the $\Lambda_c(2940)^{+} \rightarrow \Sigma_c(2455)^{0}\;\pi^{+}$
and  $\Lambda_c(2940)^{+} \rightarrow \Sigma_c(2455)^{++}\;\pi^{-}$ decay modes.
It is important to remove the background contribution to determine the mass and 
width of  $Y_c^{*+}$ from the mass spectrum.
The $D^*$ tagging is the most effective method for background reduction in $Y_c^{*+}$ production,
and thus we use $D^{*-}$ detection in the $\pi^-\; p$ reaction.
In $D^{*-}$ tagging, 
we selected the events which satisfy the $K^+\;\pi^-$ and $K^+\;\pi^-\pi^-$ invariant masses were 
the $\bar{D}^0$ and $D^{*-}$ masses ($Q=M_{K^+\pi^-\pi^-}-M_{K^+\pi^-}-M_{\pi^-}$ was used 
instead of the $K^+\;\pi^-\pi^-$ invariant mass in an actual analysis), 
respectively, and
obtained a background reduction factor of $2\times10^6$.
The sensitivity of the production cross section was found to be 0.1--0.2 nb.
The details of the background reduction method are described
in sub-subsection~\ref{sec:charm-spectroscopy-suppl1}.

%%%%%%%%%%%%%%%%%%%%%%%%%%%%%%%%%%%%%%%%%%%%%%%%%%%%%%%%%%%%%%%%%%%%%%%%%%%%%%%%%%%%%
\subsubsection*{Expected  $Y_c^{*+}$-mass spectra}

\begin{figure}[t]
  \begin{center}
  \includegraphics[width=15cm,keepaspectratio,clip]{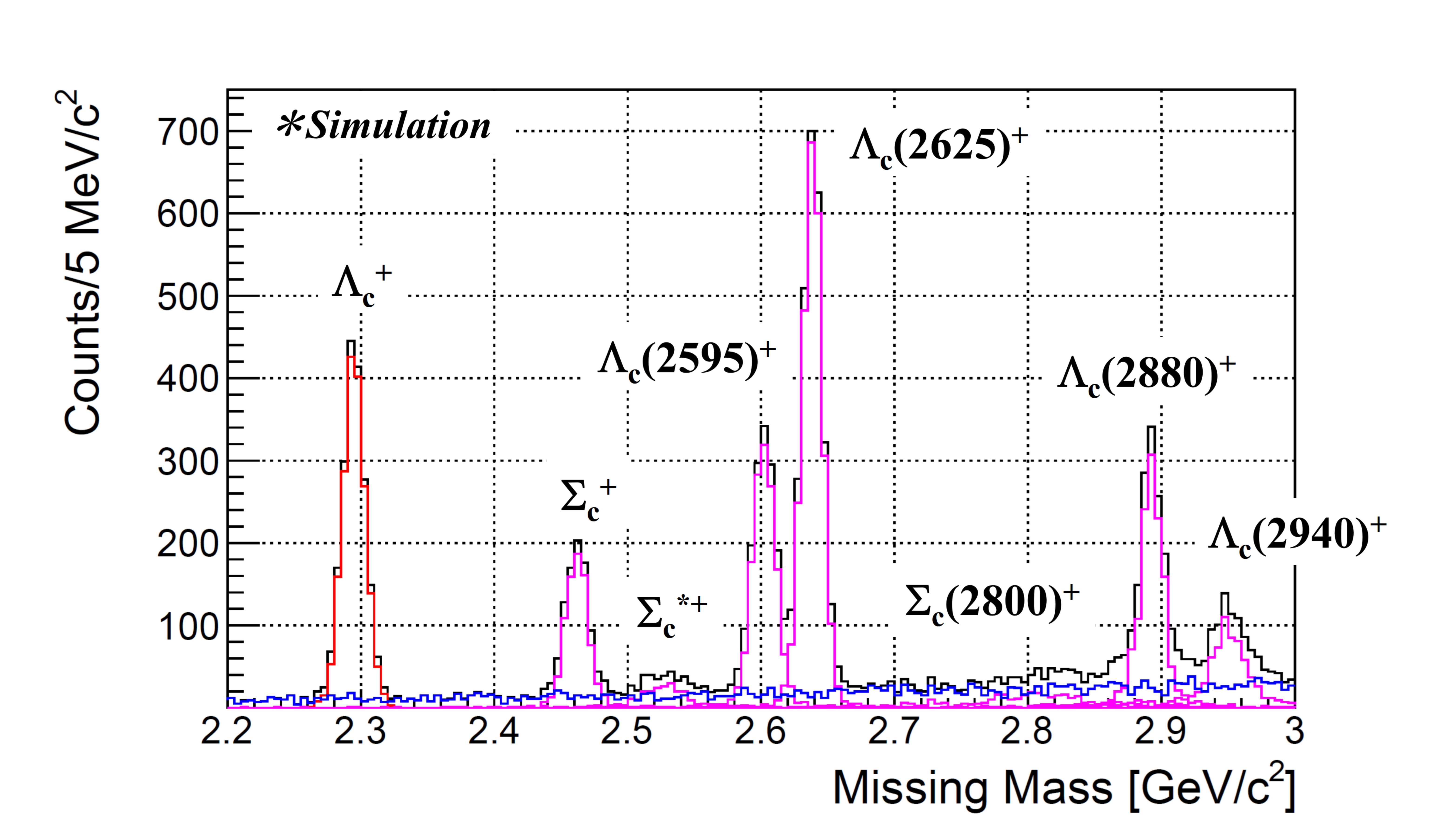}
  \end{center}
  \caption{ Expected $Y_c^*$-mass spectrum including the background contribution 
  at the incident pion momentum of 20 GeV/$c$ in a 100-day beam time. 
  The spectrum is obtained by the $p(\pi^-, D^{*-})$ missing mass for the generated $Y_c^*$-produced events and background processes.
  The background contribution shown in the blue histogram is estimated by JAM~\cite{JAM}.
  It should be noted that the background contribution is highly suppressed owing 
to $D^{*-}$ tagging.
  The contribution from the ground-state $\Lambda_c^+$ is plotted in red, 
  and that from each excited state ($\Lambda_c^*$ or $\Sigma_c^*$) is represented in magenta.
  }
  \label{k10_charm_fig2}
\end{figure}

We show here the most realistic mass spectra for $Y_c^{*+}$'s,
which are different from those in the proposal for the E50 experiment~\cite{E50exp}
since we have much progress on the theoretical studies for  $Y_c^{*+}$ production.
Figure~\ref{k10_charm_fig2} shows the expected $Y_c^*$-mass spectrum including the background contribution.
The incident pion momentum is 20 GeV/$c$, and a 100-day beam time is assumed to obtain this mass spectrum.

The production rates of $Y_c^{*+}$'s  in $\pi^-\; p \to D^{*-}\; Y_c^{*+}$ were originally
estimated using a Regge model where vector-meson ($D^*$) exchange dominated (one-quark process)~\cite{shkim14}.
In this estimation, a $\lambda$-mode excitation was likely to take place, and $\Sigma_c^*$'s were hardly observed.
The relative production rates of 
$\Lambda_c(2286)1/2^+$,
$\Sigma_c(2455)1/2^+$, and 
$\Sigma_c(2520)3/2^+$ were 1.00, 0.03, and 0.17, respectively.
Recently, a new reaction mechanism was proposed which considered diquark correlation in baryons
(two-quark process)~\cite{shim2}. This new mechanism caused both the $\lambda$- and $\rho$-mode excitations,
In this case, the relative production rates became
1.0, 2.9, and 0 for $\Lambda_c(2286)1/2^+$, $\Sigma_c(2455)1/2^+$, and  $\Sigma_c(2520)3/2^+$, respectively.
Since both the mechanisms give the extreme relative ratio of $\Sigma_c(2455)1/2^+$ to the ground-state $\Lambda_c(2285)1/2^+$,
we consider both the one- and two-quark processes are mixed in the real situation.
Here, the mixing ratio is determined so that the production rate
of  the ground-state $\Lambda$ is twice as much as that of the ground-state $\Sigma$
to reproduce the experimental data in the strangeness sector~\cite{Crennell:1972km}.

In Fig.~\ref{k10_charm_fig2},
the yield for each $Y_c^*$ state is obtained using the corresponding relative production rate.
The production cross section for the ground-state $\Lambda_c(2286)1/2^+$ is fixed at 1 nb.
The number of events expected for $\Lambda_c(2285)1/2^+$ production is 1,600 in a 100-day beam time.
It should be noted that the mass resolution is $\sim 8$ MeV ($\sigma$) for $Y_c^*$'s.
The background contribution is estimated by using the JAM code~\cite{JAM}.
We expect to observe not only $\Lambda_c^{(*)}$'s but also $\Sigma_c^{(*)}$'s.
The two heavy-quark spin doublets would appear corresponding to the $P$- and $D$-wave $\lambda$-mode excitations,
$\Lambda_c(2595)1/2^-$-$\Lambda_c(2625)3/2^-$ and $\Lambda_c(2880)5/2^+$-$\Lambda_c(2940)3/2^+$, respectively.
We can assign the excitation mode ($\lambda$ or $\rho$) for each $Y_c^*$
from the production measurement.

In addition, the differential cross section ${d\sigma}/{dt}$ corresponding to the 
transition form factor can be obtained by analyzing the events that $D^{*-}$'s are emitted at
finite angles.
The slope parameter $b$ in a form ${d\sigma}/{dt}\propto e^{-bt}$
can be measured with a statistical error of $\sim$5\% from 1,600 events.
The details of determining the differential cross section  as a function of $t$  are described in
 sub-subsection~\ref{sec:charm-spectroscopy-suppl1}.

%%%%%%%%%%%%%%%%%%%%%%%%%%%%%%%%%%%%%%%%%%%%%%%%%%%%%%%%%%%%%%%%%%%%%%%%%%%%%%%%%%%%%
\subsubsection*{Mass spectra of daughter particles from the $Y_c^{*+}$ decay}

\begin{figure}[t]
  \begin{center}
  \includegraphics[width=15cm,keepaspectratio,clip]{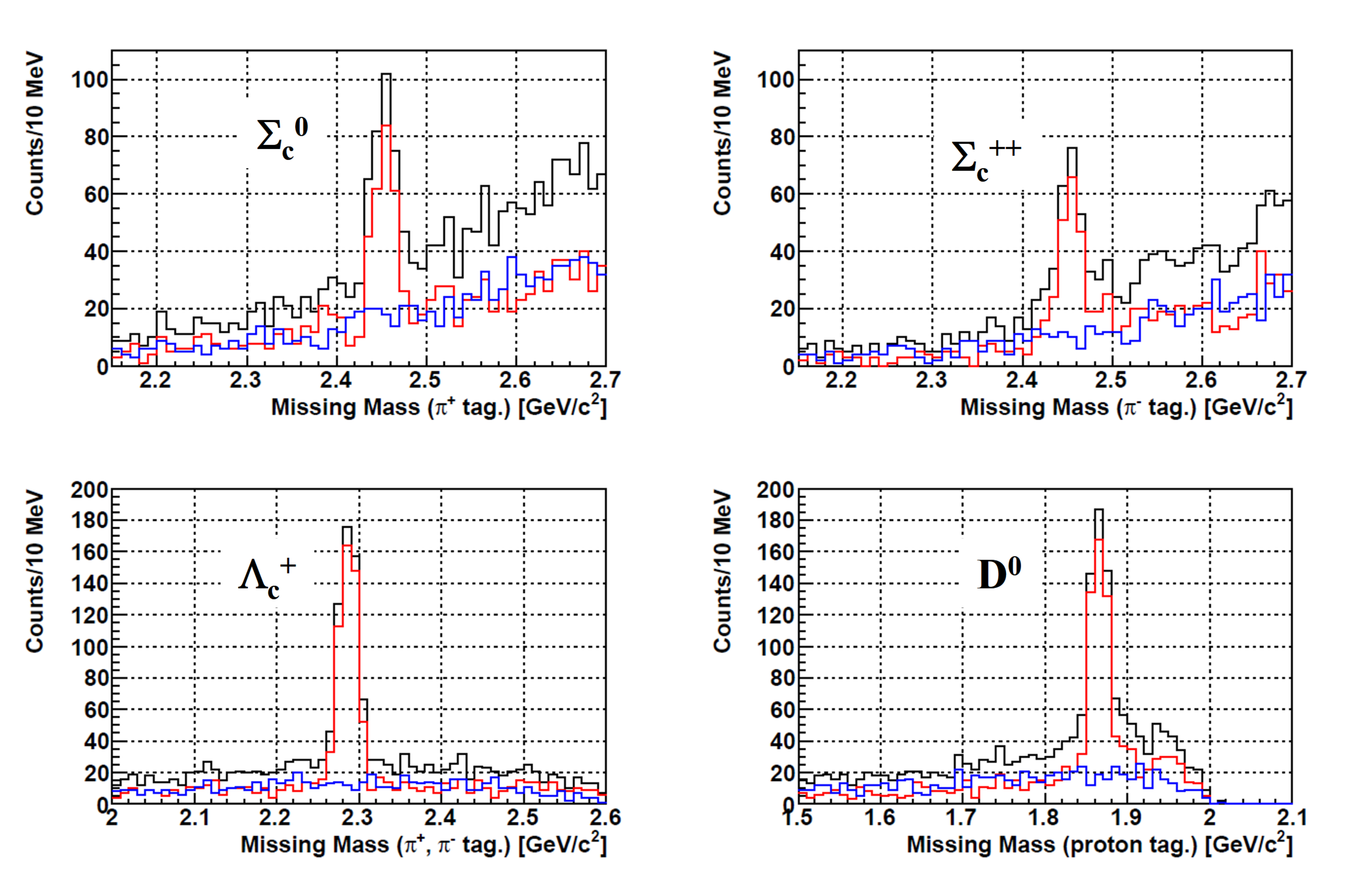}
  \end{center}
  \caption{Mass spectra for the daughter charmed baryons and $D^0$ meson 
  from the $\Lambda_{c}(2940)^+(\Lambda_{c}^{*+})$ 
  including the background contributions:
$(\Lambda_{c}^{*+},\pi^+)$ missing-mass spectrum for the  $\Lambda_{c}^{*+}\to \Sigma_c(2445)^{0}\;\pi^{+}$  decay (a);
 $(\Lambda_{c}^{*+},\pi^-)$ spectrum for $\Lambda_{c}^{*+}\to \Sigma_c(2445)^{++}\;\pi^{-}$ (b);
$(\Lambda_{c}^{*+},\pi^+\pi^-)$ spectrum for $\Lambda_{c}^{*+}\to \Lambda_c^{+}\;\pi^{+}\;\pi^{-}$ (c);
and 
$(\Lambda_{c}^{*+},p)$ spectrum for $\Lambda_{c}^{*+}\to p\; D^0$ (d).
The black spectra correspond to all the events,
the red the $Y_c^{*+}$-produced events,
and the blue the hyperon-produced events.
All the events satisfy the $D^{*-}$-tagging condition,
and all the spectra include the combinatorial background.
Peaks are formed 
only when 
the associated daughter particle(s) from $Y_c^{*+}$ written in red
is (are) correctly selected.
}
  \label{k10_charm_fig3}
\end{figure}

The decay measurement is necessary to investigate the structure of $Y_c^*$'s.
The branching ratio between $\Sigma_c\;\pi$ and $p\;D^0$ is important to assign the $\lambda/\rho$
 excitation mode for revealing  diquark correlation.
The four-momentum of the produced $Y^{*+}_c$ is known after it is identified 
by using the $p(\pi^-, D^{*-})$ missing mass.
The masses of the daughter particles,
$\Sigma_c^{0}$,
$\Sigma_c^{++}$,
and $D^{0}$
from the $Y_c^{*+}$ decay can be obtained only by detecting associated
$\pi^+$,
$\pi^-$, and
$p$,
 respectively.
The mass spectra for the daughter charmed baryons and $D^0$ meson from the $\Lambda_{c}(2940)^+$ decay were
estimated 
for some decay channels:
$\Sigma_c(2445)^{0}\;\pi^{+}$, $\Sigma_c(2445)^{++}\;\pi^{-}$, $\Lambda_c^{+}\;\pi^{+}\;\pi^{-}$ and $p\;D^{0}$.
Here, the $\Lambda_{c}(2940)^+$-produced events were selected
within a mass window $\pm30$ MeV.
Each of the mass spectra for the daughter particles has a clear peak as shown in Fig.~\ref{k10_charm_fig3}.
Here, assumed were the branching ratios of
$\mathcal{B}(\Lambda_{c}(2940)^+ \to \Sigma_c(2445)^{0} \pi^{+})=0.13$, 
$\mathcal{B}(\Lambda_{c}(2940)^+ \to \Sigma_c(2445)^{++} \pi^{-})=0.13$, 
$\mathcal{B}(\Lambda_{c}(2940)^+ \to \Lambda_c^+ \pi^+ \pi^-)=0.10$
and $\mathcal{B}(\Lambda_{c}(2940)^+ \to p D^0)=0.20$.
The source of the background contribution,
which satisfied the $D^{*-}$-tagging condition,
was a combinatorial background (wrong combinations of the daughter particles),
strangeness production,
and wrong particle identification.
In Fig.~\ref{k10_charm_fig3},
we can clearly observe peaks for the daughter particles $\Sigma_c^0$, $\Sigma_c^{++}$,
$\Lambda_c^{+}$, and $D^-$ from $Y_c^{*+}$,
and these peaks can be  formed 
only when the associated daughter $\pi^+$, $\pi^-$, $\pi^+\pi^-$,
and $p$ particles are correctly selected, respectively.
Thus,
the absolute branching ratios can be determined from the decay measurements. 
We can assign the $\lambda/\rho$ mode excitation for each produced $Y_c^{*+}$
by combining both the production rate and absolute branching ratio.

%%%%%%%%%%%%%%%%%%%%%%%%%%%%%%%%%%%%%%%%%%%%%%%%%%%%%%%%%%%%%%%%%%%%%%%%%%%%%%%%%%%%%

% flatex input end: [./k10docu_exp/k10-charm-spectroscopy_v3.tex]
\label{sec:charm-spectroscopy}
%%Experiment
% flatex input: [./k10docu_exp/k10-xi-spectroscopy_v4.tex]
%%%%%%%%%%%%%%%%%%%%%%%%%%%%%%%%%%%%%%%%%%%%%%%%%%%%%%%%%%%%%%%%%%%%%%%%%%%%%%%%%%%%%
\subsubsection{$\Xi$ baryons}

\begin{figure}[t]
  \begin{center}
  \includegraphics[width=15cm,keepaspectratio,clip]{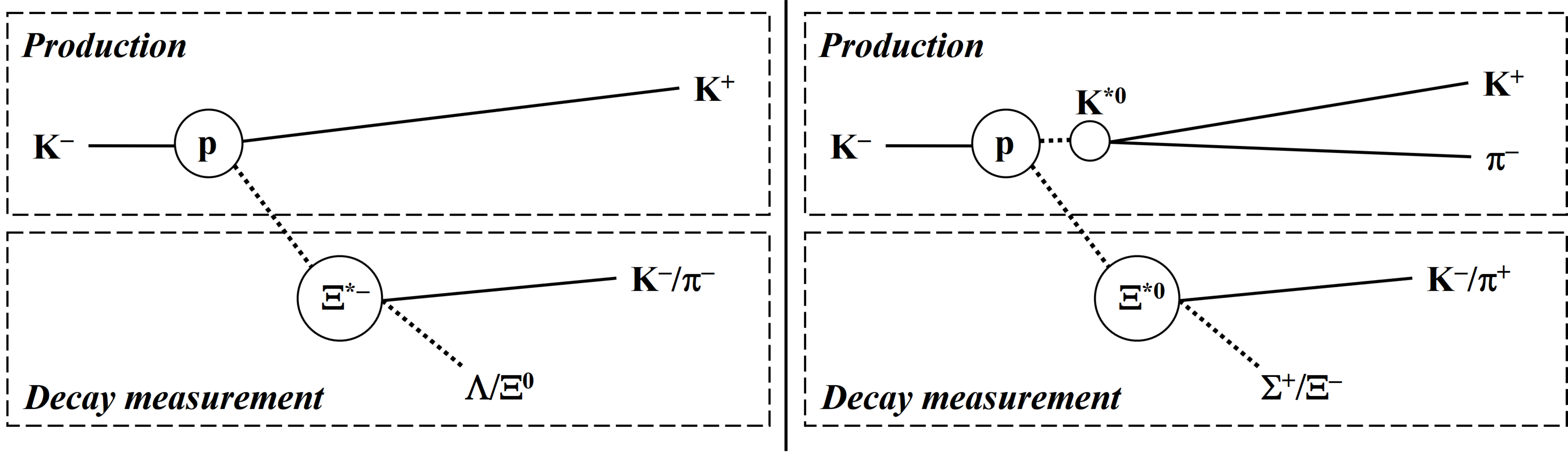}
  \end{center}
  \caption{
  Schematic view of the production mechanisms of $\Xi^*$'s in the $K^-\;p$ reaction, and their decay.
  The solid lines represent the initial- and final-state particles to be directly detected. 
  The dotted lines show unstable particles to be reconstructed from kinematic variables of the detected particles.
  $\Xi^*$'s are expected to decay into $\Lambda\;K^-$ and $\Xi^0\;\pi^-$(left), 
  and $\Sigma^+\;K^-$ and $\Xi^-\;\pi^+$ (right).
  }
  \label{k10_xi_fig1}
\end{figure}

At the $\pi20$ beam line,
we also plan to investigate the masses and decays of 
excited $\Xi$ baryons ($\Xi^*$'s) to study $su$- and $ds$-diquark correlation
in the $K^{-}\;p \rightarrow K^+\; \Xi^{*-}$ 
and $K^{-}\;p \rightarrow K^{0}\;\Xi^{*0}$ reactions at incident kaon momenta ranging from 5 to 8 GeV$/c$.
The $K^-$ beam intensity is not sufficient 
for studying the high-mass $\Xi^*$'s
since $\pi 20$ provides unseparated secondary particles which contain a small fraction of negative kaons.
Thus, construction of the K10 beam line is desired,
and highly-excited $\Xi^*$'s are intensively investigated at K10.
Figure~\ref{k10_xi_fig1} shows 
the schematic view of the production mechanisms of $\Xi^*$'s in the $K^-\;p$ reaction, and their decay.
The mass of a produced $\Xi^*$ including the ground-state $\Xi$ ($\Xi^{(*)}$)
can be determined in a missing-mass technique
using the four-momenta of the initial-state $K^-\;p$ and final-state $K^+$ and $K^+\;\pi^-$
in the $K^{-}\;p \rightarrow K^+\; \Xi^{(*)-}$ and $K^{-}\;p \rightarrow K^{0}\; \Xi^{(*)0}$ reactions, respectively.
Here, the selected events are those in which the $K^+\;\pi^-$ invariant mass should give the $K^{*0}$ mass.
Since the four-momentum of the produced $\Xi^*$ is already given in 
the missing-mass technique,
the dominant decay modes 
$\Xi^{*-} \rightarrow \Lambda\;K^-$, 
$\Xi^{*0} \rightarrow \Sigma^{+}\;K^-$,
$\Xi^{*-} \rightarrow \Xi^{0}\;\pi^{-}$,
and 
$\Xi^{*0} \rightarrow \Xi^{-}\;\pi^{+}$
can be identified only by detecting the emitted $K^-$ and $\pi^{\pm}$
and by calculating the masses of the associated 
daughter $\Lambda$, $\Sigma$ and $\Xi$ particles in a similar missing-mass technique.
Both the production and decay measurements of $\Xi^*$'s are essential in their systematic studies.

%%%%%%%%%%%%%%%%%%%%%%%%%%%%%%%%%%%%%%%%%%%%%%%%%%%%%%%%%%%%%%%%%%%%%%%%%%%%%%%%%%%%%
\subsubsection*{Expected $\Xi^*$-mass spectra}

\begin{figure}[t]
  \begin{center}
  \includegraphics[width=15cm,keepaspectratio,clip]{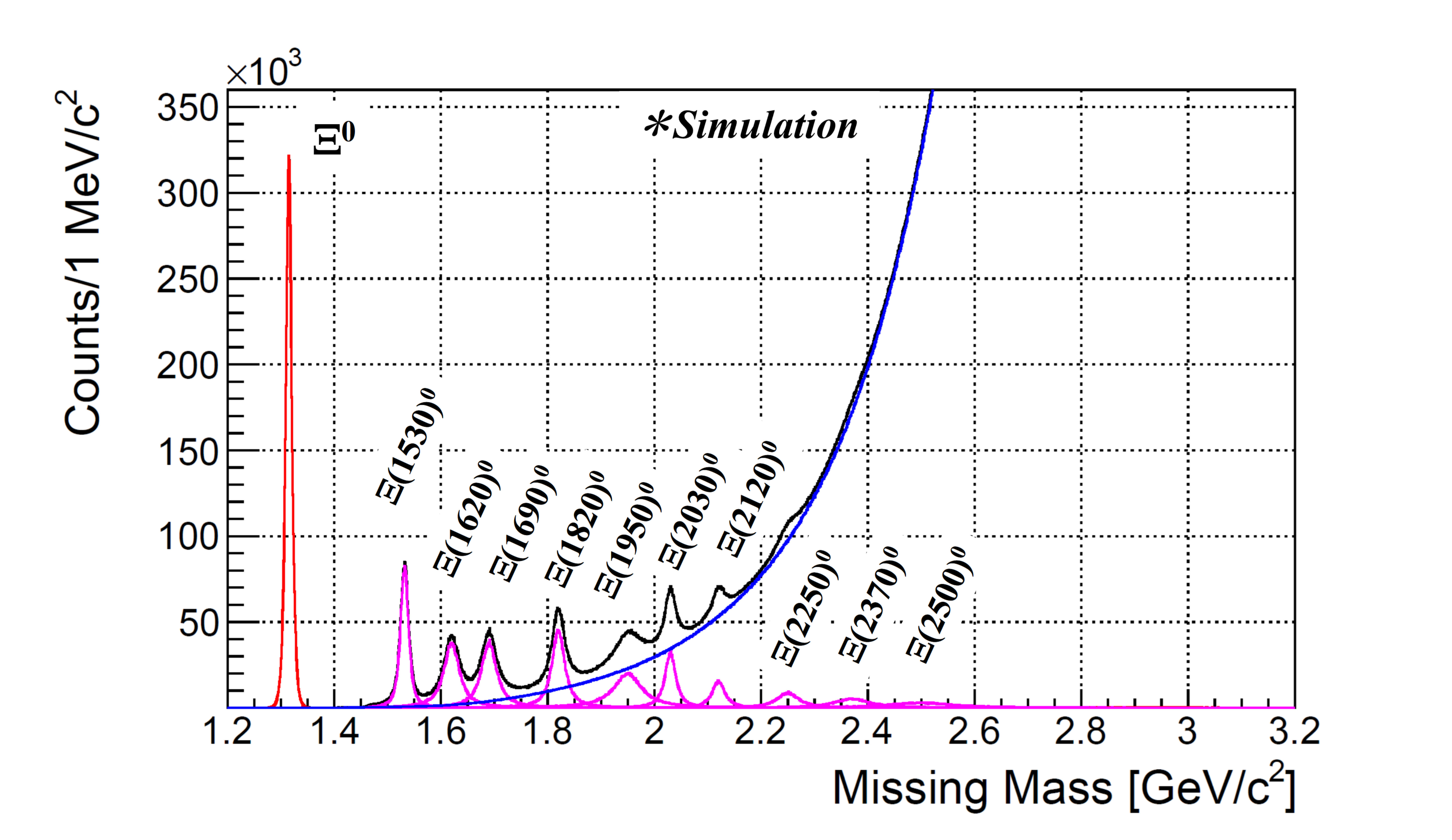}
  \end{center}
  \caption{ Expected $\Xi^*$-mass spectrum including the background contribution 
  at an incident kaon momentum of 8 GeV/$c$ in a 30-day beam time. 
  The spectrum is obtained by the $p(K^-, K^{*0})$ missing mass 
  for the generated $\Xi^*$-produced events and background processes. 
  The background contribution shown in the blue histogram is estimated by JAM~\cite{JAM}.
  The contribution from the ground-state $\Xi^0$ is plotted in red, 
  and that from each excited state ($\Xi^{*0}$) is represented in magenta.
  }
  \label{k10_xi_fig2}
\end{figure}

We will investigate the masses and decays of 
$\Xi^*$'s in the $K^-\;p \to K^{*0}\;\Xi^{*0}$
reaction at incident kaon momenta ranging from 5 to 8 GeV/$c$.
The $\Xi^*$ mass is determined by a $p(K^-, K^{*0})$ missing mass.
Figure~\ref{k10_xi_fig2} shows the expected $\Xi^*$-mass spectrum including 
the background contribution.
Here, a one-day beam time is assumed at an incident kaon momentum of 8 GeV/$c$.
In the simulation, we took the masses and widths of $\Xi^*$'s
described in the Review of Particle Physics
%~\cite{ref:PDG}.
~\cite{Zyla:2020zbs}
In Fig.~\ref{k10_xi_fig2},
the total cross section of 2 $\mu$b was assumed for the ground-state $\Xi$ production
in the $K^-\;p \to K^{*0}\;\Xi^0$ at the incident momentum of 8 GeV$/c$.
The cross section of the ground-state $\Xi$ production
was 7.2 $\mu$b  
in $K^-\;p \to K^{-}\;\Xi^-$ at an incident momentum of 5 GeV$/c$~\cite{ref:Jenkins}.
We adopted a smaller cross section of 2 $\mu$b since
the reaction of interest ($K^-\;p \to K^{*0}\;\Xi^0$) is different 
and incident momentum (8 GeV$/c$) is also different from the available data.
We obtained the production cross section for the excited states
by scaling those reported in Ref.~\cite{ref:Jenkins} 
to give 2.0 $\mu$b for the ground-state $\Xi^0$.
The kaon beam intensity assumed was  $7.0\times 10^6$ in a spill expecting the K10 beam line was ready.
The number of events expected for the $K^-\;p\to K^{*0}\;\Xi^{0}$ 
reaction is $5.3\times 10^6$ in a 30-day beam time as shown in Fig.~\ref{k10_xi_fig2}.
It is $1.8\times 10^7$ in a 100-day beam time, which is required for spectroscopy of $\Omega^*$'s
discussed in sub-subsection \ref{sec:omega-spectroscopy}.
Estimated by a Geant4-based simulation were
the acceptance of the $\Xi^*$-produced events,
and mass resolution for $\Xi^*$, $\sim 6.6$ MeV ($\sigma$), including the straggling effects of the energy loss in the target material.
The background contribution which mainly distributed to the higher mass region,
$\sim$500~$\mu$b at 8 GeV/$c$, was again estimated by using the JAM code~\cite{JAM}.
The $\Xi^*$'s are effectively produced at K10 
since the kaon beam intensity at K10 is 12 times higher than that at $\pi 20$.

\begin{figure}[htbp]
  \begin{center}
  \includegraphics[width=15cm,keepaspectratio,clip]{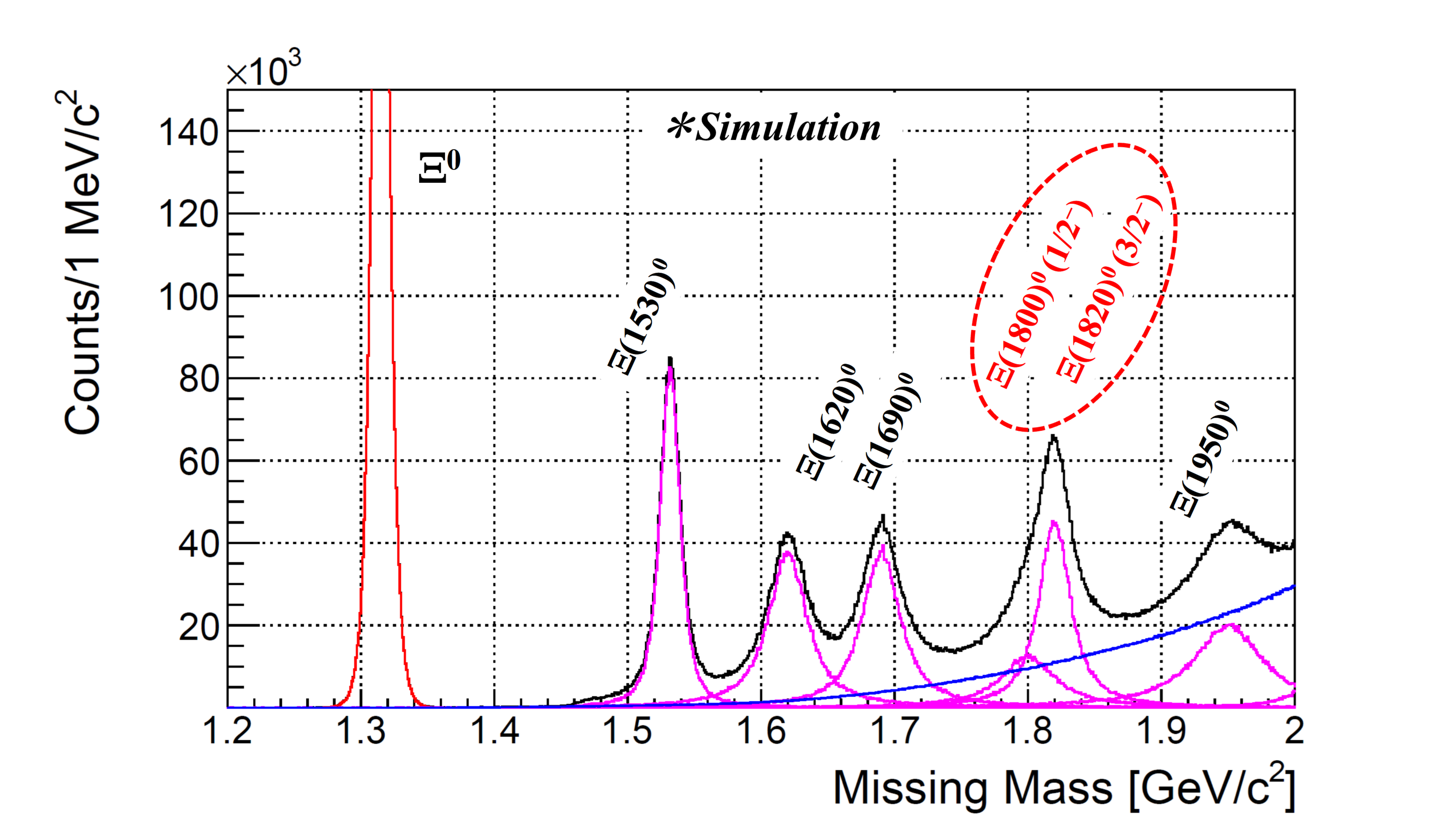}
  \end{center}
  \caption{ Close-up of Fig.~\ref{k10_xi_fig2}
in the mass region from 1.2 to 2 GeV. 
  Here, the $\rho$-mode excited states of $1/2^-$ and $3/2^-$ are highlighted.
  Just for demonstration, we assume that $\Xi(1800)1/2^-$ is the $LS$ partner of $\Xi(1820)3/2^-$ 
  and its mass and width are 1800 MeV and 46 MeV (twice the width of $\Xi(1820)3/2^-$).
  }
  \label{k10_xi_fig3}
\end{figure}

Figure~\ref{k10_xi_fig3} shows a close-up of Fig.~\ref{k10_xi_fig2} in the mass region from 1.2 to 2 GeV. 
Here, the  $1/2^-$ and $3/2^-$ states 
corresponding to the $\rho$-mode excitation
are highlighted, a similar situation observed in $\Omega^*$'s 
will be discussed in sub-subsection~\ref{sec:omega-spectroscopy}.
Just for demonstration, we assume that $\Xi(1800)1/2^-$ is an $LS$ partner of $\Xi(1820)3/2^-$ 
and its mass and width are 1800 and 46 MeV (twice the width of $\Xi(1820)3/2^-$).
The production cross section of $\Xi(1800)1/2^-$ is assumed to be a half that of $\Xi(1820)3/2^-$.
Although we cannot separately observe the $\Xi(1800)1/2^-$ peak in the mass spectrum, 
it distorts the observed $\Xi(1820)3/2^-$ peak in the lower-tail region.
By analyzing the asymmetric $\Xi(1820)3/2^-$-peak  structure, 
we may extract the mass and width of $\Xi(1800)1/2^-$.
The $\Xi(1820)3/2^-$ state is expected to correspond to the  $\rho$-mode excitation. 
Thus $\lambda$ and $\rho$ mode assignment would be checked by observing
the $LS$ partners, 
and determining their production ratios, and decay branching ratios. 
Owing to their large yields, we could determine their spin-parities from the decay angular distributions. 
Clarified would be  $su$- and $ds$-diquark correlation 
and the origin of the spin-dependent interaction 
combining information from 
spectroscopy of  $\Omega^*$'s where no diquark correlation is expected, as discussed in sub-subsection~\ref{sec:omega-spectroscopy}.

\begin{figure}[t]
  \begin{center}
  \includegraphics[width=15.5cm,keepaspectratio,clip]{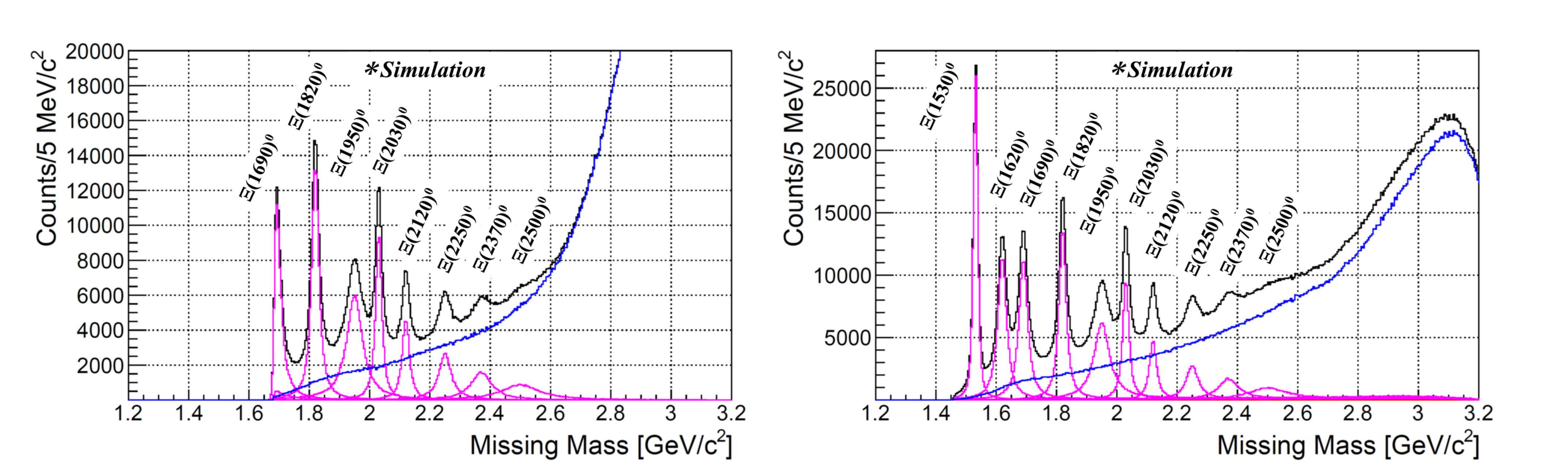}
  \end{center}
  \caption{ Expected $\Xi^*$-mass spectra by identifying the $\Xi^{*0}$ decay into $\Sigma^+\;K^-$ (left) 
  and $\Xi^-\;\pi^+$ (right)
  at an incident kaon momentum of 8 GeV/$c$ in a 30-day beam time.
  The S/N ratios are improved by a factor of more than 200 for highly-excited $\Xi^*$'s above 2.4 GeV.
  }
  \label{k10_xi_fig4}
\end{figure}

The large background contribution prevents us from observing highly-excited $\Xi^*$'s above 2.4 GeV.
In this case, an additional method is necessary to reduce the background contribution.
The S/N ratio is expected to be significantly improved by identifying the $\Xi^{*0}$ decay
into $\Sigma^+\;K^-$ or $\Xi^-\;\pi^+$ as shown in Fig~\ref{k10_xi_fig4}.
If the branching ratios are 0.1 for the $\Xi^{*0} \rightarrow \Sigma^+\;K^-$ 
and $\Xi^{*0} \rightarrow  \Xi^-\;\pi^+$ decays,
we can obtain a reduction factor of more than 200 for the high-mass $\Xi^*$'s.
Although the expected number of the $\Xi^*$-produced events reduced to
$\sim10^4$ from $5.3\times 10^6$ in a 30-day beam time by identifying the $\Xi^{*0}$ decay,
we could determine their spin-parities from the decay angular distributions. 
It is concluded that we can obtain the $\Xi^{(*)}$-mass spectrum with a high S/N ratio at K10.
The details of the background reduction are described in sub-subsection~\ref{sec:xi-spectroscopy-suppl1}.

%%%%%%%%%%%%%%%%%%%%%%%%%%%%%%%%%%%%%%%%%%%%%%%%%%%%%%%%%%%%%%%%%%%%%%%%%%%%%%%%%%%%%

% flatex input end: [./k10docu_exp/k10-xi-spectroscopy_v4.tex]
\label{sec:xi-spectroscopy}
%%Experiment
% flatex input: [./k10docu_exp/k10-omega-spectroscopy_v4.tex]
%%%%%%%%%%%%%%%%%%%%%%%%%%%%%%%%%%%%%%%%%%%%%%%%%%%%%%%%%%%%%%%%%%%%%%%%%%%%%%%%%%%%%
\subsubsection{$\Omega$ baryons}

\begin{figure}[t]
  \begin{center}
  \includegraphics[width=15cm,keepaspectratio,clip]{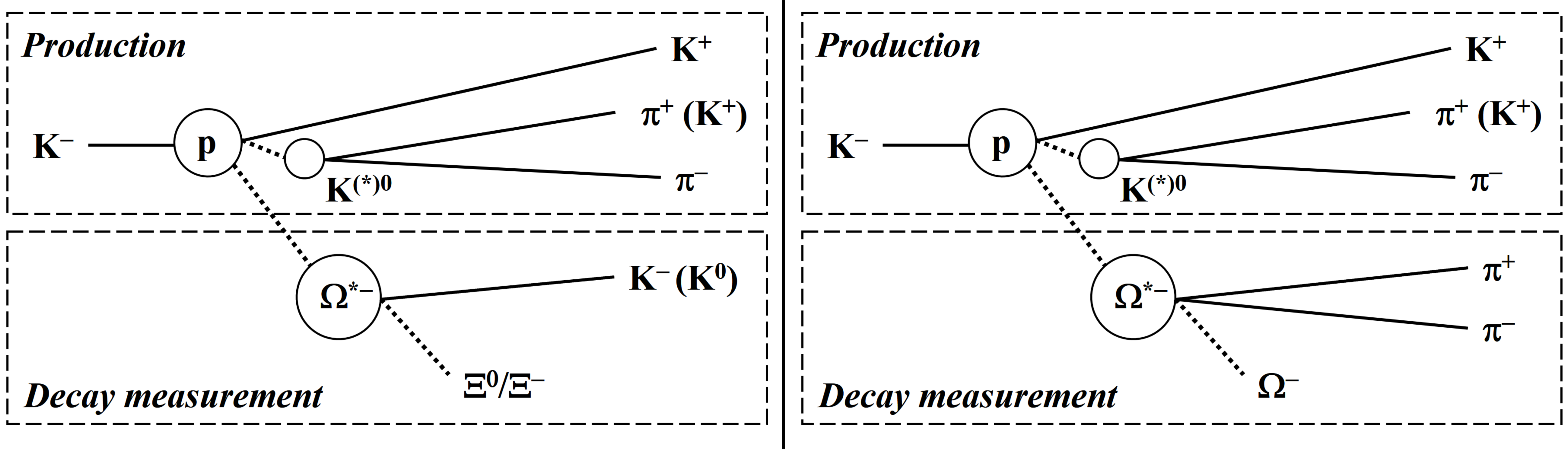}
  \end{center}
  \caption{
  Schematic view of the production mechanisms of $\Omega^*$'s in the $K^-\;p$ reaction, and their decay.
  The solid lines represent the initial- and final-state particles to be directly detected. 
  The dotted lines show unstable particles to be reconstructed from kinematic variables of the detected particles.
  $\Omega^*$'s are expected to decay into $\Xi^0\;K^-$ (left) and $\Omega^-\;\pi^+\;\pi^-$ (right).
  }
  \label{k10_omega_fig1}
\end{figure}

At the K10 beam line,
we plan to investigate the masses and decays of excited $\Omega$ baryons ($\Omega^*$s) 
in the $K^{-}\;p \rightarrow \Omega^{*-}\;K^+\;K^{*0}$ 
and $K^{-}\;p \rightarrow \Omega^{*-}\;K^+\;K^{0}$ reactions at incident kaon momenta ranging from 7 to 10 GeV$/c$.
Spectroscopy of $\Omega^*$'s,
which K10 is the only facility to perform,
 is a unique testing ground to reveal diquark
correlation 
by comparing other sectors
since neither diquark correlation nor pion cloud is expected 
in an $\Omega^*$ system.
Figure~\ref{k10_omega_fig1} shows 
the schematic view of the production mechanisms of $\Omega^*$'s in the $K^-\;p$ reaction, and their decay.
The mass of a produced $\Omega^*$ including the ground-state $\Omega$ ($\Omega^{(*)}$)
can be determined by a missing-mass technique
using the four-momenta of the initial-state $K^-\;p$ and final-state  $K^+\;K^+\;\pi^-$ ($K^+\; \pi^+\; \pi^-$)
in the $K^{-}\;p \rightarrow \Omega^{(*)-}\;K^+\;K^{*0}$ ($K^{-}\;p \rightarrow \Omega^{(*)-}\;K^+\;K^{0}$) reaction.
Here, the selected events are those in which the $K^+\;\pi^-$ ($\pi^+\; \pi^-$) invariant mass 
should give the $K^{*0}$ ($K^0$ or $K_S$) mass.
Since the four-momentum of the produced $\Omega^*$ is already given in 
the missing-mass technique,
the dominant decay mode $\Omega^{*-} \rightarrow \Xi^0\;K^-$ ($\Omega^{*-} \rightarrow \Omega^-\;\pi^+\;\pi^- $)
can be identified only by detecting the emitted $K^-$ ($\pi^+ \; \pi^-$)
and calculating the mass of the associated 
daughter $\Xi^0$ ($\Omega^-$) particle by a similar missing-mass technique.
Both the production and decay measurements of $\Omega^*$'s are essential in their systematic studies.

The performance of the E50 spectrometer for the $\Omega^{(*)}$-produced
events has been estimated by a Monte Carlo simulation based on Geant4.
The details of observing  $\Omega^*$'s
are described in
sub-subsection~\ref{sec:omega-spectroscopy-suppl1}.

%%%%%%%%%%%%%%%%%%%%%%%%%%%%%%%%%%%%%%%%%%%%%%%%%%%%%%%%%%%%%%%%%%%%%%%%%%%%%%%%%%%%%
\subsubsection*{Expected $\Omega^*$-mass spectra}

\begin{figure}[t]
  \begin{center}
  \includegraphics[width=16cm,keepaspectratio,clip]{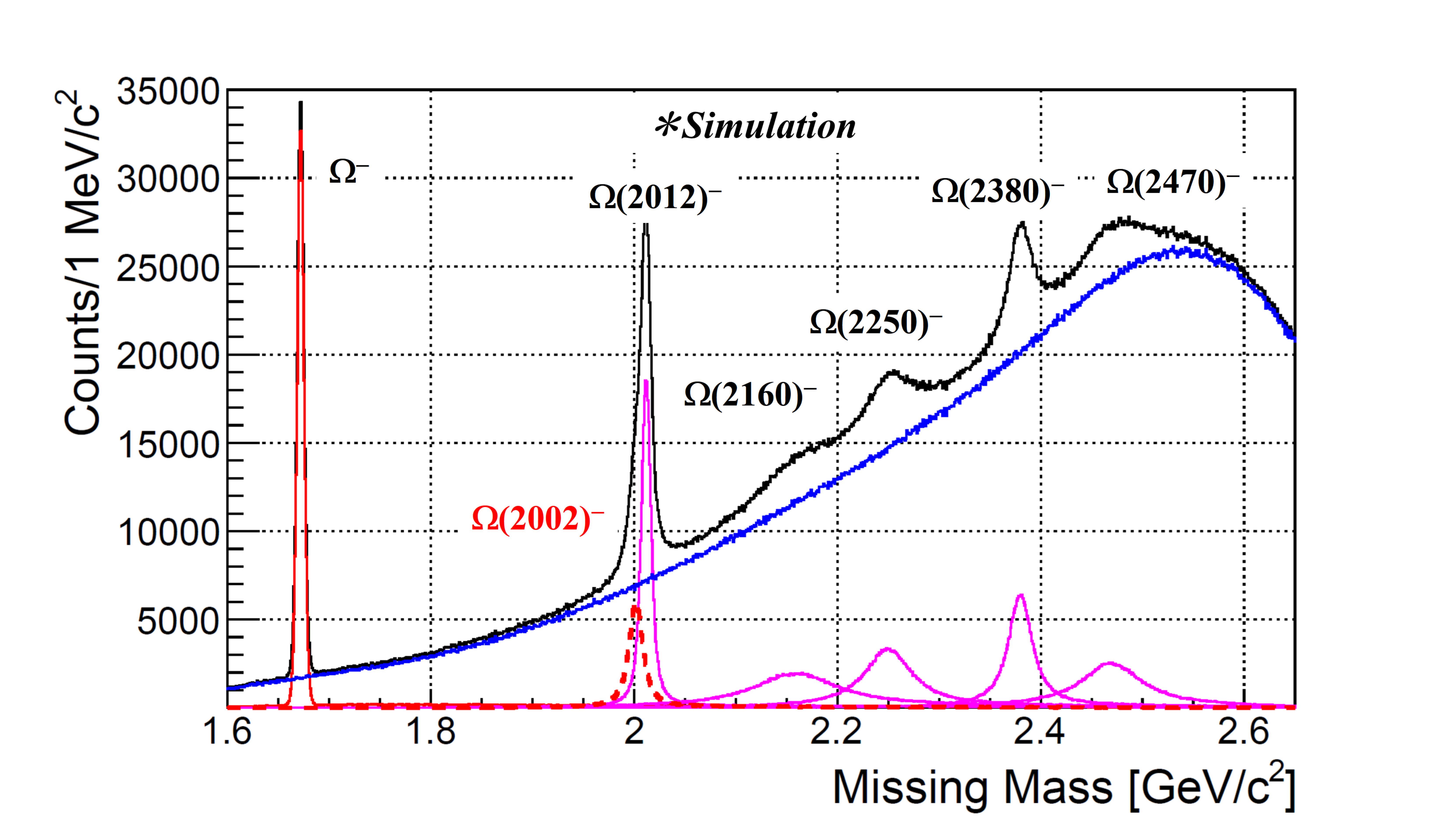}
  \end{center}
  \caption{ 
  Expected $\Omega^{(*)-}$-mass spectrum including the background contribution
  at the incident kaon momentum of 8 GeV$/c$ in a 100-day beam time.
  The spectrum is obtained by the $p(K^-,K^+K^{*0})$ missing mass 
  for the generated $\Omega^{(*)-}$-produced events and background processes.
  The smooth blue curve represent the background contribution estimated by JAM, 
  and the contributions from the ground-state $\Omega^-$ 
  and $\Omega^{*-}$'s are also plotted in red and in magenta, respectively.
  The $p(K^-,K^+K^{*0})$  missing system gives $S=-3$, which eliminates huge background 
  from $\Sigma^*$-produced events.
  }
  \label{k10_omega_fig2}
\end{figure}

Figure~\ref{k10_omega_fig2} shows the expected $\Omega^{(*)}$-mass
spectrum including the background contribution at an incident kaon momentum of 8 GeV$/c$.
The background contribution is estimated by using the JAM code~\cite{JAM}.
The $\Omega^{(*)-}$ mass is calculated by the $p(K^-,K^+K^{*0})$ missing mass, 
and the cross section of $K^-\;p\to \Omega^{(*)-}\;K^+\; K^{0*}$ for each $\Omega^{(*)-}$ is assumed to be 63 nb.
The number of events for each peak is $\sim 3.3\times 10^5$, which can be obtained in a 100-day beam time. 
Since the number of the events is large, 
the precision of the mass determination in this spectrum is better than 1 MeV depending 
on the width of the corresponding $\Omega^{(*)-}$.
To get accuracy of the mass determination,
the absolute momentum scale in the spectrometer is required to be calibrated carefully.
We can use various peaks corresponding to hyperons and other baryons in similar mass spectra.
Thus, the accuracy would be also better than 1 MeV after the calibration.
It is difficult to accurately determine the width of an $\Omega^{(*)-}$.
The observed peak must be modified by the experimental mass resolution of 2.5--4.5 MeV ($\sigma$).
The peak may be different in shape from the Breit-Wigner function
owing to the overlap with other peaks and background processes including the interference effects.
The accuracy of the width determination expected is better than 1 MeV 
for an isolated Breit-Wigner peak with a width of 10 MeV.
We expect the accuracy of the width determination 
is better than a few MeV for high-mass $\Omega^*$'s, which are overlapped with other contributions.

In Fig.~\ref{k10_omega_fig2},  $\Omega(2002)^{-}$ is also shown 
as the $LS$ partner of $\Omega(2012)^{-}$ 
with a mass of 2.002 GeV and a width of 12 MeV (twice the  $\Omega(2012)^{-}$  width).
The production cross section of $\Omega(2002)^{-}$ is assumed to be a half that of $\Omega(2012)^{-}$.
Although we cannot clearly observe the peak of $\Omega(2002)^{-}$ in the expected mass spectrum, 
the $\Omega(2002)^{-}$ contribution appears in distortion 
of the $\Omega(2012)^{-}$ peak in the lower tail region.
By analyzing the asymmetric $\Omega(2012)^{-}$-peak structure, we may extract the mass and width of $\Omega(2002)^{-}$
even if the $LS$ partner has the same mass as $\Omega(2012)^{-}$ with a different width.
The high statistic data would enable us to reveal the $LS$ partner of $\Omega(2012)^{-}$.
As for the Roper-like $\Omega^*$, $\Omega(2160)^{-}$, having 
a broad width around 100 MeV,
the corresponding events are distributed in a wide range, just forming  a shoulder.
A peak or bump would be clearly observed corresponding to each of the other $\Omega^*$'s with narrower widths.

If the $\Omega^*$-production cross section is smaller by 1/10 than that in the original assumption (63 nb),
it is difficult to observe $\Omega^*$ peaks except for the ground-state $\Omega^-$ and first-excited $\Omega(2012)^-$.
In this case, an additional method is necessary to reduce the background contribution
for finding an $\Omega^*$ with a broad width.
Selecting the events containing $\Xi^0$ as a daughter particle from the $\Omega^*$ decay
is effective at background reduction in the $\Omega^{*}$-mass spectrum.
If the branching ratio is $\sim 0.3$ for the $\Omega^{*-} \rightarrow \Xi^0\;K^-$ decay,
the S/N ratio is improved by a factor of 10 for the high-mass $\Omega^*$s.
We can recognize $\Omega^*$'s with a broad width 
even when the cross section is smaller than that originally assumed.
We can also find the $\Omega(2160)^-$ or the Roper-like state having a broad width around 100 MeV
even if the cross section is smaller by 1/10 than that in the original assumption.
It is concluded that we can measure the $\Omega^{(*)}$-mass spectrum with high S/N ratio at K10.
In sub-subsection~\ref{sec:omega-spectroscopy-suppl2},
we have estimated how $\Omega^{(*)}$'s are observed in the $p(K^{-},K^+K^{*0})$ missing-mass spectrum.
We also show the expected $\Omega^{(*)}$-mass spectrum without and with 
the background contribution estimated by JAM.

The spin-parity assignment of a produced $\Omega^{*-}$ provides
crucial insight into the internal quark motion.
Suppose the branching ratio is 0.3 for the two-body $\Omega^{*-} \rightarrow \Xi^0\;K^-$ decay,
the expected yield is several thousands in a 100-day beam time, giving 
a statistical error of $\sim$1\%,
for each bin of $\Xi^0$ emission angles divided into 20.
We may identify $J$ when the angular distribution shows some structure.
We also perform more detailed analysis for determining the spin
for the $\Omega^*$ of interest by combining other information 
such as the branching ratio.
The high statistic data are necessary to determine the spin-parities 
of observed $\Omega^*$'s. 
The details of spin-parity assignment of a produced $\Omega^{*-}$ are described
in sub-subsection~\ref{sec:omega-spectroscopy-suppl3}.

%%%%%%%%%%%%%%%%%%%%%%%%%%%%%%%%%%%%%%%%%%%%%%%%%%%%%%%%%%%%%%%%%%%%%%%%%%%%%%%%%%%%%
% flatex input end: [./k10docu_exp/k10-omega-spectroscopy_v4.tex]
\label{sec:omega-spectroscopy}
%%Experiment

\clearpage
%%OmegaN scattering
\subsection{Study of an \boldmath$\Omega N$ Bound State}\label{sec:omega-n}
% flatex input: [./k10docu_exp/k10-int_v3.tex]
%\clearpage
%\section{$\Omega N$ scattering}

%\subsection{Introduction}
In addition to the $\Omega^*$-mass spectrum, 
scattering of the ground-state $\Omega^-$ on the nucleon $N$ is 
of fundamental importance to be studied.
The $\Omega^- N$ system belongs to the 
octet as well as $\Delta N$ with $J=2$
among the decuplet-baryon and octet-baryon systems:
\begin{equation}
10 \otimes 8 = 35 \oplus 8 \oplus 10 \oplus 27.
\end{equation}
Recently,
a lattice QCD calculation predicts an bound state in the $\Omega N$ system 
owing to absence of a repulsive core in its $^5S_2$ state,
and its weak absorption to another channel.

Currently, the low-energy $\Omega N$ scattering is studied 
using the momentum correlation function of the $\Omega^- p$ pairs 
in relativistic heavy-ion collisions (so called the femtoscopy)~\cite{lqcd-omega-n}.
The function shows a depletion below 1 around 20--40 MeV
corresponding to attraction with a positive scattering length,
suggesting a shallow bound state.
The existence of an $\Omega^- p$ bound state
cannot be concluded
since information on the $J = 1$ $\Omega^-p$ potential is not sufficient.
Therefore,
desired is a direct $\Omega N$ scattering experiment,
and it can be realized at the K10 beam line thanks to a rather stable nature of $\Omega^-$ (only it decays in the weak interaction).

% flatex input end: [./k10docu_exp/k10-int_v3.tex]

%%OmegaN scattering
% flatex input: [./k10docu_exp/k10-omega-n_v3.tex]
%%==================================================
%\input{k10omega-n}
%%==================================================

\subsubsection{$\Omega^-N$ interaction}

The deuteron is the simplest nucleus of a proton-neutron bound system. It has provided a lot of information on the nuclear force. One of important goals in strangeness nuclear physics at J-PARC is to clarify the baryon-baryon interactions in the framework of the flavor $SU(3)$ symmetry. 
Experimental data for hypernuclear structure as well as hyperon-nucleon scattering  are thus of particular importance to extract the hyperon-nucleon interactions. 

Recently, the correlation function of $\Omega^-$-$p$ has been extracted as a function of the relative momentum between the two particles in high-energy nucleus-nucleus collisions by the ALICE collaboration at LHC \cite{femtospectroscopy}. 
The observed distribution is consistent with a lattice QCD calculation which predicts an attractive potential and a bound state with the binding energy of 1.54(0.3)($^{+0.04}_{-0.10}$) MeV in the $^5S_2$ spin state of $\Omega N$ \cite{lqcd-omega-n}. The $\Omega N$ bound state has yet to be observed to provide fruitful information on the interaction between decuplet-octet baryons in the flavor $SU(3)$ framework. 

Sekihara {\it et al.} have provided the $\Omega N$ potential to simulate the lattice QCD calculation within a meson exchange model \cite{sekihara-omega-n}. The authors reported a narrow state at a pole position of $2611.3-0.7i$ MeV, suggesting 
the existence of a shallow $\Omega N({}^5S_2)$ bound state with a weak
binding energy of $\sim 0.1$ MeV and a narrow width of $\Gamma\sim 1.5$ MeV.

\begin{figure}[htbp]
\centerline{\includegraphics[width=0.6\textwidth]{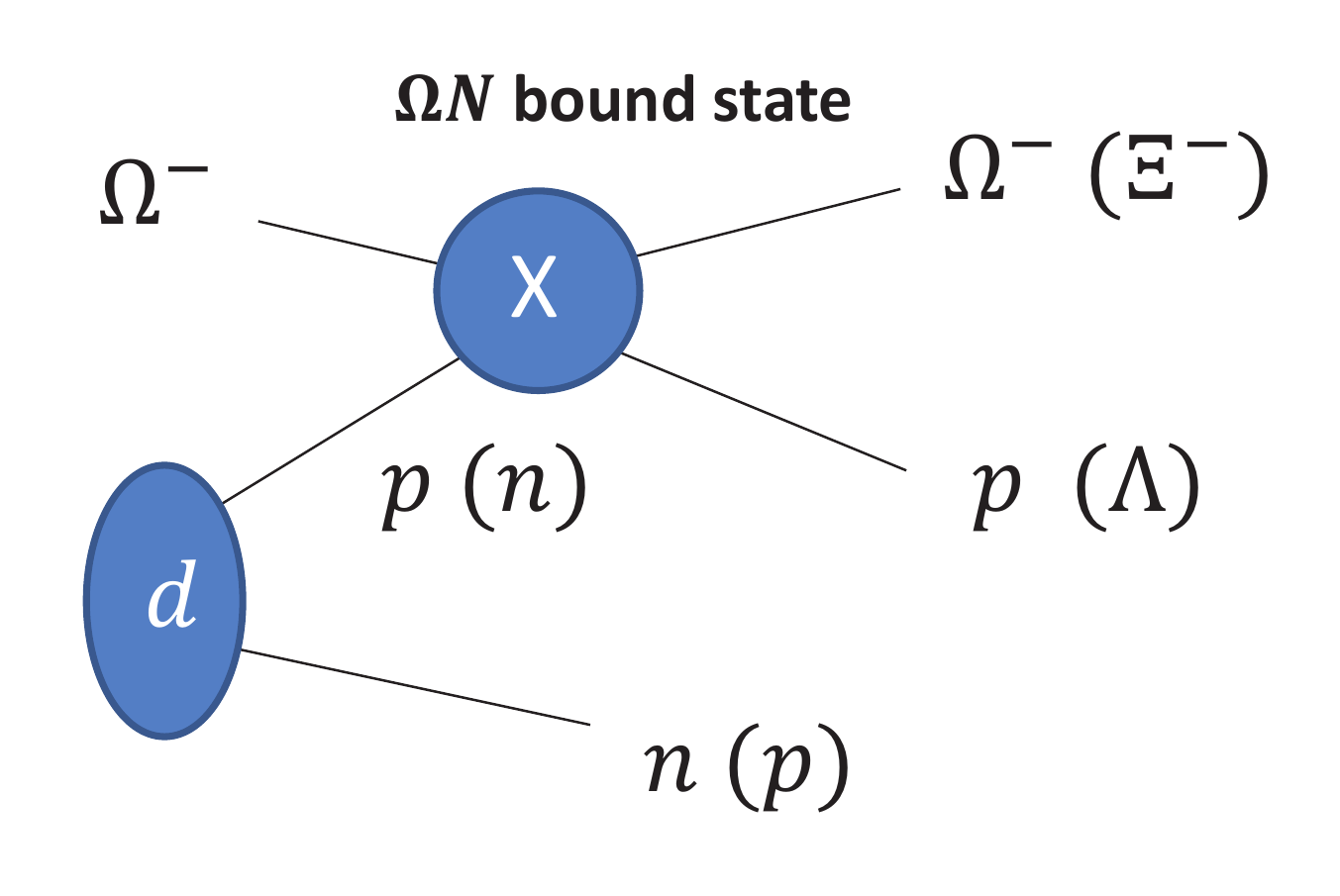}}
 \caption{Diagram of the $\Omega^\;-d$ reaction 
to produce an $\Omega^- N$ bound/resonance state near the threshold.
A possible shallow $\Omega^-N$ bound state can be produced
in the $\Omega^-\; n$ collision, 
decaying into $\Xi^-\; \Lambda$.
\label{fig:omega-d-diagram}}
\end{figure}

J-PARC provides a unique opportunity to investigate the $\Omega N$ bound state. The $\Omega^-$ baryon will be produced via the $K^-\; p\rightarrow 
\Omega^-\; \bar{K}^{(*)0}\; K^+$ reaction. 
The typical momentum of $\Omega^-$ is around 3 GeV$/c$
distributing from 2 to 4 GeV$/c$ at an incident kaon momentum of 5 GeV/c. 
It would be difficult to slow down the $\Omega$ to realize
the $S$-wave scattering with a nucleon at rest,
and to produce a possible bound state.
Thus, we consider to use an off-shell nucleon,
and discuss the $\Omega^-d\rightarrow\Xi^-\Lambda p$ reaction as illustrated in Fig.~\ref{fig:omega-d-diagram}.

%------------------------------------
%------------------------------------

The $\Omega^-$ reacts with a bound neutron ($n_b$) in the deuteron ($d$) to produce $\Xi^-\Lambda$, where the residual proton acts as a spectator.  Since the bound neutron ($n_b$) is an off-shell particle, 
the collision energy or the $\Omega^-n_b$-CM energy can be smaller and reach even below the $\Omega^-n$ threshold. 

\begin{figure}[htbp]
\centerline{\includegraphics[width=0.95\textwidth]{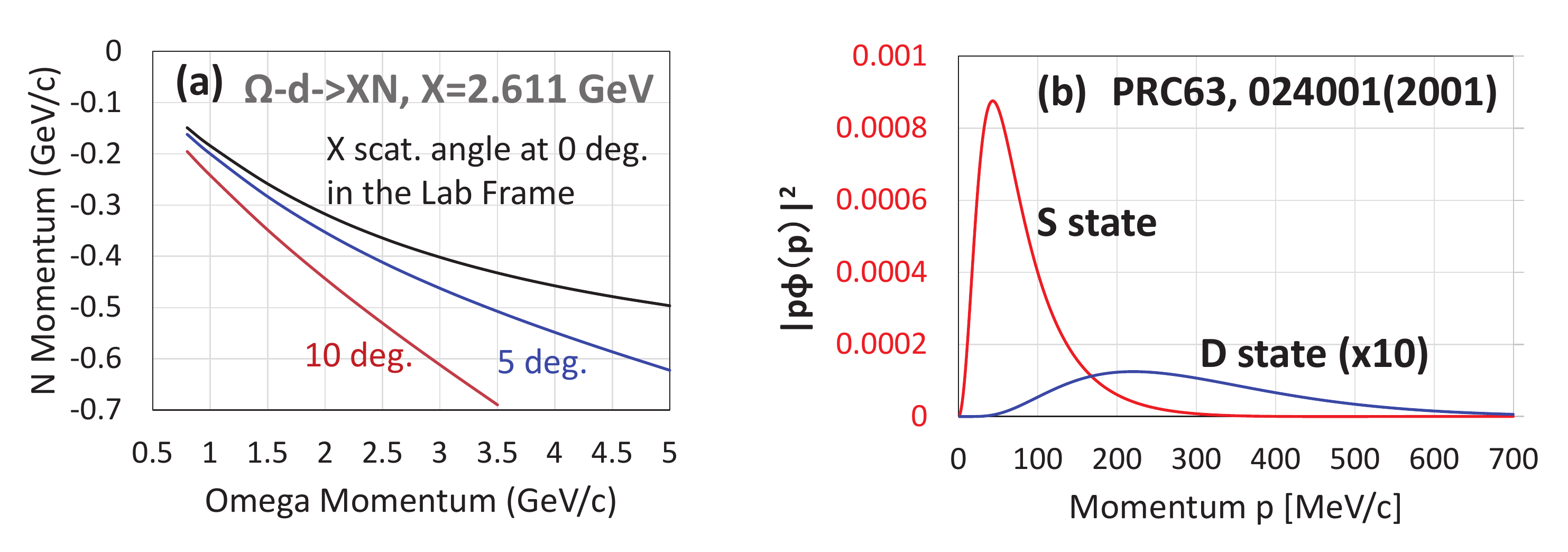}}
 \caption{(a) Calculated momentum of the spectator nucleon ($N_s$) as a function of the incident $\Omega^-$ momentum in the $\Omega^\;-d\rightarrow X\; N_s$ process.(b) Momentum distributions of the bound nucleon in the deuteron 
for the $S$- (red) and $D$-wave (blue) components.
The magnitude of the $D$-wave component is enhanced by a factor of ten 
for easy comparison. \label{fig:fermi-momentum}}
\end{figure}

Let us consider the two-body reaction, $\Omega^-d\rightarrow X N_s$. Here, $N_s$ denotes a spectator nucleon in the reaction diagram shown in Fig.~\ref{fig:omega-d-diagram}. Putting the (invariant) mass of $X$ at the mass of 
2.611 GeV as for the predicted shallow $\Omega^-N$ bound state, one can calculate the momentum of $N_s$, which tells us the momentum of a participant nucleon
to produce the bound state. Fig.~\ref{fig:fermi-momentum}(a) shows 
the  $N_s$ momentum as a function of the incident $\Omega^-$ momentum 
at the $X$ emission angle of $0^\circ$, $5^\circ$, and $10^\circ$
 in the laboratory frame.
A negative $N_s$ momentum means that 
the head-tail collisions take place between $\Omega^-$ and the participant nucleon.
The absolute momentum of the participant nucleon becomes higher 
to several hundred MeV$/c$ as the incident $\Omega^-$ momentum increases. 
The deuteron is known as a loosely-bound two-nucleon system.
However, it is well known that a bound nucleon has a $D$-wave component
making the momentum of the bound nucleon higher in the deuteron
owing to the tensor interaction of the nuclear force, 
as illustrated in Fig.~ \ref{fig:fermi-momentum}(b). 
Thanks to the higher momentum component, 
we can access the $\Omega^-N$-CM energy near the $\Omega^-N$ threshold 
in a collision of an energetic $\Omega^-$ to a bound nucleon in the deuteron.

%------------------------------------
%------------------------------------

The cross section of the $\Omega^-\;n_b\rightarrow\Xi^-\;\Lambda$ is expected to be enhanced at the pole energy, which would be observed in the $\Xi^-\Lambda$ invariant-mass ($M_{\Xi^-\Lambda}$) spectrum.
The cross section as a function of $M_{\Xi^-\Lambda}$
is given by
\begin{equation}
\frac{d\sigma}{dM_{\Xi^-\Lambda}}=\int{|T_{\Omega^-n_b\rightarrow\Xi^-\Lambda}(s)\psi(\mbox{\boldmath$p$})|^2\delta(P_{\Xi^-\Lambda}-p_\Omega-p_{n_b})}d\mbox{\boldmath$p$},\label{eq:cs}
\end{equation}
where $P_{\Xi^-\Lambda}=(E_{\Xi\Lambda},\mbox{\boldmath$p$}_{\Xi\Lambda})$
denotes the four-momentum of the ${\Xi^-\Lambda}$ system with
$E_{\Xi\Lambda}=\sqrt{M_{\Xi\Lambda}^2+\mbox{\boldmath$p$}_{\Xi\Lambda}^2}.$
The four-momentum of the bound neutron $p_{n_b}=(M_d-E_p,\mbox{\boldmath$p$})$ 
is obtained by assuming that the spectator proton is free and on-shell 
($E_p=\sqrt{m_p^2+p^2}$).
The total energy of $n_b$ is thus $E_b=M_d-\sqrt{M_p^2+p^2}$.
The wave function of $n_b$ in momentum space, $\psi(\mbox{\boldmath$p$})$,
is  normalized so that 
\begin{eqnarray}
 \int{|\psi(\mbox{\boldmath$p$})|^2}d\mbox{\boldmath$p$}=1.
\end{eqnarray}
The scattering (transition) amplitude 
$T_{\Omega^-n_b\rightarrow\Xi^-\Lambda}(s)$ for the $S$-wave $\Omega^-n_b\rightarrow\Xi^-\Lambda$ reaction
will be described in the next sub-subsection.

\subsubsection{$\Xi^-\Lambda$ invariant mass spectrum in the $\Omega^- d\rightarrow\Xi^-\Lambda p$ reaction}

\begin{figure}[htbp]
\centerline{\includegraphics[width=\textwidth]{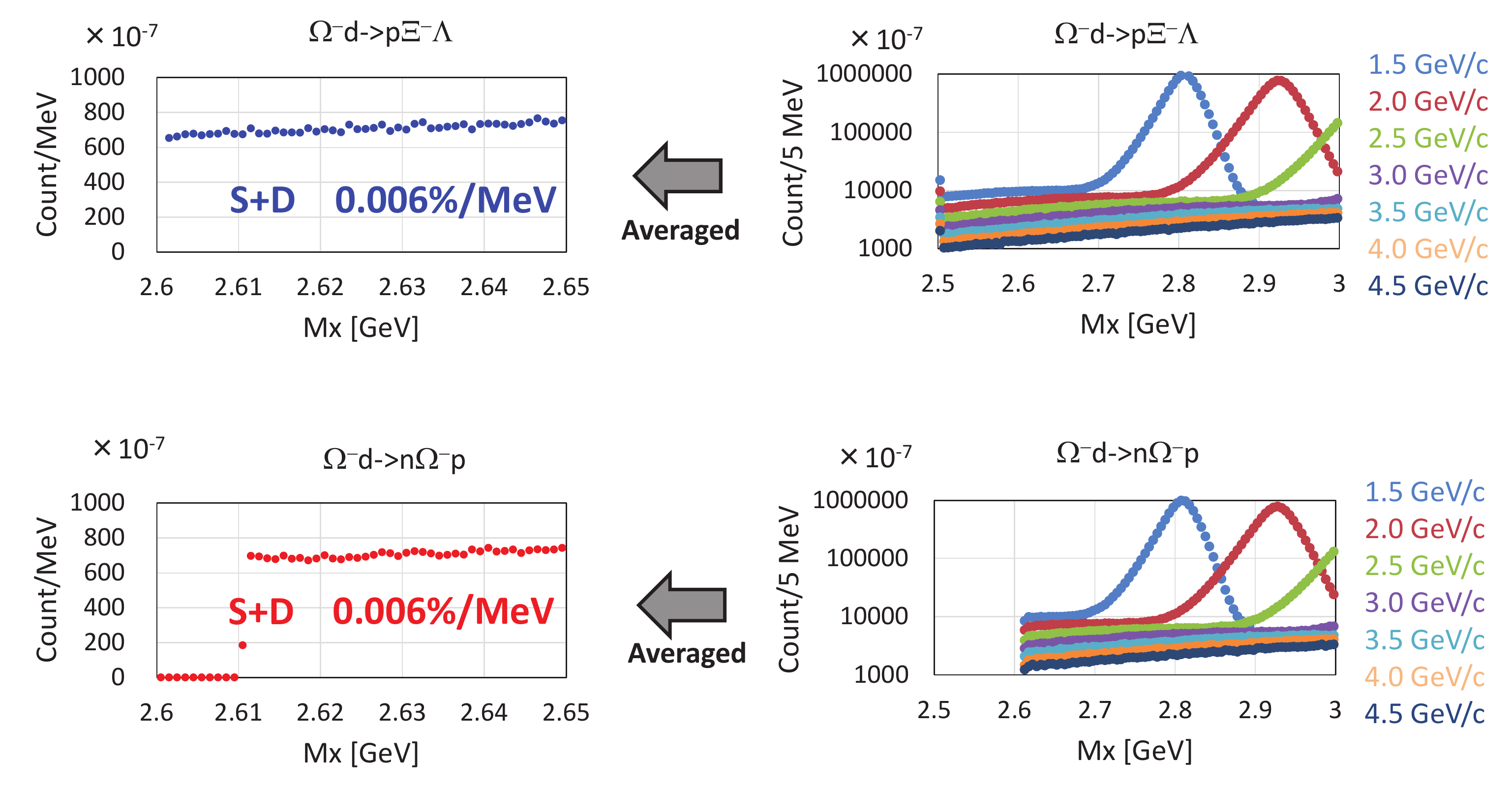}}
 \caption{
Response functions, or the probability of $\Omega^-$ colliding to a bound nucleon with a momentum 
of $p_{n_b}$ at a collision energy of $\sqrt{s}$.
The upper panels shows the response function for $\Omega^-\;n_b\to\Xi^-\;\Lambda$,
and the lower that for $\Omega^-\;p_b\to\Omega^-\;p$,
where the subscript $b$ stands for the bound nucleon.
\label{fig:response}}
\end{figure}

To estimate the spectral shape in Eq.~(\ref{eq:cs}) for $M_{\Xi^-\Lambda}$, 
we applied a factorization approximation:
\begin{equation}
\frac{d\sigma}{dM_{\Xi^-\Lambda}}\sim|T_{\Omega^-n_b\rightarrow\Xi^-\Lambda}(s)|^2\int{|\psi(\mbox{\boldmath$p$})|^2\delta(P_{\Xi^-\Lambda}-p_\Omega-p_{n_b})}d\mbox{\boldmath$p$}.
\end{equation}
We evaluated the integral that represents a response function, 
or a probability of $\Omega^-$ colliding to a bound nucleon with a momentum 
of $p_{n_b}$ at a collision energy of $\sqrt{s}$, as shown in Fig.~\ref{fig:response}.

%------------------------------------
%------------------------------------

A major component of the fermi momentum of the participant nucleon makes 
a bump far above the $\Omega^- N$ mass threshold in the response function, 
corresponding to the incident $\Omega^-$ momentum. 
One finds a long tail to the region around the $\Omega^- N$ mass threshold with a small but a finite strength of $\sim 0.006\%$/MeV in total. It is notable that the distribution is almost uniform and no artificial structure would be expected 
in this response function.

Sekihara $et\ al.$ provided the scattering length, $a=5.3-4.3i$ fm, and the effective range,
 $r_{\rm eff}=0.74+0.04i$ fm, for the 
$^5S_2$ $\Omega^- N$ channel. The $S$-wave scattering amplitude in the elastic channel, $\Omega^- N\rightarrow\Omega^- N$, can be written as
\begin{equation}
f(s,\theta)= \frac{1}{D(k_2)} {\rm\ with\ }
D(k_2)=\displaystyle{\frac{1}{a}}- ik_2+\frac{1}{2}r_{\rm eff}k_2^2, 
\end{equation}
where $k_2$ is the momentum of $\Omega^-$ in the 
$\Omega^- N$-CM frame. Solving a two-channel coupled problem in the $S$-wave, the scattering amplitude in the conversion channel, $\Omega^- N\rightarrow\Xi \Lambda$, can be written as
\begin{eqnarray}
g(s,\theta)= \frac{1}{\sqrt{k_1}}e^{i\delta_0}\frac{\sqrt{{\rm Im}\left[a\right]-
\frac{1}{2}\left|a\right|^2\,{\rm Im}\left[r_{\rm eff}\right]k_2^2}}{D(k_2)},
\end{eqnarray}
where $k_1$ is the CM momentum of $\Xi$ and $\delta_0$ is a phase parameter.
%The cross section of this channel is $\sigma=4\pi|g(\theta)|^2$. 
It should be noted that $\sqrt{4\pi}g(s,\theta)$ corresponds to $T_{\Omega n_b\rightarrow\Xi^-\Lambda}(s)$ in Eq.(\ref{eq:cs}). 

\begin{figure}
\centerline{\includegraphics[width=0.8\textwidth]{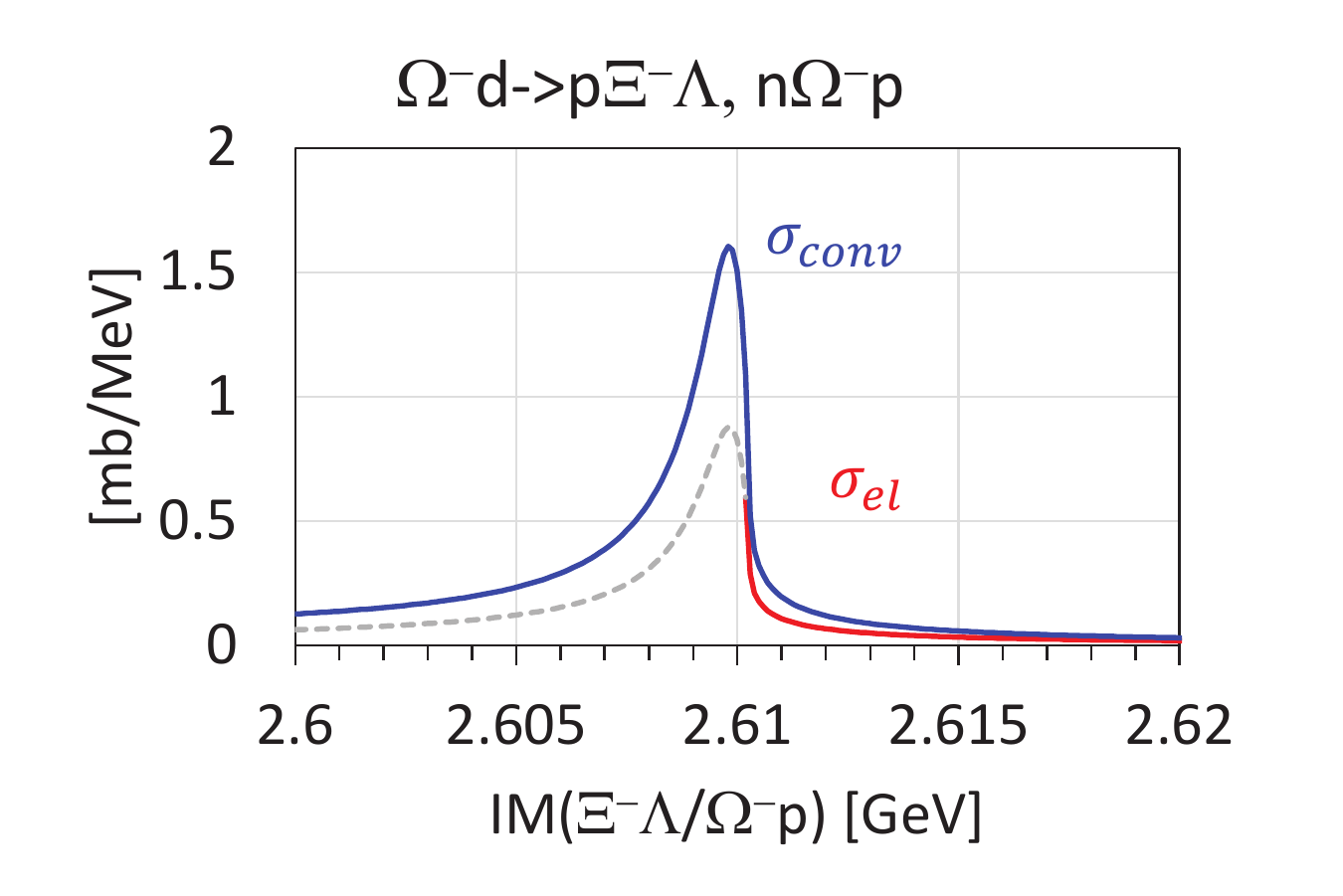}}
 \caption{ Expected $\Xi^-\Lambda$ ($\Omega^-p$) invariant-mass spectrum. \label{fig:xilam-im}}
\end{figure}

The total cross sections of the elastic channel
is obtained 
\begin{equation}
\sigma(\Omega N\rightarrow\Omega N)=4\pi|f(s,\theta)|^2,\\
\end{equation}
and that of the conversion channel is given by
\begin{equation}
\sigma(\Omega N\rightarrow\Xi\Lambda)= 4\pi|g(s,\theta)|^2.
\end{equation}
Multiplying a factor of 0.006\% corresponding to 
the strength of the response function that we have discussed in the previous 
sub-subsection, 
we can obtain the $\Xi^-\;\Lambda$ ($\Omega^-p$) 
invariant-mass spectrum
as shown in Fig.~\ref{fig:xilam-im}.

%------------------------------------
%------------------------------------

A clear peak with a narrow width of $\Gamma\sim$1.5 MeV is located just below the $\Omega^- N$ threshold in the $\Omega^-\;d\rightarrow\Xi\;\Lambda\; N$ channel.
The cross section around the peak is as large as 3 mb.
It is worth challenging a direct observation of the $\Omega^-N$ bound state 
via $\Omega^-$-induced $\Xi\;\Lambda$ production off the deuteron at K10.

% flatex input end: [./k10docu_exp/k10-omega-n_v3.tex]

%%OmegaN scattering
% flatex input: [./k10docu_exp/k10-omega-n-spinparity_v2.tex]
%========================================
% Local definitions
%
\newcommand{\be}{\begin{eqnarray}}
\newcommand{\ee}{\end{eqnarray}}
\newcommand{\ket}{\rangle}
\newcommand{\bra}{\langle}
\newcommand{\del}{\partial}
\newcommand{\tr}{\, {\rm tr}\, }
\newcommand{\pslash}{{p\hspace{-5pt}/}}
\newcommand{\dslash}{{\del \hspace{-5pt}/}}
\newcommand{\zslash}{{z\hspace{-5pt}/}}
\newcommand{\kslash}{{k\hspace{-6pt}/}}
\newcommand{\calL}{{\cal L}}
\newcommand{\calM}{{\cal M}}
\newcommand{\calO}{{\cal O}}

%-----
\newcommand{\Slash}[1]{\ooalign{\hfil/\hfil\crcr$#1$}}

\newcommand{\half}{\frac{1}{2}}

\def\gsim{\displaystyle\mathop{>}_{\sim}}
\def\lsim{\displaystyle\mathop{<}_{\sim}}
%========================================

%\title{\bf How can we determine spin and parity?}
%\author{Atsushi Hosaka\\
%{\it Research Center for Nuclear Physics (RCNP)}\\
%{\it Osaka University, Ibaraki, 567-0047, Japan}}

%\date{}
%\maketitle

%
%%%======================
%%% Abstract
%\begin{center}
%{\bf Abstract}
%\vspace*{1cm}\\
%\begin{minipage}{14cm}
%{\small 
%We compute decays of charmed baryons via one-step process.}
%\end{minipage}
%\end{center}
%%======================

%%=====================
%\section{Structure}
%%=====================
%
%Excited states of charmed baryons are classified by the orbital motions of 
%the $\rho$ and $\lambda$ modes.  

%=====================
\subsubsection{Determination of spin and parity}
%=====================

In sub-subsection~\ref{sec:angulardist}, we will briefly look
 at how angular correlation 
extracts the information of spin of $\Omega^*$ through the decay into $\Xi\; \bar K$.  
In this sub-subsection we extend it to the decay of $\Omega N$ quasi-bound state decaying 
into $\Xi\; \Lambda$.  
In addition to spin we also discuss how parity can be determined.  

%=====================
\subsubsection*{Spin}
%=====================

Let us assume that we can fix an axis along which the spin of $\Omega N$
is quantized in its rest frame.  
Then we measure the distribution of the $\Xi$ and $\Lambda$ decaying to 
direction of angle $(\theta, \phi)$.  
The method of the helicity amplitudes
is applied~\cite{Jacob:1959at,Copley:1969ft}.  
Assuming that the initial state of $\Omega N$ is given by an ensemble average of 
a spin $JM$ state, the decay angle dependence is given by the formula
\be
F(\Omega) 
&\sim&
\sum_{M\mu}
\rho_{MM}
D^{J *}_{\mu, M}(\Omega)D^{J}_{\mu, M}(\Omega)
\ee
where the sum over $\mu = -1, 0, +1$ is for the unobserved helicities 
of the final state $\Xi$ and  $\Lambda$, 
and $\rho_{MM}$ is the spin density matrix characterized by the polarization 
of the initial state and actual interaction that causes the decay.  
In experimental analysis, they are treated as free parameters to be fitted to the observed 
angular correlations.  

For $J=0,1$ after summing up the finial helicity state with equal weight, the resulting 
angular correlation is 
\be
F(\Omega)
&\sim&
\sum_{M\mu}
\rho_{MM}
|d^{2 }_{\mu, M}(\theta)|^2
\nonumber \\
&=&
\rho_{0 0}
\sum_\mu |d^{2 }_{\mu, 0}(\theta)|^2 
+
2 \rho_{1, 1}
\sum_\mu |d^{2 }_{\mu, 1}(\theta)|^2 
+
2\rho_{2 2}
\sum_\mu |d^{2 }_{\mu, 2}(\theta)|^2 
\nonumber \\
&\sim&
\rho_{0 0} 
\left( - \frac{3}{4}c^4 + \frac{3}{2} c^2 + \frac{1}{4} \right)
+
\rho_{1 1} 
\left(c^4 + 2 \right)
+
\rho_{2 2} 
\left( - \frac{1}{4} c^4 - \frac{3}{2} c^2 + \frac{7}{4} \right)
\ee
where $c = \cos \theta$.
In the second line the factor 2 before $\rho_{11}, \rho_{22}$ is to count
$\pm M$.
Therefore, if initial state is at least partially polarized (not all of the 
density matrices are equal), the angular correlations will follow 
a fourth order even power polynomial of $\cos \theta$.  

Similarly, for $J = 3$, we find
\be
F(\Omega) &=&
\frac{\rho_{00}}{16}
\left( 6 - 30c^2 + 90 c^4 - 50 c^6 \right)
\nonumber \\
&+&
 \frac{\rho_{11}}{16}
 \left( 7 + 45c^2 - 95 c^4 + 75 c^6 \right)
\nonumber \\
&+&
 \frac{\rho_{22}}{16}
 \left( 10 +30c^2 - 10 c^4 -30 c^6 \right)
\nonumber \\
&+&
 \frac{\rho_{33}}{16}
 \left( 25 - 45 c^2 +15  c^4+ 5 c^6 \right).
\ee
Thus the angular correlation follows an even power polynomials of the
sixth power or less.  
For $J = 1$ and 0, if the final state helicity is equally summed, 
the angular correlation will be flat.  
If, however, we can fix either the initial state $M$ or the final state $\mu$, then 
the angular distribution of $J = 1$ will depend on $\cos^2\theta$, while 
that of $J = 0$ remains flat.  

%------------------
\subsubsection*{Parity}
%------------------

To know the parity of the system, we need to extract the information 
of the orbital angular momentum $\bm L$.
We may also use the information of the internal spin $\bm S$.  
%This cannot be done by the helicity method, when the total system of 
%$\bm L$ and $\bm S$ are rotated simultaneously.  

Let us assume that a spin-quantization axis is determined,
and consider the decay of an $\Omega N$ bound of spin parity 
$J^P = 2^{\pm}, 1^{\pm}$ into $\Xi \;\Lambda$ of total spin $J^P$ (conserved) which is composed of the
orbital angular momentum $ L_f$ and spin $ S_f$ ($\bm S_f = \bm S_\Xi + \bm S_\Lambda$)
\be
(\Omega N): J^P \ \ \to \ \ \Xi\; N: L_f \ S_f .
\ee
Hence the total spin $J$ state is given by 
\be
|JM\ket = \sum_\mu (L_f \mu, S_f M\!\! - \!\! \mu|JM) |L_f \mu\ket |S_f M\!\! -\!\! \mu\ket
\ee
and the parity $P$ is given by 
$
P = (-1)^{L_f}
$
when the parities of $\Xi$ and $\Lambda$ are positive.  
In the momentum and spin representation (along the initially defined quantization axis), 
we find
\be
\bra \Omega |JM\ket = \sum_\mu (L_f \mu, S_f M\!\! - \!\! \mu|JM) Y_{L_f, \mu}(\Omega) \chi_{S_f, M-\mu}
\label{eq:16}
\ee
where 
\be
\chi_{1,1} =\ \uparrow \uparrow, \ \
\chi_{1,0} =\ \frac{1}{\sqrt{2}}(\uparrow \downarrow + \downarrow \uparrow), \ \ 
\chi_{1,-1} =\ \downarrow \downarrow; \ \ \ \
\chi_{0,0} =\ \frac{1}{\sqrt{2}}(\uparrow \downarrow - \downarrow \uparrow).
\ee

For a given $J^P$ of the $\Omega N$, we find possible values of 
$L_f$ and $S_f$  as summarized in Table~\ref{table_Sf}.
We have tabulated all possible cases where the internal orbital angular momentum 
of the $\Omega N$ is $L_i = 0$ or 1.
The states of $L_i = 2$ are or larger $L_i$ are not tabulated because they are 
highly excited and is not likely to be  a candidate of the $\Omega N$ bound state.  

%For $S_f = 0$ the spins of $\Xi$ and $\Lambda$ must be anti-parallel.
%But $S_f = 0$ is always accompanied by $S_f = 1$ which appears in all cases.  
%Therefore, the signal $S_f = 0$ may not be used. 
From this table, the angular correlation of the final state $\Xi \;\Lambda$ with $S_f = 1$ may be used.  
The parity is related to orbital angular momentum, and hence to 
even or odd power of the $Y$-function that determines the angular dependence.  
For instance, for the $J^P = 2^+$ the $\uparrow \uparrow \sim \chi_{1,1}$ component in Eq.~(\ref{eq:16}) has $Y_{2 \mu}(\Omega)$ in its coefficient, which shows one-period dependence
while $0 \le \theta \le \pi$.
Contrary for $J^P = 2^-$ the coefficient of $\uparrow \uparrow$ is $Y_{1 \mu}(\Omega)$, 
and so shows half-period dependence on $\theta$.  
%The result in Table~\ref{table_Sf}
%shows that if these spin-angular correlations are measured
%the parity is determined.  

\begin{table}
\centering
\caption{Orbital angular momentum $L_f$ and spin $S_f$ in the final state $\Xi;\Lambda$.}
\vspace*{2mm}
\label{table_Sf}
\begin{tabular}{  c c c c c c c c}
\hline
$J^P$\ \   & $2^+$ & $2^-$  & $1^+$ & $1^-$ & $0^-$ & $3^-$ & \\
\hline
$L_f$\ \   & 2 & 1, 3 & 0,\ 2 & 1 & 1 & 3 & \\
$S_f$\ \   & 0,\ 1 & 1 & 1 & 0,\ 1 & 1 & 0, 1&\\
\hline
\end{tabular}
\end{table}

% flatex input end: [./k10docu_exp/k10-omega-n-spinparity_v2.tex]

%%OmegaN scattering
% flatex input: [./k10docu_exp/k10-omega-n-exp_v4.tex]
\subsubsection{Production of the $\Omega N$ bound state}
\begin{figure}[t]
  \begin{center}
  \includegraphics[width=15cm,keepaspectratio,clip]{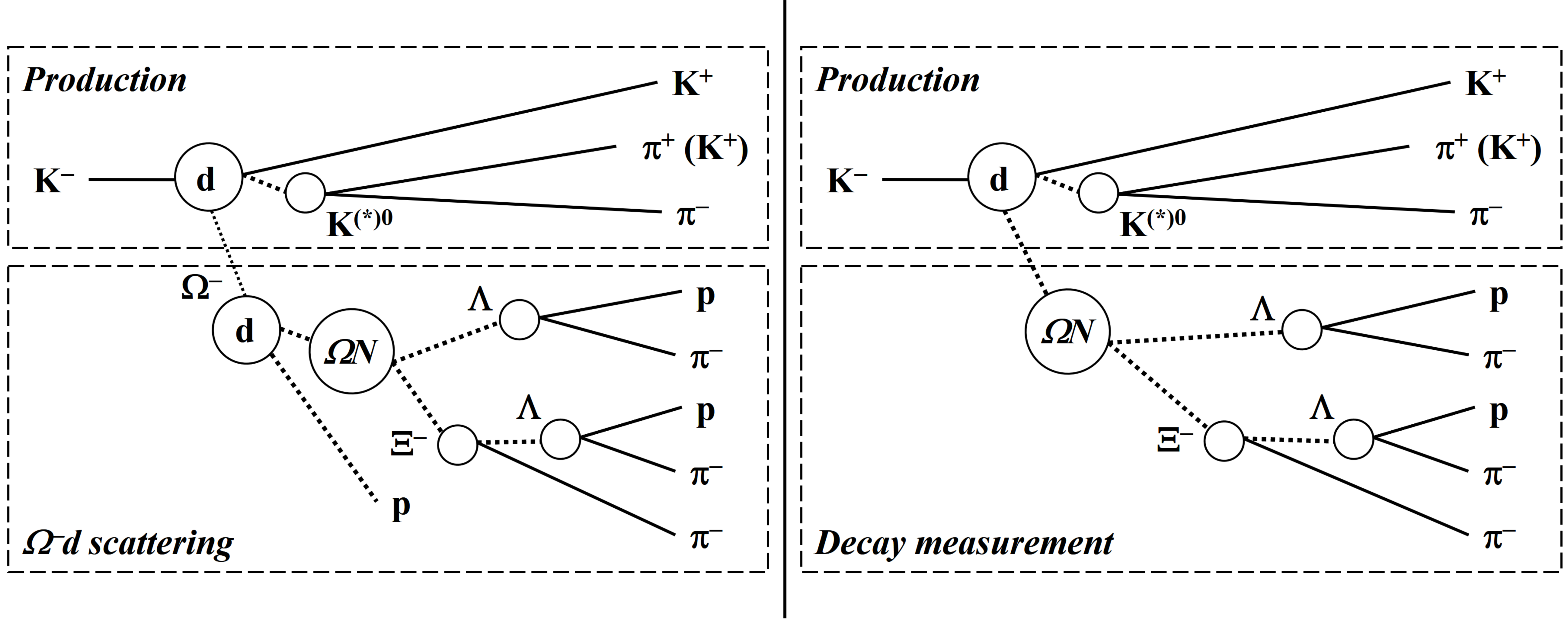}
  \end{center}
  \caption{Diagrams for producing the low-energy  $\Omega\; N$  system
including the $\Omega N$ bound state.
(a) $\Omega^-$ is produced from the deuteron in the $K^-\;d\to \Omega^-\;K^+\; K^{(*)0}\; n$ reaction, 
followed by the $\Omega^-\; n$ reaction on another deuteron
($\Omega^-\; d$ scattering).
(b) the $\Omega^-\; n$ system is directly produced
from the deuteron in the $K^-\;d\to \Omega^-\; n\; K^+\; K^{(*)0}$ reaction
(direct $\Omega^-\; N$ production).
The solid lines represent the initial- and final-state particles to be directly 
detected. 
The dotted lines show unstable particles to be reconstructed from kinematic
variables of the detected particles,
and spectator nucleons difficult to be detected.
  }
  \label{k10exp_fig106}
\end{figure}

Of particular importance for the $\Omega N$ interaction are the measurement of 
the low-energy $\Omega{N}$ scattering cross section,
and the determination of the mass and width of a possible $\Omega{N}$ 
bound state.
The two reactions are considered for these purposes here:
\begin{enumerate}
\item $\Omega^-$ is produced from the deuteron
in the $K^-\;d\to \Omega^-\;K^+\; K^{(*)0}\; n$ reaction,
followed by the $\Omega^-\; n$ reaction on another deuteron
($\Omega^-\; d$ scattering), and
\item $\Omega^- n$ is directly produced from the deuteron 
in the $K^-\;d\to \Omega^- \;n\; K^+\; K^{(*)0}$ reaction 
(direct $\Omega^-\; N$ production).
\end{enumerate}
Figure~\ref{k10exp_fig106} shows the diagrams of these reactions 
for producing the low-energy $\Omega\; N$ system including the $\Omega N$
bound state.
In $\Omega^-\; d$ scattering,
we select the $\Omega^{-}\; n \rightarrow \Xi^{-}\;\Lambda$ events
by detecting the final-state $p\;\pi^-$ particles from the $\Lambda$ decay
and $p \;\pi^- \;\pi^-$ from $\Xi^-\to \Lambda \;\pi^-$ followed by 
$\Lambda\to p\;\pi^-$.
It should be noted that 
a finite distance is observed in this case
between the production vertex of $\Omega^-$
and decay (or scattering) vertex of $\Omega \;N\to \Xi^-\;\Lambda$.
In direct $\Omega^-\; N$ production,
we measure the final-state $K^+\;K^{(*)0}$ particles for $\Omega\; N$ 
production, and the $\Xi^-\;\Lambda$ for the $\Omega\; N$ decay.
The analysis procedure is very similar to that for $\Omega^-\;d$ scattering
but the production vertex of $\Omega^-$ and decay 
vertex of $\Omega \;N\to \Xi^-\;\Lambda$ is common in direct $\Omega^- \;N$ production.
The total number of emitted particles  is 8 in the events to be analyzed.

\subsubsection{Requirements for the spectrometer}
To produce the low-energy $\Omega\; N$ system,
we need high-intensity high-momentum negative-kaon beam similarly
to $\Omega^*$ production.
At such incident kaon momenta, almost all the particles are likely to 
be emitted at forward angles in the laboratory frame.
A spectrometer system with a dipole magnet covering the forward direction 
is also suitable to detect the final-state particles in 
$\Omega\; N$ production.
The requirements of the spectrometer for the experiment 
producing $\Omega^*$'s is common to $\Omega\; N$ production.
However, the number of the final-state particles to be detected 
in $\Omega\; N$ production is larger than that in $\Omega^*$ production.
Additional items should be considered as follows:
\begin{itemize}
 \item the target is located inside the gap of the dipole magnet 
for maximizing the acceptance of the multiple-particle detection
with a wide angular coverage,
 \item additional detectors are required for surrounding the target 
to detect the daughter $\pi^-$'s effectively from the $\Omega^-\; n$ decay 
emitted at backward angles up to $\sim 90^\circ$ with a momentum lower than 1 GeV/$c$,
 \item the amount of substance in the additional detectors should be small
to avoid the multiple scattering and nuclear reactions of the slow $\pi$'s,
 \item precise vertex determination is required for 
recognizing finite flight lengths of $\Omega^-$, $\Xi^-$, and $\Lambda$
by detecting their daughter particles to reduce a combinatorial
background,
 \item the momentum resolution ($\Delta p/p$) is better than $0.2\%(\sigma)$
to achieve enough $\Omega N$-mass resolution of $\sim 1$ MeV$(\sigma)$
(comparable or better than the $\Omega^- n$ width), and
 \item large acceptance as long as possible and high momentum-resolution of a few $10^{-3}(\sigma) $
for the other high-momentum final-state $K^+\;K^+\;\pi^-$ ($K^+\;\pi^+\;\pi^-$) emitted 
at forward angles to give an $S=-3$ 
condition.
\end{itemize}
 
\begin{figure}[t]
  \begin{center}
  \includegraphics[width=15.5cm,keepaspectratio,clip]{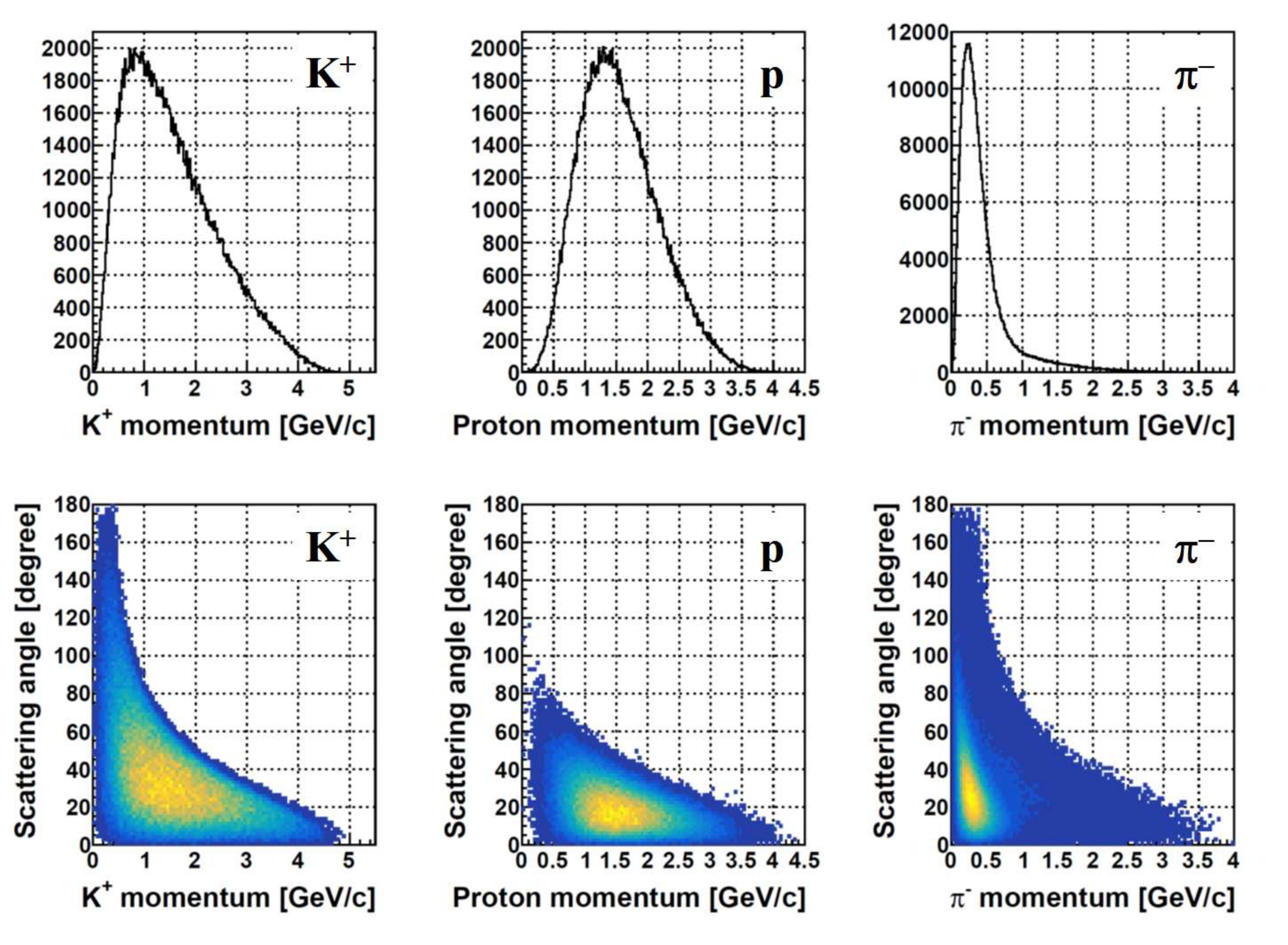}
  \end{center}
  \caption{Momentum distributions of the
final-state $K^+$, $p$ and $\pi^-$ particles 
from the left to right panels (top).
Correlation plots between the emission angle and momentum 
for $K^+$, $p$ and $\pi^-$ from the left to right panels (bottom).
}
  \label{k10exp_fig107}
\end{figure}

\subsubsection{Yield for the $\Omega N$ bound state}
As already discussed,
we select the $\Omega^{-}\;n \rightarrow \Xi^{-}\;\Lambda$ events
by detecting the final-state $p\;\pi^-$ particles from the $\Lambda$ decay
and $p\; \pi^-\; \pi^-$ from $\Xi^-\to \Lambda\; \pi^-$ followed by 
$\Lambda\to p\;\pi^-$ 
both in $\Omega^-\;d$ scattering and direct $\Omega^- \;N$ production.
A difference of two reactions are 
observed in the distance between
the production vertex of $\Omega^-$ and decay 
vertex of $\Omega N\to \Xi^-\Lambda$.
It is crucial to confirm each decay vertex
of $K^{(*)0}$, $\Xi^-$, and $\Lambda$ is consistent with a composite 
three-momentum and previous decay (or production) vertex
for reducing the combinatorial background.

We estimate the $\Omega N$-mass spectrum in the direct 
$\Omega N$-production case
at an incident kaon momentum of 7 GeV$/c$
in a similar simulation using the E50 spectrometer to that for $\Omega^*$ production.
Here, it is assumed that 
the possible $\Omega^-n$ bound state takes a Breit-Wigner shape, and 
that its mass and width is 2611 and 1.4 MeV, respectively,
Both the angular distributions assumed are isotropic 
in the CM frame for $\Omega N$ production and 
in the rest frame of $\Omega N$ for its decay into $\Xi\;\Lambda$.
The acceptance  estimated is $\sim 25\%$
for detecting $K^+\;K^{(*)0}$ associated with $\Omega\; N$ production,
and that is $\sim 42\%$ for detecting $\Xi^-\;\Lambda$ in the $\Omega\; N$ decay.
Therefore, the acceptance is $\sim$10\%  for detecting 
all the particles from $K^+\;K^{(*)0}$ and  $\Xi^-\;\Lambda$.
Figure~\ref{k10exp_fig107} shows the 
momentum distributions of the final-state 
$K^+$, $p$ and $\pi^-$ particles, and 
correlations between the emission angle and momentum for them.
The $K^+$'s and $\pi^-$'s are distributed in a wide angle range,
and those emitted at backward angles ($\theta_{\rm LAB}> 40^\circ$) 
are not detected with the E50 spectrometer.
The E50 spectrometer is not the most suitable for the study of the 
possible  $\Omega N$ bound state since the spectrometer
does not cover backward angles.
Desired is construction of a new spectrometer dedicated to this study
with additional detectors surrounding the target to increase the 
acceptance for the final-state particles emitted at backward angles.
In the yield estimation,
it is assumed that both the acceptances are 80\%
using a dedicated new spectrometer 
for detecting $K^+\;K^{(*)0}$ and for detecting $\Xi^-\;\Lambda$.

\begin{table}[t]
\caption{Different factors from Table~\ref{factors1}
in sub-subsection~\ref{sec:omega-spectroscopy-suppl1} (supplemental information)
for estimating the $\Omega^- d$-scattering events
at the incident beam momentum of 7 GeV/c.
}
\begin{center}
\begin{tabular}{lcc} \hline \hline
Reaction 
&  $K^{-}\;p \rightarrow \Omega^{*-}\;K^+\;K^{0}$
&  $K^{-}\;p \rightarrow \Omega^{*-}\;K^+\;K^{*0}$ \\  
Cross section & 2.0 $\mu$b & 0.05 $\mu$b \\
Acceptance &  0.80 &  0.80 \\
$\Omega^-$ rate in a spill & 10.0 & 0.48  \\ \hline
Total $\Omega^-$ rate in a spill
&  \multicolumn{2}{c}{10.5} \\
Cross section ($\Omega^-\; N$ scattering) &  \multicolumn{2}{c}{3.6 mb} \\
Branching ratio ($\Omega^-\;N \rightarrow \Xi^{-}\;\Lambda$) & \multicolumn{2}{c}{0.80} \\
Branching ratio($\Lambda \rightarrow p\;\pi^-$) & \multicolumn{2}{c}{0.64} \\
branching ratio ($\Xi \rightarrow \Lambda\;\pi^-$) & \multicolumn{2}{c}{0.99} \\
Average $\Omega^-$ flight length & \multicolumn{2}{c}{5.2 cm} \\
Effective target thickness & \multicolumn{2}{c}{ 0.8 g/cm$^2$} \\
Acceptance ($\Omega^-\; N \rightarrow \Xi^{-}\;\Lambda$)  & \multicolumn{2}{c}{0.80} \\
Total efficiency &\multicolumn{2}{c}{ 0.50 (five-track events)} \\ \hline
Expected yield & &\\
in a 100-day beam time & \multicolumn{2}{c}{2.0$\times$10$^3$} \\ \hline
\end{tabular}
\label{factors102}
\end{center}
\end{table}

The yield is estimated for the events that the $\Omega^-n$ bound state is produced
at the incident beam momentum of 7 GeV/c.
The numbers of the $\Omega^-$-produced events
are 10.0 and 0.48 in a spill
in the $K^{-}\;p \rightarrow \Omega^{-}\;K^+\;K^{0}$ and 
$K^{-}\;p \rightarrow \Omega^{-}\;K^+\;K^{*0}$ reactions
with cross sections of 
2.0 and 0.05 $\mu$b, respectively.
Once the produced $\Omega^-$'s are identified, they can be used as incident particles
with an intensity of 10.5/spill
to produce the $\Omega^-n$ bound state.
The total cross section assumed is 3.6 mb
for the $\Omega^-n$ reaction.
The branching ratio (fraction) of $\Omega^-\;n\to \Xi^-\;\Lambda$
assumed is 0.80.
The average flight length of the produced $\Omega^-$'s 
is $\beta \gamma c \tau \sim 5.2$ cm
at an incident kaon momentum of 7 GeV/$c$
since their average momentum is 3.5 GeV/$c$.
Thus, the effective mass-thickness of the deuteron target is 0.8 g/cm$^2$.
The total efficiency is 0.50 for five-track events where 
the in-flight decay of $\pi^-$'s is not considered.
The production rate is found to be $\sim 20$ a day
 for the events that the $\Omega^-n$ bound state is produced.
We can detect $\sim 2,000$ events in a 100-day beam time.
Additional factors are listed in Table~\ref{factors102}
for estimating the yield for the events that
the $\Omega^-n$ bound state is produced.
In this estimation, direct $\Omega^-n$ production is not included.
If this cross section is $\sim$0.3 nb, the yield of $\sim$1,000
can be additionally obtained.

\subsubsection{Mass resolution for the $\Omega N$ bound state}
\begin{figure}[t]
  \begin{center}
  \includegraphics[width=13cm,keepaspectratio,clip]{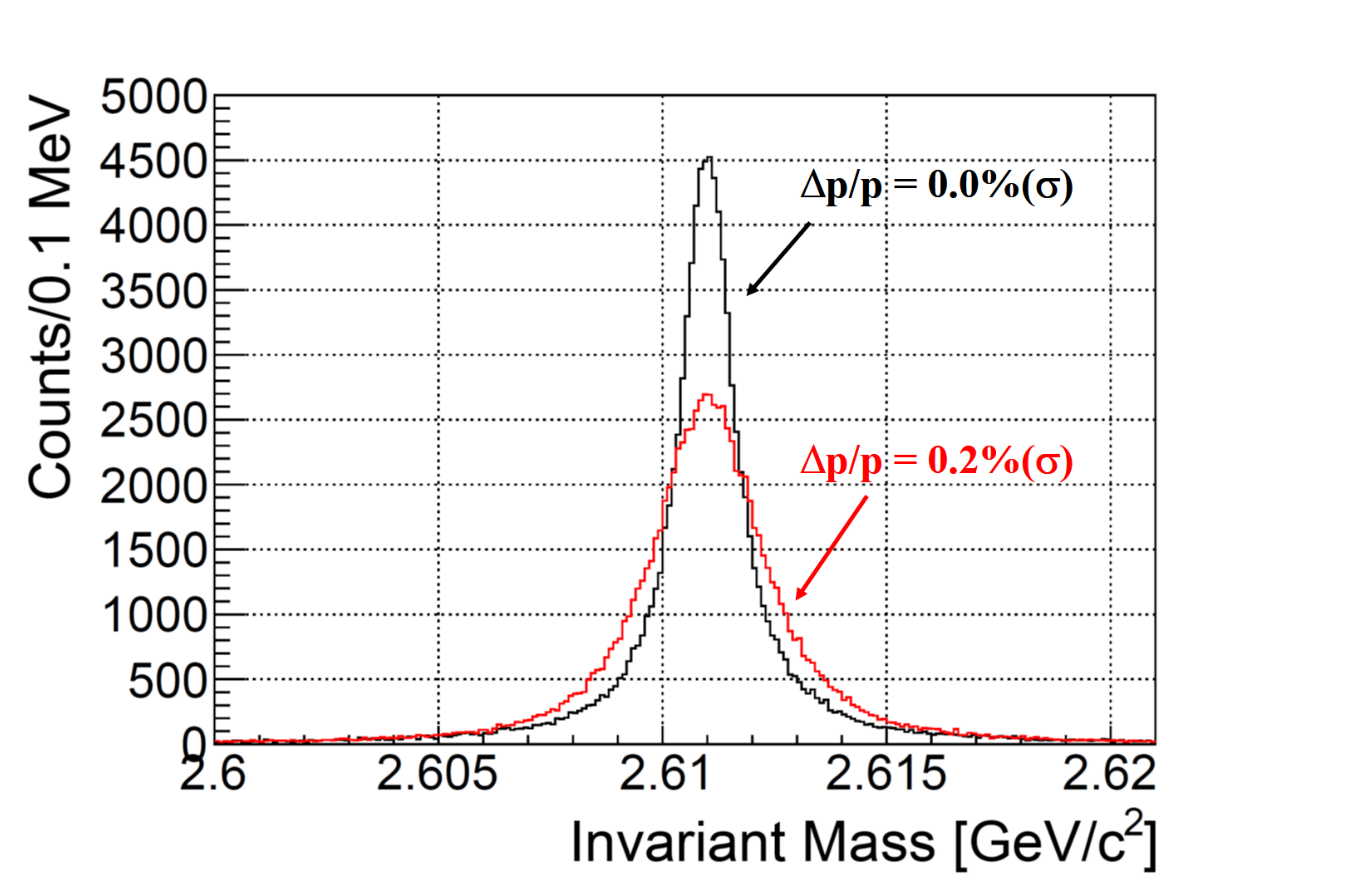}
  \end{center}
  \caption{$\Xi^-\;\Lambda$ invariant-mass spectrum for 
the $\Omega^- n$ bound state. The black curve corresponds to the spectrum
for the expected $\Omega^- n$ bound state,
and the red one shows that with incorporating the experimental mass resolution.}
  \label{k10exp_fig108}
\end{figure}

\begin{figure}[t]
  \begin{center}
  \includegraphics[width=15.5cm,keepaspectratio,clip]{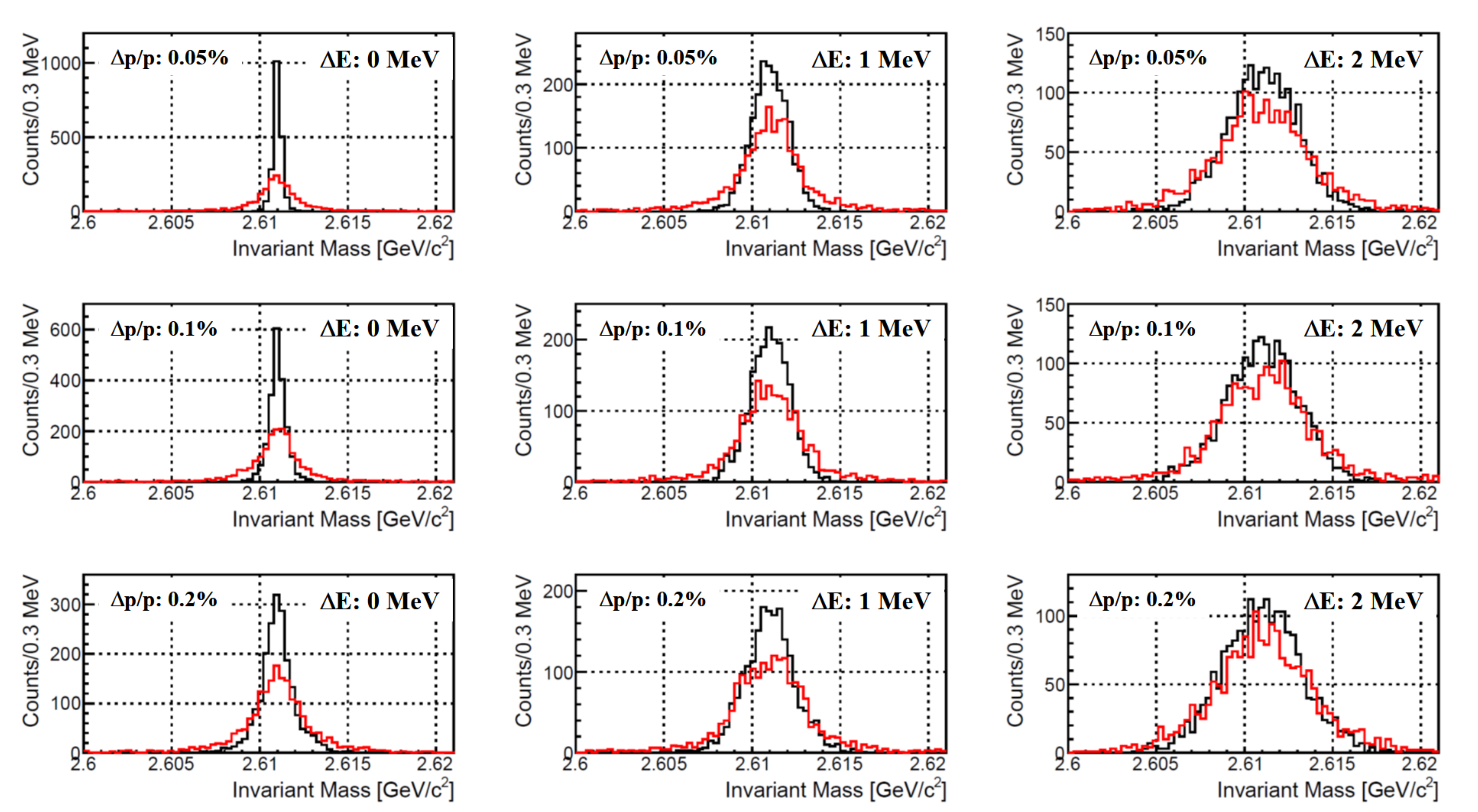}
  \end{center}
  \caption{
$\Xi^-\;\Lambda$ invariant-mass spectra for 2,000 events of the $\Omega^- n$ bound state with different momentum resolution ($\Delta p_{\rm spec}/p_{\rm spec}=
0.05\%$, 0.1\%, and 0.2\% $(\sigma)$ from the top to the bottom)
and 
different straggling effects of the energy loss in the target 
material (0, 1, and 2 MeV($\sigma$) from the left to the right).
In each panel, the black curve shows the measured spectrum for the 
$\Omega N$ bound state with a width of 0,
and the red for the state taking a Breit-Wigner shape with a width of 1.4 MeV.
}
  \label{k10exp_fig109}
\end{figure}

The $\Omega^-n$-mass resolution  is estimated assuming 
the $\Omega^-n$ bound state is observed in the $\Xi^-\Lambda$ invariant mass
with a momentum resolution of $\Delta p_{\rm spec}/p_{\rm spec}=0.2\%$($\sigma$) for
the final-state particles.
Figure~\ref{k10exp_fig108} shows the $\Xi^-\;\Lambda$ invariant-mass spectrum for 
the $\Omega^- n$ bound state.
Here, $10^4$ events are generated for the $\Omega^- n$ bound state.
The combinatorial background coming from wrong combinations of the final-state 
particles for reconstructing unstable particles.
The mass resolution estimated for the $\Omega^-n$ state is $\sim 1$ MeV ($\sigma$) when  the momentum resolution is $\Delta p_{\rm spec}/p_{\rm spec}=
0.2\%(\sigma)$ for the final-state particles.
The observed width in the mass spectrum becomes 1.6 MeV($\sigma$)
by incorporating the 1.4-MeV width of the $\Omega^-n$ bound state.

The straggling effects of the energy loss in the target material also deteriorate
the $\Omega^- n$-mass resolution.
Figure~\ref{k10exp_fig109} shows
the 
$\Xi^-\;\Lambda$ invariant-mass spectra for 2,000 events of the $\Omega^- n$ bound state with different straggling effects and 
different momentum resolutions.
Plotted are the spectra with widths of 0 and 1.4 MeV
for the  $\Omega^- n$ bound state.
The broadening by the momentum resolution
of  $\Delta p_{\rm spec}/p_{\rm spec}=0.2\%(\sigma)$ 
is equivalent to that by the $\Omega^-n$ width of 1.4 MeV.
The momentum resolution is required to be better than 
 $\Delta p_{\rm spec}/p_{\rm spec}=
0.2\%(\sigma)$ for the final-state particles 
to recognize the broadening of the $\Omega^-n$ peak owing to the 
width of the $\Omega^- n$ bound state.
The broadening of the $\Omega^-n$ peak caused by
the width of the $\Omega^- n$ bound state
can be observed 
even if the straggling effect is 1 MeV($\sigma$) of the energy loss.
However, it is difficult to determine the width of the $\Omega^-n$ bound state
if  the straggling effect becomes 2 MeV($\sigma$) of the energy loss.
The straggling effect should be lower than 1 MeV$(\sigma)$.

\begin{figure}[t]
  \begin{center}
  \includegraphics[width=14cm,keepaspectratio,clip]{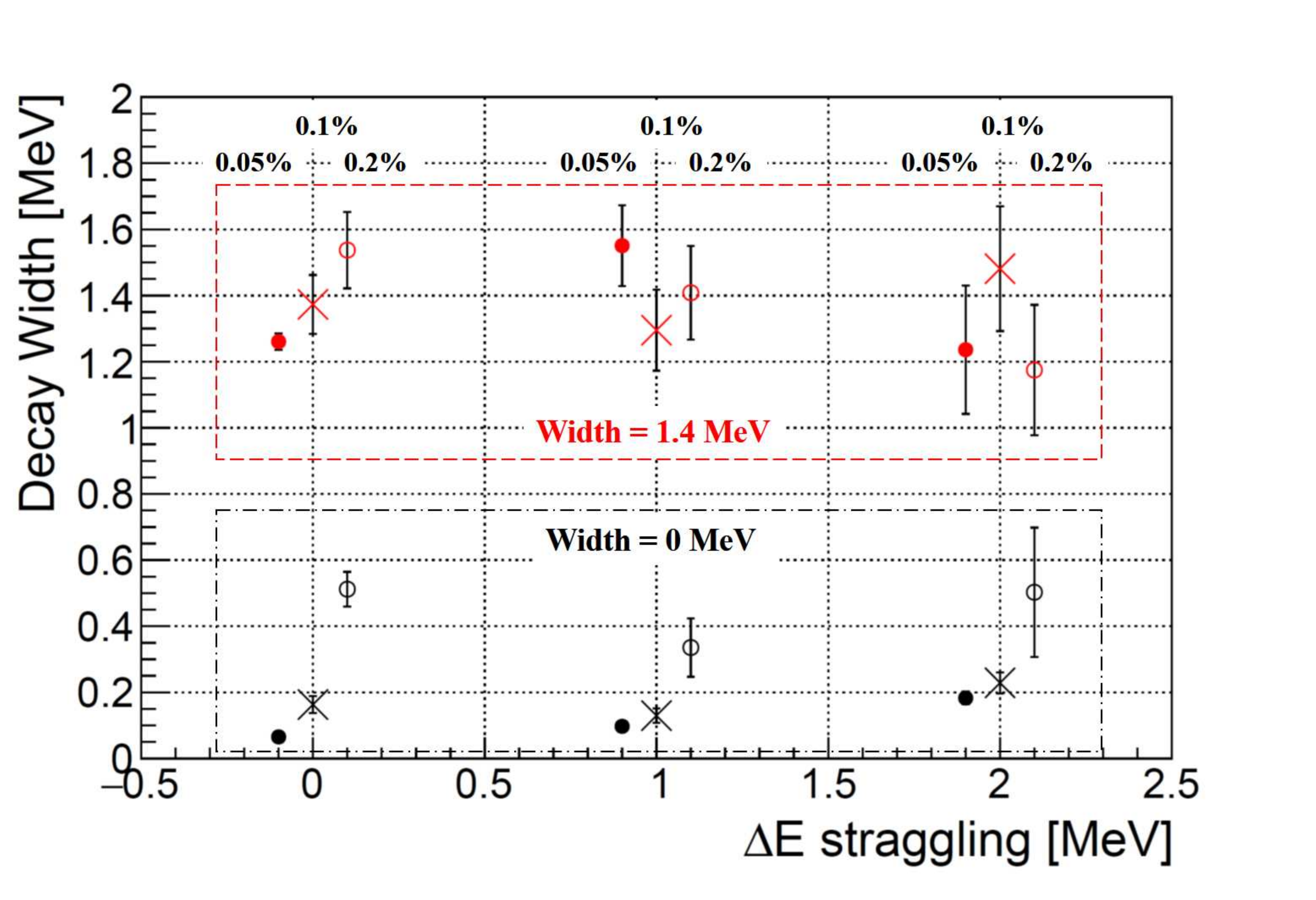}
  \end{center}
  \caption{Measured width for the $\Omega^-n$ bound state 
as a function of the straggling effect of the energy loss in the target material
(0, 1, and 2 MeV ($\sigma$)).
The filled circles, X markers, and open circles 
represent the data for $\Delta p_{\rm spec}/p_{\rm spec}=0.05\%$, 
0.1\%, and 0.2\% $(\sigma)$, respectively. 
The black and red makers represent the data for the state with widths 
of 0 and 1.4 MeV.  }
  \label{k10exp_fig110}
\end{figure}

By fitting a Breit-Wigner function convoluted by a Gaussian 
to the  $\Omega^-n$-spectra
shown in Fig.~\ref{k10exp_fig109},
the measured widths for the $\Omega^-n$ bound state
are  estimated for different straggling effects of the energy loss,
and for different momentum resolutions for the final-state particles.
Figure~\ref{k10exp_fig110} shows 
the measured width for the $\Omega^-n$ bound state 
as a function of the straggling effect of the energy loss.
Required are high momentum-resolution better than $\Delta p_{\rm spec}/p_{\rm spec}=0.2\%$
($\sigma$)  and small straggling effect of the energy loss less than 1 MeV($\sigma$).
When the momentum resolution is  $\Delta p_{\rm spec}/p_{\rm spec}=0.2\%$
and the straggling effect of the energy loss is 1 MeV($\sigma$),
the accuracy of the width determination is found to be $\pm0.1$ MeV ($\sigma$).

%%%%%%%%%%%%%%%%%%%%%%%%%%%%%%%%%%%%%%%%%%%%%%%%%%%%%%%%%%%%%%%%%%%%%%%%%%%%%%%%%%%%%

% flatex input end: [./k10docu_exp/k10-omega-n-exp_v4.tex]

%%OmegaN scattering

\clearpage
% flatex input: [./k10docu/k10world-situation.tex]
\subsection{World Situation}

{\bf Collider Experimental Facility}\\

From a global perspective, the hadron spectroscopy at the collider experiment has played an important role in hadron physics.
Data collection is over, but SLAC National Accelerator's PEP-II / BaBar experiment and KEKB / Belle experiments have discovered many excited states of hadrons including some exotic hadron candidates.
In addition,  the LHCb experiment which is one of the proton-proton collision experiments at the European Organization for Nuclear Research (CERN), is currently in operation. Many excited and exotic hadrons such as Pentaquark state including charm quark, $\Omega_c $ baryon excited state, and $\Xi_{cc} $ with two charm quarks have been discovered. Also, in the Beijing BES III experiment, many interesting results have been obtained not only exotic hadrons but also more exotic states, such as the antiproton-proton bound states.
Nowadays, the next-generation electron-positron collider experiment, i.e., the SuperKEKB/Belle II  experiment,  has started to take data. Reports of the discovery of new hadrons including exotic hadrons will expect to be coming near future.\\
\

\noindent
{\bf Jefferson National Accelerator Laboratory(JLab)  at US }\\

Nowadays, the upgraded project of the JLab, increasing the electron beam energy up to 12 GeV, has been completed. This upgrade of the facility opens a new scientific project on the JLab physics programs, i.e., the further elucidation of the nucleon structure and photo-production of hadrons with charm quarks. The three existing experimental facilities (Hall-A, Hall-B, Hall-C), new experimental apparatus, Glux at Hall-D, have been constructed.
One of the main goals for the Glue-X experiment is to search for exotic hadrons, such as glueballs and hybrid mesons.  Also, in the Hall-D, a large amount of $ K ^ 0_ {L} $ meson production using high-intensity gamma rays generated at the upstream of Hall-D has been planned. The experimental proposal of the spectroscopy of baryons with strangeness $ S = -1 $ and $ S = -2 $ by utilized high intensity $ K ^ 0_ {L} $ meson has been discussed.\\
\

\noindent
{\bf International Antiproton Heavy Ion Research Facility} \\

In Germany, the construction of the International Anti-proton Heavy Ion Research Facility (FAIR) is underway, led by the German National Institute for Heavy Ion Research (GSI).
The beam will be accelerated and delivered to the experimental hall after 2025, and the experiment will start.\\
\

\noindent
{\bf PANDA}\footnote{The antiProton ANnihilations at DArmstadt}\\

The PANDA experiment is aiming to reveal quarks and gluons dynamics that exhibit strong interaction non-perturbative behavior.
A ring HESR that cools and stores antiprotons from 1.5 GeV/$c$ to 15 GeV/$c$ generated from a proton beam that is delivered from the SIS100 accelerator. By using a cluster or pellet targets placed on HESR, hadron spectroscopy with high resolution (Beam momentum resolution $ 0.4-2.0 \times 10^{-4}$) and  high luminosity ($ <~ 2 \times 10^{32}$/cm$^2$/s) have expected to be achieved.
A multi-purpose detector system consists of two parts: a barrel detector and a forward detector. The barrel detector consists of
a Solenoid magnet installed with cylindrical charged particles and a photon-detector. The forward detector has a magnetic spectrometer in front of the muon detector. The spectrometer system can operate at reaction rates up to 20 MHz and measure a wide range of final states.  The main topics are the production of hadrons, including exotic hadrons, and the elucidation of their internal structure. Nucleon structure through the measurement of form factors with a time-like electromagnetic probe. Moreover, Hadron formation and hadron spectroscopy in the nuclear matter to investigate the partial restoration of chiral symmetry in the nucleus.
Besides, hypernucleus generation and its spectroscopic studies are planned. 

% flatex input end: [./k10docu/k10world-situation.tex]

%%OmegaN scattering

\clearpage
%Supplemental information
\subsection{Supplemental Information}\label{sec:suppl}
% flatex input: [./k10docu/k10SpinDepInt.tex]
%--------------------------------------
\subsubsection{Spin dependent interactions\label{sec:spin-dep-int}}
%--------------------------------------

One of the longstanding mysteries in hadron physics is in the (hyper-)fine structure of spectrum
associated with the spin-spin ($ss$) and spin-orbit ($LS$) interaction.  
The difficulty in the  spin-spin interaction has been in the attempt to explain the decuplet-octet mass difference such as 
$N\Delta$ mass splitting, $\Delta M = M_\Delta - M_N \sim 300$ MeV.  
%Having the ``ordinary path", one would expect 
%the one-gluon exchange (OGE) force between the constituent quarks 
%as one possible source of the splitting.  
One would expect a source of this splitting 
is 
the one-gluon exchange (OGE) force between the constituent quarks.
It is possible but
%But then one need to set 
requires too large a coupling constant $\alpha_S > 1$, 
%and if so
preventing from 
perturbation calculations.
%is no longer available.  

The difficulty in the spin-orbit interaction is in the almost degenerate  $LS$ multiplets such as 
$S_{11}(1535)$ and $D_{13}(1520)$
--- denoted by $N(1535)1/2^-$ and $N(1520)3/2^-$, respectively, in Review of Particle Physics~\cite{Zyla:2020zbs}. 
Here the notation is $L_{2I\, 2J}$ where $L$ is the orbital angular momentum of the scattering pseudoscalar meson and baryon which form the resonant state, 
$I$ is the isospin and $J$ is the total angular momentum
(For hyperons with an integer $I$, the notation is $L_{I\, 2J}$).
The situation is similar in other expected multiplets as shown in Fig.~\ref{fig_systematics}.
However, the use of the $LS$ interaction derived from the OGE
leads to large splittings of order 100 MeV or more~\cite{Takeuchi:1998mv}. 

%{\color{red}
%（ここらに、メソンとバリオンのSS力のフレーバー（クォーク質量）依存性の図とバリオンにおけるLS力のフレーバー（クォーク質量）依存性の図（←メソン系ではどうなっているのだろうか）を配置してはどうか。）
%}

Let us look at the above situation in a little bit more detail.  
As in QED, the interaction mediated by OGE decomposes into Coulomb (color-electric), 
spin-spin (color-magnetic) and spin-orbit interaction terms 
(and further terms~\footnote{
In addition to these terms, there are antisymmetric $LS$ and tensor terms.}) 
in the non-relativistic expansion for the interaction.  
The corresponding potentials are written for the color $\bar 3$ quark pair ($ij$)~\cite{Yoshida:2015tia}, 
\begin{eqnarray}
V^{\rm Coul}_{ij}(r_{ij}) &=& - \alpha_S^{\rm Coul} \frac{2}{3r_{ij}}
\nonumber \\
V^{ss}_{ij}(r_{ij}) &=& \alpha_S^{ss} \frac{16\pi}{9m_i m_j} \delta(r_{ij}) \vec s_i \cdot \vec s_j
\nonumber \\
V^{LS}_{ij}(r_{ij}) &=& \alpha_S^{LS} \frac{1}{3r_{ij}} 
\left(
\frac{1}{m_i^2} + \frac{1}{m_j^2}  + \frac{4}{m_i m_j} 
\vec L_{ij} \cdot (\vec s_i + \vec s_j)
\right)
\end{eqnarray}
A single color coupling constant would have to determine the strengths of the three terms, 
$\alpha_S^{\rm Coul} = \alpha_S^{ss} = \alpha_S^{LS}$, 
but actually phenomenological fitting requires that they are different, 
e.g., 
$\alpha_S^{\rm Coul} \sim 0.6$, 
$\alpha_S^{ss} \sim 1.2$ and 
$\alpha_S^{LS} \sim 0.08$.
The strength for the Coulomb part is qualitatively consistent with the running coupling constant 
$\alpha_S(Q)$ at $Q \sim 1$ GeV while the other two are considerably different from it.
These values indicate the anticipated problems; 
the spin-spin interaction $\alpha_S^{ss}$ is too large, while the spin-orbit interaction 
is suppressed.  
It was argued that the instanton induced interaction (III) can help to fill the discrepancy.
The III contributes to the spin-spin interaction with the same sign as that of the OGE.
It was shown that when the strength of III was fixed to reproduce the $\eta$-$\eta^\prime$ mass difference, 
about 40 \% of the phenomenological strength of 
$\alpha_S^{ss}$ was brought by  III.  
The III also contributes to the spin-orbit interaction which is now with opposite sign to that of OGE.  
Using the above $\alpha_S$ values, with the 40 \% of $\alpha_S^{ss} \sim 1.2$ added to 
$\alpha_S^{\rm Coul} \sim 0.6$, we would find $\alpha_S^{\rm Coul} - 0.4 \alpha_S^{ss} \sim 0.1$ 
which agree qualitatively with the phenomenologically determined value $\alpha_S^{LS} \sim 0.08$. 

The above examples are the evidences of the III interaction that may resolve the mystery 
in the fine structure of baryon spectrum.
It is shown that the OGE and III depend differently on quark flavors.
Therefore, systematic studies over the wide range of flavors including $\Omega$ baryons
will solve the unsolved/unconfirmed issues.  

\subsubsection{Baryon with a single heavy quark and two-light quarks \label{sec:baryonshf}}
\begin{figure}[htbp] 
 \centerline{\includegraphics[width=0.8\textwidth]{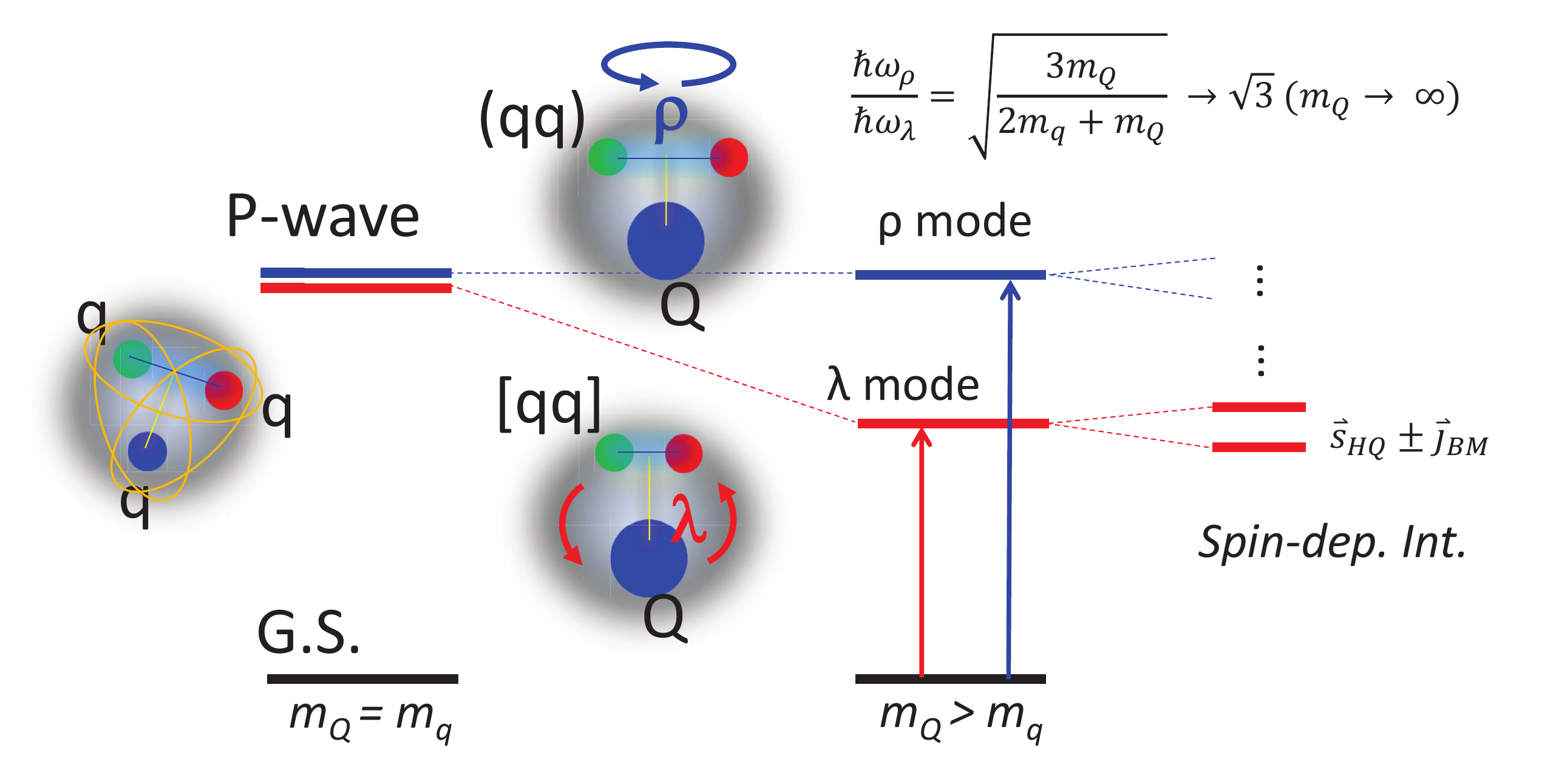}}
 \caption{Schematic illustration of the orbital excitations for the $\lambda$ and $\rho$ modes in a baryon with a heavy quark. The $\lambda$ and $\rho$ modes are degenerated in a light baryon. \label{fig:hq}}
\end{figure}

%----------------------------------------------------
To study dynamics of light degrees of freedom such as the light constituent quarks and diquarks, it is helpful to introduce a heavy flavor in a baryon, where the heavy quark behaves as an inert particle. In excited states, a collective motion of two light quarks to a heavy quark ($\lambda$ mode) is separated in kinematics from a relative motion between the two light quarks ($\rho$ mode), as shown in Fig~\ref{fig:hq}. The orbital excitation energy of the $\lambda$ mode becomes lower than that of the $\rho$ mode by a factor $[3m_Q/(2m_q+m_Q)]^{1/2}$ approximately, where $m_Q$ and $m_q$ are masses of the heavy and light quark, respectively. It is known as the so-called isotope shift. 

%{\color{red}[$qq$](0$^+$)と($qq$)(0$^-$)が$U_A$(1) anomalyを反映する状態であることに関した記述を加える。}

%--------------------------------------
%--------------------------------------

It is known that the QCD has the so-called heavy quark spin symmetry. Namely, the heavy quark spin is conserved in the heavy quark mass limit ($m_Q\rightarrow\infty$).
%, where the heavy quark spin operator ($s_Q$) and the Hamiltonian of a hadronic system including heavy quarks commutes. 
Since the color-magnetic (color-spin) interaction between quarks is proportional to the inverse of the relevant quark masses, spin-dependent interactions to the heavier quark become weaker as the quark mass increases, being zero in the heavy quark mass limit. As a result, the heavy-quark spin ($s_Q$=1/2) as well as a baryon spin ($J$), and hence the rest spin (the so-called Brown-muck spin $j_{b} = J-s_Q$), become ``good" quantum numbers. Then, the heavy-quark spin doublet states are formed, which are degenerated in the heavy quark limit. These behaviors of baryons characterized by heavy quarks should be reflected in the level structure, the production rates, and the decay properties. Through spectroscopy of excited baryons, we could investigate the nature of baryons over various flavors with respect to the quark masses. 

\begin{figure}[htbp] 
 \centerline{\includegraphics[width=0.9\textwidth]{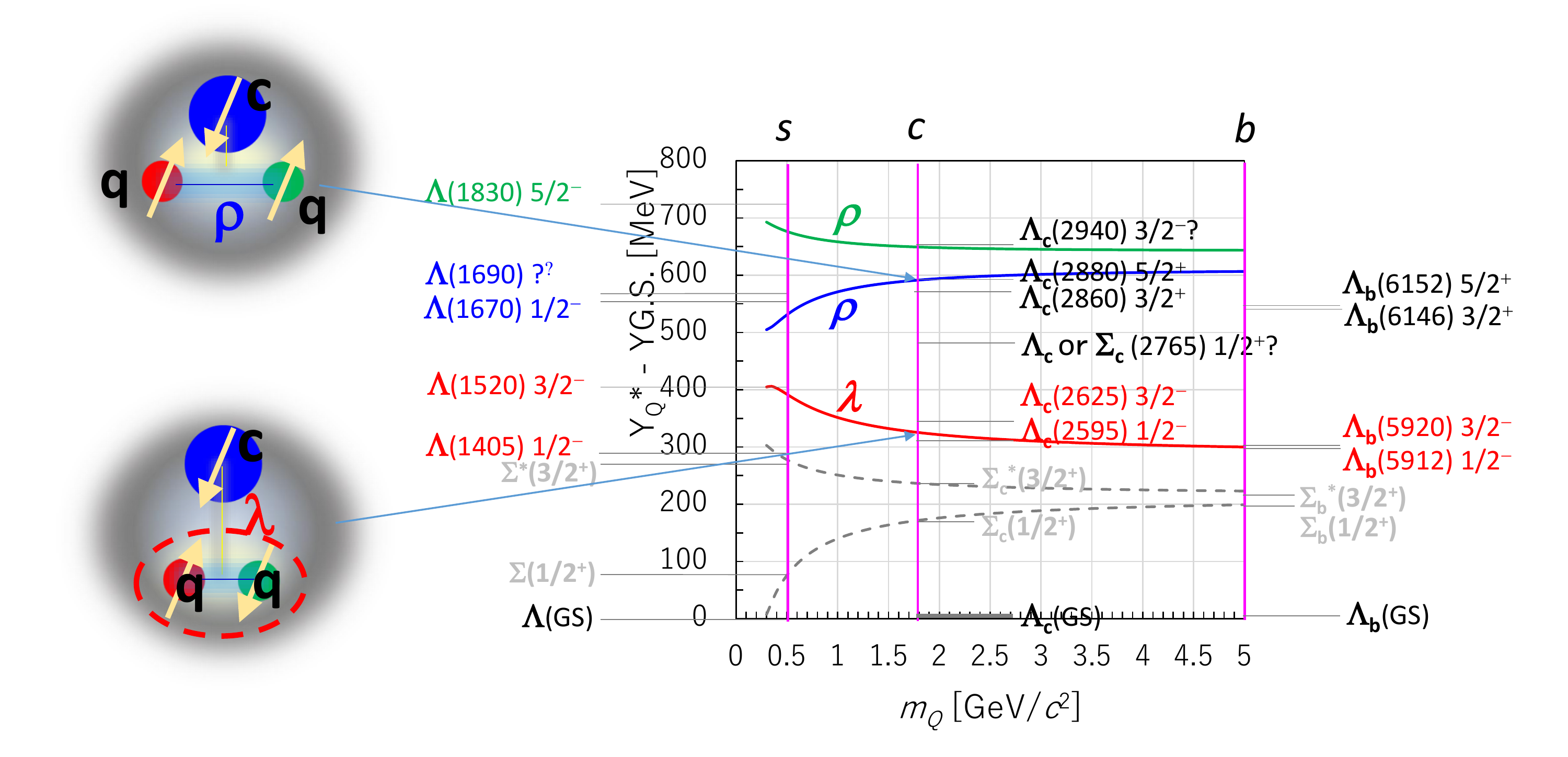}}
 \caption{Excitation energy measured from the ground state $\Lambda_Q$ in the cases of $\Lambda_Q$ baryons based on a constituent quark model (curves) are illustrated together with the observed states reported in PDG. \label{fig:ex-lambda}}
 \vspace{5mm}
 \centerline{\includegraphics[width=0.9\textwidth]{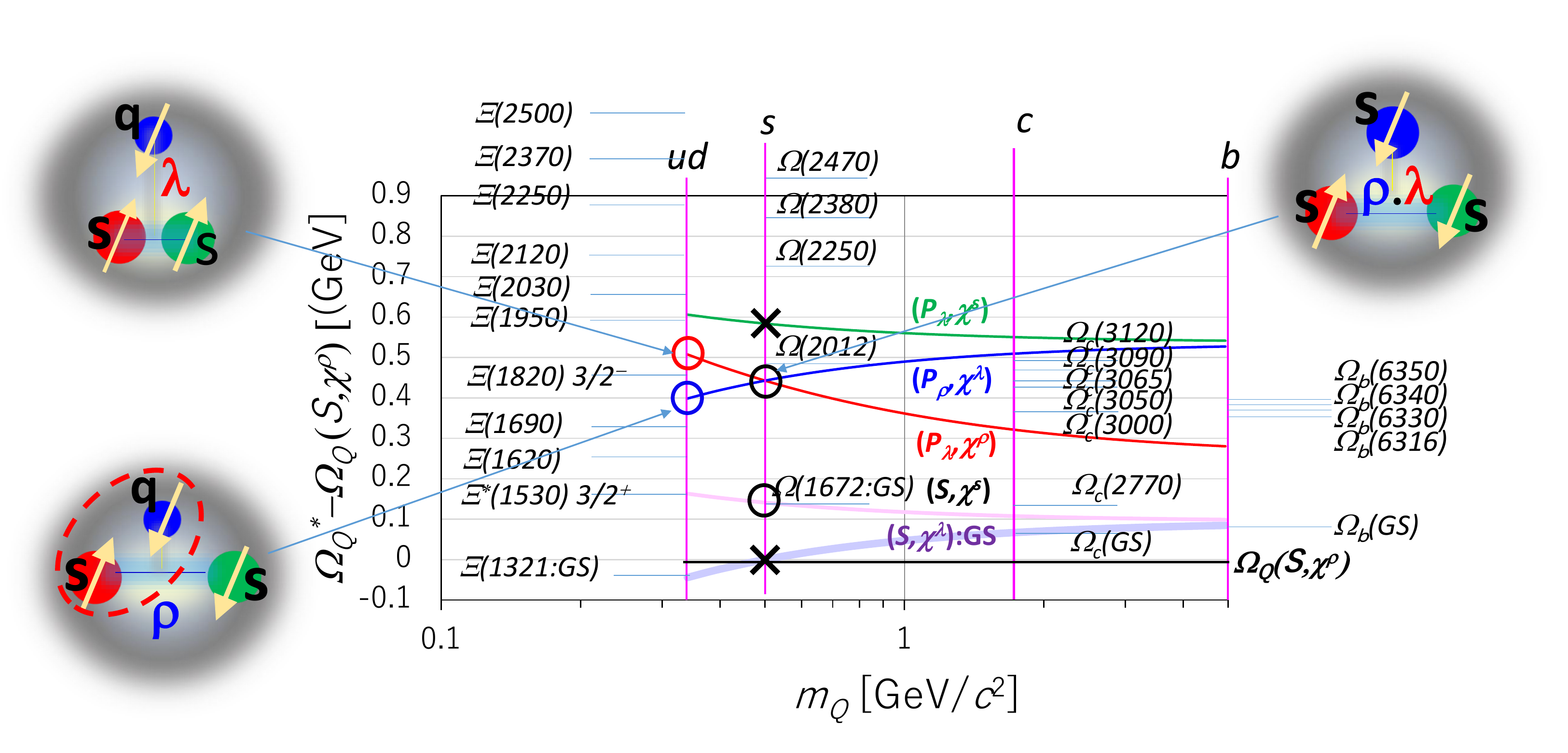}}
 \caption{Excitation energy measured from the forbidden states indicated as $\Omega_Q(S,\chi^\rho)$ in the cases of $\Omega_Q$ baryons based on the constituent quark model (curves) are illustrated together with the observed states reported in PDG. Note that in the cases of $Q=u$ or $=d$ $\Omega_Q$ is nothing but $\Xi$. \label{fig:ex-omega}}
\end{figure}

We have shown schematically how the excitation energies of the $\lambda$ and $\rho$ modes change as a function of the mass of the heavy quark $Q$ in  Fig.~\ref{fig_baryonspecies}. Here again, we demonstrate a bit more in detail how the level structure of baryons behaves as the quark masses. Fig.~\ref{fig:ex-lambda} shows a level scheme of the excitation energies in the cases of $\Lambda$ baryons as a function of the heavy quark mass ($m_Q$),  where a $\Lambda$ baryon can be described as a system of a single heavy quark and the two light quarks, $Qqq$, with the isospin equal to zero. 
In the figure, the excitation energies are plotted as mass differences to the ground states of $\Lambda_Q$'s.
Lines are calculated by a non-relativistic quark model \cite{
%prd92:yoshida
Yoshida:2015tia} as changing $m_Q$, each of which shows the average of the spin-orbit splitting states. 
The $\lambda$ mode is lower as $m_Q$ increases.
Two $\rho$-mode lines are indicated. The lower (blue) line corresponds to the spin-antisymmetric $ud$ state with its relative angular momentum of  1 ($^1P_0$). The other (green) line is for the state that all the quark spins are symmetric and the $ud$ relative angular momentum is 1 ($^3P_0$), and thus the $ud$ isospin is antisymmetric.
Dashed (grey) lines are the cases of the $ud$ isospin symmetric state in S-wave, which correspond to octet $\Sigma_Q(1/2^+)$ and decuplet $\Sigma_Q(3/2^+)$, respectively.
The energy splittings between them are due to the spin-spin interaction between the quarks.
As the spin-spin interaction are proportional to the inverse of the relevant quarks masses, the dashed lines are degenerated in the limit of $m_Q\rightarrow\infty$ (heavy quark limit). This is due to the heavy quark symmetry.
The two $\rho$-mode lines are also degenerated in the heavy quark limit due to the heavy quark symmetry.

%{\color{red} Here, only two parameters for the confinement and spin-spin interaction potential are determined by the $\Lambda$ and $\Sigma$ baryons. }
They are good guide lines to see relations between excited states with the same spin-parity over the 
These two $\rho$-mode lines are degenerated due to the heavy quark symmetry in the limit of $m_Q\rightarrow\infty$.
various flavors as compared with observed states reported in PDG~\cite{Zyla:2020zbs} as shown in the figure.

One finds that the $\lambda$-mode line fits well the averaged masses of $[M(\Lambda_c(2595)1/2^-) + 2M(\Lambda_c(2625)3/2^-)]/3$ and $[M(\Lambda_b(5912)1/2^-)+2M(\Lambda_c(5920)3/2^-)]/3$, where $M$ represents the mass of the state indicated in parenthesis. They are believed to be the so-called $LS$ partners with their internal configurations of an orbital angular momentum of 1 between $c$ and $ud$ and $ud$ is the so-called "good" diquark with both spin and isospin equal to 0. In strangeness sector, the $\Lambda(1405)1/2^-$ state, which is thought to be a $\bar{K}N$ bound state, is lower than the $\lambda$-mode line. In E50, we will measure the production cross sections of $\Lambda_c(2595)$ and $\Lambda_c(2625)$ in the $\pi^-p\rightarrow D^{*-}Y_c^*$ reactions, as shown in Fig.~\ref{k10_charm_fig2}. The cross section ratios are expected to be $L:L+1=1+2$ if they are the $\lambda$-mode states. By measuring the cross sections, the spin-parities of these states are experimentally determined.

%--------------------------------------
%--------------------------------------

% flatex input end: [./k10docu/k10BaryonHF.tex]

%Supplemental information
% flatex input: [./k10docu/k10BaryonW2strange.tex]
%----------------------------------------------------
%\subsubsection{Baryon with a single heavy quarks with two-strange quarks}
%----------------------------------------------------
\subsubsection{Baryon with a single heavy quark and two-strange quarks \label{sec:baryonsqss}}

Now let us turn to the systems of $Qss$, $\Omega_Q$ baryons.
In Fig.~\ref{fig:ex-omega}, we  draw 5 lines for excited S- and P-states, connecting the states with having the same spatial (parity) and spin configuration but different masses (flavors). 
Excitation energies for $\Xi$, $\Omega$, $\Omega_c$, and $\Omega_b$ reported by the Particle Data Group (PDG) \cite{Zyla:2020zbs} are overlaid for reference. 
Most of their quantum numbers have not been determined. 
The behaviors of 5 lines are similar to the cases in the $Qqq$ system. The internal configuration of the $Qss$ system is constraint due that the $ss$ pair must be flavor-symmetric. In fact, the S-wave $Qss$ systems with the $SS$ spin-singlet is forbidden, however their masses can be numerically evaluated. It should be noted that the excitation energies shown in Fig.~\ref{fig:ex-omega} are relative to the forbidden states (indicated as $\Omega_Q(S,\chi^\rho)$ in the figure) as one can recognize the similarity of $Qss$ to $Qqq$.  
From Fig.~\ref{fig:ex-omega}, we observe unique features.
The excitation energy of the $\rho$ mode is lower than that of the $\lambda$ mode in $\Xi$. The $\Xi(1820)3/2^-$ state is a candidate of the $\rho$-mode excited states. If so, its decay into $\bar{K}\Lambda$ or $\bar{K}\Sigma$ is expected to be dominant.
One of the $Qs$ pair in the $\rho$-mode $\Xi$ state, where $Q$ is $u$ or $d$, is expected to be the so-called "good" diquark, the spin singlet state, and thus the state reflects the $ds/su$ diquark correlation.
Several $\Xi$ states are observed but little of spin-parities and decay branching ratios are known. 
At K10, we will populate the $\Xi$ states in the $K^-p\rightarrow\Xi^*\bar{K}^{(*)}$ reactions, as shown in Fig.~\ref{k10_xi_fig2}. We will determine the excitation modes and properties of the excited $\Xi$ states by measuring their decay branching ratios and angular distributions.

$\Omega$ is made of only $s$ quarks. The level structure of $\Omega$ baryons becomes much simpler due to flavor symmetry. 
The $\lambda$ and $\rho$ modes are degenerated 
(the point of the upper circle at $Q = s$), and  forbidden modes at $Q = s$ 
due to the Pauli exclusion principle (crosses at $Q = s)$.  
These observations are not yet established experimentally that can be primarily studied in the K10 project. 
In addition, there is further unique features 
in the $Qss$ $\Omega_Q$ baryons as we discuss below.%in the next subsection.  

The systems $Qss$ with no $u$, $d$ quarks that we have discussed so far are considered very different from the protons and neutrons, where pion clouds surround the nucleon forming the tail of the structure. Hence we have a picture for the nucleon, the quark core and the pion cloud around it. The pion of light mass is the necessity of spontaneous symmetry breaking of chiral symmetry and leads to various model independent relations as the low energy theorems. 
The pion also gives the most important component of the nuclear force at long distances. In spite of these good features, the pion may complicate the structure of the nucleon and makes it difficult to establish a unified description including its resonances. 
Moreover, the light pions are easily emitted from the resonances, that makes theoretical analysis involved.

In the standard quark model where the pion cloud is not taken into account, baryons are $qqq$ and their resonances are described as single particle excitations of the constituent quarks. 
In this word, the quark model explains properties of many low lying hadrons qualitatively with various flavors including heavy quarks. 
It also describes the short range part of the nuclear force which has been supported by the recent lattice calculations. 
A very opposite description of the nucleon emerges when maximally utilizing the pion cloud. This is the skyrmion for the nucleon. 
In this picture the nucleon excitations emerge as collective motion of the pionic soliton. The model is good at describing the phenomena where the pion is involved. The nucleons or in general hadrons containing light quarks share both features quark dynamics and pion dynamics in some way or another.

The two models reflect important aspects of QCD, but their advantages are somewhat different. In some cases they make very different predictions. 
Under such situation, we propose to explore the study of baryons removing pion clouds, ``pionless spectroscopy". This should be useful to unveil the genuine role of the constituent quarks, and in turn that of the pions. Our eventual goal is then to establish an effective method keeping its applicability to explain and predicts various hadron phenomena.

%{\color{red} $\Omega_{css}$や$\Omega_{bss}$については、今後, HL-LHCにおいて新しいデータの登場を期待する。$\Omega$や$\Xi$についてはJ-PARCやBelle-IIが役割を果たす。}

% flatex input end: [./k10docu/k10BaryonW2strange.tex]

%Supplemental information
% flatex input: [./k10docu/k10sigma.tex]
\subsubsection{High-energy hyperon and proton scattering\label{sec:ynscattering}}
\begin{figure}[hthb]
  \begin{center}
  \includegraphics[width=0.75\textwidth]{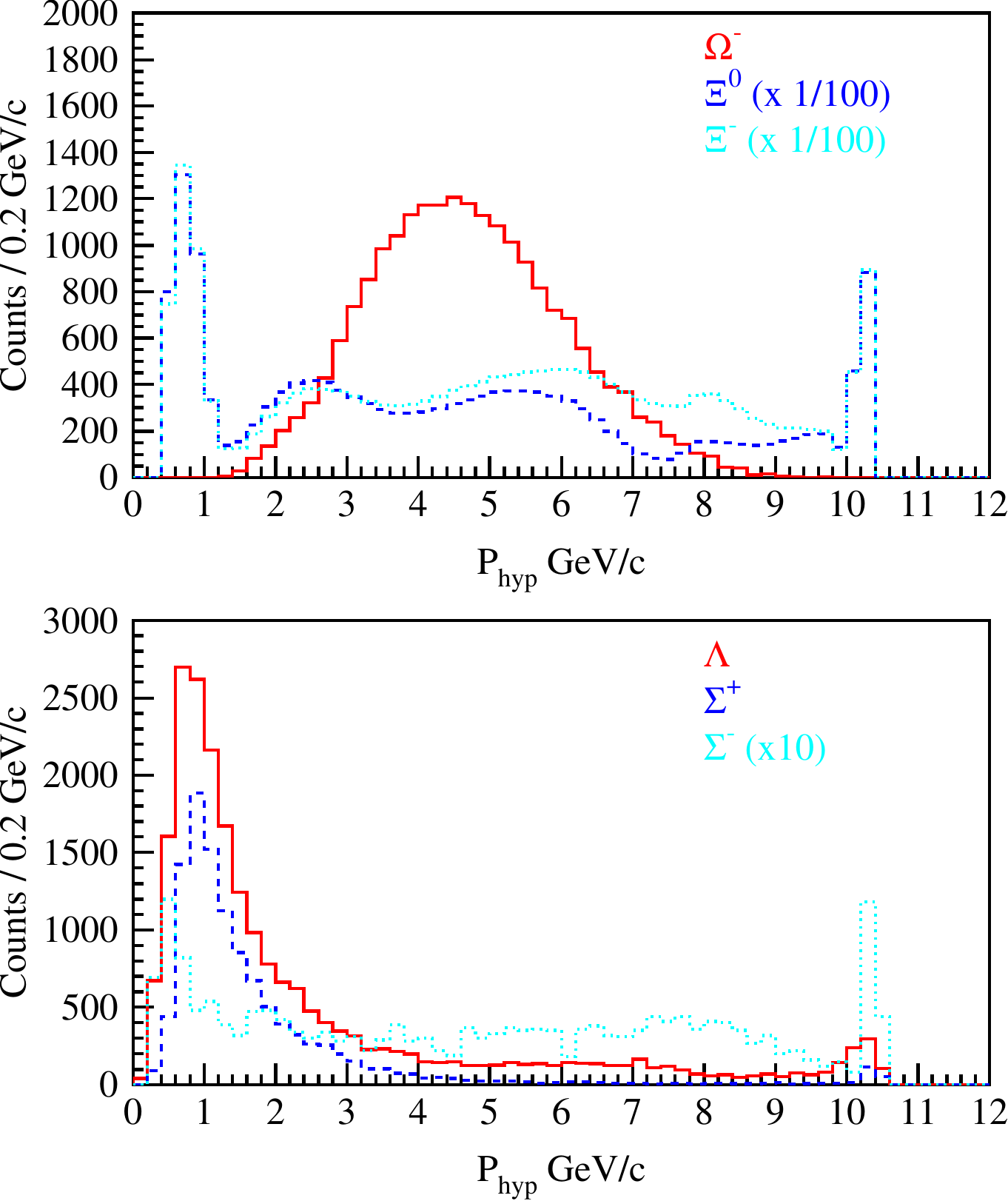}
  \end{center}
  \caption{
Momentum distributions for $\Xi$ and $\Omega^-$ (top),
and those for $\Lambda$ and $\Sigma$ (bottom).
The yield for each of $\Xi$'s is scaled by 1/100,
and that for $\Sigma^-$ is scaled by 10.
  }
  \label{k10-mom-k10}
\end{figure}

At the K10 beam line, a huge amount of free ground-state hyperons 
are expected to be produced such as $\Lambda$,
$\Sigma^\pm$, $\Xi^{0/-}$, and $\Omega^-$.
These hyperons are rather stable or long-lived since they can decay only with 
the weak interaction.
Of primary importance for the interaction between a hyperon and the nucleon would be 
the $S$-wave (and $P$-wave) low-energy scattering.
However, scattering on the nucleon at higher energies would  provide
information on the size of hyperons, and interaction mechanism 
(what the exchange particle is).
We have estimated the expected momentum distributions for the free hyperons 
produced at the K10 beam line
by using the JAM code~\cite{JAM}.
Fig.~\ref{k10-mom-k10} shows the momentum distributions 
for the produced hyperons. 
Among the ground-state hyperons, 
it is difficult to use $\Sigma^0$'s since they decay promptly after production
not with the weak interaction but with the electromagnetic interaction.
A missing-mass technique enables us to obtain the momentum-tagged hyperons.
The momentum of the produced $\Omega^-$ 
shows a bump at 4.5 GeV/c, covering a wide range from 2 to 8 GeV/$c$.
The momentum distributions are rather flat from the minimum (1 GeV$/c$) to the maximum
(10 GeV/$c$) for $\Xi^0$ and $\Xi^-$.
The two peaks at the minimum and maximum momenta come from 
the backward and forward direct $\Xi$ production in the $K^-\;p \to K\; \Xi$,
respectively.
Almost all the $\Lambda$'s and $\Sigma$'s have
the momenta below 4 GeV/$c$ but the momenta reach 10 GeV$/c$.
We can study the hyperon and nucleon collisions at 
CM energies $W \sim 3.2$ GeV and up to $W=4.7$ GeV
where the reaction cross section are not affected by 
appearance of baryon resonances.

\begin{figure}[htb]
  \begin{center}
  \includegraphics[width=0.85\textwidth]{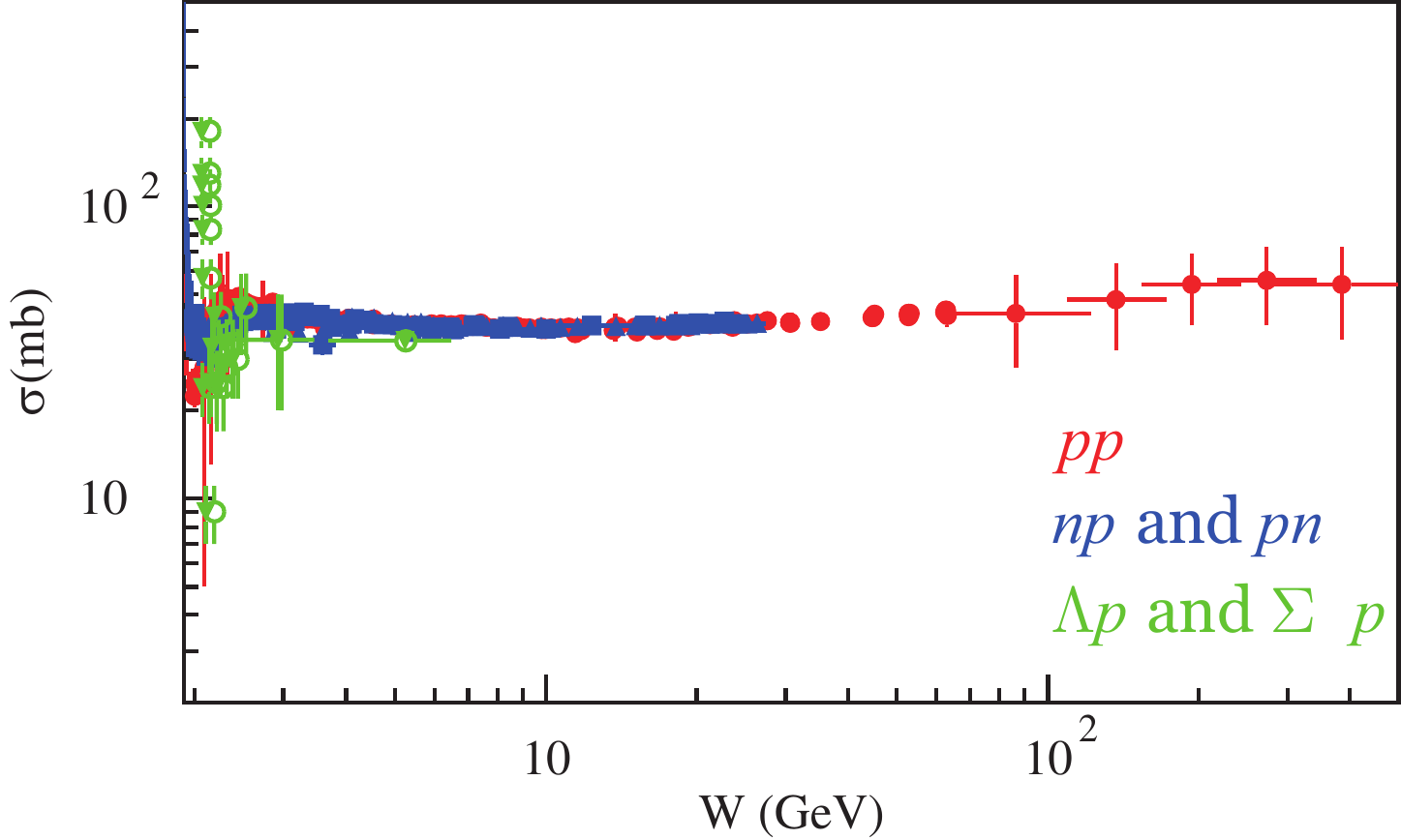}
  \end{center}
  \caption{
Total cross section $\sigma$ as a function of the CM energy $W=\sqrt{s}$.
The $\sigma$'s for $p\;p$,
for $p\;n$ and $n\;p$,
and for $\Lambda\; p$ and $\Sigma^-\; p$
are plotted by the the red, blue, and green makers,
respectively.
The data are taken from the Review of Particle Physics~\cite{Zyla:2020zbs}.
  }
  \label{k10-tcs1}
\end{figure}

The total cross section provides an input to the size of baryons.
In a classical situation,
one obtains the total cross section of $\sigma =\pi \left(R_1+R_2\right)^2$ 
for the two balls with radii of $R_1$ and $R_2$.
Given the transverse position of the first ball, the second one must be within 
an area of $\sigma$ to make the two collide.
The total cross section in the two-baryon collision
must include information on the sum of their radii.
Figure~\ref{k10-tcs1} shows the total cross section $\sigma$ as a function of 
the CM energy $W=\sqrt{s}$.
It shows a prominent enhancement in each of $p\;p$, $p\;n$, and $n\;p$ collisions
near the threshold owing to strong attraction in the corresponding $S$-wave channel.
It should be noted that an enhancement in the $p\;n$ and $n\;p$ collisions is larger than 
that in $p\;p$ because of the existence of a $p\;n$ or $n\;p$ bound state, or the deuteron.
After that $\sigma$ shows a complicated structure below a few GeV 
having several enhancements corresponding to the appearance of baryon resonances.

At high energies,
$\sigma$ increases as the square of a logarithm (relativistic rise of $\sigma$): 
\begin{equation}
\sigma\sim \log^2\left(W/W_0\right).
\end{equation}
With greater energy, a baryon can excite the other baryon and produce
one or several mesons on it from ever greater distances.
Since the radius of interaction is thus growing, so is the total cross section.
In the intermediate energies from a few to several tens of GeV,
$\sigma$ is almost constant, $\sim 38$ mb, independently of the kind of baryons,
indicating a core of a baryon measures $\sim 0.55$ fm in radius.
Since we do not expect a pion cloud in $\Omega^-$,
a smaller value than $\sim 38$ mb would be obtained for $\sigma$ in
the $\Omega^- p$ collision.
The relativistic rise of $\sigma$ would be also highly suppressed for $\Omega^- p$,
resulting $\sigma$ is constant in a wide energy region.

Another important feature of the hyperon and proton collision would appear
in diffractive elastic scattering (elastic scattering at small angles)
which is directly connected to the longstanding problem of QCD at large distances.
Especially, $\Lambda p$ and $\Omega^-p$ scatterings are of particular interest.
In these scatterings, a single pion exchange is forbidden owing to the 
isospin conservation between the initial and final isoscalar $\Lambda$'s or
$\Omega^-$'s.
Owing to the $\Lambda N$-$\Sigma N$ coupling as well as $\Lambda$-$\Sigma^0$ mixing coming 
from the isospin symmetry breaking,
$\Lambda p$ scattering takes place more likely than $\Omega^- p$ scattering.
The $\Lambda p$-$\Sigma N$ coupling enables a sequential $\Lambda\;p \to \Sigma\; N \to \Lambda\; p$ transition,
and a pion exchange takes place in each step. It is difficult to consider a 
similar sequential transition for $\Omega^-\; p$ scattering.
The $\Lambda p$ scattering also takes place with a single pion exchange at 
a few percent level isospin breaking. It is also difficult to consider a similar situation 
for $\Omega^-\; p$ scattering.
In elastic $\Omega^-\;p$ scattering,
only $\sigma$, $\eta$ and $\eta'$ exchanges are allowed in the first order.
The systematic studies of high-energy $Y\;N$ scatterings allow us to study the reaction mechanisms
of diffractive process (what the exchange particle is).

% flatex input end: [./k10docu/k10sigma.tex]

%Supplemental information

%\clearpage
% flatex input: [./k10docu_exp/k10-charm-suppl_v2.tex]
%%%%%%%%%%%%%%%%%%%%%%%%%%%%%%%%%%%%%%%%%%%%%%%%%%%%%%%%%%%%%%%%%%%%%%%%%%%%%%%%%%%%%
\subsubsection{Differential cross section as a function of $t$ for $Y_c^*$ production}

\begin{figure}[t]
  \begin{center}
  \includegraphics[width=15cm,keepaspectratio,clip]{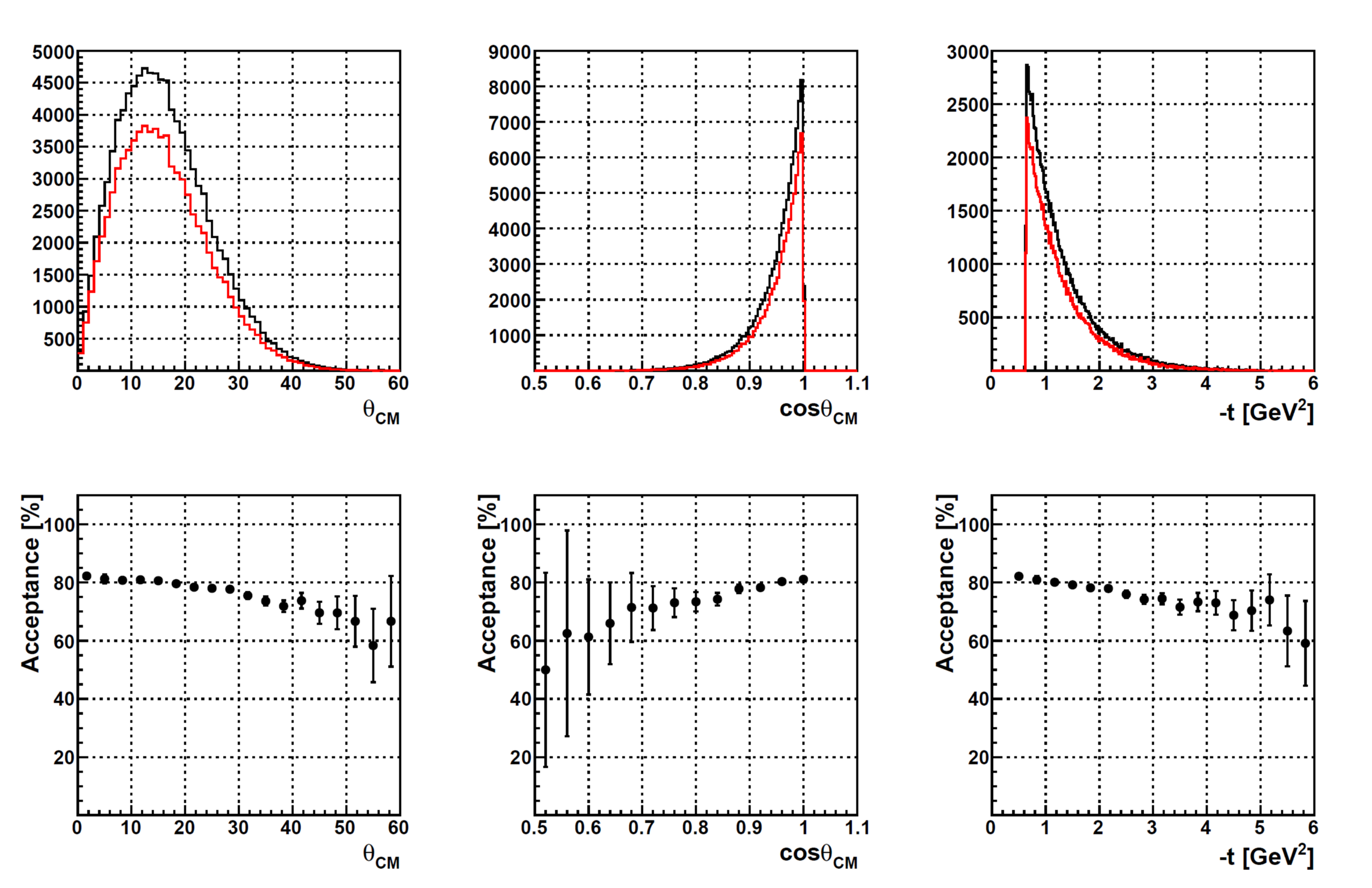}
  \end{center}
  \caption{
Angular distribution of $D^{*-}$ emission (top)
and acceptance of the $D^{*-}$-produced events (bottom)
as a function of the emission angle in the CM frame,
$\theta_{CM}$ (left) and $\cos\theta_{CM}$ (central)
together with those as a function of $-t$ (right).
Here,
the differential cross section is assumed to take an exponential form expressed by
Eq.~(\ref{eq:fit-dsdt}) 
with a slope parameter $b=1.40 {\rm\ GeV}^{-2}/c^{-2}$.
The back and red histograms are the distributions for 
the generated and accepted events, respectively.
  }
  \label{k10_charm_fig4}
\end{figure}

\begin{figure}[t]
  \begin{center}
  \includegraphics[width=12cm,keepaspectratio,clip]{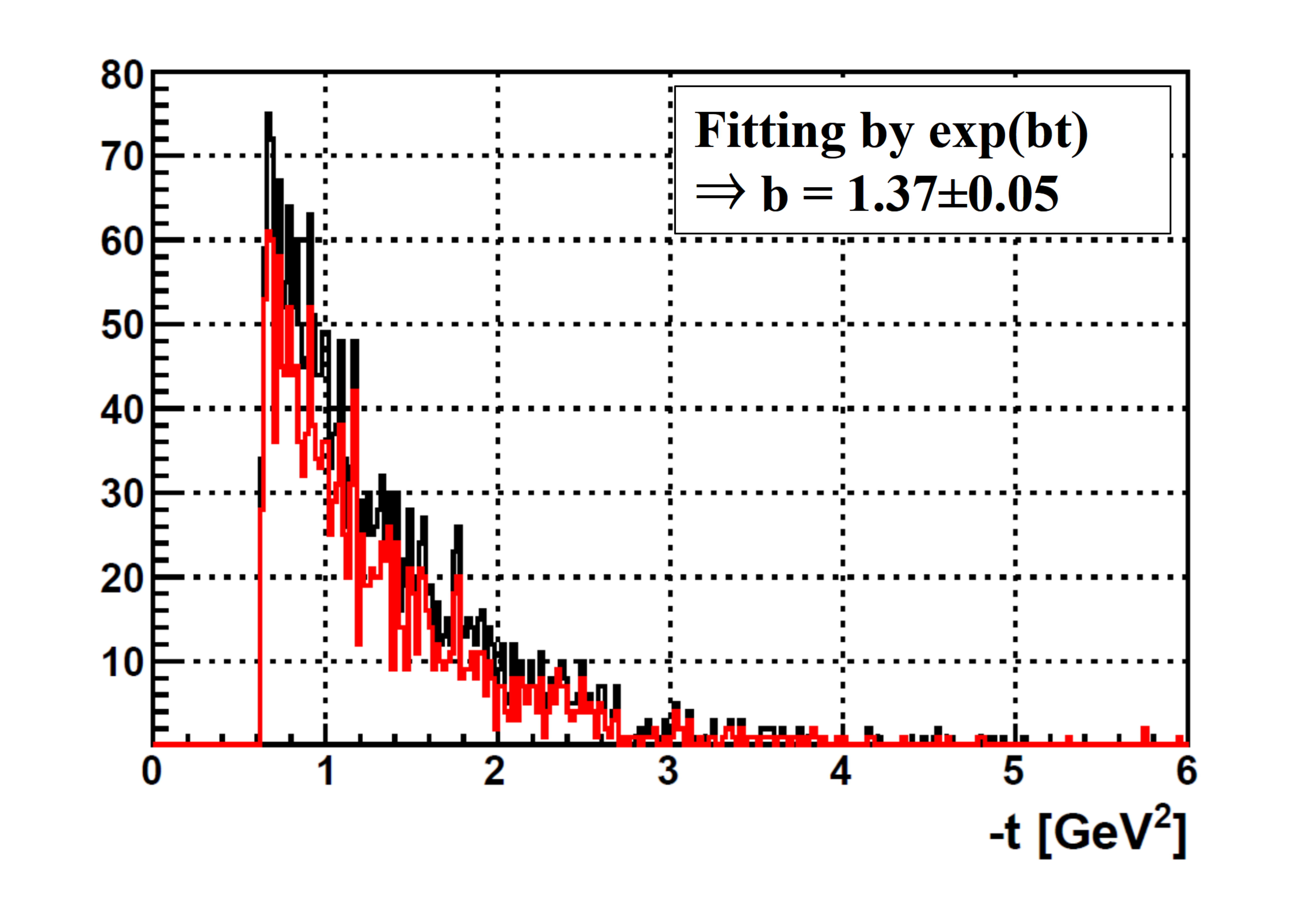}
  \end{center}
  \caption{ $-t$ distributions for the generated and accepted events.
  The number of accepted events is 1,600 as expected.
Here,
the differential cross section is assumed to take an exponential form expressed by
Eq.~(\ref{eq:fit-dsdt}) 
with a slope parameter $b=1.40 {\rm\ GeV}^{-2}/c^{-2}$.
The back and red histograms are the distributions for 
the generated and accepted events, respectively.
A value of $b=1.37 \pm 0.05 {\rm\ GeV}^{-2}/c^{-2}$
is obtained for the slope parameter by fitting an function (\ref{eq:fit-dsdt}) 
to the distribution without any acceptance correction for the accepted events.
  }
  \label{k10_charm_fig5}
\end{figure}

The differential cross section $d\sigma/dt$ for charmed-baryon
production,
corresponding to the transition form factor from the nucleon to the  
charmed baryon,
includes information on the internal structure of the charmed baryon.
The $d\sigma/dt$  can be determined from the events 
where the $D^{*-}$ mesons are emitted at finite angles.
Figure~\ref{k10_charm_fig4} shows 
the angular distribution of $D^{*-}$ emission 
and acceptance of the $D^{*-}$-produced events
as a function of the emission angle in the CM frame
($\theta_{CM}$ and $\cos\theta_{CM}$)
together with those as a function of $-t$.
Here, 
the differential cross section was assumed to take a 
form 
\begin{equation}
\frac{d\sigma}{dt}\propto e^{-bt} \label{eq:fit-dsdt}
\end{equation}
with a slope parameter $b=1.40 {\rm\ GeV}^{-2}/c^{-2}$.
Acceptance of the $D^{*-}$-produced events detected 
with the E50 spectrometer
was found to be  $\sim$80\% almost independently of the 
emission angle or $-t$.
Figure~\ref{k10_charm_fig5} shows the $-t$ distributions for the 
generated and accepted events
(the number of accepted events is $\sim 1,600$).
A value of $b=1.37 \pm 0.05 {\rm\ GeV}^{-2}/c^{-2}$
was obtained for the slope parameter
by fitting an function (\ref{eq:fit-dsdt}) to the $-t$ distribution for the 
accepted events.
To determine the slope parameter correctly,
we need acceptance correction as a function of $-t$
but the fitting result of $b$ indicates
that we can extract the slope parameter with a statistical error of 3.6\%.

%%%%%%%%%%%%%%%%%%%%%%%%%%%%%%%%%%%%%%%%%%%%%%%%%%%%%%%%%%%%%%%%%%%%%%%%%%%%%%%%%%%%%
\subsubsection{Background reduction for $Y_c^*$ production}

\begin{figure}[t]
  \begin{center}
  \includegraphics[width=15cm,keepaspectratio,clip]{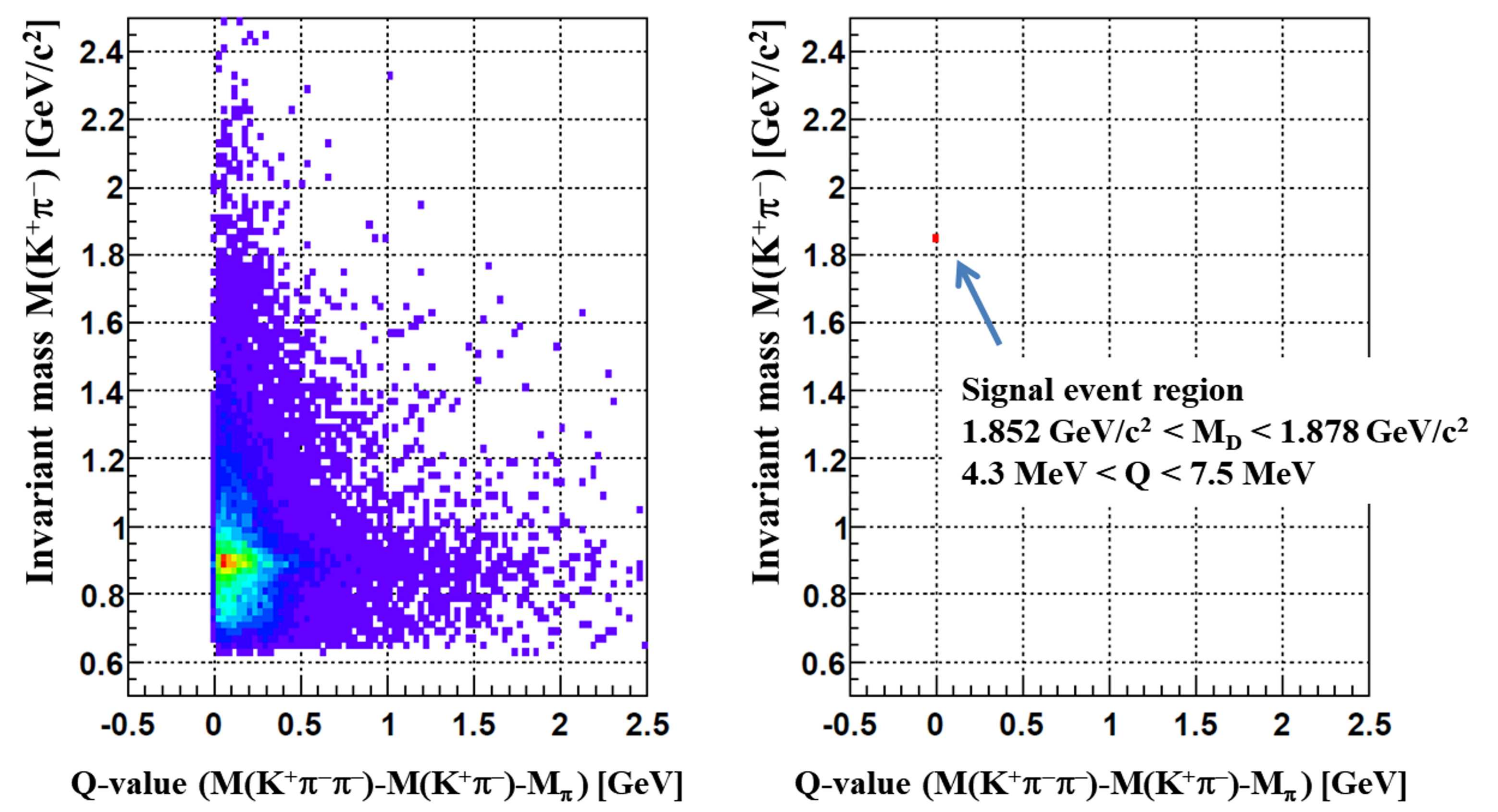}
  \end{center}
  \caption{ Correlation plots between $M_{K^+\pi^-}$ and $Q=M_{K^+\pi^-\pi^-}-M_{K^+\pi^-}-M_{\pi^-}$ before (a) and after (b) selecting the events that satisfy
$M_{K^+\pi^-}=1.865\pm 0.013$ GeV ($\bar{D}^0$ mass)
and $Q=5.9 \pm 1.6$ MeV.
Note that the $Q$ value corresponds to the excitation energy of 
$\bar{D}^0$ ($M_{D^{*-}}-M_{\bar{D}^0}$).
  }
  \label{k10_charm_fig6}
\end{figure}

The  estimated cross section of $Y_c^{*+}$ production 
is 10$^{-4}$ smaller 
in the $\pi^-\; p \to D^{*-}\; Y_c^{*+}$ reaction
than that of hyperon ($Y^*$) production (10--100 $\mu$b for  $\pi^{-}\;p \rightarrow K^{*}\;Y^{*}$).
Thus, it is necessary 
to reduce the background contribution significantly
for observing $Y_c^*$'s and determine their masses and widths.
To optimize the background reduction, 
we investigate three kinds of background processes:
1) strangeness production,
2) wrong particle identification, and 
3) $D^{*-}$ production associated without the $Y_c^*$ of interest.
Although strangeness production is the main contributor to the background,
the corresponding events are effectively reduced by using a $D^*$ tagging method.
Figure~\ref{k10_charm_fig6} shows the 
correlation plots between 
$M_{K^+\pi^-}$ and $Q=M_{K^+\pi^-\pi^-}-M_{K^+\pi^-}-M_{\pi^-}$
where the $Q$ value corresponds to the excitation energy of 
$\bar{D}^0$ ($M_{D^{*-}}-M_{\bar{D}^0}$).
A reduction factor  of 2$\times$$10^6$ was obtained
by selecting the events that satisfied 
$M_{K^+\pi^-}=1.865\pm 0.013$ GeV ($\bar{D}^0$ mass)
and $Q=5.9 \pm 1.6$ MeV.
Although the reduction factor of $D^*$ tagging is quite high,
we have still a sizable background contribution.
Other possible methods for further reducing the background 
contribution are
to use the opening angle in the CM frame 
between $K^+$ and $\pi^-$ from $\bar{D}^0$,
and to use the kinematic condition corresponding to the
$t$-channel $Y_c^{*+}$ in the $\pi^-\; p \to D^{*-}\; Y_c^{*+}$ reaction.
Additional reduction factor of $\sim$15 can be achieved from these 
although survival ratio of the events of interest is 0.54.
The final background level is expected to 5 counts/MeV in average
at $Y_c^*$ masses ranging from 2.2 to 3.4 GeV.

Wrong particle identification also deteriorates the background level.
We also estimated its effect taking into account the limited performance of the 
PID detector, and obtained a similar level of 6 counts/MeV to that 
of strangeness production.
The $D^{*-}$-produced events associated without 
the $Y_c^*$ of interest.
were also investigated by using 
the Geant4 simulation with JAM.
The contribution was minor compared with the strangeness production and the wrong particle identification events.
Finally, the the average background level was found to be 11 counts/MeV 
at $Y_c^*$ masses ranging from 2.2 to 3.4 GeV.
It is difficult to estimate the systematic uncertainty of the background level
since we do not have any experimental data currently.
We just compared the estimated background levels obtained from 
the JAM~\cite{JAM} and PYTHIA~\cite{PYTHIA} codes,
and considered the systematic uncertainty was a factor of 2.
The sensitivity of the production cross section was found to be 0.1--0.2 nb.

%%%%%%%%%%%%%%%%%%%%%%%%%%%%%%%%%%%%%%%%%%%%%%%%%%%%%%%%%%%%%%%%%%%%%%%%%%%%%%%%%%%%%
% flatex input end: [./k10docu_exp/k10-charm-suppl_v2.tex]
\label{sec:charm-spectroscopy-suppl1}
%\clearpage
% flatex input: [./k10docu_exp/k10-xi-suppl_v2.tex]
%%%%%%%%%%%%%%%%%%%%%%%%%%%%%%%%%%%%%%%%%%%%%%%%%%%%%%%%%%%%%%%%%%%%%%%%%%%%%%%%%%%%%
\subsubsection{Possible background reduction for observing highly-excited $\Xi^*$s}

\begin{figure}[htpb]
  \begin{center}
  \includegraphics[width=15.5cm,keepaspectratio,clip]{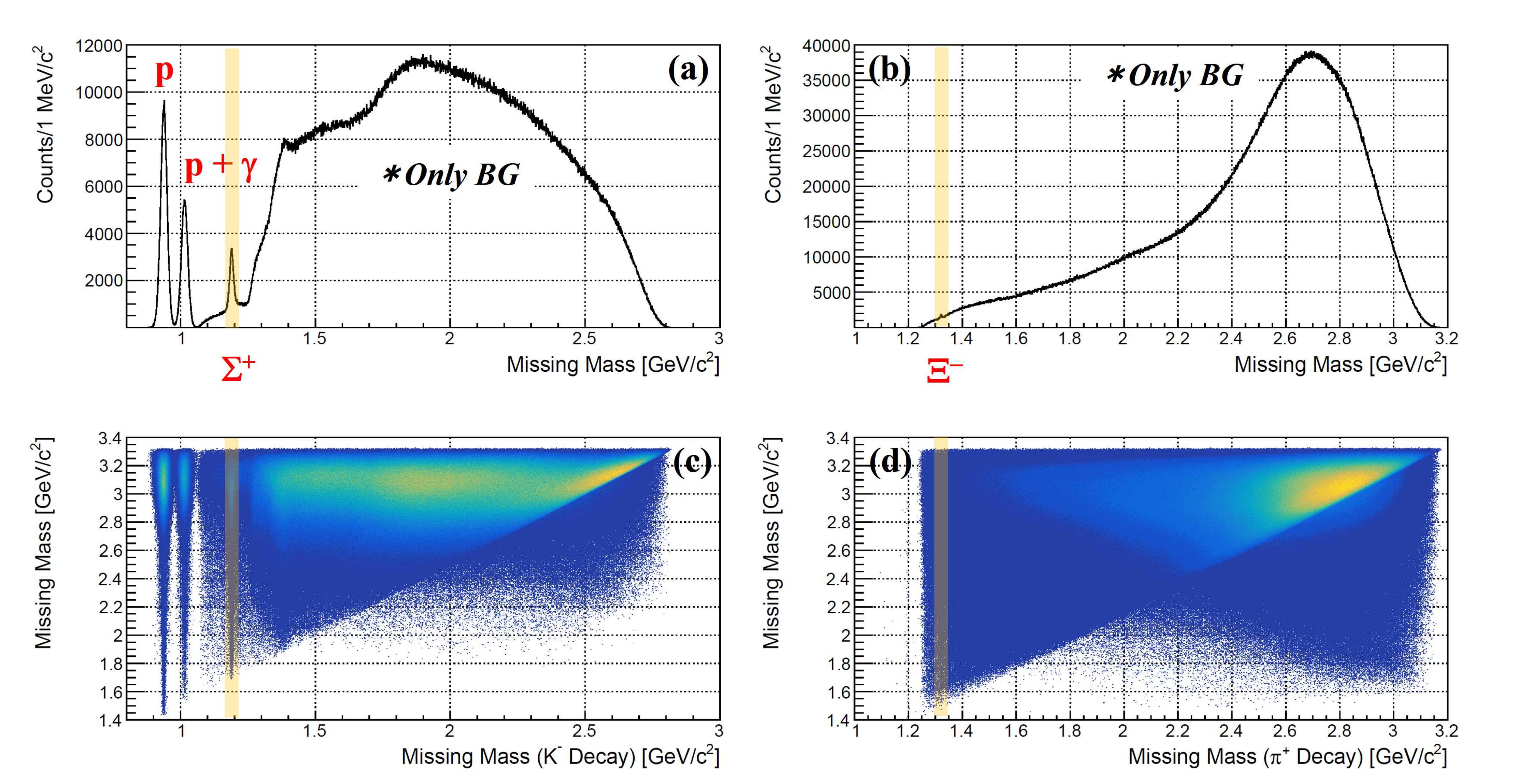}
  \end{center}
  \caption{ 
  (a) $p(K^-,K^{*0}K^-)$ missing-mass spectrum for the background events generated by JAM.
  (b) $p(K^-,K^{*0}\pi^+)$  missing-mass spectrum for the background events.
  (c) Correlation between the the $p(K^-,K^{*0})$ and the $p(K^-,K^{*0}K^-)$ missing masses for the background events.
  (d) Correlation between the the $p(K^-,K^{*0})$ and the $p(K^-,K^{*0}\pi^+)$ missing masses for the background events.
  }
  \label{k10_xi_fig5}
\end{figure}
\begin{figure}[htbp]
  \begin{center}
  \includegraphics[width=15.5cm,keepaspectratio,clip]{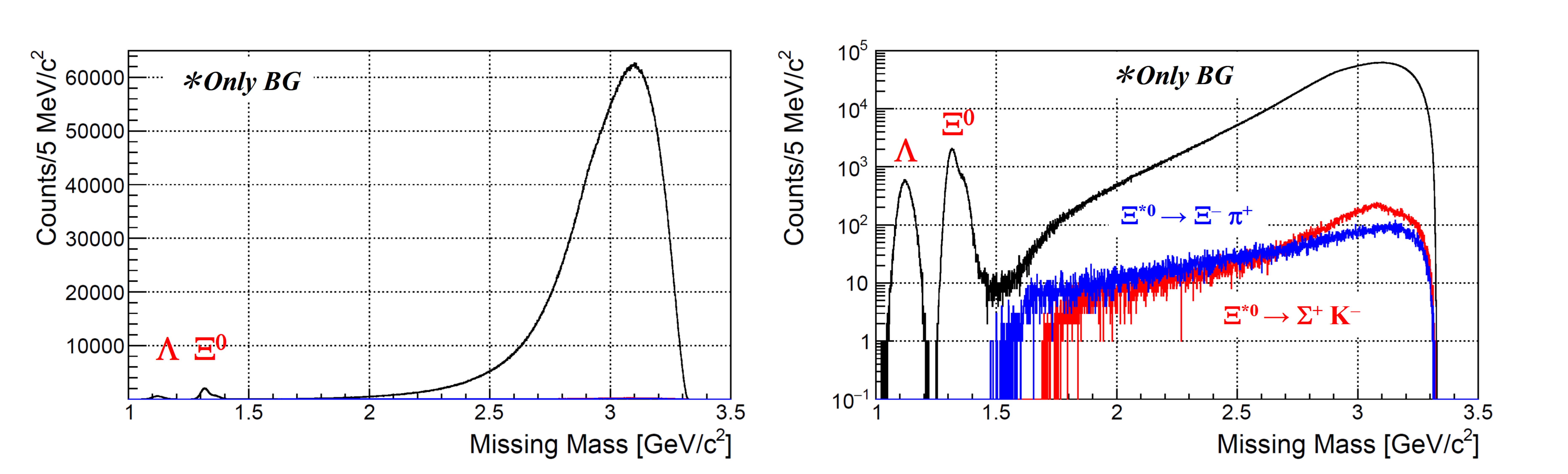}
  \end{center}
  \caption{ $p(K^-,K^{*0})$ missing-mass spectra for the background events
   in the linear scale (left) and in the logarithmic scale (right)
  after selecting the events that contain $\Sigma^+$ 
  from the $\Xi^0\to\Sigma^+\;K^-$ decay or 
$\Xi^0$ from $\Xi^0\to\Xi^-\;\pi^+$.
  }
  \label{k10_xi_fig6}
\end{figure}

The background contribution gradually increases with increase 
of the $p(K^-, K^{(*)0})$ missing mass, and 
the $\Xi^*$ peaks are difficult to be observed 
in the mass region above 2.2 GeV.
It is crucial to make the S/N ratio higher for observing highly-excited $\Xi^*$'s.
Selecting the events that contain the daughter particles from $\Xi^*$'s is a possible 
way to get the higher S/N ratio.
From now on, we focus on the $\Xi^{*0} \rightarrow \Sigma^+\;K^-$ and $\Xi^{*0} \rightarrow \Xi^-\;\pi^+$ decay channels since their branching ratios are expected to be large.
It should be noted that the produced $\Xi^{*0}$'s can decay into not only the 
channels described above but also other channels
in the simulation depending on the JAM parameters.
Figure~\ref{k10_xi_fig5}(a) shows the $p(K^-,K^{*0}K^-)$ missing-mass spectrum 
for the background events generated by JAM.
Three peaks are clearly observed at masses of 0.94, 1.02, and 1.19 GeV.
The 0.94-GeV (1.19-GeV) peak corresponds to the events that the proton
 ($\Sigma^+$) is missing.
The 1.02-GeV peak is made by the proton and $\gamma$ ($p+\gamma$) from 
the $\Sigma^0\to \Lambda\; \gamma$ decay.
The generated events in JAM include $\Xi^*$'s from $K^-\;p\to K^+\;\Xi^{*-}$, 
which decay into  $\Lambda\;K^-$ and $\Sigma^0\;K^-$.
Since $\Lambda$ decays into $p\; \pi^-$, the $p$ and $p+\gamma$ peaks are generated.
It should be noted that the final-state
$K^+\; \pi^-$ can be identified as $K^{*0}$ 
because of its broad width.
A detailed analysis of the production and decay vertices is expected to remove 
these background processes.
Figure~\ref{k10exp_fig10}(b) shows
the$p(K^-,K^{*0}\pi^+)$ missing-mass spectrum for the background events.
A peak corresponding to $\Xi^-$ can be observed with a small yield.
Figures~\ref{k10_xi_fig5}(c) and (d) show the correlation plots
between the $p(K^-,K^{*0})$ and the $p(K^-,K^{*0}K^-)$  missing masses, 
and between $p(K^-,K^{*0})$ and the $p(K^-,K^{*0}\pi^+)$, respectively.
To remove the background contribution to the $\Xi^0$-produced events 
with the $\Xi^0\to\Sigma^+\;K^-$ decay,
we select the events that the $p(K^-,K^{*0}K^-)$ missing mass is consistent 
with the $\Sigma^+$ mass.
Similarly, the $\Xi^0\to\Xi^-\;\pi^+$ events can be selected 
with background suppression in the high-mass region
by applying the condition that the $p(K^-,K^{*0}\pi^-)$ missing mass
 is consistent with the $\Xi^-$ mass.
Figure~\ref{k10_xi_fig6} shows the $p(K^-,K^{*0})$ missing-mass spectra for the background events
after selecting the events that contain $\Sigma^+$ 
  from the $\Xi^0\to\Sigma^+\;K^-$ decay or 
$\Xi^0$ from $\Xi^0\to\Xi^-\;\pi^+$, showing 
a reduction factor of more than 200 in the high-mass region.
A significant improvement of the S/N ratios are expected 
only if the branching ratios are $\sim 0.1$ or larger
for the $\Xi^{*0} \rightarrow \Sigma^+\;K^-$ and $\Xi^{*0} \rightarrow \Xi^-\;\pi^+$  decays.

%%%%%%%%%%%%%%%%%%%%%%%%%%%%%%%%%%%%%%%%%%%%%%%%%%%%%%%%%%%%%%%%%%%%%%%%%%%%%%%%%%%%%

% flatex input end: [./k10docu_exp/k10-xi-suppl_v2.tex]
\label{sec:xi-spectroscopy-suppl1}
%\clearpage
% flatex input: [./k10docu_exp/k10-omega-exp1_v2.tex]
%%%%%%%%%%%%%%%%%%%%%%%%%%%%%%%%%%%%%%%%%%%%%%%%%%%%%%%%%%%%%%%%%%%%%%%%%%%%%%%%%%%%%
\subsubsection{Performance of the spectrometer for observing $\Omega^*$s}

The performance of the E50 spectrometer for the $\Omega^{(*)}$-produced
events has been estimated by a Monte Carlo simulation based on Geant4.
In this sub-subsection, described are
the acceptance of the $\Omega^{(*)}$-produced events,
the experimental $\Omega^{(*)}$-mass resolution,
and their yields.

%%%%%%%%%%%%%%%%%%%%%%%%%%%%%%%%%%%%%%%%%%%%%%%%%%%%%%%%%%%%%%%%%%%%%%%%%%%%%%%%%%%%%
\subsubsection*{(a) Acceptance of the $\Omega^{(*)}$-produced events}

\begin{figure}[htbp]
 \begin{center}
  \begin{minipage}[t]{0.49\hsize} \centering
   \includegraphics[width=7.5cm,keepaspectratio,clip]{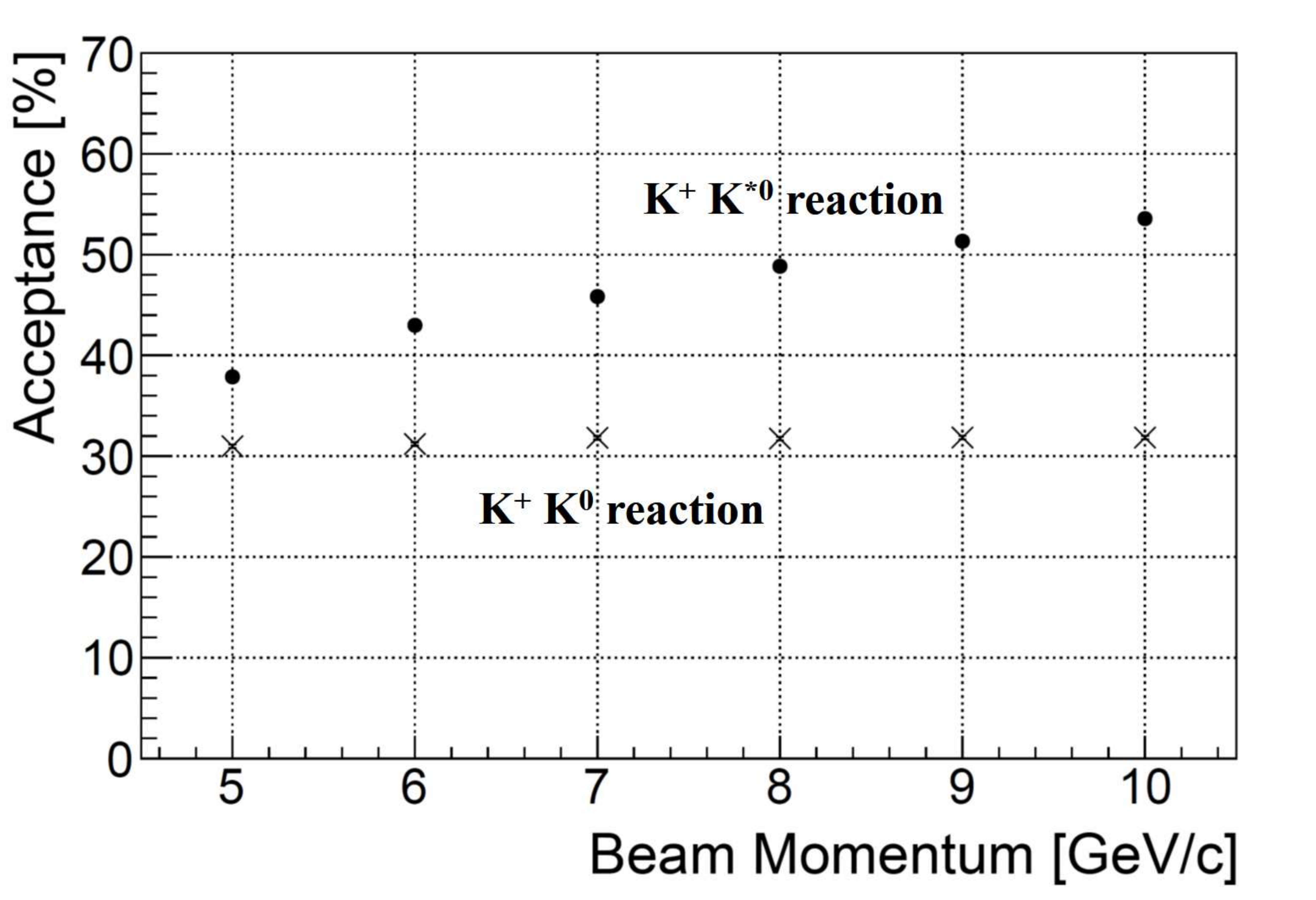}
  \end{minipage}
  \begin{minipage}[t]{0.49\hsize} \centering
   \includegraphics[width=7.5cm,keepaspectratio,clip]{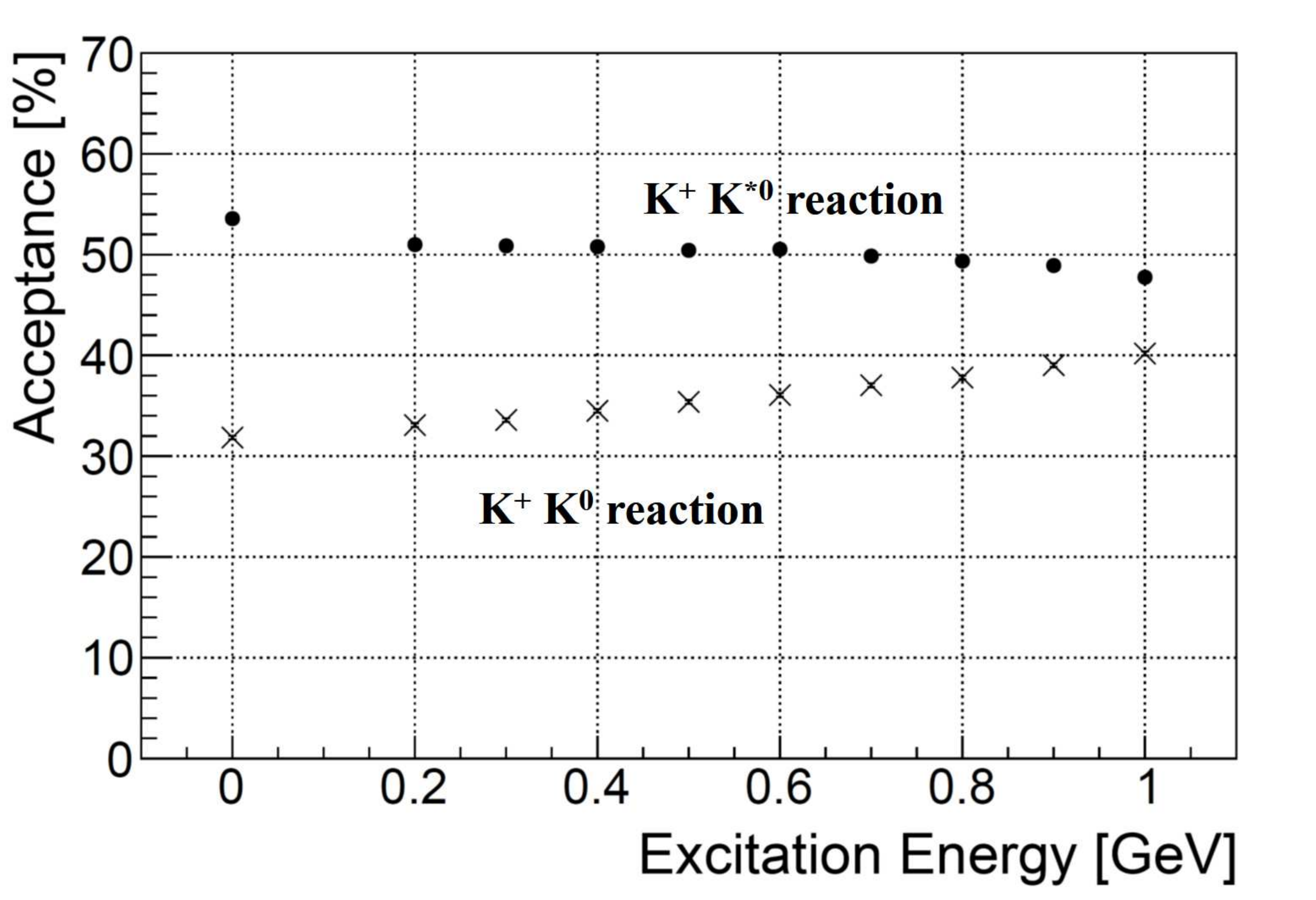}
  \end{minipage}
  \caption{
  Detector acceptance for the $\Omega^-$-produced events (the ground state) as a function of the incident kaon momentum (left),
  and that for the $\Omega^*$-produced events as a function of the excitation energy  at the incident momentum of 10 GeV/$c$ (right).
  The filled circles and X markers represent the acceptances 
  in the $K^{-}\;p \rightarrow \Omega^{(*)-}\;K^+\;K^{*0}$ and $K^{-}\;p \rightarrow \Omega^{(*)-}\;K^+\;K^{0}$ reactions, respectively.
  Here, detection of daughter particles are not required from the $\Omega^{(*)}$ decays. 
  }
  \label{k10exp_fig2}
 \end{center}
\end{figure}

The $\Omega^{(*)}$-produced events are selected 
by using the $p(K^-,K^+K^{(*)0})$ missing mass.
It is necessary to detect the final-state 
$K^+\;K^+\;\pi^-$ ($K^+\;\pi^+\;\pi^-$) particles from
$K^+\;K^{(*)0}$ emitted at forward angles.
Estimated is the acceptance for the $\Omega^{(*)}$-produced events,
or a fraction of the events that all the 
$K^+\;K^+\;\pi^-$ ($K^+\;\pi^+\;\pi^-$) particles are detected.
Fig.~\ref{k10exp_fig2}(left) shows the detector acceptance 
for the $\Omega^-$-produced events  (the ground state) as a function of the incident kaon momentum.
The detector acceptance for the $\Omega^-$-produced events are found to be $30\%$--$50\%$ and $\sim 30\%$ 
in the $K^{-}\;p \rightarrow \Omega^{-}\;K^+\;K^{*0}$ and $K^{-}\;p \rightarrow \Omega^{-}\;K^+\;K^{0}$ reactions, respectively.
Here, the angular distribution of $\Omega^-$ production is assumed to be isotropic in the $K^-\;p$-CM frame.
It should be noted that the acceptance includes the effects of final-state $K^+$ and $\pi^\pm$ decays in flight.
The acceptance gradually increases with increase of the incident kaon momentum in the $K^{-}\;p \rightarrow \Omega^{-}\;K^+\;K^{*0}$ reaction,
while it is almost independent of the incident momentum below 10 GeV/$c$ in $K^{-}\;p \rightarrow \Omega^{-}\;K^+\;K^{0}$.
Figure~\ref{k10exp_fig2}(right) shows the acceptance for the $\Omega^*$-produced events
below the excitation energy of 1 GeV at the incident kaon momentum of 10 GeV/$c$.
The angular distribution of $\Omega^*$ production is also assumed to be isotropic in the $K^-p$-CM frame.
The acceptance becomes higher with increase of the excitation energy both 
in the two reactions.
This is because $K^+\;K^{*0}$ ($K^+\;K^0$) are likely to be emitted at forward angles 
where the E50 spectrometer has high acceptance in case of producing high-mass $\Omega^*$'s.

\begin{figure}[htbp]
  \begin{center}
  \includegraphics[width=15cm,keepaspectratio,clip]{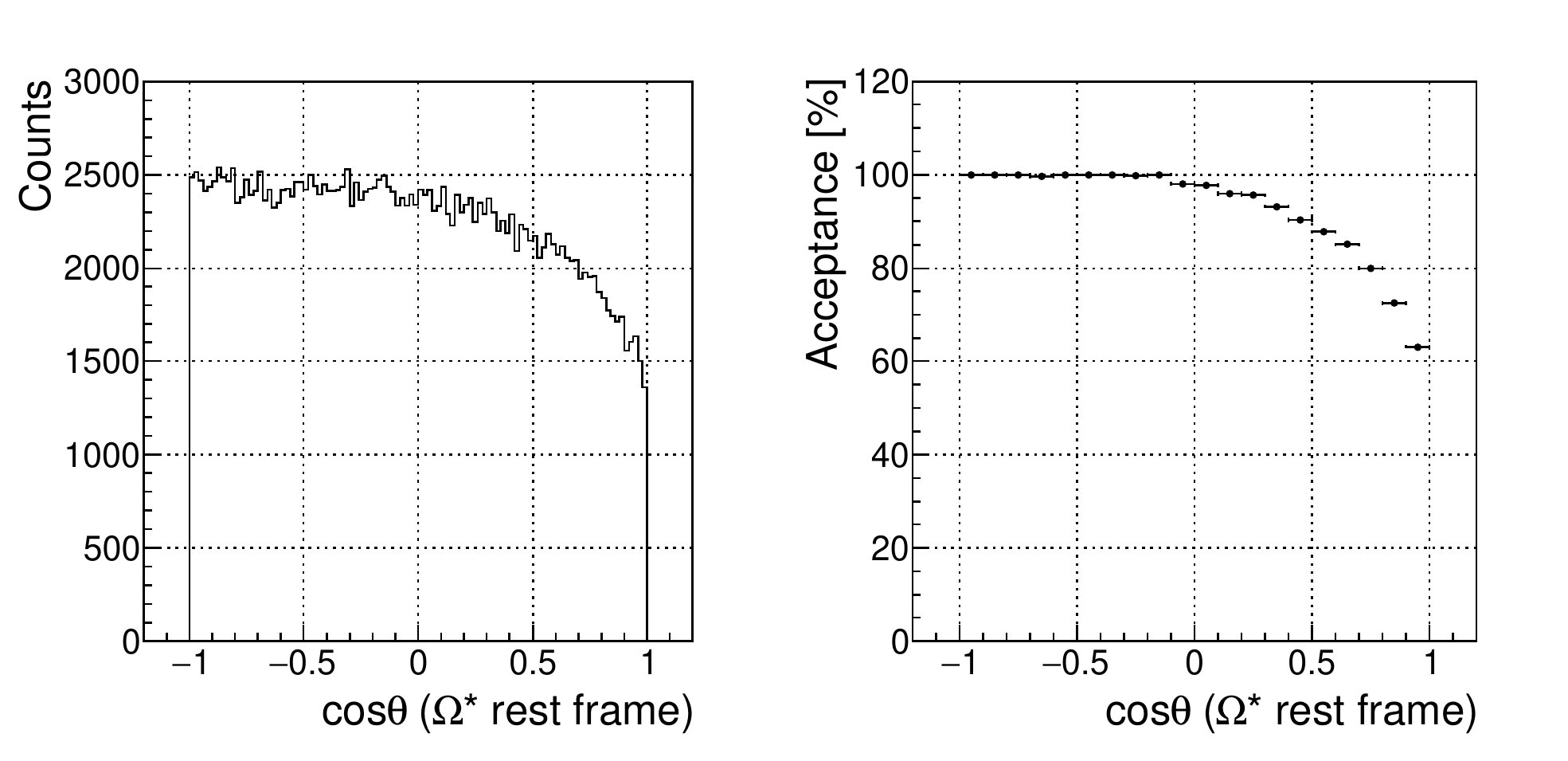}
  \end{center}
  \caption{ 
  Angular distribution of $\Xi^0$ emission (left)
  and acceptance as a function of the $\Xi^0$ emission angle (right)
  in the rest frame of the produced $\Omega^{*-}$ 
  ($z$-axis: opposite to the direction of the $K^+\;K^{(*)0}$ composite system produced together with $\Omega^*$).
  The two-body $\Omega^{*-}\to\Xi^0\;K^-$ decay can be identified by additionally detecting the daughter $K^-$ from the 
  $\Omega^{*-}$ decay.
  }
  \label{k10exp_fig3}
\end{figure}

The daughter particles from the $\Omega^*$ decays are covered by the internal detectors. 
The horizontally-emitted particles finally comes to the detectors in front of the yoke, and the vertically-emitted to those on the pole pieces.
Owing to these internal detectors, a wide coverage $\cos\theta_{\Xi^0}>-0.9$ 
is obtained for the polar angles of $\Xi^0$ emission 
(opposite to the direction of $K^-$ emission)
in the $\Omega^*$ rest frame from the two-body decay of $\Omega^*$ ($\Omega^{*-}\to \Xi^0\;K^-$).
The acceptance of the events detecting $K^-$ ($\pi^+\;\pi^-$) is found to be $\sim$95\% ($\sim$90\%) 
in the two-body $\Omega^{*-}\to \Xi^0\;K^-$ (three-body $\Omega^{*-}\to \Omega^-\;\pi^+\;\pi^-$) decay.
It should be noted that the acceptance described above 
does not include the effects of final-state $K^-$ and $\pi^\pm$ decays in flight.
The effects of the in-flight decay of $K^-$'s deteriorate the acceptance by 10\%.
Figure~\ref{k10exp_fig3} shows the acceptance of detecting the two-body $\Omega^{*-} \rightarrow \Xi^0\;K^-$ decay 
as a function of the $\Xi^0$ emission angle in the $\Omega^{*-}$ rest frame.
Here, $z$-axis is defined to be opposite to the direction of the $K^+\; K^{(*)0}$ composite system produced in association with $\Omega^*$.
The daughter $K^-$ from the $\Omega^{*-}$ decay is emitted at forward angles in the laboratory frame,
and the acceptance is high for detecting the $\Omega^{*-} \rightarrow \Xi^0\;K^-$ decay.
Some acceptance drop is observed at the $\Xi^0$ emission angle of $\cos\theta_{\Xi^0} \sim 1.0$.
In this case, the associated $K^-$'s are emitted at backward angles, 
and some of them are out of the angular coverage of the E50 spectrometer.
Owing to high acceptance of the events with identifying the $\Omega^*$ decay,
we can obtain the decay angular distribution,
and determine the corresponding branching ratio.

%%%%%%%%%%%%%%%%%%%%%%%%%%%%%%%%%%%%%%%%%%%%%%%%%%%%%%%%%%%%%%%%%%%%%%%%%%%%%%%%%%%%%
\subsubsection*{(b) $\Omega^{(*)}$-mass resolution}

\begin{figure}[t]
 \begin{center}
  \begin{minipage}[t]{0.49\hsize} \centering
   \includegraphics[width=7.5cm,keepaspectratio,clip]{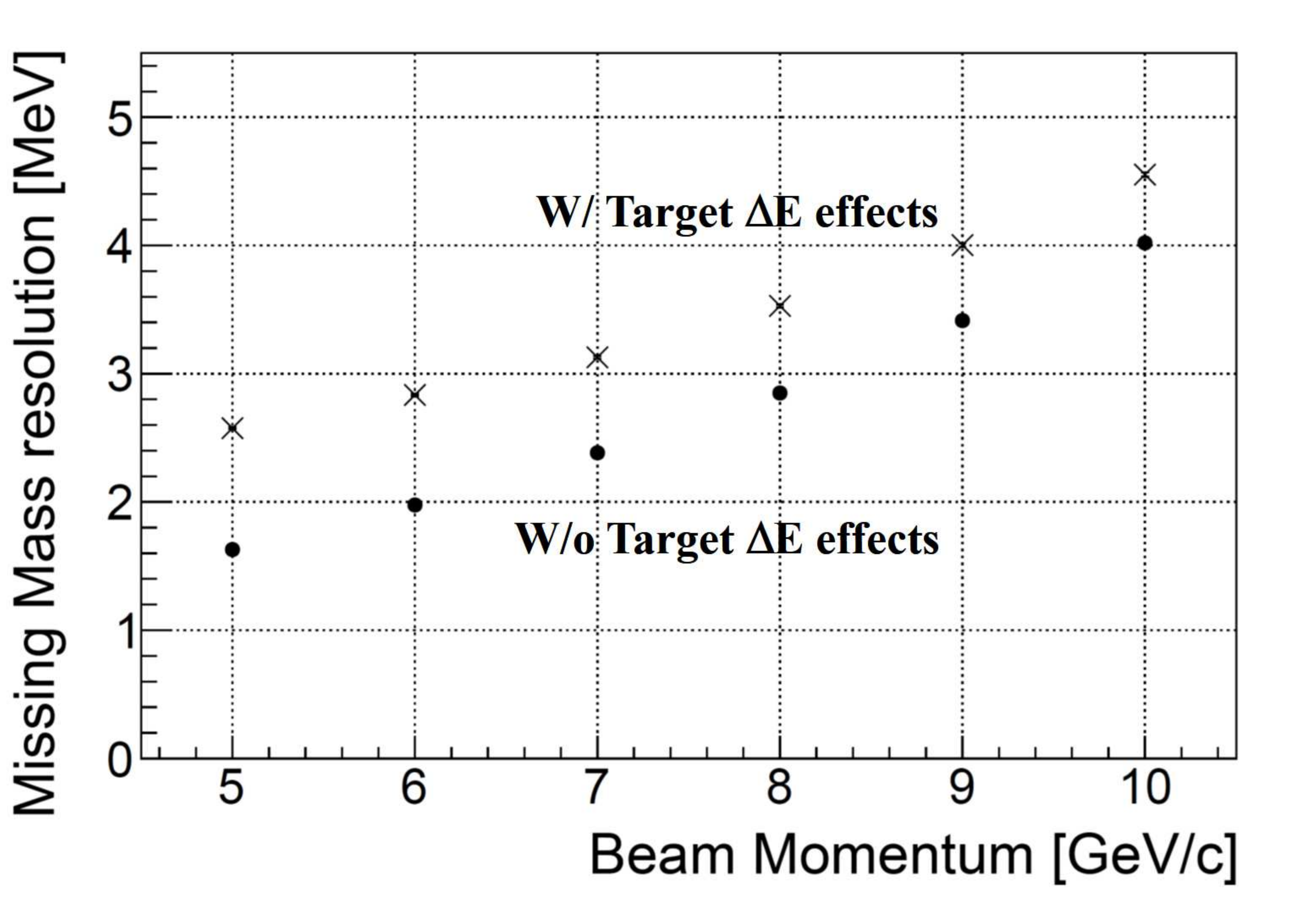}
  \end{minipage}
  \begin{minipage}[t]{0.49\hsize} \centering
   \includegraphics[width=7.5cm,keepaspectratio,clip]{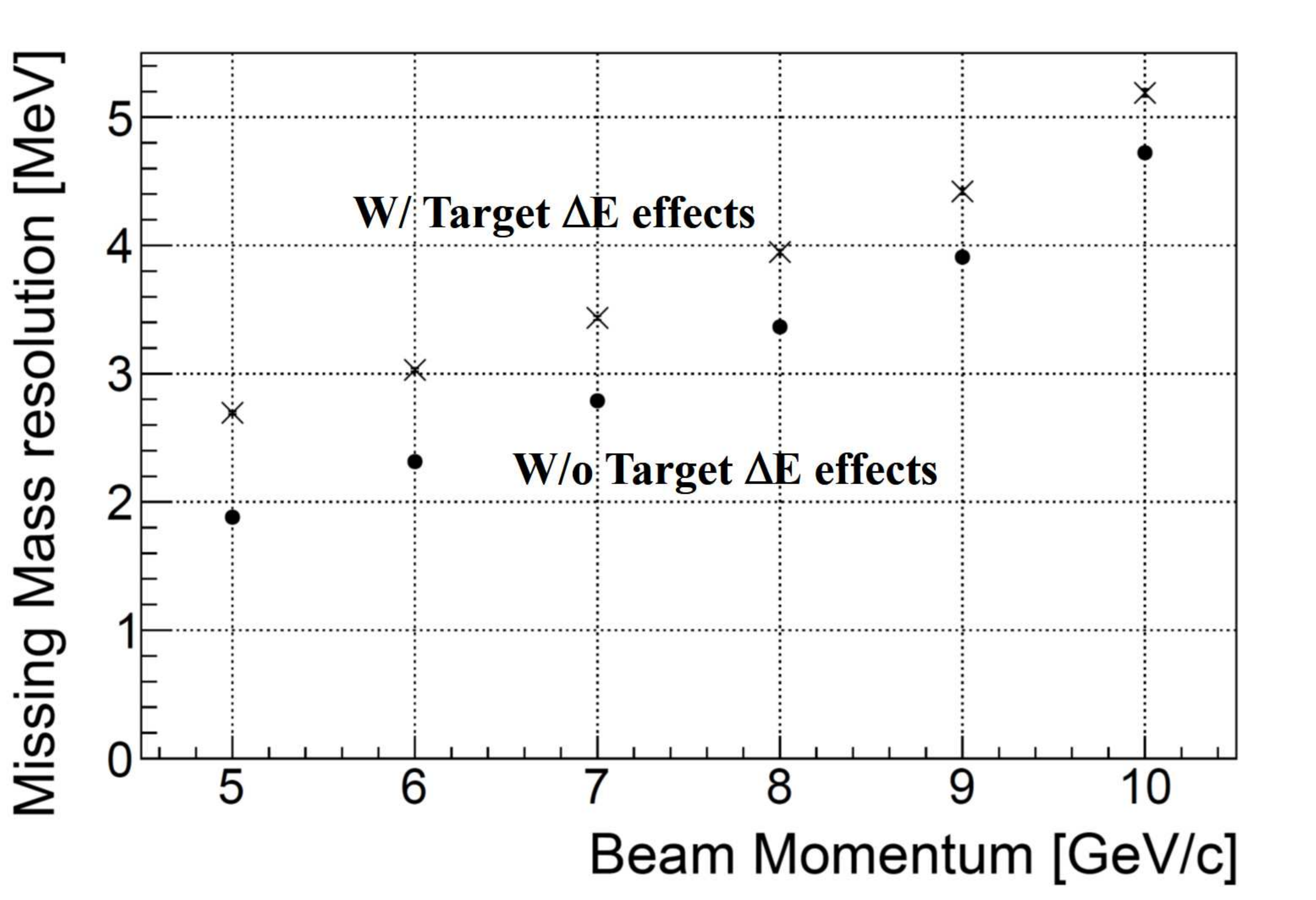}
  \end{minipage}
  \caption{
  Resolution ($\sigma$) of the $\Omega^-$ (ground state) mass as a function of the incident kaon momentum. 
  The $\Omega^-$-mass spectrum is obtained from the $p(K^{-},K^+K^{*0})$ missing mass 
  in the $K^{-}\;p \rightarrow \Omega^{-}\;K^+\;K^{*0}$ reaction (left),
  and from $p(K^{-},K^+K^{0})$ in $K^{-}\;p \rightarrow \Omega^{-}\;K^+\;K^{0}$ (right).
  The filled circles and X markers show the resolutions without and with the straggling effect of the energy loss in the target material, respectively.
  }
  \label{k10exp_fig4}
 \end{center}
\end{figure}

\begin{figure}[t]
 \begin{center}
  \begin{minipage}[t]{0.49\hsize} \centering
   \includegraphics[width=7.5cm,keepaspectratio,clip]{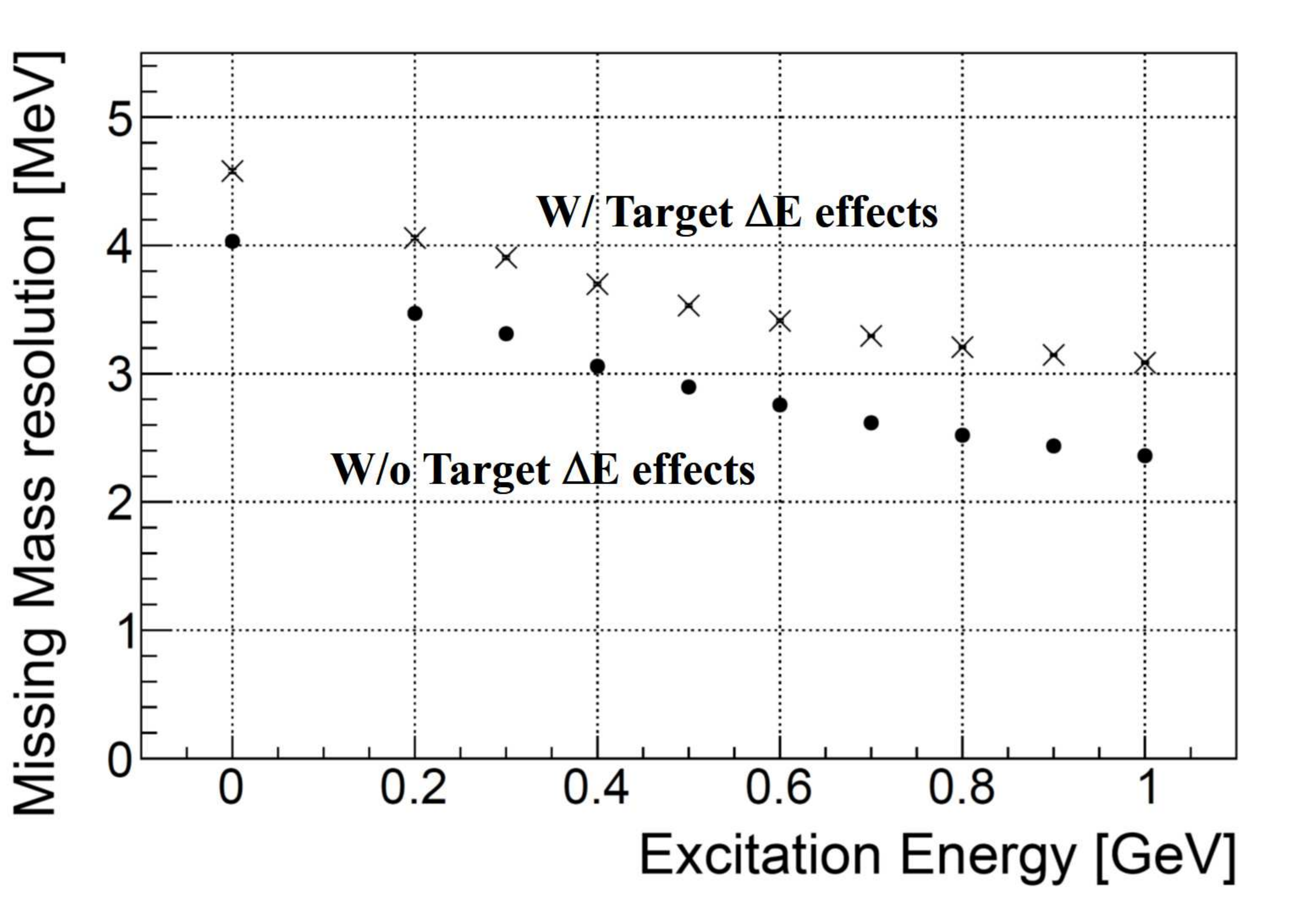}
  \end{minipage}
  \begin{minipage}[t]{0.49\hsize} \centering
   \includegraphics[width=7.5cm,keepaspectratio,clip]{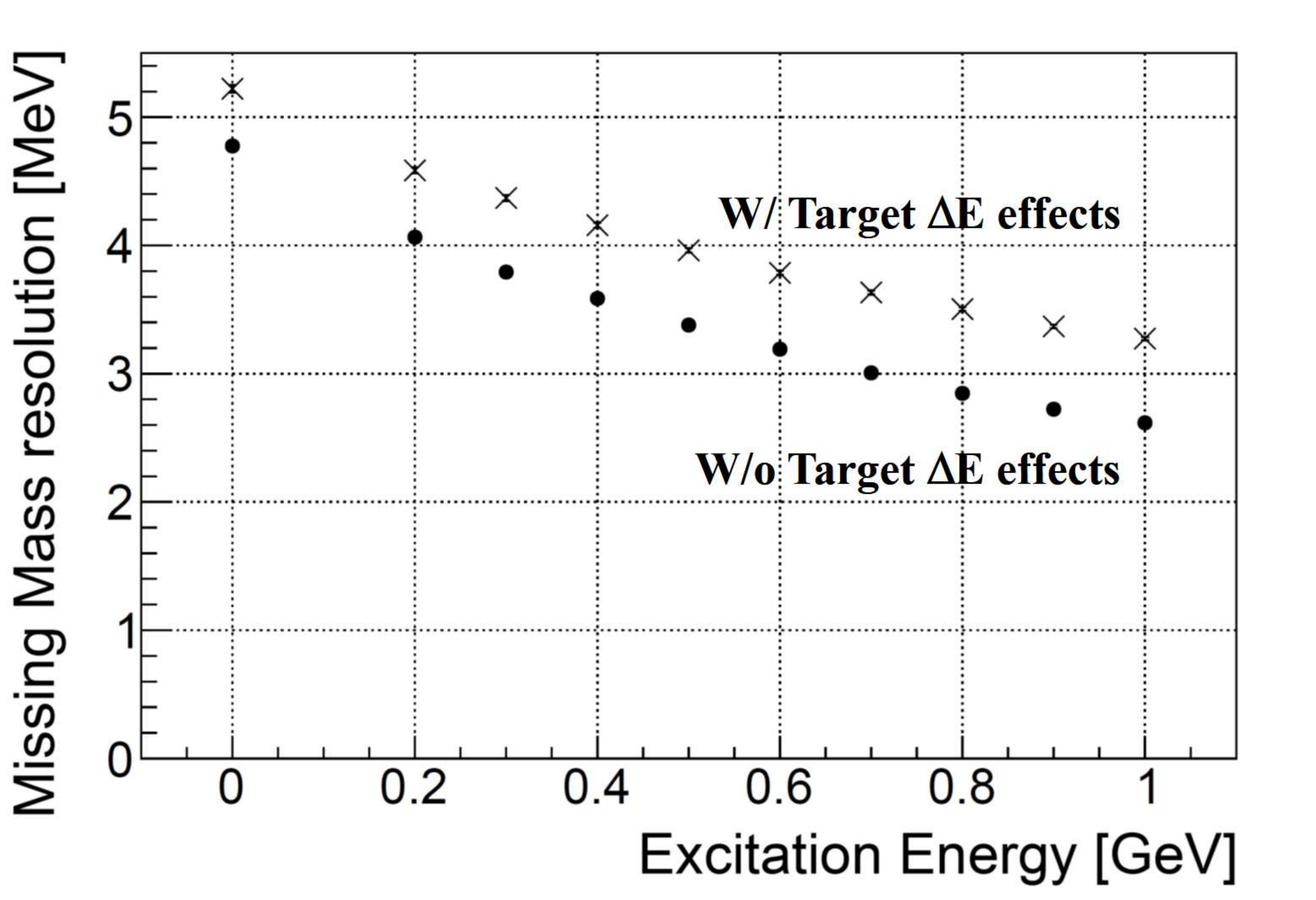}
  \end{minipage}
  \caption{
  Resolution ($\sigma$) of the $\Omega^*$ mass as a function of the excitation energy at the incident kaon momentum of 10 GeV/$c$. 
  The $\Omega^{*}$ mass is obtained from the $p(K^{-},K^+K^{*0})$ missing mass 
  in the $K^{-}\;p \rightarrow \Omega^{*-}\;K^+\;K^{*0}$ reaction (left),
  and from $p(K^{-},K^+K^{0})$ in $K^{-}\;p \rightarrow \Omega^{*-}\;K^+\;K^{0}$ (right).
  The filled circles and X markers show the resolutions without and with the straggling effect of the energy loss in the target material, respectively.
  }
  \label{k10exp_fig5}
 \end{center}
\end{figure}

The resolution is estimated for the $\Omega^{(*)}$ mass, which is calculated as the $p(K^{-},K^+K^{*0})$ missing mass 
in the $K^{-}\;p \rightarrow \Omega^{(*)-}\;K^+\;K^{*0}$ reaction, 
or the $p(K^{-},K^+K^{0})$ missing mass in the $K^{-}\;p \rightarrow \Omega^{(*)-}\;K^+\;K^{0}$ reaction.
Here, the momentum resolutions are assumed as same as those in the planned E50 experiment for the incident kaon momentum,
and for each of the final-state $K^+\;K^+\;\pi^-$ and $K^+\;\pi^+\;\pi^-$ particles detected with the E50 spectrometer 
in the $K^{-}\;p \rightarrow \Omega^{(*)-}\;K^+\;K^{*0}$, and $K^{-}\;p \rightarrow \Omega^{(*)-}\;K^+\;K^{0}$ reactions, respectively.
The incident momentum resolution assumed is a fixed value of $\Delta p_{\rm beam}/p_{\rm beam}=0.1\%$ ($\sigma$).
The momentum resolution assumed is $\Delta p_{\rm spec}/p_{\rm spec}=0.2\%$ ($\sigma$)
for 5-GeV/$c$ particles detected with the E50 spectrometer,
and is given by $\Delta p_{\rm spec}/p_{\rm spec}= 0.2\% \times p_{\rm spec}/\left(5\ {\rm GeV}/c\right)$
 at a certain momentum $p_{\rm spec}$.
The 2-MeV ($\sigma$) straggling is incorporated for the energy loss in the target material, 
which corresponds to the 4-g/cm$^2$ thickness of the hydrogen target.
The target thickness should be decided 
in the actual experiment considering the balance between the mass resolution and yield for $\Omega^*$'s.

Figure~\ref{k10exp_fig4} shows 
the resolution ($\sigma$) of the $\Omega^{-}$ (ground state)  mass as a function of the incident kaon momentum in the
$K^{-}\;p \rightarrow \Omega^{*-}\;K^+\;K^{*0}$ (left)
and 
$K^{-}\;p \rightarrow \Omega^{*-}\;K^+\;K^{0}$ (right)
reactions.
The momentum becomes higher for each of the final-state $K^+\;K^+\;\pi^-$ ($K^+\;\pi^+\;\pi^-$) with increase of the incident momentum,
deteriorating the resolution of the $\Omega^{(*)-}$ mass as well as that of the momentum of each final-state particle.
Figure~\ref{k10exp_fig5} shows the resolution ($\sigma$) of the $\Omega^{*-}$ mass
as a function of the excitation energy at the incident kaon momentum of 10~GeV/$c$.
The mass resolution becomes higher with increase of the excitation energy
since production of high-mass $\Omega^*$ makes the momentum lower 
for each final-state particle out of  $K^+\;K^+\;\pi^-$ ($K^+\;\pi^+\;\pi^-$).
The mass resolution is found to be 2.5--4.5 MeV ($\sigma$) depending on the incident kaon momentum and excitation energy.
The high mass-resolution and narrow width of $\Omega^{*}$ 
($\Gamma$ is expected from several MeV to several tens of MeV for $\Omega^{*}$'s)
enable us to observe $\Omega^{*-}$s separably.
Additionally, the $\Omega^{*-}$ width is expected to be measured directly since the mass resolution is comparable or smaller than the width.

%%%%%%%%%%%%%%%%%%%%%%%%%%%%%%%%%%%%%%%%%%%%%%%%%%%%%%%%%%%%%%%%%%%%%%%%%%%%%%%%%%%%%
\subsubsection*{(c) Yield of the $\Omega^*$-produced events}

We estimate the expected yields of the $\Omega^*$-produced events detected with the E50 spectrometer at the K10 beam line.
The existing data show that the total cross section ranges from 2.0 to 3.5 $\mu$b 
at incident kaon momenta ranging from 7 to 10 GeV/$c$
for inclusive $\Omega^-$ production in the $K^-\;p$ reaction:
$0.5\pm 0.1$ $\mu$b at 4.2 GeV$/c$~\cite{gang77,hemi78},
$1.4\pm 0.6$ $\mu$b at 6.5 GeV$/c$~\cite{hass81},
$2.1\pm 0.3$ $\mu$b at 8.25 GeV$/c$~\cite{baub81},
$3.7\pm 0.9$ $\mu$b at 10 GeV$/c$~\cite{sixe79}, and
$3.9\pm 0.6$ $\mu$b at 11 GeV$/c$~\cite{asto85}.
The $K^{-}\;p \rightarrow \Omega^{-}\;K^+\;K^{0}$ reaction dominates 
in inclusive $\Omega^-$ production
at incident kaon momentum below 10 GeV/$c$.
Thus, we simply assume the cross sections are 2.0 and 3.5 $\mu$b for the $K^{-}\;p \rightarrow \Omega^{-}\;K^+\;K^{0}$ reaction
at incident kaon momenta of 7 and 10 GeV$/c$, respectively.
Since the total cross section increases rather linearly with increase of the incident kaon momentum,
the total cross section is given by interpolating the 
7- and 10-GeV$/c$ cross sections to give 
a cross section 
at a certain incident momentum.
Among the inclusive $\Omega^-$-produced events at 4.2~GeV/$c$ in Ref.~\cite{hemi78},
the numbers of events containing $K^0$ and $K^+\;\pi^-$ in the final state are 39 and 1, respectively.
We assume that the total cross section of the $K^{-}\;p \rightarrow \Omega^{-}\;K^+\;K^{*0}$ reaction
is 1/40 that of $K^{-}\;p \rightarrow \Omega^{-}\;K^+\;K^{0}$ at a fixed incident kaon momentum between 7 and 10 GeV/$c$.
This assumption might not be realistic 
since a branching ratio  of 50\% for the $K^0\to \pi\;\pi$ decay is ignored here,
and since the $K^-\;p\to \Omega^-\;K^+\;K^{*0}$ reaction
is suppressed owing to the limited phase space at 4.2 GeV/$c$.
We suppose these two effects are canceled out at incident kaon momenta of 7--10 GeV/$c$.
Additionally, the production cross section of an $\Omega^{*-}$ is assumed to be the same as that of 
$\Omega^-$ depending on the reaction and incident kaon momentum.

\begin{table}[t]
\caption{ 
Factors assumed for the yield estimation of $\Omega^*$ production, 
and expected yield of the $\Omega^*$-produced events at the incident beam momentum of 8 GeV/$c$. 
}
\begin{center}
\begin{tabular}{lcc} \hline \hline
Reaction
& $K^{-}\;p \rightarrow \Omega^{*-}\;K^+\;K^{*0}$
& $K^{-}\;p \rightarrow \Omega^{*-}\;K^+\;K^{0}$
\\
Cross section  
& 0.06 $\mu$b
& 2.50 $\mu$b
\\
Branching ratio
& 0.67 ($K^{*0}\to K^+\pi^-$)
& 0.35 ($K^{0}\to\pi^+\pi^-$) \\
Beam intensity & \multicolumn{2}{c}{$7.0\times 10^6$/spill (2-s duration in a 5.2-s cycle)}  \\
Target thickness & \multicolumn{2}{c}{4.0 g/cm$^2$ (57-cm-thick hydrogen)} \\
Acceptance ($\Omega^*$ production)& 0.48 & 0.23 \\
\hline
Tracking efficiency & \multicolumn{2}{c}{0.90 (each particle)} \\
Particle identification & \multicolumn{2}{c}{0.97 (each particle)} \\
Data acquisition efficiency & \multicolumn{2}{c}{0.99 (Streaming DAQ)} \\
Total efficiency & \multicolumn{2}{c}{0.66 (three-track events)} \\
\hline
$\Omega^*$ yield in a day &
$3.3\times 10^3$ &
$4.6\times 10^4$ \\ 
$\Omega^*$ yield in a 100-day beam time &
$3.3\times 10^5$ &
$4.6\times 10^6$ \\ 
\hline\hline
\end{tabular}
\label{factors1}
\end{center}
\end{table}

\begin{table}[t]
\caption{ 
Additional factors assumed for the yield estimation of the $\Omega^*$ decay angular distribution,
and expected yield in each bin of the $\Xi^0$ emission angle in the rest frame of $\Omega^*$ divided into 20. 
The factors listed in this table are additional ones to those in Table~\ref{factors1}.
}
\begin{center}
\begin{tabular}{llc} \hline \hline
Acceptance ($\Omega^{*-} \rightarrow \Xi^0\;K^-$ decay)  & \multicolumn{2}{c}{0.86} \\
(including in-flight decay of $K^-$) & \\
Decay branching ratio ($\Omega^{*-} \rightarrow \Xi^0\;K^-$ decay) &  \multicolumn{2}{c}{0.30} \\
Total efficiency for detecting $K^-$ & \multicolumn{2}{c}{0.75} \\
\hline
$\Omega^{*-}\to\Xi^0\;K^-$ yield in a 100-day beam time &  \multicolumn{2}{c}{$\sim 4000$}  \\
&  (each bin out of 20 angle bins)
 \\
\hline
\hline 
\end{tabular}
\end{center}
\label{factors2}
\end{table}

The expected yield of the $\Omega^*$-produced events is estimated in the following assumptions:
the cross section of $\Omega^*$ production is 0.06 $\mu$b and 2.50 $\mu$b at the incident kaon momentum of 8 GeV/$c$
in the $K^{-}\;p \rightarrow \Omega^{*-}\;K^+\;K^{*0}$ and $K^{-}\;p \rightarrow \Omega^{*-}\;K^+\;K^{0}$ reactions, respectively, 
the tracking efficiency is 0.90 for each particle;
the particle-identification efficiency is 0.97 for each final-state particle from 
$K^+\;K^{(*)0}$;
DAQ efficiency is 0.99;
and the total efficiency is $\sim 0.66$ for three-track events.
The acceptance of the events for each reaction includes in-flight decay of the final-state
 particles to be detected (both $K^+$ and $\pi^{\pm}$).
Table~\ref{factors1} summarizes the factors assumed for the yield estimation of $\Omega^*$ production.
The expected yields of $\Omega^*$-produced events are $3.3\times10^3$ and $4.6\times 10^4$ a day 
in the $K^{-}\;p \rightarrow \Omega^{*-}\;K^+\;K^{*0}$ and $K^{-}\;p \rightarrow \Omega^{*-}\;K^+\;K^{0}$ reactions, respectively.
To obtain several $10^5$ of the $\Omega^*$-produced events, we need a 100-day beam time.
In analysis of the decay angular distribution of $\Omega^*$,
the $\Omega^*$-produced events may be combined 
for the $K^{-}\;p \rightarrow \Omega^{*-}\;K^+\;K^{*0}$ and $K^{-}\;p \rightarrow \Omega^{*-}\;K^+\;K^{0}$ reactions,
and more than $10^5$ $\Omega^*$-produced events are available in a 100-day beam time.
Suppose the branching ratio is 0.3 for the two-body $\Omega^{*-} \rightarrow \Xi^0\;K^-$ decay,
the expected yield is several thousands for each bin of $\Xi^0$ emission angles divided into 20.
Additional factors are listed in Table~\ref{factors2} to get the expected yields in the decay angular distribution.

%%%%%%%%%%%%%%%%%%%%%%%%%%%%%%%%%%%%%%%%%%%%%%%%%%%%%%%%%%%%%%%%%%%%%%%%%%%%%%%%%%%%%

% flatex input end: [./k10docu_exp/k10-omega-exp1_v2.tex]
\label{sec:omega-spectroscopy-suppl1}
%\clearpage
% flatex input: [./k10docu_exp/k10-omega-exp2_v2.tex]
%%%%%%%%%%%%%%%%%%%%%%%%%%%%%%%%%%%%%%%%%%%%%%%%%%%%%%%%%%%%%%%%%%%%%%%%%%%%%%%%%%%%%

\clearpage
\subsubsection{Expected $\Omega^{(*)}$-mass spectrum}

%%%%%%%%%%%%%%%%%%%%%%%%%%%%%%%%%%%%%%%%%%%%%%%%%%%%%%%%%%%%%%%%%%%%%%%%%%%%%%%%%%%%%
\subsubsection*{(a) Spectrum without the background contribution}

\begin{figure}[t]
  \begin{center}
  \includegraphics[width=15cm,keepaspectratio,clip]{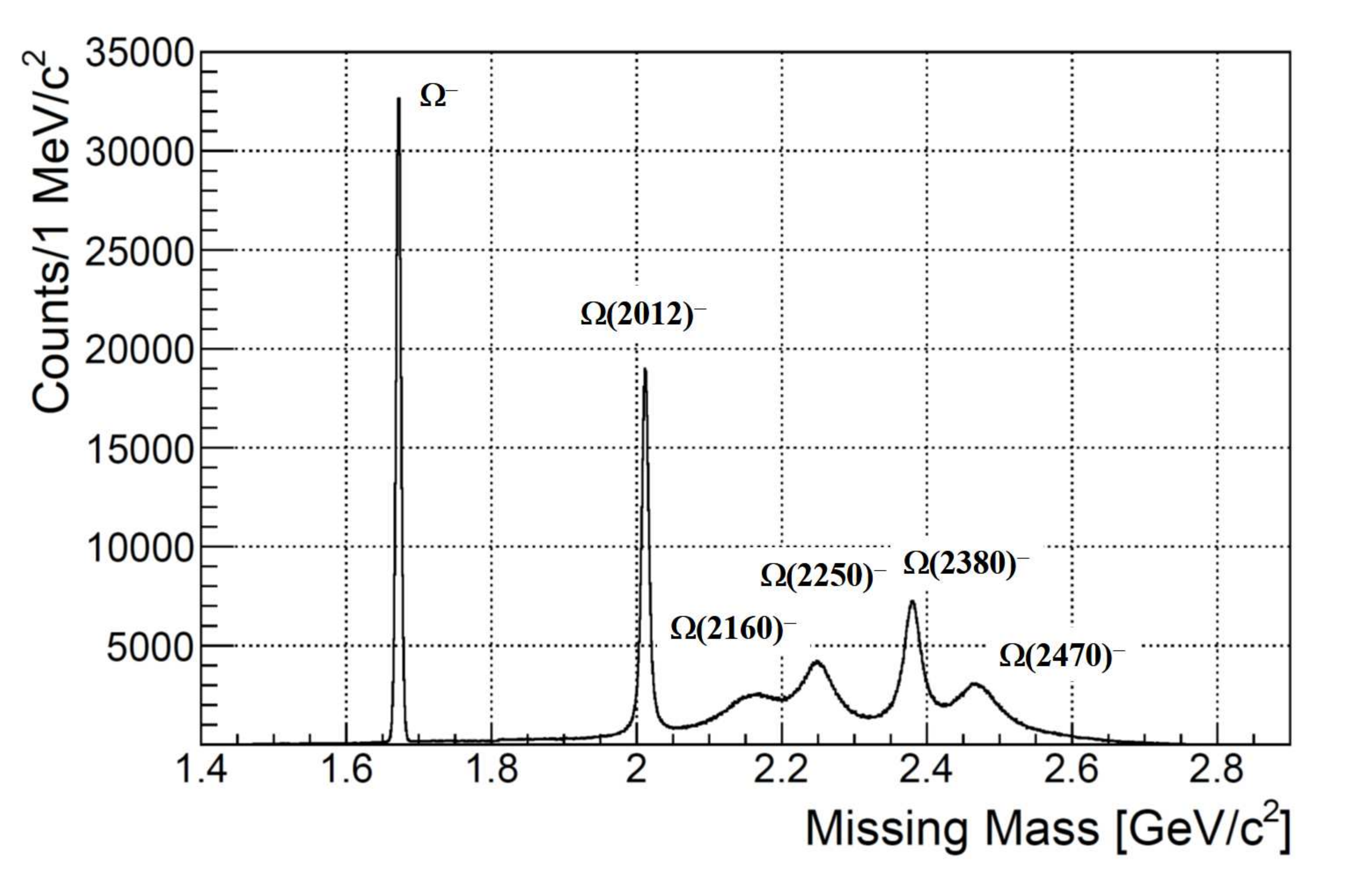}
  \end{center}
  \caption{
  Expected $\Omega^{(*)}$-mass spectrum without any background processes 
  at the incident kaon momentum of 8 GeV$/c$  in a 100-day beam time.
  The spectrum is obtained by the $p(K^-,K^+K^{*0})$ missing mass for the generated $\Omega^{(*)}$-produced events.
  }
  \label{k10exp_fig7}
\end{figure}

At first, we show the expected $\Omega^{(*)}$-mass spectrum without any background processes.
The $\Omega^{(*)}$'s considered are summarized in Table~\ref{OmegaPDG}, 
which are taken from the Review of Particle Physics~\cite{Zyla:2020zbs}.
Additionally, we incorporate a Rope-like resonance $\Omega(2160)^-$ with the parameters described in Table~\ref{OmegaPDG}.
\begin{table}[b]
\caption{
Mass and width parameters of $\Omega^{(*)}$'s for estimating the 
$\Omega^{(*)}$-mass spectrum in the simulation.
We assume the parameters for $\Omega(2160)$ as expected,
and take those for the others as listed in Review of Particle Physics~\cite{Zyla:2020zbs}.
}
\begin{center}
\begin{tabular}{ccc} \hline \hline
$\Omega^{(*)}$ & Mass [MeV] & Width [MeV] \\ \hline
$\Omega(2470)^-$ & 2470 & 72 \\
$\Omega(2380)^-$ & 2380 & 26 \\
$\Omega(2250)^-$ & 2250 & 55 \\
$\Omega(2160)^-$ & 2160 & 100 \\
$\Omega(2012)^-$ & 2012 & 6.4 \\
$\Omega^-$         & 1672 & --- \\
\hline
\hline 
\end{tabular}
\end{center}
\label{OmegaPDG}
\end{table}
Figure~\ref{k10exp_fig7} shows the expected $\Omega^{(*)}$-mass spectrum 
without any background processes at the incident kaon momentum of 8 GeV$/c$.
Here, the $\Omega^{(*)}$ mass is given by the $p(K^-,K^+K^{*0})$ missing mass,
and the number of produced $\Omega^{(*)}$'s is fixed at $3.3\times10^5$ for each $\Omega^{(*)}$ corresponding to a 100-day beam time.
Peaks are separably observed for narrow $\Omega^{(*)}$'s with widths narrower
than 30 MeV,
 and those are clearly distinguished from others even for rather wide ones.
It should be noted that all the experimental resolutions are included:
momentum resolution, angular resolution, and straggling of the 
energy loss in the target material.

Since the number of events is $\sim1.0\times10^5$ in each peak, 
the precision of the mass determination in this spectrum is better than 1 MeV depending 
on the width of the corresponding $\Omega^{(*)}$.
To get accuracy of the mass determination,
the absolute momentum scale in the spectrometer is required to be calibrated carefully.
We can use various peaks corresponding to hyperons and other baryons in similar mass spectra.
Thus, the accuracy expected is also better than 1 MeV after the calibration.

It is difficult to accurately determine the width of an $\Omega^{(*)}$.
The observed peak must be modified by the experimental mass resolution of 2.5--4.5 MeV ($\sigma$).
The peak may be different in shape from the Breit-Wigner function
owing to the overlap with other peaks and background contributions, and interference with other processes.
The accuracy of the width determination expected is better than 1 MeV 
for an isolated Breit-Wigner peak with a width of 10 MeV.
We expect the accuracy of the width determination 
is better than a few MeV for high-mass $\Omega^*$'s, which are overlapped with other contributions.

%%%%%%%%%%%%%%%%%%%%%%%%%%%%%%%%%%%%%%%%%%%%%%%%%%%%%%%%%%%%%%%%%%%%%%%%%%%%%%%%%%%%%
\subsubsection*{(b) Background contribution}

It is difficult to estimate the background contribution correctly 
in the $\Omega^{(*)}$-mass spectrum obtained from the $K^-\;p$ reaction for several-GeV/$c$ incident kaons.
Available experimental data are quite limited in this momentum range,
and we cannot estimate inclusive cross sections or multiplicities of generated particles 
even if we find some exclusive cross sections in a particular reaction.
Thus, we use a hadron-reaction generating code called JAM (version 1.90597)~\cite{JAM} for the background estimation as usual.
JAM  includes many elementary reaction processes at wide CM energies,
covering the resonance region ($\sqrt{s}<4$ GeV), 
the string region ($4<\sqrt{s}<10$ GeV),
and the perturbative-QCD (pQCD) region ($\sqrt{s}>10$ GeV).

\begin{table}[t]
\caption{
$K^-\;p$ cross sections for the background contributions estimated by JAM
together with those assumed for the signal $\Omega^*$-produced processes.
The ratios of the cross sections of the $\Omega^*$-produced 
$K^{-}\;p \rightarrow \Omega^{*-}\;K^+\;K^{0}$ and $K^{-}\;p \rightarrow \Omega^{*-}\;K^+\;K^{*0}$ reactions
to the background $K^-\;p\to K^+\;\pi^-\;X$  and $K^-p \to K^+\;K^+\;\pi^-\;X$ are also listed, respectively.
}
\begin{center}
\begin{tabular}{cccccccc} \hline \hline
Beam & 
$\sigma_{\rm tot}$
& 
$\sigma_{K^+\;\pi^-}$ 
& 
$\sigma_{K^+\;K^+\;\pi^-}$
& 
$\sigma_{\Omega^*\;K^+\;K^0}$
& 
$\sigma_{\Omega^*\;K^+\;K^{*0}}$
& Ratio [\%] & Ratio [\%]  \\
\mbox{}[GeV/$c$] &
[mb] & 
[$\mu$b] & 
[$\mu$b] & 
[$\mu$b] & 
[$\mu$b] & 
$\frac{\sigma_{\Omega^*\;K^+\;K^0}}{\sigma_{K^+\;\pi^-}}$
& 
$\frac{\sigma_{\Omega^*\;K^+\;K^{*0}}}{\sigma_{K^+\;K^+\;\pi^-}}$
\\ \hline
7.0  & 25.6 & 463 & 1.80 & 2.00 & 0.050 & 0.43 & 2.8 \\
8.0  & 23.6 & 503 & 2.46 & 2.50 & 0.063 & 0.50 & 2.6 \\
9.0  & 23.2 & 548 & 3.16 & 3.00 & 0.075 & 0.55 & 2.4 \\
10.0 & 22.6 & 585 & 4.22 & 3.50 & 0.088 & 0.60 & 2.0 \\
\hline
\hline 
\end{tabular}
\end{center}
\label{JAMBG}
\end{table}

Of interest in the proposed experiment is the string region 
where hadrons are generated 
mainly in the string-string scattering processes.
The hadronization process is described in the Lund string model~\cite{Lund},
which is also adopted in another hadron-reaction generating code called PYTHIA~\cite{PYTHIA}.
Adjusted in JAM to reproduce the existing experimental data
are the conditions of the string-string scattering and hadronization processes:
no string-string scattering before hadronization, 
no color-flow during the string generation, 
and the production ratios of the generated hadron resonances.
Those conditions and hadron resonances considered are different from those in PYTHIA.
Here, we do not use PYTHIA but JAM 
since we have found JAM reproduces the experimental data better than PYTHIA 
at incident momenta of several GeV/$c$\footnote{
We have investigated both the codes, and compared with the experimental data
for the background studies in the E50 experiment for spectroscopy
of charmed baryons~\cite{E50exp}.}.

Table~\ref{JAMBG} shows the $K^-\;p$ cross sections for the background contributions 
estimated by JAM together with those assumed for the signal $\Omega^*$-produced processes.
We investigate $K^+$ and $K^+\;K^+\;\pi^-$ production as 
the background contribution.
The cross section for the $K^+K^+\pi^-$-produced events 
is found to be two-order smaller than that for $K^+$-produced events.
Thus, the $\Omega^{(*)}$-mass spectrum obtained in the $K^{-}\;p \rightarrow \Omega^{(*)-}\;K^+\;K^{*0}$ reaction
is expected to have low background as compared with that in the $K^{-}\;p \rightarrow \Omega^{(*)-}\;K^+\;K^{0}$.
Additionally, some of $K^0$ candidates reconstructed from $\pi^+\pi^-$ can be $\bar{K}^0$
so that the $\Omega^{(*)}$-mass spectrum includes the $\Lambda^*$- and $\Sigma^*$-produced events.
It is necessary to detect not $K^0$ but $K^{*0}$ for reducing the 
background contribution in the $\Omega^{(*)}$-mass spectrum.
We can obtain the $\Omega^{(*)}$-mass spectrum with high S/N ratio 
from the $p(K^{-},K^+K^{*0})$ missing mass which gives only an $S=-3$ system.

%%%%%%%%%%%%%%%%%%%%%%%%%%%%%%%%%%%%%%%%%%%%%%%%%%%%%%%%%%%%%%%%%%%%%%%%%%%%%%%%%%%%%
\subsubsection*{(c) Smaller $\Omega^*$-production cross sections}

\begin{figure}[htpb]
  \begin{center}
  \includegraphics[width=15.5cm,keepaspectratio,clip]{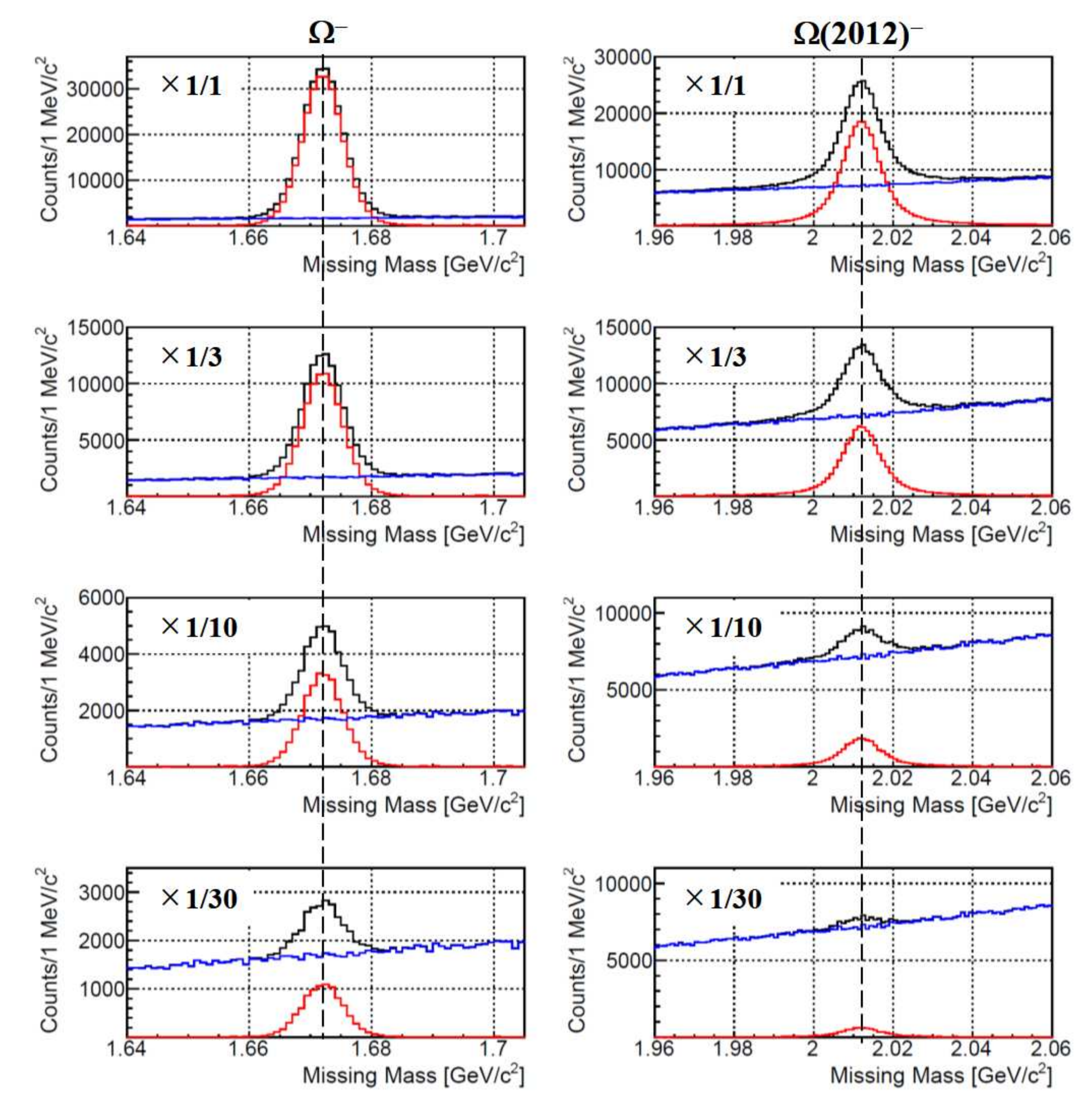}
  \end{center}
  \caption{ 
  Expected $\Omega^{(*)}$-mass spectra,
  or the $p(K^-,K^+K^{*0})$ missing-mass spectra,
  including the background contribution 
  at the incident kaon momentum of 8 GeV$/c$
    in a 100-day beam time
  by reducing $\Omega^*$-production cross sections by a factor 
  of 1/1 (original), 1/3, 1/10 and 1/30 from the top to bottom panels.
  The mass spectra are expanded 
  for the ground-state $\Omega^-$ and $\Omega(2012)^-$ in the left and right panels, respectively.
  }
  \label{k10exp_fig9_1}
\end{figure}

\begin{figure}[htpb]
  \begin{center}
  \includegraphics[width=15.5cm,keepaspectratio,clip]{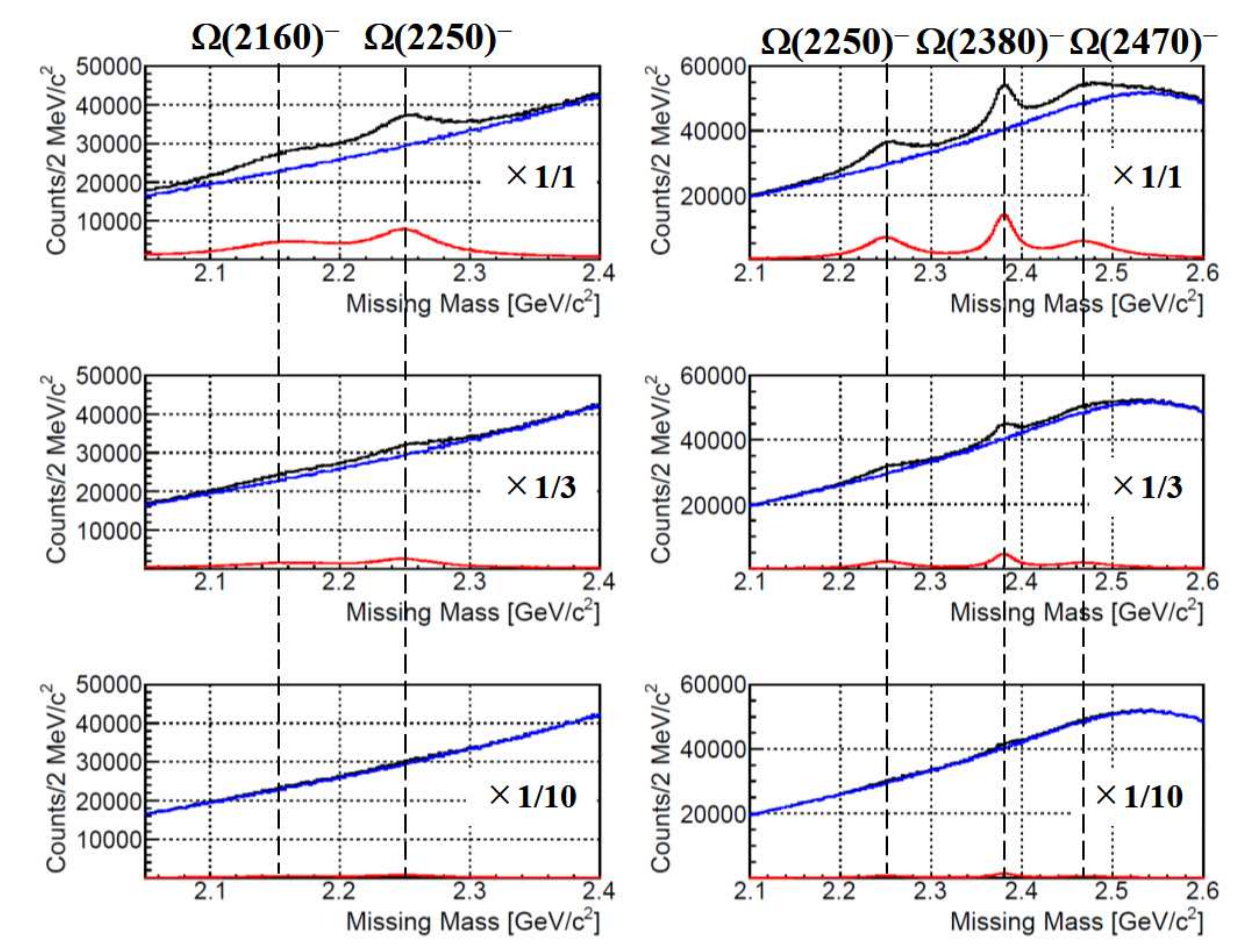}
  \end{center}
  \caption{
  Expected $\Omega^{*}$-mass spectra, or the
$p(K^-,K^+K^{*0})$ missing-mass spectra,
including the background contribution 
  at the incident kaon momentum of 8 GeV$/c$
  in a 100-day beam time
  by reducing $\Omega^*$-production cross sections by a factor 
  of 1/1 (original), 1/3 and 1/10 from the top to bottom panels.
  The mass spectra are expanded 
  for $\Omega(2160)^-$ and high-mass $\Omega^*$'s in the left and right panels, respectively.
  }
  \label{k10exp_fig9_2}
\end{figure}

Even if the cross section is smaller by a factor of 1/3, 
we can still recognize the $\Omega^-$, $\Omega(2012)^-$, $\Omega(2250)^-$, and $\Omega(2380)^-$ peaks.
Figures~\ref{k10exp_fig9_1} and~\ref{k10exp_fig9_2} show the expected $\Omega^{(*)}$-mass spectra  
including the background contribution at the incident kaon momentum of 8 GeV$/c$
by reducing $\Omega^*$-production cross sections by a factor of 1/1, 1/3, 1/10 and 1/30.
If the cross section is smaller by 1/10 than that in the original assumption (63 nb),
it is difficult to identify $\Omega^*$ peaks except for the ground-state $\Omega^-$ and first-excited $\Omega(2012)^-$.
In this case, additional background reduction 
is required 
for finding an $\Omega^*$ with a broad width.

%%%%%%%%%%%%%%%%%%%%%%%%%%%%%%%%%%%%%%%%%%%%%%%%%%%%%%%%%%%%%%%%%%%%%%%%%%%%%%%%%%%%%
\subsubsection*{(d) Possible background reduction}\label{sec:omega-improve-sn}

\begin{figure}[t]
  \begin{center}
  \includegraphics[width=15.5cm,keepaspectratio,clip]{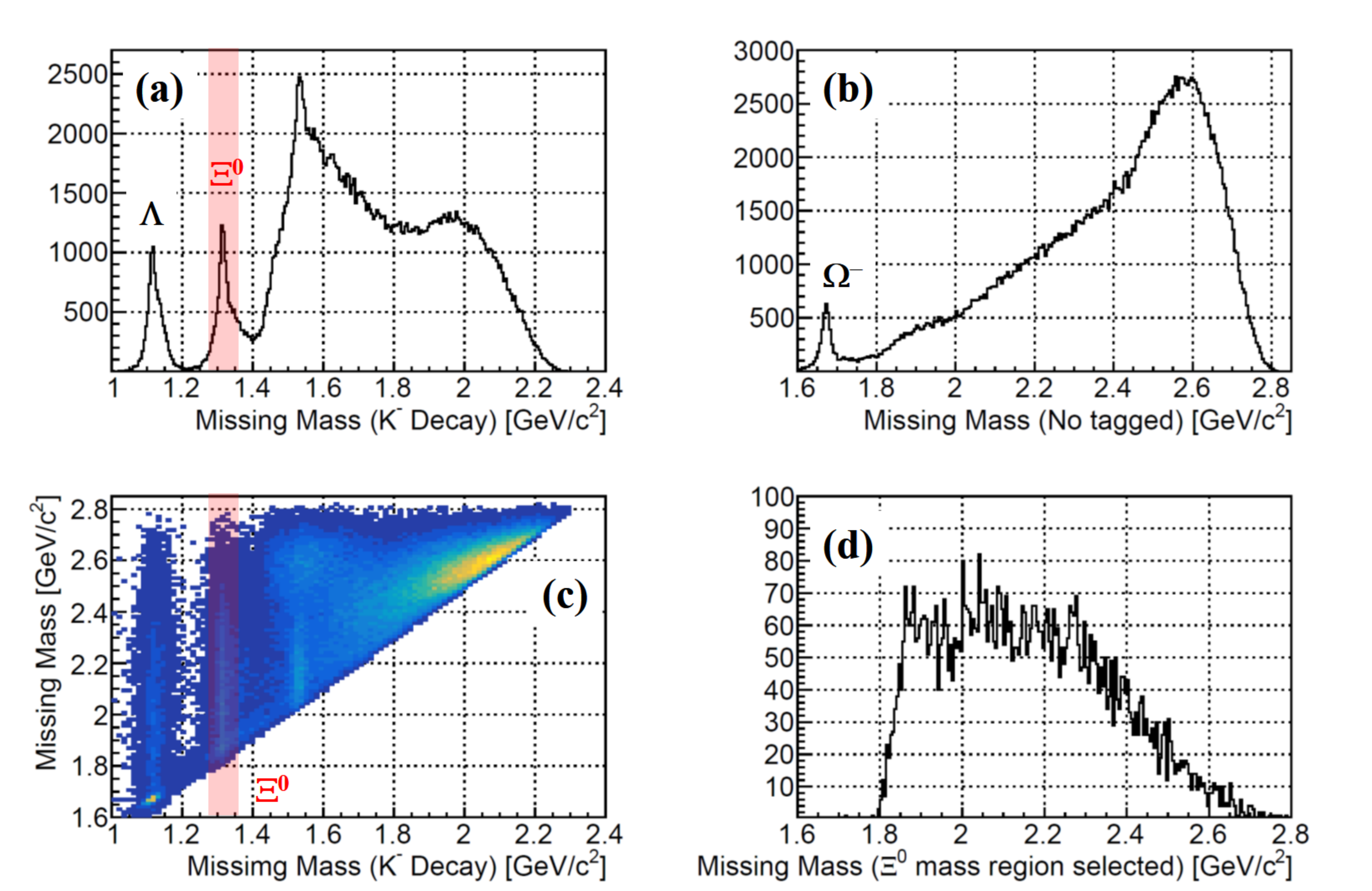}
  \end{center}
  \caption{ 
  (a) $p(K^-,K^+K^{*0}K^-)$ missing-mass spectrum for the background events generated by JAM.
  (b) $p(K^-,K^+K^{*0})$  missing-mass spectrum for the background events.
  (c) Correlation between the the $p(K^-,K^+K^{*0})$ and the $p(K^-,K^+K^{*0}K^-)$ missing masses for the background events.
  (d) $p(K^-,K^+K^{*0})$ missing-mass spectrum for the background events
  after selecting the events near $\Xi^0$ mass in the $p(K^-,K^+K^{*0}K^-)$ missing-mass spectrum.
  }
  \label{k10exp_fig10}
\end{figure}

We plan to determine the mass spectrum of $\Omega^{*-}$'s originally without 
using any information on their decay.
Selecting the events that contain the daughter particles from $\Omega^{*-}$'s is a possible 
way to get the higher S/N ratio.
From now on, we focus on 
the $\Omega^{*-} \rightarrow \Xi^0\;K^-$ decay channel 
since its branching ratio is expected to be large.
It should be noted that the produced $\Omega^{*-}$'s 
can decay into not only the 
channel described above but also other channels
in the simulation depending on the JAM parameters.
Figure~\ref{k10exp_fig10}(a) shows the $p(K^-,K^+K^{*0}K^-)$ missing-mass spectrum 
for the background events generated by JAM.
In Fig.~\ref{k10exp_fig10}(a), 
peaks corresponding to $\Lambda$, $\Xi^0$, and $\Xi(1530)^{0}$ are observed. 
The generated events in JAM include those from $K^-\;p\to K^+\;K^{*0}\;K^-\;\Xi^{0(*)}$, 
showing the $\Xi^0$ and $\Xi^{*0}$ peaks.
The events forming the $\Lambda$ peak come mainly 
from the decay of the ground-state $\Omega^-$ which is included in JAM.
It also comes from  the $K^-\;p \to \Xi^- K^+ \phi$ reaction
followed by the $\Xi^-\to \Lambda\pi^-$ and $\phi \to K^+\;K^-$ decays.
Since the width of $K^{*0}$ is rather broad,
$K^+$ and $\pi^-$ in the final-state of this reaction sequence can be identified as $K^{*0}$.
Detailed analysis of the production and decay vertices is expected to remove this background process.
Figure~\ref{k10exp_fig10}(b) shows
the $p(K^-,K^+K^{*0})$ missing-mass spectrum for the background events.
A peak corresponding to $\Omega^-$ can be observed clearly.
Above this peak, a huge amount of the continuum background events are observed
which prevent us from observing highly-excited $\Omega^*$'s.
Figure~\ref{k10exp_fig10}(c) shows the correlation plot
between the the $p(K^-,K^+K^{*0})$ and the $p(K^-,K^+K^{*0}K^-)$ missing masses.
By selecting the events that the $p(K^-,K^+K^{*0}K^-)$ missing masses 
are consistent with the  $\Xi^0$ mass,
we can remove huge background concentrated as shown in
Fig,~\ref{k10exp_fig10}(c) at a $p(K^-,K^+K^{*0})$ missing mass of $\sim 2.6$ GeV and
a $p(K^-,K^+K^{*0}K^-)$ missing mass of $\sim 2.1$ GeV.
Figure~\ref{k10exp_fig10}(d) shows the $p(K^-,K^+K^{*0})$ missing-mass spectrum for the background events
after selecting the events 
that the $p(K^-,K^+K^{*0}K^-)$ missing masses
are consistent with the $\Xi^0$ mass.
As compared with Fig.~\ref{k10exp_fig10}(b), the background level has been reduced by a factor of /10 or 1/100.
To get a significantly improved S/N ratio,
the branching ratio must be $\sim 0.3$ or higher 
for the $\Omega^{*-} \rightarrow \Xi^0\;K^-$ decay.

\begin{figure}[htpb]
  \begin{center}
  \includegraphics[width=15.5cm,keepaspectratio,clip]{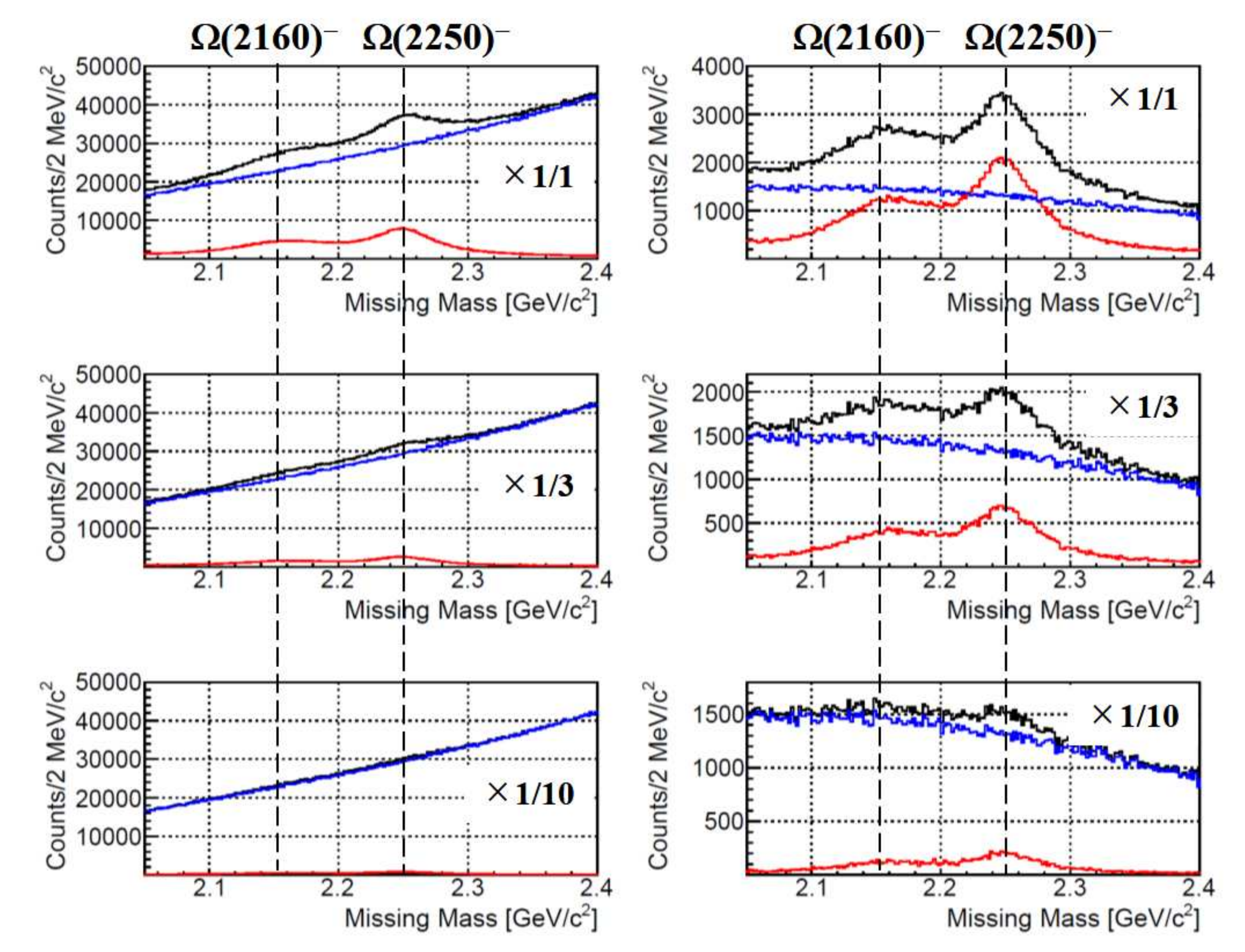}
  \end{center}
  \caption{
  Expected $\Omega^{*}$-mass spectra, or the $p(K^-,K^+K^{*0})$ missing mass spectra including the background contribution
  at the incident kaon momentum of 8 GeV$/c$
    in a 100-day beam time.
Here, the expected $\Omega^*$-production cross sections 
are reduced by a factor 
  of 1/1, 1/3 and 1/10 from the top to bottom panels.
  The left panel corresponds to the spectra that any requirement is applied for the $p(K^-,K^+K^{*0}K^-)$ missing mass.
  The right panel corresponds to the spectra that 
  the events are selected near the $\Xi^0$ mass in the $p(K^-,K^+K^{*0}K^-)$ missing-mass spectrum.
  The mass spectra are expanded for $\Omega(2160)^-$.
  }
 \label{k10exp_fig11_1}
\end{figure}

\begin{figure}[htpb]
  \begin{center}
  \includegraphics[width=15.5cm,keepaspectratio,clip]{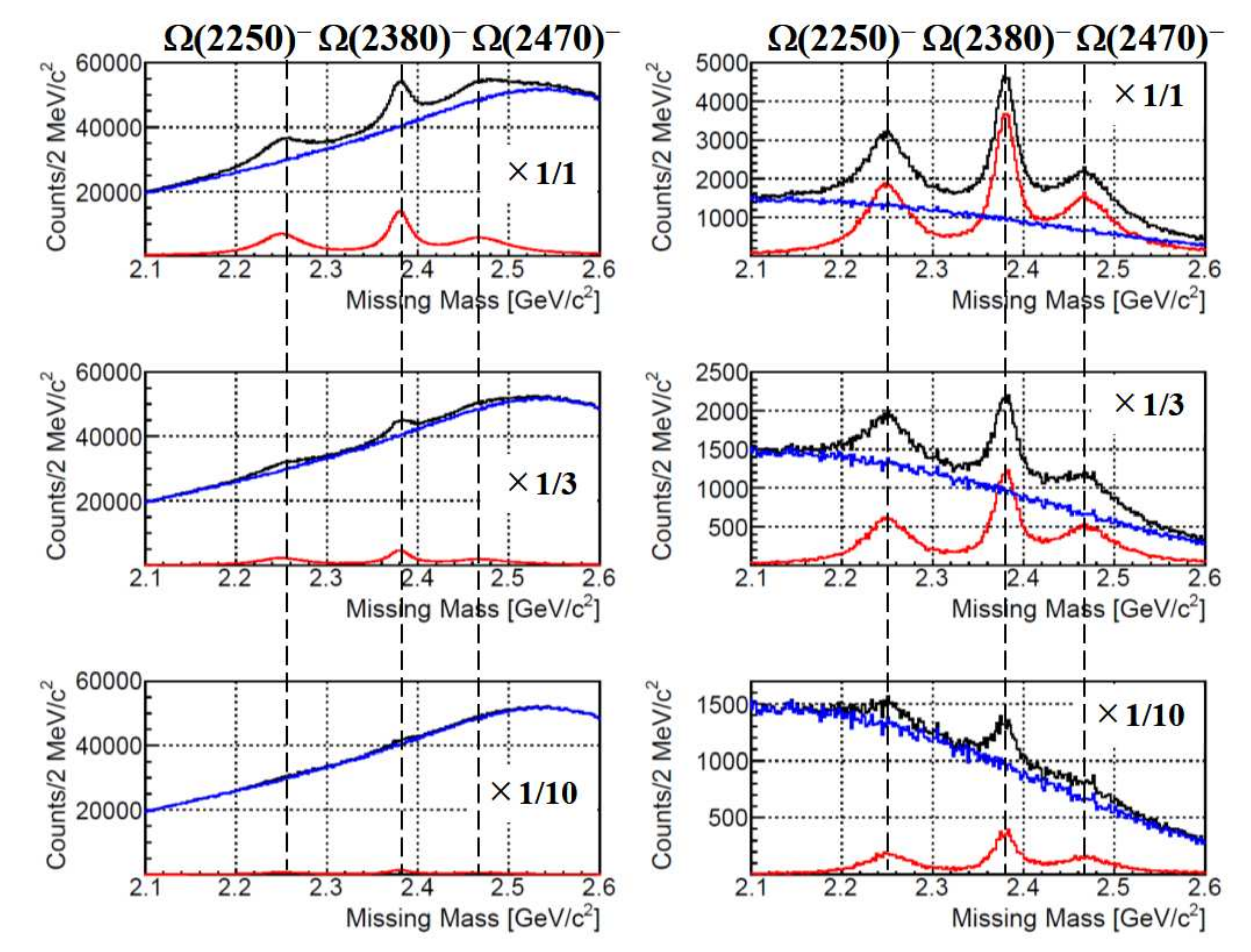}
  \end{center}
  \caption{
  Expected $\Omega^{*}$-mass spectrum, or the $p(K^-,K^+K^{*0})$ missing mass spectra  including  thebackground contribution
  at the incident kaon momentum of 8 GeV$/c$ 
  in a 100-day beam time.
Here, the expected $\Omega^*$-production cross sections 
are reduced by a factor 
  of 1/1, 1/3 and 1/10 from the top to bottom panels.
  The left panel corresponds to the spectra that any requirement is applied for the $p(K^-,K^+K^{*0}K^-)$ missing mass.
  The right panel corresponds to the spectra that 
  the events are selected near the $\Xi^0$ mass in the $p(K^-,K^+K^{*0}K^-)$ missing-mass spectrum.
  The mass spectra are expanded for high-mass $\Omega^*$'s.
  }
 \label{k10exp_fig11_2}
\end{figure}

Selecting the events containing $\Xi^0$ as a daughter particle from the $\Omega^*$ decay
is effective at background reduction in the $\Omega^{*}$-mass spectrum
if the branching ratio is $\sim 0.3$ for the $\Omega^{*-} \rightarrow \Xi^0\;K^-$ decay.
As shown in Figs.~\ref{k10exp_fig11_1} and~\ref{k10exp_fig11_2},
the S/N ratio is improved by a factor of 1/10 for the high-mass $\Omega^*$'s
so that we can recognize $\Omega^*$'s with a broad width 
even when the cross section is smaller than that originally assumed.
We can also find the $\Omega(2160)^-$ or the Roper-like state having a broad width of 100 MeV
even if the cross section is smaller by 1/10 than 
that in the original assumption.
Since $\Xi^0$ is a rather stable particle, we can recognize its flight path before decaying.
Thus, detailed analysis of the production and decay vertices of an $\Omega^*$ 
would help us to get the further higher S/N ratio.

\clearpage
%%%%%%%%%%%%%%%%%%%%%%%%%%%%%%%%%%%%%%%%%%%%%%%%%%%%%%%%%%%%%%%%%%%%%%%%%%%%%%%%%%%%%
\subsubsection*{(e) Different incident kaon momenta}

\begin{figure}[t]
  \begin{center}
  \includegraphics[width=16cm,keepaspectratio,clip]{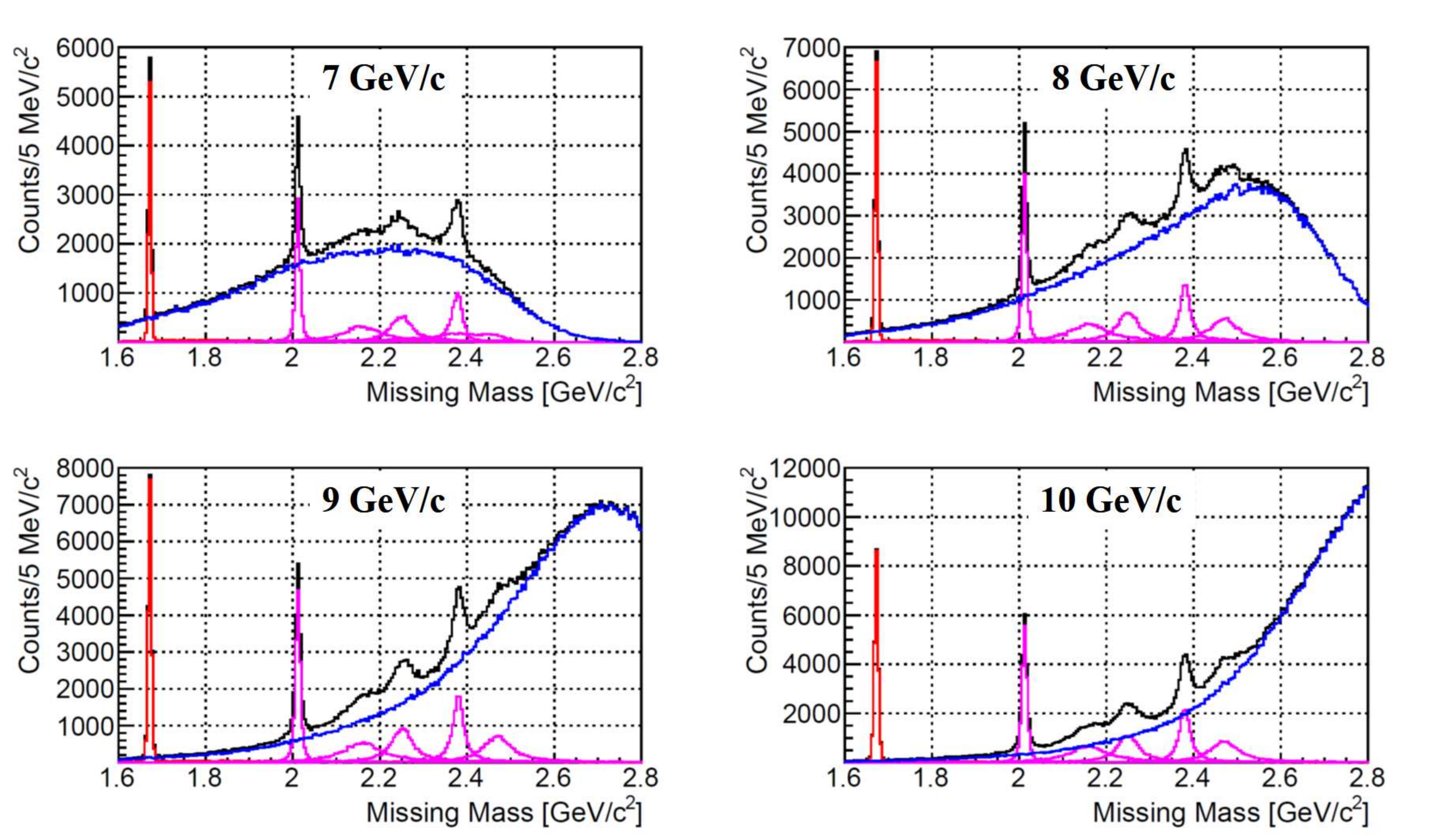}
  \end{center}
  \caption{ 
  Expected $\Omega^{(*)}$-mass spectra, or the $p(K^-,K^+K^{*0})$ missing mass spectra,  including the background contribution
  at the incident kaon momenta of 7 GeV$/c$ (top-left), 8 GeV$/c$ (top-right), 
  9  GeV$/c$ (bottom-left), and 10 GeV$/c$ (bottom-right)
  in a 3-day beam time.
  The smooth blue curve represent the background contribution estimated by JAM, 
  and the contributions from the ground-state $\Omega$ and $\Omega^*$'s are also plotted in red and in magenta.
  }
  \label{k10exp_fig12}
\end{figure}

Thus far, shown are only the expected $\Omega^{(*)}$-mass spectra at the incident kaon momentum of 8 GeV/$c$.
We also estimate those at several different incident kaon momenta.
Figure~\ref{k10exp_fig12} shows the expected $\Omega^{(*)}$-mass spectra
at incident kaon momenta of 7, 8, 9, and 10 GeV$/c$.
The excitation-energy range covered in the measurement becomes wider with increase of the incident kaon momentum,
and high-mass $\Omega^*$'s are likely to be observed 
up to the excitation energy of 1.5 GeV at the highest incident kaon momentum of 10 GeV/$c$.
Within the excitation-energy coverage,
all the peaks corresponding to $\Omega^*$'s are clearly observed with high S/N ratios.
It is important to measure the $\Omega^*$-mass spectra at different 
incident kaon momenta.
A fake structure may appear caused by some kinematic effects
since the measured spectrum reflects the appearance of a resonance 
in other system of the final-state particles than that corresponding to
$\Omega^*$'s.
This kind of kinematic effects are observed in a different shape 
at a different incident kaon momentum.
The mass spectra with several incident momenta
enable us to identify fake structures coming from the kinematic effects.
The mass spectra with different incident
momenta are also useful for determination
of the the $\Omega^{*-}$ widths.
The optimum incident kaon momentum is different 
for different excitation energies since the sensitivity
of the width determination highly depends on 
the background level and background shape.

%%%%%%%%%%%%%%%%%%%%%%%%%%%%%%%%%%%%%%%%%%%%%%%%%%%%%%%%%%%%%%%%%%%%%%%%%%%%%%%%%%%%%

% flatex input end: [./k10docu_exp/k10-omega-exp2_v2.tex]
\label{sec:omega-spectroscopy-suppl2}
%\clearpage
% flatex input: [./k10docu_exp/k10-omega-spinparity_v2.tex]
%%%%%%%%%%%%%%%%%%%%%%%%%%%%%%%%%%%%%%%%%%%%%%%%%%%%%%%%%%%%%%%%%%%%%%%%%%%%%%%%%%%%%
\subsubsection{Spin-parity determination of $\Omega^{*}$s \label{sec:angulardist}}

The spin-parity assignment of a produced $\Omega^{*-}$ provides
crucial insight into the internal quark motion.
Here, we discuss possible ways to determine the spin-parity of 
$\Omega^{*-}$.
First, 
we determine the spin of $\Omega^{*-}$
from the angular distribution of $\Omega^{*-}\to \Xi^0\;K^-$.
Since the spin of $\Xi^0$ is 1/2 and that of $K^-$ is 0,
the distribution of $K^-$ ($\Xi^0$) emission 
in the rest frame of $\Omega^{*-}$
is expressed using
the spin-density matrix (SDM) of $\Omega^{*-}$ with a spin of $J$ as 
\begin{equation}
\begin{array}{lll}
W(\theta)\propto &1 
& {\rm for\ } J=1/2,\\
W(\theta)\propto & 3\rho_{33}\sin^2\theta+\rho_{11}(1+3\cos^2\theta) 
& {\rm for\ } J=3/2, {\rm and}\\
W(\theta)\propto &5\rho_{55}(1-\cos^2\theta)^2 +\rho_{33}(1+14\cos^2\theta-15\cos^4\theta)\\
&+ 2\rho_{11}(1-2\cos^2\theta+5\cos^4\theta) 
& {\rm for\ } J=5/2.\\
\\
\end{array}
\label{eq:sdm}
\end{equation}
Here,  $\theta$ is the emission angle of the daughter particle  ($\Xi^0$ or $K^-$)
with respect to the $z$ axis in the rest frame of $\Omega^{*-}$,
and $\rho_{2J_z2J_z}$ denotes the diagonal elements of SDM.
It should be noted that $\rho_{-2J_z-2J_z}=\rho_{2J_z2J_z}$.
We have several choices for the $z$-axis.
Suppose $z$-axis is defined to be opposite to the direction of the $K^+\; K^{(*)0}$ composite system produced together with $\Omega^{*-}$.
In this case, $J_z$ is equivalent to the helicity of $\Omega^{*-}$,
and non-uniform helicity distribution enables us to determine the spin of $\Omega^*$.
However, we obtain the uniform angular distribution 
independently of the spin of $\Omega^{*-}$ when the helicity distribution is uniform.
Taking into account the $K^-\;p\to K^+ \;\Xi^{*0}$ reaction followed by 
the $\Xi^{*0}\to  \Omega^{*-}\;K^{(*)0}$ decay alternatively,
we can define the $z$ axis is 
the normal vector of the $\Xi^{*0}$-production plane containing 
the initial $K^-$- and final $K^+$-momentum vectors.
The $\Xi^{*0}$ must be polarized and $\Omega^{*-}$ is also expected to be polarized in some fraction with respect to this $z$ axis.

\begin{figure}[t]
  \begin{center}
  \includegraphics[width=15.5cm,keepaspectratio,clip]{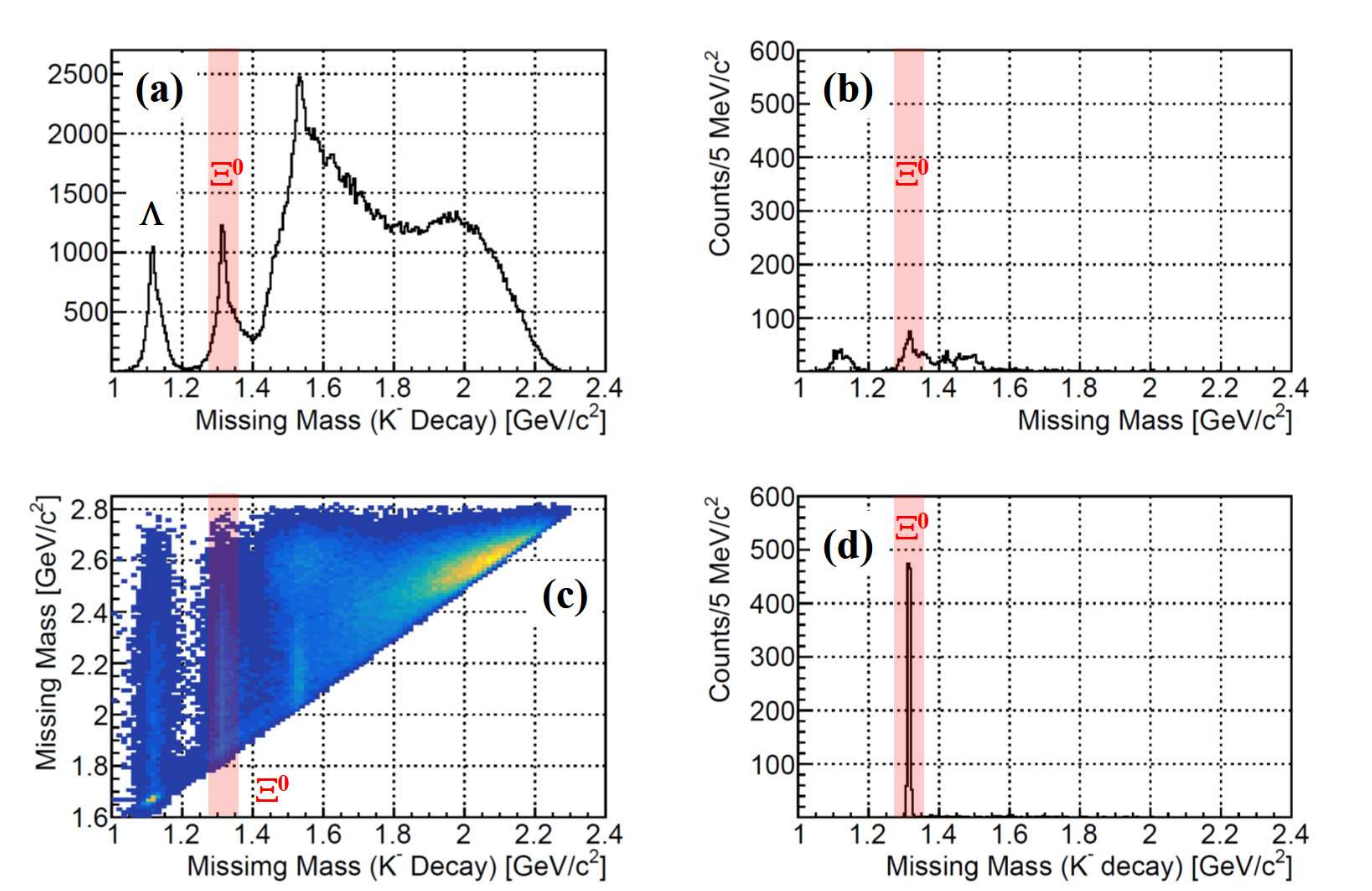}
  \end{center}
  \caption{ 
  (a) $p(K^-,K^+K^{*0}K^-)$ missing-mass spectrum for the background events estimated by JAM (same as Fig.~\ref{k10exp_fig10}(a)).
  (b) $p(K^-,K^+K^{*0}K^-)$ missing-mass spectrum for the background events
  after selecting the events near $\Omega^-(2012)$ mass in the $p(K^-,K^+K^{*0})$ missing mass spectrum.
  (c) Correlation between the the $p(K^-,K^+\;K^{*0})$ and the $p(K^-,K^+K^{*0}K^-)$ missing masses for the background events.
  (d) $p(K^-,K^+K^{*0})$ missing-mass spectrum for the $\Omega(2012)^-$-produced events
  after selecting the events near the $\Omega(2012)^-$ mass in the $p(K^-,K^+K^{*0})$ missing mass spectrum.
  }
  \label{k10exp_fig13}
\end{figure}

We show here how to obtain the angular distribution 
of $\Xi^0$ emission (decay angular distribution)
of an $\Omega^{*-}$ in the two-body
$\Omega^{*-} \rightarrow \Xi^0\;K^-$ decay.
The branching ratio assumed is 0.3 for this decay channel.
In principle, the $\Omega^{*-}$-produced events are identified by using the $p(K^-,K^+ K^{*0})$ missing mass. 
The decay angular distribution can be given additionally detecting the daughter $K^-$ from the $\Omega^{*-}$ decay.
Here, we take an example for the decay angular distribution 
for $\Omega(2012)^-$ as an $\Omega^{*-}$.
We select the events that the $p(K^-,K^+K^{*0})$ and $p(K^-,K^+K^{*0}K^-)$ missing masses 
are close to the $\Omega(2012)^-$ and $\Xi^0$ masses, respectively.
The selected region is indicated in Fig.~\ref{k10exp_fig13}(c).
Figure~\ref{k10exp_fig13}(b) shows the $p(K^-,K^+K^{*0}K^-)$ missing-mass spectrum for the background events
after selecting the events near the 
$\Omega(2012)^-$ mass in the $p(K^-,K^+K^{*0})$ missing-mass spectrum,
and Figure~\ref{k10exp_fig13}(d) for the $\Omega(2012)^-$-produced events.
Let us look at the $p(K^-,K^+K^{*0}K^-)$ missing-mass spectrum
as shown in Fig.~\ref{k10exp_fig13} (b) for the background contribution
and in Fig.~\ref{k10exp_fig13} (d) for $\Omega(2012)^-$ production.
The background contribution is well suppressed and distributed 
in a wider range as compared with $\Omega(2012)^-$ production.
Thus, a side-band subtraction method can be used to extract the $\Omega(2012)^-$-produced
events for deduction of the decay angular distribution.
The decay angular distribution can be obtained similarly to $\Omega(2012)^-$ for the other $\Omega^*$'s.

\begin{figure}[t]
  \begin{center}
  \includegraphics[width=10.5cm,keepaspectratio,clip]{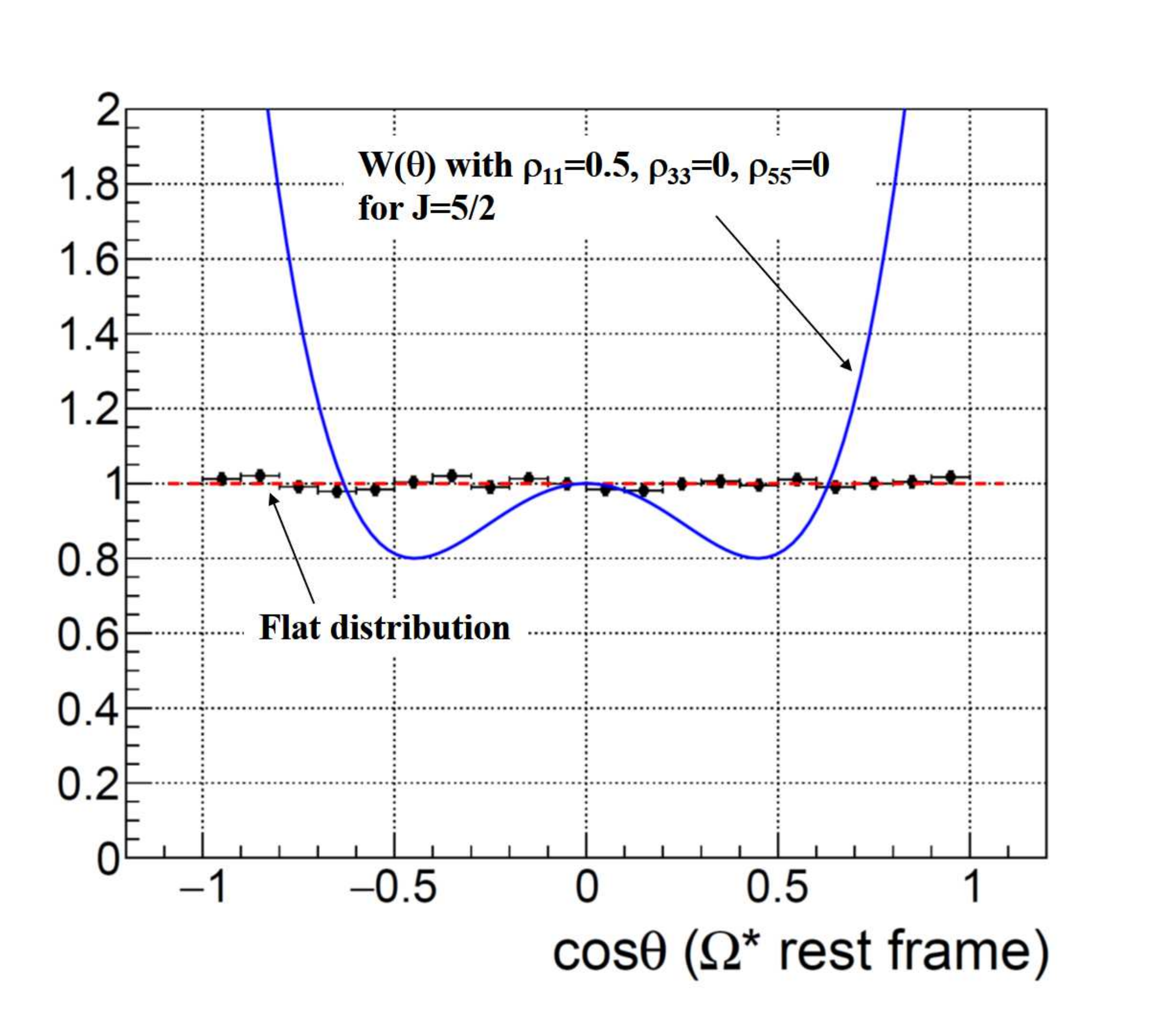}
  \end{center}
  \caption{
  Acceptance-corrected decay angular distribution of $\Xi^0$ emission in the reset frame of $\Omega^{*-}$
  ($z$-axis: opposite to the direction of the $K^+\; K^{*0}$ composite system).
  Isotropic distributions corresponding to the event generation are obtained correctly.
  The data are plotted only with statistical errors.
  The $W(\theta)$ functions corresponding to the isotropic generation
  and the case with $\rho_{11}=0.5,\;\rho_{33}=0,\;\rho_{55}=0$ for $J=5/2$
  are also plotted in the red dashed and blue curves, respectively.
  }
  \label{k10exp_fig14}
\end{figure}

We estimate the decay angular distribution in the rest frame of the produced $\Omega^{*-}$.
Here, we take the $z$-axis along the opposite direction to the momentum 
of the $K^+ K^{(*)0}$ composite system produced together with $\Omega^*$.
Figure~\ref{k10exp_fig14} shows the acceptance-corrected decay angular distribution.
The isotropic decay angular distributions are obtained according to isotropic event generation 
of the $\Omega^{*-} \rightarrow \Xi^0\;K^-$ decay in the simulation.
The azimuthal-angle asymmetry is simply ignored here,
the effect of which is canceled out in the polar angle distribution.
The $\rho_{2J_z2J_z}$ can be determined by fitting a function expressed 
by Eq.~(\ref{eq:sdm}) to the observed angular distribution.
In Fig.~\ref{k10exp_fig14}, 
the $W(\theta)$ functions corresponding to the isotropic generation 
(uniform $J_z$ sub-population for a certain $J$)
and the case with $\rho_{11}=0.5,\;\rho_{33}=0,\;\rho_{55}=0$ for $J=5/2$
are plotted.
%An isotropic distribution can be explained by any $J$ with uniform $J_z$ 
%sub-population:
%always for $J=1/2$, $\rho_{11}=\rho_{33}=1/4$ for $J=3/2$, and
%$\rho_{11}=\rho_{33}=\rho_{55}=1/6$ for $J=5/2$.
When the angular distribution shows some structure,
we may identify  $J$ and determine $\rho_{2J_z2J_z}$'s with an error of $\sim$1\%.
We can perform more detailed analysis for determining the spin for the $\Omega^{*-}$
of interest by combining other information such as the branching ratio.

\begin{figure}[t]
  \begin{center}
  \includegraphics[width=7.cm,keepaspectratio,clip]{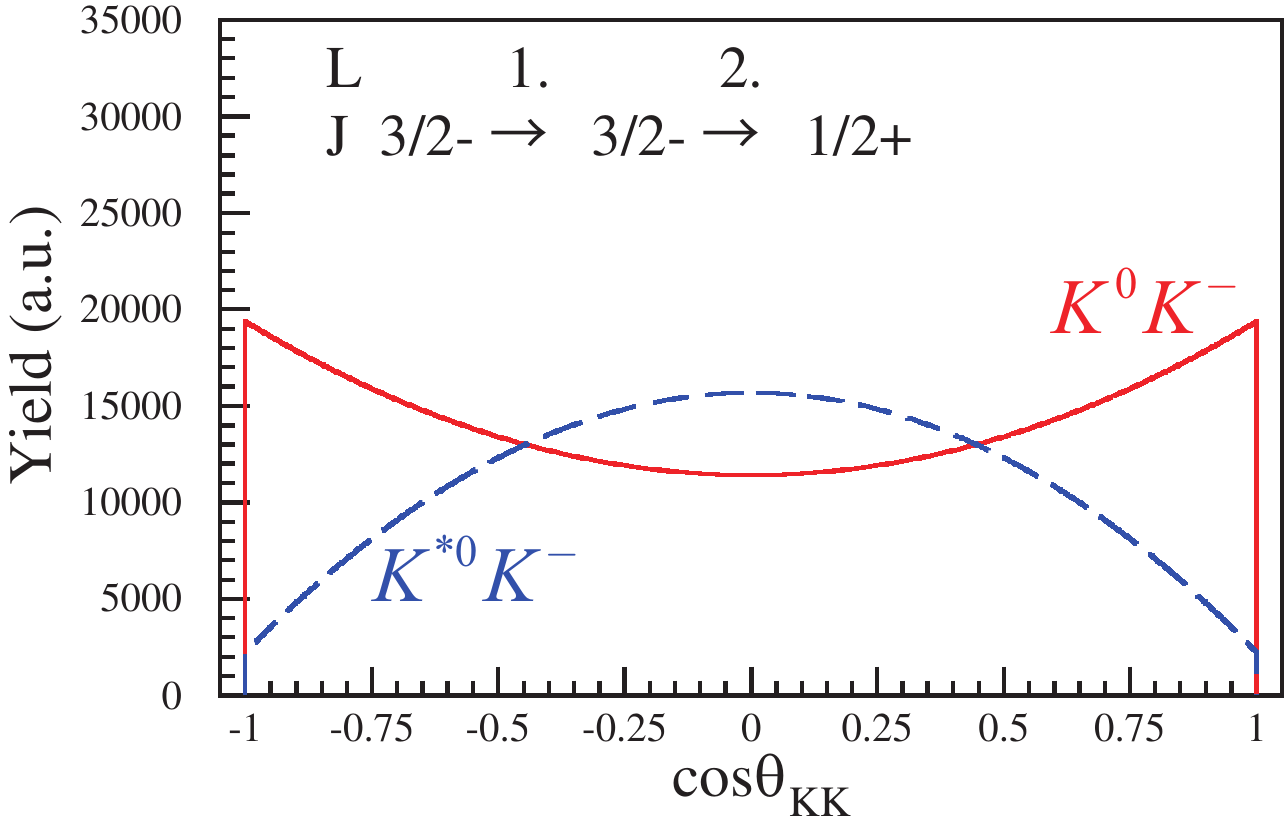}
  \includegraphics[width=7.cm,keepaspectratio,clip]{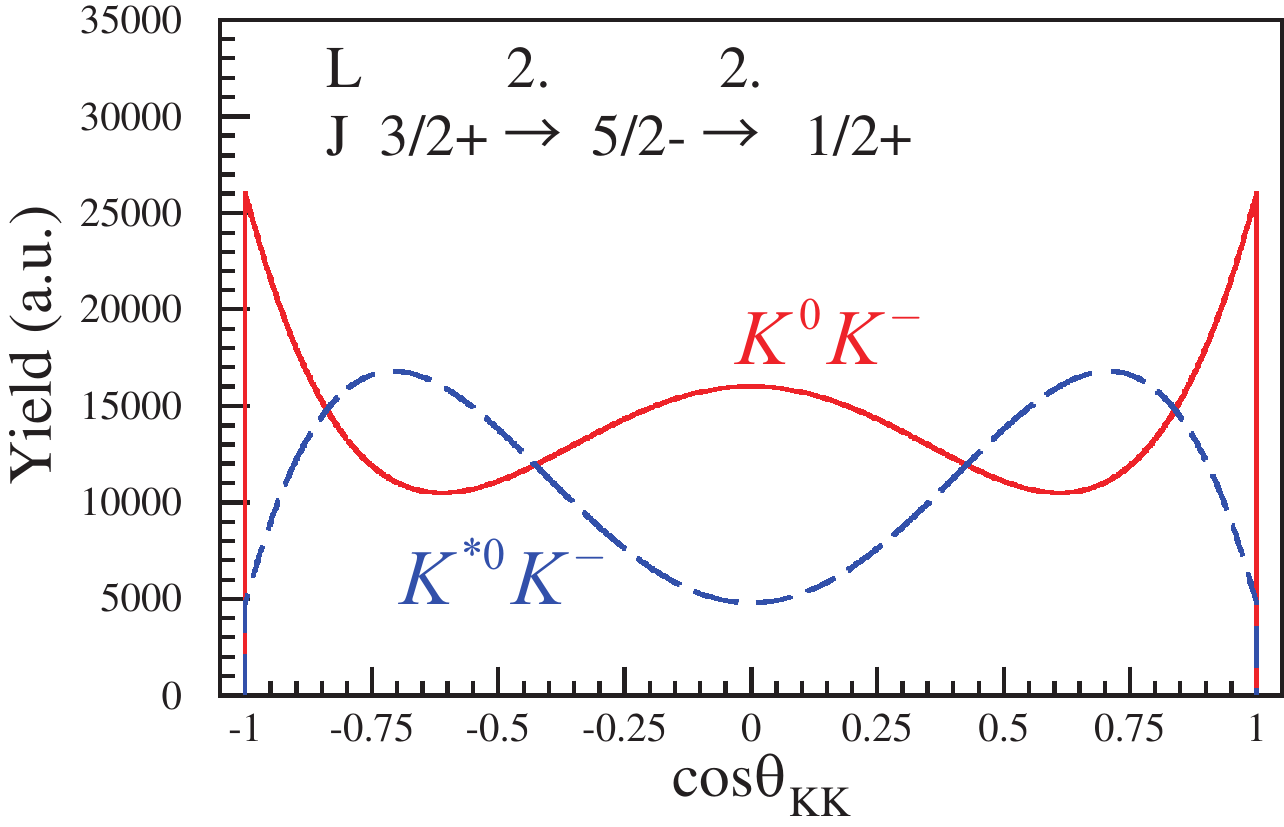}
  \end{center}
  \caption{Typical angular correlations for the 
$\Xi^{*-} \to \Omega^{*-}\to \Xi^0$ transitions.
In the left (right) panel, spin-parities of $\Xi^{*-}$ and $\Omega^{*-}$ are 
$3/2^-$ 
($3/2^+$)
and $3/2^-$ ($5/2^-$), respectively.
The angular momenta carried by meson emissions from 
$\Xi^{*-}$ and $\Omega^{*-}$ are
1 (2) and 2 (2), respectively.
The $\Omega^{*-}$ is assumed in its decay to emit $K^-$,
and both $K^0$ and $K^{*0}$ emissions are considered for the $\Xi^{*-}$ decay.
The red solid curves show the angular correlations between $K^{0}$ and 
$K^-$, and the blue dashed show those between $K^{*0}$ and $K^-$.
  }
  \label{k10exp_fig_angcor}
\end{figure}

%\bibitem{bied53} L.C.~Biedenharn, and M.E.~Rose, Rev.\ Mod.\ Phys.\ {\bf 25},  729 (1953).
%\bibitem{mori14} K.~Moriya {\it et al.} (CLAS Collaboration), Phys.\ Rev.\ Lett.\ {\bf 112}, 082004 (2014).
When we observe an isolated $\Xi^{*-}$ state with a certain spin-parity
in the $K^-\; p \rightarrow K^+\; \Xi^{*-}$ reaction followed 
by $\Xi^{*-} \rightarrow \Omega^{*-}\;K^{(*)0}$
and  $\Omega^{*-}\to \Xi^0\;K^- $ decays,
we can determine the spin-parity of $\Omega^{*-}$ 
from the angular correlation between $K^{(*)0}$ and $K^-$.
Figure~\ref{k10exp_fig_angcor}
shows the typical angular correlations for the 
$\Xi^{*-} \to \Omega^{*-}\to \Xi^0$ transitions for
certain 
 spin-parities of $\Xi^{*-}$ and $\Omega^{*-}$,
and angular momenta carried by meson emissions.
Here, we have calculated the $K^{(*)0}$ and $K^-$ angular correlations  
for the sequential decay 
of $\Xi^{*-} \to \Omega^{*-}\to \Xi^0$
using the density matrix (statistical tensor) formalism~\cite{bied53}.
It should be noted that 
the spin-parity of $\Xi^{*-}$ can be determined in its decay 
into the ground-state $\Omega^-$ with a spin-parity of $3/2^+$,
$\Xi^{*-} \rightarrow \Omega^-\;K^{(*)0}$.
The angular correlation is isotropic (uniform) 
%and the parity cannot be determined 
when the spin of $\Omega^{*-}$ is 1/2.
In this case, the parity of  $\Omega^{*-}$ is determined 
by using the polarization transfer from $\Omega^{*-}$ to $\Xi^0$
similarly to the parity determination of $\Lambda(1405)1/2^-$
by the CLAS collaboration~\cite{mori14}.
When the spin-parity of $\Omega^{*-}$ is $1/2^-$ ($1/2^+$),
the  $\Omega^{*-}\to\Xi^0\; K^-$ decay takes place in the 
$S$ ($P$) wave.
The polarization direction of $\Xi^0$,
which can be obtained from the decay asymmetry of $\Xi^0 \to \Lambda\; \pi^0$,
is parallel to that of $\Omega^{*-}$ in the $S$-wave case.
On the hand, the polarization direction is
expressed by $\cos\theta_\Xi$ where 
$\theta_\Xi$ denotes the $\Xi^0$ emission angle with respect 
to the polarization direction of $\Omega^{*-}$.

\begin{figure}[htbp] 
 \centerline{\includegraphics[width=0.95\textwidth]{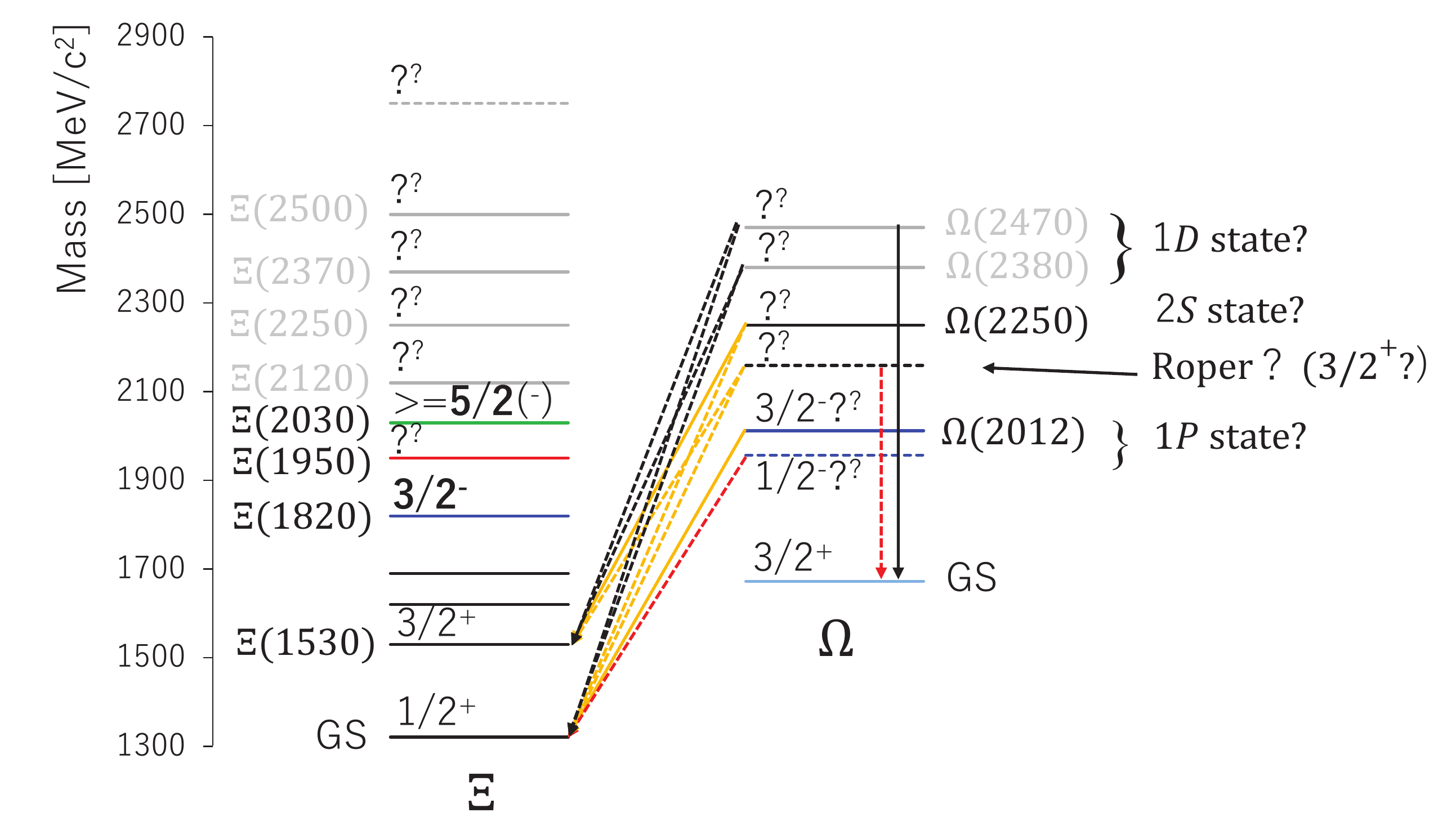}}
 \caption{Possible decays from $\Omega^{*-}$ to $\Xi^{(*)}$ by emitting 
$\bar{K}^{(*)}$.
The observed and expected decays are 
represented by the solid and dashed arrows, respectively.
\label{fig:level-omega1}}
 \centerline{\includegraphics[width=0.8\textwidth]{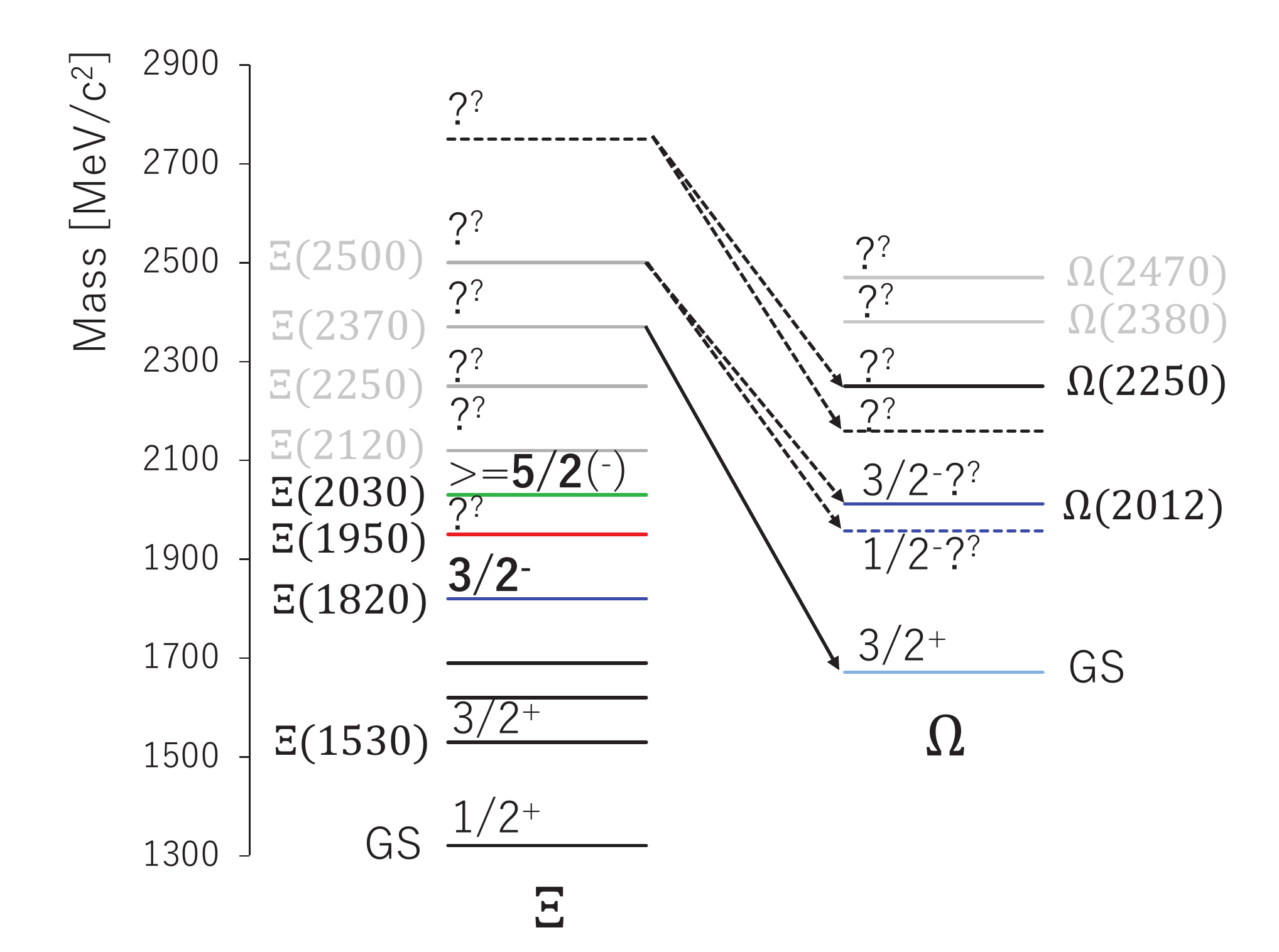}}
 \caption{Possible decays from $\Xi^{*-}$ to $\Omega^{*-}$ by emitting 
${K}^{(*)}$.
There are several predicted highly-excited $\Xi^*$'s,
which can play a role as a doorway to $\Omega^{*-}$ production.
\label{fig:level-omega2}}
\end{figure}

It is important to find a sequential $\Xi^{*-} \to \Omega^{(*)-} \to \Xi^0$
decay for $\Omega^{(*)-}$ with a known spin-parity
for the spin-parity determination of  $\Xi^{*-}$.
Figure~\ref{fig:level-omega1} shows possible decays from 
$\Omega^{*-}$ to $\Xi^{(*)}$.
The solid arrows indicate the observed decays
listed in Review of Particle Physics~\cite{Zyla:2020zbs}.
The $\Omega^{*-}$'s are expected to decay directly into the ground-state $\Xi$ 
by emitting $\bar{K}$. Even if this direct decay is not observed,
we can expect the decay into a low-lying $\Xi^*$ state of which 
spin-parity is known, $\Xi(1530)3/2^+$ for example.
On the other hand,
it is uncertain to find an isolated $\Xi^{*-}$ state decaying into $\Omega^{*-}$.
Figure~\ref{fig:level-omega2} shows
possible decays from $\Xi^{*-}$ to $\Omega^{*-}$.
There are several predicted highly-excited $\Xi^*$s,
which can play a role as a doorway to $\Omega^{*-}$ production.
These $\Xi^*$'s decay into $\Lambda^*$ or $\Sigma^*$ by emitting 
$\bar{K}^{(*)}$, and they must also decay into $\Omega^{*-}$ by emitting
${K}^{(*)}$.
Since the statistical tensor formalism can treat a mixed state for each 
of intermediate states and meson emissions,
we can also estimate the interference effects 
between several $\Xi^{*-}$ states and those with background contributions.

%--------------------------------------
%--------------------------------------

%\subsubsection{Decay partial width (branching ratio)}
We also expect to determine the parities from the decay widths for some $\Omega^*$'s.
Let us consider an $\Omega^{*-}$ with a spin of 1/2 decaying into 
$\Xi^0 \;K^-$.
The angular momentum $\lambda$ carried by $K^-$ emission is 
0 and 1 for the negative and positive parities of the $\Omega^{*-}$, respectively.
The decay width is proportional to $p^{2\lambda+1}$
where $p$ denotes the relative momentum between the daughter particles.
Thus the branching ratio of  $\Omega^{*-}\to \Xi^0 \;K^-$ with respect
to $\Omega^{*-}\to \Omega^-\; \pi\;\pi$ 
would be useful for the parity determination.
%However, the two body decay with larger $\lambda$ is suppressed 
%by a factor of $p/\Lambda$ 
%when $p$ becomes a few hundred MeV or higher.
%Here, $\Lambda \sim 0.8$ GeV$/c$ denotes the cut-off parameter 
%corresponding to the typical hadron size $\sim 0.5$ fm.
% flatex input end: [./k10docu_exp/k10-omega-spinparity_v2.tex]
\label{sec:omega-spectroscopy-suppl3}
%\clearpage

\clearpage

\clearpage

% flatex input end: [./k10docu/k10main.tex]

%==================================================================%

\printbibliography[segment=\therefsegment,heading=subbibliography]

\clearpage

%==================================================================%
\section{\centering Physics and Experiment at KL2 Beam Line}
\chapterauthor{
H.~Nanjo, T.~Nomura, K.~Shiomi, and G.~Y.~Lim,\\
for the KOTO collaboration\\
}
%==================================================================%

% flatex input: [./KLdocu/KLmain.tex]
% flatex input: [KLdocu/common/std_defs.tex]
\newcommand{\memo}[1]{\textcolor{red}{\textbf{#1}}}
\newcommand{\ie}{\textit{i}.\textit{e}.}

\newcommand{\kl}{K_L}
\newcommand{\klpionn}{K_L \to \pi^0 \nu \overline{\nu}}
\newcommand{\kpinn}{K \to \pi \nu \overline{\nu}}
\newcommand{\klgg}{K_L \to 2\gamma}
\newcommand{\klpiopio}{K_L \to \pi^0 \pi^0}
\newcommand{\klpiopiopio}{K_L \to \pi^0 \pi^0 \pi^0}
\newcommand{\klppm}{K_L \to \pi^+ \pi^- \pi^0}

\newcommand{\kpluspnn}{K^{+} \to \pi^+ \nu \overline{\nu}}

\newcommand{\pt}{p_{\mathrm{T}}}
\newcommand{\zvtx}{z_{\mathrm{vtx}}}

\newcommand{\kpien}{K_L \to \pi^\pm e^\mp \nu}
\newcommand{\kpimun}{K_L \to \pi^\pm \mu^\mp \nu}

\makeatletter
\newcommand\subsubsubsection{\@startsection{paragraph}{4}{\z@}%
                                     {-3.25ex\@plus -1ex \@minus -.2ex}%
                                     {1.5ex \@plus .2ex}%
                                     {\normalfont\normalsize\bfseries}}
\makeatother

% flatex input end: [KLdocu/common/std_defs.tex]

\newcounter{secnumdepthsave}
\setcounter{secnumdepthsave}{\value{secnumdepth}}
\setcounter{secnumdepth}{4}

%\section{Physics and Experiment at KL2 Beam Line}
% flatex input: [KLdocu/physics/physics.tex]
%\section{Physics Motivation}
\subsection{Physics Motivation}
\label{chap:physics}
\begin{figure}[tbph]
 \begin{center}
  \includegraphics[width=0.25\textwidth]{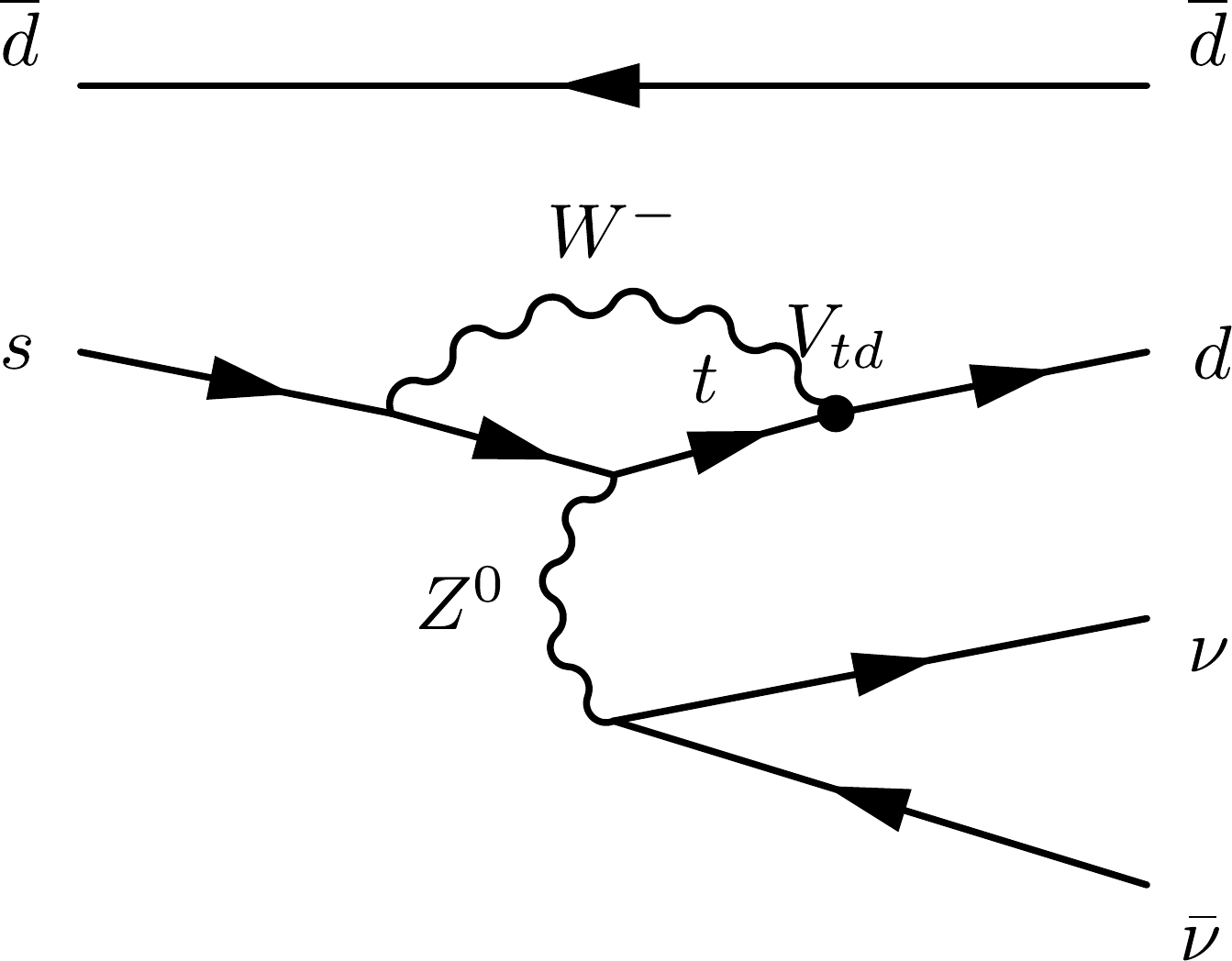}
  \hspace*{3mm}
  \includegraphics[width=0.25\textwidth]{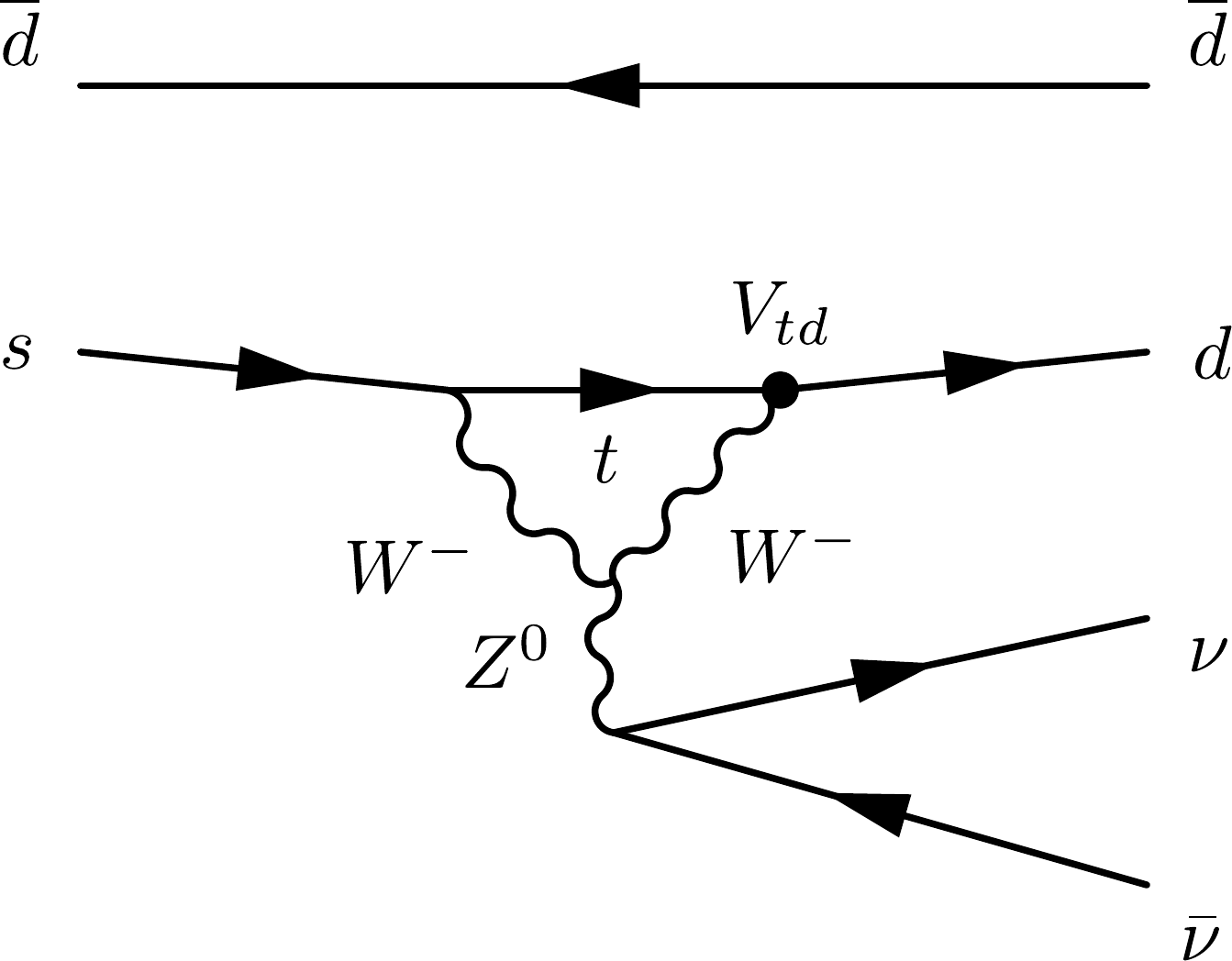} 
  \vspace*{3mm}
  \includegraphics[width=0.25\textwidth]{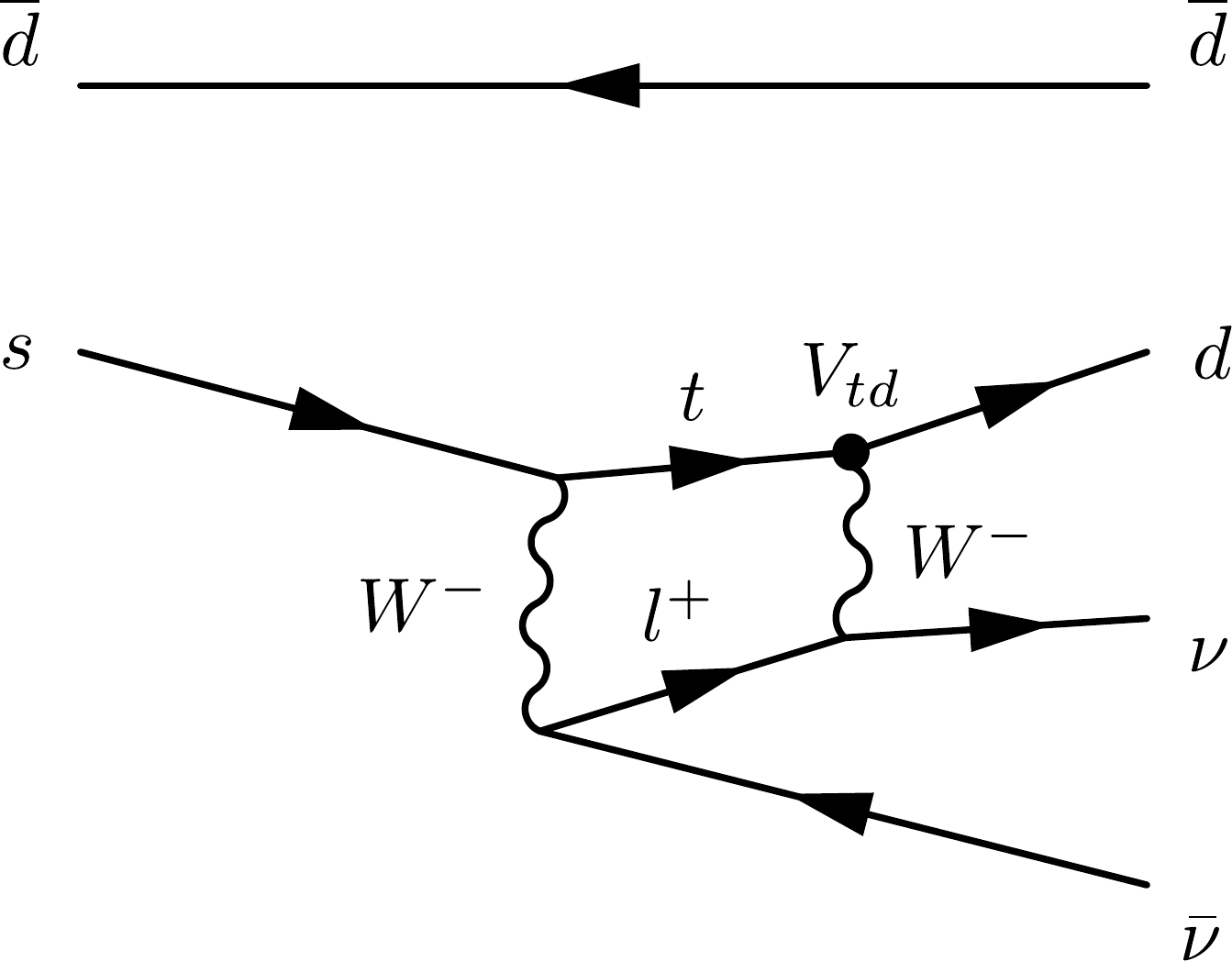}
  \caption{
  Feynman diagrams of the $\klpionn$ decay in the Standard Model. }
  \label{fdiagram}
 \end{center}
\end{figure}

The kaon rare decay $\klpionn$ provides a unique opportunity to search for new physics beyond the
Standard Model (SM) in particle physics.
The decay proceeds by a Flavor Changing Neutral Current (FCNC) from a strange to a down quark (s$\rightarrow$d transition) through loop effects 
expressed by the electroweak penguin and box diagrams shown in Fig.~\ref{fdiagram}.  
The s$\rightarrow$d transition is most strongly suppressed in the SM among other FCNC transitions such as b$\rightarrow$d and b$\rightarrow$s transitions 
due to the Glashow-Iliopoulos-Maiani (GIM) mechanism and the hierarchical structure of the Cabibbo-Kobayashi-Maskawa (CKM) matrix.
On the other hand, the flavor structure on new physics does not exhibit the CKM hierarchies in general.  
The contribution of new physics in the loop could be observed as the deviation of the branching ratio from the SM prediction
even if the energy scale of new physics is more than 100 TeV, which can not be reached by the LHC~\cite{ref:Zepto}.

The SM prediction of the branching ratio (BR) on the $\klpionn$ decay is $(3.00\pm0.30)\times10^{-11}$~\cite{ref:kpnnBR}. 
The uncertainties are dominated by the parametric uncertainties on the CKM elements, and the theoretical uncertainties 
are only 2.5\%~\cite{ref:kpnnBRError}, because the decay is entirely governed by short-distance physics involving the top quark.
The long-distance interaction intermediated by photons does not exist, and the hadronic matrix element can be precisely estimated 
by $K \rightarrow\pi e\nu$ data.  Therefore we can extract small effect of new physics from the SM contribution.

\begin{figure}[tbph]
 \begin{center}
  \includegraphics[width=12cm]{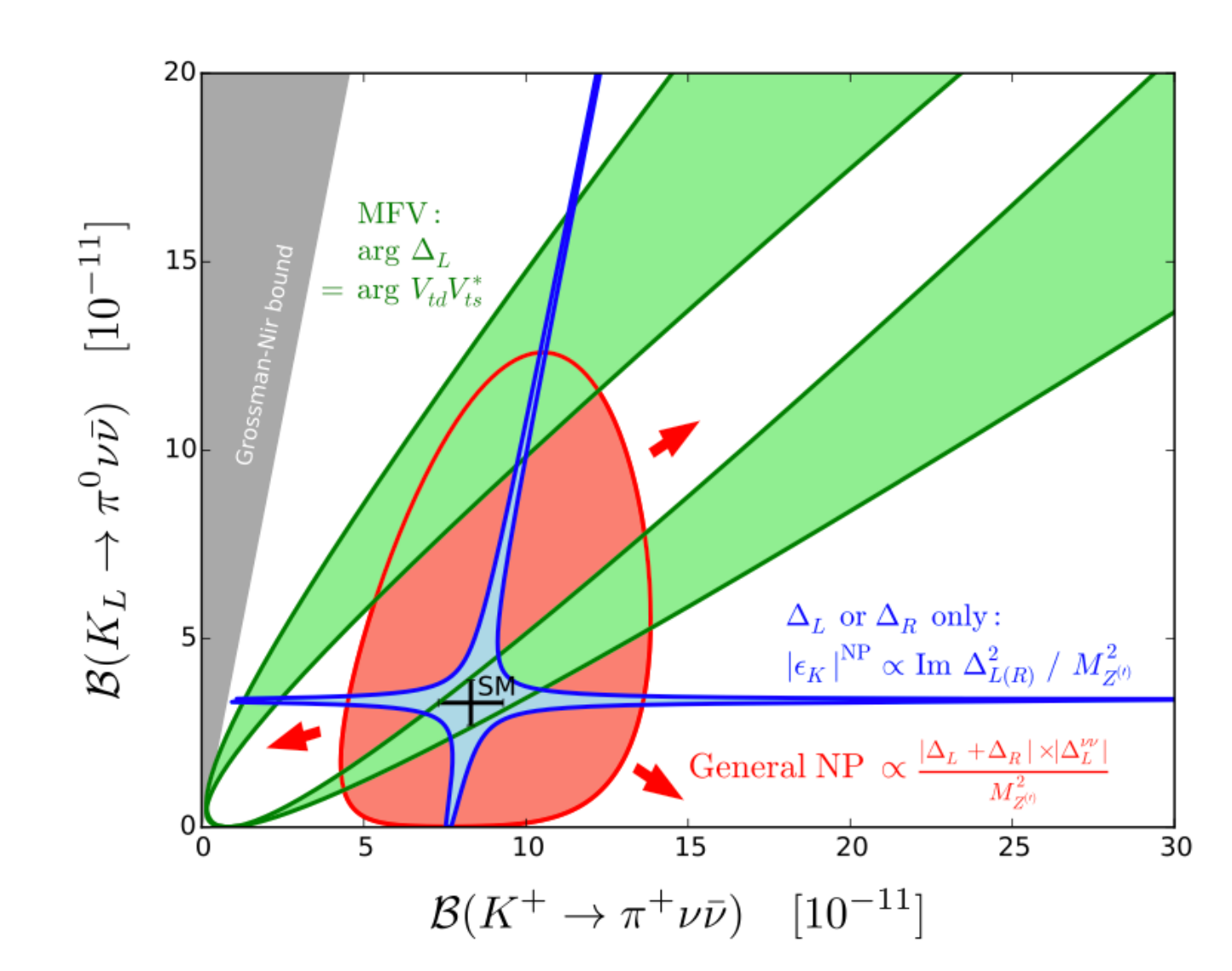}
  \end{center}
    \caption{The correlation between BR($\klpionn$) and BR($\kpluspnn$) under the various new physics models.
    The blue region shows the correlation coming from the constraint by the $K$-$\overline{K}$ mixing parameter $\epsilon_{K}$ if
    only left-handed or right-handed couplings are present. The green region shows the correlation for models having
    a CKM-like structure of flavor interactions.  The red region shows the lack of correlation for models with general left-handed and
     right-handed couplings~\cite{ref:newphysics}. }
  \label{newphysics}
\end{figure}

In addition, the $\klpionn$ decay is sensitive to new physics that violates CP symmetry 
because the $K_{L}$ is mostly $CP$-odd and $\pi^{0}\nu\bar{\nu}$ is $CP$-even.
On the other hand, the charged mode decay, $\kpluspnn$, has both the contribution of $CP$-violating and $CP$-conserving process.
The BRs between the $\klpionn$ and $\kpluspnn$ decay are connected under the isospin symmetry 
by the model-indepdendent bound BR($\klpionn$)$\leq$ 4.4$\times$BR($\kpluspnn$)~\cite{ref:Grossman}. Figure~\ref{newphysics} reproduced from \cite{ref:newphysics} shows how the BRs of the $\klpionn$ and $\kpluspnn$ would be affected due to new physics effects.  In some physics scenarios, those BRs have a strong correlation and we can distinguish the physics scenarios by measuring both branching ratios.

The BRs of the $\kpinn$ decay would be related to other flavor observables such as 
lepton-flavour-universality-violation in the $B$ sector.  
Most theoretical models to explain such phenomena have strong couplings to the third generation fermions
which cause significant effects on the BRs of the $\kpinn$ decay through couplings to $\tau$ neutrinos~\cite{ref:Banomalies}.
Together with observables in the $B$ sector, measurements of BRs on the $\kpinn$ decays give us crucial 
information to investigate the flavor structure in new physics.

% flatex input end: [KLdocu/physics/physics.tex]

%\section{Physics and Experiment at KL2 Beam Line}
\clearpage
% flatex input: [KLdocu/concept/concept.tex]
%\section{Basic Concept of KOTO Step-2}
\subsection{Basic Concept of KOTO Step-2}
\label{chap:concept}
The KOTO experiment, which searches for the $\klpionn$ decay and has been running in the Hadron Experimental Facility, will reach a sensitivity level better than $10^{-10}$ in 3--4 years but would take longer time toward the sensitivity predicted by the Standard Model,  $3\times 10^{-11}$,
considering the operation plan of the Main Ring accelerator and the expected running time in future.
%, and extrapolating from the achieved sensitivity so far.
% It is desirable to 
We thus should have a new experiment that can discover and observe 
%a certain amount 
a large number 
of $\klpionn$ events and measure its branching ratio.
%Such a next-generation experiment, called KOTO step-2 is being discussed and designed.
An idea of such a next-step experiment, called KOTO step-2, was already mentioned in the KOTO proposal in 2006~\cite{KOTOproposal}.
Now we have 
%learned many things from 
gained
experiences in the KOTO experiment and are ready to consider and design the next-generation experiment more realistically.
%We are eagerly considering 
We eagerly consider
the realization of KOTO step-2 in the early phase of the extension of the Hadron Experimental Facility.

\begin{figure}[H]
\begin{minipage}{0.5\linewidth}
\includegraphics[width=\linewidth]{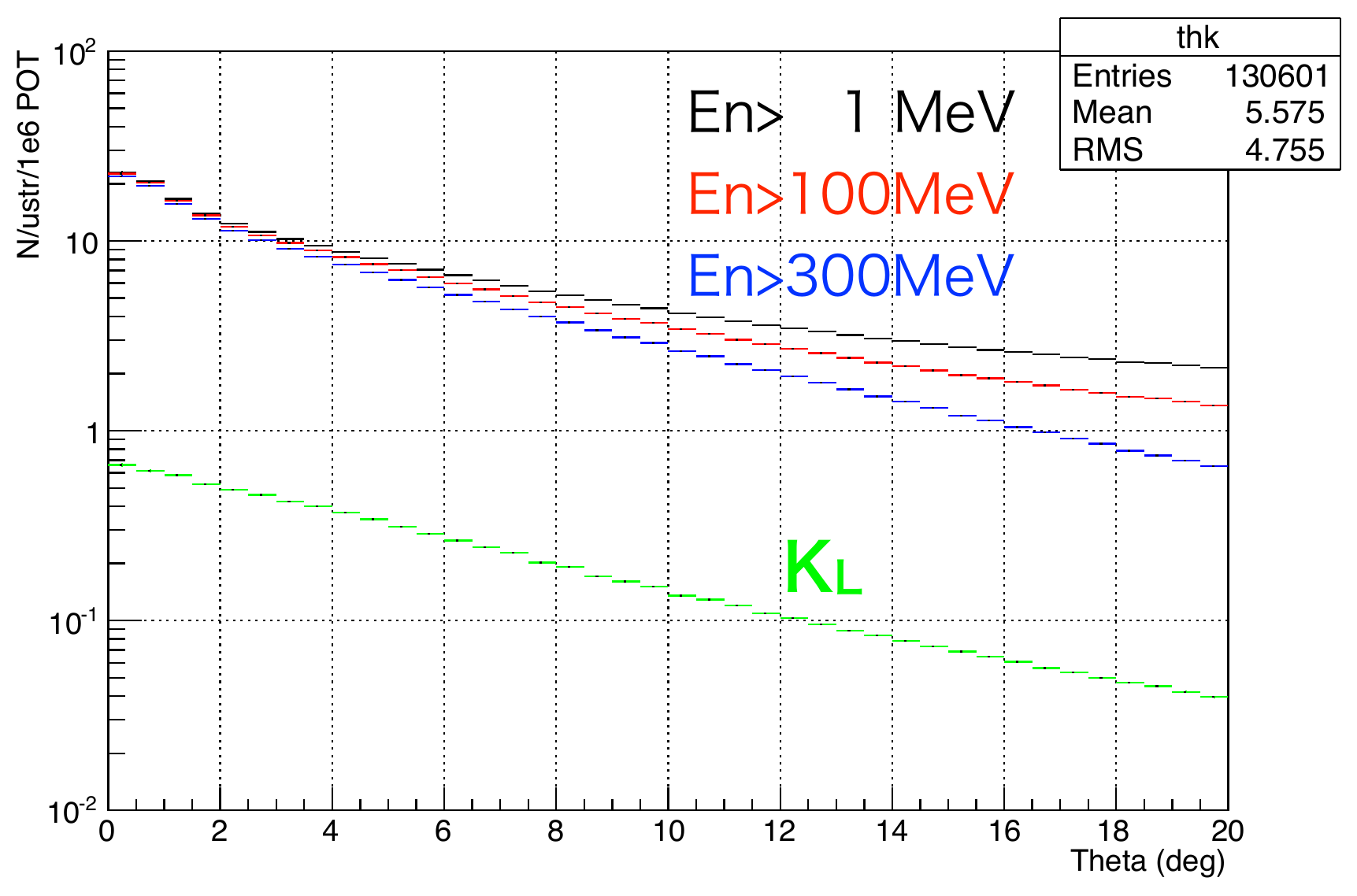}
\end{minipage}
%\hspace{0.1\linewidth}%
\begin{minipage}{0.5\linewidth}
\includegraphics[width=\linewidth]{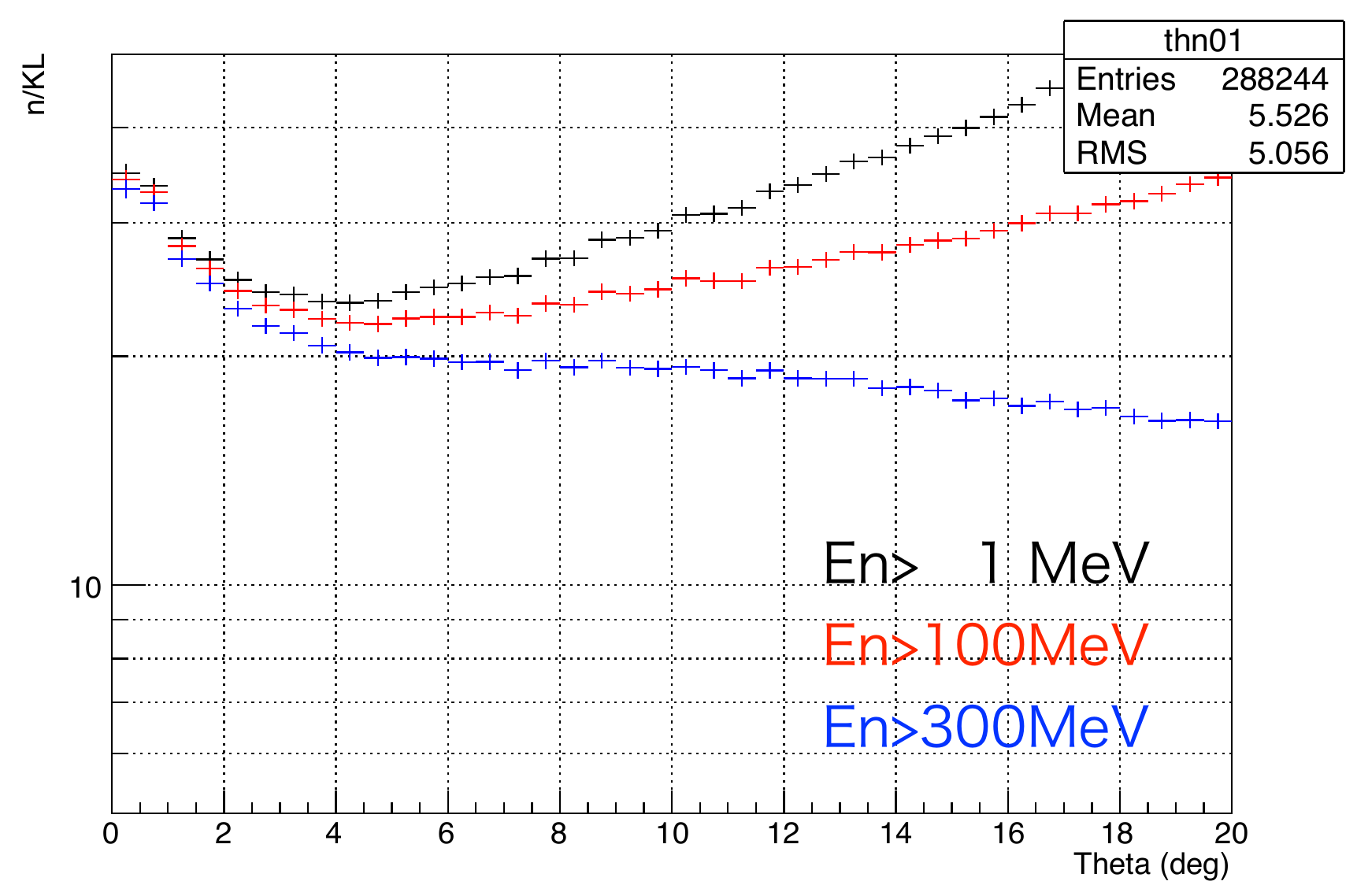}
\end{minipage} 
\caption{\label{fig:vsangle}Simulated $K_L$ and neutron yields (left) and their ratio (right) as functions of the production angle~\cite{ref:kaon2019}. 
The yields were evaluated at 1~m downstream of the target, 
normalized by the solid angle ($\mu$str).
Black, red, and blue points indicate the results when selecting neutrons with their energies of more than 1, 100, and 300~MeV, respectively.}
\end{figure}

\begin{figure}[H]
\centering
\includegraphics[width=0.8\linewidth]{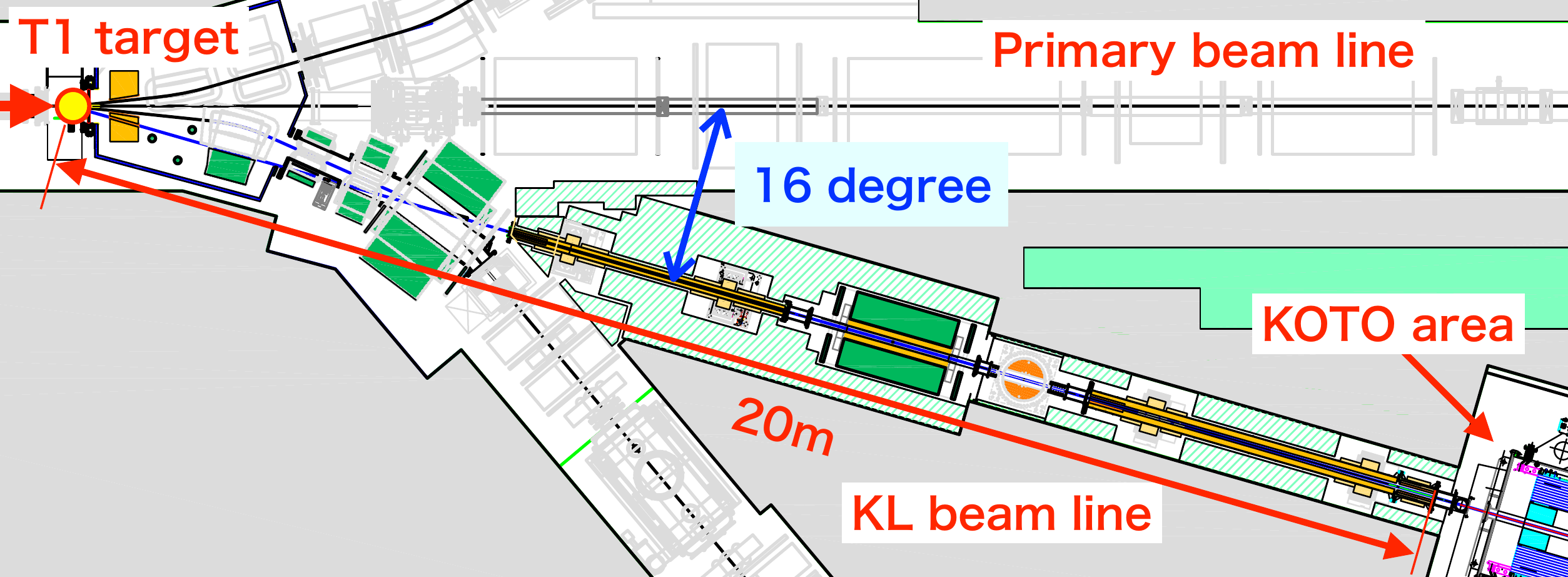}
\caption{\label{fig:kotobl}Schematic drawing of the KL beam line for the KOTO experiment in the current Hadron Experimental Facility.}
\end{figure}

\begin{figure}[h]
\centering
\includegraphics[width=\linewidth]{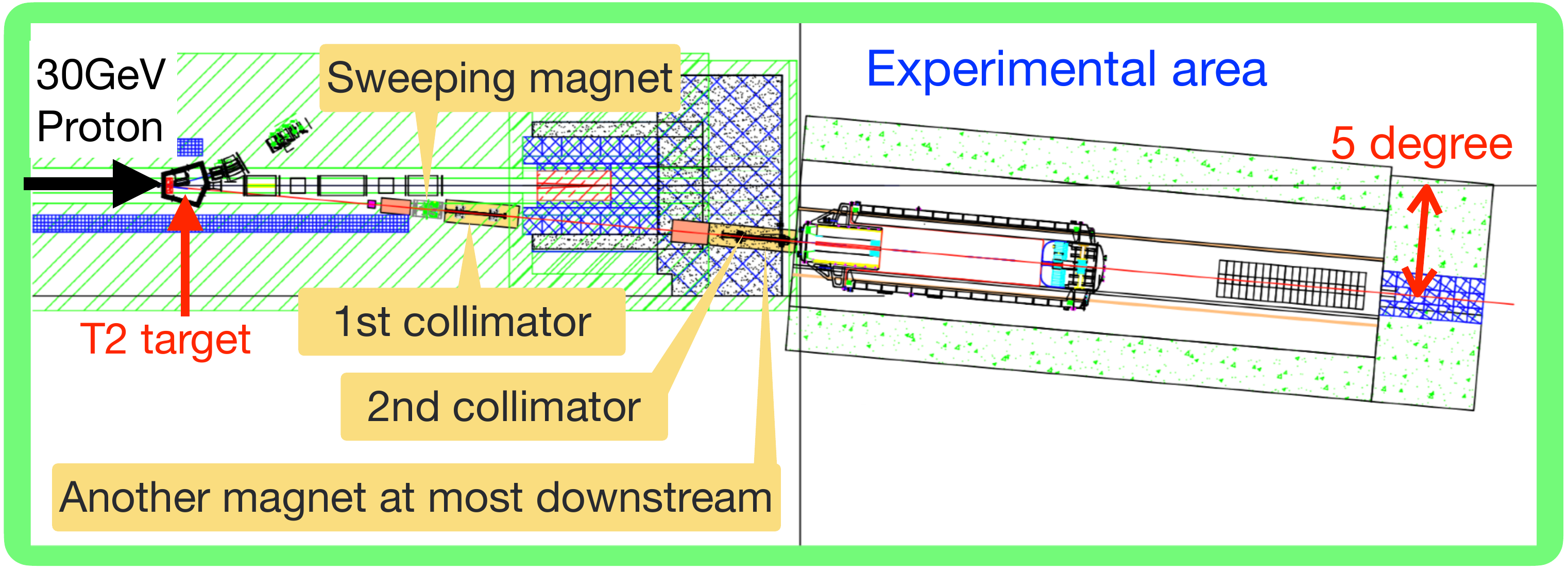}
\caption{\label{fig:koto2bl}Schematic drawing of the KOTO step-2 beam line in the Extended Hadron Experimental Facility.
The experimental area is located behind the beam dump. Distance between the T2 target and the experimental area is assumed to be 43~m in the present studies.
A modeled cylindrical detector with 20~m in length and 3~m in diameter is described in the experimental area as a reference.}
\end{figure}

%Toward a higher experimental sensitivity in the $\klpionn$ measurement, 
To achieve a higher experimental sensitivity for the $\klpionn$ measurement, 
 we must consider to maximize the $\kl$ flux, the detection acceptance of the signal, and the signal-to-background ratio.
The $\kl$ flux is determined by the production angle and the solid angle of the secondary neutral beam.
Other parameters are the achievable intensity of the primary proton beam and the target properties (material, thickness, etc.).
The production angle is defined as the angle between the primary proton beam and secondary neutral beam directions. 
Figure~\ref{fig:vsangle} shows the $\kl$ and neutron yields and the neutron-to-$\kl$ flux ratio as functions of the production angle when a 102-mm-long gold target is used.
KOTO step-2 chooses the production angle of 5~degrees as an optimum point for a higher $\kl$ flux with a smaller neutron fraction in the beam. 
%In case of the KOTO experiment, the 16-degree production angle was chosen, as shown in Fig.~\ref{fig:kotobl}, 
%in order to utilize the T1 target with the experimental area away from the primary beam.
In case of the KOTO experiment, the production angle is 16-degree, as shown in Fig.~\ref{fig:kotobl}, 
which was chosen to utilize the T1 target with the experimental area away from the primary beam.
In order to realize the 5-degree production while keeping the solid angle of the neutral beam as large as possible,
$\ie$ with the shortest beam line, a new experimental area behind the primary beam dump and a new target station close to the dump are necessary.
Figure~\ref{fig:koto2bl} shows a possible configuration in the Extended Hadron Experimental Facility, utilizing the second target (T2).

Here we introduce the KOTO step-2 detector base design.
The acceptance of the signal is primarily determined by the detector geometry.
The basic detector configuration is same as the current KOTO experiment, a cylindrical detector system with an electromagnetic calorimeter to detect two photons from the $\pi^0$ decay at the end cap. 
A longer $\kl$ decay region and a larger diameter of the calorimeter are being considered to obtain a larger acceptance.
The 5-degree production also provides a benefit in view of the signal acceptance; a harder $\kl$ spectrum than KOTO is expected, and two photons are boosted more it the forward direction, and thus the acceptance gain by a longer decay region can be utilized.

The ability of the signal discovery and the precision of the branching ratio measurement depend on the signal-to-background ratio, as well as the expected number of observed events.
The background level is affected by many factors such as the beam size at the detector, the flux of beam particles (neutrons, $\kl$) leaking outside the beam (beam-halo),
%to the halo region,
 charged kaons in the neutral beam, and detector performances.

In the following two sections, Sections~\ref{chap:beamline} and \ref{chap:detector}, a modeled beam line and a conceptual design of the detector are described.
Discussions of the sensitivity and background follows in Section~\ref{chap:sensitivity}, with parametrized detector performance.

% flatex input end: [KLdocu/concept/concept.tex]

%\section{Physics and Experiment at KL2 Beam Line}
\clearpage
% flatex input: [KLdocu/beamline/beamline.tex]
%\section{Beam Line}
\subsection{Beam Line}
\label{chap:beamline}

%\subsection{Performance of the beam line}
\subsubsection{Performance of the beam line}
Table~\ref{tab:beampar} summarizes the beam parameters for KOTO step-2
%, comparing with 
and those in the current KOTO experiment. 
\begin{table}[h]
\caption{\label{tab:beampar}Beam parameters for KOTO step-2 and the current KOTO experiment.}
\begin{center}
\begin{threeparttable}
\begin{tabular}{lll}
\hline
&KOTO step-2\tnote{*}&KOTO\\
\hline
Beam power&100~kW&64~kW (100~kW in future)\\
Target&102-mm-long gold&60-mm-long gold\\
Production angle&5$^{\circ}$&16$^{\circ}$\\
Beam line length&43~m&20~m\\
Solid angle&4.8~$\mu$sr&7.8~$\mu$sr\\
\hline
\end{tabular}
\begin{tablenotes}\footnotesize
\item[*] 
Note the parameters for step-2 are tentative for this study.
\end{tablenotes}
\end{threeparttable}
\end{center}
\end{table}

\begin{figure}[h]
\centering
\begin{minipage}{0.45\linewidth}
\includegraphics[width=\linewidth]{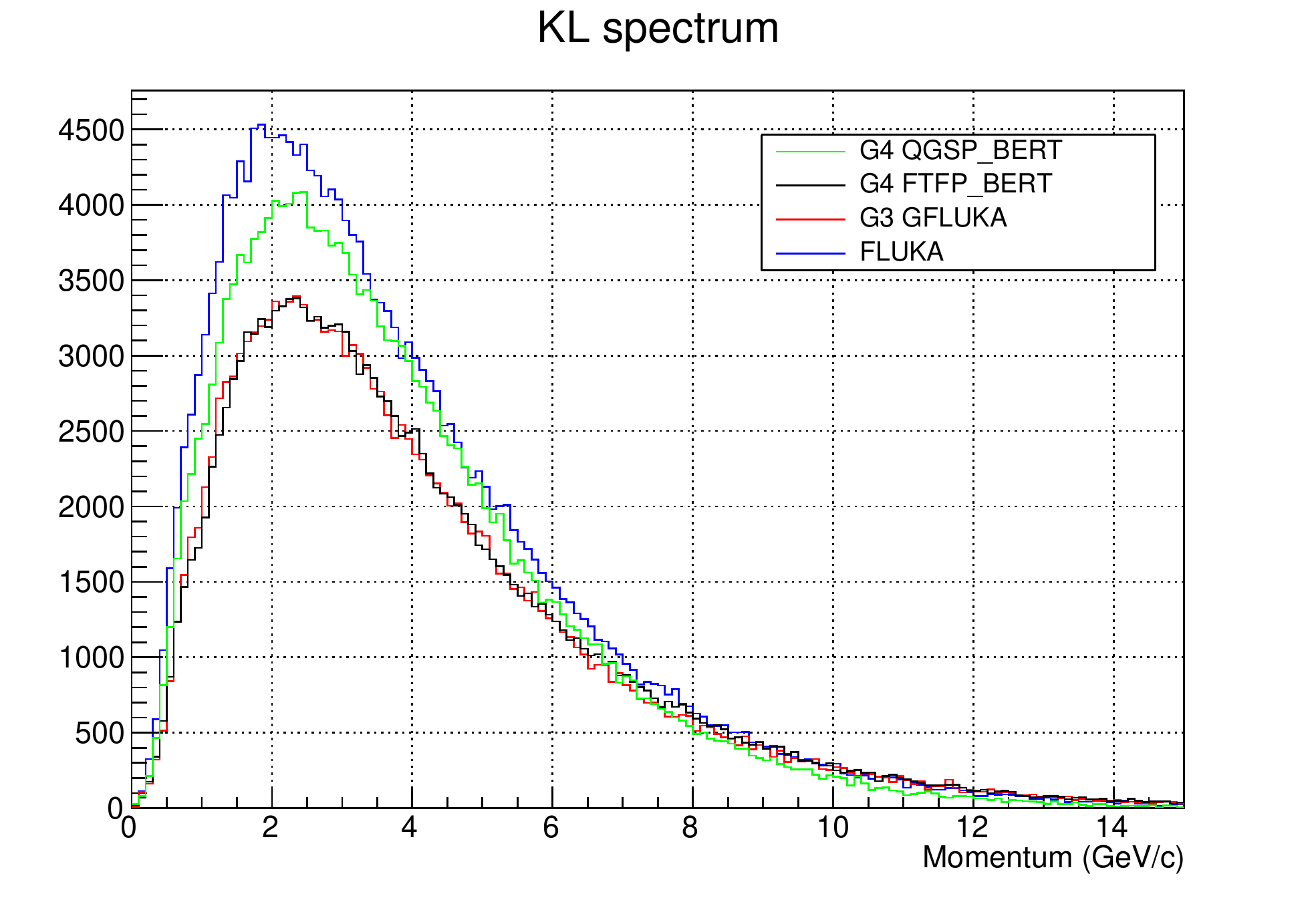}
\end{minipage}
\begin{minipage}{0.45\linewidth}
\includegraphics[width=\linewidth]{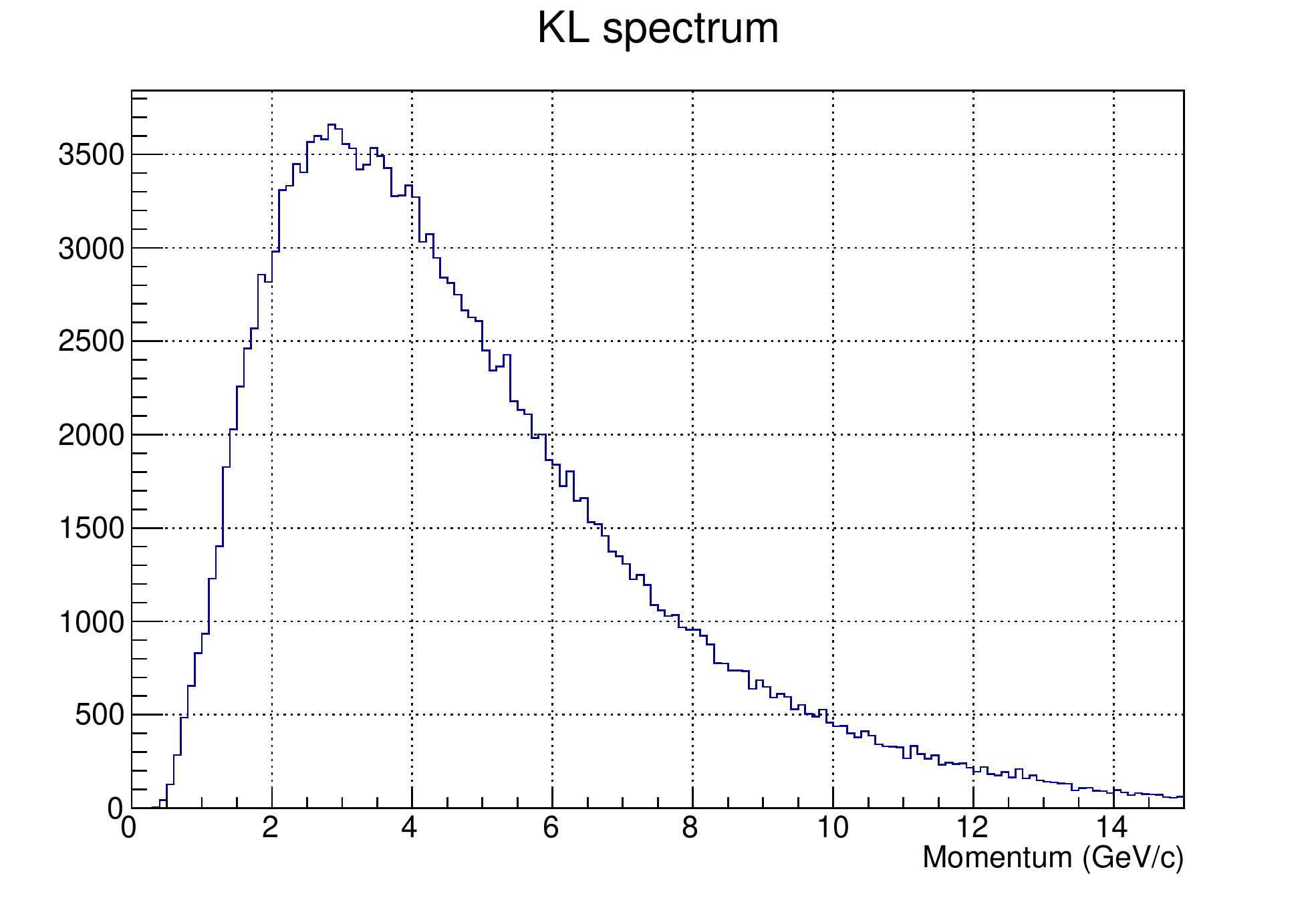}
\end{minipage}
\caption{\label{fig:klmom}
%Simulated $\kl$ spectrum 
$\kl$ spectra at 1~m from the T2 target by the target simulation (left) and at the exit of the KL2 beam line at 43~m from the T2 target by the beam line simulation (right). In the left plot, the results by using various simulation packages are also shown, as well as the result by the GEANT3-based simulation (labeled ``G3 GFLUKA'') which is our default in this study.}
\end{figure}

To evaluate the performance of the beam line for KOTO step-2 (KL2 beam line), the target and beam line simulations were conducted. The target in the study was chosen to be a simple cylindrical rod made of gold with its diameter of 10~mm and length of 102~mm, which corresponds to 1$\lambda_{I}$ (interaction length). The 30~GeV primary protons were injected to the target with the beam size ($\sigma$) of 1.6~mm in both horizontal and vertical directions. No beam divergence was considered in the simulation.
The secondary particles which went in the direction of 5~degree (within $\pm$0.3~degree) to the primary beam direction were recorded at 1~m downstream from the target 
%for use 
to be used in the following beam line simulation as inputs.
For the simulation of the particle production at the target, we used the GEANT3-based simulation as a default, and GEANT4 (10.5.1 with a physics list of QGSP\_BERT or FTFP\_BERT) and FLUKA (2020.0.3) for comparison, as shown in Fig.~\ref{fig:klmom} (left).
The resultant $\kl$ fluxes were found to agree with each other within 30\%. GEANT3 provided 
%the least result
the smallest $\kl$ yield
 and thus is considered to be a conservative choice in the discussion of the sensitivity.

In designing the KL2 beam line, we first follow the design strategy of the KOTO beam line (KL beam line)~\cite{ref:kotobl}.
The KL2 beam line consists of two stages of 5-m-long collimators, a photon absorber in the beam, and a sweeping magnet to sweep out charged particles from the target.
%Although an additional sweeping magnet is needed at the end of the beam line in order to sweep out charged kaons which are produced by interactions of neutral particles with the collimators, its effect was evaluated independently from the base design beam line.
The photon absorber, made of 7-cm-long lead, is located at 7~m downstream of the target.
The first collimator, starting from 20~m from the target, defines the solid angle and shape of the neutral beam.
The solid angle is set to be 4.8~$\mu$str.
The second collimator, starting from 38~m from the target, cut the particles coming from the interactions at the photon absorber and the beam-defining edge of the first collimator. The bore shape of the second collimator is designed not to be seen from the target so that particles coming directly from the target do not hit the inner surface and thus do not generate particles leaking 
%to the halo region.
outside the beam.
The first sweeping magnet is located upstream of the first collimator in this study.
Although an additional sweeping magnet is needed at the end of the beam line in order to sweep out charged kaons which are produced by interactions of neutral particles with the collimators, its effect was evaluated independently from the base design beam line, as discussed later.

\begin{figure}[h]
\centering
\begin{minipage}{0.45\linewidth}
\includegraphics[width=\linewidth]{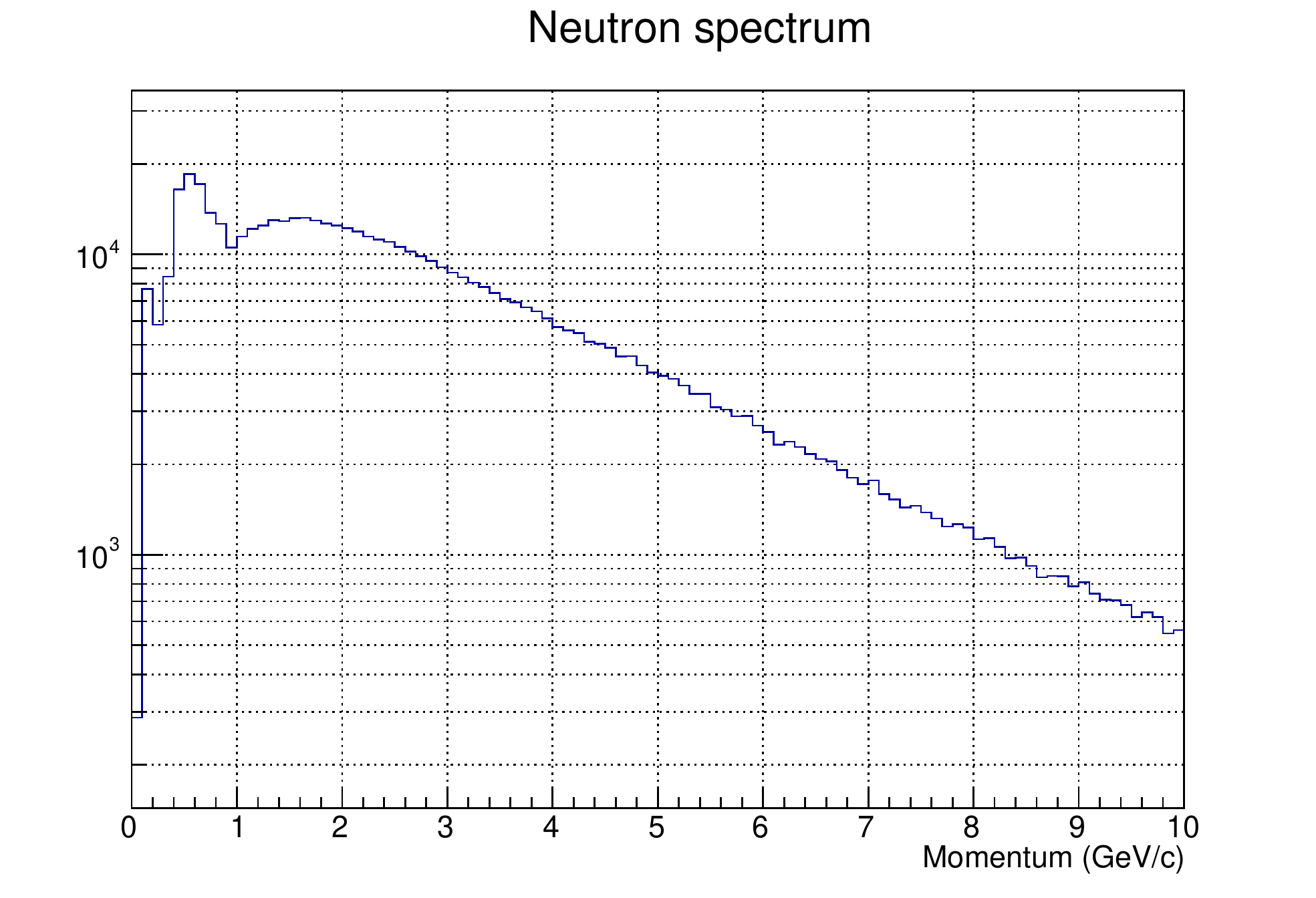}
\end{minipage}
\begin{minipage}{0.45\linewidth}
\includegraphics[width=\linewidth]{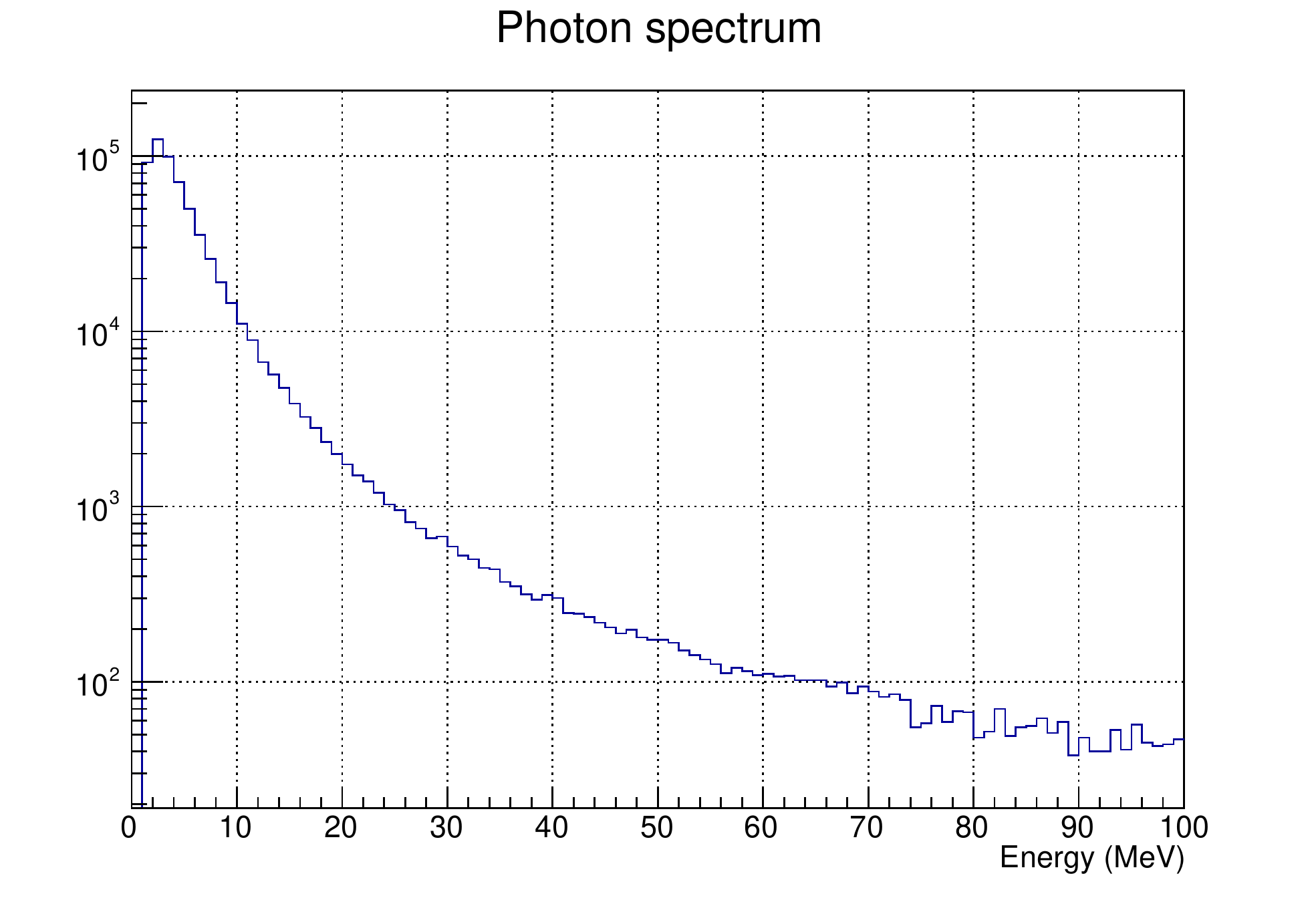}
\end{minipage}
\caption{\label{fig:ngspectra}Simulated neutron (left) and photon (right) spectra at the exit of the beam line.}
\end{figure}

Figures~\ref{fig:klmom} (right) and \ref{fig:ngspectra} show the simulated spectra of $\kl$, neutrons, and photons at the exit of the KL2 beam line, respectively.
The $\kl$ yield was evaluated to be $1.1\times 10^7$ per $2\times 10^{13}$ protons on the target (POT).
Note that the beam power of 100~kW corresponds to $2\times 10^{13}$~POT per second with 30~GeV protons.
%The resultant $\kl$ flux is 2.6 times higher than that of the current KOTO experiment, comparing the POT-normalized values.
The resultant $\kl$ flux per POT is 2.6 times higher than that of the current KOTO experiment.
The $\kl$ spectrum peaks at 3~GeV/c, while it is 1.4~GeV/c in the current KOTO experiment.
The simulated particle fluxes are summarized in Table~\ref{tab:yield}.
\begin{table}[h]
\caption{\label{tab:yield}Expected particle yields estimated by the simulations.}
\begin{center}
\begin{threeparttable}
\begin{tabular}{cccc}
%\br
\hline
\multirow{2}{*}{Particle}&\multirow{2}{*}{Energy range}&Yield & On-spill rate\\
&&(per $2\times 10^{13}$~POT) & (MHz)\\
\hline
$\kl$&&$1.1\times 10^7$&24\\
%\mr
\hline
\multirow{2}{*}{Photon}&$>$10~MeV&$5.3\times10^7$&110\\
&$>$100~MeV&$1.2\times10^7$&24\\
%\mr
\hline
\multirow{2}{*}{Neutron}&$>$0.1~GeV&$3.1\times10^8$&660\\
&$>$1~GeV&$2.1\times10^8$&450\\
%\br
\hline
\end{tabular}
\begin{tablenotes}\footnotesize
\item[]
The beam power of 100~kW corresponds to $2\times 10^{13}$~POT/s with 30~GeV protons.
The on-spill rate means the instantaneous rate during the beam spill, assuming 2-second beam spill every 4.2 seconds. 
\end{tablenotes}
\end{threeparttable}
\end{center}
\end{table}

\begin{figure}[H]
\centering
\centering
\begin{minipage}{0.45\linewidth}
\includegraphics[width=\linewidth, height=\linewidth]{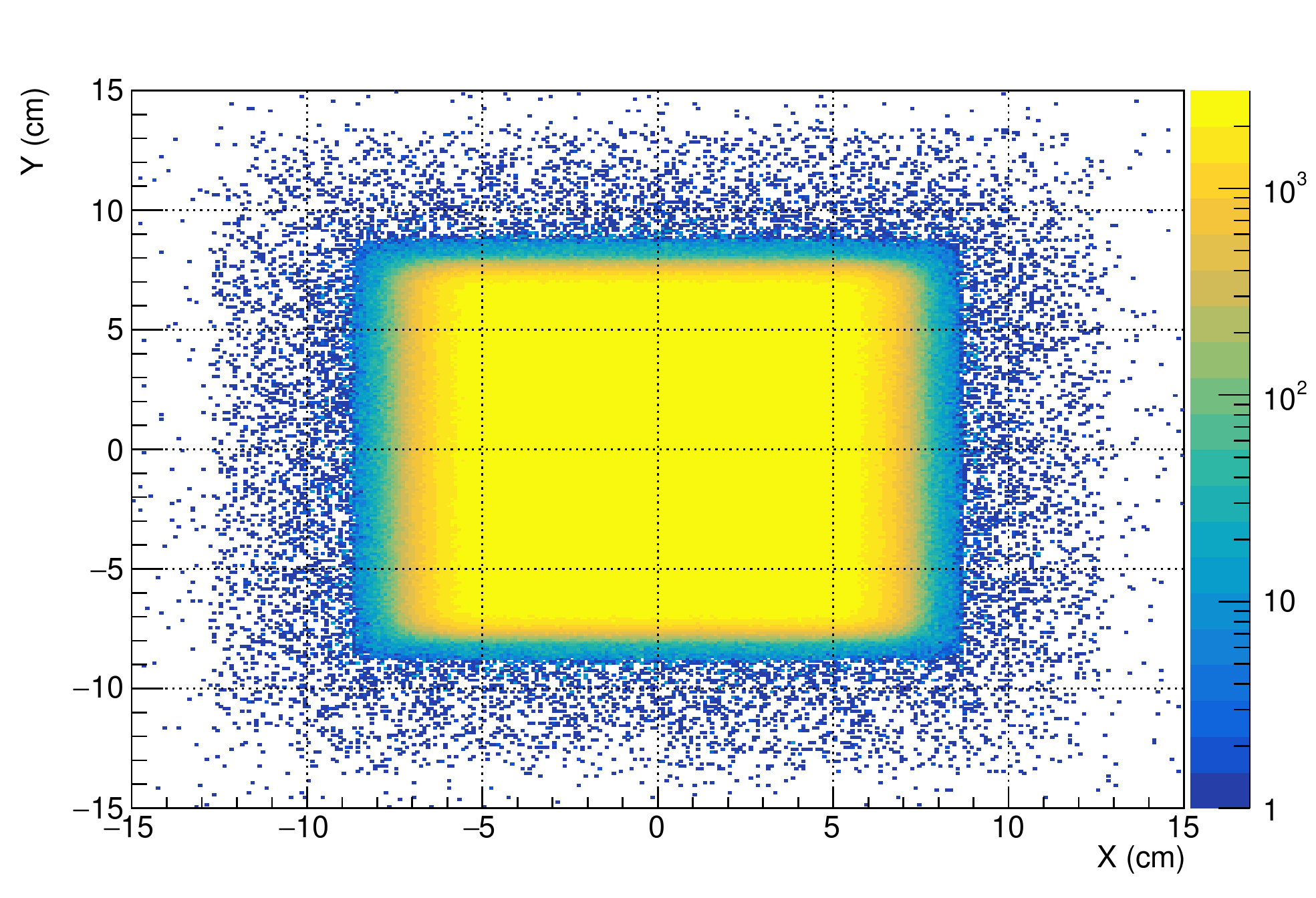}
\end{minipage}
\begin{minipage}{0.45\linewidth}
\includegraphics[width=\linewidth]{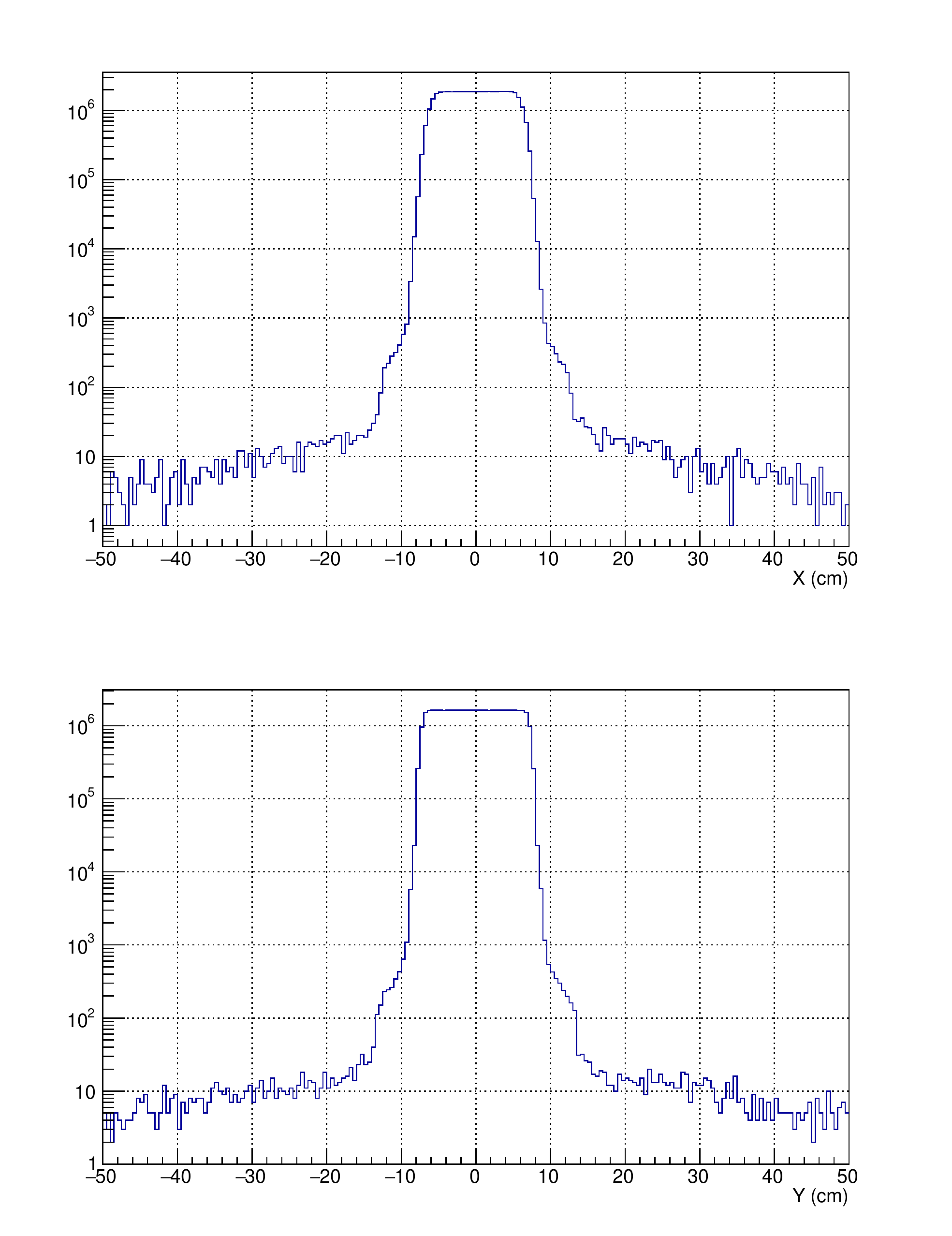}
\end{minipage}
\caption{\label{fig:nprof}Beam shape at the end-cap plane, represented by the neutron profile.
The left figure shows the distribution in $\pm 15$~cm of the beam center, and the right top (bottom) histogram indicates the horizontal (vertical) distribution in $\pm 50$~cm of the beam center.}
\end{figure}

Figure~\ref{fig:nprof} shows the neutron profile at the assumed calorimeter location, 64~m from the T2 target. As shown in the figure, the neutral beam is shaped so as to be a square at the calorimeter location.
Evaluation of neutrons spreading to the beam halo region, called ``halo neutron'', is important since they are potential sources of backgrounds due to their interaction with the detector materials.
%The flux evaluation depends on the definition of the halo region; 
Here we define the core and halo neutrons as those inside and outside the $\pm$10~cm region at the calorimeter, respectively.
The ratio of the halo neutron yield to the core yield was found to be $1.8 \times 10^{-4}$.

%\subsubsection{Charged kaons in the neutral beam}
\subsubsubsection{Charged kaons in the neutral beam}
The contamination of the charged kaons in the neutral beam line is harmful, since $K^\pm$ 
decays such as $K^\pm \to \pi^0 e^\pm \nu$
 in the detector region can mimic the signal, which were pointed out in the analysis of the KOTO experiment
 ~\cite{KOTO:2020prk}.
 %~\cite{KOTO2016-18}.
The major production point of charged kaons is the second collimator. 
Neutral particles ($\kl$ and neutrons) hit the inner surface of the collimator and the produced charged kaon in the interactions can reach the end of the beam line. Charged pions from $\kl$ decays hitting the collimator also can produce charged kaons.
According to the beam line simulation, 
the flux ratio of the charged kaon and $\kl$ entering the decay region is $R(K^\pm/\kl)=4.1\times 10^{-6}$.
%We confirmed that the additional sweeping magnet can reduce the ratio to be $R<1.1\times 10^{-6}$, limited by the simulation statistics.
To evaluate the reduction by an additional sweeping magnet, 
we conducted another beam line simulation with a sweeping magnet that provides a magnetic field of 2~Tesla in 1.5~m~long at the end of the beam line.
We confirmed that it can reduce the ratio to $R<1.1\times 10^{-6}$, which is limited by the simulation statistics.
%Further optimization continues.

\begin{figure}[h]
\centering
\includegraphics[width=0.7\linewidth]{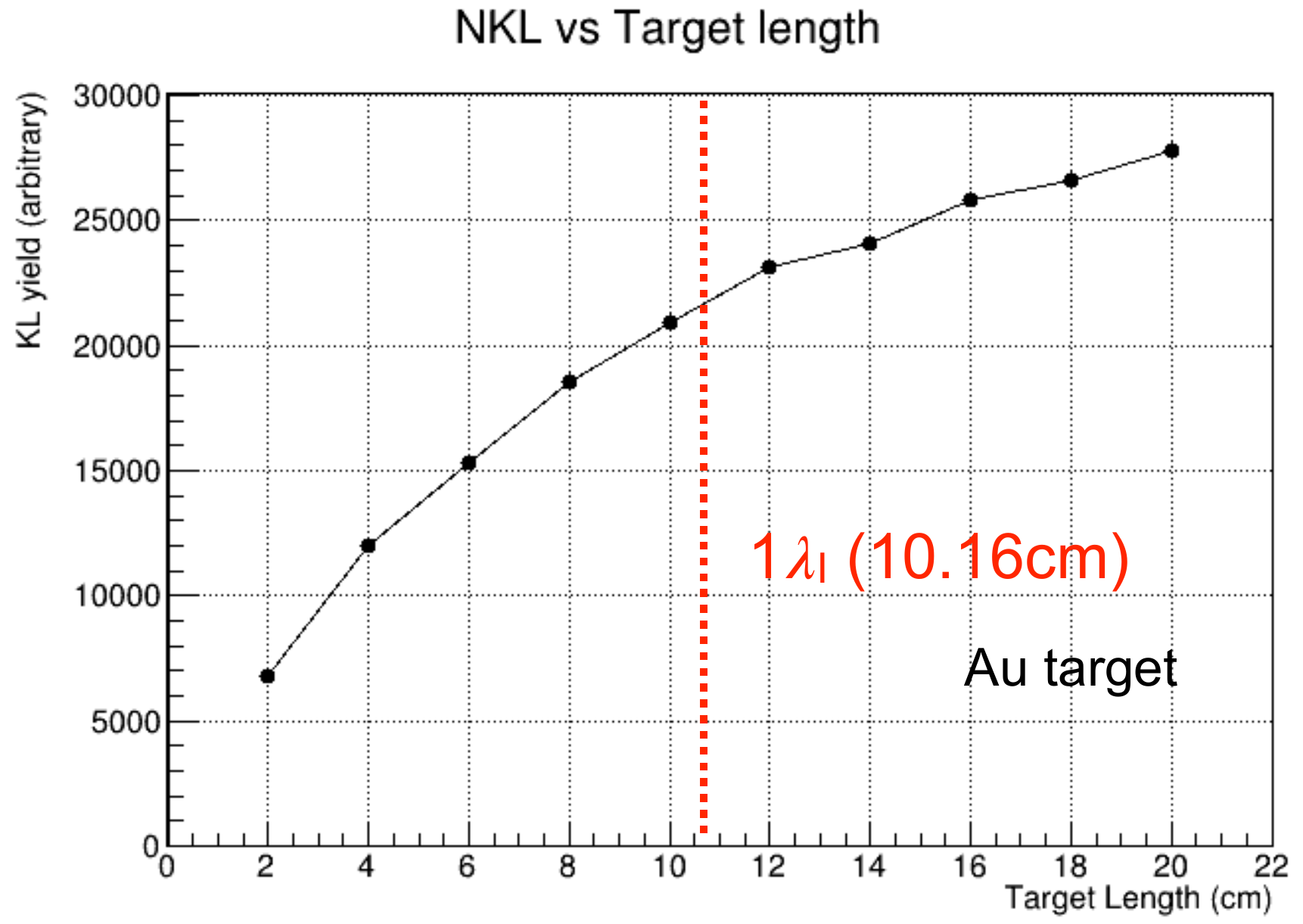}
\caption{\label{fig:tlen}$\kl$ yield as a function of the target length. The yield is evaluated at 1~m from the T2 target. A FLUKA-based simulation is used for this study.}
\end{figure}

%\subsubsection{Discussion on the target length}
\subsubsubsection{Discussion on the target length}
%In the baseline simulation studies, 
In the target simulation,
the gold target was assumed to be 102~mm~long, while the T1 target used in the current Hadron Experimental Facility is 60~mm~long.
There might be technical difficulties to use a thicker target from the view point of cooling.
Figure~\ref{fig:tlen} indicates the relative $\kl$ yield as a function of the target length.
%As can be seen, our expected $\kl$ yield with a 102-mm-long gold target provided 40\% more than that with a 60-mm-long gold, while the physical length is 70\% longer.
As can be seen, a 102-mm-long gold target provided 40\% more $\kl$ yield than a 60-mm-long gold target.
%while the physical length is 70\% longer.
%
%

%\subsection{Activities in the experimental area behind the beam dump}
\subsubsection{Activities in the experimental area behind the beam dump}
\begin{figure}[h]
\centering
\includegraphics[width=\linewidth]{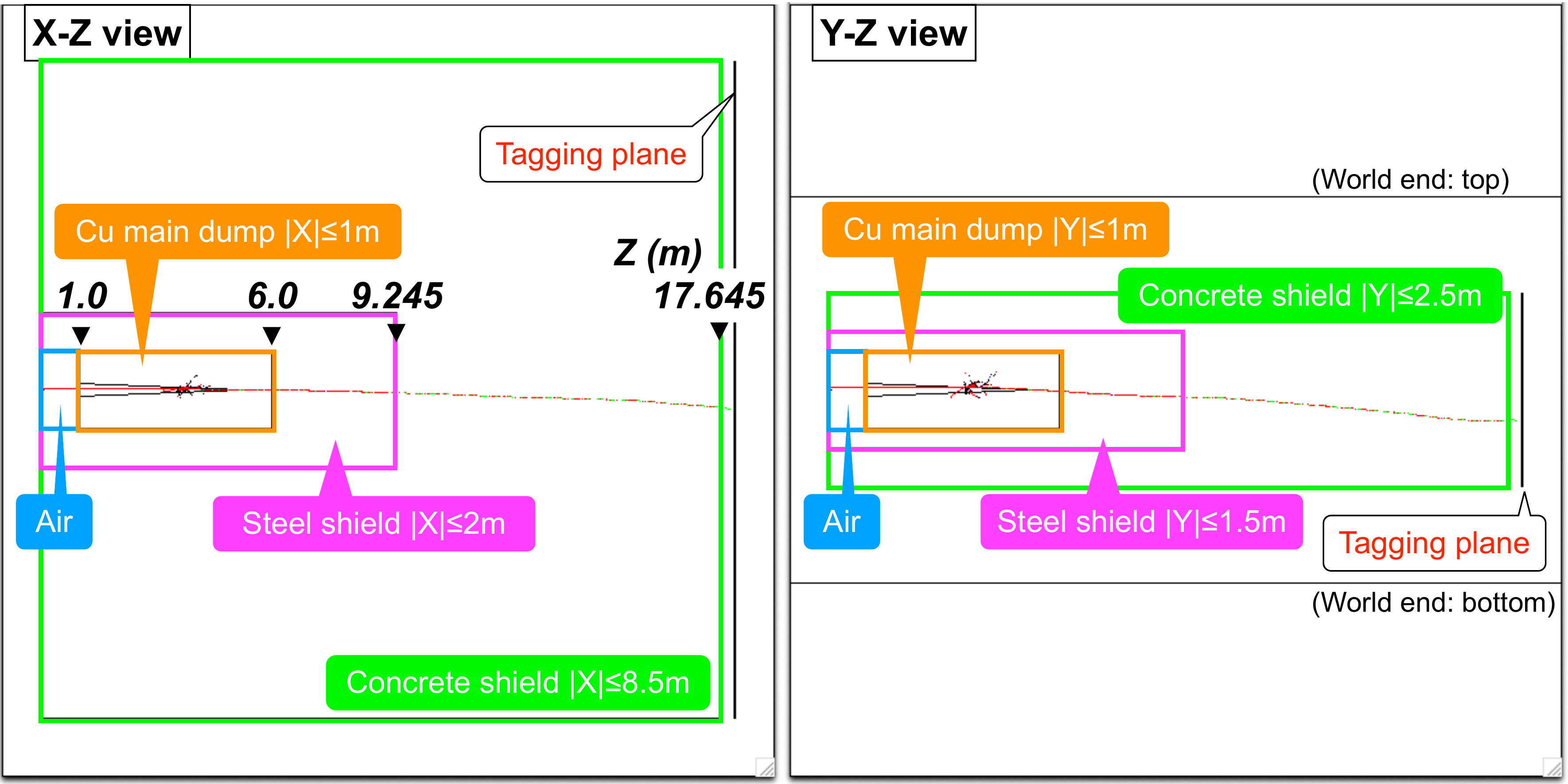}
\caption{\label{fig:bd}Model of the current primary beam dump used in the simulation study
in horizontal (left) and vertical cutaway views.}
\end{figure}

To realize the production angle of 5 degree, the experimental area must be located behind the primary beam dump.
There is a concern that many particles from interactions in the beam dump penetrate the shield and reach the experimental area, which cause a high rate of accidental hits in the detector.
To evaluate the flux in the experimental area, a GEANT3-based simulation of the beam dump was conducted.
Figure~\ref{fig:bd} illustrates the model of the current beam dump.
The main body of the dump is made of copper, on which a tapered hole exists to spread the proton hit position along the beam direction and thus distribute the heat dissipation. Steel and concrete shields follow behind the copper dump.
In this simulation, the proton beam with the size ($\sigma$) of 3~cm is injected into the dump, which roughly indicate the parameter in the current operation.

\begin{figure}[h]
\centering
\includegraphics[width=0.7\linewidth]{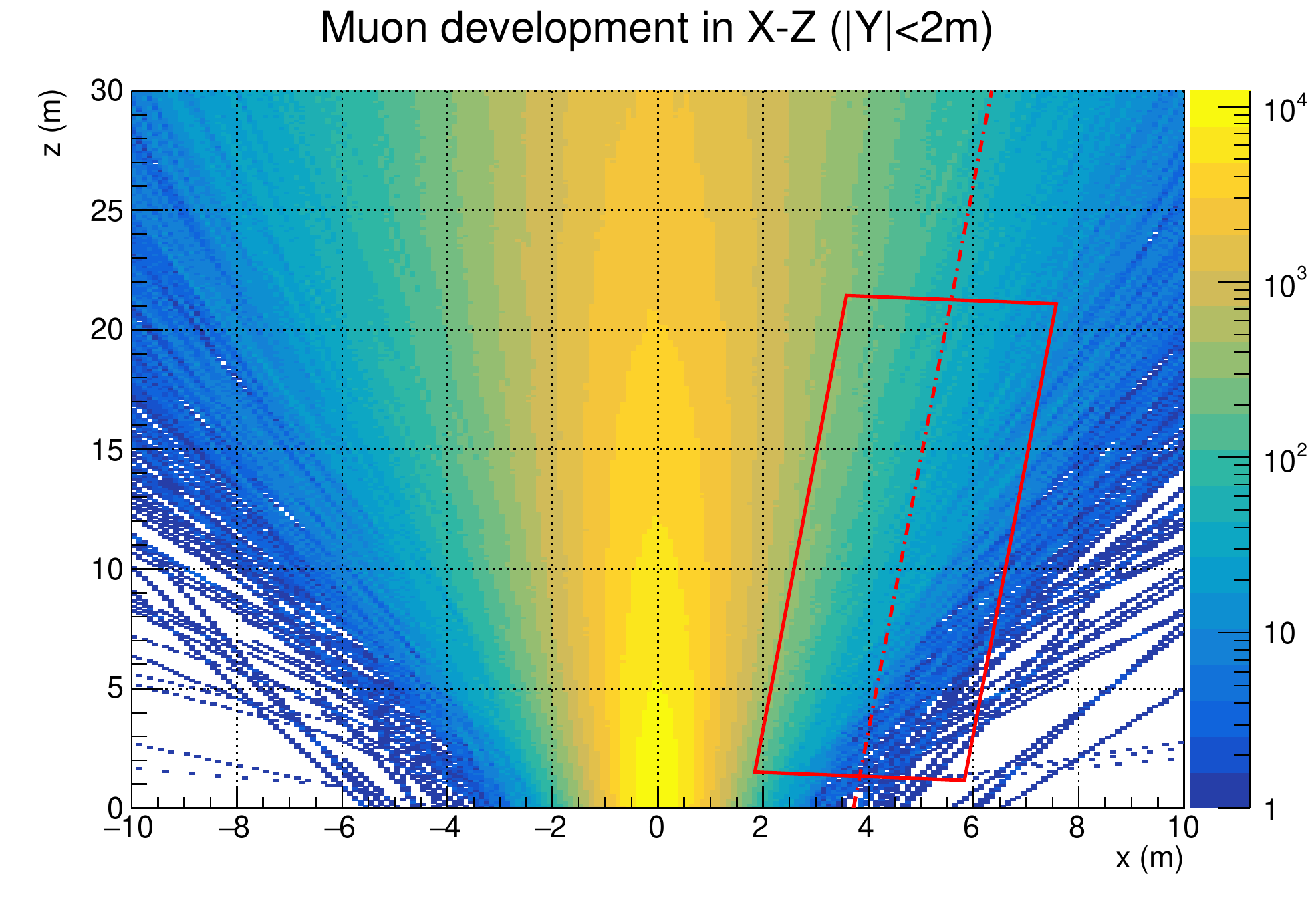}
\caption{\label{fig:bd-mu-develop}Simulated development of the muon tracks behind the current beam dump, distributed in the horizontal (x) and the beam direction (z). The red box indicates the assumed location of the KOTO step-2 detector, and the red dashed line indicates the center of the KL2 beam line.}
\end{figure}

Almost all the particles entering the area were muons. 
As shown in Fig.~\ref{fig:bd-mu-develop}, the KOTO step-2 detector is not in the most intense region but still in the region where the flux is high.
The on-spill counting rate due to the muons which cross the detector region was estimated to be 15~MHz when the 100~kW beam is fully injected into the dump.
The contribution of other particles such as neutrons were found to be small, about 25~kHz even integrated over the whole region behind the dump.

\begin{figure}[h]
\centering
\includegraphics[width=0.7\linewidth]{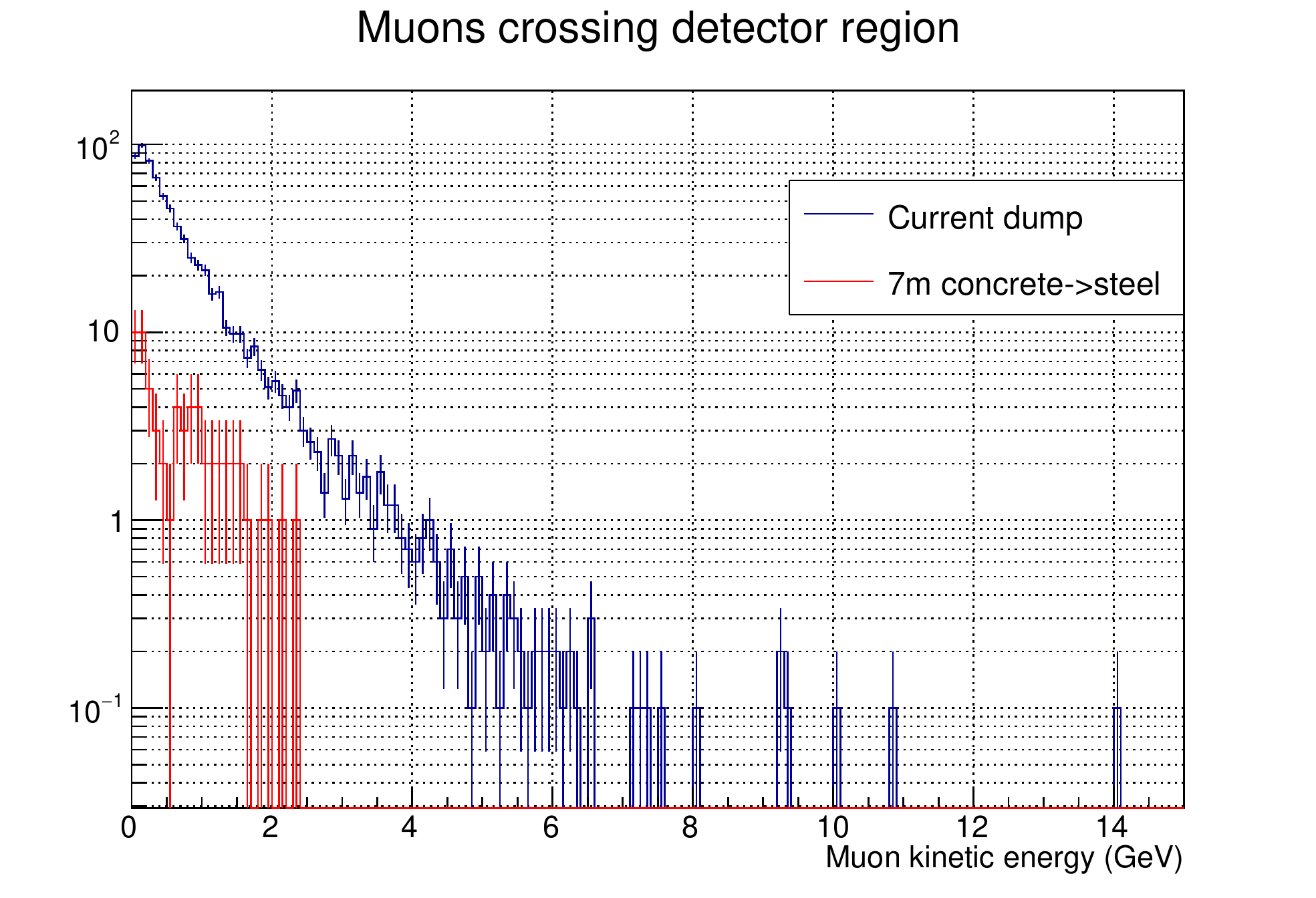}
\caption{\label{fig:bd-emu}Energy distribution of muons whose extrapolated trajectory crosses the KOTO step-2 detector region in case of the current dump (blue) and the modified dump (red).}
\end{figure}

To reduce the muon flux, a part of the concrete shield must be replaced with steel shield.
Replacing the material of the 4~m (horizontal) $\times$ 3~m (vertical) $\times$ 7~m (beam direction) volume from concrete to steel reduces the flux by an order of magnitude to 1.3~MHz.
Figure~\ref{fig:bd-emu} shows the energy distribution of the muons crossing the detector region before and after the replacement.

% flatex input end: [KLdocu/beamline/beamline.tex]

%\section{Physics and Experiment at KL2 Beam Line}
\clearpage
% flatex input: [KLdocu/detector/detector.tex]
%\section{Detector Model}
\subsection{Detector Model}
\label{chap:detector}
%\subsection{Concept of detector}
\subsubsection{Concept of detector}
%One signature of the signal from $K_L\to \pi^0\nu\overline{\nu}$ is
The signature of the $K_L\to \pi^0\nu\overline{\nu}$ decay is
two photons from the $\pi^0$
without any other detectable particles.
%Another signature is 
In addition,
a large transverse momentum ($\pt$) of the $\pi^0$
is expected
due to the kinematics of the decay.

The detector concept is the same as in the KOTO step-1 detector:
An endcap calorimeter is used to detect two photons. 
An evacuated decay region at the upstream of the calorimeter is surrounded
by a hermetic veto-detector system to ensure no other detectable particles.
The decay vertex of the $\pi^0$ is reconstructed on the beam axis by
assuming the nominal $\pi^0$ mass for the invariant mass of the two photons.
With the reconstructed decay vertex, 
$\pt$ of the $\pi^0$ can be calculated.

Backgrounds are categorized into three:
$K_L$ decay, $K^\pm$ decay, and halo-neutron backgrounds.
\begin{itemize}
 \item $K_L$ decay\\
       $K_L$ decays with branching fractions
       larger than $10^{-4}$ are listed in Table~\ref{tab:KLdecay}.
       $\kpien$ (Ke3),
       $\kpimun$ (K$\mu$3),
       $K_L\to\pi^+\pi^-$,
       and
       $K_L\to 2\gamma$ have only two observable particles in the final state.
       The Ke3, K$\mu$3, and $K_L\to\pi^+\pi^-$ can be reduced by identifying the charged particles.
       The $K_L\to 2\gamma$ decay can be reduced by requiring large $\pt$
       for the reconstructed $\pi^0$,
       although a fake $\pi^0$ is reconstructed
       from the two clusters in the calorimeter.
%       $K_L$ in the beam scatters at the beam line components, and exists in the
%       beam halo region.
%      We call it ``halo-$K_L$''.
       A $K_L$ which spreads out to the beam halo region is called ``halo $K_L$''.
       When such a halo $K_L$ decays into two photons
       reconstructed $\pt$ can be larger
       because the vertex is assumed to be on the beam axis.
       This halo $K_L\to2\gamma$ background can be reduced with
       incident-angle information on the photons at the calorimeter.
       The other $K_L$ decays have more than two particles
       in the final state,
       and
       extra particles which are not used to reconstruct a $\pi^0$
       can be used to veto the events.
 \item $K^\pm$ decay\\
       $K^\pm$ is generated from the interaction of $K_L$,
       neutron, or $\pi^\pm$
       at the collimator in the beam line.
       The second sweeping magnet
       near the entrance of the detector
       will reduce the contribution.
       Some $K^\pm$ can pass through
       the second magnet,
       and $K^\pm\to\pi^0 e^\pm\nu$
       decay occurs in the detector.
       This becomes a background if $e^\pm$ is undetected.
       The kinematics of the $\pi^0$ is similar to
       $K_L\to\pi^0\nu\overline{\nu}$, and thus
       this decay is one of the serious backgrounds.
       Detection of $e^\pm$ is one of the keys to reduce the background.
 \item Halo-neutron background\\
       Neutrons in the beam halo (halo neutrons) interact with the detector material
       and produce $\pi^0$ or $\eta$, which can decay into two photons
       with large branching fractions (98.8\% for $\pi^0$, 39.4\% for $\eta$).
       Minimizing 
       material near the beam is essential to reduce these backgrounds.
       Fully active detector will reduce the background by
       efficiently detecting
       other particles generated in the $\pi^0$ or $\eta$ production.
       Another type of halo-neutron background is
       ``hadron cluster background'':
       A halo neutron hits the calorimeter to produce a first hadronic shower,
       and another neutron in the shower travels inside the calorimeter,
       and produces a second hadronic shower apart from the first one.
       These two hadronic clusters can mimic the signal.
       
\end{itemize}
\begin{table}[h]
 \centering
 \caption{$K_L$ decay of the signal and those with the branching fraction larger than $10^{-4}$.}
 \label{tab:KLdecay}
 \begin{tabular}{llll}\hline
  Decay mode                 &branching fraction &$\pi^0$ maximum $\pt$
	   & key to reduce background \\
  \hline
  $\pi^0 \nu\overline{\nu}$  &$3\times 10^{-11}$ (in SM) &$230~\mathrm{MeV}/c$ &\\
  \hline
  $\pi^\pm e^\mp\nu$         &40.6\% & &charged particle ID\\
  $\pi^\pm \mu^\mp\nu$       &27.0\% & &charged particle ID\\
  $3\pi^0$                   &19.5\% & $139~\mathrm{MeV}/c$&extra-photon veto\\
  $\pi^+\pi^-\pi^0$          &12.5\% & $133~\mathrm{MeV}/c$&charged-particle veto\\
  $\pi^+\pi^-$               &$1.97\times 10^{-3}$  & &charged particle ID\\
  $2\pi^0$                   &$8.64\times 10^{-4}$  &$209~\mathrm{MeV}/c$&extra-photon veto\\
  $\pi^\pm e^\mp\nu\gamma$ &$3.79\times 10^{-3}$  & & extra-particle veto\\
  $\pi^\pm \mu^\mp\nu\gamma$ &$5.65\times 10^{-4}$  & &extra-particle veto\\
  $2\gamma$                  &$5.47\times 10^{-4}$  & &$\pt$ of reconstructed $\pi^0$ \\
  \hline
 \end{tabular}
\end{table}

%\subsection{Conceptual detector for the base design}
\subsubsection{Conceptual detector for the base design}
\begin{figure}[h]
 \includegraphics[width=\textwidth]{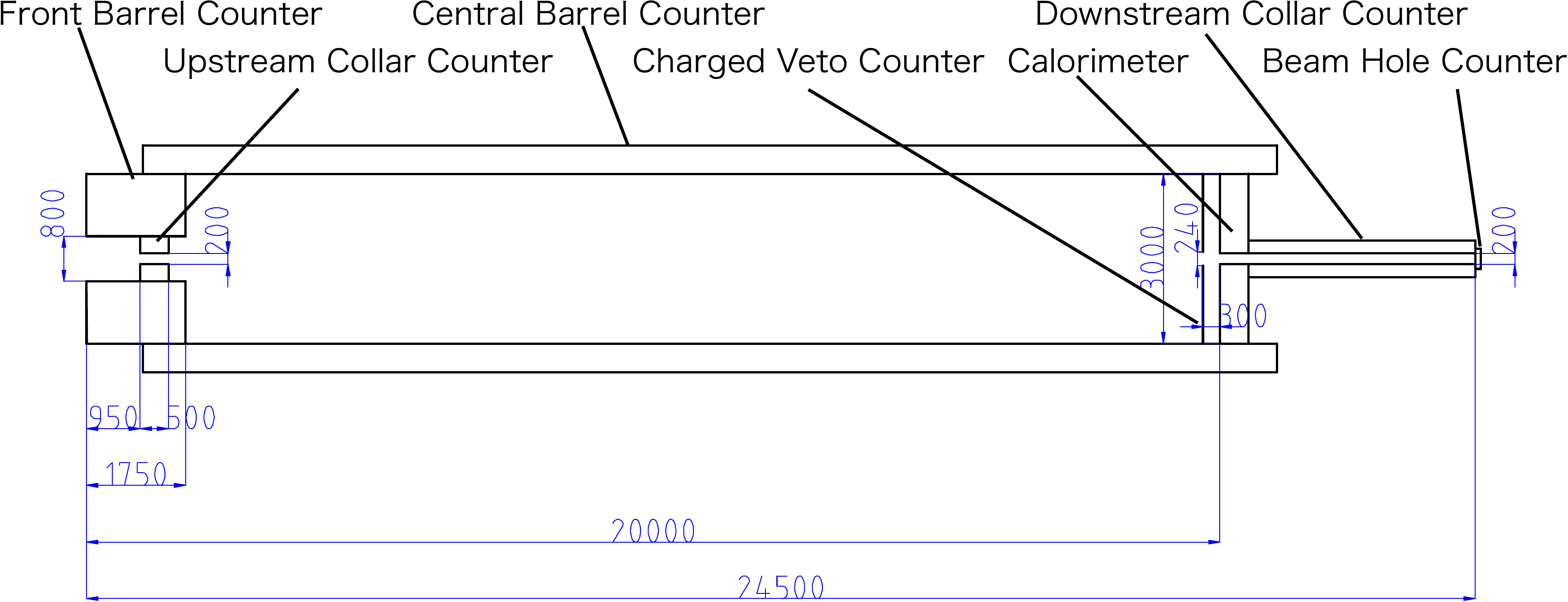}
 \caption{Conceptual detector.
 The upstream edge of the Front Barrel Counter is 44~m from T2 target.}
 \label{fig:conceptualDetector}
\end{figure}

A conceptual detector used in the base design is shown in Fig.~\ref{fig:conceptualDetector}.
We define the $z$ axis on the beam axis pointing downstream
with the origin at the upstream surface of the Front Barrel Counter,
which is 44~m from the T2 target (43-m long beam line and 1-m long space).
We use the following conceptual detector;
the diameter of the calorimeter is 3~m to gain the signal acceptance.
%The size is limited by the realistic design of the vacuum tank
%to contain the calorimeter and the barrel detector (Central Barrel Counter).
The beam hole in the calorimeter is 20~cm $\times$ 20~cm to accept
the 15~cm $\times$ 15~cm beam size.
The $z$ position of the calorimeter is 20~m with
%a larger decay volume to enhance the signal acceptance
%for higher $K_L$ momentum.
a larger decay volume to enhance the $K_L$ decay.
The larger decay volume is effectvie in the KOTO step-2,
because the higher $K_L$ momentum keeps the signal acceptance.

The Charged Veto Counter
is a charged-particle veto counter at 30~cm upstream of the calorimeter
to veto $K_L\to \pi^{\pm} e^{\mp}\nu$, $K_L\to \pi^{\pm} \mu^{\mp}\nu$,
or $K_L\to \pi^+\pi^-\pi^0$.
The beam hole at the Charged Veto Counter is 24~cm $\times$ 24~cm to avoid
$\pi^0$ or $\eta$ produced by
the interaction with neutrons.
The Front Barrel Counter is 1.75-m long and
the Upstream Collar Counter is 0.5-m long
to veto $K_L\to3\pi^0$ decay at upstream of and inside the Front Barrel Counter.
The Central Barrel Counter is 20-m long mainly to veto $K_L\to 2\pi^0$
by detecting extra photons from the decay.
The Downstream Collar Counter is 4-m long to veto particles passing
through the beam hole in the calorimeter but going
outside the beam region.
The Beam Hole Counter covers in-beam region starting from 24.5~m
to veto particles escaping through the calorimeter beam hole.
The Front Barrel Counter, the Upstream Collar Counter,
the Central Barrel Counter,
and the Downstream Collar Counter,
act as either photon and charged-particle veto
for the conceptual design.
For the Beam Hole Counter,
we introduce two separate counters, 
a beam-hole charged-veto counter and 
a beam-hole photon-veto counter.
The signal region is defined in a 12-m region from 3~m to 15~m in $z$.

We use the conceptual detector in the following sections
to evaluate the signal acceptance and 
the background contributions, and the hit rates.

%\subsection{Modeling of detector response}
\subsubsection{Modeling of detector response}
The interaction of particles in the calorimeter can be modeled
in terms of the {\bf energy / position resolutions},
and {\bf two-photon fusion probability}
(a probability to identify two incident photons nearby
as a single cluster).
The interaction of particles with respect to the veto performance 
is modeled by {\bf inefficiency} as a function of the particle type,
the incident energy, and the incident angle.
The timing smearing of the Central Barrel Counter is applied
for some studies.
Other energy or timing smearing
is not applied for the veto counters.

The photon-detection inefficiency of the barrel counter
is estimated 
based on the study performed
for the newly installed 
barrel photon counter in KOTO step-1~\cite{Murayama:2020mcp}.
The inefficiency of the in-beam detector
(the Beam Hole Counter in the conceptual detector)
is based on the performance of the current detector in the KOTO step-1.
The other modelings are the same as in the proposal~\cite{KOTOproposal}.

For the inefficiency, here we just introduce models with
detection thresholds.
These relate to the background estimation in 
Sec.~\ref{chap:sensitivity}.
The thresholds relate to
the detection rate of the counters to be described in
Sec.~\ref{sec:accidentalLoss}.
We decided the thresholds considering both.

%\subsubsection{Energy / position resolutions of the calorimeter}
\subsubsubsection{Energy / position resolutions of the calorimeter}
%The reconstruction of $\pi^0$ is affected with
%the energy and position resolutions.
The $\pt$ and $\zvtx$ resolutions of 
the reconstructed $\pi^0$ is affected by
the energy and position resolutions of the calorimeter.

\begin{figure}[h]
 \centering
 \includegraphics[bb=0 0 216 215,clip,width=0.45\textwidth]
 {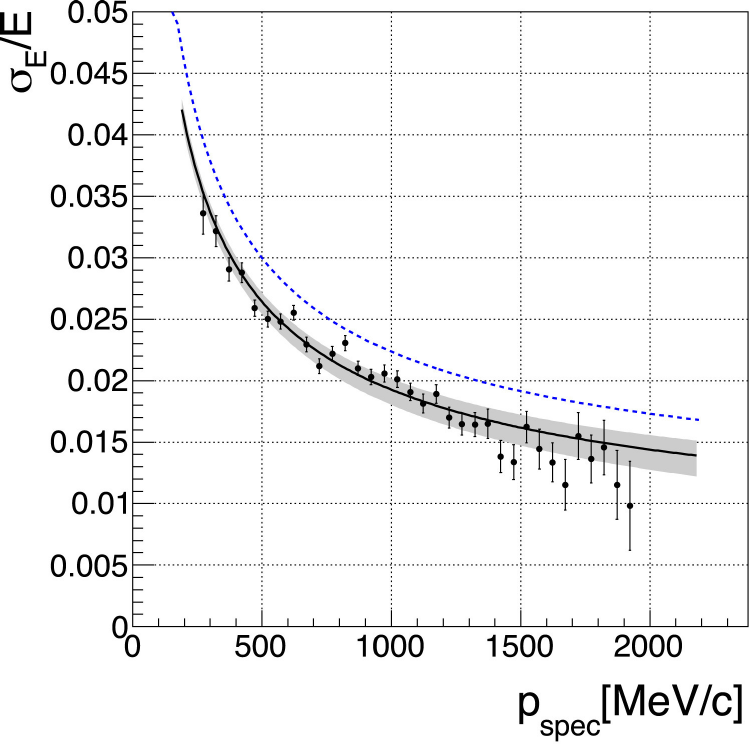}
 \includegraphics[bb=0 0 201 198,clip,width=0.45\textwidth]{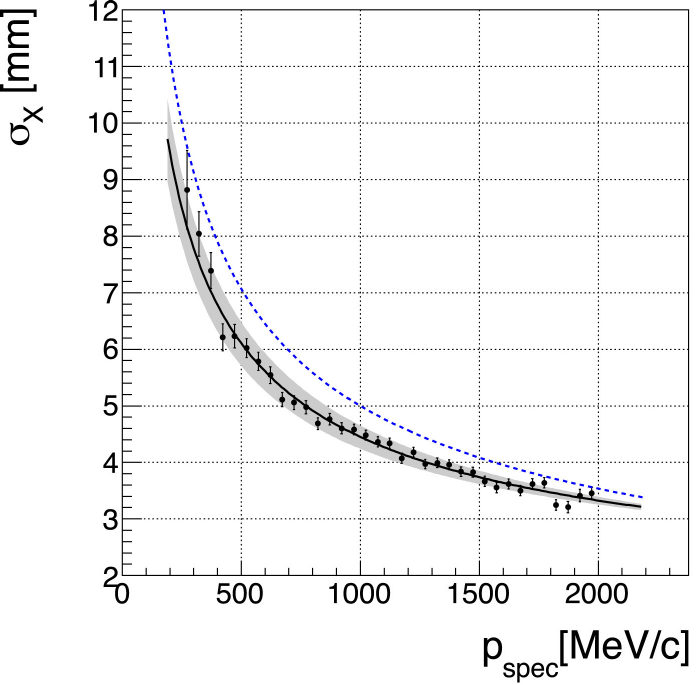}
 \caption{Energy (left) and position (right) resolutions
 for the central region of the calorimeter~\cite{Sato:2020kpq}.
 The points with error bars show the measured data with electron,
 the solid line shows
 a fit with a function,
 and the filled area shows the combined statistical and systematic errors.
 The dashed line shows the model used in this study.
 }\label{fig:calreso}
\end{figure}

The energy resolution is modeled as follows:
\begin{align*}
 \frac{\sigma_E}{E}=&\left(1\oplus \frac{2}{\sqrt{E(\mathrm{GeV})}}\right)\%.
\end{align*}
The position resolution is modeled as follows:
\begin{align*}
 \sigma_x=&\frac{5}{\sqrt{E(\mathrm{GeV})}}\left(\mathrm{mm}\right).
\end{align*}
The comparisons between the model and actual measurements in the KOTO step-1 calorimeter are shown in Fig.~\ref{fig:calreso}~\cite{Sato:2020kpq}.
The model is more conservative
than the resolutions in the inner region of the KOTO step-1 calorimeter.
\footnote{
The modeled energy resolution is also more conservative
than the actual measurements in the outer region.
The modeled position resolution 
is better by at most 3.2~mm 
than the actual resolution
in the outer region of the current calorimeter
for the incident energy smaller than 2~GeV.
5-cm-square CsI crystals 
are used in the outer region instead of 2.5-cm-square ones
in the inner region.}

%\paragraph{Two-photon fusion probability in the calorimeter}
\subsubsubsection{Two-photon fusion probability in the calorimeter}
The two-photon fusion contributes to the $K_L\to 2\pi^0$ background.
Missing two of four photons from the decay causes the background.
Fusion is one of the mechanisms to miss a photon.

\begin{figure}[h]
 \centering
 \includegraphics[bb=0 0 535 363,clip,width=0.5\textwidth]{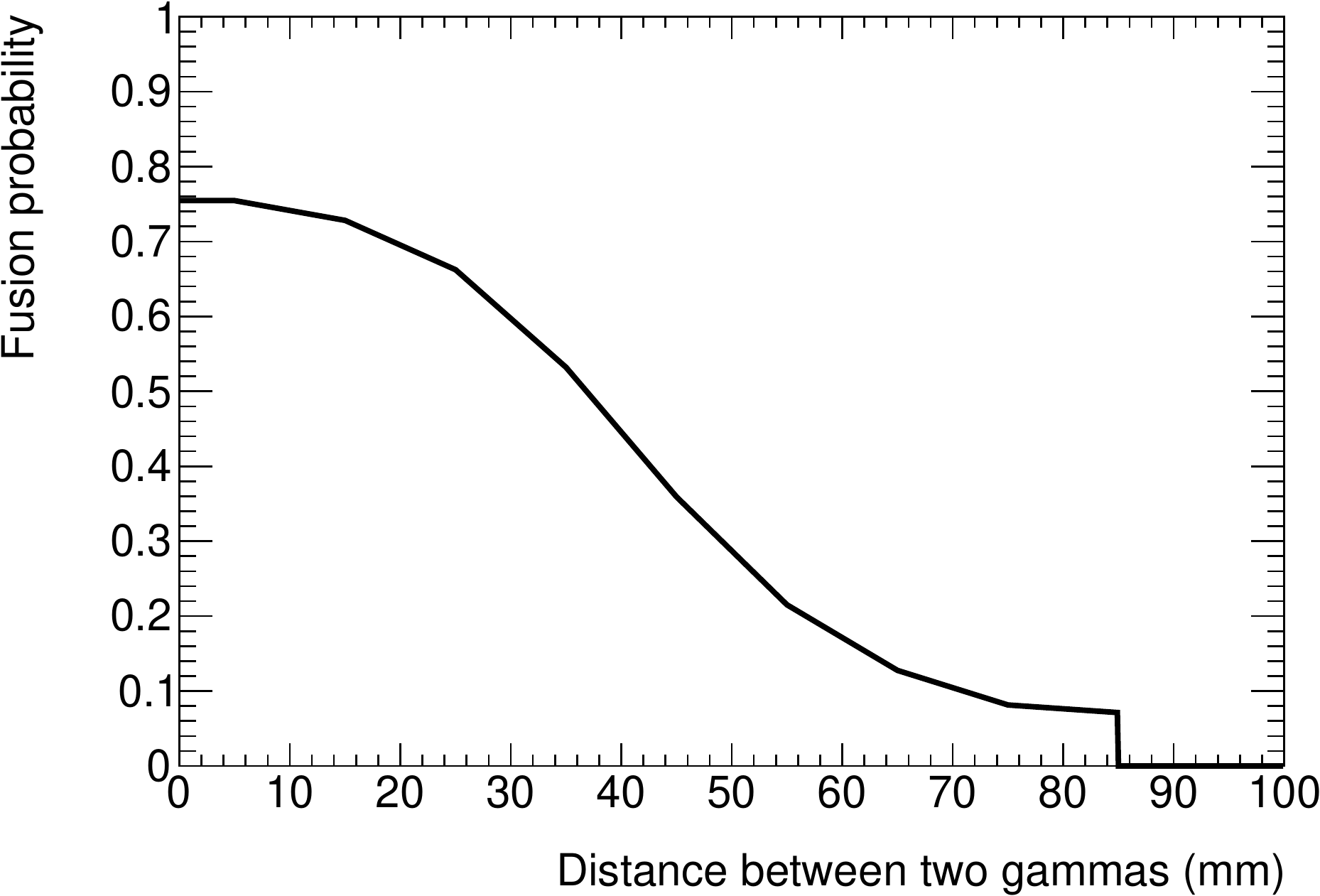}
 \caption{Fusion probability as a function of two-gamma distance on the calorimeter.\label{fig:fusionProb}}
\end{figure}

The model of the fusion probability is shown in Fig.~\ref{fig:fusionProb}
as a function of the distance between two-gamma incident-positions on the calorimeter.
This model was prepared with a MC study using
the calorimeter in KOTO step-1.

\begin{figure}[h]
 \centering
 \includegraphics[bb=0 0 540 363,clip,width=0.5\textwidth]{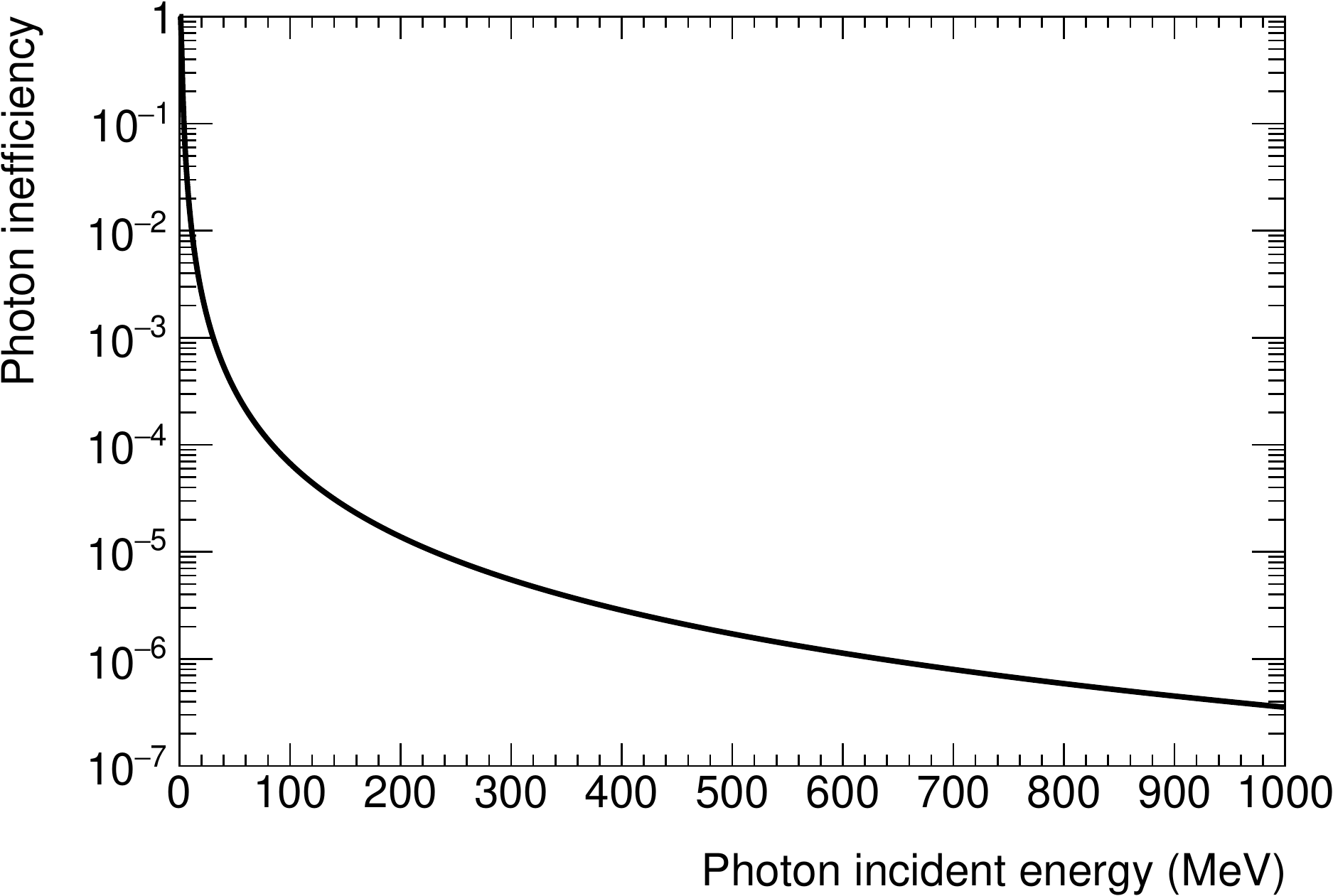}
 \caption{Photon inefficiency of the calorimeter.}\label{fig:csiIneffi}
\end{figure}

\subsubsubsection{Inefficiency of the particle veto}
\subparagraph{Calorimeter photon inefficiency}
The photon inefficiency of the calorimeter contributes
the $K_L\to2\pi^0$ background.
The modeled inefficiency is shown in Fig.~\ref{fig:csiIneffi},
which is the same as in the proposal and was obtained with a MC study.

\begin{figure}[h]
 \centering
 \subfloat[]{
 \includegraphics[page=1,width=0.32\textwidth]
 {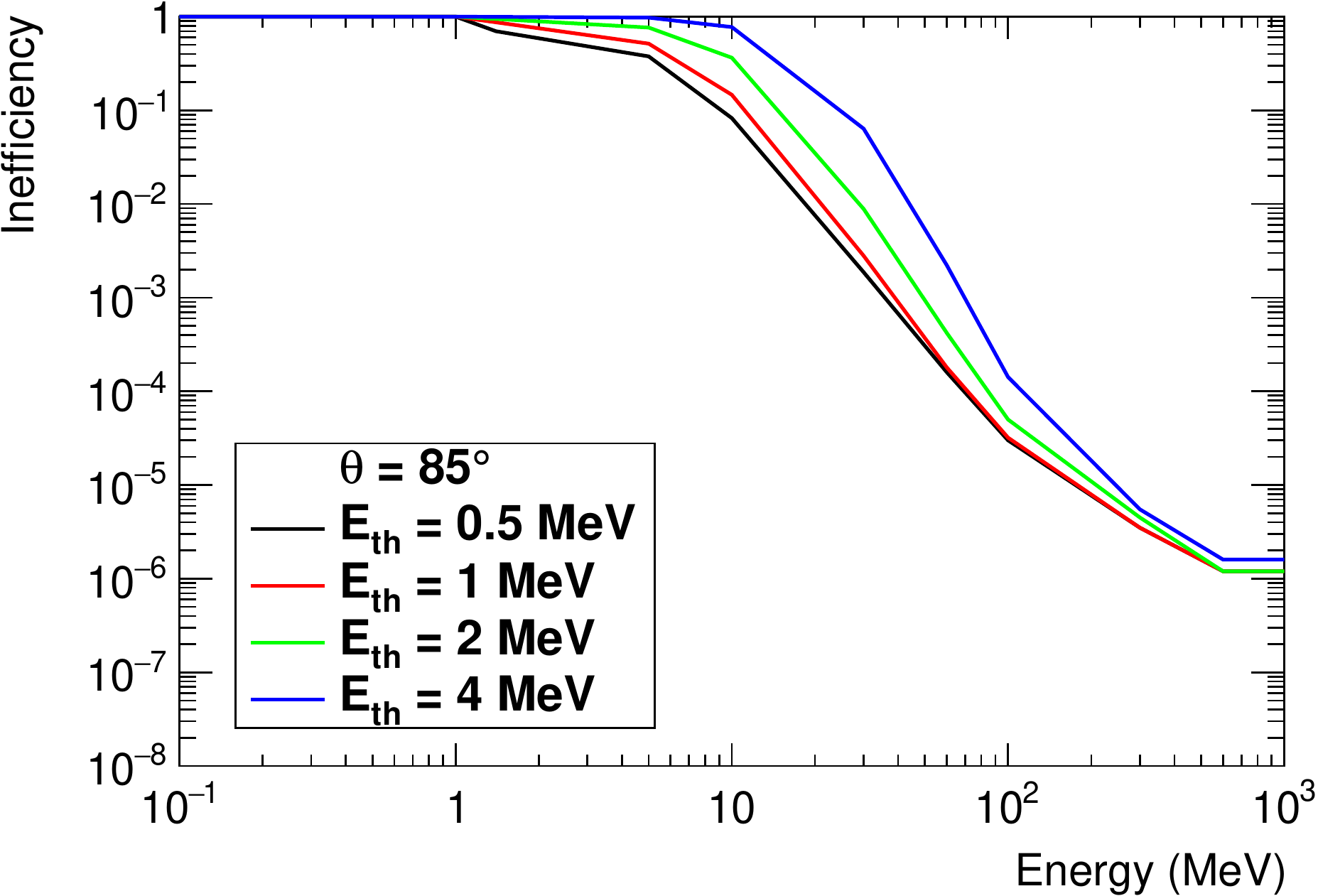}
 }
 \subfloat[]{
 \includegraphics[page=2,width=0.32\textwidth]
 {KLdocu/detector/figure/drawIBIneffi2.pdf}
 }
 \subfloat[]{
 \includegraphics[page=3,width=0.32\textwidth]
 {KLdocu/detector/figure/drawIBIneffi2.pdf}
 }
 
 \subfloat[]{
 \includegraphics[page=4,width=0.32\textwidth]
 {KLdocu/detector/figure/drawIBIneffi2.pdf}
 }
 \subfloat[]{
 \includegraphics[page=5,width=0.32\textwidth]
 {KLdocu/detector/figure/drawIBIneffi2.pdf}
 }
 \caption{Photon inefficiency of the Central Barrel Counter
 for the photon incident-angles
 $85^{\circ}$ (a),
 $45^{\circ}$ (b),
 $25^{\circ}$ (c),
 $15^{\circ}$ (d), and
 $5^{\circ}$ (e).
 When the photon direction is perpendicular to the barrel detector surface,
 the incident angle is defined to be $90^\circ$.
 }
 \label{fig:ibIneffi}
\end{figure}

\subparagraph{Barrel photon inefficiency}
The photon inefficiency of the barrel counter
is shown in Fig.~\ref{fig:ibIneffi},
which was prepared for the
barrel detector upgrade in KOTO step-1.
The barrel detector in KOTO step-1
is composed of 1-mm-thick or 2-mm-thick lead plates
and 5-mm-thick plastic scintillator plates.
The inefficiency was prepared with a full shower simulation
by applying some energy thresholds in the energy deposit.
In this report,
we use the energy threshold of 1 MeV.

%\subparagraph{Collar-counter photon inefficiency}

\subparagraph{Charged Veto Counter inefficiency for penetrating charged particles}
The inefficiency of the Charged Veto Counter contributes to
the backgrounds from 
the $\kpien$ and $\kpimun$ decays.
The two charged particles could make two clusters on the calorimeter,
which mimics the signal if these are not detected with the Charged Veto Counter.

In the conceptual design, the reduction of $10^{-12}$ for
these backgrounds with the Charged Veto Counter is required.
%This is the requirement on the detector design.
For example, 
two planes of the Charged Veto Counter with $10^{-3}$ reduction
of a single charged particle per plane
will reduce the background
with the two charged particles by $10^{-12}$.
In KOTO step-1,
we achieved the $10^{-5}$ reduction of a single charged particle with one plane~\cite{Naito:2015vrz}.
Therefore, the $10^{-3}$ reduction with one plane is achievable.

%\subparagraph{Charged Veto Counter inefficiency for $\pi^\pm$}
%\subparagraph{Collar counter charged-particle inefficiency}

\subparagraph{Beam Hole Counter charged-particle inefficiency}
In KOTO step-1, we are operating a MWPC-type gas-wire chamber,
and achieved $5\times 10^{-3}$ inefficiency
for charged particles~\cite{Kamiji:2017deh}.
Based on the result,
we assume the same $5\times 10^{-3}$ inefficiency
for the beam-hole charged-veto counter
in the KOTO step-2 design.

\subparagraph{Beam Hole Counter photon inefficiency}
In KOTO step-1, we are operating 16 modules of
a lead-aerogel Cherenkov counter
~\cite{Maeda:2014pga}
as the in-beam photon veto counter.
It is insensitive to beam neutrons,
because protons or charged pions generated from the neutron-interaction
tend to be slow and emit less Cherenkov radiation.
By taking three-consecutive coincident hits in the modules,
electromagnetic shower is efficiently detected,
because it develops in the forward direction and is laterally well collimated.

\begin{figure}[h]
 \centering
 \includegraphics[page=2,width=0.5\textwidth]{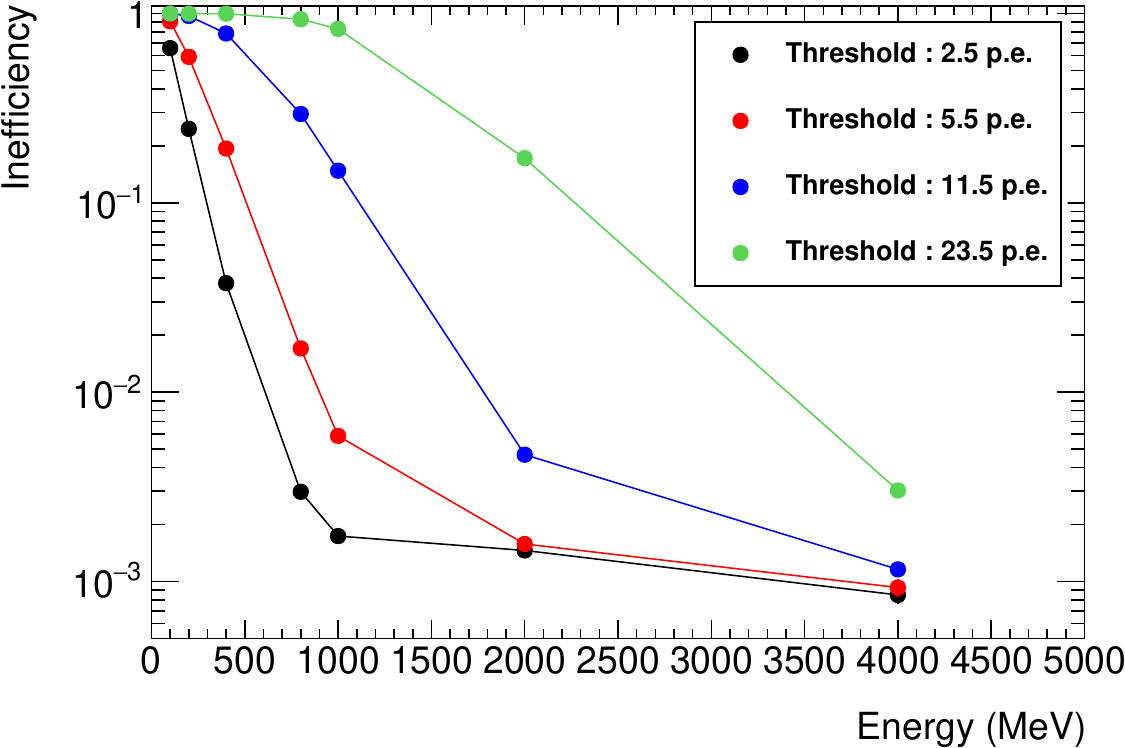}
 \caption{Photon inefficiency of the Beam Hole Counter.
 }\label{fig:bhpvinef}
\end{figure}

We assume 25 modules of such a counter for the 
beam-hole photon-veto counter.
The photon detection performance was studied with
a reliable full-shower simulation developed in KOTO step-1.
Inefficiencies as a function of the incident-photon energy are
shown in Fig.~\ref{fig:bhpvinef} for several detection
thresholds on the number of observed photoelectrons.
In this report,
we use the threshold of 5.5 photoelectrons.

\begin{figure}[h]
 \centering
 \includegraphics[width=0.5\textwidth]{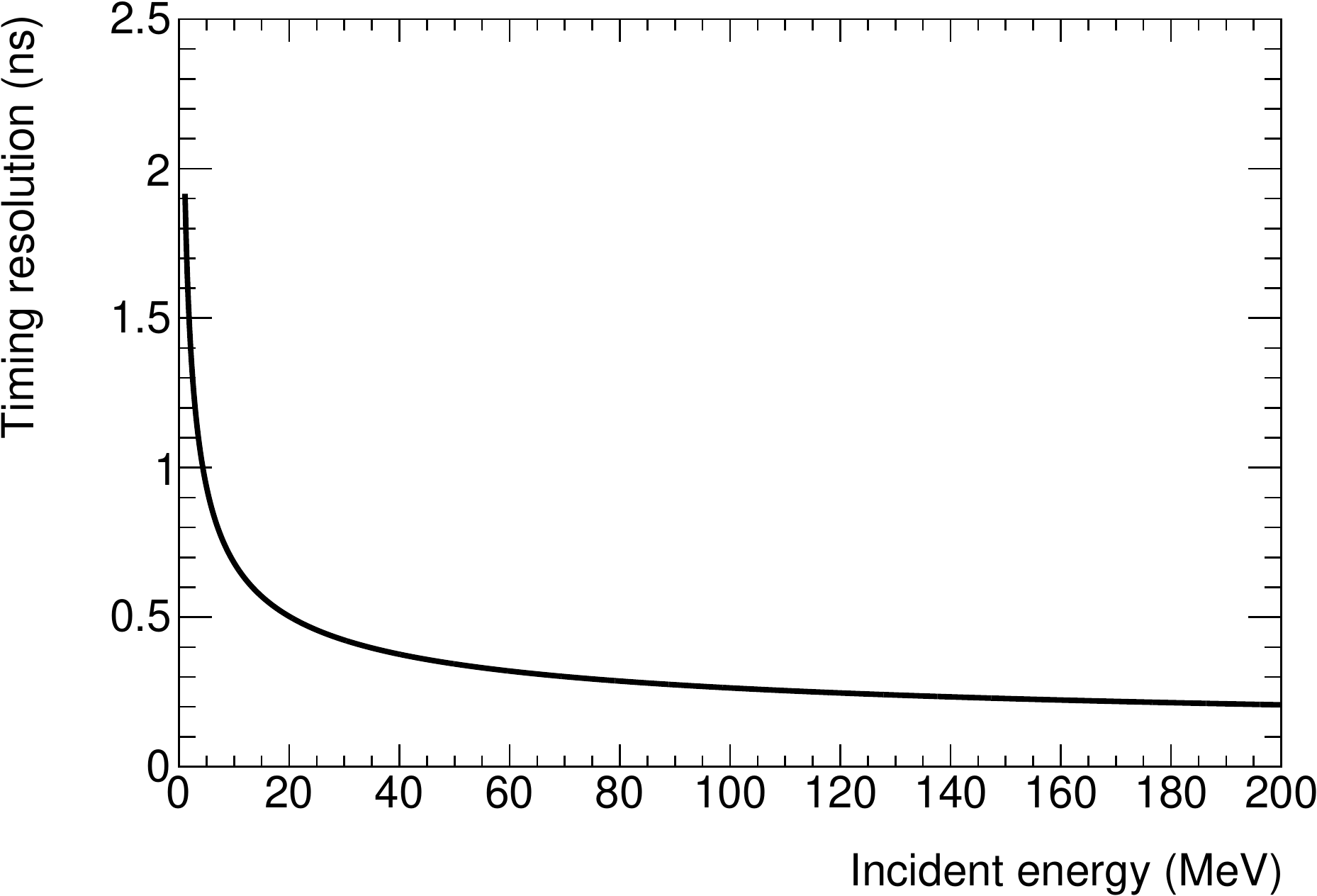}
 \caption{Assumed timing resolution of the Central Barrel Detector.
 }\label{fig:barrelTreso}
\end{figure}

\subsubsubsection{Timing resolution of the Central Barrel Counter}
\label{sec:barrelTreso}
We assume the timing resolution of the  Central Barrel Counter
as shown in Fig.~\ref{fig:barrelTreso}
based on the study performed
for the new 
barrel photon counter installed in KOTO step-1
~\cite{Murayama:2020mcp}.
The resolution is modeled as a function of the incident energy.
Is is 2~ns for the incident energy of 1~MeV, for example.

% flatex input end: [KLdocu/detector/detector.tex]

%\section{Physics and Experiment at KL2 Beam Line}
\clearpage
% flatex input: [KLdocu/sensitivity/sensitivity.tex]
%\section{Sensitivity and Background Estimation}
\subsection{Sensitivity and Background Estimation}
\label{chap:sensitivity}

%\subsection{Beam conditions}
\subsubsection{Beam conditions}
We assume the beam condition and running time as in Table~\ref{tab:beam}.
\begin{table}[h]
 \centering
 \caption{Assumed beam and running time.}\label{tab:beam}
 \begin{tabular}{lll}\hline
  Beam power & 100~kW &(at 1-interaction-length T2 target) \\
  && ($1.1\times 10^7 K_L/2\times 10^{13}~\mathrm{POT}$)\\
  Repetition cycle &4.2~s  & \\
  Spill length &2~s  & \\
  Running time &$3\times 10^7$~s  & \\\hline
 \end{tabular}
\end{table}

%\subsection{Reconsturction}
\subsubsection{Reconstruction}
We evaluated yields of the signal and backgrounds
with Monte Carlo simulations.
The calorimeter response was simulated either with model responses
as explained in Section~\ref{chap:detector}
or with shower simulations in the calorimeter.
We assumed 50-cm-long CsI crystals for the calorimeter material in the shower simulations.

A shower is generated by a particle incident on the calorimeter.
%A cluster is formed from multiple hits of the calorimeter channels.
A cluster is formed based on energy deposits
in the calorimeter segmented in x-y directions.
Assuming the incident particle to be a photon,
the energy and position are reconstructed.
We treat it as a photon in the later analysis
regardless of the original particle species.

\begin{figure}[h]
 \centering
 \includegraphics[width=0.5\textwidth]{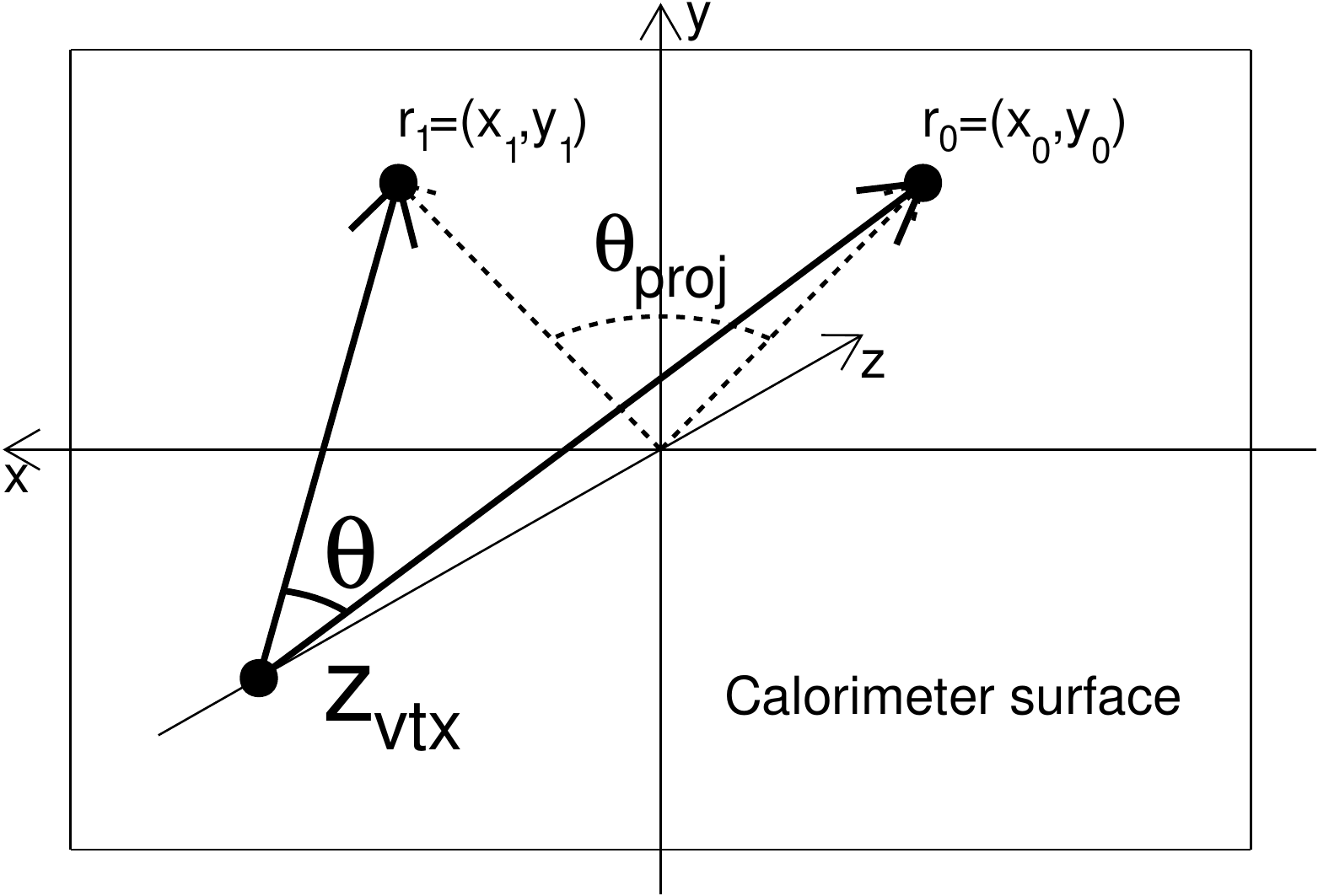}
 \caption{Geometrical relation in the vertex reconstruction.}\label{fig:zvert}
\end{figure}

A $\pi^0$ is reconstructed from the two photons on the calorimeter;
The opening angle of two photon-momentum directions ($\theta$)
can be evaluated with the energies of the two photons ($E_0, E_1$)
from 4-momentum conservation:
\begin{align*}
 {p}_{\pi^0}=&{p}_0+{p}_1,\\
 m_{\pi^0}^2=& 2 E_0 E_1 (1-\cos\theta).
\end{align*}
${p}_{\pi^0}$ is four-momentum of the $\pi^0$.
${p}_0$ and ${p}_1$ are four-momenta
of two photons. $m_{\pi^0}$ is the nominal mass of $\pi^0$.
The vertex position of the $\pi^0$ is assumed to be on the $z$ axis
owing to the narrow beam, and
the $z$ vertex position ($\zvtx$) is calculated from the
geometrical relation among $\theta$ and hit positions
${\bf r}_0=(x_0, y_0)$, ${\bf r_1}=(x_1, y_1)$ on the calorimeter
as shown in Fig.~\ref{fig:zvert}.
The $\zvtx$ gives the momenta of two photons,
and the sum of the momenta gives momentum of $\pi^0$.
Accordingly, the $\pi^0$ transverse momentum ($\pt$) is obtained.

%\subsection{Fiducial decay and event selection}\label{sec:cuts}
\subsubsection{Event selection}\label{sec:cuts}
\begin{figure}[h]
 \centering
 \includegraphics[width=0.5\textwidth]{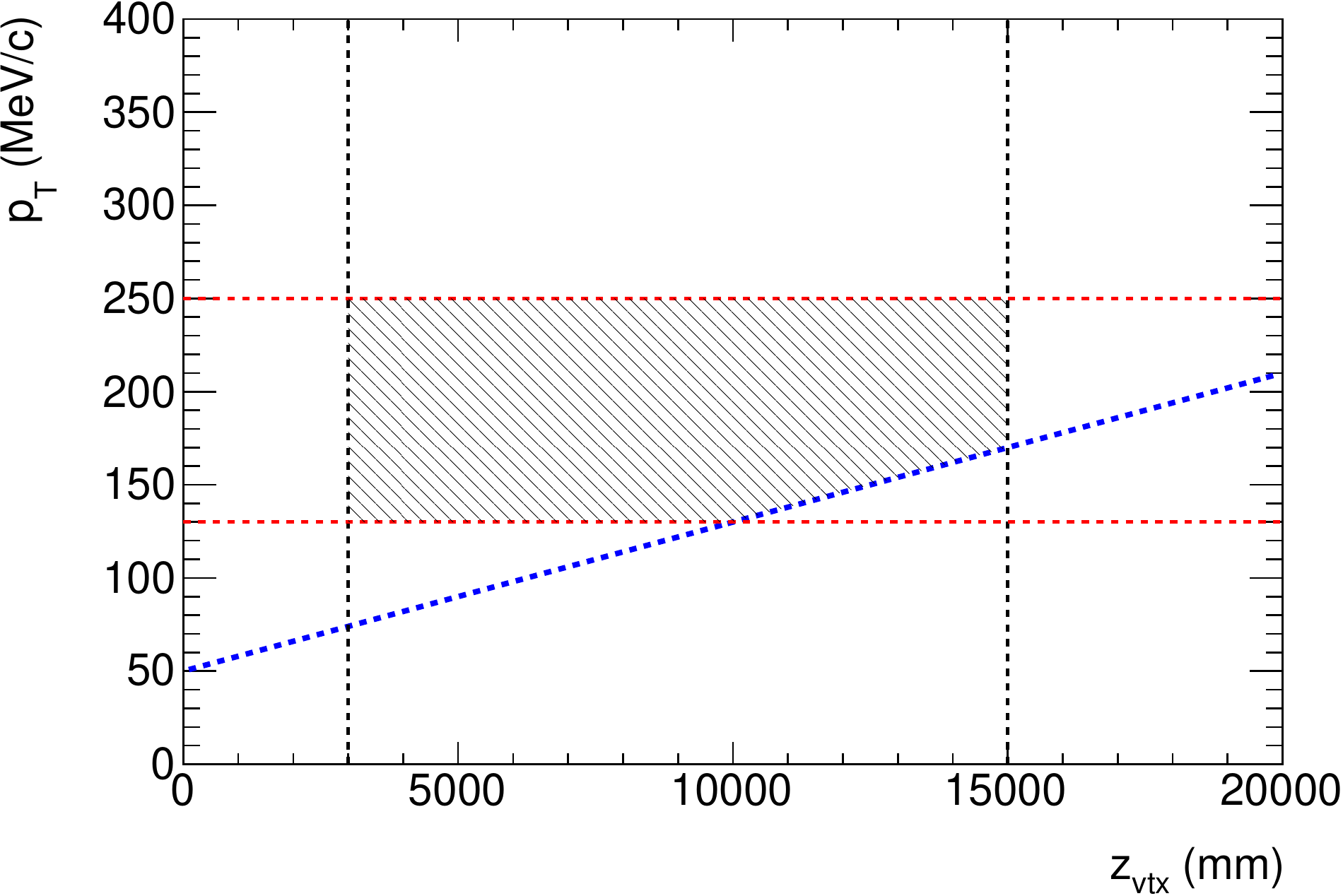}
 \caption{The shaded area shows the $\pt$ criteria in the $\zvtx$-$\pt$ plane.
 The blue doted line
 shows the tighter $\pt$ criteria in the downstream region.}
 \label{fig:hexiagonal}
\end{figure}

%\paragraph{Fiducial decay definition}
%\subparagraph{Fiducial decay definition}
%We require the decay $z$ position in the following range,
%and require the two photons from the decay entering the calorimeter.
%\begin{itemize}
% \item Decay $z$ position : $3~\mathrm{m}< z <15~\mathrm{m}$.
% \item Two photons in the calorimeter.
%\end{itemize}
%\paragraph{Event selection} We use the following event selections.
We use the following event selections
for the events with two clusters in the calorimeter.
\begin{enumerate}
 \item Sum of two photon energies : $E_0+E_1>500~\mathrm{MeV}$.
 \item Calorimeter fiducial area :
       $\sqrt{x_0^2+y_0^2}<1350~\mathrm{mm}$,
       $\sqrt{x_1^2+y_1^2}<1350~\mathrm{mm}$.
 \item Calorimeter fiducial area :
       $\mathrm{max}(|x_0|, |y_0|)>175~\mathrm{mm}$,
       $\mathrm{max}(|x_1|, |y_1|)>175~\mathrm{mm}$.
 \item Photon energy :
       $E_0>100~\mathrm{MeV}$,
       $E_1>100~\mathrm{MeV}$.
 \item Distance between two photons : $|{\bf r}_1-{\bf r}_0|>300~\mathrm{mm}$.
 \item Projection angle ($\theta_{\mathrm{proj}}$ as shown in Fig.~\ref{fig:zvert}) : $\theta_{\mathrm{proj}}\equiv \mathrm{acos}\left( \frac{{\bf r}_0\cdot{\bf r}_1}{|{\bf r}_0||{\bf r}_1|}\right)<150^{\circ}$.
 \item $\pi^0$ decay vertex : $3~\mathrm{m}<\zvtx <15~\mathrm{m}$.
 \item $\pi^0$ transverse momentum : $130~\mathrm{MeV}/c<\pt <250~\mathrm{MeV}/c$.
 \item Tighter $\pi^0$ $\pt$ criteria in the downstream (Fig.~\ref{fig:hexiagonal}):
       $\frac{\pt}{(\mathrm{MeV}/c)}>\frac{\zvtx}{(\mathrm{mm})}\times 0.008+50$.
 \item Selection to reduce hadron cluster background\\
       In order to reduce neutron clusters,
       cluster shape, pulse shape, and 
       depth information of the hits in the calorimeter
       are used as in the analysis of the KOTO step-1.
       The signal selection efficiency of $0.9^3=0.73$ is assumed.
       The reduction of the background is discussed in Sec.~\ref{sec:hadronCluster}.
 \item Selection to reduce halo $K_L\to 2\gamma$  background\\
       The photon incident-angle information is used to reduce the halo $K_L\to 2\gamma$  background
       as in the KOTO step-1. The signal selection efficiency of $0.9$ is assumed.
       The reduction of the background is discussed in Sec.~\ref{sec:haloKL}.
\end{enumerate}
The first 5 selections ensure the quality of the photon cluster;
The sum of the photon energies is useful to reduce a trigger bias,
because we plan to use the sum of the calorimeter energy for the trigger.
The edge region of the calorimeter is avoided to reduce the energy leak outside the calorimeter.
Higher energy photons give good resolution.
Large distance between the two photons
reduces the overlap of two clusters.

The next four are kinematic selections.
The projection angle selection requires
no back-to-back configuration of the two photons to reduce $K_L\to2\gamma$.
Larger $\pi^0$ $\pt$ is required
to match the kinematics of the signal.
The tighter $\pt$ selection is required in the downstream region,
because the reconstructed $\pt$ tends to be larger due to 
worse $\pt$ resolution for the decay near the calorimeter.

The last two are the identification criteria with the calorimeter
to discriminate photon and neutron clusters, or
to discriminate correct and incorrect photon-incident angles.

%\subsection{Signal yield}
\subsubsection{Signal yield}
The yield of $\klpionn$ can be factorized into
the decay probability within the $z$ region
from 3~m to 15~m,
the geometrical acceptance to have 2 photons in the calorimeter,
and the cut acceptance described in the following subsections.
The signal losses
called  ``accidental loss'' and ``shower-leakage loss''
will be introduced and discussed in the later subsections.

%\subsubsection{Decay probability and geometrical acceptance of two photons at the calorimeter}
\subsubsubsection{Decay probability and geometrical acceptance of two photons at the calorimeter}

\begin{figure}[h]
 \centering
 \subfloat[]{
 \includegraphics[page=1,width=0.45\textwidth]{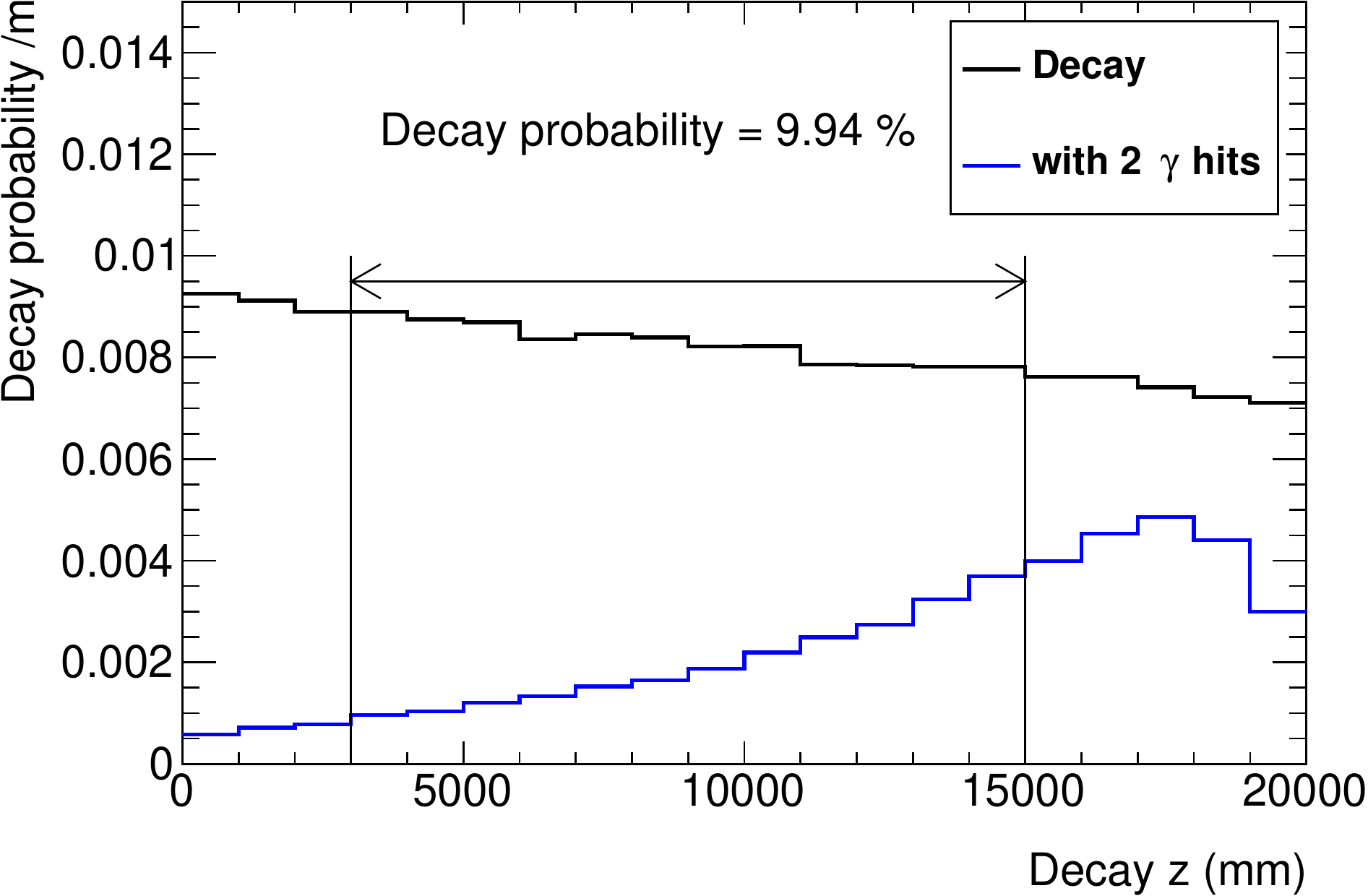}
 }
 \subfloat[]{
 \includegraphics[page=2,width=0.45\textwidth]{KLdocu/sensitivity/figure/decayGeom.pdf}
 }
 \caption{Decay probability (a) and geometrical acceptance (b).}
 \label{fig:decayGeom}
\end{figure}

The decay probability ($P^{\mathrm{truth}}_{\mathrm{decay}}$) is defined:
\begin{align*}
 P^{\mathrm{truth}}_{\mathrm{decay}}=&
 \frac{\text{Number of $K_L$'s that decayed in $3~\mathrm{m}<z<15~\mathrm{m}$}}
 {\text{Total number of $K_L$'s at $z=-1$~m}}.
\end{align*}
It is evaluated with a MC simulation as in Fig.~\ref{fig:decayGeom}(a)
to be $9.9$\%.
% We can roughly calculate it with the peak $K_L$ momentum 
% $p_{\mathrm{peak}}(=2.9~\mathrm{GeV}/c)$, $K_L$ mass $m_{K_L}(=0.497~\mathrm{GeV}/c^2)$,
% and $K_L$ lifetime $c\tau(=15.3~\mathrm{m})$:
% \begin{align*}
%  &
%  \exp\left(-\frac{(3~\mathrm{m}-(-1)~\mathrm{m})}
%  {(p_{\mathrm{peak}}/m_{K_L}) c\tau}\right)
%  \left[
%  1-\exp\left(-\frac{(15~\mathrm{m}-3~\mathrm{m})}
%  {(p_{\mathrm{peak}}/m_{K_L}) c\tau}\right)
%  \right]\\
%  \sim&
%  \exp\left(-\frac{4~\mathrm{m}}{90~\mathrm{m}}\right)
%  \left[
%  1-\exp\left(-\frac{12~\mathrm{m}}
%  {90~\mathrm{m}}\right)
%  \right]\\
%  \sim & 96\% \times 12\%\sim 12\%.
% \end{align*}

The geometrical acceptance ($A^{\mathrm{truth}}_{\mathrm{geom}}$) is defined:
\begin{align*}
 A^{\mathrm{truth}}_{\mathrm{geom}}=&
 \frac
 {\text{Number of $K_L$'s with 2$\gamma$'s in the calorimeter that decayed in $3<z<15$~m}}
 {\text{Number of $K_L$'s that decayed in $3~\mathrm{m}<z<15~\mathrm{m}$}}.
\end{align*}
% {\text{Number of $K_L$ decays with 2$\gamma$'s in the calorimeter within the denominator}}
It is also evaluated with a MC simulation 
to be 24\%
as shown in Fig.~\ref{fig:decayGeom}(b).

The
$P^{\mathrm{truth}}_{\mathrm{decay}}$ and 
$A^{\mathrm{truth}}_{\mathrm{geom}}$
relate to the true decay $z$ position. 
In the following sections, reconstructed $\zvtx$ is used
to give a realistic evaluation.
$P^{\mathrm{truth}}_{\mathrm{decay}}\times A^{\mathrm{truth}}_{\mathrm{geom}}$
can be compared with $A_{2\gamma}$ which is defined with
reconstructed $\zvtx$:
\begin{align*}
 A_{2\gamma}=&
 \frac{\text{Number of events with 2$\gamma$ hits with $3~\mathrm{m}<\zvtx <15~\mathrm{m}$}}
 {\text{Total number of $K_L$ at $z=-1$~m}}.
\end{align*}
With a MC simulation,
$A_{2\gamma}$ of 2.4\% is obtained,
which is consistent with
$P^{\mathrm{truth}}_{\mathrm{decay}}\times A^{\mathrm{truth}}_{\mathrm{geom}}=2.4\%$.

\begin{figure}[h]
 \centering
 \includegraphics[page=21,width=0.5\textwidth]{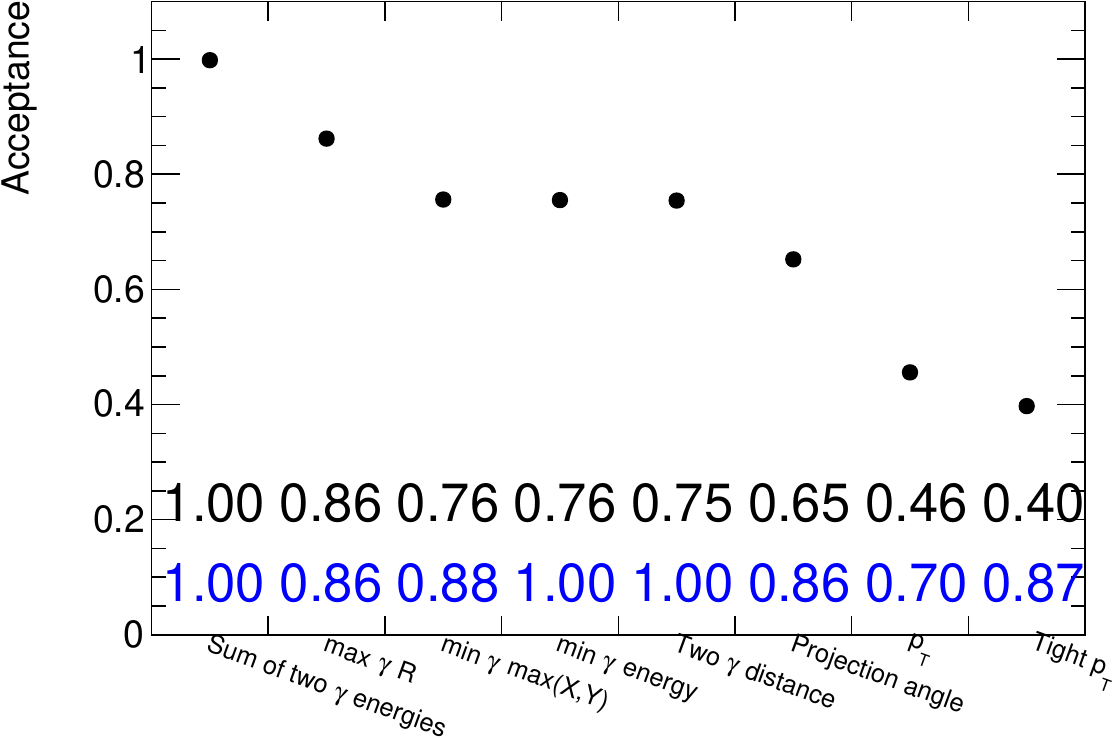}
 \caption{Cut acceptance. The black numbers in the figure show cumulative acceptances,
 the blue numbers show individual acceptances.
 }
 \label{fig:cutAccSummary}
\end{figure}

\begin{figure}[h]
 \centering
 \subfloat[]{
 \includegraphics[page=12,width=0.32\textwidth]{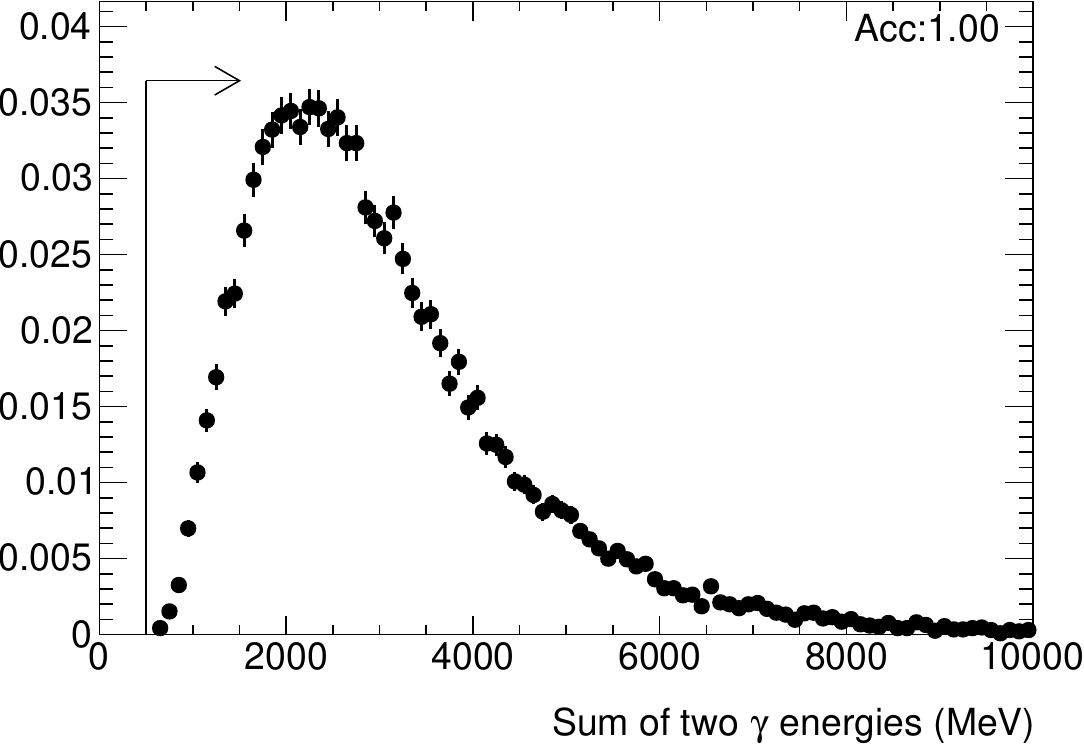}
 }
 \subfloat[]{
 \includegraphics[page=13,width=0.32\textwidth]{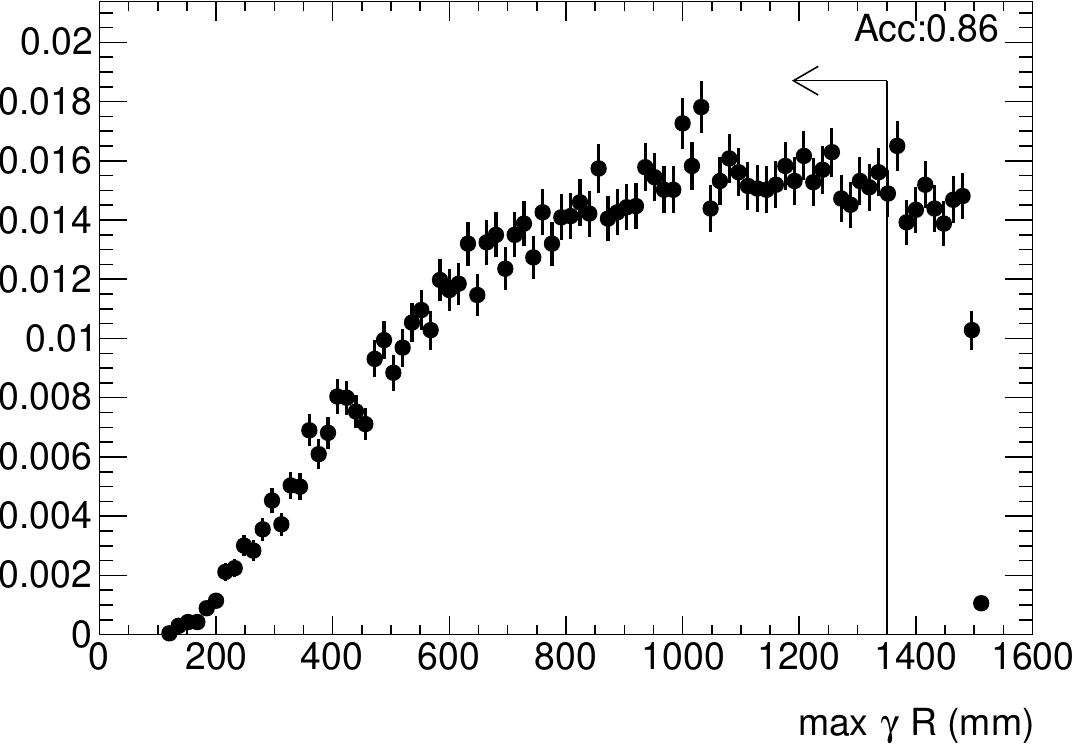}
 }
 \subfloat[]{
 \includegraphics[page=14,width=0.32\textwidth]{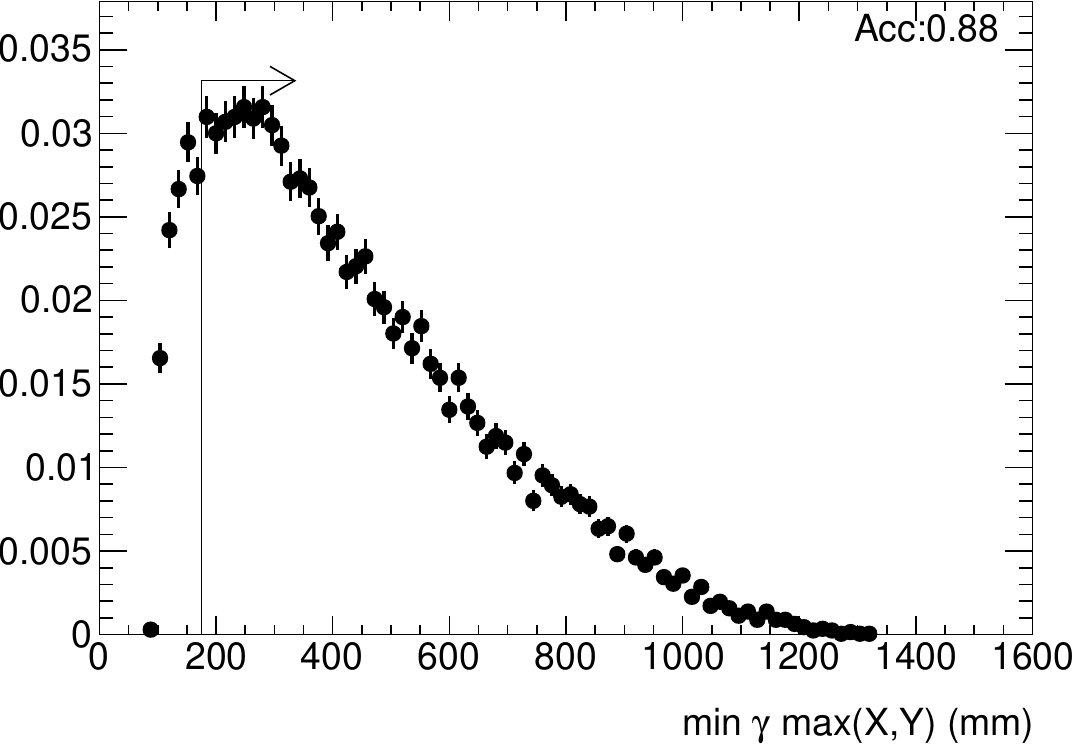}
 }
 
 \subfloat[]{
 \includegraphics[page=16,width=0.32\textwidth]{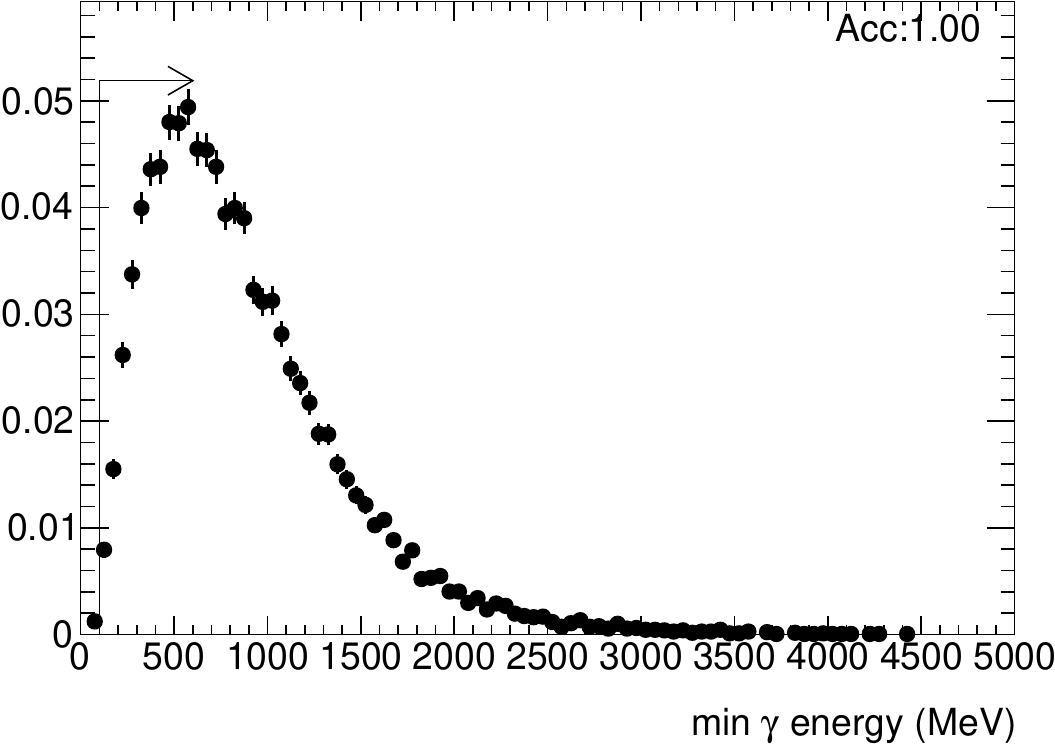}
 }
 \subfloat[]{
 \includegraphics[page=17,width=0.32\textwidth]{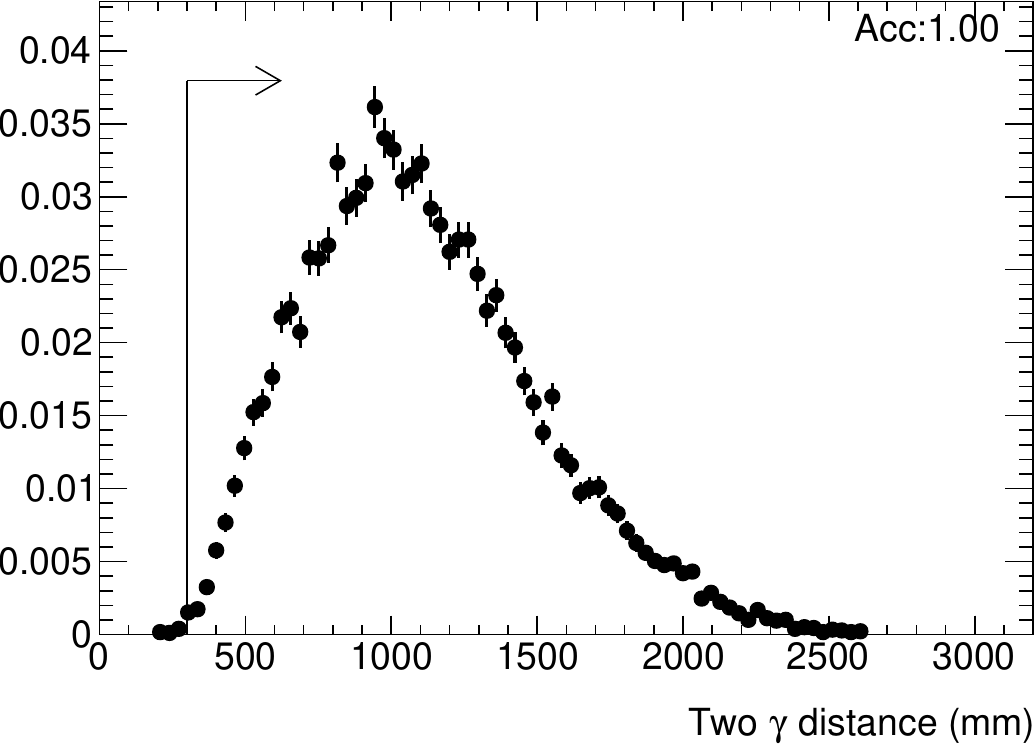}
 }
 \subfloat[]{
 \includegraphics[page=18,width=0.32\textwidth]{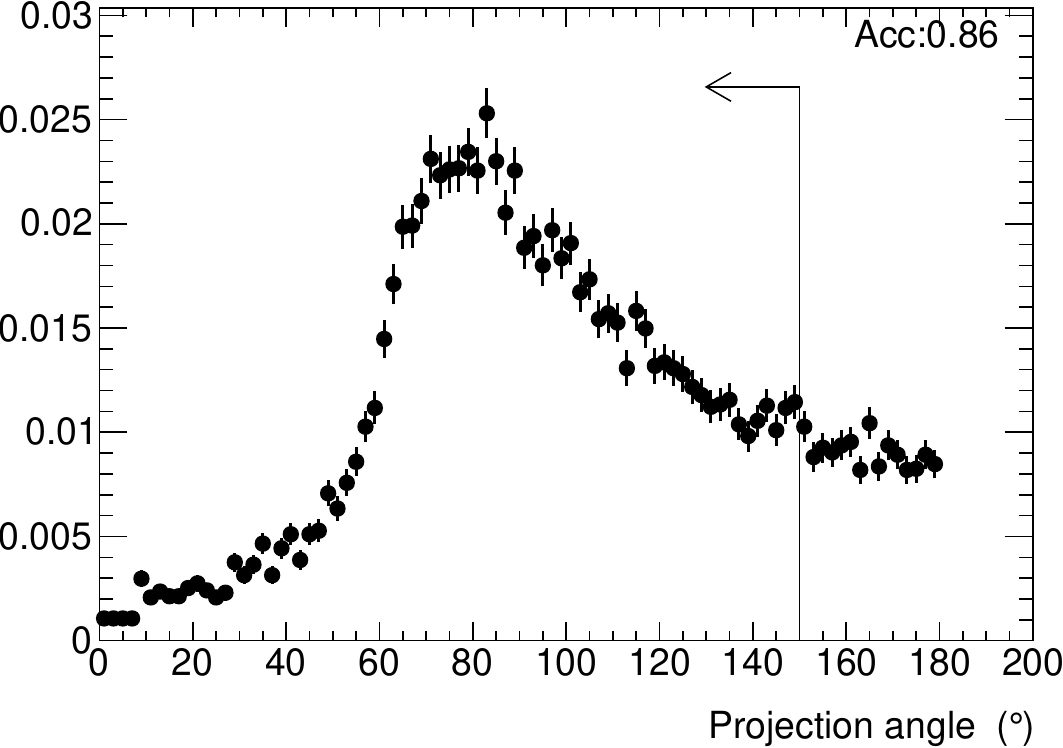}
 }
 
 \subfloat[]{
 \includegraphics[page=19,width=0.32\textwidth]{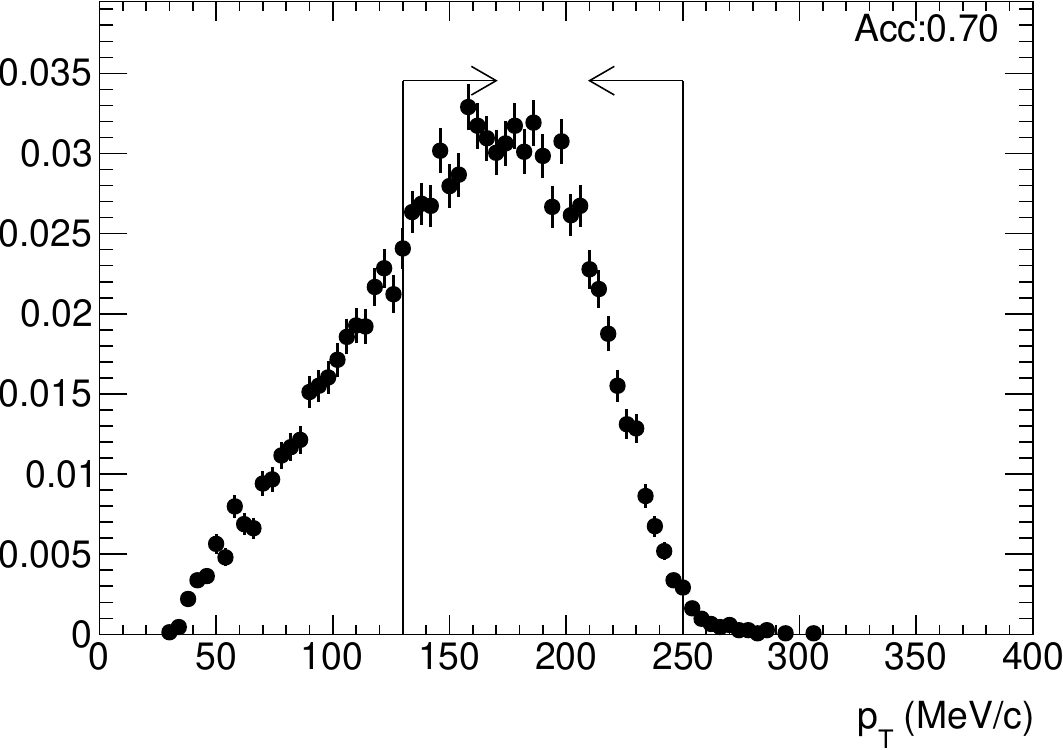}
 }
 \subfloat[]{
 \includegraphics[page=20,width=0.32\textwidth]{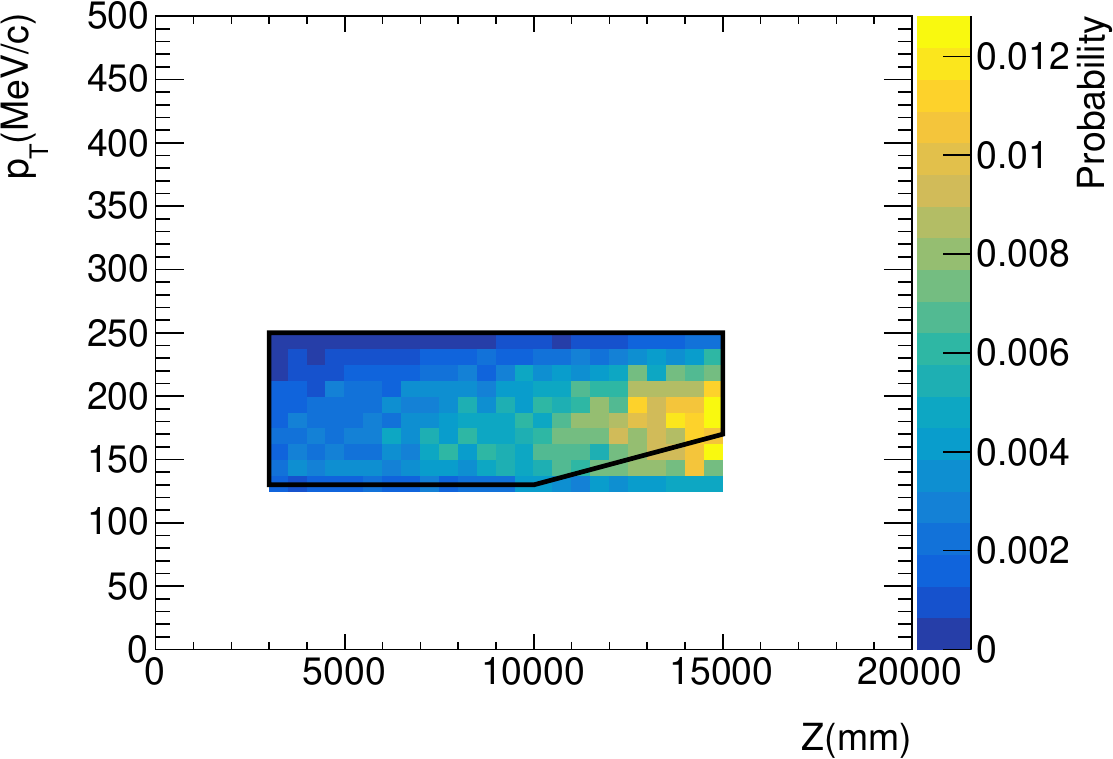}
 }
 \caption{Distributions of variables used in the event selections:
 (a) sum of two photon energies, (b) radial hit position,
(c) inner hit positions (d) minimum photon energy,
 (e) distance between two photons,
 (f) projection angle, (g) $\pt$, and (h) tighter $\pt$ selection in the downstream
 }
 \label{fig:cutAcc}
\end{figure}

\subsubsubsection{Cut acceptance}
The cut acceptances for the cuts from 1-6 and 8-9
\footnote{The seventh cut, $\zvtx$ selection, is already treated in the previous section.}
listed in Sec.~\ref{sec:cuts}
are summarized in Fig.~\ref{fig:cutAccSummary}.
The overall cut acceptance
after applying all those cuts is 40\%.
The distributions of cut variables are shown in Fig.~\ref{fig:cutAcc}.
The assumed acceptance for all the additional cuts
to reduce the hadron-cluster background
and halo $K_L\to 2\gamma$ is $0.9^4=66\%$.
Including all the above, the cut acceptance is 26\%.

\subsubsubsection{Accidental loss}
\label{sec:accidentalLoss}
In order to veto background events,
we set a timing window (veto window)
to detect extra particles 
with respect to the two-photon hit timing at the calorimeter.
The width of the veto window is set to 40 ns for the Central Barrel Counter,
30 ns for the beam-hole charged-veto counter,
6 ns for the beam-hole photon-veto counter,
and 20 ns for the other counters
throughout this report.

When the $\klpionn$ signal is detected with the calorimeter,
another $K_L$ might decay accidentally
and its daughter-particle may hit a counter at the same time.
Similarly, a photon or neutron in the beam might hit the Beam Hole Counter
at the same time.
These accidental hits will veto the signal if the hit timing is within the veto window.
We call this type of signal loss as ``accidental loss''.

First, we explain the accidental loss from the detectors
other than the Beam-Hole Counter.
Next, we explain the accidental loss
of the Beam Hole Counter.
%using a result from a beam line simulation which includes
%in-beam low-energy photons and neutrons.

%\paragraph{Detectors other than Beam-Hole Counter}
\subparagraph{Detectors other than Beam-Hole Counter}
% In principle, all the $K_L$ decays inside the detector contribute
% the accidnetal hits, because our detector is hermetic.
% Sometimes the energy of incident particle is $O(1~mathrm{MeV})$
% and not detected, but such contribution is small.
% Therefore, we can use the $K_L$-decay rate in the detector
% for the estimation of the hit rate.
% Here we assume the hits from the multiple particles in a single decay
% happen in the same veto window.

% The $K_L$ decay rate is shown in Fig.~\ref{fig:KdecayRate} (a)
% at each $K_L$ decay position.

% The $K_L$ decay at $z$ roughly from -1~m to 23~m alwasy
% makes hits at the detector.
% In total, $K_L$ decay rate with detector hits is 5.12~MHz,
% and the accidental loss probability within the 20-ns veto window
% is 9.73\%.
%%Table~\ref{tab:rateWidth} shows the detector-hit rates
%%obtained from a $K_L$-decay simulation.
The hit rate of each detector and the veto width are summarized in Table~\ref{tab:rateWidth}. 
%%%In this simulation,
%%%a 4-m long collimator with a 20-cm-square beam hole
%%%was located in $z=-5$~m,
%%%and $K_L$'s were generated at $z=-10~\mathrm{m}$.
The detector rates were obtained by a $K_L$-decay simulation. 
In the simulation,
$K_L$'s were shot from $z=-10~\mathrm{m}$, 
and a 4-m-long modeled collimator with a 20-cm-square beam hole was located at $z$ from $-5$~m to $-1$~m.
The $K_L$ decay rate with detector hits was found to be 5.1~MHz,
to which the $K_L$-decay at $z$ from $-1$~m to 23~m mainly contributes.
% The sum of the rates is 7.6~MHz, which is larger than
% the $K_L$ decay rate because part of the multiple hits in a decay
% are counted separately in each detector.
%This rates gives the accidental loss of 14.1\%. (20ns window)
%%This gives the accidental loss of 17.9\%.
We evaluated the accidental loss by the detectors other than the beam-hole counters to be 17.9\% in total.
This is a conservative number, because
the two or more counters can have coincident hits
from the same $K_L$-decay, but 
these hits are counted separately in different counters.

%%%The $K_L$ decay rate with detector hits is 5.12~MHz,
%%%where the $K_L$-decay at $z$ from -1~m to 23~m mainly contributes.
%%%We can evaluate the accidental loss with a common 40-ns veto
%%%window to be 18.5\%, just for a reference.
%
%\begin{figure}[h]
% \centering
 % \subfloat[]{
 % \includegraphics[page=1,width=0.5\textwidth]{KLdocu/sensitivity/figure/accidentalKdecay.pdf}
 % }
 %\subfloat[]{
% \includegraphics[page=23,width=0.5\textwidth]{KLdocu/sensitivity/figure/accidentalKdecay.pdf}
 %}
% \caption{
 % (a) $K_L$ decay rate in each decay $z$ psotiion.
 % The hollow one shows the $K_L$ decay positions, and
 % the filled one shows those with hits at the detector 
 % other than Beam-Hole Counter.
% Hit rates for each detector.
% }\label{fig:KdecayRate}
%\end{figure}

\begin{figure}[h]
 \centering
 \includegraphics[page=40,width=0.5\textwidth]{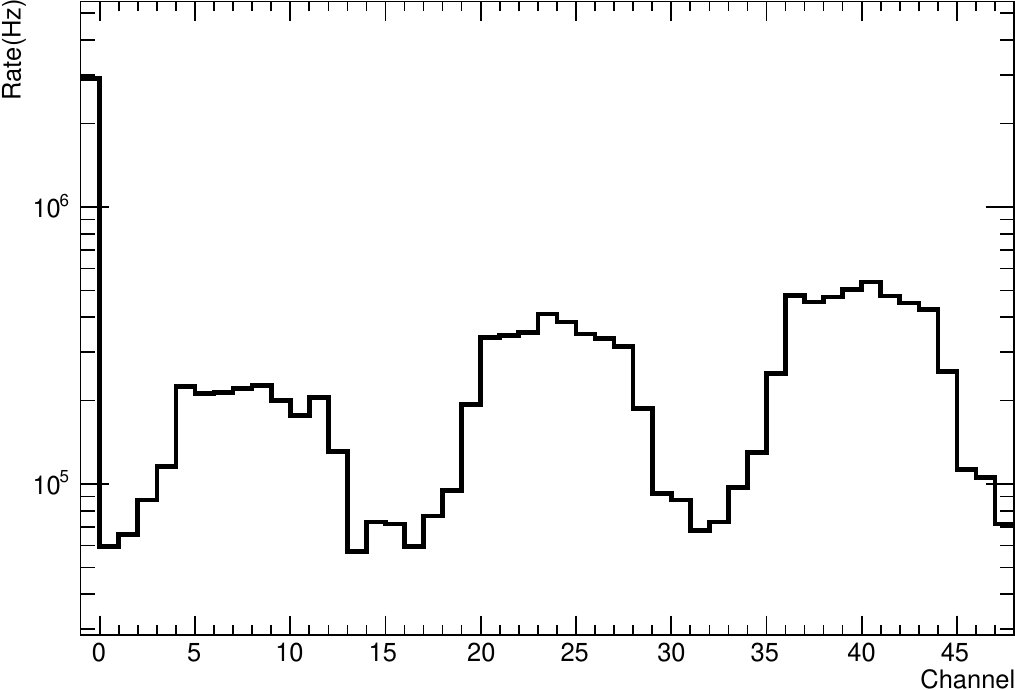}
 \caption{Expected hit rate of the beam-hole charged-particle veto counter at KOTO step-2
 with the 1/4 minimum-ionizing-particle peak threshold.
 It consists of three layers of a MWPC-type wire chamber with 16-channel readout for each layer.
 The 0-47 channels in the x-axis corresponds to all the readout channels.
 The channel -1 in the x-axis corresponds to the detector rate
 decided from two coincident hits out of three layers.}\label{fig:newBHCV}
\end{figure}

%\paragraph{Beam-Hole Counter}
\subparagraph{Beam-hole charged-veto counter}
\label{sec:rateBHCV}
We evaluated the hit rate of the beam-hole charged-veto counter
by using the current detector design in the KOTO step-1.
It consists of three layers of a MWPC-type wire chamber~\cite{Kamiji:2017deh}
with a small amount of material:
the thickness of the gas volume for each layer is 2.8 mm, and
the cathode plane is a 50-$\mu$m-thick graphite-coated polyimide film.
This design reduces the hit rate from neutral particles,
such as photons, neutrons, and $K_L$.
The layer-hit is
defined as
the energy deposit larger than
1/4 of the minimum-ionizing-particle peak.
The counter-hit is
defined as
two coincident layer-hits out of three layers,
which maintains the charged-particle efficiency
to be better than 99.5\%
with less contribution from neutral particles.
The width of the veto window is 30 ns
to cover the drift time of the ionized electrons in the chamber.
The particles in the beam simulated with the beam line simulation
were injected into the beam-hole charged-particle veto counter.
Figure~\ref{fig:newBHCV} shows the hit rate of each readout channel.
The counter-hit rate
with the two-out-of-three logic is 2.9~MHz
as shown at the channel -1 in the figure.
The accidental loss with this counter is 8.3\% with a 30-ns veto window.

\subparagraph{Beam-hole photon-veto counter}
We evaluated the hit rate of the beam-hole photon-veto counter
based on the current detector design in the KOTO step-1,
which consists of 16 modules of
lead-aerogel Cherenkov counters~\cite{Maeda:2014pga}.
A high-energy photon generates an $e^+e^-$ pair in the lead plate,
and these generate Cherenkov light in the aerogel radiator.
The Cherenkov light is guided by mirrors to a PMT.
In this report, we use 25 modules for KOTO step-2.
The detail of the lead or aerogel thickness for each module
is described in Sec.~\ref{sec:beamHolePhotonVetoCounter}.

\begin{figure}[h]
 \centering
 \includegraphics[page=28,width=0.5\textwidth]
 {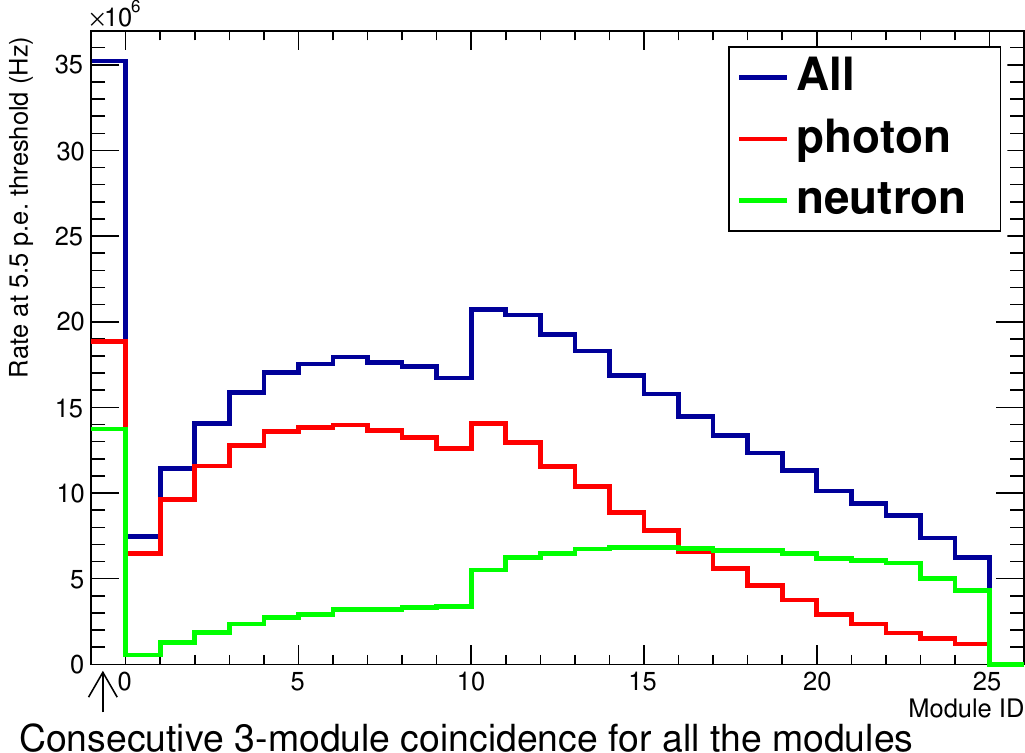}
 \caption{Expected hit rate of the beam-hole photon-veto counter
 at KOTO step-2.
 The rates for the channels from 0 to 24 in the x-axis show
 the module-hit rates with 5.5-p.e.-threshold.
 The rate at the channel -1 shows
 the counter-hit rate (consecutive three-module coincidence).
 }
 \label{fig:bhpvrate}
\end{figure}

We injected all the particles collected in the beam line simulation
to the beam-hole photon-veto counter with 25 modules.
A full-shower simulation and optical-photon tracking to the PMTs were performed,
and the observed number of photoelectrons was recorded.
The individual module-hit is defined with 
a 5.5-photoelectron (p.e.) threshold.
The counter-hit is defined with the 
consecutive three-module coincidence.
Those module-hit rates and the counter-hit rate
are shown in Fig.~\ref{fig:bhpvrate}.
Of the counter-rate of $35.2~\mathrm{MHz}$,
60\% comes from the beam-photon, and 30\% comes from the beam-neutron.
The accidental loss with this counter is 19\% with a 6-ns veto window.

%\paragraph{Conclusion on the accidental loss}
\subparagraph{Conclusion on the accidental loss}
The rate ($r_i$), the veto width $(w_i)$, and
the individual loss $(1-\exp\left(- w_i r_i\right))$
for each detector are summarized
in Table~\ref{tab:rateWidth}.

The total accidental loss is evaluated:
\begin{align*}
 \text{Accidental loss}=&1-\exp\left(-\sum_i w_i r_i\right)\\
 =&39\%.
\end{align*}
\begin{table}[h]
 \centering
 \caption{Summary of rate, veto width, and individual accidental loss.}\label{tab:rateWidth}
 \begin{tabular}{llll}
  Detector&Rate(MHz) & Veto width (ns) & Individual loss (\%)\\ \hline
  Front Barrel Counter&0.18 &20 & 0.4\\
  Upstream Collar Counter&0.80 &20 & 1.6\\
  Central Barrel Counter&2.21 &40 & 8.5\\
  Calorimeter&3.45 &20 & 6.7\\
  Downstream Collar Counter&0.97 &20 & 1.9\\\hline
  Beam-hole charged-veto&2.9 &30 & 8.3\\
  Beam-hole photon-veto&35.2 &6 & 19\\
  \hline
 \end{tabular}
\end{table}

%\subsubsection{Shower-leak loss}\label{sec:backsplash}
\subsubsubsection{Shower-leakage loss}\label{sec:backsplash}
When two photons from the $\klpionn$ decay are detected
in the calorimeter,
the shower can leak both the downstream and upstream of the calorimeter,
and make hits on the other counters such as the Central Barrel Counter.
Such hits will veto the signal if the hit timing is within the veto window.
We call this signal loss ``shower-leakage loss''.
In particular,
we call the loss caused by
shower leakage toward the upstream
``backsplash loss''.

\subparagraph{Downstream shower-leakage loss}
The downstream shower-leakage affects the shower-leakage loss in 
the Downstream Collar Counter and the Central Barrel Counter.

The Downstream Collar Counter is hit by a shower leakage passing through
the 50-cm long (27 radiation-length) CsI crystal of the calorimeter.
%The larger photon-energy compared to the KOTO step-1 causes it
%a higher occurrence.
This signal loss of 8\% is neglected, because
this effect can be mitigated with
hit-position information in the Downstream Collar Counter,
or an absorber upstream of the Downstream Collar Counter.
This is one of the requirements on the design of the Downstream Collar Counter.

For the Central Barrel Counter,
a shower-leakage at the outer edge of the calorimeter makes
hits in the barrel counter,
because the barrel counter covers the side of the calorimeter.
This makes 3.4\% loss of the signal; however we allow it because
the barrel coverage is also effective to reduce the background from $K_L\to2\pi^0$.

\subparagraph{Backsplash loss at the Charged Veto Counter}
We will discuss the backsplash loss (upstream shower-leak loss).
The Charged Veto Counter covers
the upstream side of the calorimeter, and would
suffer from the backsplash.
Because the Charged Veto Counter is located 30-cm upstream of the calorimeter,
the timing of the Charged Veto Counter
defined with respect to the calorimeter-timing
is different between the $K_L$-decay particles
and the backsplash particles by 2 ns at least;
the $K_L$ decay particles give hits at-least 1-ns earlier
than the calorimeter timing,
and the backsplash particles give hits at-least 1-ns later. 
We require the timing resolutions of
the Charged Veto Counter
and the calorimeter to resolve the two
in order not to contribute to the backsplash loss.
This is one of the requirements on the detector design.
For example,
a 300~ps resolution with the calorimeter and the Charged Veto Counter
would work.
Reducing the material of the Charged Veto Counter will also
reduce the backsplash loss, because photons are dominant
(The fraction of photons in the backsplash particles is 95\%).

\begin{figure}[h]
 \centering
 \includegraphics[width=\textwidth]{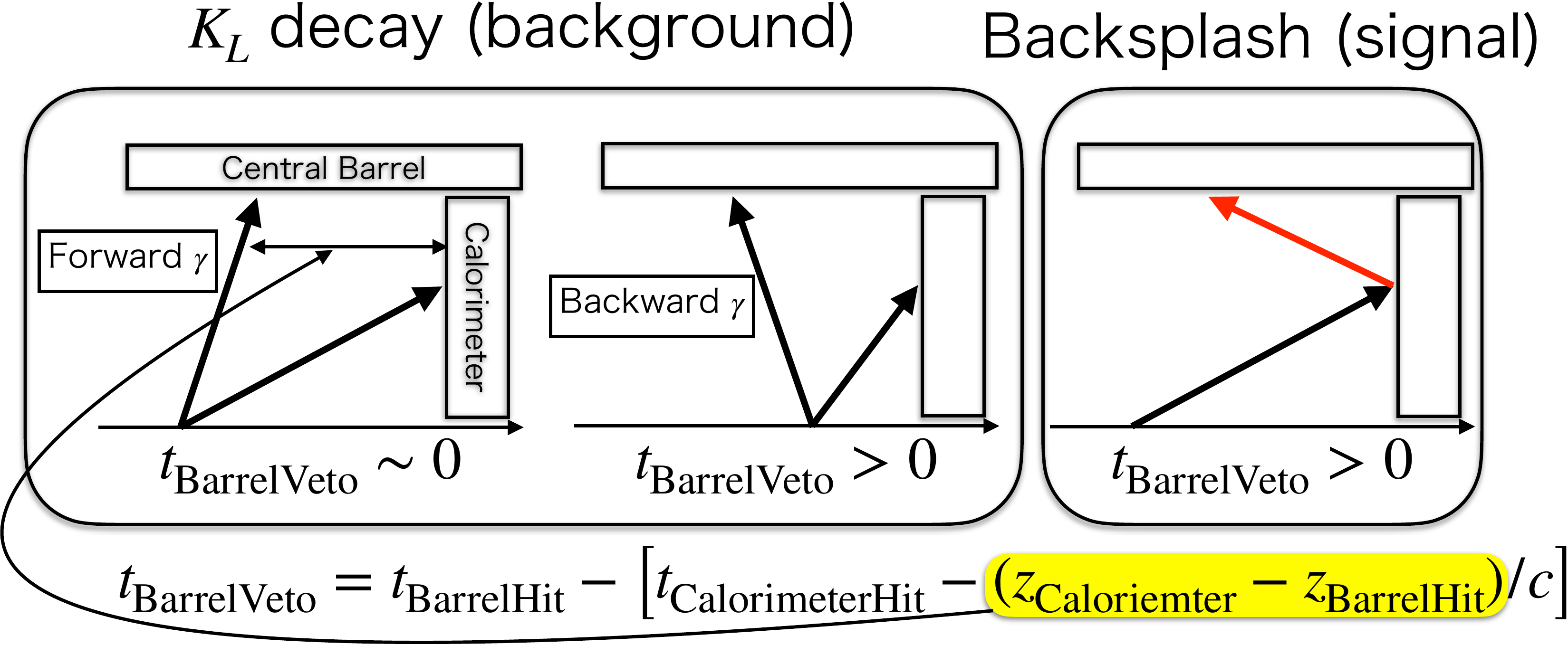}
 \caption{Configurations of barrel hits.}\label{fig:barrelTiming}
\end{figure}

\subparagraph{Backsplash loss and  barrel-timing definition}
The large coverage of the Central Barrel Counter gives significant effect
on the backsplash loss. First, we introduce a timing definition
of the Central Barrel Counter ($t_{\mathrm{BarrelVeto}}$):
\begin{align*}
 t_{\mathrm{BarrelVeto}}=&
 t_{\mathrm{BarrelHit}}-
 \left[t_{\mathrm{CalorimeterHit}}
 -\frac{z_{\mathrm{Calorimeter}}-z_{\mathrm{BarrelHit}}}{c}
 \right].
\end{align*}
The concept of this definition is illustrated in Fig.~\ref{fig:barrelTiming}.
The calorimeter hit timing
($t_{\mathrm{CalorimeterHit}}$)
is corrected with
the expected time of flight
from the barrel hit $z$ position ($z_{\mathrm{BarrelHit}}$) and
the calorimeter $z$ position ($z_{\mathrm{Calorimeter}}$).
$t_{\mathrm{BarrelVeto}}$ is the relative timing of
the actual barrel hit timing ($t_{\mathrm{BarrelHit}}$)
from the corrected calorimeter timing.
This reduces the barrel timing fluctuation in $K_L$ decays,
and the  veto window can be shortened.
The backward-going particle in the $K_L$
shown
in the middle of Fig.~\ref{fig:barrelTiming}
makes larger  $t_{\mathrm{BarrelVeto}}$, which requires the veto window
from $-5$~ns to $35$~ns.
The backsplash particles give also larger
$t_{\mathrm{BarrelVeto}}$, and the 40-ns veto-timing
requirement reduces the
backsplash loss for $t_{\mathrm{BarrelVeto}}>35~\mathrm{ns}$.

\begin{figure}[h]
 \centering
 \subfloat[]{
 \includegraphics[page=2,width=0.33\textwidth]{KLdocu/sensitivity/figure/anaBSL.pdf}
 }
 \subfloat[]{
 \includegraphics[page=1,width=0.33\textwidth]{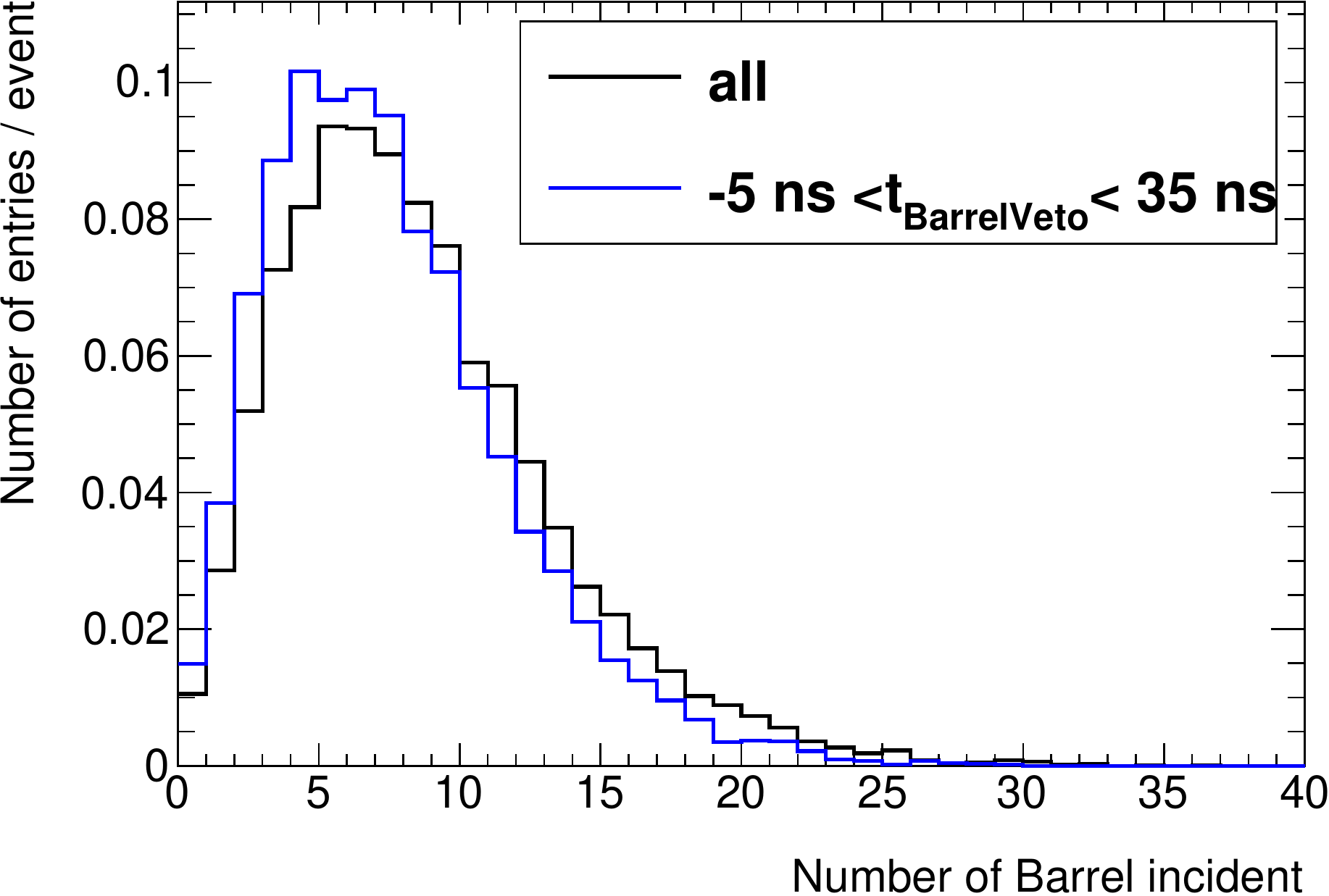}
 }
 \subfloat[]{
 \includegraphics[page=4,width=0.33\textwidth]{KLdocu/sensitivity/figure/anaBSL.pdf}
 }
 \caption{
 Distributions of 
 the incident timing of particles on the barrel from the shower-leakage (a),
 the multiplicity of the barrel-incident particles (b),
 and the maximum energy of the barrel-incident photons in a event
 within the veto window (c).
 The left-right arrow in (a) shows the veto window.
 }\label{fig:BShit}
\end{figure}

\subparagraph{Characteristics of backsplash}
The barrel incident timing of the shower-leakage particles is
shown in Fig.~\ref{fig:BShit}(a).
The incident-particle  multiplicity is shown in Fig.~\ref{fig:BShit}(b).
The mean multiplicity is 8.0 without any timing requirement.
This large multiplicity
makes the reduction of the backsplash loss difficult,
because only a single-particle detection from these multiple particles
will kill the signal.
The timing requirement of the veto window reduces the mean multiplicity
to be 7.1.

Most of the incident particles are low energy photons,
and the fraction of electrons or positrons is 4.8\%.
The distribution of the maximum photon energy 
in a event
within the veto window
is shown in Fig.~\ref{fig:BShit}(c).
The probability of containing one or more photons with
the energy larger than 1~MeV (3~MeV) is still 84\% (35\%).

Events survive after the veto
if all the backsplash particles are not detected.
The probability not to detect a particle is inefficiency
($=1-\text{efficiency}$), therefore
the event survival probability is obtained
by multiplying all the inefficiencies of the
incident particles.
The inefficiency for the 3-MeV photon is roughly 60\%, for example.
These contribute to the backsplash loss.
At this stage, the survival probability of a event
after vetoing the shower-leak particles is 57\%,
which will be improved in the next paragraph.

\begin{figure}[h]
 \centering
 \includegraphics[page=2,width=0.5\textwidth]{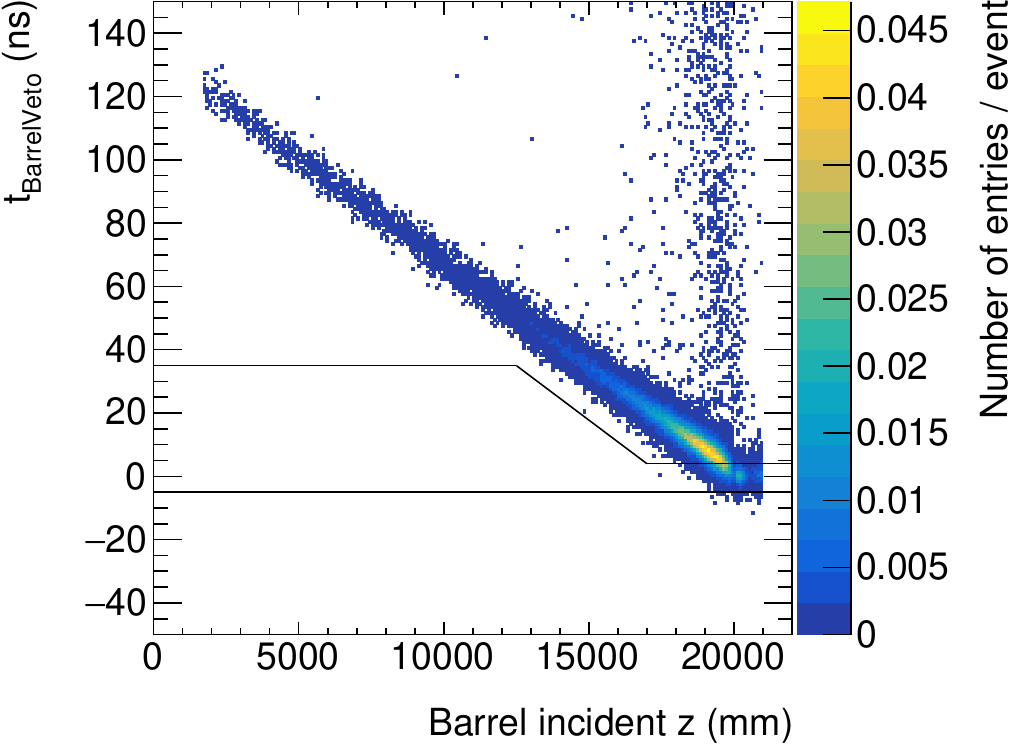}
 \caption{Incident timing of the backsplash particles
 on the Central Barrel Counter
 as a function of the incident $z$ position.
 The smearing with the timing resolution is applied
 depending on the incident energy.
 }\label{fig:BStiming}
\end{figure}

\subparagraph{Veto window depending on hit-position}
Figure~\ref{fig:BStiming} shows the incident timing of the
backsplash particles on the Central Barrel Counter
as a function of the incident $z$ position.
The timing-smearing is applied with the timing resolution
depending on the incident energy
as shown in Sec.\ref{sec:barrelTreso}.
Events with larger $t_{\mathrm{BarrelVeto}}$ at $z\sim 20000~\mathrm{mm}$
are generated with neutrons from the electromagnetic shower.
Events with smaller $z$ tend to have larger $t_{\mathrm{BarrelVeto}}$
due to longer flight distance.
This clear correlation can be used to exclude the backsplash particles
from the veto to reduce the backsplash loss.
We loosen the veto criteria
\footnote{
The loose veto criteria could reduce also the accidental loss
described in Sec.~\ref{sec:accidentalLoss}.
In this report, we did not change the accidental loss
obtained in Sec.~\ref{sec:accidentalLoss},
which can be improved with this loose veto.}
at the downstream region:
for $z$ from 12.5~m to 17~m, the largest timing of the veto window
is changed from 35~ns to 4~ns linearly, and 
for $z>17~\mathrm{m}$, it is 4~ns.
The region within the two lines in the figure shows the 
new veto region, which gives the survival probability
of $91\%$, equivalently the backsplash loss of 9\%.

In addition, if we added $z$
smearing using a Gaussian with the $\sigma$ of 500~mm,
the same survival probability is obtained.

\begin{figure}[h]
 \centering
 \includegraphics[page=8,width=0.5\textwidth]{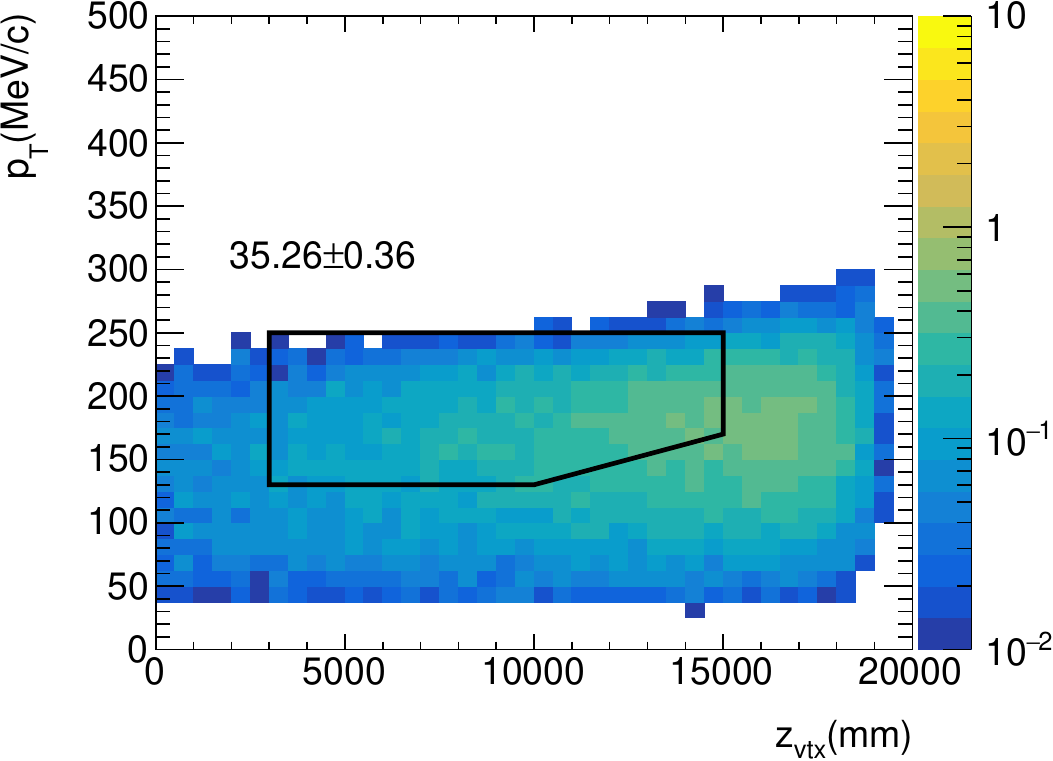}
 \caption{Distribution in the $\zvtx$-$\pt$ plane for $\klpionn$
 for the running time of $3\times 10^7$~s.
 All the cuts other than $\pt$ and $\zvtx$ cuts are applied.
 }
 \label{fig:ptz_pi0nn}
\end{figure}

%\subsubsection{Signal yield includging the signal losses}
\subsubsubsection{Signal yield including the signal losses}
We evaluate the number of the signal ($S$) in $3\times 10^7$ s running time
with $\mathrm{BR}_{\mathrm{\klpionn}}=3\times 10^{-11}$:
%a branching fraction of $\klpionn$ $3\times 10^{-11}$:
\begin{align*}
 S=&
 \frac{(\text{beam power})\times (\text{running time})}
 {(\text{beam energy})}
 \times (\text{number of }K_L\text{/POT})\\
 &\times P_{\mathrm{decay}}
 \times A_{\mathrm{geom}}
 \times A_{\mathrm{cut}}
 \times (\text{1-accidental loss})
 \times (\text{1-backsplash loss}) \times \mathcal{B}_{\klpionn}\\
 =&
 \frac{
 (100~\mathrm{kW})\times (3\times 10^7~\mathrm{s})
 }
 {
 (30~\mathrm{GeV})
 }
 \times
 \frac{(1.1\times 10^7 K_L)}
 {(2\times 10^{13}~\mathrm{POT})}\\
 &\times 9.9\% \times 24\% \times 26\% \times (1-39\%) \times 91\%
 \times (3\times 10^{-11})\\
 =&35.
\end{align*}
Here, $P_{\mathrm{decay}}$ is the decay probability,
$A_{\mathrm{geom}}$ is the geometrical acceptance for the two photons
entering the calorimeter,
$A_{\mathrm{cut}}$ is the cut acceptance, and
$\mathcal{B}_{\klpionn}$ is the branching fraction of $\klpionn$.
Distribution in the $\zvtx$-$\pt$ plane is shown in Fig.~\ref{fig:ptz_pi0nn}.

\subsubsection{Background estimation}
\begin{figure}[h]
 \centering
 \includegraphics[page=12,width=0.5\textwidth]{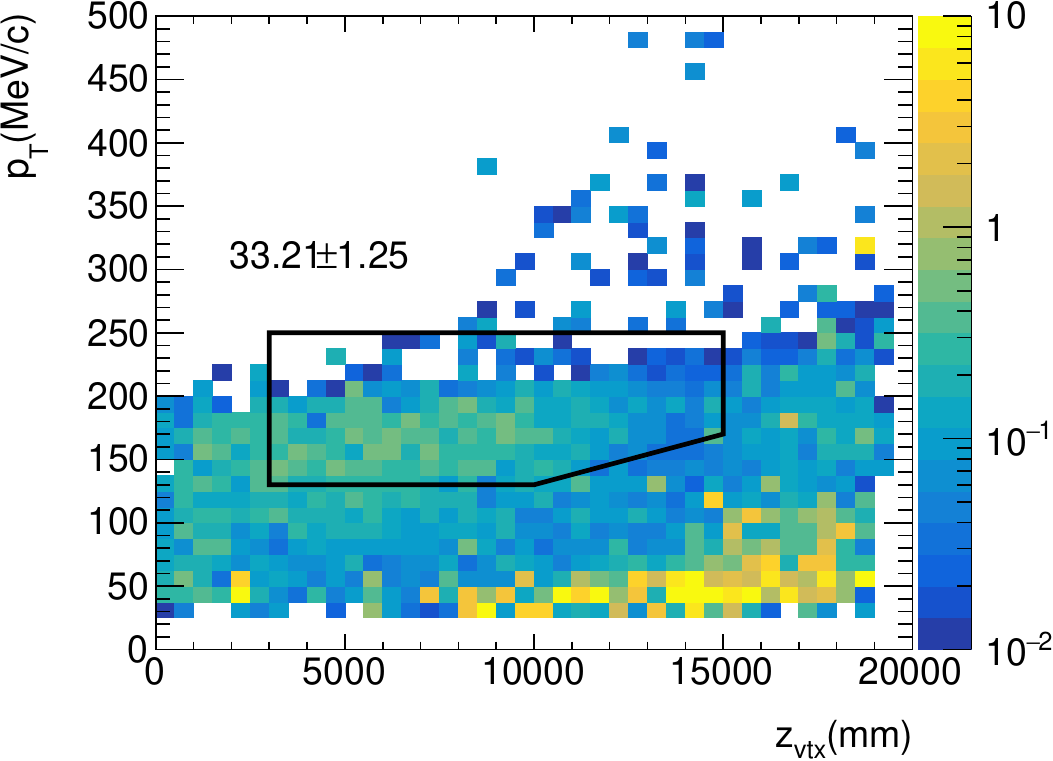}
 \caption{Distribution in the $\zvtx$-$\pt$ plane for $\klpiopio$
 background events
 for the running time of $3\times 10^7$~s.
 All the cuts other than $\pt$ and $\zvtx$ cuts are applied.
 }
 \label{fig:ptz_pi0pi0}
\end{figure}

\subsubsubsection{$\klpiopio$}
$\klpiopio$ becomes a background
when two clusters are formed at the calorimeter and
the other photons are missed in the following cases.
\begin{enumerate}
 \item Fusion background: Three photons enter the calorimeter,
       and two of them are fused into one cluster.
       The other one photon is missed due to the detector inefficiency.
 \item Even-pairing background: Two photons from a $\pi^0$-decay
       form two clusters in the calorimeter. Two photons from
       the other $\pi^0$ are missed due to the detector inefficiency.
 \item Odd-pairing background : One photon from a $\pi^0$ and
       one photon from the other $\pi^0$ form two clusters in the calorimeter.
       The other two photons are missed due to the detector inefficiency.
\end{enumerate}
The number of this background is evaluated to be 33.2,
in which 2 event come from the fusion, 27 events from the even-pairing,
4 events from the odd-pairing background.
Fig.~\ref{fig:ptz_pi0pi0} shows the distribution 
in the $\zvtx$-$\pt$ plane with all the cuts
other than the $\zvtx$ and $\pt$ selections.

Among the 33.2 background events,
both two photons are missed in the Central Barrel Counter
for 29 events, which gives the dominant contribution.

\begin{figure}[h]
 \centering
 \subfloat[]{
 \includegraphics[page=3,width=0.45\textwidth]{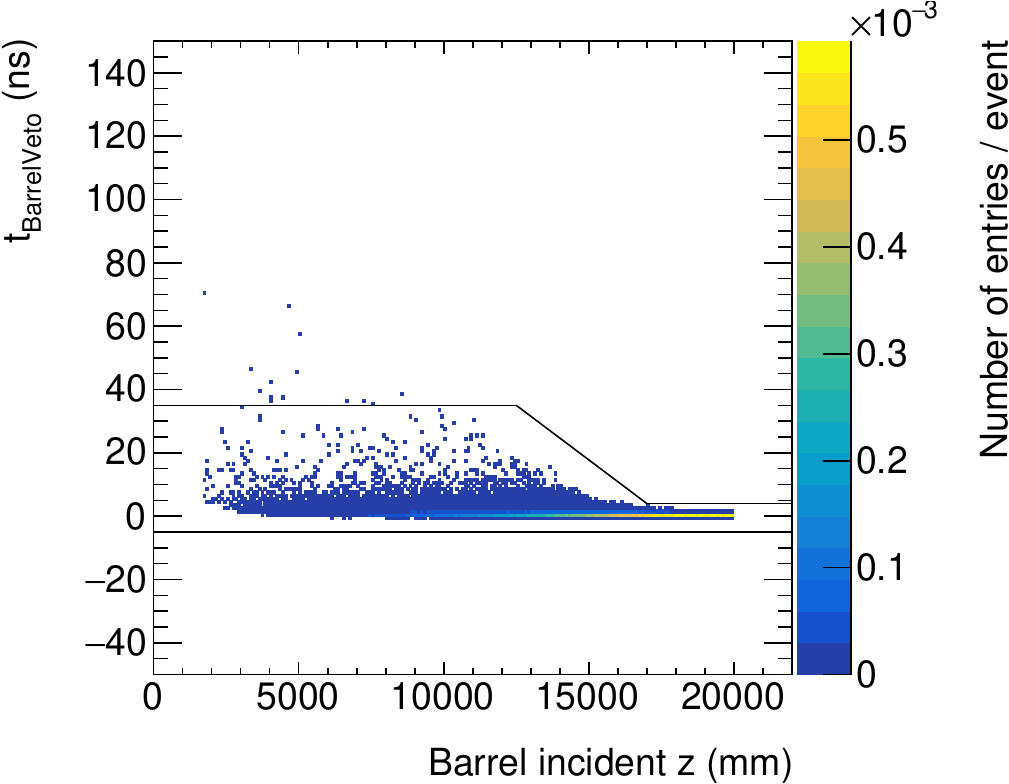}
 }
 \hspace{0.3cm}
 \subfloat[]{
 \includegraphics[page=6,width=0.45\textwidth]{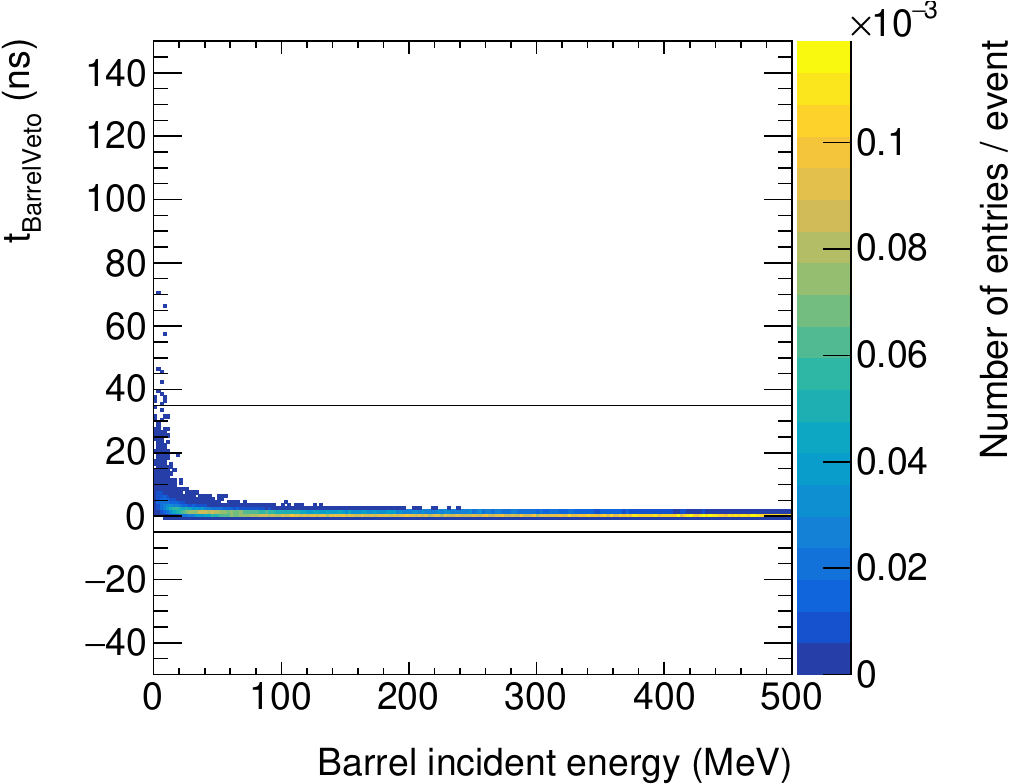}
 }
 \caption{Correlation between
 $t_{\mathrm{BarrelVeto}}$ and the barrel incident $z$ position (a), and
 correlation between
 $t_{\mathrm{BarrelVeto}}$ and the barrel incident energy (b)
 for $\klpiopio$.
 Both are after imposing all the cuts, where 
 the timing-smearing is applied with the timing resolution
 depending on the incident energy.
 }
 \label{fig:timing2pi0}
\end{figure}

For the barrel veto, the timing definition of $t_{\mathrm{BarrelVeto}}$,
and the veto window depending on the barrel hit position 
are introduced in Sec.~\ref{sec:backsplash}.
Fig.~\ref{fig:timing2pi0} shows the correlation between
$t_{\mathrm{BarrelVeto}}$ and the barrel incident $z$ position
after imposing all the cuts.
The timing-smearing is applied with the timing resolution
depending on the incident energy
as described in Sec.\ref{sec:barrelTreso}.

For the barrel incident $z$ position downstream of 15~m,
the photon direction tends to be forward
because the signal region is upstream of 15~m.
Therefore, the timing fluctuation is small, and the timing resolution
is better due to the higher energy photons.
The number of background events does not change by introducing
the veto window that depends on the barrel hit position.

For the barrel incident position upstream of 15~m,
the timing fluctuation is larger, and 
events with $t_{\mathrm{BarrelVeto}}>35~\mathrm{ns}$ exist.
For the events with larger $t_{\mathrm{BarrelVeto}}$,
the energy is small 
as shown in Fig.~\ref{fig:timing2pi0} (b),
because of the backward photon.
The low energy photons are originally less detectable,
and therefore missing them outside the veto window
has less impact on the increase of the $\klpiopio$ background.
Actually, 
the number of events
with $t_{\mathrm{BarrelVeto}}>35~\mathrm{ns}$
is small (0.1 events).

If we use the selection $t_{\mathrm{BarrelVeto}}<15~\mathrm{ns}$
instead of the base design $t_{\mathrm{BarrelVeto}}<35~\mathrm{ns}$,
the number of background events increases by 2.7 events.

If we added $z$
smearing using a Gaussian with the $\sigma$ of 500~mm
on the distribution in Fig.~\ref{fig:timing2pi0},
the same number of background event is obtained.
%There is a room to optimize the veto condition.

%not timing cut
%../g2ana/step2geom_step2mome15cm_pi0pi0_1MeV_6.root     1.78981 0.298835        4.31395 0.527486        26.9675 1.08731 33.0713 1.2449
%35ns
%../g2ana/step2geom_step2mome15cm_pi0pi0_1MeV_6.root     1.78981 0.298835        4.31417 0.527486        27.1063 1.09406 33.2103 1.2508
%15ns
%../g2ana/step2geom_step2mome15cm_pi0pi0_1MeV_6.root     1.78981 0.298835        4.32917 0.527574        29.8569 1.27791 35.9759 1.41446 

\begin{figure}[h]
 \centering
 \includegraphics[page=20,width=0.5\textwidth]{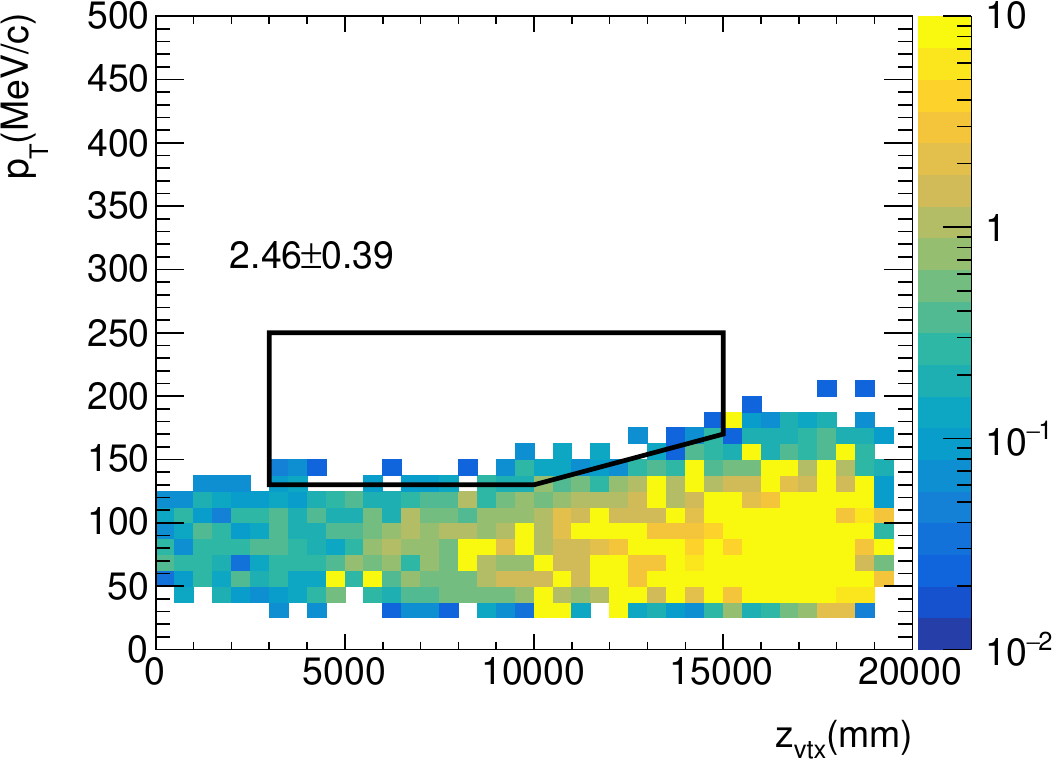}
 \caption{Distribution in the $\zvtx$-$\pt$ plane for $\klppm$
 background
 for the running time of $3\times 10^7$~s.
 All the cuts other than $\pt$ and $\zvtx$ cuts are applied.}
 \label{fig:ptz_pipipi0}
\end{figure}

\subsubsubsection{$\klppm$}
$\klppm$ becomes a background when $\pi^+$ and $\pi^-$ are not detected.
The number of this background events
is evaluated to be 2.5
as shown in Fig.~\ref{fig:ptz_pipipi0},
where one charged pion is lost in the Charged Veto Counter,
and the other is lost in the Beam Hole Counter.
The maximum $\pt$ of the reconstructed $\pi^0$ is
limited with the kinematics and the $\pt$ resolution.
The tighter $\pt$ selection in the downstream
makes the pentagonal cut in the $\zvtx$-$\pt$ plane
as shown in the figure.
This cut reduces the background
because the $\pt$ resolution is worse in the downstream.

\begin{figure}[h]
 \centering
 \includegraphics[page=2,width=0.5\textwidth]{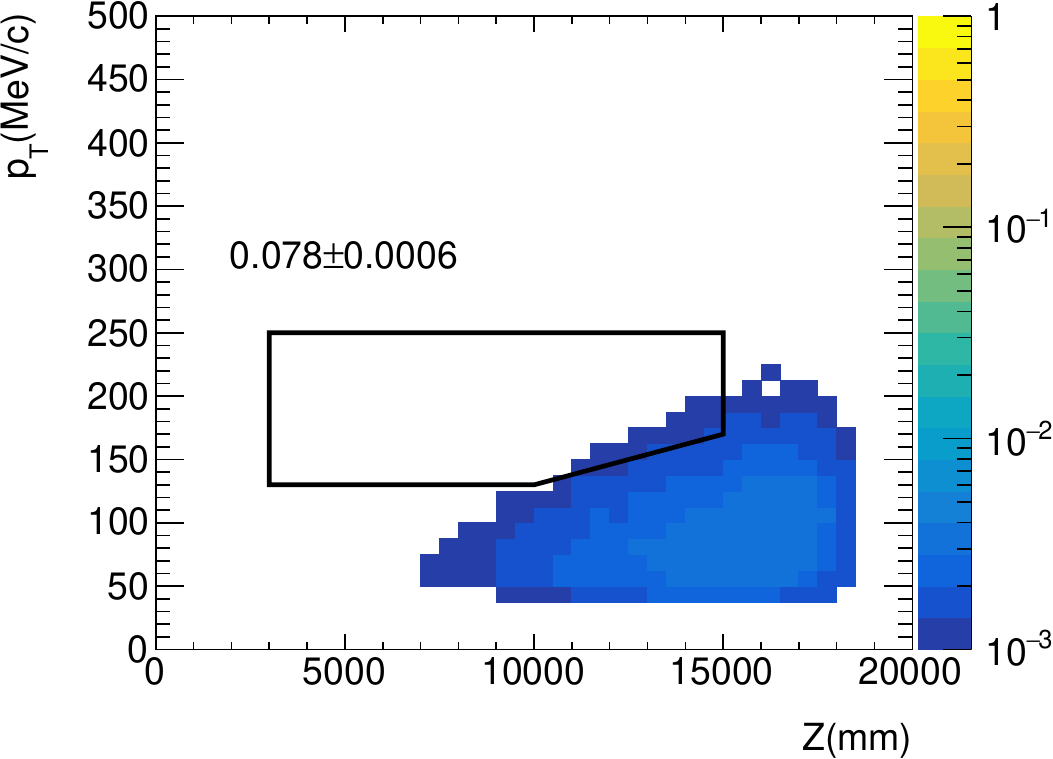}
 \caption{Distribution in the $\zvtx$-$\pt$ plane for Ke3
 background
 for the running time of $3\times 10^7$~s.
 All the cuts other than $\pt$ and $\zvtx$ cuts are applied.}
 \label{fig:ptz_ke3}
\end{figure}

\subsubsubsection{$\kpien$ (Ke3)}
The Ke3 background happens when the electron and the charged pion are not
identified with the Charged Veto Counter.
The number of background events is evaluated to be
0.08
with $10^{-12}$ reduction with the Charged Veto Counter 
as shown in Fig.~\ref{fig:ptz_ke3}.

\begin{figure}[h]
 \centering
 \includegraphics[page=2,width=0.5\textwidth]{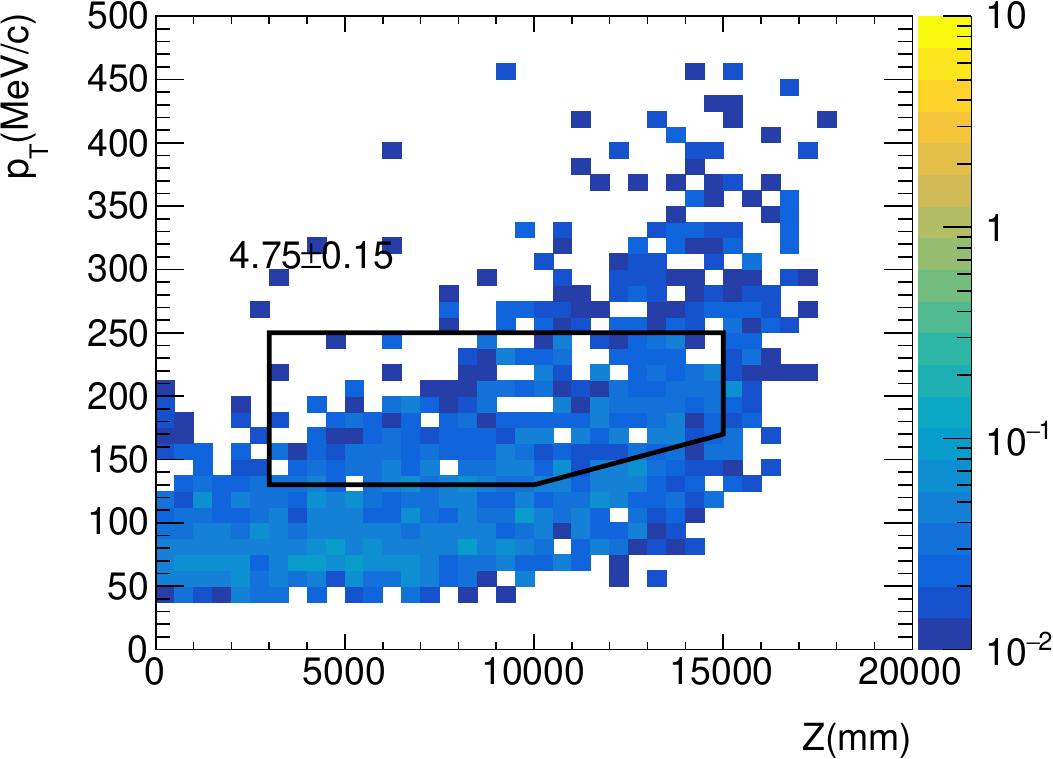}
 \caption{Distribution in the $\zvtx$-$\pt$ plane
 for halo $K_L\to2\gamma$ background
 for the running time of $3\times 10^7$~s.
 All the cuts other than $\pt$ and $\zvtx$ cuts are applied.}
 \label{fig:ptz_haloK}
\end{figure}

%\subsubsection{$\klgg$ for halo $K_L$} \label{sec:haloKL}
\subsubsubsection{$\klgg$ for halo $K_L$} \label{sec:haloKL}
$K_L$ in the beam scatters at the beam line components, and exists in the
beam halo region. When such a halo $K_L$ decays into two photons,
larger $\pt$ is possible
due to the assumption of the vertex on the $z$-axis.
The decay vertex is wrongly reconstructed
with the nominal pion mass assumption.
This fake vertex gives a wrong photon-incident angle.
Therefore, this halo $K_L\to2\gamma$ background can be reduced with
incident-angle information at the calorimeter.
We can reconstruct
another vertex with the nominal $K_L$ mass assumption,
which gives a correct photon-incident angle.
By comparing the observed cluster shape to 
those from the incorrect and correct photon-incident angles,
this background is reduced to be 10\% in the KOTO step-1, 
while keeping 90\% signal efficiency.
In this report, we assume that the background is reduced
to be 1\%,
because the higher energy photon in the step-2 will give
a better resolution in the photon-incident angle.
We will study it more in the future.
The number of this background is evaluated to be 4.8
as shown in Fig.~\ref{fig:ptz_haloK}.
For the halo $K_L$ generation in this simulation,
the core $K_L$ momentum spectrum and 
the halo neutron directions were used.
%The flux and spectrum of the halo $K_L$ are obtained
%from the beam line simulation.
Systematic uncertainties on the flux and spectrum
are also one of the future studies.

%\subsubsection{$K^\pm\to\pi^0 e^\pm\nu$}
\subsubsubsection{$K^\pm\to\pi^0 e^\pm\nu$}
$K^\pm$ is generated in the interaction of $K_L$,
neutron, or $\pi^\pm$
at the collimator in the beam line.
Here we assume that 
the second sweeping magnet
near the entrance of the detector
will reduce the contribution to be 10\%.

Higher momentum $K^\pm$ can survive
in the downstream of the second magnet,
and $K^\pm\to\pi^0 e^\pm\nu$
decay occurs in the detector.
This becomes a background if $e^\pm$ is undetected.
The kinematics of $\pi^0$ is similar to
$K_L\to\pi^0\nu\overline{\nu}$, therefore
this is one of the serious backgrounds.
Detection of $e^\pm$ is one of the keys to reduce the background.

\begin{figure}[h]
 \centering
 \includegraphics[page=2,width=0.5\textwidth]{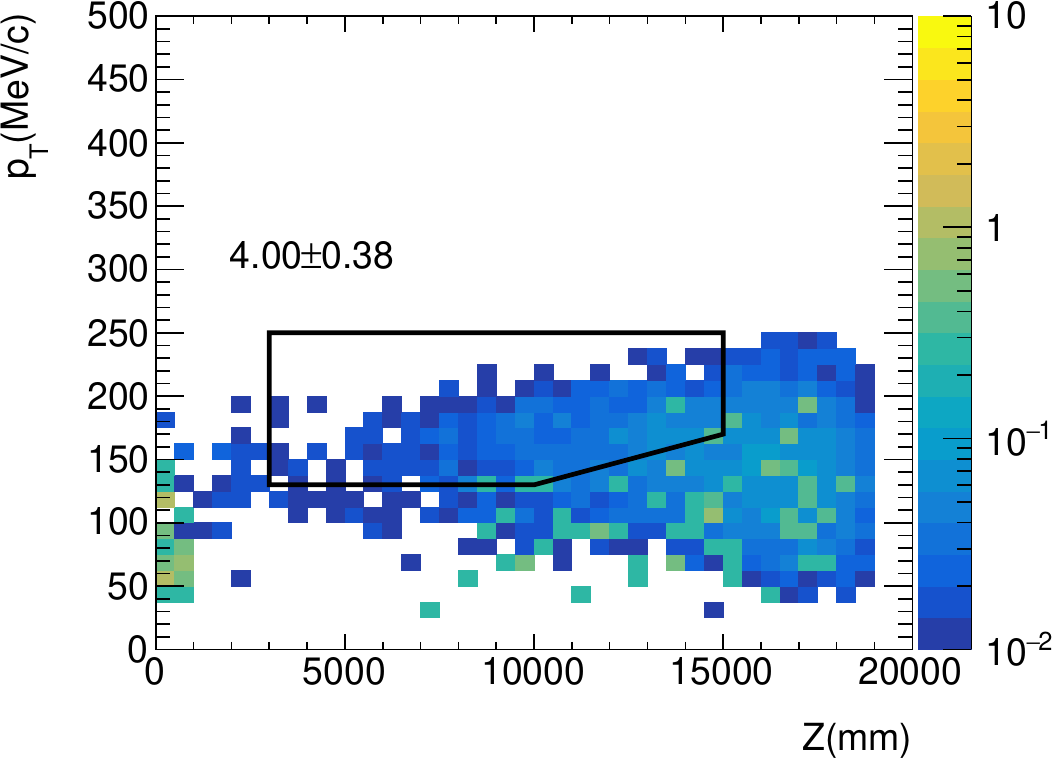}
 \caption{Distribution in the $\zvtx$-$\pt$ plane
 for $K^\pm\to\pi^0 e^\pm\nu$
 for the running time of $3\times 10^7$~s.
 All the cuts other than $\pt$ and $\zvtx$ cuts are applied.}
 \label{fig:ptz_Kplus}
\end{figure}

We evaluated the number of the background events to be 4.0 as shown
in Fig.~\ref{fig:ptz_Kplus}.
In the current beam line simulation, statistics is not large enough.
%The number of $K^\pm$ with the momentum direction toward the decay volume
%through the beam hole of the Upstream Collar Counter
%is 8. The second magnet reduces it to be 0.
We use the $K_L$ momentum spectrum and directions
%halo-neutron directions
for the $K^\pm$ generation.
We will evaluate these with more statistics with the beam line simulation in the future.

\begin{figure}[h]
 \centering
 \includegraphics[page=3,width=0.45\textwidth]{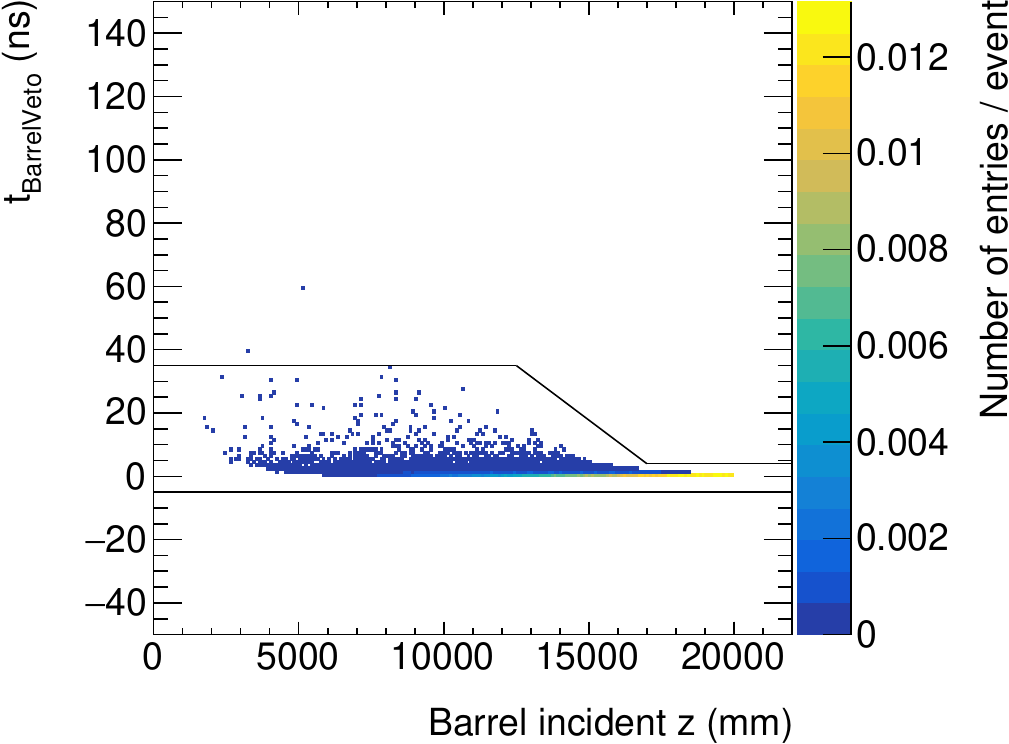}
 \caption{
 Correlation between
 $t_{\mathrm{BarrelVeto}}$ and the barrel incident $z$ position
 for $K^\pm\to\pi^0 e^\pm\nu$ background
 after imposing all the cuts.
 The timing-smearing is not applied.
 }
 \label{fig:timing_Kplus}
\end{figure}

%../g2ana/step2geom_step2mome15cm_Kplus_1MeV_100.root    0       0       0       0       3.48675 0.0839693       3.48675 0.0839693       
%../g2ana/step2geom_step2mome15cm_Kplus_1MeV_100.root    0       0       0       0       322.48  35.1124 322.48  35.1124
%../g2ana/step2geom_step2mome15cm_Kplus_1MeV_100.root    0       0       0       0       4.00163 0.37747 4.00163 0.37747 
The veto timing of the barrel detector is essential also for this decay.
Figure~\ref{fig:timing_Kplus} shows
the correlation between the barrel hit-$z$-position and
$t_{\mathrm{BarrelVeto}}$.
The lower momentum electrons or positrons contribute to the events with larger $t_{\mathrm{BarrelVeto}}$
due to the backward-going configuration similarly to $\klpiopio$.
Unlike the photon detection,
the detection efficiency is high because
a few-MeV electron is still a minimum-ionizing particle.
Therefore, the loss of the low-momentum particles outside the veto window
could give a large impact to increase the number of background events.
The 40-ns veto window from $-5$~ns to $35$~ns is adopted
because the number of events with 
$t_{\mathrm{BarrelVeto}}>35~\mathrm{ns}$ is small (0.5 events).
The number of background events increases to 322, for example,
if we set a 20-ns veto window 
($-5~\mathrm{ns}<t_{\mathrm{BarrelVeto}}<35~\mathrm{ns}$)
instead of the 40-ns veto window.

In this report, we use the same veto timing on this charged veto as in the photon veto.
There is a room to optimize the veto timings independently.

%\subsubsection{Hadron cluster}\label{sec:hadronCluster}
\subsubsubsection{Hadron cluster}\label{sec:hadronCluster}
A halo neutron hits the calorimeter to produce a first hadronic shower,
and another neutron in the shower travels inside the calorimeter,
and produces a second hadronic shower apart from the first one.
These two hadronic clusters mimic the signal.

We evaluated this background using halo neutrons prepared
with the beam line simulation,
and with a calorimeter composed of 50-cm-long CsI crystals.
A full-shower simulation was performed with those neutrons.

\begin{figure}[h]
 \centering
 \includegraphics[page=2,width=0.5\textwidth]{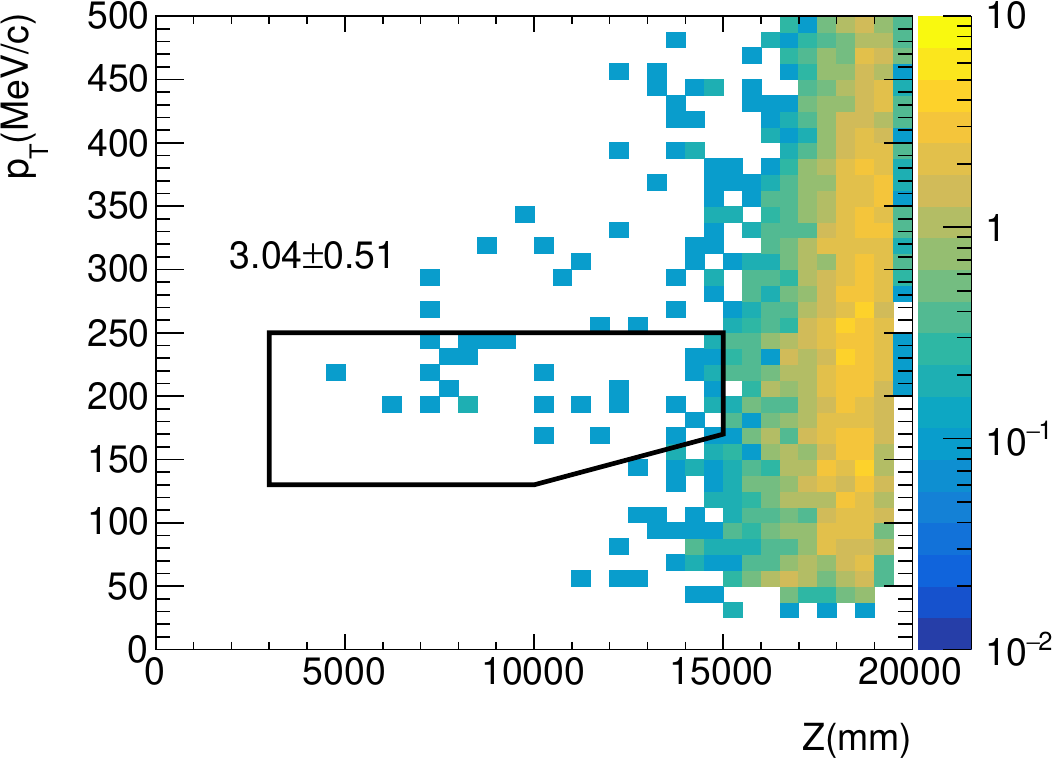}
 \caption{Distribution in the $\zvtx$-$\pt$ plane
 of the hadron cluster background
 for the running time of $3\times 10^7$~s.
 All the cuts other than $\pt$ and $\zvtx$ cuts are applied.}
 \label{fig:ptz_NN}
\end{figure}

We evaluated the number of background events to be 3.0
as shown in Fig.~\ref{fig:ptz_NN}.
Here we assume $10^{-7}$ reduction with
the cluster-shape information, pulse-shape information,
and shower depth information in the calorimeter.
In the KOTO step-1, we achieved to reduce the background to be
$\times ((2.5 \pm 0.01)\times 10^{-6})$
with 72\% signal efficiency
by using cluster and pulse shapes.
By using the shower depth information,
we also achieved to reduce it to be  $\times (2.1\times 10^{-2})$
with 90\% signal efficiency with small correlation to
the cluster and pulse shape cuts.
In total, $10^{7}$ reduction is feasible.
This reduction power is one of the requirements on the calorimeter design.

%\subsubsection{$\pi^0$ production at the Upstream Collar Counter}
\subsubsubsection{$\pi^0$ production at the Upstream Collar Counter}
If a halo neutron hits the Upstream Collar Counter, and produces a $\pi^0$,
which decays into two photons, it mimics the signal.

\begin{figure}[h]
 \centering
 \includegraphics[page=2,width=0.5\textwidth]{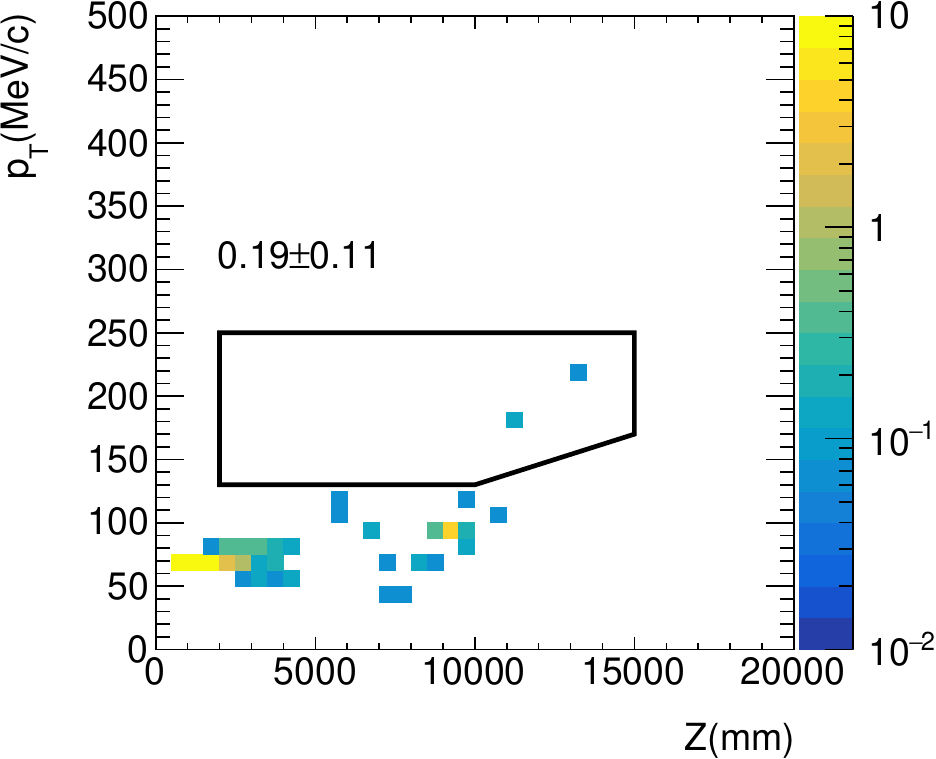}
 \caption{Distribution in the $\zvtx$-$\pt$ plane
 of the background from $\pi^0$ production at the
 Upstream Collar Counter
 for the running time of $3\times 10^7$~s.
 All the cuts other than $\pt$ and $\zvtx$ cuts are applied.}
 \label{fig:ptz_NCCpi0}
\end{figure}

Halo neutrons obtained from the beam line simulation are used
to simulate the $\pi^0$ production in the Upstream Collar Counter.
We assume fully-active CsI crystals for the detector.
Other particles produced in the $\pi^0$ production can deposit energy
in the detector, and can veto the event.
In the simulation, such events were discarded at first.
In the next step, only the $\pi^0$-decay was generated
in the Upstream Collar Counter.
Two photons from the $\pi^0$ can also interact 
the Upstream Collar Counter, and deposit energy in the counter.
Such events were also discarded.
The $\pi^0$ production near the downstream surface of the
Upstream Collar Counter mainly survives.
Finally when the two photons hit the calorimeter,
a full shower simulation was performed.
In this process, photon energy can be mis-measured due to
photo-nuclear interaction.
Accordingly the distribution of the events in the $\zvtx$-$\pt$ plane
was obtained as shown in Fig.~\ref{fig:ptz_NCCpi0}.
We evaluated the number of background events to be 0.19.

%\subsubsection{$\eta$ production at the Charged Veto Counter}
\subsubsubsection{$\eta$ production at the Charged Veto Counter}
A halo neutron hits the Charged Veto Counter, and produces a $\eta$.
The $\eta$ decays into two photons with the branching fraction of 39.4\%,
which can mimic the signal.
The decay vertex will be reconstructed at the upstream of the
Charged Veto Counter because the $\eta$ mass is
four times larger than the $\pi^0$ mass.

\begin{figure}[h]
 \centering
 \includegraphics[page=2,width=0.5\textwidth]{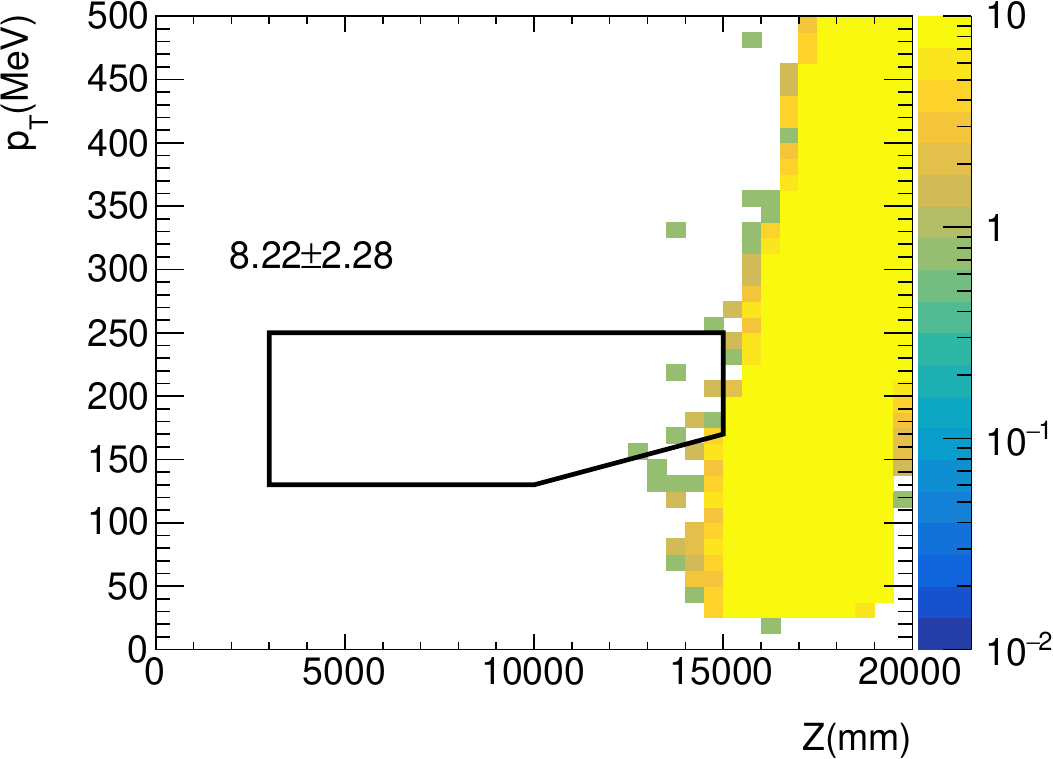}
 \caption{Distribution in the $\zvtx$-$\pt$ plane
 of the background from $\eta$ production at the Charged Veto Counter
 for the running time of $3\times 10^7$~s.
 All the cuts other than $\pt$ and $\zvtx$ cuts are applied.}
 \label{fig:ptz_CVeta}
\end{figure}

Halo neutrons obtained from the beam line simulation are used
to simulate the $\eta$ production in the Charged Veto Counter.
We assume a 3-mm-thick plastic scintillator at 30-cm upstream of
the calorimeter.
Other particles produced in the $\eta$ production can deposit energy in the
Charged Veto Counter, and can veto the event.
In the simulation, such events were discarded at first.
In the next step, only the $\eta$-decay was generated.
When two photons from the $\eta$ hit the calorimeter,
a full-shower simulation was performed.
Two clusters were formed, and 
the distribution of the events in the $\zvtx$-$\pt$ plane
was obtained as shown in Fig.~\ref{fig:ptz_CVeta}.
We evaluated the number of background events to be 8.2.

%\subsubsection{Summary of the background estimation}
\subsubsubsection{Summary of the background estimation}
A summary of  the background estimations is shown in Table~\ref{tab:bg}.
The total number of background events is
$56.0\pm 2.8$, where $33.2\pm 1.3$ comes from
the $\klpiopio$ decay.
\begin{table}[H]
 \centering
 \caption{Summary of background estimations.}\label{tab:bg}
 \begin{tabular}{lrr}\hline
  Background&Number&\\\hline
  $\klpiopio$&33.2 &$\pm$1.3 \\
  $\klppm$&2.5 &$\pm$0.4 \\
  $\kpien$& 0.08 & $\pm$0.0006\\
  halo $K_L\to 2\gamma$&4.8 &$\pm$0.2 \\
  $K^\pm\to\pi^0 e^\pm\nu$&4.0 &$\pm$0.4 \\
  hadron cluster&3.0 &$\pm$0.5 \\
  $\pi^0$ at upstream&0.2 &$\pm$0.1 \\
  $\eta$ at downstream&8.2 &$\pm$2.3 \\ \hline
  Total& 56.0 & $\pm$2.8 \\\hline
 \end{tabular}
\end{table}

%\clearpage
%\subsection{Sensitivity and the impact}
\subsubsection{Sensitivity and the impact}
We assume $3\times 10^7$ s running time with
100 kW beam on a 1-interaction-length T2 target,
where the $K_L$ flux is
$1.1\times 10^7$ per $2\times 10^{13}$ protons on target.

The sensitivity and the impacts are calculated and 
summarized in Table~\ref{tab:sensitivity}.
Here we assume
that 
the statistical uncertainties in the numbers of events
are dominant.

The single event sensitivity is evaluated to be 
$8.5\times 10^{-13}$.
The expected number of background events is 56.
With the SM branching fraction of $3\times 10^{-11}$,
35 signal events are expected with a signal-to-background ratio (S/B) of 0.63.
 \begin{itemize}
  \item $4.7$-$\sigma$ observation is expected
	for the signal branching fraction
	$\sim 3\times 10^{-11}$.
  \item It indicates new physics at the 90\% confidence level (C.L.) if
	the new physics gives
	$44\%$ deviation on the BR from the SM prediction.
  \item It corresponds to $14\%$ measurement of the CP-violating CKM parameter $\eta$ in the SM (The branching fraction is proportional to $\eta^2$). 
 \end{itemize}

\begin{table}[h]
 \centering
 \caption{Summary of the sensitivity and the impact.}
 \label{tab:sensitivity}
 \begin{threeparttable}
 \begin{tabular}{lcc}\hline
 & Formula & Value \\
 \hline
  Signal (branching fraction : $3\times 10^{-11}$) & $S$ & $35.3 \pm 0.4$\\
  Background & $B$ & $56.0 \pm 2.8$ \\
  \hline
  Single event sensitivity & $(3\times 10^{-11})/S$ & $8.5\times 10^{-13}$\\
  Signal-to-background ratio & $S/B$ & 0.63 \\
  Significance of the observation&  $S/\sqrt{B}$ & $4.7 \sigma$\\
  90\%-C.L. excess / deficit & $1.64\times \sqrt{S+B}$ & 16 events \\
  & $1.64\times \sqrt{S+B}/S $ & 44\% of SM\\
  Precision on branching fraction & $\sqrt{S+B}/S$ & 27\%\\
  Precision on CKM parameter $\eta$ &$0.5\times \sqrt{S+B}/S$ &14\%\\
  \hline
 \end{tabular}
 \begin{tablenotes}\footnotesize
 \item[*] Running time of $3\times 10^7$~s is assumed in the calculation.
 \end{tablenotes}
 \end{threeparttable}
\end{table}

% flatex input end: [KLdocu/sensitivity/sensitivity.tex]

%\section{Physics and Experiment at KL2 Beam Line}
\clearpage
% flatex input: [KLdocu/improvement/improvement.tex]
\subsection{Discussion on Sensitivity Improvement}
\label{chap:improvement}
In the previous section, we described the baseline evaluation of the sensitivity and the background estimation.
However, it is not final and we should continue considering improvements to achieve a better sensitivity.
Here we discuss possible examples to increase the signal acceptance.

\subsubsection{Extension of the signal region}
\begin{figure}[h]
\centering
\begin{minipage}{0.45\linewidth}
\includegraphics[width=\linewidth]{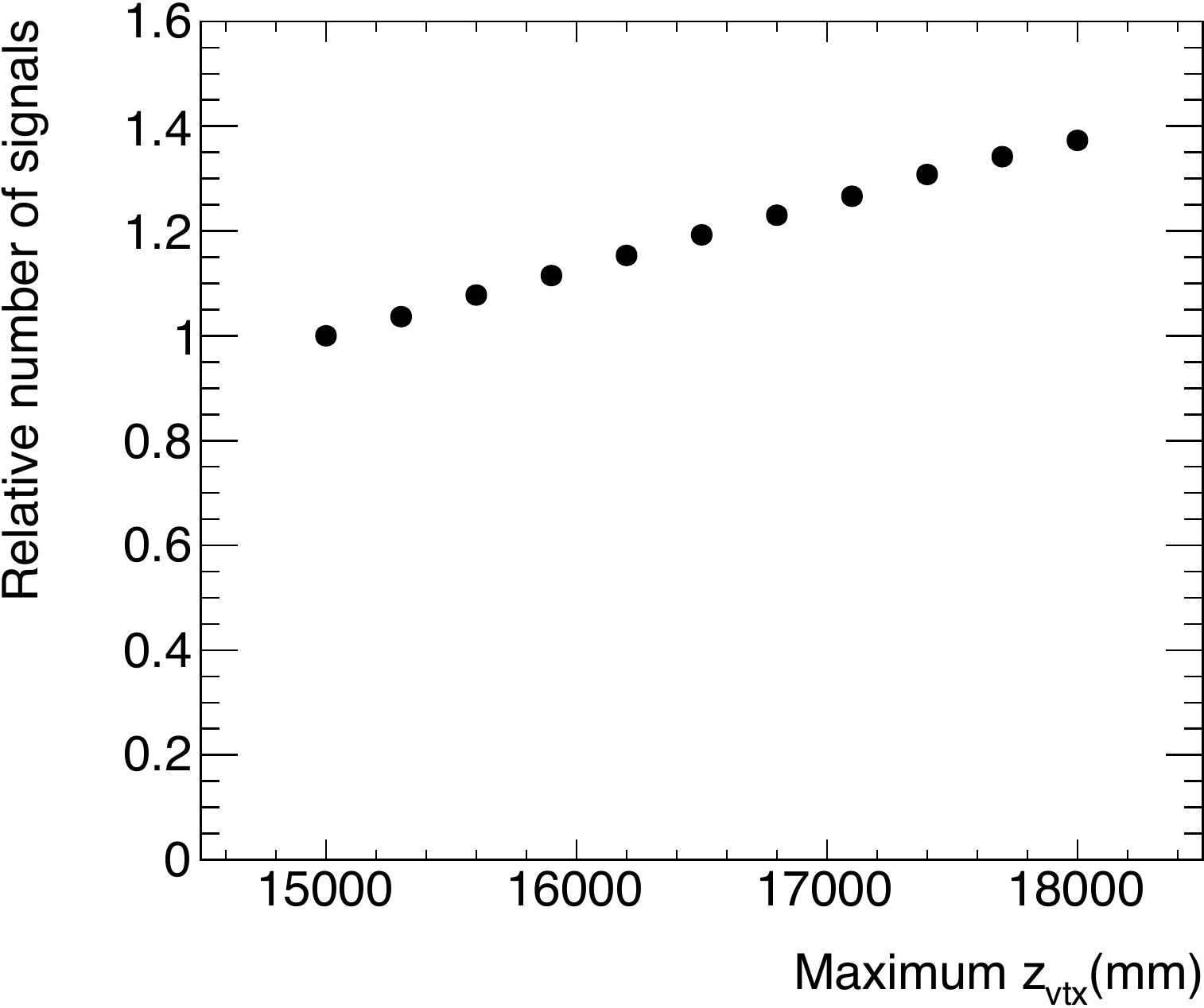}
\end{minipage}
\hspace{0.05\linewidth}
\begin{minipage}{0.45\linewidth}
\includegraphics[width=\linewidth]{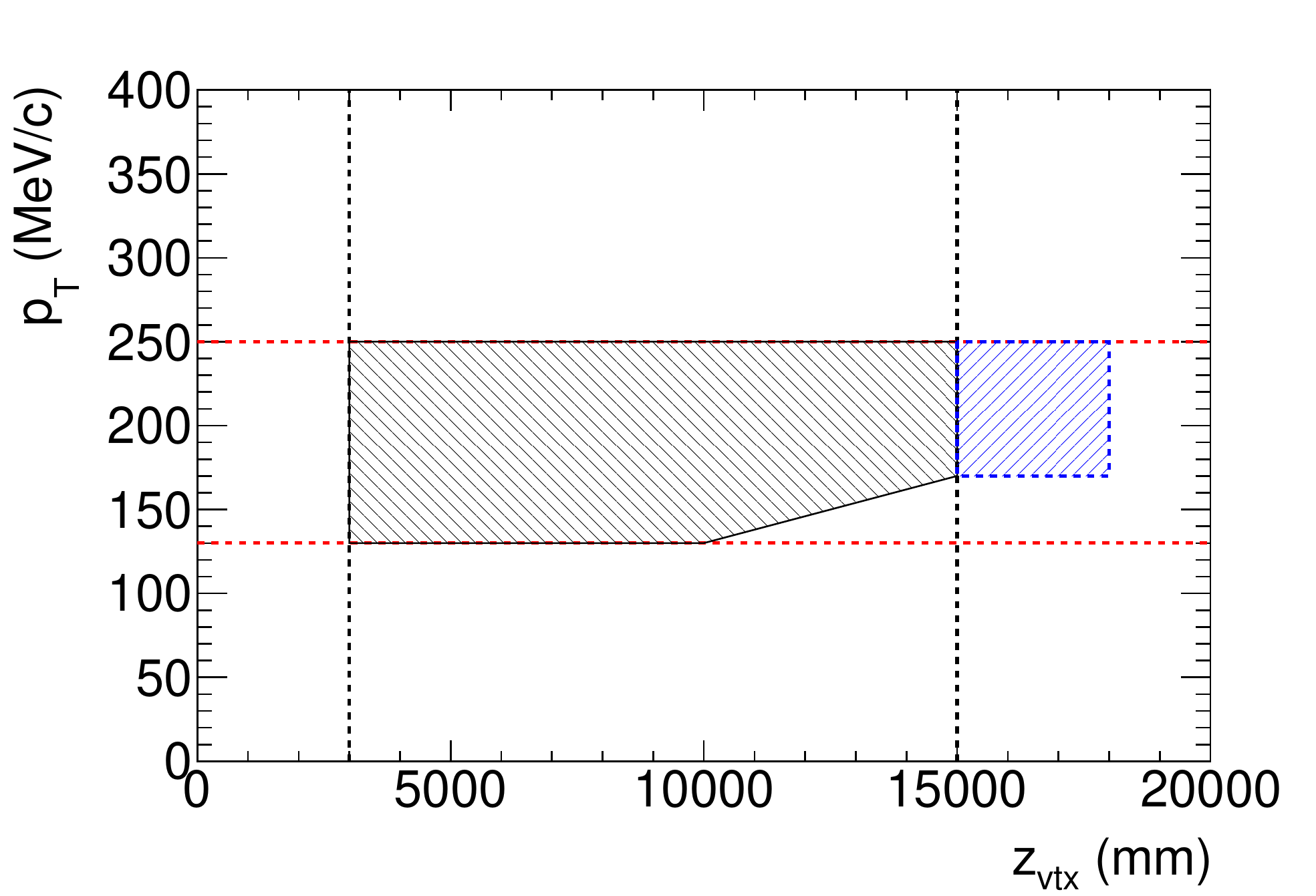}
\end{minipage}
\caption{\label{fig:ExtendSigR}(Left) Gain of the number of signals by extending the signal region.
The horizontal axis indicates the downstream end of the extended signal region.\\
(Right) Example of the extended signal region when the maximum $\zvtx$ is 18~m.}
\end{figure}

As can be seen in Fig.~\ref{fig:ptz_pi0nn}, the current definition of the signal region loses the signal in the downstream region ($\zvtx>15000$~mm), where the signal distribution is prominent.
%
%Table~\ref{tab:boxext} summarizes the possible increase of the number of signals as the signal region is extended.
%
% \begin{table}[h]
 %\centering
 %\caption{Expected number of the $\klpionn$ signals as a function of the downstream end of the signal region.}
 %\label{tab:boxext}
 %\begin{tabular}{lccc}\hline
%Downstream end of the signal region (Z; m) & 15 & 16 & 17\\
%Number of events & \memo{20} & \memo{25} & \memo{30}\\
%\hline
% \end{tabular}
%\end{table}
%
%
Figure~\ref{fig:ExtendSigR} shows the expected gain of the number of signals when the signal region is extended downstream.
Here, the $\pt$ range in the extended region was set to be $170 < \pt < 250$~MeV/c.
In the baseline sensitivity estimation, we avoid this region since large contaminations of halo neutron backgrounds (hadron cluster and $\eta$ production backgrounds) are expected, as shown in Figs.~\ref{fig:ptz_NN} and \ref{fig:ptz_CVeta}.
In order to realize the extension of the signal region, we must develop a method to reduce these backgrounds.
This is one of the motivations to develop the detector that can measure the photon incident angle to the calorimeter, described in the following section, which can remove the events with wrong vertex reconstruction.

% flatex input end: [KLdocu/improvement/improvement.tex]

%\section{Physics and Experiment at KL2 Beam Line}
\clearpage
% flatex input: [KLdocu/dfeasibility/dfeasibility.tex]
\subsection{Detector Feasibility Study}
% flatex input: [KLdocu/dfeasibility/photon-angle.tex]
%\subsection{Angle measurement of photon}
\subsubsection{Angle measurement of photon}
%In addition to the current reconstruction of $\pi^0$ by using energies and positions of two photons detected in a calorimeter, we can extract its decay position independently when we can measure incident angle of the photons. By requiring a consistency between two vertices reconstructed from different observables, we can select genuine single $\pi^0$ decay rejecting various kinds of wrong pair of photons such as odd-pairing of $K_L \rightarrow \pi^0 \pi^0$ decays, neutron induced two clusters, $\eta$ decays, and so on.
In addition to the current reconstruction of $\pi^0$ by using energies and positions of two photons detected in a calorimeter, we can extract its decay position independently when we can measure incident angle of the photons. By requiring a consistency between two vertices reconstructed from different observables, we can select genuine single $\pi^0$ decay and reject various kinds of wrong pair of photons such as odd-pairing $K_L \rightarrow \pi^0 \pi^0$ backgrounds and neutron induced two clusters. It also enables us to remove $\pi^0$ and $\eta$ decays produced by halo neutrons because they occur far from the beam axis. We can expect similar rejection against $K_L \rightarrow 2\gamma$ decays when the $K_L$ deviate from the beam center due to scattering. 

\begin{figure}[h]
\includegraphics[width=\textwidth]{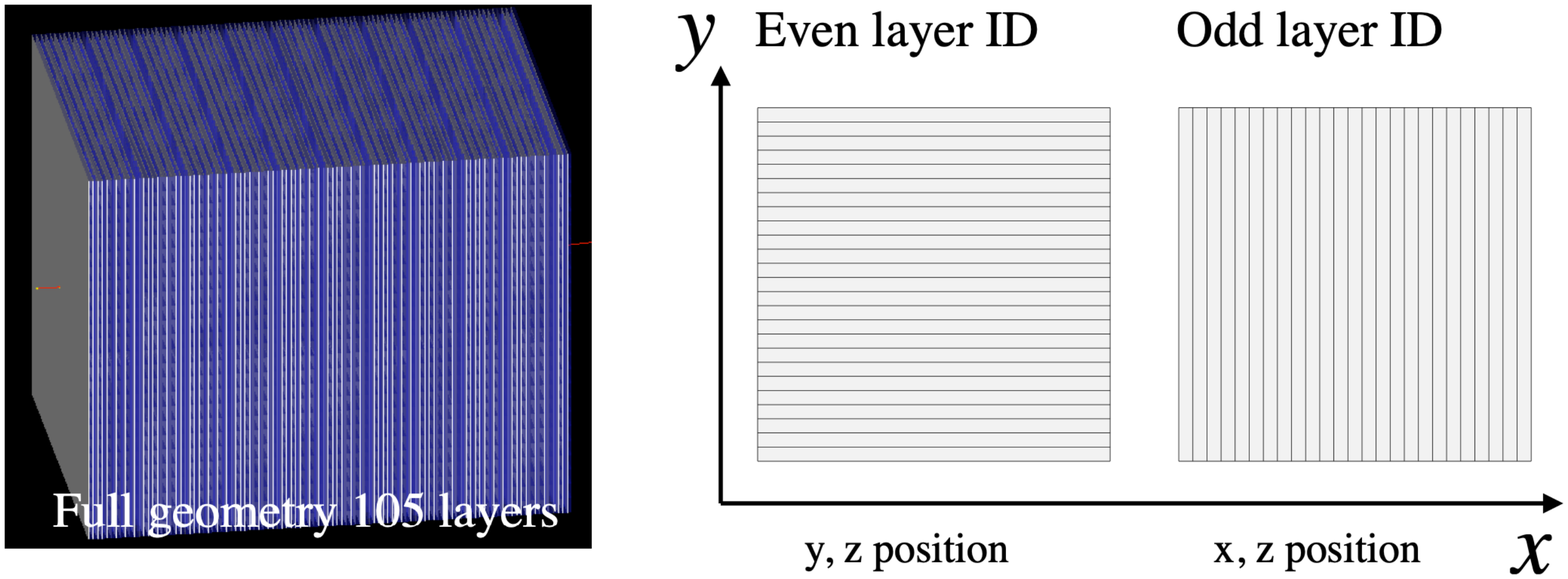}
 \caption{Simulation setup for studying on angle measurement. The detector is a sampling calorimeter consisting of alternating lead sheets and strips of plastic scintillator.}
 \label{fig:ang_conf}
\end{figure}

For the angle measurement of photon, we should get information about the spatial profile of shower particles generated in the electromagnetic calorimeter. This will be realized by recording energy deposits in a detector finely segmented in three dimensions. 
%Since the angular resolution of this measurement would suffer an intrinsic limitation due to stochastic process in the electromagnetic shower generation, we need to understand an achievable pesrformance of the measurement and  and an experimental goal. The number of read-out channels is also an important factor in the detector design.
We started the study with a setup of sampling calorimeter consisting of alternating lead sheets and strips of plastic scintillator as shown in Fig.~\ref{fig:ang_conf}. 
%The thickness of lead and scintillator is 1~mm and 5~mm, respectively. The lead sheet has a cross-section of  500 mm $\times$ 500 mm and the scintillator strip of 500 mm $\times$ 15 mm. 
The dimensions of a lead sheet and a plastic scintillator strip are $500 \times 500 \times 1$~mm$^3$ and $500 \times 15 \times 5$~mm$^3$, respectively.
By arranging the strips in x- and y-direction alternatively along z-direction, we can get shower profiles in x-z and y-z planes in turn.

\begin{figure}[h]
\begin{minipage}{0.5\linewidth}
\includegraphics[width=\linewidth]{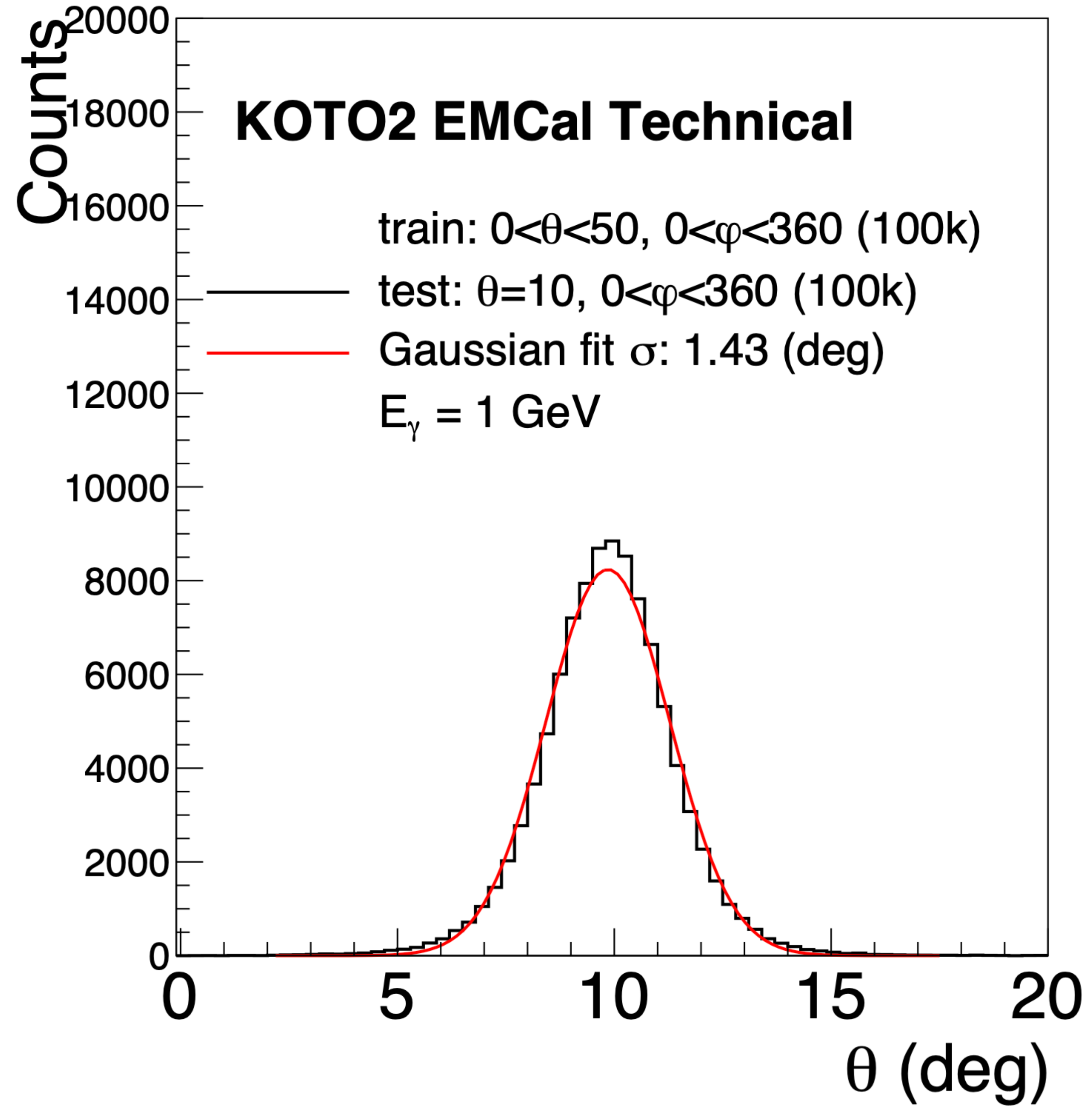}
%\caption{Reconstructed angle for 1 GeV photon entering with an angle of 10 degree.}
 %\label{fig:ang_rec}
\end{minipage}
%\hspace{0.1\linewidth}%
\begin{minipage}{0.5\linewidth}
\includegraphics[width=\linewidth]{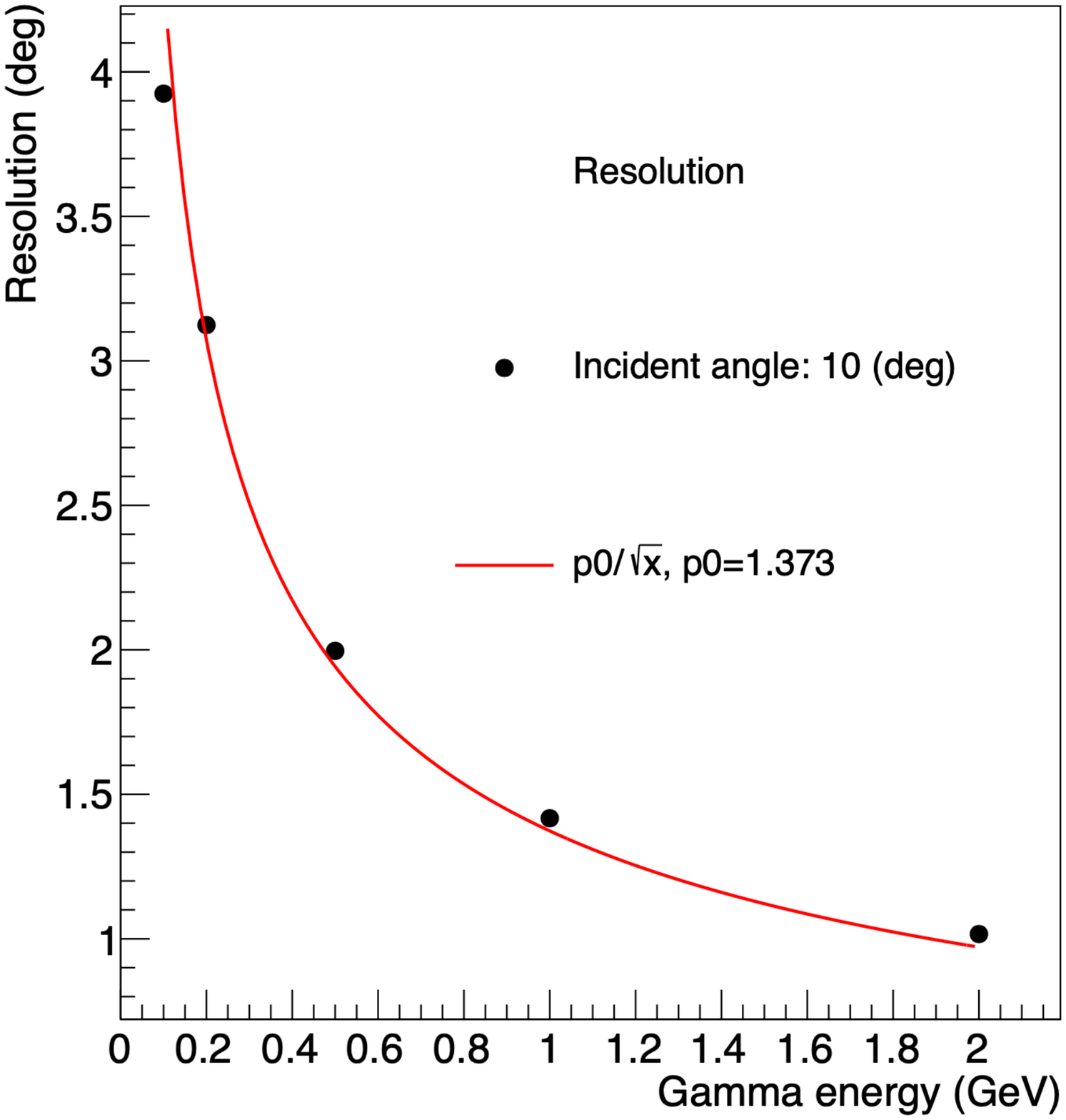}
\end{minipage} 

\caption{Left: Reconstructed angular distribution for 1~GeV photon entering with an incident angle of 10~degrees.  Right: Energy dependency of angular resolution.}
\label{fig:ang_resol}
\end{figure}

Figure~\ref{fig:ang_resol} (left panel)  shows a distribution of reconstructed angle for 1 GeV photon entering the calorimeter with an angle of 10 degrees. Showers were generated by using GEANT4 package and the angle was reconstructed by XGboost package for the machine learning. As a training sample, we used $10^5$ photons generated uniformly in the range of 0 to 50 degrees in polar angle and 0 to 2$\pi$ in azimuthal angle. We evaluated angular resolution 
for photons with an incident angle of 10~degrees to be in 
1.4~degrees based on $10^5$ testing sample.
The angular resolution depends on photon's energy
 %as expressed by $1/\sqrt{E_\gamma}$, 
in proportion to $1/\sqrt{E_\gamma}$, 
where $E_\gamma$ is energy of photon in GeV (right panel of Fig.~\ref{fig:ang_resol} ).

%\begin{figure}[h]
% \includegraphics[bb=14 362 1011 745,clip,width=\textwidth]{./figure/det_conf.pdf}
%\includegraphics[width=\textwidth]{./AngleMeas/figure/ang_rec.pdf}
 %\caption{Reconstructed angle for 1 GeV photon entering with an angle of 10 degree.}
 %\label{fig:ang_rec}
%\end{figure}

%\begin{figure}[h]
% \includegraphics[bb=14 362 1011 745,clip,width=\textwidth]{./figure/det_conf.pdf}
%\includegraphics[bb=14 362 1011 745,clip,width=\textwidth]
 %\caption{Energy dependency of angular resolution.}
 %\label{fig:ang_resol}
%\end{figure}

\begin{figure}[h]
\begin{center}
\includegraphics[width=0.5\textwidth]{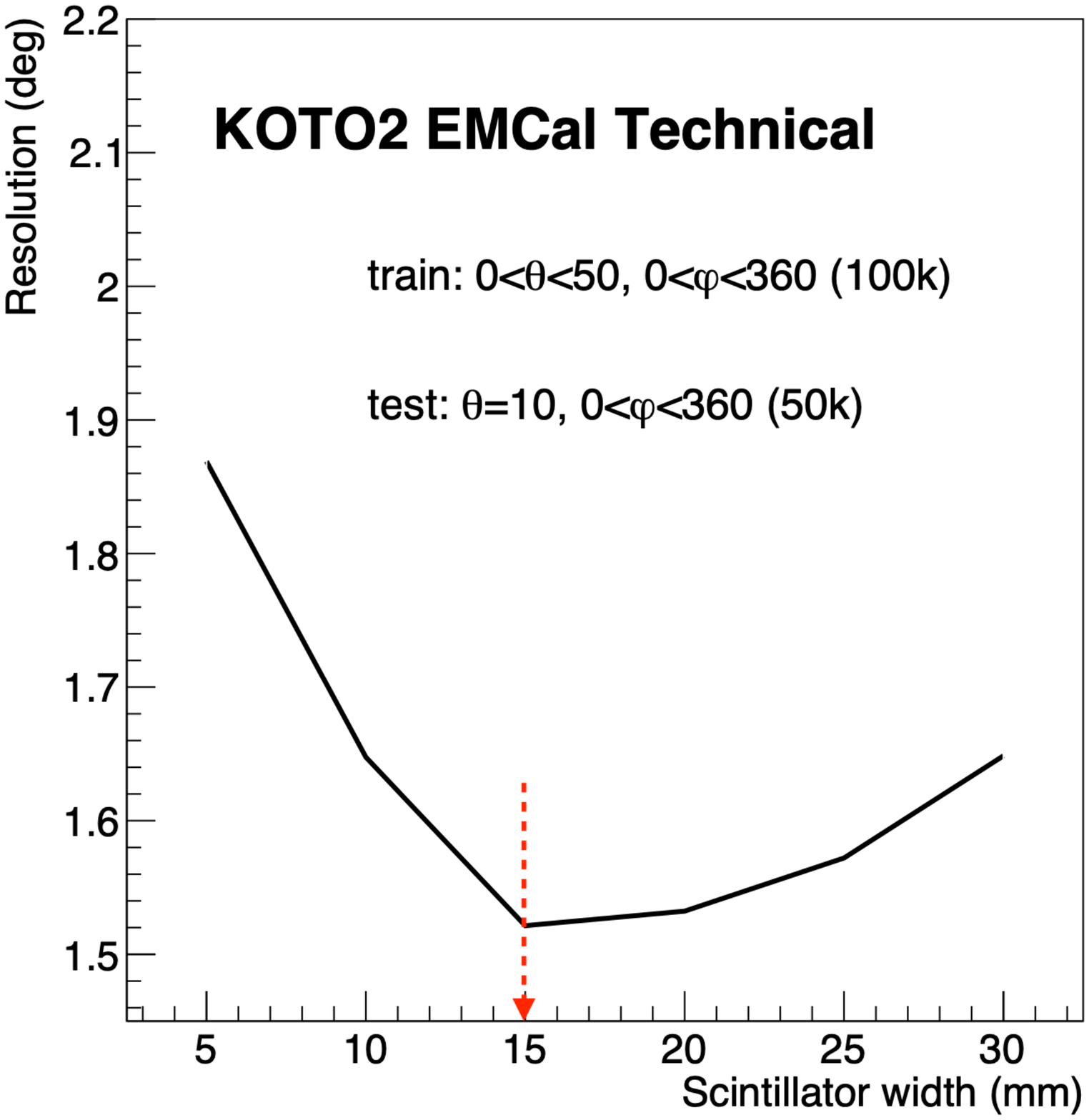}
 \caption{Angular resolution according to the width of scintillator strips.}
 \label{fig:ang_width}
 \end{center}
\end{figure}

%We have many study items for optimal detector configuration such as wid
The angular resolution depends on the width of scintillator strips as shown in Fig.~\ref{fig:ang_width}. However, the result implies that the narrower width does not give a better resolution
Its reason could be due to intrinsic shower property or training process of the machine running, and is under study.
We will optimize the detector configuration based on these studies. 

When we focus on the angle measurement (pre-shower detector), we might place the detector of five radiation length (5~$X_0$) in depth in front of the main calorimeter. This configuration will provide the same angular resolution with a 97\% detection efficiency. 
We plan to further optimize the detector configuration to achieve better angular resolution in particular for low energy photons. 
Also, we plan to make a prototype to test the simulation results and to examine the fabrication process in detail.

% flatex input end: [KLdocu/dfeasibility/photon-angle.tex]

% flatex input: [KLdocu/dfeasibility/cal-perform.tex]
\subsubsection{Discussion on the calorimeter performance}
In the background estimation, we assumed that the new calorimeter for the KOTO step-2 experiment
consisted of 50-cm-long CsI crystals the same as the KOTO experiment.
However, the cost to make the whole new calorimeter of CsI crystals is high because
the diameter of the new calorimeter is 3m and the area is 2.2 times larger than that of the KOTO experiment. 
It is thus important to consider an alternative option which reduce the cost while keeping the signal and noise ratio.
In this study, we evaluated the numbers of the $\klpiopio$ and $\klppm$ background events
by changing the energy resolution used in the MC simulation. 
We considered sampling calorimeters such as the calorimeters used in the KLOE experiment at DA$\Phi$NE,
and the calorimeter designed for the KOPIO experiment at BNL. The basic parameters on those calorimeters are listed in Table~\ref{table:EnergyTimeReso}.
The results are shown in Table~\ref{table:RelativeBG}.  We can keep the current background level by using a calorimeter having energy
and position resolutions achievable by the sampling calorimeters.  The effect on other background events will be investigated in the future.

The new calorimeter is also required to distinguish photon clusters from hadron clusters to reduce the hadron cluster background
events. In the KOTO experiment, the discrimination is achieved by using the information of cluster shapes, pulse shapes and
the depth of the interaction. The depth of the interaction is estimated by using the timing difference between the photo sensors attached 
at the upstream and downstream surfaces of the crystals. The timing resolution is a key to get a better discrimination from the depth 
information. As shown in Table~\ref{table:EnergyTimeReso}, sampling calorimeters can achieve the similar timing resolution to that of the calorimeter 
used in the KOTO experiment and thus same discrimination power.
The discrimination power of a sampling calorimeter by using the information of cluster shapes and pulse shapes
still need to be studied.  We plan to make a prototype detector and
evaluate the performance of the discrimination between photon clusters and neutron clusters with electron and neutron beams.

\begin{table}[htbp]
\centering
  \caption{Energy and timing resolutions of sampling calorimeters together with the technologies and the depth of the calorimeters.
  The performance of the CsI calorimeter used in KOTO is also shown as a reference. $E$ is in GeV. The energy and timing resolutions of the 
  calorimeters are referred from \cite{
  %ref:PDG,
  Zyla:2020zbs,
  ref:KOPIO,ref:Calorimeter,ref:CsITime}. }
  \label{table:EnergyTimeReso}
\begin{tabular}{llll}
\hline \hline
  Technology (Experiment) & Depth & Energy resolution & Timing resolution\\
 \hline 
   CsI (KOTO) & 27$X_{0}$  &   $2\%/\sqrt{E}\oplus1\%$  & $115~\mathrm{ps}/\sqrt{E}\oplus5~\mathrm{ps}/E$ \\
   & & & $\oplus130~\mathrm{ps}$ \\
   \hline
   Scintillator/Pb (KOPIO) & 16$X_{0}$  &   $3\%/\sqrt{E}$ & $90~\mathrm{ps}/\sqrt{E}$ \\
    \hline
   Scintillator fiber/Pb & 15$X_{0}$  &   $5.7\%/\sqrt{E}\oplus0.6\%$ & $54~\mathrm{ps}/\sqrt{E}\oplus140~\mathrm{ps}$  \\
   spaghetti (KLOE)    &                   &                         &                \\    
   \hline
  \end{tabular} 
\end{table}

\begin{table}[htbp]
\centering
  \caption{ The relative numbers of background events by changing the energy resolution assumed in the simulation.
  The numbers in the table are normalized by the numbers of expected background events  with the default calorimeter consisting 
  of CsI crystals.
}
\label{table:RelativeBG}
\begin{tabular}{lll}
\hline \hline
  The type of the calorimeter & \# of $\klpiopio$ BG &  \# of $\klppm$ BG  \\
 \hline 
  KOPIO & 1.01  &   0.96 \\
  KLOE   & 0.99  &   1.02 \\
   \hline
  \end{tabular}  
\end{table}

%%% ==========
%%% sample description how to designate the image file
%%% ==========
%%%\begin{figure}[h]
%%%\centering
%%%%\includegraphics{KLdocu/dfeasibility/figs/aaa.pdf}
%%%\caption{\label{fig:aaa}This is a sample description.}
%%%\end{figure}

% flatex input end: [KLdocu/dfeasibility/cal-perform.tex]

% flatex input: [KLdocu/dfeasibility/bhc.tex]
\subsubsection{Beam hole photon veto counter}
\label{sec:beamHolePhotonVetoCounter}
A beam-hole photon-veto counter detects a photon from the $\klpiopio$ decay
in the neutral beam, and reduces the background.
To operate it in the neutral beam,
it should be less sensitive to neutrons, and photons with energies
lower than those from the $\klpiopio$ decay.
In this section, we discuss the design and rate of the beam-hole photon-veto counter.

\begin{figure}[h]
 \includegraphics[width=0.9\textwidth]
 {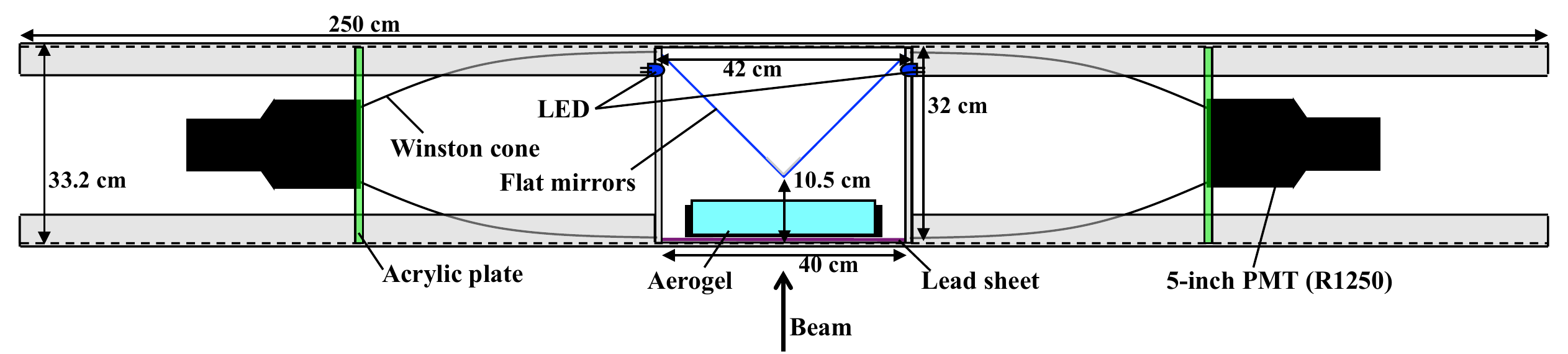}
 \caption{Structure of a module for the beam-hole photon-veto counter in KOTO step-1
 ~\cite{Maeda:2014pga}.}\label{fig:BHPVmodule}
\end{figure}

\subsubsubsection{The design of beam hole photon veto counter}
The beam-hole photon-veto counter consists of an array of a module
with a lead photon-converter and aerogel Cherenkov radiators.
The module structure which is the same as that in KOTO-step1 
is shown in Fig.~\ref{fig:BHPVmodule}.
A high-energy photon is converted to $e^+$ and $e^-$,
and these generate Cherenkov light in the aerogel radiator.
The Cherenkov light is guided by mirrors to a PMT 5-inch in diameter.
The Cherenkov light is generated less for slow particles
such as protons or charged pions from the hadronic interaction
of beam neutrons.  Several modules are lined up along the beam
and three-consecutive coincident hits in the modules are required,
which keep high efficiency against high energy photons while causing 
low energy photons and neutrons to be less sensitive.  
This is because electro-magnetic showers of high energy photons
develop in the forward direction while hadronic showers  develop 
isotropically and electro-magnetic showers of low energy photons cannot
develop in three or more modules.

\begin{figure}[h]
 \includegraphics[width=0.9\textwidth]
 {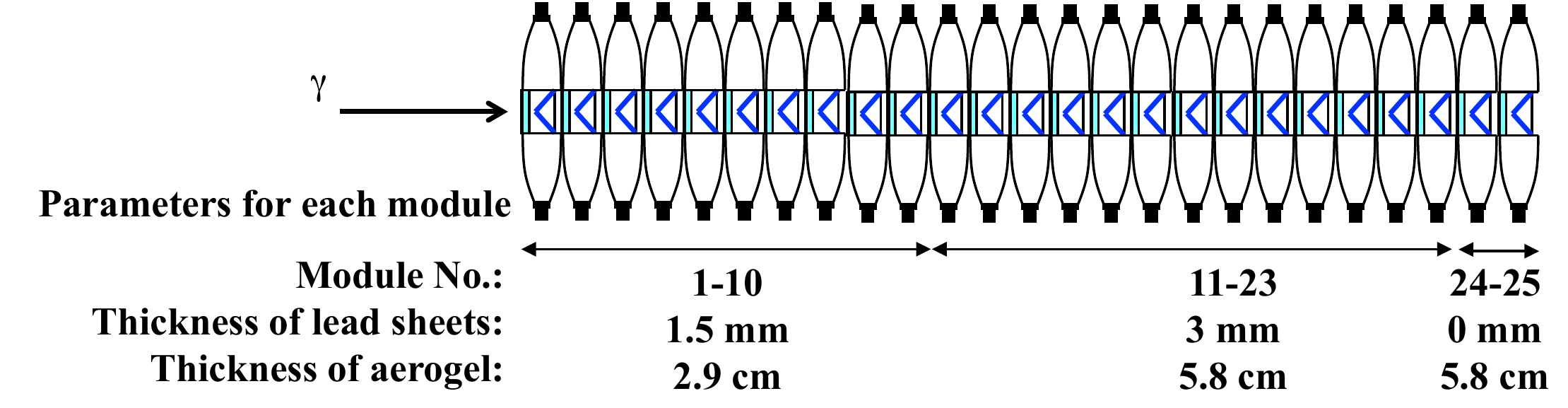}
 \caption{Configurations of the modules of the beam-hole photon-veto counter.}
 \label{fig:BHPVconfig}
\end{figure}

The configuration of the beam-hole photon-veto counter in KOTO step-2 is shown in  Fig~\ref{fig:BHPVconfig}.
The thickness of lead and aerogel sheets  is changed according to the location of the module 
to balance the hit rates among the modules. The total thickness of the lead sheets
is 54~mm, which corresponds to 9.6 $X_{0}$, to keep the photon punch-through inefficiency to be less than $10^{-3}$.

We evaluated the inefficiency of the beam-hole photon-veto counter against photons as a function of the incident energy by using a full-shower simulation
as already shown in Fig.~\ref{fig:bhpvinef}.
We can achieve $10^{-3}$ inefficiency for 2000~MeV photons with a threshold of 5.5 p.e.
which is considered as the default threshold.


\subsubsubsection{Expected counting rate and accidental loss in KOTO step-2}
A single module is sensitive to 1-MeV photons.
The incident rate of photons is 0.75~GHz for the energy larger than 1~MeV.
If the detection efficiency is $\sim 5$\%, the hit rate is $O(10)$~MHz.

\begin{figure}[h]
 \centering
 \includegraphics[page=26,width=0.45\textwidth]
 {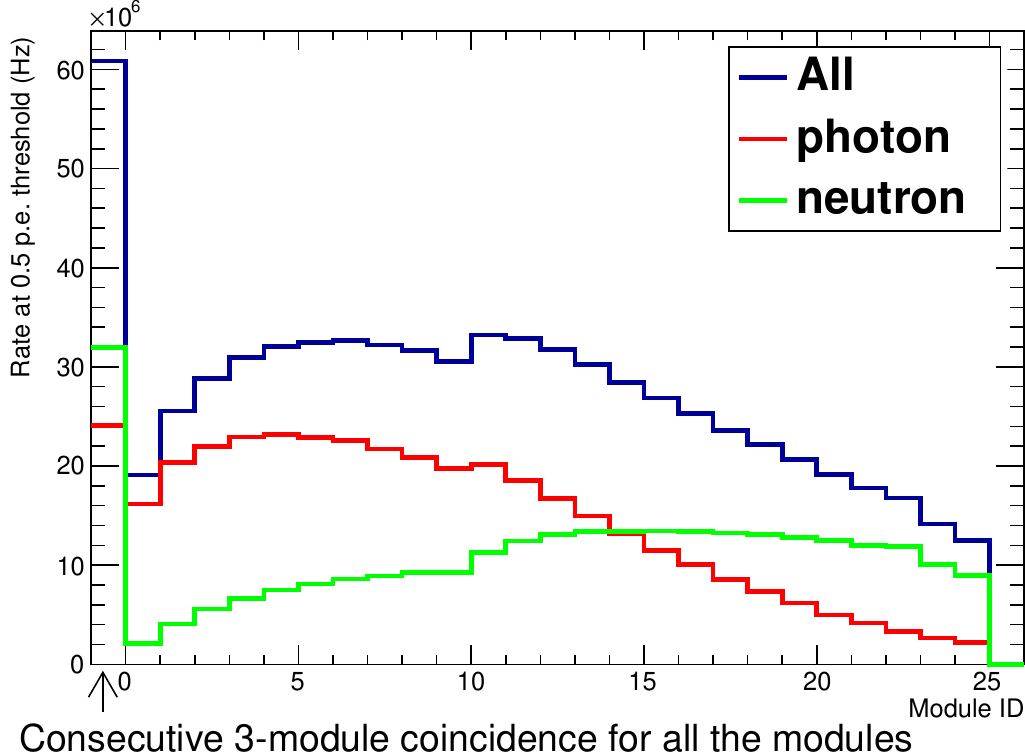}
 \caption{Hit rate of each module and counter (consecutive three-module coincidence)
 are shown with 0.5-p.e.-threshold. 
 The first bin of the histogram corresponds to the counter rate.}
 \label{fig:BHPVrate}
\end{figure}

We injected all the particles collected in the beam-line simulation
to the beam-hole photon-veto counter with 25 modules.
A full shower simulation and optical-photon tracking to the PMTs were performed,
and the observed number of photoelectrons was recorded.
For the module hit, ``OR'' of the PMTs on both sides is used
for a certain photoelectron (p.e.) threshold.
Consecutive three-module coincidence with a certain p.e. threshold
(counter rate) was also calculated.
Those module hit-rate and the consecutive three-module-coincidence rate (counter rate)
are already shown in Fig.~\ref{fig:bhpvrate} for the 5.5-p.e.\ threshold as a base design.
Those for the 0.5-p.e.\ threshold 
are shown in Fig.~\ref{fig:BHPVrate}.
Average module hit-rate in 2-sec spill is more than 30~MHz.
The counter rate at the default threshold is $35.2~\mathrm{MHz}$.
The resultant accidental loss is 19\% with the 6-ns veto window, which correspond to 
$\pm 5 \sigma$ of the timing resolution of 0.59~ns.
It was achieved with the beam-hole photon-veto
counter at KOTO step-1 by requiring the three-module coincidence with the 2.5-p.e.\ threshold~\cite{Maeda:2014pga}.

\begin{figure}[h]
 \centering
 \subfloat[]{
 \includegraphics[page=36,width=0.45\textwidth]
 {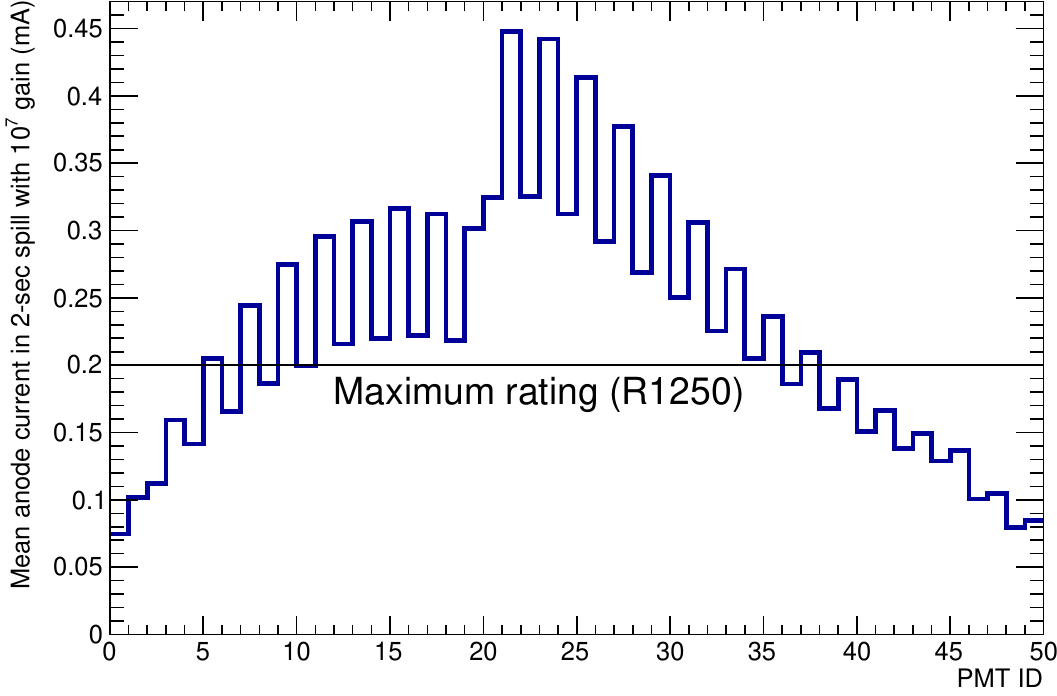}
 }
 \subfloat[]{
 \includegraphics[page=11,width=0.45\textwidth]
 {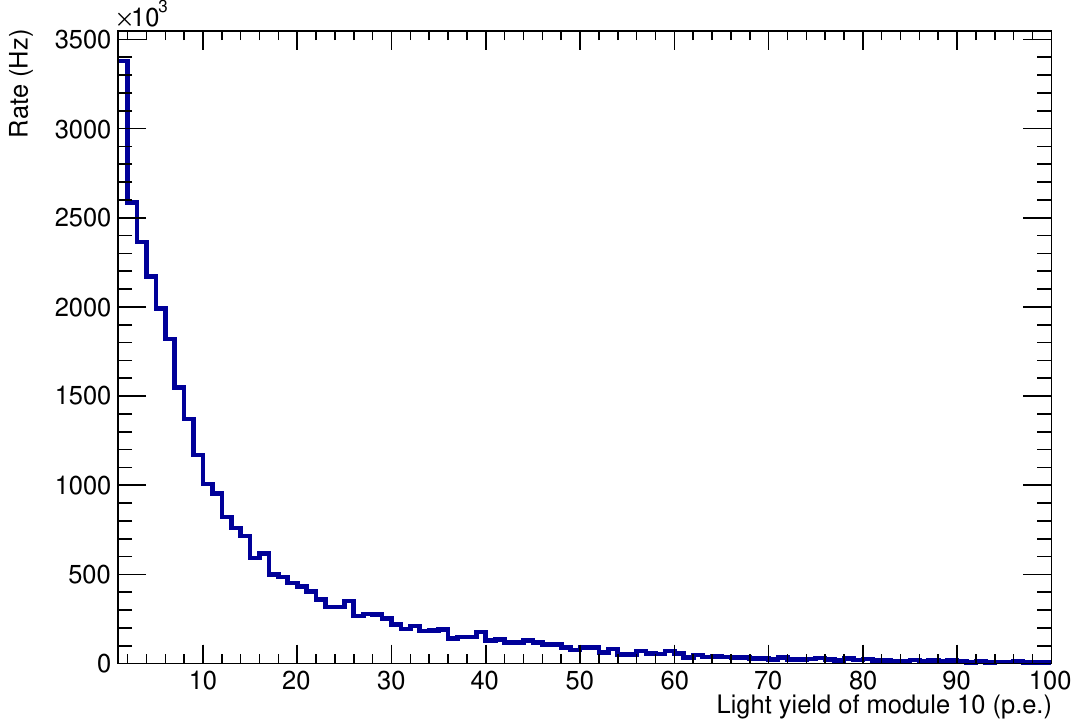}
 }
 \caption{(a)Expected average anode-current for each PMT
 with the gain of $10^7$. (b)Distribution of number of p.e. for a module (ID:10).}
 \label{fig:AnodeCurrent} 
\end{figure}

\subsubsubsection{PMT operation and expected waveforms}
We evaluated the average PMT-anode current
in Fig.~\ref{fig:AnodeCurrent}(a) based on the p.e. yields
from the full simulation with the particles in the beam
(Module ID 10 for example in Fig.~\ref{fig:AnodeCurrent}(b)).
A $10^7$ gain of the PMT is assumed.
In more than half of PMTs, the anode currents exceed
the maximum rating (0.2~mA) of the current PMT
used in KOTO step-1.
We may use a pre-amplifier to reduce the gain of the PMT
and keep the single-p.e.sensitivity.

\begin{figure}[h]
 \centering
 \subfloat[]{
 \includegraphics[page=101,width=0.32\textwidth]
 {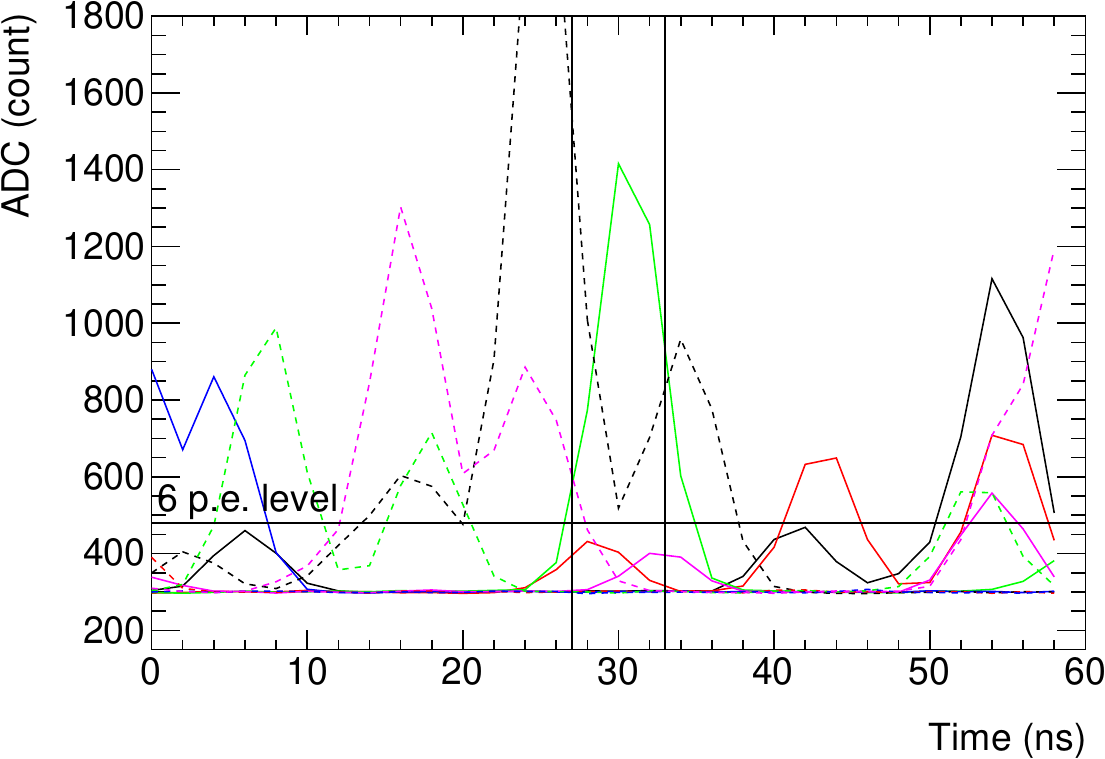}
 }
 \subfloat[]{
 \includegraphics[page=102,width=0.32\textwidth]
 {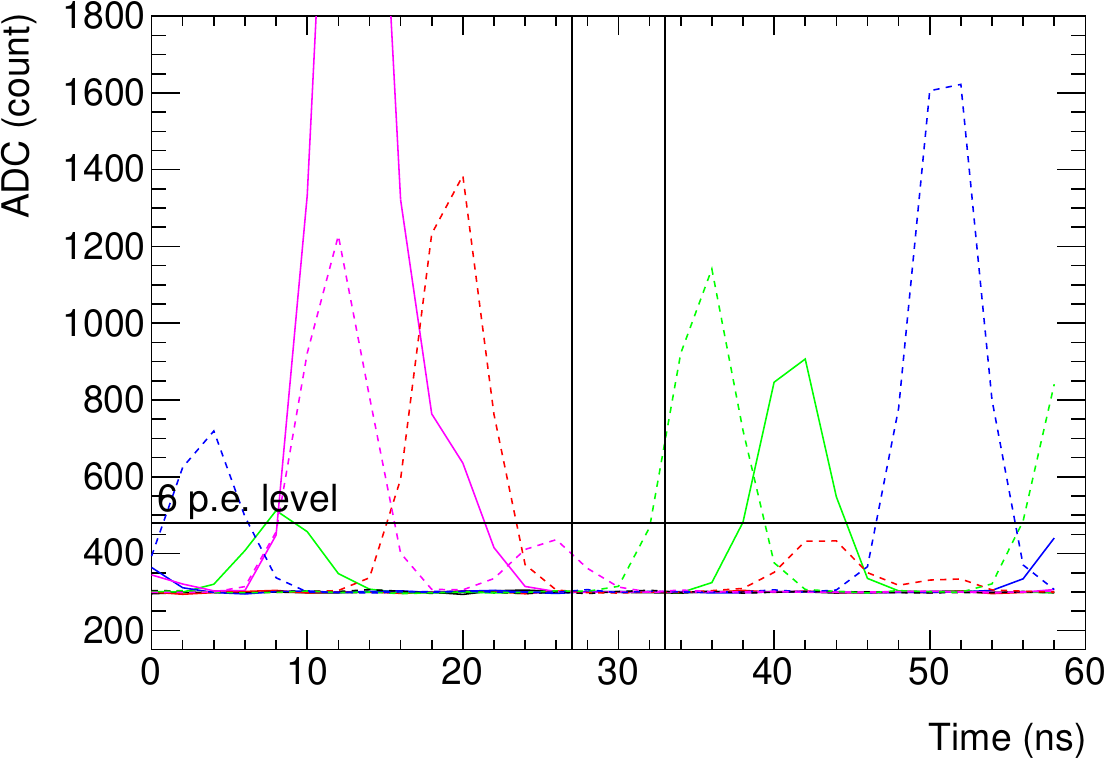}
 }
 \subfloat[]{
 \includegraphics[page=103,width=0.32\textwidth]
 {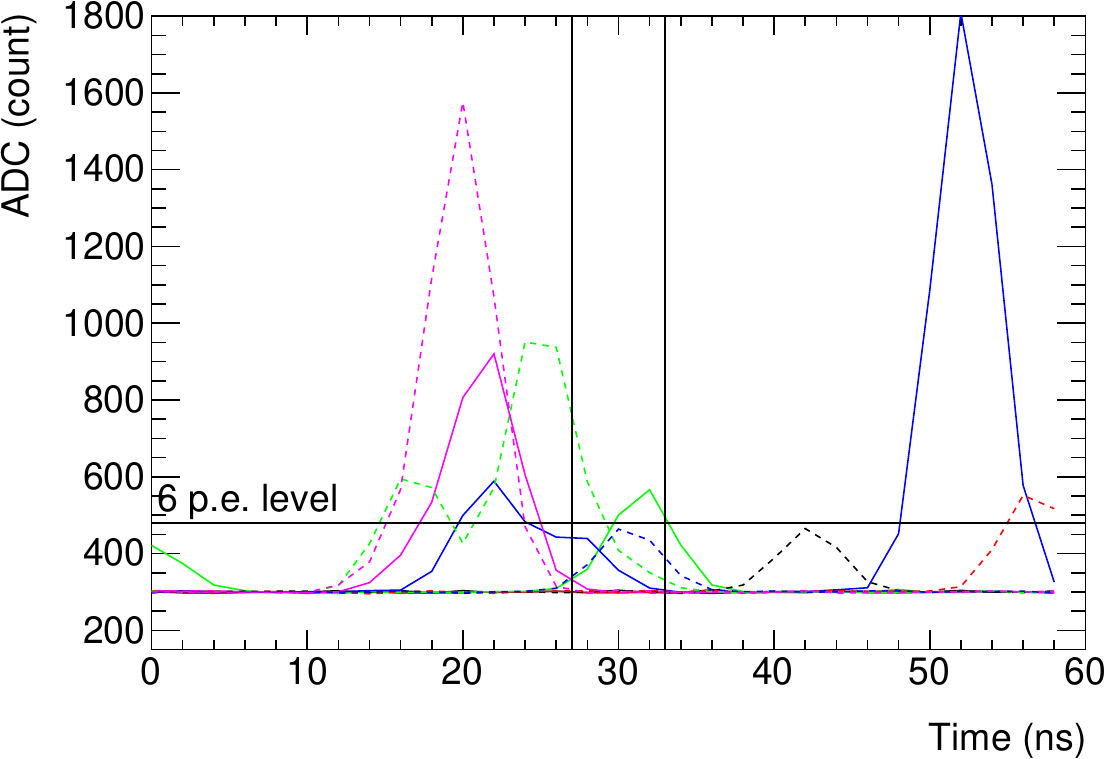}
 }

 \caption{Ten expected waveforms are overlaid in each panel. 
 Two vertical lines indicate the 6-ns veto window.
 A horizontal line shows 6-p.e.-level pulse height.}
 \label{fig:waveform}
\end{figure}

Expected waveforms after summing up the PMTs on both sides
are shown in Fig.~\ref{fig:waveform}.
In each panel, 10 waveforms are overlaid.
The waveforms do overlap, but peaks
can still be separated thus photons can be vetoed.
Thus, we can still operate the beam-hole photon-veto counter  in KOTO step-2 despite the harsh environment.

%For another example,
%efficient photon detection with less material,
%therefore less hadron interaction
%is discussed in the KLEVER collaboration~\cite{Moulson:2019ifj}
%by using coherent interaction of photons in the crystalline tungsten.

%%% ==========
%%% sample description how to designate the image file
%%% ==========
%%%\begin{figure}[h]
%%%\centering
%%%%\includegraphics{KLdocu/dfeasibility/figs/aaa.pdf}
%%%\caption{\label{fig:aaa}This is a sample description.}
%%%\end{figure}

% flatex input end: [KLdocu/dfeasibility/bhc.tex]

% flatex input end: [KLdocu/dfeasibility/dfeasibility.tex]

%\section{Physics and Experiment at KL2 Beam Line}
\clearpage
% flatex input: [KLdocu/conclusion/conclusion.tex]
\subsection{Conclusion}
Measurement of the CP-violating $\kl$ rare decay mode $\klpionn$ plays an important role in the study of the flavor physics. 
%It is one of the good probes of physics beyond the Standard Model. 
It is one of the best probes for physics beyond the Standard Model. 
Following the ongoing search by the KOTO experiment,
%in the J-PARC Hadron Experimental Facility, 
which will reach the sensitivity of better than $10^{-10}$ in 3--4 years,
KOTO step-2 is %being
discussed and designed aiming to observe several tens of $\klpionn$ signals.
To achieve more than 100 times better sensitivity than KOTO step-1, a new beam line with the production angle of 5 degrees
%are necessary, which is 
is needed and could be accommodated in the Hadron Experimental Facility Extension project.

We evaluate the experimental sensitivity and the background level with a modeled beam line and detector.
%%With a total of $1.5\times 10^{21}$ protons on target, equivalent to $7\times 10^7$ seconds of running with the 100~kW proton beam power,
%%observation of $\sim$60 Standard Model (SM) events is expected, corresponding to the single event sensitivity of $4.7\times 10^{-13}$, with a signal-to-background ratio of 0.4.
%%Even with 3 snow-mass-year running ($3\times 10^7$~s), 3$\sigma$ discovery can be achieved, with $\sim$30 SM events observation over $\sim$70 background events.
%With a total of $6.3\times 10^{20}$ protons on target, equivalent to $3\times 10^7$ seconds of running with the 100~kW proton beam power,
%observation of 22 Standard Model (SM) events is expected, corresponding to the single event sensitivity of $1.4\times 10^{-12}$.
%The estimated number of backgrounds is 35 events, which corresponds to the signal-to-background ratio of 0.63.
With a total of $6.3\times 10^{20}$ protons on target, which is equivalent to $3\times 10^7$ seconds of running with the 100~kW proton beam power,
observation of 35 Standard Model (SM) events is expected.
The corresponding single event sensitivity is $8.5\times 10^{-13}$.
The estimated number of background events is 56, which corresponds to the signal-to-background ratio of 0.63.
The significance of the observation for SM events is expected to be 4.7~$\sigma$.

%%There is room to 
%%achieve a better sensitivity by improving the signal acceptance.
%%We will continue optimization, including analysis methods and R\&D of the detectors, toward the realization of the experiment.
Moving forward beyond the discussed base design for better sensitivity and signal acceptance, 
we continue to study and optimize many known effects and techniques including R\&D of possible new detectors and analysis methods.
This importance of the physics deserves a world class experiment (KOTO step-2) at the world class accelerator facility (J-PARC and Extended Hadron Experimental Facility).
% flatex input end: [KLdocu/conclusion/conclusion.tex]

%\section{Physics and Experiment at KL2 Beam Line}
\clearpage

\setcounter{secnumdepth}{\value{secnumdepthsave}}

% flatex input end: [./KLdocu/KLmain.tex]

%==================================================================%

\printbibliography[segment=\therefsegment,heading=subbibliography]

\clearpage

%\bibliographystyle{utphys}
%\bibliography{bibliography,KLdocu/common/koto2}

\end{document}